\documentclass[a4paper, 10pt, twoside]{book}
\usepackage[english,italian]{babel}
\usepackage{epsfig}
\usepackage[draft]{hyperref}
\usepackage{amssymb}
\usepackage{amsmath}
\usepackage{epsf}
\usepackage{subfigure}
\usepackage{rotating}
\usepackage{graphicx}
\usepackage{amsmath}
\usepackage{fancyhdr}
\usepackage{lineno}
\usepackage{graphics}
\usepackage{geometry}
\usepackage{slashed}
\usepackage{color}

\makeatletter
\def\cleardoublepage{\clearpage\if@twoside
\ifodd\c@page
\else\hbox{}\thispagestyle{empty}\newpage
\if@twocolumn\hbox{}\newpage\fi\fi\fi}
\makeatother

\let\a=\alpha   \let\b=\beta   \let\g=\gamma   \let\d=\delta
\let\e=\epsilon \let\z=\zeta   \let\h=\eta    
    \let\k=\kappa  \let\l=\lambda  
                 \let\r=\rho
\let\s=\sigma        \let\f=\phi
\let\c=\chi         \let\w=\omega

\newcommand{\bea}{\begin{eqnarray}}
\newcommand{\eea}{\end{eqnarray}}
\def\a{\alpha}
\def\b{\beta}
\def\g{\gamma}
\def\d{\delta}
\def\w{\omega}

\def\nn{\nonumber}
\def\beq{\begin{equation}}
\def\eeq{\end{equation}}
\def\ba{\begin{eqnarray}}
\def\lag{\left\langle}
\def\rag{\right\rangle}
\def\veps{\varepsilon}

\newcommand{\sq}{\square}
\newcommand{\bmi}{\begin{minipage}}
\newcommand{\emi}{\end{minipage}}
\newcommand{\beqa}{\begin{eqnarray}}
\newcommand{\eeqa}{\end{eqnarray}}
\newcommand{\bann}{\begin{eqnarray*}}
\newcommand{\eann}{\end{eqnarray*}}
\newcommand{\eps}{\epsilon}

\newcommand{\si}{\sigma}
\newcommand{\ksl}{k \! \! \!  /}
\newcommand{\psl}{p \! \! \!  /}
\newcommand{\qsl}{q \! \! \!  /}
\newcommand{\lsl}{l \! \! \!  /}

\newcommand{\pd}{\partial}

\newcommand{\mD}{\mathcal{D}}

\newcommand{\bh}{\bar{\h}}

\def\th{\theta}

\renewcommand{\Im}{\mathrm{Im}}
\newcommand{\bq}{\bar{q}}

\newcommand{\muu}{\mu_{1}}
\newcommand{\nuu}{\nu_{1}}
\newcommand{\ku}{\vec{k_1}}
\newcommand{\xu}{x_{1}}

\newcommand{\mud}{\mu_{2}}
\newcommand{\nud}{\nu_{2}}
\newcommand{\kd}{\vec{k_2}}
\newcommand{\xd}{x_{2}}

\newcommand{\mut}{\mu_{3}}
\newcommand{\nut}{\nu_{3}}
\newcommand{\kt}{\vec{k_3}}
\newcommand{\xt}{x_{3}}

\newcommand{\muq}{\mu_{4}}
\newcommand{\nuq}{\nu_{4}}
\newcommand{\kq}{\vec{k_4}}
\newcommand{\xq}{x_{4}}

\newcommand{\tRE}{\widetilde{\rm Re}}

\newcommand{\bes}{\begin{subequations}}
\newcommand{\ees}{\end{subequations}}

\newcommand{\HRule}{\rule{\linewidth}{0.5mm}}

\begin{document}
\thispagestyle{empty}

\begin{titlepage}
\begin{center}
\includegraphics[scale=.15]{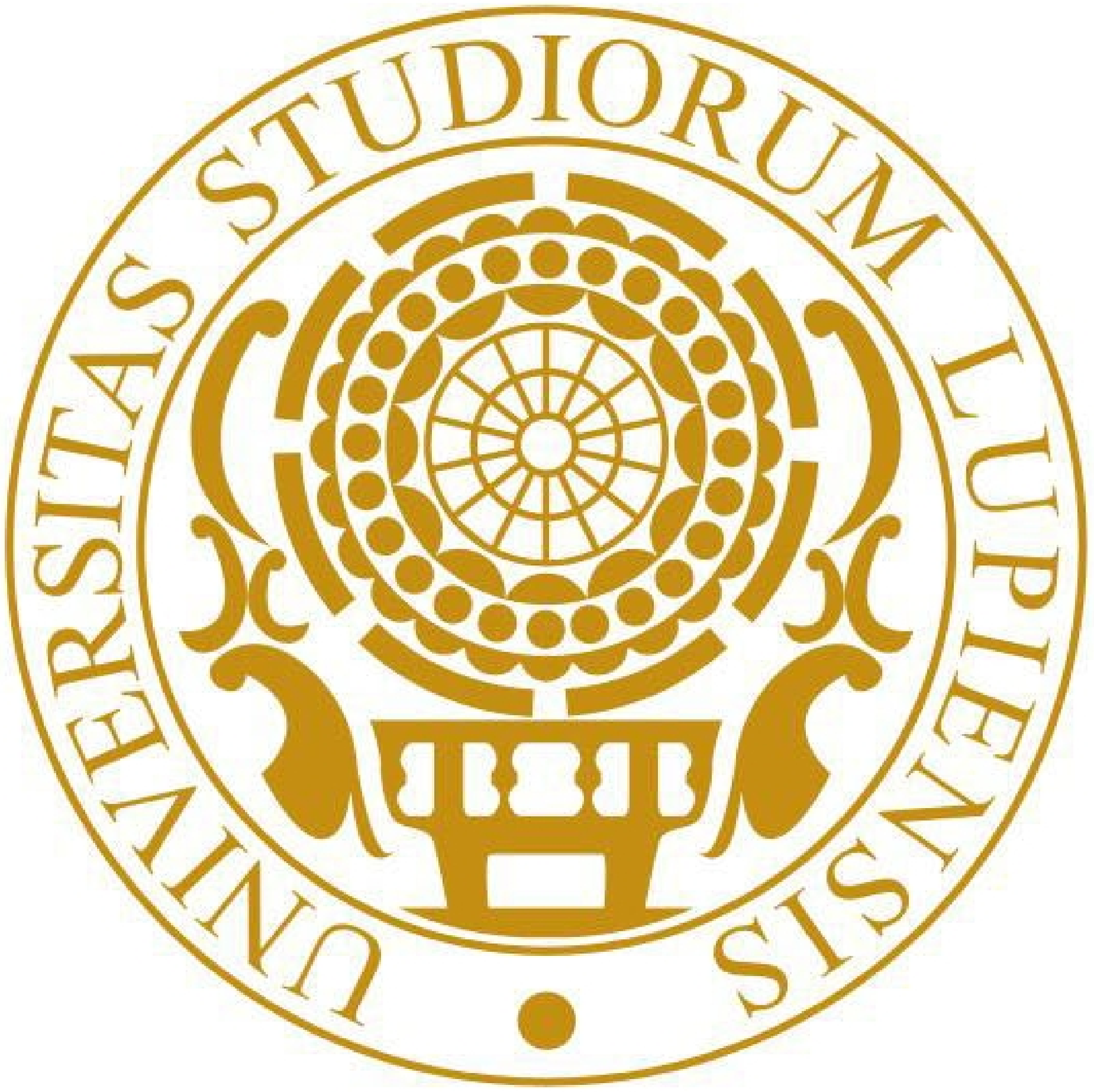}\\
\HRule\\[0.1cm]
\textrm{\LARGE Universit\`a del Salento}\\[0.3cm]
\textrm{\Large Facolt\`a di Scienze Matematiche, Fisiche e Naturali}\\[0.3cm]
\textrm{\Large Dipartimento di Matematica e Fisica ``Ennio De Giorgi''}\\[0.1cm]

\HRule \\[3cm]

{\Large { \bf{The Standard Model in a Weak Gravitational Background}}}\\[0.5cm]
{\large \bf {\em Dilatons, Scale Anomalies and Conformal Methods }}\\[5cm]

\begin{minipage}{0.4\textwidth}
\begin{flushleft}\large
\textit{Supervisor}\\
Prof.~Claudio Corian\`o
\end{flushleft}
\end{minipage}
\begin{minipage}{0.4\textwidth}
\begin{flushright} \large
\textit{Candidate}\\
Luigi Delle Rose
\end{flushright}
\end{minipage}\\[2cm]
Tesi di Dottorato di Ricerca in Fisica -- XXV ciclo
\vfill

\end{center}
\end{titlepage}

\frontmatter
\newpage
%
%
\selectlanguage{english}%
\tableofcontents
%
%
\chapter{List of publications}
The publications discussed in this thesis are 

\begin{itemize}
\item 
Roberta Armillis, Claudio Corian\`o, Luigi Delle Rose, Marco Guzzi  \\
\textbf{\emph{"Anomalous U(1) Models in Four and Five Dimensions and their Anomaly Poles"}} \\
Published in JHEP 0912 (2009) 029 \\
e-Print: arXiv:0905.0865 [hep-ph]
\item
Roberta Armillis, Claudio Corian\`o, Luigi Delle Rose \\
\textbf{\emph{"Anomaly Poles as Common Signatures of Chiral and Conformal Anomalies"}} \\
Published in Phys.Lett. B682 (2009) 322-327 \\
e-Print: arXiv:0909.4522 [hep-ph]
\item
Roberta Armillis, Claudio Corian\`o, Luigi Delle Rose \\
\textbf{\emph{"Conformal Anomalies and the Gravitational Effective Action: The TJJ Correlator for a Dirac Fermion"}} \\
Published in Phys.Rev. D81 (2010) 085001 \\
e-Print: arXiv:0910.3381 [hep-ph]
\item
Roberta Armillis, Claudio Corian\`o, Luigi Delle Rose \\
\textbf{\emph{"Trace Anomaly, Massless Scalars and the Gravitational Coupling of QCD"}} \\
Published in Phys.Rev. D82 (2010) 064023 \\
e-Print: arXiv:1005.4173 [hep-ph]
\item
Claudio Corian\`o, Luigi Delle Rose, Antonio Quintavalle, Mirko Serino \\
\textbf{\emph{"The Conformal Anomaly and the Neutral Currents Sector of the Standard Model"}} \\
Published in Phys.Lett. B700 (2011) 29-38 \\
e-Print: arXiv:1101.1624 [hep-ph]
\item
Claudio Corian\`o, Luigi Delle Rose, Mirko Serino \\
\textbf{\emph{"Gravity and the Neutral Currents: Effective Interactions from the Trace Anomaly"}} \\
Published in Phys.Rev. D83 (2011) 125028 \\
e-Print: arXiv:1102.4558 [hep-ph]
\item
Claudio Corian\`o, Luigi Delle Rose, Antonio Quintavalle, Mirko Serino \\
\textbf{\emph{"Dilaton Interactions and the Anomalous Breaking of Scale Invariance of the Standard Model"}} \\
Published in JHEP 1306 (2013) 077 \\
e-Print: arXiv:1206.0590 [hep-ph] 
\item
Claudio Corian\`o, Luigi Delle Rose, Mirko Serino \\
\textbf{\emph{"Three and Four Point Functions of Stress Energy Tensors in D=3 for the Analysis of Cosmological Non-Gaussianities"}} \\
Published in JHEP 1212 (2012) 090 \\
e-Print: arXiv:1210.0136 [hep-th]
\item
Claudio Corian\`o, Luigi Delle Rose, Emidio Gabrielli, Luca Trentadue \\
\textbf{\emph{"One loop Standard Model corrections to flavor diagonal fermion-graviton vertices"}} \\
Published in Phys.Rev. D87 (2013) 054020 \\
e-Print: arXiv:1212.5029 [hep-ph]
\item
Claudio Corian\`o, Luigi Delle Rose, Emidio Gabrielli, Luca Trentadue \\
\textbf{\emph{"Mass Corrections to Flavor-Changing Fermion-Graviton Vertices in the Standard Model"}} \\
e-Print: arXiv:1303.1305 [hep-th]
\item
Claudio Corian\`o, Luigi Delle Rose, Emil Mottola, Mirko Serino \\
\textbf{\emph{"Solving the Conformal Constraints for Scalar Operators in Momentum Space and the Evaluation of Feynman's Master Integrals"}}\\
Published in JHEP 1307 (2013) 011
e-Print: arXiv:1304.6944 [hep-th]

\end{itemize}

\section*{Publications not included}
The following is a list of related papers which are part of the thesis work but are not discussed in detail 
in this manuscript
\begin{itemize}
\item
Roberta Armillis, Claudio Corian\`o, Luigi Delle Rose, Luigi Manni \\
\textbf{\emph{"The Trace Anomaly and the Gravitational Coupling of an Anomalous U(1)"}} \\
Published in Int.J.Mod.Phys. A26 (2011) 2405-2435 \\
e-Print: arXiv:1003.3930 [hep-ph]
\item
Roberta Armillis, Claudio Corian\`o, Luigi Delle Rose, A.R. Fazio \\
\textbf{\emph{"Comments on Anomaly Cancellations by Pole Subtractions and Ghost Instabilities with Gravity"}} \\
Published in Class.Quant.Grav. 28 (2011) 145004 \\
e-Print: arXiv:1103.1590 [hep-ph]
\item
Claudio Corian\`o, Luigi Delle Rose, Emil Mottola, Mirko Serino \\
\textbf{\emph{"Graviton Vertices and the Mapping of Anomalous Correlators to Momentum Space for a General Conformal Field Theory"}} \\
Published in JHEP 1208 (2012) 147 \\
e-Print: arXiv:1203.1339 [hep-th]

\item
Claudio Corian\`o, Luigi Delle Rose, Carlo Marzo, Mirko Serino \\
\textbf{\emph{"Higher Order Dilaton Interactions in the Nearly Conformal Limit of the Standard Model"}} \\
Published in Phys.Lett. B717 (2012) 182-187 \\
e-Print: arXiv:1207.2930 [hep-ph]

\item
Claudio Corian\`o, Luigi Delle Rose, Carlo Marzo, Mirko Serino \\
\textbf{\emph{"Conformal Trace Relations from the Dilaton Wess-Zumino Action"}} \\
e-Print: arXiv:1306.4248 [hep-ph]
\end{itemize}

\section*{Proceedings}
\begin{itemize}
\item
Roberta Armillis, Claudio Corian\`o, Luigi Delle Rose \\
\textbf{\emph{"Trilinear Gauge Interactions in Extensions of the Standard Model and Unitarity"}} \\
Published in Nuovo Cim. C32N3-4 (2009) 261-264 \\
e-Print: arXiv:0905.4410 [hep-ph]
\item
Roberta Armillis, Claudio Corian\`o, Luigi Delle Rose, Marco Guzzi, Antonio Mariano \\
\textbf{\emph{"The Effective Actions of Pseudoscalar and Scalar Particles in Theories with Gauge and Conformal Anomalies"}} \\
Published in Fortsch.Phys. 58 (2010) 708-711 \\
e-Print: arXiv:1001.5240 [hep-ph]
\item
Roberta Armillis, Claudio Corian\`o, Luigi Delle Rose \\
\textbf{\emph{"The Trace Anomaly and the Couplings of QED and QCD to Gravity"}} \\
Published in AIP Conf.Proc. 1317 (2011) 185-190 \\
e-Print: arXiv:1007.2141 [hep-ph]
\item
Luigi Delle Rose, Mirko Serino \\
\textbf{\emph{"Massless Scalar Degrees of Freedom in QCD and in the Electroweak Sector from the Trace Anomaly"}} \\
Published in AIP Conf.Proc. 1492 (2012) 205-209 \\
e-Print: arXiv:1208.6425 [hep-ph]
\item
Luigi Delle Rose, Mirko Serino \\
\textbf{\emph{"Dilaton Interactions in QCD and in the Electroweak Sector of the Standard Model"}} \\
Published in AIP Conf.Proc. 1492 (2012) 210-213 \\
e-Print: arXiv:1208.6432 [hep-ph]

\end{itemize}

\newpage

\thispagestyle{plain}

\centerline{\bf Preface }

The principal goal of the physics of the fundamental interactions is to provide a consistent description of the nature of the subnuclear forces, which manifest in our universe, together with the gravitational force, in a unified framework. This attempt, which is far from being complete, is characterized by two milestones, which we identify with the Standard Model of the elementary particles and with Einstein's theory of General Relativity. The Standard Model is formulated as a renormalizable Yang-Mills (quantum) gauge theory which embodies the mechanism of mass generation via a scalar sector, the so called "Higgs sector". 
It has even raised to the laymans' attention thanks to the recent observation of a Standard Model Higgs-like resonance at the Large Hadron Collider at CERN.
Einstein's theory, on the other hand, describes the large scale structures of the universe, 
and is a geometrical theory which is not consistent at quantum level, preventing us to achieve a simple unification with the Standard Model using the basic principles of quantum field theory.

Quantum field theory is extremely effective in the description of the phenomenology of the Standard Model, but much less reliable if applied to the gravitational interactions. 
Nevertheless, the coupling of an ordinary field theory
to a weak gravitational background provides significant information concerning the coupling of matter 
to gravity and, as we are going to discuss in the various chapters of this work, allows to study in a systematic way the origin of the conformal anomaly.
For this reason, the analyses presented here are of double significance since they characterize the interactions which are part of the effective action describing gravity and matter at leading order in the gravitational coupling, while, at the same time, provide the basis for a discussion of the mechanism of generation of the conformal anomaly in quantum field theory. 
For instance, in a classical conformal invariant theory, such as massless Quantum Electrodynamics (QED), the dilatation is a tree-level symmetry of the Lagrangian, broken by the renormalization procedure. After quantization, the symmetric and conserved energy momentum tensor (EMT) of the theory, which appears in the definition of the Noether's current of scale transformations, acquires a non-vanishing trace characterizing the effective action with an anomalous contribution, the scale (or trace) anomaly. 

To appreciate how the source of this anomaly can be studied with the help of the background gravity, it is sufficient to observe that a symmetric and conserved energy momentum tensor of a given theory can be computed by coupling the corresponding Lagrangian to a curved background. This procedure can be seen both as a formal trick, since the EMT obtained by this prescription is automatically equivalent to that provided by the Belinfante's technique of symmetrization of its canonical form, but also as a way to study the coupling of gravity to the theory for its own sake. We recall, in fact, that in the weak field limit, gravity couples to matter via the EMT.

Therefore, the breaking of the conformal symmetry can indeed be viewed both as a phenomenon which 
is justified {\em per se} with no relation to gravity, being the EMT related to the dilatation current, but also in a gravitational context as a consequence of an anomaly 
generated by Green's functions with one or more gravitons on the external lines. 
For this reason, the computation of correlation functions of a theory such as the Standard Model in a weak gravitational background is of remarkable interest and, as we are going to see, the consequences of this analysis are also of phenomenological relevance. 
For instance, they concern the 
effective interaction of the dilatation currents of the Standard Model. These are affected by the appearance in the spectrum of the theory of a {\em composite state}, the dilaton,  which is identified, in perturbation theory, by an {\em infrared coupled} anomaly pole.

Our analysis will then cover the perturbative description of anomalous correlators involving EMT insertions in several cases, as QED, QCD, 
and the electroweak theory. In the final chapter of the first part of this thesis we will discuss the implications related to the appearance of a dilaton state in a scale invariant extension of the Standard Model. In this case, we propose that the Standard Model Higgs may be accompanied by an extra composite scalar singlet, the {\em effective dilaton}, which can be extensively studied at the LHC.

The study of some applications of the methods discussed so far, which are in the second part of this thesis, can be read quite independently from the other chapters and deserves a special comment. \\
The analysis of correlation functions of EMT's in odd dimensions, which are not affected by the conformal anomaly, is relevant also in different contexts. 
For instance, it has been shown that the theory of the classical metric perturbations in gravity can be related to some specific field theories via a holographic mapping. This requires the computation of correlation functions of EMT's in $D=3$ spacetime dimensions. 
In this scenario, the (scalar and tensor) gravitational perturbations can be expressed in terms of 
these three dimensional correlators and play a role in the study of the non-gaussianities of the cosmic background radiation in the early universe.

Finally, we mention an application in momentum space of the constraints derived from the conformal Ward identities in the computation of some correlation functions. This relies on the solution of a system of second order partial differential equations which introduces a certain class of generalized hypergeometric functions of two variables. This method is employed in the calculation of some nontrivial Feynman (master) integrals appearing in perturbation theory at higher orders and allows to bypass the use of 
Mellin-Barnes transforms. At the same time, it is shown that some recursion relations which are typical of such integrals correspond, in the conformal language, to the requirement of scale invariance.  \\
These two applications exploit all the methodologies developed in the first chapters of this manuscript and, as such, are part of this thesis work. 

\thispagestyle{plain}

\chapter{Acknowledgements}

I would like to express my sincerest gratitude to my advisor Prof. Claudio Corian\`{o} for his encouragements and immense patience during these years.
His unconditional passion for Physics guided my postgraduate education and contributed to my personal and professional growth. \\
I am also indebted to all the people that make this work possible. In particular I would like to thank 
Profs. Emil Mottola, Luca Trentadue, Emidio Gabrielli, Pietro Colangelo and Angelo Raffaele Fazio for their physical insights and support.
I thank Prof. Paolo Ciafaloni for a reading of this manuscript and comments. \\
A special acknowledgement goes to all my colleagues of the Physics Department in Lecce, to Roberta Armillis for the nice time spent together, 
for being a precious friend and an invaluable collaborator, to Mirko Serino for his genuine friendship,
to Antonio Mariano, the computer expert of our working group, to Alfredo Urbano, Ylenia Maruccia, Marco Guzzi and Antonio Quintavalle. \\
I cannot forget to thank all my friends of the "Collegio Fiorini" for the great moments had together in these years. 
Among them Enza Marra and Luigi Manni with whom I shared a lot of fun.\\
Finally, I would like to thank my love Annalisa De Lorenzis, always by my side along this path, for being a constant inspiration in life and physics.
Her understanding and encouragement have been very important for me.

\chapter[Introduction]{General Introduction}  

Understanding the interaction of gravity with the particle content of the Standard Model is one of the most challenging topics in quantum field theory, both from the theoretical and the phenomenological side. At classical level the theory of gravitation is well established and explained in the framework of general relativity through a geometrical formulation.  The Einstein's field equations, which represent the core of the classical theory of gravity, relate geometrical quantities, built from the curved metric, to a rank-two tensor, the energy momentum tensor of the matter fields. A complete quantum field theory version of gravity, instead, is still lacking of a consistent description, and one must resort to string theory in order to reconcile general relativity with quantum physics. Without having to rely on the formalism of string theory, we can anyway incorporate the leading quantum effects of the gravity-matter system by promoting the classical energy momentum tensor to an operator, and retaining the gravitational field at the classical level. In this way we can construct an effective field theory where the quantum fluctuations of the matter sector are coupled to a classical curved gravitational background.

One of most important features of an effective field theory is that of incorporating the decoupling of the microscopic degrees of freedom from the macroscopic ones. Accordingly, the effective theory becomes quite insensitive to the underlying short distance physics, whose effects are only parametrized by the coefficients of the local operators which are generated in the effective action. \\
%
%
%
Violation of the decoupling mechanism occurs when the quantum corrections break a classical symmetry, giving rise to an anomaly.
Such contributions, being not suppressed by any mass scale, survive in the decoupling limit, leaving their imprints at all scales and, sometimes, even dominating on the large distance physics. These effects are not captured by the usual local effective field theory and must be explicitly introduced into the effective action in the form of nonlocal terms. Their coefficients are entirely determined by the microscopic degrees of freedom and by their quantum numbers. The contributions from the anomaly can then be identified by a detailed analysis of the quantum corrections of the corresponding anomalous correlators, and appear in the form of effective massless degrees of freedom. We will extensively elaborate on this point in the following chapters, bringing perturbative evidence of the appearance of massless poles in the structure of the effective Lagrangian of the Standard Model, due to the emergence of an anomalous symmetry. Understanding the significance of these degrees of freedom and their connection with a possible scale invariant extension of the Standard Model is an important component of our analysis that will be developed in this work. 
We recall, in fact, that the Standard Model is not scale invariant, due to the mass term of the Higgs field, but would be such if an extra field, the dilaton, is suitably introduced in the scalar potential. 
Therefore, the appearance of an effective scalar degree of freedom, related to the anomalous breaking of the conformal symmetry, lead us to a plausible interpretation of this perturbative result as a hint of the 
existence of a conformal phase of the Standard Model, where the dilaton plays the role of a composite Nambu-Goldstone mode. 
We will analyze the implications of this hypothesis in the last chapter of the first part of this work, before our conclusions.   
\begin{itemize} 
\item{\bf \large The energy momentum tensor and the conformal anomaly} 
\end{itemize} 

In a scale invariant theory, when the energy momentum tensor is promoted to an operator it becomes impossible to maintain at the quantum level both the invariance under general coordinate transformations and the conformal symmetry. Because the conservation of the energy momentum tensor is an essential requirement in order to satisfy the equivalence principle, we are forced to give up on conformal invariance. This implies that the energy momentum tensor acquires a non-vanishing anomalous trace which depends on the number of degrees of freedom and describes the quantum breaking of the dilatation symmetry. It is worth to note that, except for some rare cases, the invariance under Poincar\'{e} and scale transformations automatically implies the invariance under the full conformal group, including the special conformal transformations. Therefore, for our purpose, the trace anomaly is equivalent to a conformal anomaly and vice versa.

The interplay between two symmetries and the impossibility to maintain both of them at quantum level is also a characteristic of another manifestation of the anomaly in quantum field theory, the chiral anomaly. In this case one chooses to preserve the conservation of the vector current, associated with the gauge invariance, sacrificing the axial-vector one.
The latter is the Noether's current of a global symmetry and therefore its conservation can be violated without affecting the consistency of the theory. 
Moreover, as shown by a direct perturbative analysis, the chiral anomaly contribution is defined in terms of the microscopic degrees of freedom, represented by the quark fields, but its effects extend to low energy as well. The non-decoupling property of the anomaly, for instance, manifests as an infrared effect, enhancing the decay rate of the pion into two photons, as observed experimentally.

The conformal anomalous effective action is a significant feature of the quantum gravitational theories. Its expression was obtained by Riegert for the first time \cite{Riegert:1987kt} as a variational solution of the trace anomaly equation, generalizing 
previous results in two-dimensional gravity and string theory. Riegert's action only accounts for the anomaly contributions and misses completely all the non-anomalous terms. 
To account for all such contributions, one may turn to a perturbative approach, as we will do in this thesis work, performing an analysis of the quantum gravitational effective action at one-loop order. In fact, the leading quantum corrections, which describe both anomalous and non-anomalous effects, can be extracted by one-loop perturbative computations of a given set of correlation functions. As we have mentioned above, they are characterized by the insertion of the energy momentum tensor operator and define the interaction of gravity with matter at leading order in the gravitational coupling.
\begin{itemize}
\item{\bf \large Connection with conformal field theories}
\end{itemize}

The analysis of the correlation functions with insertions of the energy momentum tensor has found widespread interest over the years \cite{Fradkin:1997df} also in $d$ dimensional quantum field theory possessing conformal invariance. Given the infinite dimensional character of the conformal algebra in $d=2$ dimensions, conformal field theories (CFT's) in $d=2$ have received the most attention. Indeed one can show that, in $d=2$, the $n$-point correlation functions can be computed exactly relying only on symmetry arguments. Nevertheless, higher dimensional CFT's have also been studied, motivated by renormalization group analyses \cite{Komargodski:2011vj, Elvang:2012yc, Elvang:2012st, Luty:2012ww}. In fact, a quantum field theory is expected to become conformally invariant (at the quantum level) at the renormalization group fixed points where the beta functions vanish. Unlike the $d=2$ case, in CFT's with $d>2$ dimensions the structure of generic conformal correlators is not entirely fixed by conformal symmetry, but for 2- and 3-point functions built out of conserved currents the situation is rather special and these can be significantly constrained, up to a small number of constants. 

In \cite{Osborn:1993cr, Erdmenger:1996yc}, exploiting the constraints of the conformal group, a general procedure for constructing conformally invariant 2- and 3-point functions in position space and for arbitrary dimensions was presented. In particular, the authors considered correlation functions with a scalar operator $O$, a conserved vector current $J_\mu$ and a symmetric, conserved and traceless energy momentum tensor $T_{\mu\nu}$. In this approach the anomaly is introduced through a regularization procedure of the singular behaviour of the same correlators in the limit of coincidence points.   
In a recent paper \cite{Coriano:2012wp}, we have re-investigated this procedure in momentum space. 
The work has not been included in this thesis but we will refer the reader to the original literature for further details.  
The use of the conformal Ward identities directly in momentum space remains quite challenging because
they amount to partial differential equations, rather than to the algebraic constraints of the position space, and, as such, are far more difficult to solve. 
For this reason, the literature on conformal symmetry in momentum space has received a minor attention compared to the position space approach. 

It should be stressed, moreover, that dynamical information on the emergence of effective degrees of freedom related to the conformal anomaly 
are not extracted from position space, but require a parallel analysis in momentum space. For this reason a systematic analysis of a certain class of anomalous correlators in momentum space is in order.  

We remark that, in general, a CFT does not always possess a Lagrangian description. In fact, the use of symmetry principles in order to infer the general solution of conformal Ward identities, for some specific correlation functions, allows to gather information about a conformal theory even when a Lagrangian formulation is not readily found or may not exist at all. 
However, when the conformal symmetry is only approximate, such as that of the Standard Model, due to the breaking of the electroweak symmetry, a formal approach, based on characterization of the correlators through the solution of the conformal constraints, would not be helpful, except in the very high energy limit, when all the mass-dependent terms can be dropped. 
Among all the possible vertex functions that can be studied in a CFT, a particular role is taken by those involving the energy momentum tensor, as for example the $TT$, $TJJ$ and $TTT$, being the insertion of a $T$ operator responsible of the appearance of the conformal anomaly. These will play a key role in our analysis.
\begin{itemize}
\item{\bf \large The anomaly supermultiplet}
\end{itemize}
In our work all of our analysis deals with non-supersymmetric theories, but it is important to point out that some of our results could be 
generalized to the supersymmetric case. 
In fact, it is worth to recall that in a supersymmetric Yang-Mills theory (take for instance $SU(N)$ with $\mathcal N=1$ supersymmetric charges) the trace of the energy momentum tensor, the gamma-trace of the supersymmetric current and the divergence of the chiral $U(1)_R$ current lie in the same anomalous supermultiplet $({T^{\mu}}_{\mu}, \gamma \cdot s, \partial \cdot J_5)$ \cite{Ferrara:1974pz} which describes the radiative breaking of the superconformal symmetry. In this context, chiral and conformal anomalies, that we have separately investigated in our works, are entangled and are likely to play a unified role. In particular, the appearance of an anomaly pole in the chiral $U(1)_R$ current \cite{Dolgov:1971ri} suggests that a similar structure should manifest also in the correlation functions of either the energy momentum tensor or of the supersymmetric current. This result is necessary for a consistent formulation of the anomalous effective action in superspace. This is indeed the case, as we are going to show in the following, at least for the correlators ($TJJ$) with an energy momentum tensor insertion on the $JJ$ 2-point function of vector currents $J$. The emergence of a massless pole in the $TJJ$ vertex, identified in the spectral density of the same correlator by a dispersive approach, is far from being obvious, since its identification requires a rather involved computation, not carried out until recently \cite{Giannotti:2008cv, Armillis:2009pq, Armillis:2010qk, Coriano:2011zk}. It is then interesting to observe that the lifting to superspace of the chiral anomaly pole of the $U(1)_R$ current induces a similar pole in the correlator responsible for the trace anomaly. In this respect it would be relevant to confirm the appearance of these structures also in the supersymmetric current. \\
Although the presence of anomalies in the conservation of global symmetries does not rise to any problem (it can even be predictive), it is instead harmful for the consistency of the theory when the same symmetries are promoted to the local level. Indeed, when the supersymmetric multiplet is embedded in a supergravity context, the anomalous currents of the anomaly multiplet are gauged. The energy momentum tensor couples to the graviton ($g_{\mu\nu}$), the supersymmetric current couples to the gravitino ($\psi_{\mu}$), and the $U(1)_R$ current couples to the axial-vector gauge boson $B_\mu$. The invariance under general coordinate and supersymmetric transformations gives the standard conservation conditions for $T_{\mu\nu}$ and for the spinor current $s_\mu$, but the super-Weyl and $U(1)_R$ symmetries are radiatively broken (see \cite{Freedman:1976uk, Castano:1995ci, Chamseddine:1995gb} for related studies). Therefore, a mechanism of anomaly cancellation must be introduced to restore the predictability of the theory. For this purpose Ovrut and Cardoso \cite{LopesCardoso:1991zt, LopesCardoso:1992yd} developed a procedure which is a realization of the Green-Schwarz mechanism of anomaly cancellation of string theory \cite{Green:1984sg} in a supergravity context. It amounts to the subtraction, in the superspace formalism, of the anomaly poles from the quantum effective action. This nonlocal subtraction implicitly assumes that poles {\em should be present} in the other components of the anomaly supermultiplet, when the 
corresponding currents are inserted in appropriate correlation functions. 
It is then clear that also in the context of supergravity theories, an in-depth comprehension of the structure and of the features of both the chiral and conformal anomaly actions is required.
\begin{itemize}
\item{\bf \large Conformal anomalies and the physics beyond the Standard Model}
\end{itemize}

Conformal anomalies play a role also in several unification contexts, which are of direct phenomenological relevance at the electroweak scale, 
currently under exploration at the LHC. Historically, the interest of the high energy physics community has promoted the study of different extensions of the Standard Model, which is not a complete theory, in several new directions. Among these, some of the most popular have been supersymmetry, technicolor, theories with extra dimensions and scale invariant extensions. 

One of the main motivations for the introduction of supersymmetry in gauge theories is the solution of the hierarchy problem in the Standard Model (SM). Indeed, the electroweak scale is not stable under radiative corrections because the mass-squared $m_H^2$ of the Higgs field gets quadratic contributions from its self-interactions ($\sim \lambda_H$) and from its coupling with the massive fermions ($\sim \lambda_F$) as 
\bea
m_H^2(\textrm{phys}) \simeq m_H^2(\textrm{tree}) + c (\lambda_H - \lambda_F^2) \Lambda^2 \,.
\eea
In the previous equation $m_H(\textrm{phys})$ is the physical Higgs mass ($\sim 10^2$ GeV), $m_H(\textrm{tree})$ the tree level mass and $\Lambda$ is the ultraviolet cut-off. Therefore, regarding the Standard Model as a low-energy effective theory of a larger, fundamental theory, $\Lambda$ represents the scale at which the unknown physics should appear. If we assume that the SM is not affected by any new physics effect until quantum gravity contributions become important, the natural scale to cut-off the quadratic corrections to the Higgs mass is the Planck scale ($\Lambda \sim M_P \sim 10^{19}$ GeV). Now, given that the one-loop corrections are much larger then the electroweak scale, an extraordinary and unpleasant fine-tuning is necessary in order to get down to $10^2$ GeV. \\
A simple way to overcome this problem, as mentioned above, is to assume the existence of a boson-fermion (super)symmetry which relates the scalar and the fermion couplings ($\lambda_H = \lambda_F^2$) in order to cancel the quadratic sensitivity to $\Lambda$ in the structure of the radiative corrections, and reduce it to a harmless logarithmic dependence.

Another interesting possibility to overcome the same hierarchy problem, is to assume that the Planck scale might not be fundamental, and that the scale of quantum gravity could indeed be much lower \cite{ArkaniHamed:1998rs, Antoniadis:1998ig, ArkaniHamed:1998nn, Randall:1999ee} than $10^{19}$ GeV. This mechanism is realized by the introduction of $n$ compactified extra dimensions in which gravity can propagate while the Standard Model fields remain localized on a four-dimensional brane. In this scenario the true fundamental scale $M_*$ would be related to the Planck scale through the equation $M_P^2 = M_*^{2 + n} V_n$, and, therefore, it could be lowered to the TeV scale, solving the electroweak hierarchy problem, by a suitable choice of the compactification radius and of the number of extra dimensions. \\
In this kind of scenarios, the four dimensional effective theory is characterized by the appearance of an infinite tower of spin-2 and spin-0 Kaluza-Klein states, namely the graviton and the graviscalar (also called the dilaton or the radion in the literature) excitations \cite{Han:1998sg, Giudice:1998ck}. They couple to the Standard Model fields through the energy momentum tensor and its trace respectively, the latter being anomalous due to the quantum breaking of the conformal symmetry. Therefore, a more satisfactory understanding of these interactions requires the analysis, at least at one-loop level, of the insertions of the energy momentum tensor on the correlation functions of the Standard Model fields. 
In this respect, the conformal anomaly contribution could play a central role in the characterization of the graviscalar interactions since it 
determines a strong enhancement of its decay and production channels with two photons and two gluons respectively \cite{Giudice:2000av, Csaki:2007ns, Toharia:2008tm, Cheung:2011nv, Grzadkowski:2012ng}.

Moreover, perturbative studies of the conformal anomalies show that the Standard Model effective action is characterized by the appearance of an effective degree of freedom, the {\em effective dilaton}, which describes the anomalous breaking of the conformal symmetry. We discuss the dynamics of this state interpreted as the Nambu-Goldstone mode of such a breaking.
In support of this interpretation we follow, as close as possible, the analogy with the physics of the strong interactions where the chiral anomaly pole of the $AVV$ diagram finds its natural identification as the pion field, the Goldstone boson of the spontaneous breaking of the chiral symmetry. This interpretation is in line with the recently proposed scenarios of scale invariant extension of the Standard Model \cite{Goldberger:2007zk, Fan:2008jk} in which a light scalar field can appear in combination of the Higgs sector.

\begin{itemize} 
\item{\bf \large Cosmological applications: the analysis of CMB non-gaussianities} 
\end{itemize}
Besides these intriguing gravitational effects in high-energy particle physics, the theory of gravity certainly plays a paramount role in the cosmological context. Among all the interesting subjects in the standard and particle cosmology, the analysis of the primordial gravitational fluctuations provides clues on the physics of the early universe, leaving an imprint on the cosmic microwave background (CMB) and on the formation and evolution of large scale structures \cite{Mukhanov:1990me, Bartolo:2010qu}. Their origin is well understood as a quantum effect in the context of cosmological inflation, which is considered the leading theory for explaining the flatness of the universe and the seeds of the density fluctuations which gave rise, in the full nonlinear regime, to the complex structures that we observe nowadays at large scales. This picture has been recently confirmed by the cosmological data extracted from the analysis of the CMB temperature anisotropies provided by 
the Wilkinson Microwave Anisotropy Probe (WMAP) experiment \cite{2009ApJS..180..225H}. Moreover, the Planck experiment \cite{PlanckExpurl} may be also sensitive to non-linear effects in the cosmological perturbations which manifest themselves through a non-gaussian behaviour in the CMB \cite{Bartolo:2004if}. The gaussian character of these fluctuations is parameterized by the 2-point correlation function of the gravitational perturbations, also called the "power spectrum", whose measurements alone are, unfortunately, not able to describe completely the physics of the inflationary epoch. Higher-point correlation functions on the celestial sphere, as the bispectrum or the trispectrum, are required to discriminate among various cosmological models, because different realizations of the inflationary mechanism predict different amounts of non-gaussianities (NG). For instance, in the single field inflationary scenario, it was proved \cite{Maldacena:2002vr} that the non-gaussian fluctuations are vey tiny and then a possible experimental detection of a certain amount of NG would allow to rule out this particular model.

Another interesting proposal relies on the idea that a quantum gravitational system in $n$ dimensions is described by a dual quantum field theory (QFT) without gravity in $n-1$ dimensions \cite{'tHooft:1993gx, Susskind:1994vu}. This holographic principle emerged in the past to explain why the entropy of a black hole scales like the area of its horizon and not like its volume, as we would naively expect. Concrete realizations of this new paradigm have been found in string theory \cite{Maldacena:1997re} through strong-weak coupling duality, and then applied, as we have mentioned, also to describe the early cosmological evolution of our universe \cite{McFadden:2009fg}. In this framework, the inflationary epoch has a holographic description in terms of the dynamics of a three dimensional dual QFT. Indeed the cosmological observables, like the power spectrum, the bispectrum and the trispectrum, are related to correlation functions of two, three and four energy momentum tensor operators, which can be computed on the dual field theory side using ordinary perturbation theory \cite{McFadden:2009fg, McFadden:2010vh, McFadden:2011kk, Bzowski:2011ab, Coriano:2012hd}. Interestingly, the analysis of the power spectrum is found to be compatible \cite{Easther:2011wh} with the current data from WMAP.

 \mainmatter

\chapter*{Part one}
{\LARGE{ \textbf {Conformal anomalies in the Standard Model \\ \\ and the effective dilaton}}}

\vspace{2cm}

\chapter{Polology and the example of the chiral anomaly}
\label{Chap.AnomalyPoles}
%
 \section{Introduction}
 
This chapter is a basic introduction to polology, showing as a typical application the case of the anomaly pole present in the $AVV$ diagram which is responsible for the chiral anomaly. 
The first section of this chapter consists of a brief review of the quantum field theory connection between the analytical structure of the poles in a given correlation function $G$, defined by a set of operators $\mathcal O_i$, and the implications that such a structure has for the physical spectrum of the theory. We follow the line of reasoning addressed in \cite{WeinbergBook1}, where it is shown that a pole, in the kinematical invariants of $G$, appears only when a given one-particle state interpolates with $\mathcal O_i$ and the vacuum. This singularity, which can be either massive or massless, may be an elementary particle associated to a field operator appearing in the Lagrangian or even a composite one. For instance, the old-known massless pole of the chiral anomalous $AVV$ diagram \cite{Dolgov:1971ri, Achasov:1992bu, Horejsi:1994aj} surely belongs to the latter case, being interpreted as the pion state interpolating between the axial current ($A$) and two on-shell photons (described by the action of the vector currents $(V)$ on the vacuum).\\
Moreover we review some of the characteristics of the chiral anomaly, summarizing part of the work presented in \cite{Armillis:2009sm} which has not been included in this dissertation. There we have shown that the chiral anomaly pole can be identified in three possible ways: by a direct perturbative computation, by a solution of the anomalous Ward identity satisfied by the $AVV$ diagram, or, finally, by a solution of a variational equation satisfied by the anomalous effective action. 
As we have mentioned, one feature of the chiral anomaly, analyzed in a perturbative framework, is the appearance of massless poles which account for it. They can be extracted by a spectral analysis and are usually interpreted as being of an infrared origin. Nevertheless their presence is not just confined in the infrared, but they appear in the effective action under the most general kinematical condition \cite{Armillis:2009sm}.
They are also responsible for the non-unitary behaviour of these theories in the ultraviolet (UV) region \cite{Armillis:2009sm}. This analysis can be extended to the case of the gravitational conformal anomaly, showing that the effective action describing the interaction of gauge fields with gravity is characterized by anomaly poles that give the entire anomaly and that are decoupled in the infrared (IR) in the presence of fermion mass corrections, in complete analogy with the chiral case. In a related analysis on the trace anomaly in gravity \cite{Giannotti:2008cv}, confined in the IR, an anomaly pole has been identified in the corresponding correlator using dispersion theory. Our extension to the most general kinematical configuration is based on an exact computation of the off-shell correlation function, responsible for the appearance of the anomaly, involving an energy-momentum tensor and two vector currents (the gauge-gauge-graviton vertex). The analysis of the conformal anomaly will be presented in the next chapters while here we will focus on the chiral case. \\

\section{Polology}

The only way a physical amplitude can develop a pole in one of the momenta of the external lines, or in a combination of them, is through the coupling of a particle in the physical spectrum of the theory to the operators taken into account. Therefore, poles do not appear in correlation functions for no reason. Usually these singularities match the elementary particles which correspond to the fields appearing in the Lagrangian, however this is not the only possibility. A pole term can emerge also when the exchanged particle in the correlation function is a bound state of the elementary fields that do appear in the defining theory. This situation is most frequently encountered in theories with a confining phase of which QCD is an enlightening example. For instance, the axial anomalous $AVV$ amplitude, where $A$ and $V$ are the axial and vector currents respectively, exhibits a pole term which completely saturates the anomaly coefficient. This structure clearly describes the pseudoscalar pion $\pi$ which interpolates between the axial and the two vector currents and it is essential to correctly reproduce the decay rate of $\pi \rightarrow \gamma\gamma$. We remark that it was exactly the understanding of this process that led to the discovery of the anomaly in the theory of the strong interactions.

It is then clear that the analysis of the kinematical singularities of a certain class of correlators is of a paramount importance in order to improve our knowledge on the physical spectrum of a theory, in particular on the composite particles, to better understand their interactions and, in principle, even to predict the existence of new states. In this chapter we briefly illustrate how and under which circumstances a pole structure can emerge from a correlation function. 

For this purpose, let's review the proof given in \cite{WeinbergBook1}, which is not restricted to perturbation theory, but relies on \emph{non-perturbative} methods. Such a proof relies only on symmetry arguments and on the existence of a completeness relation.

We consider a generic correlator $G$ as a function of the momenta of the external lines. In coordinate space it is given by
\bea
G(x_1, \ldots, x_n) = \langle 0 | T \left\{ A_1(x_1) \ldots A_n (x_n) \right\} | 0 \rangle 
\eea 
where $|0\rangle$ is the vacuum, the symbol $T$ denotes the time order product and $A_n$ are arbitrary operators. They may be the fields appearing in the Lagrangian or even composite local operators. Although the analysis can be developed in full generality with an arbitrary number of operator insertions, we specialize to a simpler case with just three operators, say $\mathcal O(z), A_1(x_1)$ and $A_2(x_2)$. This is the situation encountered in our studies on chiral and conformal anomalies. The general proof can be found in \cite{WeinbergBook1} and follows the same steps presented here. \\
Then we move to momentum space and consider the correlator
\bea
G(k, p_1, p_2) = \int d^4 z d^4 x_1 d^4 x_2 \, e^{- i k z - i p_1 x_1 - i p_2 x_2} \langle 0 | T \left\{ \mathcal O(z) A_1(x_1)  A_2 (x_2) \right\} | 0 \rangle
\eea
as a function of the virtuality of $\mathcal  O$, namely $k^2 = (- p_1 - p_2)^2$. Notice that the virtualities of the external momenta $p_1^2$ and $p_2^2$ are not fixed by any on-shellness condition and can be arbitrary. \\
In order to proceed with the analysis we isolate the operator $\mathcal O$ from the T product and retain only the term in which $\mathcal O$ appears to the far left
\bea
G(k, p_1, p_2) &=& \int d^4 z d^4 x_1 d^4 x_2 \, e^{- i k z - i p_1 x_1 - i p_2 x_2} \nn \\
&\times& \bigg\{ \theta(z^0 - \max\{x_1^0, x_2^0\}) \, \langle 0 | \mathcal O(z) T \left\{ A_1(x_1)  A_2 (x_2) \right\} | 0 \rangle + \ldots \bigg\} \,,
\eea
where the ellipsis stand for the other time ordering products which we have ignored. They do not contribute with any pole structure to the correlator.  \\
Now we insert a complete set of intermediate states between the operator $\mathcal O$ and the other ones, isolating only single particle states with a specific mass $m$. We discard the other single particle states with different masses (they will contribute with poles but at other kinematical positions) and multi particle states (which appear as branch cuts). We obtain
\bea
G(k, p_1, p_2) &=& \sum_\sigma \int d^4 z d^4 x_1 d^4 x_2 d^3 \vec{p} \, e^{- i k z - i p_1 x_1 - i p_2 x_2} \nn \\
&\times& \bigg\{ \theta(z^0 - \max\{x_1^0, x_2^0\}) \langle 0 | \mathcal O(z) | \vec{p}, \sigma \rangle \langle \vec{p}, \sigma | T \left\{ A_1(x_1)  A_2 (x_2) \right\} | 0 \rangle + \ldots \bigg\}
\eea
where $|\vec{p},\sigma \rangle$ is a single particle state with mass $m$ ($p^2  = m^2$) and with quantum numbers collectively identified by $\sigma$. \\
In order to easily perform the integration over the spacetime coordinates, it will be useful to extract the $z$ and $x_1$ dependences from the matrix elements appearing in the previous equation, and to introduce the new integration variable $y =  x_1 -  x_2$ in place of $x_2$. Finally we insert the integral representation of the step function $\theta(t)$ given by
\bea
\theta(t) = \frac{i}{2 \pi} \int_{- \infty}^{+ \infty} d \omega \frac{e^{- i \omega \, t}}{\omega + i \epsilon}
\eea
where $\epsilon$ is an infinitesimal and positive constant. We have 
\bea
G(k, p_1, p_2) &=& \frac{i}{2 \pi} \sum_\sigma \int d^4 z \, d^4 x_1 \, d^4 y \, d^3 \vec{p} \, \frac{d \omega}{\omega + i \epsilon}  \, e^{- i k z - i (p_1 + p_2) x_1 - i p_2 y } \nn \\
&& \hspace{-2cm} \times \, \, e^{ - i \omega (z^0 - x_1^0 - \max\{0, y^0\})} e^{i p z - i p x_1}  \langle 0 | \mathcal O(0) | \vec{p}, \sigma \rangle \langle \vec{p}, \sigma | T \left\{ A_1(0)  A_2 (y) \right\} | 0 \rangle + \ldots  \,,
\eea
where the integration over $z$ and $x_1$ is straightforward and gives only delta functions
\bea
&& \hspace{-0.7cm} G(k, p_1, p_2) = \frac{i}{2 \pi} \sum_\sigma \int  d^4 y \, d^3 \vec{p} \, \frac{d \omega}{\omega + i \epsilon}  \, e^{ - i p_2 y + i \omega \max\{0, y^0\} }  
\langle 0 | \mathcal O(0) | \vec{p}, \sigma \rangle \langle \vec{p}, \sigma | T \left\{ A_1(0)  A_2 (y) \right\} | 0 \rangle \nn \\
&& \hspace{-0.7cm} \times (2 \pi)^8 \delta^{(3)}(\vec{k} - \vec{p}) \, \delta(k^0 - \sqrt{\vec{p}^2 + m^2}  + \omega) \delta^{(3)}(\vec{p_1} + \vec{p_2} + \vec{p}) \, \delta(p_1^0 + p_2^0 + \sqrt{ \vec{p}^2 + m^2} - \omega) + \ldots \,. 
\eea
The integrations over the momenta $\vec{p}$ and $\omega$ are now trivial due to the delta functions and lead to
\bea
&& G(k, p_1, p_2) = (2 \pi)^4 \delta^{(4)}(k + p_1 + p_2) i \frac{(2 \pi)^3}{\sqrt{\vec{k}^2 + m^2} - k^0 + i \epsilon} \nn \\
&& \times \sum_\sigma \int d^4 y  \, e^{  i \left( \sqrt{\vec{k}^2 + m^2} - k^0  \right) \max\{0, y^0\}} e^{ - i p_2 y}
\langle 0 | \mathcal O(0) | \vec{k}, \sigma \rangle \langle \vec{k}, \sigma | T \left\{ A_1(0)  A_2 (y) \right\} | 0 \rangle \,.
\eea
The appearance of the pole in the limit $k^0 \rightarrow \sqrt{\vec{k}^2 + m^2}$ in the correlation function is now explicitly manifest and originates from the massless pole in $\omega$, which comes, in turn, from the integral parameterization of the step function. In order to make the pole structure more clear we notice that near the pole
\bea
\frac{1}{\sqrt{\vec{k}^2 + m^2} - k^0 + i \epsilon}  \sim  \frac{2 k^0  }{k^2 - m^2 - i \epsilon}
\eea
while the exponential function under integration goes to unity. This allows us to define the matrix elements
\bea
(2 \pi)^4 \delta^{(4)}(k - p) \, \mathcal M_{0 | (k,\sigma)}(k) &\equiv& \int d^4 z e^{-i p z} \langle 0 | \mathcal O(z) | \vec{k}, \sigma \rangle \, \\
(2 \pi)^4 \delta^{(4)}(k + p_1 + p_2) \, \mathcal M_{(k,\sigma) | 0}(k, p_1, p_2) &\equiv& \int d^4 x_1 d^4 x_2 e^{-i p_1 x_1 - i p_2 x_2} \langle \vec{k}, \sigma | T \left\{ A_1(x_1)  A_2 (x_2) \right\} | 0 \rangle \,.\nn \\
\eea
With these definitions and simplifications the pole behaviour of the correlator is now explicit and reads as
\bea
\label{polebehaviour}
G(k, p_1, p_2) & \stackrel{k^2 \rightarrow m^2}{\longrightarrow} & (2 \pi)^4 \delta^{(4)}(k + p_1 + p_2) \nn \\
&\times&
 \sum_\sigma   \sqrt{2 (2 \pi)^3 k^0} \mathcal M_{0 | (k,\sigma)}(k)      \frac{ i }{k^2 - m^2 - i \epsilon}    \sqrt{2 (2 \pi)^3 k^0}  \mathcal M_{(k,\sigma) | 0}(k, p_1, p_2) \,. 
\eea
\begin{figure}[t]
\begin{center}
\includegraphics[scale=0.5]{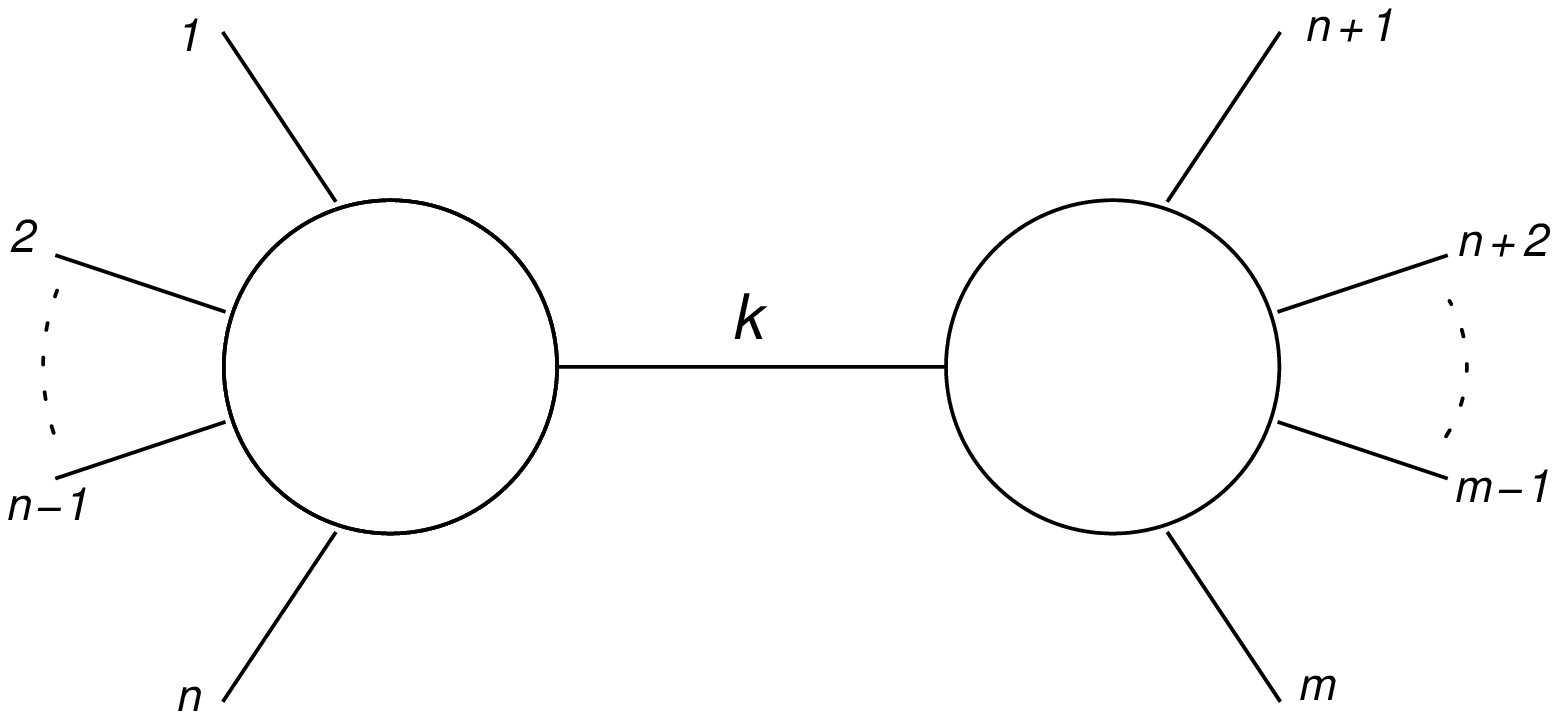}
\caption{\small A correlation function exhibits a pole exchange with momentum $k$ corresponding to an elementary particle which appears in the Lagrangian.}
\label{Polology.Elementary}
\end{center}
\begin{center}
\includegraphics[scale=0.5]{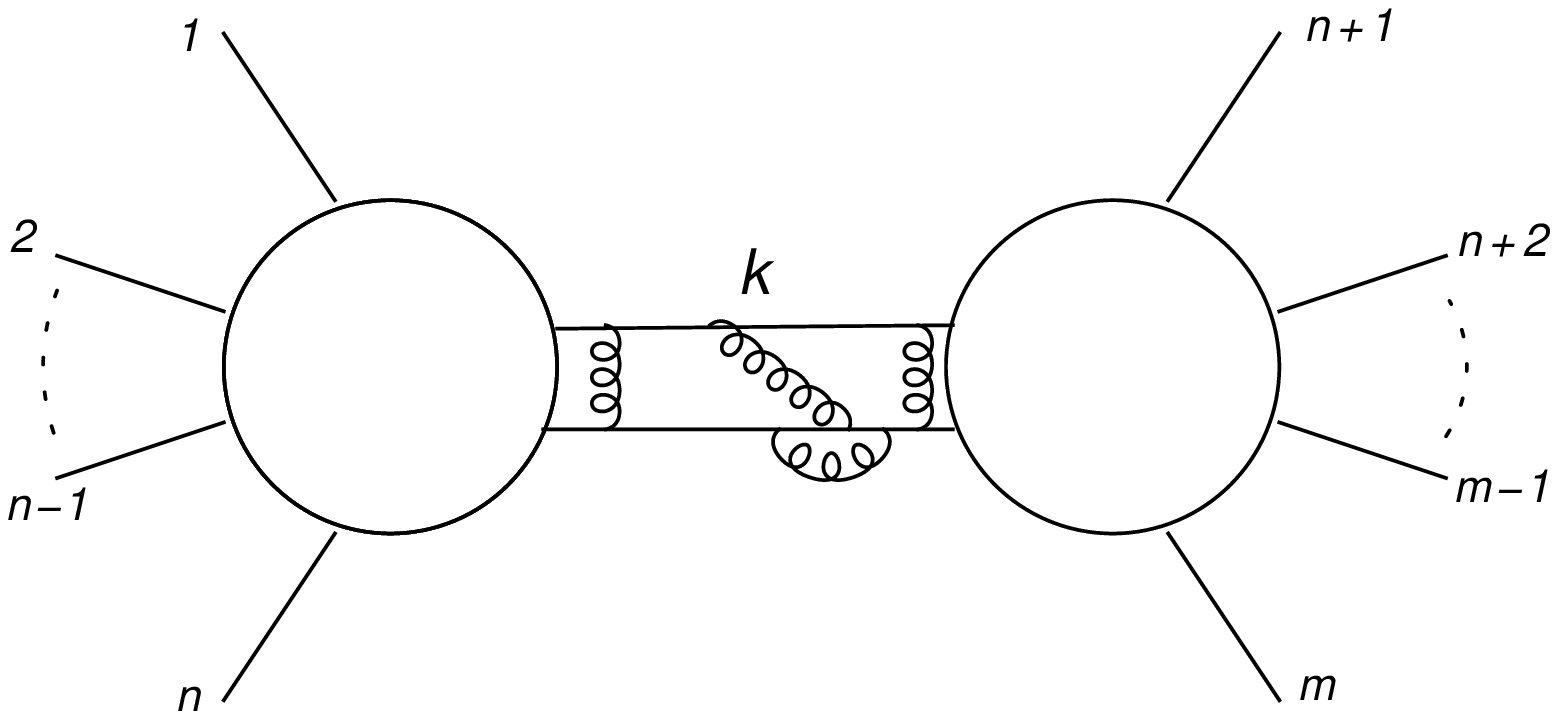}
\caption{\small A bound state interpolates between the two subamplitudes in a given correlation function. In this case the pole corresponds to a composite, a bound state of two elementary particles represented by straight lines, which interact together by the exchange of other elementary states (curly lines).}
\label{Polology.Composite}
\end{center}
\end{figure}
Notice that the term $\sqrt{2 (2 \pi)^3 k^0}$ is just a kinematical factor which ensures the correct normalization of the $| k, \sigma \rangle$ state. \\
As we can see form Eq.(\ref{polebehaviour}), the correlation function in momentum space exhibits a pole behaviour in the external kinematical invariant $k^2$ near $m^2$ if the operators under consideration have non-vanishing transition amplitudes between the vacuum and a single particle state with mass $m$. In other words, the matrix elements defined above must be non zero. \\
We remark again that this state can be created by a field operator that appears in the Lagrangian of the theory, Fig.\ref{Polology.Elementary}, but the most interesting case certainly occurs when the pole is associated with a bound state, Fig.\ref{Polology.Composite}, namely, a composite particle of the elementary fields defined in the Lagrangian. 

One of the goal of the work presented in this thesis is to investigate the existence of massless pole contributions in correlation functions with a single insertion of the energy momentum tensor and to understand their connection to the trace anomaly. The correlator responsible for the appearance of this massless degree of freedom has been analyzed in a generic kinematical configuration and in a reliable physical theory like the Standard Model. Complete analytic results have been obtained allowing for an unambiguous identification of these singularities without resorting to dispersion theory.

\section{The chiral anomaly poles}
In the case of chiral (and anomalous) theories, the corresponding anomalous Ward identities, which are at the core of the quantum formulation, have a natural and obvious solution, which can be written down quite straightforwardly, in terms of anomaly poles. This takes place even before that any direct computation of the anomaly diagram allows to really identify the presence (or eventually the absence) of such contributions in the explicit expression of an anomalous correlator of the type $AVV$ (A= Axial-Vector, V=Vector) or $AAA$. 

\begin{figure}[t]
\begin{center}
\includegraphics[scale=1.0]{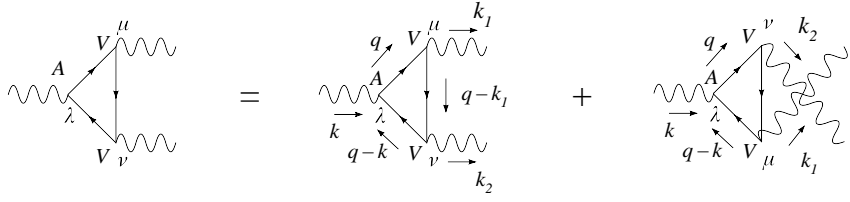}
\caption{\small Triangle diagram and momentum conventions for an AVV correlator.}
\label{P1AVV}
\end{center}
\end{figure}

To state it simply, the pole appears by solving the anomalous Ward identity for the corresponding $AVV$ amplitude $\Delta^{\lambda\mu\nu} (k_1,k_2) $  of a massless theory (we use momenta as in Fig.~\ref{P1AVV}  with $k=k_1+k_2$)
\beq
k_\lambda \Delta^{\lambda\mu\nu} (k_1,k_2)= a_n \epsilon^{\mu\nu\alpha\beta}\, k_{1\alpha} \, k_{2\beta}
\eeq
rather trivially, using the longitudinal tensor structure
\beq
\Delta^{\lambda\mu\nu}= a_n \, \frac{k^{\lambda}}{k^2} \, \epsilon^{\mu\nu\alpha\beta}\, k_{1\alpha} \, k_{2\beta}.
\label{P1IRpole}
\eeq
In the expression above the coefficient $a_n = -i / 2 \pi^2$ denotes the anomaly contribution.
The presence of this tensor structure with a $1/k^2$ behaviour is the signature of the anomaly. 
This result holds for an $AVV$ graph, but can be trivially generalized to more general anomalous correlators, such as the
$AAA$ functions, by adding poles in the invariants of the remaining external lines, i.e. $1/k_1^2$ and $1/k_2^2$
\beqa
\Delta^{\lambda \mu \nu}_{AAA}(k,k_1,k_2) &=& \frac{1}{3} \bigg[
\frac{a_n}{k^2} \, k^\lambda \, \epsilon[\mu, \nu, k_1, k_2]
+ \frac{a_n}{k_1^2} \, k_1^\mu \, \epsilon[\lambda, \nu, k, k_2]
+ \frac{a_n}{k_2^2} \, k_2^\nu \, \epsilon[\lambda, \mu, k, k_1]\bigg],
\eeqa
and imposing an equal distribution of the anomaly on the three axial-vector legs \footnote{The symbol $\epsilon[\mu,\nu,k_1,k_2]$ is defined as $\epsilon^{\mu\nu\alpha\beta} k_{1\, \alpha} k_{2 \,\beta}$}.

The same Ward identity can be formulated also as a variational equation on the quantum effective action. For this purpose consider the simplest case of a theory describing a single anomalous spin 1, $B_\mu$, coupled to the axial current, with a Lagrangian defined as
\beq
\mathcal{L}_{B}= \overline{\psi} \left( i \, \partial \! \! \!  / + e B \! \! \! \! /  \gamma_5\right)\psi - \frac{1}{4} F_B^2 \,.
\label{P1count0}
\eeq
The anomalous variation $(\delta B_\mu=\partial_\mu\theta_B)$ of Eq. (\ref{P1count0}) implies for the quantum effective action $\Gamma_B$
\beq
\delta \, \Gamma_B = \frac{ i \, e^3 \, a_n}{24} \, \int d^4 x \, \theta_B(x) \, F_B\wedge F_B \,.
\label{P1var1}
\eeq
This can be reproduced by the nonlocal action
\beq
{\Gamma}_{pole}= \frac{e^3}{48 \, \pi^2} \, \langle \partial B(x) \square^{-1}(x-y) F_B(y)\wedge F_B(y) \rangle \,,
\label{P1var2}
\eeq
which is the coordinate space analogous of Eq. (\ref{P1IRpole}).
Given a solution (\ref{P1var2}) of the variational equation (\ref{P1var1}), it is mandatory to check whether the $1/\square$ (nonlocal) structure is indeed justified by a perturbative computation. \\
\begin{figure}[t]
\begin{center}
\includegraphics[scale=0.8]{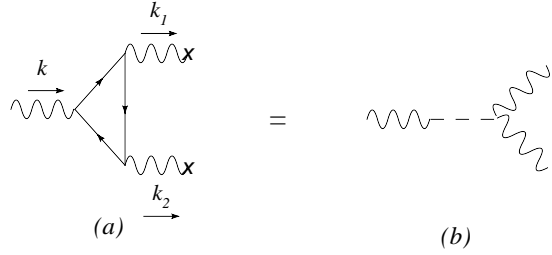}
\caption{\small The amplitude $\Delta^{\lambda\mu\nu}(k_1,k_2)$ shown in $a)$ for the kinematical configuration $k_1^2=k_2^2=0$ reduces to the polar form depicted in $b)$ and given by Eq.~(\ref{P1IRpole}).}
\label{P1pole}
\end{center}
\end{figure}
The analysis shows that the kinematical configuration responsible for the appearance of the pole can be depicted as in Fig.~\ref{P1pole}. In this graph, containing the mixing of a spin $1$ with a spin $0$, the anomalous current interpolates with the two photons via an intermediate massless state. This can be interpreted as describing a collinear 
fermion-antifermion pair (a pseudoscalar composite state) coupled to the two on-shell photons (see also the discussion in \cite{Giannotti:2008cv}). The anomaly graph is characterized, in this limit, by a nonzero spectral density proportional to $\delta(k^2)$ \cite{Dolgov:1971ri}. This kinematical configuration, in which the two photons are on-shell and the fermions are massless, is entirely described by the anomaly pole, which has a clear infrared (IR) interpretation \cite{Coleman:1982yg}. The IR ($k^2 \rightarrow 0$) coupling of the pole present in the correlator is, in this case, rather obvious since the limit
\beq
\lim_{k^2\rightarrow 0} \, k^2 \, \Delta^{\lambda\mu\nu}=k^\lambda \,  a_n \, \epsilon^{\mu\nu\alpha\beta} \, k_{1\alpha} \, k_{2\beta}
\eeq
allows to attribute to this amplitude a non-vanishing residue.

The infrared analysis sketched above is well suited for the identification of anomaly poles which have a rather clear interpretation in this region, but does not allow to identify other similar pole terms which might emerge in far more general kinematical configurations (for instance with massive fermions). In \cite{Armillis:2009sm} we have shown that only a complete and explicit computation of the anomalous effective action allows the identification of the extra anomaly poles present in an $AVV$ correlator, that otherwise would escape detection. For this purpose we will consider also, unless otherwise stated, a fermions mass $m$. These have been identified\footnote{A single pole term for an AVV and three pole terms for an AAA diagram.}  using a special representation of the anomaly amplitude developed in \cite{Knecht:2003xy,Knecht:2002hr} (that we have called the ``Longitudinal/Transverse" or L/T parameterization), based on the general solution of an anomalous Ward identity. This parameterization takes the form
\beq
  \Delta^{\lambda\mu\nu} (k_1,k_2) = \mathcal \, W ^{\lambda\mu\nu}= \frac{1}{8 \,\pi^2} \left [  \mathcal \, W^{L\, \lambda\mu\nu} -  \mathcal \, W^{T\, \lambda\mu\nu} \right]
\label{P1long}
\eeq
where the longitudinal component ($W_L$) has a pole contribution ($w_L=-4 i/s$) plus mass corrections 
($\mathcal{F}$) computed in \cite{Armillis:2009sm}
\beq
 \mathcal \, W^{L\, \lambda\mu\nu}= \left(w_L  - \mathcal{F}(m, s,s_1,s_2)\right)  \, k^\lambda \veps[\mu,\nu,k_1,k_2]
\label{P1brokenW}
\eeq
with
\beq
\mathcal{F}(m, s,s_1,s_2)= \frac{8 \, m^2}{\pi ^2 \, s} \, C_0(s, s_1,s_2,m^2) \,,
\label{P1due}
\eeq
and where $C_0$ is the standard three-point scalar integral with equal masses, while $s = k^2, \, s_1 = k_1^2, \, s_2 =k_2^2$.
The transverse form factors appearing in $W_T$ contribute homogeneously to the anomalous Ward identity. They have been given in the most general case in \cite{Armillis:2009sm}.

Obviously, some doubts concerning the correctness of this parameterization may easily arise, especially if one is accustomed 
to look for anomaly poles using a standard infrared analysis. 
For this reason and to dissolve any possible doubt, a direct computation shows 
that the L/T representation introduced in \cite{Knecht:2003xy,Knecht:2002hr} is, indeed, completely equivalent to the other existing parameterization in the literature \cite{Rosenberg:1962pp}, even though no poles apparently come to the surface when using this alternative description of the anomaly graph.
In \cite{Armillis:2009sm} one can find an extension of the same parameterization to the massive fermion case. \\
We conclude this section with some comments about the IR decoupling of the anomaly pole in the presence of massive fermions. Indeed it can be seen that, for fixed $m^2 \neq 0$, the residue of the $AVV$ diagram in the limit $k^2 \rightarrow 0$ is vanishing 
\beq
\lim_{k^2\rightarrow 0} \, k^2 \, \Delta^{\lambda\mu\nu} = 0 \,,
\eeq
and therefore the anomalous diagram with mass corrections does not exhibit the pole singularity for $k^2 \rightarrow 0$. This behaviour may be understood as a consequence of the decoupling theorem. The limit $k^2 \rightarrow 0$ can be equivalently reproduced by $k^2/m^2 \rightarrow 0$ because no other mass scale appears in the computation (we are considering the on-shell case $k_1^2 = k_2^2 = 0$), and therefore it is the same as the large mass limit $m^2 \rightarrow \infty$ with $k^2$ fixed. In this case the loop contribution should vanish since the momentum $k$ is not large enough to resolve the quantum fluctuations. For more details on this argument we refer to \cite{Giannotti:2008cv, Armillis:2009sm}, where the IR pole decoupling has been examined also for off-shell anomalous amplitudes. 
Finally, we have shown that the anomaly pole is indeed decoupled in the IR under general kinematic conditions, unless $k_1^2 = k_2^2 = m^2 = 0$.
\section{Pole-dominated amplitudes} 
\begin{figure}[t]
\begin{center}
\includegraphics[scale=0.8]{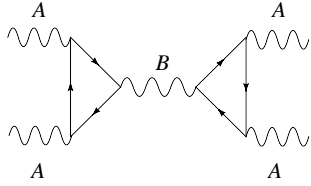}
\caption{\small A ``two-triangles" anomaly amplitude in the $s$-channel which is pole-dominated. In this case we have assumed $A$ to be a non-anomalous boson while $B$ is anomalous.}
\label{P1BIM}
\end{center}
\end{figure}
A useful device  to investigate the meaning of these new anomaly poles is provided by a class of amplitudes \cite{Bouchiat:1972iq} 
which connect initial and final state via anomalous correlators, one example of them being shown in Fig.~(\ref{P1BIM}). These amplitudes are unitarily unbound in the UV \cite{Coriano:2008pg}. This property can be easily derived by considering the scattering of external {\em massless} spin-1 fields coupled via a longitudinal exchange of an anomalous boson. The amplitude in the s-channel is shown in Fig.~\ref{P1BIM}. 
In the case of scattering of massless bosons, the pole of Eq.~(\ref{P1IRpole}) saturates each of the two subamplitudes (i.e. for $m=s_1=s_2=0$) and provides a cross section which grows quadratically with the energy scale.
This is an obvious manifestation of the fact that an anomaly 
pole has dangerous effects in the UV because of the violation of the Froissart bound at high energies. This behaviour is retained also under general kinematics, for instance in the scattering of massive bosons, when each of the two triangle 
subdiagrams of Fig.~(\ref{P1BIM}) takes the more general form given by Eqs. (\ref{P1long}), (\ref{P1brokenW}). Interestingly enough, if we subtract the pole component contained in $w_L$  
the quadratic growth of the amplitude disappears \cite{Armillis:2009sm}. Therefore the manifestation of the anomaly and the breaking of unitarity in the UV is necessarily attributed to the $w_L$ component, even if it is decoupled in the IR. After the subtraction, the Ward identity used in the computation of the amplitude is not anomalous anymore, even if it could remain broken (due to the massive fermions). The apparent breaking of unitarity in the UV is not ameliorated by a more complete analysis of this S-matrix amplitude involving the Higgs sector, since a massless fermion in each of the two anomaly loops would not allow the exchange of a Higgs in the s-channel. Explicit computations show that the corresponding amplitude would still have the same asymptotic behaviour even for a massive fermion. 

The only possible conclusion extrapolated from this example is that amplitudes which are dominated by anomaly poles in the UV region demonstrate the inconsistency of an anomalous theory, as expected by common lore.
We conclude that unitarity provides a hint on the UV significance of the anomaly poles of the anomaly graphs surfacing in the L/T parameterization, poles which escape detection in the usual IR analysis. Even if they have been identified in the UV, this does not necessarily exclude a possible (indirect) role played by these contributions in the IR region.

The formal solution of the Ward identity \cite{Knecht:2003xy} that takes to the L/T parameterization and to the isolation of an anomaly pole is indeed in agreement with what found in a direct computation. 
These results, as we are going to show, emerge also from the perturbative analysis of the effective action for the conformal anomaly and are likely to correspond to a generic feature of other manifestations of the anomalies in quantum field theory. 

\section{The complete anomalous effective action and its expansions in the chiral case}
The point made in \cite{Armillis:2009sm} is that the anomaly is always completely given by $w_L$, under {\em any} kinematical conditions, while the mass corrections (generated, for instance, by spontaneous symmetry breaking) are clearly (and separately) identifiable as extra terms which contribute to the broken anomalous Ward identity satisfied by the correlator. It is important that these two 
sources of breaking of the gauge symmetry (anomalous and spontaneous) be thought of as having both an independent status. For this reason one can provide several organizations of the effective actions of anomalous theories, with similarities that cover both the case of the chiral anomaly and of the conformal anomaly, as we will discuss next. 

The complete effective action, in the chiral case, can be given in several forms. The simplest, valid for any energy range, is the full one
 \beq
 \Gamma^{(3)}=  \Gamma^{(3)}_{pole} + \tilde{\Gamma}^{(3)}
 \eeq
 with the pole part given by (\ref{P1var2}) and the remainder ( $\tilde{\Gamma}^{(3)}$) given by a complicated nonlocal expression which contributes homogeneously to the Ward identity of the anomaly graph
 \beqa
  \tilde{\Gamma}^{(3)}&=&  -  \frac{e^3}{48 \pi^2} \int d^4 x \, d^4 y \, d^4 z \,  
  \partial \cdot B(z) F_B (x) \wedge F _B (y ) \nn \\
&\times&  \int \frac{d^4  k_1 \, d^4 k_2}{(2 \pi)^8} \, 
 e^{-i k_1 \cdot (x-z) - i k_2 \cdot (y-z)} \mathcal F (k, k_1, k_2, m)  \nn \\
&& -\frac{e^3}{48 \pi^2} \int d^4 x \, d^4 y \, d^4 z \, B_\lambda (z) \, B_\mu (x) \, B_\nu (y) \nn \\
&\times&  \int  \frac{d^4  k_1 \, d^4 k_2}{(2 \pi)^8} \, e^{-i k_1 \cdot (x-z) - i k_2 \cdot (y-z)} \, W_T ^{\lambda \mu \nu} (k, k_1,k_2,m). 
\label{P1gammafull}
 \eeqa
 The analytic expressions of these form factors can be found in \cite{Armillis:2009sm}. This (rather formal) expression is an exact result, but becomes more manageable if expanded in the fermion mass (in $1/m$ or in $m$) 
 (see for example \cite{Bastianelli:2007jv, Bastianelli:2004zp}).

For instance, let's consider the $1/m$ case. One of the shortcomings of this expansion, because of the IR decoupling property of anomaly poles in the presence of fermion mass corrections, is that it does not do full justice 
of the presence of massless degrees of freedom in the theory (anomaly poles do not appear explicitly in this expansion) which, as discussed in \cite{Giannotti:2008cv} might instead be of physical significance. We refer to \cite{Armillis:2009sm} for the details.

A second expansion of the effective action 
Eq.~(\ref{P1gammafull}) can be given for a small mass $m$ (in $m^2/s$), with $s = k^2$. In this formulation the action is organized in the form of a pole contribution plus $O(m^2/s)$ corrections. Although this case it is not suitable to describe the heavy fermion limit, the massless pseudoscalar degrees of freedom introduced by the anomaly in the effective theory can be clearly identified from it. \\
For the longitudinal form factor, this expansion gives ($s<0$) 
\beq
W_L =  - \frac{4 i}{s}  - \frac{ 4 \, i \, m^2}{s^2} \log \left( - \frac{s}{m^2}\right) + O (m^3)
\label{P1smooth}
\eeq
which has a smooth massless limit. The expansion of the transverse form factor $W_T$ can be found in \cite{Armillis:2009sm}. This form of the effective action is the most suitable for the study of the UV behaviour of an anomalous theory, in the search, for instance, of a possible UV completion. Notice that the massless limit of this action reflects (correctly) the pole-dominance present in the theory in the UV region 
of $s\to \infty$, since the mass corrections are suppressed by $m^2/s$.

\chapter{The $TJJ$ vertex in QED and the conformal anomaly}
\label{Chap.TJJQED}

\section{Introduction}
Investigations of conformal anomalies in gravity (see \cite{Duff:1993wm} for an historical overview and references) \cite{Deser:1976yx} and in gauge theories \cite{Adler:1976zt,Collins:1976yq,Freedman:1974ze} as well as in string theory, have been of remarkable significance along the years. In cosmology, for instance, \cite{Starobinsky:1980te} (see also \cite{Shapiro:2008sf} for an overview)  the study of the gravitational trace anomaly has been performed in an attempt to solve the problem of the ``graceful exit" (see for instance \cite{Fabris:2000gz, Shapiro:2001rd, Shapiro:2001rh, Pelinson:2002ef}). In other analysis it has been pointed out that the conformal anomaly may prevent the future singularity occurrence in various dark energy models \cite{Nojiri:2004pf,Nojiri:2005sx}. 

In the past the analysis of the formal structure of the effective action for gravity in four dimensions,  obtained by integration of the trace anomaly \cite{Riegert:1984kt, Fradkin:1983tg}, has received a special attention, showing that the variational solution of the anomaly equation, which is nonlocal, can be made local by the introduction of extra scalar fields. The gauge contributions to these anomalies are identified at one-loop level from a set of diagrams - involving fermion loops with two external gauge lines and one graviton line - and are characterized, as shown recently by Giannotti and Mottola in \cite{Giannotti:2008cv}, by the presence of anomaly poles. Anomaly poles are familiar from the study of the chiral anomaly in gauge theories and describe the nonlocal structure of the effective action. In the case of global anomalies, as in QCD chiral dynamics, they signal the presence of a non-perturbative phase of the fundamental theory, with composite degrees of freedom (pions) which offer an equivalent description of the fundamental Lagrangian, matching the anomaly, in agreement with 't Hooft's principle. Previous studies of the role of the conformal anomaly in cosmology concerning the production of massless gauge particles and the identification of the anomaly pole in the infrared are those of Dolgov \cite{Dolgov:1981nw,Dolgov:1971ri}, while a discussion of the same pole from a dispersive derivation is contained in \cite{Horejsi:1997yn}.

In a related work \cite{Armillis:2009sm}  we have shown that anomaly poles are typical of the perturbative description of the chiral anomaly not just in some special kinematical conditions, for instance in the collinear region, where the coupling of the anomalous axial current to two (on-shell) vector currents (for the $AVV$ diagram) involves a pseudoscalar intermediate state (with a collinear and massless fermion-antifermion pair), but under any kinematical conditions.   
They are the most direct - and probably also the most significant - manifestation of the anomaly in the 
perturbative diagrammatic expansion of the effective action. These features have been briefly reviewed in the previous chapter. On a more speculative side, the interpretation of the pole in terms of composite degrees of freedom could probably have direct physical implications, including the  condensation of the composite fields, very much like Bose Einstein (BE) condensation of the pion field, under the action of gravity. Interestingly, in a recent paper, Sikivie and Yang have pointed out that Peccei-Quinn axions (PQ) may form BE condensates \cite{Sikivie:2009qn}. \\
In this chapter, which parallels a previous 
investigation of the chiral anomaly \cite{Armillis:2009sm}, we study the perturbative structure of the off-shell effective action showing the appearance of similar singularities under general kinematic conditions. Our investigation is a first step towards the computation of the exact effective action describing the coupling of the Standard Model to gravity via the conformal anomaly, that we discuss in the following chapters.  

In our study we follow closely the work of \cite{Giannotti:2008cv}. There the authors have presented a complete off-shell classification of the invariant amplitudes of the relevant correlator responsible for the conformal anomaly, which involves the energy momentum tensor (T) and two vector currents  (J),  $TJJ$, and have thoroughly  investigated it in the QED case, drawing on the analogy with the case of the chiral anomaly. The analysis of
$\cite{Giannotti:2008cv}$ is based on the use of dispersion relations, which are sufficient to identify the anomaly poles of the amplitude from the spectral density of this correlator, but not to characterize completely the off-shell effective action of the theory and the remaining non-conformal contributions, which will be discussed in this chapter. The poles that we extract from the complete effective action include both the usual poles derived from the 
spectral analysis of the diagrams, which are coupled in the infrared (IR) and other extra poles which account for the anomaly but are decoupled in the same limit. These extra poles appear under general kinematic configurations and are typical of the off-shell as well as of the on-shell effective action, both for massive and massless fermions.

We also show, in agreement with those analysis, that the pole terms which contribute to the conformal anomaly are indeed only coupled in the on-shell limit of the external gauge lines, and we identify all the mass corrections to the correlator in the general case.  This analysis is obtained by working out all the relevant kinematical limits of the perturbative corrections. We present the complete anomalous off-shell effective action describing the interaction of gravity with two photons, written in a form in which we separate the nonlocal contribution due to the anomaly pole from the rest of the action (those which are conformally invariant in the massless fermion limit). Away from the conformal limit of the theory we present  a $1/m$ expansion of the effective action as in the Euler-Heisenberg approach. This expansion, naturally, does not convey the presence of nonlocalities in the effective action.

The computation of similar diagrams, for the on-shell photon case, appears in older contributions
by Berends and Gastmans \cite{Berends:1975ah} using dimensional regularization, in their study of the gravitational scattering of photons, and by Milton using Schwinger's methods \cite{Milton:1976jr}. The presence of an anomaly pole in the amplitude has not been investigated nor noticed in these previous analysis, since they do not appear explicitly in their results, nor the $1/m$ expansion of the three form factors of the on-shell vertex, contained in \cite{Berends:1975ah}, allows their identification in the S-matrix elements of the theory.
Two related analysis by Drummond and Hathrell in their investigation of the gravitational contribution to the self-energy of the photon \cite{Drummond:1978hh} and the renormalization of the trace anomaly \cite{Drummond:1979pp} included the same on-shell vertex. Later, the same vertex has provided the ground for several elaborations concerning a possible superluminal behaviour of the photon in the presence of an external gravitational field \cite{Shore:2003zc}.

\section{The effective actions for conformal anomalies and their variational solutions}
In this section we briefly review the topic of the variational solutions of anomalous effective actions, and on the local formulations in terms of auxiliary fields.
One well known result of quantum gravity is that the effective action of the trace anomaly is given by a nonlocal form when expressed in terms of the spacetime metric $g_{\mu\nu}$. This was obtained \cite{Riegert:1984kt} from a variational solution of the equation for the trace anomaly \cite{Duff:1977ay}
\bea
T^\mu_\mu =   b \, F + b^{\prime} \, \left( E - \frac{2}{3} \, \square \, R\right) + b'' \, \square \, R +  c\, F^{\mu \nu} F_{\mu \nu},
\label{P2var2}
\eea
(see also \cite{Deser:1999zv, Deser:1993yx}
for an analysis of the gravitational sector)
which takes in $D=4$ spacetime dimensions the form
\bea
&& \hspace{-1cm}S_{anom}[g,A] =  \nn \\
&& \hspace{-1cm} \frac {1}{8}\int d^4x\sqrt{-g}\int d^4x'\sqrt{-g'} \left(E - \frac{2}{3} \square R\right)_x
 \Delta_4^{-1} (x,x')\left[ 2b\,F
 + b' \left(E - \frac{2}{3} \square R\right) + 2\, c\, F_{\mu\nu}F^{\mu\nu}\right]_{x'}. 
 \label{P2var1}
\eea
Here, the parameters $b$ and $b'$ are the coefficients of the Weyl tensor squared
\beq
F = C_{\lambda\mu\nu\rho}C^{\lambda\mu\nu\rho} = R_{\lambda\mu\nu\rho}R^{\lambda\mu\nu\rho}
-2 R_{\mu\nu}R^{\mu\nu}  + \frac{R^2}{3}
\eeq
 and of the Euler density
\beq
E = ^*\hskip-.2cmR_{\lambda\mu\nu\rho}\,^*\hskip-.1cm R^{\lambda\mu\nu\rho} =
R_{\lambda\mu\nu\rho}R^{\lambda\mu\nu\rho} - 4R_{\mu\nu}R^{\mu\nu}+ R^2
\eeq
 respectively, characterizing the trace anomaly in a general curved spacetime background. Notice that the last term in (\ref{P2var1}) is the contribution generated in the presence of a background gauge field, with coefficient $c$.
For a Dirac fermion in a classical gravitational ($g_{\mu\nu}$) and abelian ($A_{\alpha}$) background, the values of the coefficients are $b = 1/(320\,\pi^2)$, and $b' = - 11/(5760\,\pi^2)$,
and $c= -e^2/(24\,\pi^2)$, with $e$ being the electric charge of the fermion. One crucial feature of this solution is its origin,
which is purely variational. Obtained by Riegert long ago, the action was derived by solving  the variational equation satisfied by the trace of the energy momentum tensor.
$\Delta_4^{-1}(x,x')$ denotes the Green's function of the
conformally covariant differential operator of fourth order, defined by
\beq
\Delta_4 \equiv  \nabla_\mu\left(\nabla^\mu\nabla^\nu + 2 R^{\mu\nu} - \frac{2}{3} R g^{\mu\nu}\right)
\nabla_\nu = \square^2 + 2 R^{\mu\nu}\nabla_\mu\nabla_\nu +\frac{1}{3} (\nabla^\mu R)
\nabla_\mu - \frac{2}{3} R \square\,.
\label{P2Deldef}
\eeq
Given a solution of a variational equation, it is mandatory to check whether the solution is indeed justified by a perturbative computation as we are going to show in the next sections. \\
Notice that an anomaly-induced action does not reproduce the homogeneous contributions to the anomalous trace Ward identity, which require an independent computation in order to be identified. Moreover such an action does not account for all those terms which are responsible for the explicit breaking of scale invariance. For example in the case of the Standard Model such terms are obviously present in the spontaneosly broken phase of the theory and provide important corrections to the anomalous correlators, as we will illustrate in the next chapters.
\subsection{The kinematics of an anomaly pole}
\begin{figure}[t]
\begin{center}
\includegraphics[scale=1.0]{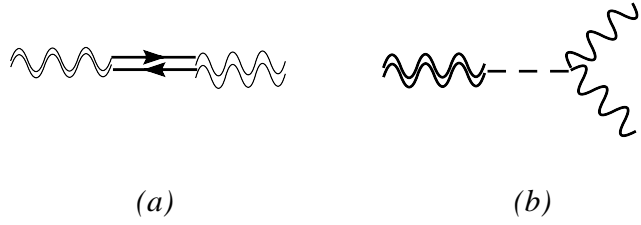}
\caption{\small The diagrams describing the anomaly pole in the dispersive approach. Fig. (a) depicts the singularity of the spectral density $\rho(s)$ as a spacetime process. Fig. (b) describes the anomalous pole part of the interaction.}
\label{P2collinear}
\end{center}
\end{figure}
In our conventions we will denote with $p$ and $q$ the outgoing momenta of the two photons and with $k$ the incoming momentum of the graviton. The kinematical invariant $s\equiv k^2 = (p + q)^2$ denotes the virtual mass of the external graviton line. A computation of the spectral density $\rho(s)$ of the $TJJ$ amplitude shows that this takes the form $\rho(s)\sim \delta(s)$ in the case of a massless fermion. The configuration responsible for the appearance of a pole is illustrated in Fig.~\ref{P2collinear} (a) and it is the analogous of the chiral anomaly situation discussed in the previous chapter. 
It describes the decay of a graviton into two on-shell photons. The decay is mediated by a collinear and on-shell fermion-antifermion pair and can be interpreted as a spacetime process. The corresponding interaction vertex, described as the exchange of a pole, is instead shown in Fig.~\ref{P2collinear} (b). The actual process depicted in Fig.~\ref{P2collinear} (a) is obtained at diagrammatic level by setting on-shell the
fermion/antifermion pair attached to the graviton line. This configuration, present in the spectral density of the diagram only for on-shell photons, generates a pole contribution which can be shown to be coupled in the infrared. This means that if we compute the residue of the amplitude for $s\to 0$ we find that it is non-vanishing. In the general expression of the vertex, a similar configuration is extracted in the high energy limit, not by a dispersive analysis, but by an explicit (off-shell) computation of the diagrams. 
Either for virtual or for real photons, a direct computation of the vertex allows to extract the pole term, without having to rely on a dispersive analysis. As we have already mentioned, this point has been illustrated in detail in the computation of the chiral anomaly vertex in \cite{Armillis:2009sm}, while the analogous analysis of the $TJJ$ vertex for QED will be discussed in this chapter. The identification of this singularity in the case of QCD and in the electroweak sector of the Standard Model is in perfect agreement with the previous results.

\subsection{The single pole from $\Delta_4$ and the local formulation}
In the case of the gravitational effective action, the appearance of the inverse of $\Delta_4$ operator seems to be hard to reconcile with the simpler $1/\square$ interaction which is predicted by the perturbative analysis of the $TJJ$ correlator, which manifests a single anomaly pole. In  \cite{Giannotti:2008cv}, Giannotti and Mottola show step by step how a single pole emerges from this quartic operator, by using the auxiliary field formulation of the same effective action.  Clearly, more computations are needed in order to show that the nonlocal effective action consistently does justice of {\em all} the poles
(of second order and higher) which should be present in the perturbative expansion. Obviously, the perturbative computations - being either based on dispersion theory or on complete evaluations of the vertices, as in our case - become rather hard as we increase the number of
external lines of the corresponding perturbative correlator. For instance, this check becomes almost impossible for correlators of the form $TTT$ or higher, due to the appearance of a very large number of tensor structure in the reduction to scalar form of the tensor Feynman integrals. In the case of $TJJ$ the computation is still manageable, since it does not require Feynman integrals beyond rank-4.

Another important issue concerns the reformulation of this action in such a way that its interactions become local. This important point 
has been analyzed in \cite{Mottola:2006ew}. The authors introduce two scalar fields $\varphi$ and $\psi$ which satisfy fourth order differential equations
\bes\bea
&& \Delta_4\, \varphi = \frac{1}{2} \left(E - \frac{2}{3} \square R\right)\,,
\label{P3auxvarphi}\\
&& \Delta_4\, \psi = \frac{1}{2}C_{\lambda\mu\nu\rho}C^{\lambda\mu\nu\rho} 
+ \frac{c}{2b} F_{\mu\nu}F^{\mu\nu} \,,
\label{P3auxvarpsi}
\eea
\label{P3auxeom}
\ees
\hspace{-.35cm}
which allow to express the nonlocal action in the local form 
\bea
S_{anom} = b' S^{(E)}_{anom} + b S^{(F)}_{anom} + \frac{c}{2} \int\,d^4x\,\sqrt{-g}\ 
F_{\mu\nu}F^{\mu\nu} \varphi\,,
\label{P3allanom}
\eea
where
\bea
&& S^{(E)}_{anom} \equiv \frac{1}{2} \int\,d^4x\,\sqrt{-g}\ \left\{
-\left(\square \varphi\right)^2 + 2\left(R^{\mu\nu} - \frac{R}{3} g^{\mu\nu}\right)(\nabla_{\mu} \varphi)
(\nabla_{\nu} \varphi) + \left(E - \frac{2}{3} \square R\right) \varphi\right\}\,,\nonumber\\
&& S^{(F)}_{anom} \equiv \,\int\,d^4x\,\sqrt{-g}\ \left\{ -\left(\square \varphi\right)
\left(\square \psi\right) + 2\left(R^{\mu\nu} - \frac{R}{3}g^{\mu\nu}\right)(\nabla_{\mu} \varphi)
(\nabla_{\nu} \psi)\right.\nonumber\\
&& \qquad\qquad\qquad + \left.\frac{1}{2} C_{\lambda\mu\nu\rho}C^{\lambda\mu\nu\rho}\varphi +
\frac{1}{2} \left(E - \frac{2}{3} \square R\right) \psi \right\}\,.
\label{P3SEF}
\eea
The equations of motion for $\psi$ and $\varphi$  (\ref{P3auxvarphi}), (\ref{P3auxvarpsi}) can be obtained by varying \ref{P3allanom} with respect to these fields. Notice that in momentum space, these equations, being quartic, show the presence of a double pole in the corresponding energy momentum tensor. This can be defined, as usual, by varying \ref{P3allanom} with respect to the background metric. The reduction of this double pole to a single pole has been discussed in \cite{Giannotti:2008cv,Mottola:2006ew}, using a perturbative formulation of the local action around the flat metric background. In particular the field $\varphi$ has to be assumed of being of first order in the metric fluctuation $h_{\mu\nu}$. \\
With this assumption, 
expanding around flat space, the nonlocal formulation of Riegert's action, as shown in \cite{Giannotti:2008cv,Mottola:2006ew}, can be rewritten in a form which manifests explicitly the appearance of the massless pole
\beq
S_{anom}[g,A]  \rightarrow  -\frac{c}{6}\int d^4x\sqrt{-g}\int d^4x'\sqrt{-g'}\, R_x
\, \square^{-1}_{x,x'}\, [F_{\alpha\beta}F^{\alpha\beta}]_{x'}\,.
\label{P2SSimple}
 \eeq
This is is valid to first order in the fluctuation of the metric around a flat background, denoted as $h_{\mu\nu}$
\beq
g_{\mu\nu}= \eta_{\mu\nu} +\kappa h_{\mu\nu}, \qquad\qquad \kappa=\sqrt{16 \pi G_N}
\eeq
with $G_N$ being the 4-dimensional Newton's constant. \\
Instead, the local formulation in terms of auxiliary fields of this action gives
\beq
S_{anom} [g,A;\varphi,\psi'] =  \int\,d^4x\,\sqrt{-g}
\left[ -\psi'\sq\,\varphi - \frac{R}{3}\, \psi'  + \frac{c}{2} F_{\alpha\beta}F^{\alpha\beta} \varphi\right]\,,
\label{P2effact}
\eeq
where $\phi$ and $\psi$ are the auxiliary scalar fields. They satisfy the equations
\bea
&&\psi' \equiv  b\, \sq\, \psi\,, \label{P2diffeq}\\
&&\square\,\psi' =  \frac{c}{2}\, F_{\alpha\beta}F^{\alpha\beta} \,,\\
&&\square\, \varphi = -\frac{R}{3}\,.
\eea

A perturbative test of the pole structure identified in the anomaly induced action is obtained by a direct computation of the correlator $TJJ$, with the insertion of the EMT on the 2-point function  $(JJ)$ at nonzero momentum transfer. The advantage of a complete computation of the correlator, respect to the variational solution found by inspection, is that it gives the possibility of extracting also the mass corrections to the pole behaviour.

In order to make contact with the $TJJ$ amplitude, one needs the expression of the energy momentum extracted from
(\ref{P2effact}) to leading order in $h_{\mu\nu}$, or, equivalently, from (\ref{P2SSimple}), that can be shown to take the form
\beq
T^{\mu\nu}_{anom}(z) =
\frac{c}{3}  \left(g^{\mu\nu}\sq - \partial^{\mu}\partial^{\nu}\right)_z \int\,d^4x'\, \sq_{z,x'}^{-1}
\left[F_{\alpha\beta}F^{\alpha\beta}\right]_{x'}.
\label{P2Tanom}
\eeq
Notice that $T^{\mu\nu}_{anom}$ is the expression of the energy momentum tensor of the theory in the background of the gravitational and gauge fields. 
Notice also that $\Gamma_{anom}$, derived either by functional differentiations of the generating functional (\ref{P2Tanom}) given by the Riegert's action or from
 the direct perturbative expansion, should nevertheless coincide in the two cases, for the pole term not to be a spurious artifact of the variational solution. In particular, as we will illustrate in great detail in this chapter, a computation performed in QED shows that the pole term extracted from $T_{anom}$ via functional differentiation
\beqa
\Gamma_{anom}^{\mu\nu\alpha\beta}(p,q) &=& i^2 \int\,d^4x\,\int\,d^4y\, e^{ip\cdot x + i q\cdot y}\,\frac{\delta^2 T^{\mu\nu}_{anom}(0)}
{\delta A_{\alpha}(x) A_{\beta}(y)} = -\frac{e^2}{18\pi^2} \frac{1}{k^2} \left(g^{\mu\nu}k^2 - k^{\mu}k^{\nu}\right)u^{\alpha\beta}(p,q)
\nonumber \\
\label{P2Gamanom}
\eeqa
with
\beq
u^{\alpha\beta}(p,q) \equiv (p\cdot q) \,  g^{\alpha\beta} - q^{\alpha} \, p^{\beta}\,,\\
\label{P2utensor}
\eeq
indeed coincides with the result of the perturbative expansion which we are going to present in the following section. Thus, the entire contribution to the anomaly is extracted form $T_{anom}$ as
\beq
g_{\mu\nu}T^{\mu\nu}_{anom} = c F_{\alpha\beta}F^{\alpha\beta} = -\frac{e^2}{24\pi^2} F_{\alpha\beta}F^{\alpha\beta}.
\eeq
As we have already mentioned, the full action (\ref{P2var1}), varied several times with respect to the background metric $g_{\mu\nu}$ and/or
the background gauge fields $A_{\alpha}$ gives those parts of the correlators of higher order, such as
$\lag TTT...JJ\rag$ and $\lag TTT...\rag$, which contribute to the trace anomaly. 

We start with an analysis of the
correlator following an approach which is close to that of \cite{Giannotti:2008cv}. The crucial point of the derivation presented in that chapter is the imposition of the Ward identity for the $TJJ$ correlator 
which allows to eliminate all the Schwinger (gradients) terms which otherwise plague any derivation based on the canonical formalism and are generated by the equal-time commutator of the energy momentum tensor with the vector currents.

\section{The construction of the $TJJ$ correlator}
\begin{figure}[t]
\begin{center}
\includegraphics[scale=0.9]{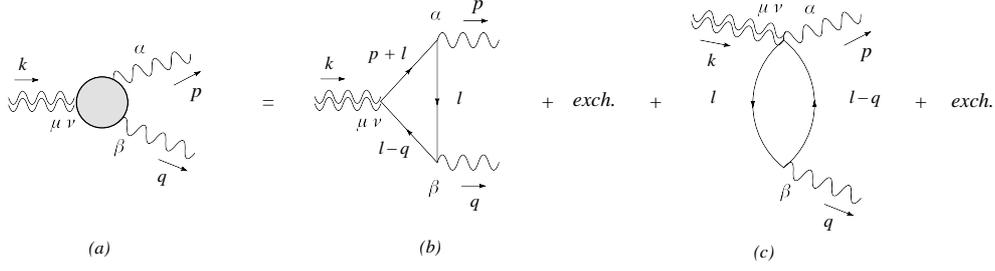}
\caption{\small The complete one-loop vertex (a) given by the sum of the 1PI contributions called $V^{\mu \nu \a \b} (p,q)$ (b) and $W^{\mu \nu \a \b} (p,q)$ (c). }
\label{PNew1vertex}
\end{center}
\end{figure}

We consider the standard QED Lagrangian
\beq
\mathcal{L}=-\frac{1}{4} F_{\mu\nu} F^{\mu\nu} + \, i \,  \bar{\psi} \gamma^\mu(\partial_\mu - i \, e \, A_\mu)\psi
- m \bar{\psi}\psi,
\eeq
with the energy momentum tensor decomposed into the free fermionic part $T_f$,  the interacting fermion-photon part
$T_{fp}$ and the photon contribution $T_{ph}$ which are given by
\beq
T^{\mu\nu}_{f} = -i \bar\psi \gamma^{(\mu}\!\!
\stackrel{\leftrightarrow}{\partial}\!^{\nu)}\psi + g^{\mu\nu}
(i \bar\psi \gamma^{\lambda}\!\!\stackrel{\leftrightarrow}{\partial}\!\!_{\lambda}\psi
- m\bar\psi\psi),
\label{PNew1tfermionic}
\eeq
\beq
T^{\mu\nu}_{fp} = -\, e J^{(\mu}A^{\nu)} + e g^{\mu\nu}J^{\lambda}A_{\lambda}\,,
\eeq
and
\beq
T^{\mu\nu}_{ph} = F^{\mu\lambda}F^{\nu}_{\ \ \lambda} - \frac{1}{4} g^{\mu\nu}
F^{\lambda\rho}F_{\lambda\rho},
\label{PNew1tphoton}
\eeq
where the current is defined as
\beq
J^{\mu}(x) = \bar\psi (x) \gamma^{\mu} \psi (x)\,.\\
\label{PNew1vectorcurrent}
\eeq
In the coupling to gravity of the total energy momentum tensor
\beq
T^{\mu\nu}\equiv T_{f}^{ \mu\nu} +T_{fp}^{ \mu\nu} + T_{ph}^{ \mu\nu}
\eeq
we keep terms linear in the gravitational field, of the form  $h_{\mu\nu} T^{\mu\nu}$, and we have
introduced some standard notation for the symmetrization of the tensor indices and left-right derivatives $H^{(\mu\nu)} \equiv (H^{\mu\nu} + H^{\nu\mu})/2$ and
$\stackrel{\leftrightarrow}{\partial}\!\!_{\mu} \equiv
(\stackrel{\rightarrow}{\partial}\!\!_{\mu} - \stackrel{\leftarrow}{\partial}\!\!_{\mu})/2$.
It is also convenient to introduce a partial energy momentum tensor $T_p$, corresponding to the sum of the Dirac and interaction terms
\beq
T_p^{\mu\nu}\equiv T_{f}^{ \mu\nu} +T_{fp}^{ \mu\nu}
\eeq
which satisfies the inhomogeneous equation
\beq
\partial_\nu T_p^{\mu\nu}= -\partial_\nu T_{ph}^{ \mu\nu}.
\eeq
Using the equations of motion for the electromagnetic (e.m.) field $\partial_\nu F^{\mu\nu}=J^{\mu}$, the inhomogeneous
equation becomes
\beq
\partial_\nu T_p^{\mu\nu}= F^{\mu\lambda} J_\lambda.
\eeq
There are two ways to identify the contributions of $T^{\mu\nu}$ and $T^{ \mu\nu}_p$  in the perturbative expansion of the effective action.
In the formalism of the background fields, the $T_pJJ$ correlator can be extracted from the defining functional integral
\beqa
\langle T_p^{\mu\nu}(z)\rangle_A &\equiv& \int D\psi D\bar{\psi} \,\,T^{\mu\nu}_p (z) \,\,e^{i \int d^4 x \mathcal{L} + \int J\cdot A(x) d^4 x}
= \langle T^{\mu\nu}_p \, e^{i \int d^4 x \, J\cdot A(x)}\rangle
\eeqa
expanded through second order in the external field $A$. The relevant terms in this expansion are explicitly given by
\beq
\langle T_p^{\mu\nu}(z)\rangle_A = \frac{1}{2!}\langle T_{f}^{\mu\nu}(z) (J\cdot A)(J\cdot A) \rangle +
\langle T_{fp}^{\mu\nu}(J\cdot A)\rangle + ... \, ,
\eeq
with  $(J\cdot A)\equiv \int d^4 x J\cdot A(x)$.
The corresponding diagrams are extracted via two functional derivatives respect to the background field $A_\mu$
and are given by
\beq
\Gamma^{\mu\nu\alpha\beta} (z; x, y) \equiv \frac{ \delta^2 \lag T_p^{\mu\nu} (z) \rag_A}
{\delta A_{\alpha}(x)\delta A_{\beta}(y)} \bigg\vert_{A=0}
= V^{\mu\nu\alpha\beta} + W^{\mu\nu\alpha\beta}
\eeq
\beq
 V^{\mu\nu\alpha\beta}=(i \, e )^2 \, \lag T_{f}^{\mu\nu} (z) J^{\alpha} (x) J^{\beta} (y) \rag_{A=0}
\eeq
\beqa
W^{\mu\nu\alpha\beta} &=&\frac{ \delta^2 \lag T_{fp}^{\mu\nu} (z) (J\cdot A) \rag}
{\delta A_{\alpha}(x)\delta A_{\beta}(y)} \bigg\vert_{A=0} \nonumber \\
&=& \delta^4(x-z)g^{\alpha (\mu} \Pi^{\nu )\beta}(z, y)
+ \delta^4 (y-z)g^{\beta(\mu} \Pi^{\nu )\alpha}(z, x)
- g^{\mu\nu}[\delta^4(x-z) -  \delta^4(y-z) ]\Pi^{\alpha\beta}(x, y). \nonumber \\
\label{PNew1tjjcorrelator}
\eea
These two contributions are of $O(e^2)$.
Alternatively, one can directly compute the matrix element
\beq
\mathcal{M}^{\mu\nu} = \langle 0| T_p^{\mu\nu}(z) \int d^4 w d^4 w' J\cdot A(w) J\cdot A(w')|\gamma \gamma \rangle,
 \eeq
 which generates the diagrams (b) and (c) shown in Fig.\ref{PNew1vertex}, respectively called the ``triangle" and the
 ``t-p-bubble" (``t-" stays for tensor), together with the two ones obtained for the exchange of
 $p$ with $q$ and $\alpha$ with $\beta$.
 The conformal anomaly appears in the perturbative expansion of $T_p$ and involves these four diagrams.

Instead, the lowest order contribution is obtained, in the background field formalism, from Maxwell's e.m. tensor, and is given by
\beqa
S^{\mu\nu\alpha\beta} &=&\frac{ \delta^2 \lag T_{ph}^{\mu\nu} (z) \rag}
{\delta A_{\alpha}(x)\delta A_{\beta}(y)} \bigg\vert_{A=0}.
\label{PNew1config}
\eeqa
Using (\ref{PNew1config}) we easily obtain
\beqa
S_{\mu\nu\alpha\beta}(z,x,y) &=& 2 g_{\alpha\beta} \partial_{\left( \right.\mu} \delta_{x z}\partial_{\nu\left.\right)}\delta_{yz} - 2 g_{\beta\left(\mu\right.}\partial_{\nu\left.\right)}\delta_{x z}\partial_\alpha\delta_{y z}  -
2 g_{\alpha\left(\nu\right.}\partial_{\left.\mu\right)}\delta_{yz}\partial_\beta\delta_{x z}\nonumber \\
&&+ g_{\alpha\mu}g_{\beta\nu} \partial_\lambda \delta_{yz} \partial^\lambda \delta_{x z}
 + g_{\alpha\nu}g_{\beta\mu} \partial_\lambda \delta_{yz} \partial^\lambda \delta_{x z}
 + g_{\mu\nu} \partial_{\beta}\delta_{x z} \partial_\alpha \delta_{y z}  
 - \partial_\rho \delta_{y z} \partial_\rho \delta_{x z} g_{\alpha\beta} g_{\mu\nu}
\nonumber \\
\eeqa
where $\partial_\mu\delta_{x z}\equiv \partial/\partial {x^\mu} \delta(x-z)$ and so on. In momentum space
this lowest order vertex is given by
\beqa
S^{\mu\nu\alpha\beta} &=&  
\big(p^{\mu} q^{\nu} + p^{\nu} q^{\mu}\big) \, g^{\alpha\beta}
+ p\cdot q\, \big(g^{\alpha\nu} g^{\beta\mu} + g^{\alpha\mu} g^{\beta\nu}\big) 
- g^{\mu\nu} \, (p\cdot q\,g^{\alpha\beta} - q^{\alpha} p^{\beta} ) \nn \\
&& - \, \big(g^{\beta\nu} p^{\mu} + g^{\beta\mu} p^{\nu} \big) \,q^{\alpha}
- \big (g^{\alpha\nu} q^{\mu} + g^{\alpha\mu} q^{\nu }\big) \, p^{\beta}.
\eeqa

Coming back to the one-loop correlator, the corresponding vertices which appear respectively in the triangle diagram and on the t-bubble
at $O(e^2)$ are given by
\beqa
 V^{\prime \, \mu\nu}(k_1, k_2)&=&\frac{1}{4} \left[\gamma^\mu (k_1 + k_2)^\nu
+\gamma^\nu (k_1 + k_2)^\mu \right] - \frac{1}{2} g^{\mu \nu}
[\gamma^{\lambda}(k_1 + k_2)_{\lambda} - 2 m]   \,,\\
 W^{\prime\,\mu\nu\alpha} &=& -\frac{1}{2} (\gamma^\mu g^{\nu\alpha}
+\gamma^\nu g^{\mu\alpha}) + g^{\mu \nu}\gamma^{\alpha}, 
\eeqa
where $k_1 (k_2)$ is outcoming (incoming).
Using the two vertices $V^{\prime \, \mu\nu}(k_1, k_2)$ and $W^{\prime\,\mu\nu\alpha} $ we obtain for the diagrams (b) and (c) of Fig.\ref{PNew1vertex}
\bea
&&V^{\mu\nu\alpha\beta}(p,q)=
- (-ie)^2 i^3  \int \frac{d^4 l}{(2 \pi)^4} 
\, \frac{{\rm tr}
\left\{V^{\prime\,\mu\nu}(l+p,l-q) (\lsl -\qsl +m)\gamma^{\beta}\,
(\lsl + m)\, \gamma^{\alpha}(\lsl +\psl + m)  \right\}}
{[l^2 - m^2] \, [(l-q)^2 - m^2] \, [(l+p)^2 - m^2] }\,,
\nn \\
\label{PNew1V}
\eea
and
\bea
W^{\mu\nu\alpha\beta}(p,q)   &=&
- (i e^2) i^2 \,  
\int \frac{d^4 l}{(2 \pi)^4} \, \frac{{\rm tr}\left\{W^{\prime \, \mu\nu\alpha}
\,(\lsl + m)\gamma^{\beta} (\lsl -\qsl +m) \right\}} {[l^2 - m^2][(l-q)^2 - m^2]},
\label{PNew1W}
\eea
so that the one-loop amplitude in Fig.~\ref{PNew1vertex} results
\bea
\Gamma^{\mu\nu\alpha\beta} (p,q) =
V^{\mu\nu\alpha\beta}(p,q) +\, V^{\mu\nu\b\a}(q,p)
+ W^{\mu\nu\alpha\beta}(p,q)+\, W^{\mu\nu\b\a}(q,p).
\label{PNew1Gamma}
\eea
The bare Ward identity which allows to define the divergent amplitudes that contribute to the anomaly in $\Gamma$ in terms of the remaining finite ones is obtained by re-expressing the classical equation
\beq
\partial_\nu T_{ph}^{\mu\nu} = - F^{\mu\nu} J_\nu
\eeq
as an equation of generating functionals in the background electromagnetic field
\beq
\partial_\nu \langle T_{ph}^{\mu\nu} \rangle_A= - F^{\mu\nu}\langle J_\nu\rangle_A,
\label{PNew1WIeq}
\eeq
which can be expanded perturbatively as
\beq
 \partial_\nu \langle T_{ph}^{\mu\nu} \rangle_A= - F^{\mu\nu}\langle J_\nu \int d^4 w (i e)J\cdot A (w)\rangle_
+ \,  ... \, .
\label{PNew1firstterm}
 \eeq
 Notice that we have omitted the first term in the Dyson's series of $\langle J_\nu\rangle_A$, shown on the r.h.s of
 (\ref{PNew1firstterm}) since $\langle J_\nu\rangle=0$.
The bare Ward identity then takes the form
\beq
 \partial_\nu  \Gamma^{\mu\nu\alpha\beta} =  \frac{ \delta^2 \left( F^{\mu\lambda}(z) \lag J_\lambda (z)\rag_A \right)}
{\delta A_{\alpha}(x)\delta A_{\beta}(y)} \bigg\vert_{A=0}
\eeq
which takes contribution only from the first term on the r.h.s of Eq. (\ref{PNew1firstterm}). This relation can be written in momentum space. For this purpose we use the definition of the vacuum polarization
\beq
\Pi^{\alpha\beta}(x,y) \equiv  - i e^2 \langle J_\alpha(x) J_\beta(y)\rangle,
\eeq
or
\bea
\Pi^{\alpha\beta} (p) &=&  - i^2 \, (-ie)^2 
\int \frac{d^4 l}{(2\pi)^4} \frac{{\rm tr}\left\{\gamma^{\alpha}
\,(\lsl + m)\gamma^{\beta} (\lsl + \psl +m) \right\}} {[l^2 - m^2] \, [(l+p)^2 - m^2]} 
=    (p^2 g^{\alpha\beta} - p^{\alpha}p^{\beta}) \,\Pi (p^2,m^2)
\label{PNew1polmom}
\eea
with
\beq
\Pi (p^2,m^2) = \frac{ e^2}{36 \, \pi^2  \, p^2 }
\biggl[ 6  \, \mathcal A_0(m^2) +  p^2 - 6 \, m^2
   - \, 3 \, \mathcal B_0(p^2,m^2) \left(2 m^2 + p^2  \right)\biggr],
\label{PNew1vacuumpol}
\eeq
which obviously satisfies the Ward identity $p_{\alpha} \, \Pi^{\alpha\beta} (p) =0$. The expressions of the $\mathcal A_0$ and $\mathcal B_0$ scalar integrals are given in Appendix \ref{P2scalars}.

Using these definitions, the unrenormalized Ward identity which allows to completely characterize the form of the correlator in momentum space becomes
\bea
k_\nu\,\Gamma^{\mu\nu\alpha\beta}(p,q)&=&  \,\,\,\,
\left(q^\mu p^\alpha p^\beta -q^\mu g^{\alpha\beta} p^2 +g^{\mu\beta}q^\alpha p^2
-g^{\mu\beta}p^\alpha p\cdot q \right)   \Pi(p^2) \nonumber\\
&&+ \left(p^\mu q^\alpha q^\beta - p^\mu g^{\alpha\beta} q^2
+g^{\mu\alpha}p^\beta q^2 - g^{\mu\alpha}q^\beta p\cdot q\right) \Pi(q^2) \,.
\label{PNew1TWard}
\eea

\subsection{Tensor expansion and invariant amplitudes}
\label{PNew1tensorexpansion}
\begin{table}
$$
\begin{array}
{|c 
 | 
 c 
 |
c 
 |
c 
|
c
|
c 
|}
\hline
& & & & & \\[-.5cm]
\begin{array}[t]{c}
p^{\mu} p^{\nu} p^{\alpha} p^{\beta}\\
q^{\mu} q^{\nu} q^{\alpha} q^{\beta}
\end{array}
&
\begin{array}[t]{c}
p^{\mu} p^{\nu} p^{\alpha} q^{\beta}\\
p^{\mu} p^{\nu} q^{\alpha} p^{\beta}\\
p^{\mu} q^{\nu} p^{\alpha} p^{\beta}\\
q^{\mu} p^{\nu} p^{\alpha} p^{\beta}
\end{array}
&
\begin{array}[t]{c}
p^{\mu} p^{\nu} q^{\alpha} q^{\beta}\\
p^{\mu} q^{\nu} p^{\alpha} q^{\beta}\\
q^{\mu} p^{\nu} p^{\alpha} q^{\beta}
\end{array}
&
\begin{array}[t]{c}
p^{\mu} q^{\nu} q^{\alpha} p^{\beta}\\
q^{\mu} p^{\nu} q^{\alpha} p^{\beta}\\
q^{\mu} q^{\nu} p^{\alpha} p^{\beta}
\end{array}
&
\begin{array}[t]{c}
p^{\mu} q^{\nu} q^{\alpha} q^{\beta}\\
q^{\mu} p^{\nu} q^{\alpha} q^{\beta}\\
q^{\mu} q^{\nu} p^{\alpha} q^{\beta}\\
q^{\mu} q^{\nu} q^{\alpha} p^{\beta}
\end{array}
&
\begin{array}[t]{c}
g^{\mu\nu}g^{\alpha\beta}\\
g^{\alpha\mu}g^{\beta\nu}\\
g^{\alpha\nu}g^{\beta\mu}
\end{array}
\\[2.1cm]
\hline
& & & & & \\[-.5cm]
\begin{array}{c}
p^{\mu} p^{\nu} g^{\alpha\beta}\\
p^{\mu} q^{\nu} g^{\alpha\beta}\\
q^{\mu} p^{\nu} g^{\alpha\beta}\\
q^{\mu} q^{\nu} g^{\alpha\beta}
\end{array}
&
\begin{array}{c}
p^{\beta} p^{\nu} g^{\alpha\mu}\\
p^{\beta} q^{\nu} g^{\alpha\mu}\\
q^{\beta} p^{\nu} g^{\alpha\mu}\\
q^{\beta} q^{\nu} g^{\alpha\mu}
\end{array}
&
\begin{array}{c}
p^{\beta} p^{\mu} g^{\alpha\nu}\\
p^{\beta} q^{\mu} g^{\alpha\nu}\\
q^{\beta} p^{\mu} g^{\alpha\nu}\\
q^{\beta} q^{\mu} g^{\alpha\nu}
\end{array}
&
\begin{array}{c}
p^{\alpha} p^{\nu} g^{\beta\mu}\\
p^{\alpha} q^{\nu} g^{\beta\mu}\\
q^{\alpha} p^{\nu} g^{\beta\mu}\\
q^{\alpha} q^{\nu} g^{\beta\mu}
\end{array}
&
\begin{array}{c}
p^{\mu} p^{\alpha} g^{\beta\nu}\\
p^{\mu} q^{\alpha} g^{\beta\nu}\\
q^{\mu} p^{\alpha} g^{\beta\nu}\\
q^{\mu} q^{\alpha} g^{\beta\nu}
\end{array}
&
\begin{array}{c}
p^{\alpha} p^{\beta} g^{\mu\nu}\\
p^{\alpha} q^{\beta} g^{\mu\nu}\\
q^{\alpha} p^{\beta} g^{\mu\nu}\\
q^{\alpha} q^{\beta} g^{\mu\nu}
\end{array}
\\ [1.2cm]\hline
\label{PNew1WI}
\end{array}
$$
\caption{\small The 43 tensor monomials built up from the metric tensor and the two independent momenta $p$ and $q$ into which a general fourth rank tensor can be expanded.
\label{PNew1monomials}}
\end{table}
The full one-loop amplitude $\Gamma^{\mu\nu\alpha\beta} (p,q)$ can be expanded on the basis provided by the 43 monomial tensors listed in Tab.\ref{PNew1monomials}
\bea
\Gamma^{\mu\nu\alpha\beta} (p,q) = \, \sum_{i=1}^{43} \, A_i(k^2,p^2,q^2) \, l_i^{\mu \nu \a \b} (p,q).
\eea
Since the amplitude $\Gamma^{\mu\nu\alpha\beta} (p,q) $ has total mass dimension equal to $2$ it is obvious that not all the coefficients $A_i$ are convergent. They can be divided into $3$ groups:
\begin{itemize}
\item[a)]  $A_1 \leq A_i \leq A_{16}$ - multiplied by a product of four momenta, they have mass dimension $-2$ and therefore are UV finite;
\item [b)] $A_{17} \leq A_i \leq A_{19}$ - these have mass dimension $2$ since the four Lorentz indices of the amplitude are carried by two metric tensors
\item [c)] $A_{20} \leq A_i \leq A_{43}$ - they appear next to a metric tensor and two momenta, have mass dimension $0$ and are divergent.
\end{itemize}
The way in which the $43$ invariant amplitudes will be managed in order to reduce them to a smaller set of independent amplitudes is the subject of this section.
The reduction is accomplished in 4 different steps and has as a guiding principle the elimination of the divergent amplitudes $A_i$ in terms of the convergent ones after imposing some conditions on the whole amplitude.
We require
\begin{itemize}
\item[a)] the symmetry in the two indices $	\mu$ and $\nu$  of the symmetric energy-momentum tensor $T^{\mu\nu}$;
\item[b)] the conservation of the two vector currents on $p^{\a}$ and $q^{\b}$;
\item[c)] the Ward identity on the vertex with the incoming momentum $k$ defined above in Eq. (\ref{PNew1WI}).
\end{itemize}
Condition a) becomes
\bea
\Gamma^{\mu\nu\alpha\beta} (p,q) = \Gamma^{\nu\mu\alpha\beta} (p,q),
\eea
giving a linear system of $43$ equations; $15$ of them being identically satisfied because the tensorial structures are already symmetric in the exchange of $\mu$ and $\nu$, while the remaining $14$ conditions are
\bea
&& \hspace{0.2cm} A_5=A_6, \hspace{1.7cm} A_8=A_9,  \hspace{1.7cm} A_{10}=A_{11},
 \hspace{1.5cm}  A_{13}=A_{14},  \hspace{1.5cm} A_{18}=A_{19}, \nn \\
&& A_{21}=A_{22},   \hspace{1.5cm} A_{24}=A_{28},  \hspace{1.5cm} A_{25}=A_{29},
 \hspace{1.5cm} A_{26} =  A_{30},  \hspace{1.5cm} A_{27} = A_{31}, \nn \\
&& A_{32} = A_{36},  \hspace{1.5cm}  A_{34} = A_{37},  \hspace{1.5cm} A_{33} = A_{38},
 \hspace{1.5cm}  A_{35} =  A_{39},
 \label{PNew1conda}
\eea
where all $A_i$ are thought as functions of the invariants $k^2, p^2, q^2$.
After substituting (\ref{PNew1conda}) into $\Gamma^{\mu\nu\alpha\beta} (p,q)$ the $43$ invariant tensors of the decomposition are multiplied by only $29$ invariant amplitudes. Condition b), which is vector current conservation on the two vertices with indices $\a$ and $\b,$ allows to re-express some divergent $A_i$ in terms of other finite ones
\bea
p_\a \, \Gamma^{\mu\nu\alpha\beta} (p,q) = q_{\b} \, \Gamma^{\mu\nu\alpha\beta} (p,q) =0.
\label{PNew1WI12}
\eea
This constraint generates two sets of $14$ independent tensor structures each. The first Ward identity leads to a linear system composed of $10$ equations
\bea
p_\a \, \Gamma^{\mu\nu\alpha\beta} (p,q) = 0 	\Rightarrow
\left \{ \begin {array}{l}
A_{19} + A_{36} \, p \cdot p + A_{37} \, p \cdot q = 0, \\
A_{38} \,  p \cdot p+A_{39}\,  p \cdot q=0, \\
A_{17} +A_{40} \,  p \cdot p+A_{42}  \,   p \cdot q=0,\\
A_{41} \, p \cdot p+A_{43}  \,   p \cdot q=0, \\
A_{20}+2 A_{28}+A_{1} \, p \cdot p+A_{4}  \,  p \cdot q=0, \\
2 \,  A_{30}+A_{3} \,  p \cdot p+A_{7}  \,  p \cdot q=0, \\
A_{22}+A_{29}+A_{6} \,  p \cdot p+A_{11} \,   p \cdot q=0, \\
A_{31}+A_{9}\,  p \cdot p+A_{14}  \,  p \cdot q=0,\\
A_{23}+A_{12} \, p \cdot p+A_{16} \,   p \cdot q=0, \\
A_{15} \, p \cdot p+A_{2} \, p \cdot q=0;
\end{array} \right.
\eea
 we choose to solve it for the set $\{A_{15}, A_{17}, A_{19}, A_{23}, A_{28}, A_{29}, A_{30}, A_{31}, A_{39}, A_{43}\}$ in which only the first one is convergent and the others are UV divergent. The set would not include
 all the divergent $A_i$ since in the last equations appear two convergent coefficients, $A_{15} $ and $A_{2}$. \\
In an analogous way we go on with the second Ward identity (WI) after replacing the solution of the previous system in the original amplitude. The new one is indicated by $\Gamma^{\mu\nu\alpha\beta}_{b} (p,q)$, where the subscript $b$ is there to indicate that we have applied condition $b)$ on $\Gamma$. The constraint gives
\bea
q_\b \, \Gamma^{\mu\nu\alpha\beta}_b (p,q) = 0 	\Rightarrow
\left \{ \begin {array}{l}
 A_{40}  \, p \cdot q+A_{41} \, q \cdot q \,=0, \\
 A_{1} \, p \cdot q+A_{3}  \, q \cdot q \,=0, \\
 A_{20}+A_{4} \, p \cdot q+A_{7} \, q \cdot q \,=0, \\
 A_{36}+A_{6} \, p \cdot q+A_{9}  \, q \cdot q \,=0,\\
 A_{22}+A_{37}+A_{11} \, p \cdot q+A_{14}  \, q \cdot q \,=0, \\
 2 A_{38}+A_{12} \, p \cdot q- A_{2}  \frac{ \, p \cdot q \, q\cdot q} {p \cdot p }=0.
 \end{array} \right.
\eea
We solve these equations determining the amplitudes in the set $\{ A_1, A_{20}, A_{22}, A_{36}, A_{38}, A_{40}\}$ in terms of the remaining ones.
The manipulations above have allowed a reduction of the number of invariant amplitudes from the initial $43$ to $13$ using the  $\{\mu,\nu\}$ symmetry ($14$ equations), the first WI on $p_\a$ ($10$ equations) and the second WI on $q_\b$ ($6$ equations). The surviving invariant amplitudes in which the amplitude $\Gamma^{\mu\nu\alpha\beta}_c (p,q)$ can be expanded using the form factors are \\
$\{ A_2, A_3, A_4, A_6, A_7, A_9, A_{11}, A_{12}, A_{14}, A_{16}, A_{37}, A_{41}, A_{42} \}$. This set still 
contains $3$ divergent amplitudes, $\{A_{37}, A_{41}, A_{42} \}$. The amplitude $\Gamma^{\mu\nu\alpha\beta}_c (p,q) $ is indeed ill-defined until we impose on it condition c), that is Eq. (\ref{PNew1TWard}). This condition gives
\bea
\textrm{Eq.} (\ref{PNew1TWard})	\Rightarrow
\left \{ \begin {array}{l}
 \vspace{0.2cm} 
 -A_{3} \, \,  \Big[1 + \frac{ \, p \cdot p \,  }{2 \, p \cdot q \, }\Big] \, \,  +A_{6}  +\frac{1}{2}  \,  A_{7}   \,  -A_{9} \,  -  \,  \frac{A_{41}    }{\, p \cdot q \, } =0, \\
 \vspace{0.2cm}
A_{37}+A_{42}+A_{4}\, \,  [ p \cdot p \, + p \cdot q] \,\,  + \,  A_{11}   \, p \cdot q \, +\frac{1}{2}  \,   \, A_{7} \,  \, q \cdot q \, + \nn \\
\vspace{0.2cm} 
\hspace{3cm} 
 +A_{11}  \,   \, q \cdot q \, +\frac{1}{2} \, A_3 \, \, \frac{ \, p \cdot p \,  \, q \cdot q \, }{   \, p \cdot q \, }=0, \\
\vspace{0.2cm} 
\frac{1}{2} A_2  \frac{\, p \cdot q   \, q \cdot q } {  \, p \cdot p}
- A_{41}  \frac{ \, p \cdot p + \, q \cdot q }{ \, p \cdot q} -
\frac{1}{2} A_3  \, p  \cdot p + A_7 ( \, p \cdot p  + \frac{1}{2} \, p \cdot q ) 
+A_6  \, p  \cdot q  \nn \\
\vspace{0.2cm} 
\hspace{3cm} 
+ A_{12} (\frac{1}{2} \, p  \cdot q +  q \cdot q ) +A_{14}
 Ê( \, p  \cdot q  +2 \, q \cdot q ) +
2 Ê A_{37} -\Pi (p^2)-\Pi (q^2)=0
  \end {array} \right.
\eea
After these steps we end up with an expression for $\Gamma$ written in terms of only $10$  invariant amplitudes, that are $\mathcal{X}\equiv\{A_2, A_3,A_4,A_6,A_7,A_9,A_{11},A_{12},A_{14},A_{16} \}$,  significantly reduced respect to the original $43$, see \cite{Armillis:2009pq} for more details. Further reductions are possible (down to $8$ independent invariant amplitudes), however, since these reductions just add to the complexity of the related tensor structures, it is convenient to select an appropriate set of reducible (but finite) components characterized by a simpler tensor structure and present the result in that form. The 13 amplitudes introduced in the final decomposition are, in this respect, a good choice since the corresponding tensor structures are rather simple. These tensors are combinations of the $43$ monomials listed in Tab.\ref{PNew1monomials}.

The set $\mathcal{X}$ is very useful for the actual computation of the tensor integrals and for the study of their reduction to scalar form. To compare with the previous study of Giannotti and Mottola \cite{Giannotti:2008cv} we have mapped the computation of the components of the set $\mathcal{X}$ into their structures $F_i$ $(i=1,2,..,13) $. Also in this case, the truly independent amplitudes are 8. One can extract, out of the 13 reducible amplitudes, a consistent subset of 8 invariant amplitudes. The remaining amplitudes in the 13 tensor structures are, in principle, obtainable from this subset.

\subsection{Reorganization of the amplitude}
Before obtaining the mapping between the amplitudes in $\mathcal{X}$ and the structures $F_i$, we briefly describe the tensor decomposition introduced in \cite{Giannotti:2008cv}.
We define the rank-2 tensors
\bea
&&u^{\alpha\beta}(p,q) \equiv (p\cdot q) \,  g^{\alpha\beta} - q^{\alpha} \, p^{\beta}\,,\nn \\
&&w^{\alpha\beta}(p,q) \equiv p^2 \, q^2 \, g^{\alpha\beta} + (p\cdot q) \, p^{\alpha} \, q^{\beta}
- q^2 \,  p^{\alpha} \, p^{\beta} - p^2 \, q^{\alpha} \, q^{\beta}\,,
\label{PNew1uwdef}
\eea
which are Bose symmetric,
\bea
&&u^{\alpha\beta}(p,q) = u^{\beta\alpha}(q,p)\,,\nn \\
&&w^{\alpha\beta}(p,q) = w^{\beta\alpha}(q,p)\,,
\eea
and conserve vector currents,
\bea
&&p_{\alpha} \, u^{\alpha\beta}(p,q)  = q_{\beta} \, u^{\alpha\beta}(p,q) = 0\,,\nn \\
&&p_{\alpha} \, w^{\alpha\beta}(p,q)  = q_{\beta} \, w^{\alpha\beta}(p,q) = 0\,.
\eea
\begin{table}
$$
\begin{array}{|c|c|}\hline
i & t_i^{\mu\nu\alpha\beta}(p,q)\\ \hline\hline
1 &
\left(k^2 g^{\mu\nu} - k^{\mu } k^{\nu}\right) u^{\alpha\beta}(p.q)\\ \hline
2 &
\left(k^2g^{\mu\nu} - k^{\mu} k^{\nu}\right) w^{\alpha\beta}(p.q)  \\ \hline
3 & \left(p^2 g^{\mu\nu} - 4 p^{\mu}  p^{\nu}\right)
u^{\alpha\beta}(p.q)\\ \hline
4 & \left(p^2 g^{\mu\nu} - 4 p^{\mu} p^{\nu}\right)
w^{\alpha\beta}(p.q)\\ \hline
5 & \left(q^2 g^{\mu\nu} - 4 q^{\mu} q^{\nu}\right)
u^{\alpha\beta}(p.q)\\ \hline
6 & \left(q^2 g^{\mu\nu} - 4 q^{\mu} q^{\nu}\right)
w^{\alpha\beta}(p.q) \\ \hline
7 & \left[p\cdot q\, g^{\mu\nu}
-2 (q^{\mu} p^{\nu} + p^{\mu} q^{\nu})\right] u^{\alpha\beta}(p.q) \\ \hline
8 & \left[p\cdot q\, g^{\mu\nu}
-2 (q^{\mu} p^{\nu} + p^{\mu} q^{\nu})\right] w^{\alpha\beta}(p.q)\\ \hline
9 & \left(p\cdot q \,p^{\alpha}  - p^2 q^{\alpha}\right)
\big[p^{\beta} \left(q^{\mu} p^{\nu} + p^{\mu} q^{\nu} \right) - p\cdot q\,
(g^{\beta\nu} p^{\mu} + g^{\beta\mu} p^{\nu})\big]  \\ \hline
10 & \big(p\cdot q \,q^{\beta} - q^2 p^{\beta}\big)\,
\big[q^{\alpha} \left(q^{\mu} p^{\nu} + p^{\mu} q^{\nu} \right) - p\cdot q\,
(g^{\alpha\nu} q^{\mu} + g^{\alpha\mu} q^{\nu})\big]  \\ \hline
11 & \left(p\cdot q \,p^{\alpha} - p^2 q^{\alpha}\right)
\big[2\, q^{\beta} q^{\mu} q^{\nu} - q^2 (g^{\beta\nu} q^ {\mu}
+ g^{\beta\mu} q^{\nu})\big]  \\ \hline
12 & \big(p\cdot q \,q^{\beta} - q^2 p^{\beta}\big)\,
\big[2 \, p^{\alpha} p^{\mu} p^{\nu} - p^2 (g^{\alpha\nu} p^ {\mu}
+ g^{\alpha\mu} p^{\nu})\big] \\ \hline
13 & \big(p^{\mu} q^{\nu} + p^{\nu} q^{\mu}\big)g^{\alpha\beta}
+ p\cdot q\, \big(g^{\alpha\nu} g^{\beta\mu}
+ g^{\alpha\mu} g^{\beta\nu}\big) - g^{\mu\nu} u^{\alpha\beta} \\
& -\big(g^{\beta\nu} p^{\mu}
+ g^{\beta\mu} p^{\nu}\big)q^{\alpha}
- \big (g^{\alpha\nu} q^{\mu}
+ g^{\alpha\mu} q^{\nu }\big)p^{\beta}  \\ \hline
\end{array}
$$
\caption{Basis of 13 fourth rank tensors satisfying the vector current conservation on the external lines with momenta $p$ and $q$. \label{PNew1genbasis}}
\end{table}
These two tensors are used to build the  set of $13$ tensors catalogued in Table \ref{PNew1genbasis}. They are linearly independent for generic $k^2, p^2, q^2$
different from zero. Five of the $13$ tensors are Bose symmetric, namely,
\bea
t_i^{\mu\nu\alpha\beta}(p,q) = t_i^{\mu\nu\beta\alpha}(q,p)\,,\qquad i=1,2,7,8,13\,,
\eea
while the remaining eight tensors form four pairs which are overall related by Bose symmetry
\bea
&&t_3^{\mu\nu\alpha\beta}(p,q) = t_5^{\mu\nu\beta\alpha}(q,p)\,,\\
&&t_4^{\mu\nu\alpha\beta}(p,q) = t_6^{\mu\nu\beta\alpha}(q,p)\,,\\
&&t_9^{\mu\nu\alpha\beta}(p,q) = t_{10}^{\mu\nu\beta\alpha}(q,p)\,,\\
&&t_{11}^{\mu\nu\alpha\beta}(p,q) = t_{12}^{\mu\nu\beta\alpha}(q,p)\,.
\label{PNew1tpairs}
\eea
The amplitude in (\ref{PNew1Gamma}) can be expanded in this basis as
\bea
\Gamma^{\mu\nu\alpha\beta}(p,q) =  \, \sum_{i=1}^{13} F_i (s; s_1, s_2,m^2)\ t_i^{\mu\nu\alpha\beta}(p,q)\,,
\label{PNew1Gamt}
\eea
where the invariant amplitudes $F_i$ are functions of the kinematical invariants $s=k^2=(p+q)^2$, $s_1=p^2$, $s_2=q^2$ and of the internal mass $m$. In \cite{Giannotti:2008cv} the authors use the
Feynman parameterization and momentum shifts in order to identify the expressions of these amplitudes in terms of parametric integrals, which was the approach followed also by Rosenberg in his original identification of the 6 invariant amplitudes of the $AVV$ anomaly diagram.
If we choose to reorganize all the monomials into the simpler set of $13$ tensor groups shown in Tab.\ref{PNew1genbasis}, then we need to map the $A_i$  in $\mathcal{\chi}$ and the $F_i$'s. Due to the lengthy expressions we do not give the complete mapping here but we refer to \cite{Armillis:2009pq} for more details.

We have shown how to obtain the $13$ $F_i$' s, starting from our derivation of the one-loop full amplitude $\Gamma^{\mu\nu\a\b} (p,q)$ leding to the ten invariant amplitudes of the set $\mathcal{X}$. Since we know the analytical expression of the $A_i$ involved, we can go one step further and give all the $F_i$' s in their analytical form in the most general kinematical configuration.
These have been obtained by re-computing the anomaly diagrams by dimensional regularization together with the tensor-to-scalar decomposition of the Feynman amplitudes. The complete expressions of the form factors $F_i$ ($i=1, \dots,13$) in the massive and then in the massless case are contained in \cite{Armillis:2009pq}. In both cases the virtualities of the external lines are generic and denoted by $s_1,s_2$. Here we give only the first invariant amplitude in the most general case.
After defining  $\gamma \equiv s -s_1 - s_2$  and $\si \equiv s^2 - 2 (s_1+s_2)\, s + (s_1-s_2)^2$ we obtain
\bea
F_1 (s;\,s_1,\,s_2,\,m^2) &=& F_{1\, pole}
  + \frac{ e^2 \,  \gamma  \,  m^2}{3 \pi^2 \, s \sigma } 
  + \frac{ e^2 m^2 \, s_2}{3 \pi^2 \,s \, \sigma^2 }  \mathcal D_2(s,s_2,m^2)\,  
  \left[s^2+4 s_1 s-2 s_2 s - 5 s_1^2+s_2^2 \right. \nn \\
&+&  \left. 4 s_1   s_2\right]  
 -  \, \frac{ e^2 \, m^2 \, s_1}{3 \, \pi^2 \,s \, \sigma^2 } \, \mathcal D_1 (s,s_1,m^2)\,
   \left[-\left(s-s_1\right)^2+5 s_2^2-4 \left(s+s_1\right)
   s_2\right]   \nn \\
&-& \, \frac{ e^2\, m^2 \, \gamma}{6 \, \pi^2 \, s \, \sigma^2 } \, \mathcal C_0 (s,s_1,s_2,m^2)\, \left[\left( s-s_1\right){}^3 - s_2^3 + \left(3 s+s_1\right)
  s_2^2 \right. \nn \\
&+& \left. \left(-3 s^2-10 s_1 s+s_1^2\right) s_2  - 4 m^2 \sigma \right], 
  \label{P1Fone}
 \eeqa
 where
\bea
F_{1\, pole} = - \frac{e^2}{18 \, \pi^2 \, s} \,,
\eea
and
\beq
 \mathcal D_i (s, s_i,  m^2) = \left[ a_i \log\frac{a_i +1}{a_i - 1}
- a_3 \log \frac{a_3 +1}{a_3 - 1}  \right],  \qquad a_i = \sqrt{ 1 - \frac{4m^2}{s_i}}.
\label{P1D_i}
\eeq
Then we discuss several kinematical limits of the $TJJ$ vertex, in particular the on-shell limit for the two vector lines ($s_1 \rightarrow 0$, $s_2  \rightarrow 0$) in order to better understand the structure of the whole correlator. The appearance of generalized anomaly poles in the correlator and their IR coupling/decoupling properties will be discussed thoroughly.

\section{Trace condition in the non-conformal case}
 
Similarly to the chiral case, we can fix the correlator by requiring the validity of a trace condition on the 
amplitude, besides the two Ward identities on the 
conserved vector currents and the Bose symmetry in their indices. This approach is alternative to the imposition of the Ward identity (\ref{PNew1TWard}) but nevertheless equivalent to it. At a diagrammatic level we obtain 
\bea
g_{\mu \nu} \Gamma^{\mu\nu\alpha\beta}(p,q) = \Lambda^{\alpha \beta}(p,q) - \, \frac{e^2}{6 \pi^2}  \, u^{\alpha\beta}(p,q).
\label{PNew1betaf}
\eea
We have also defined 
\bea
\Lambda^{\alpha \beta}(p,q) &=& -m \, (i e)^2 \int d^4x \, d^4y \, e^{i p\cdot x+iq \cdot y} \langle \bar \psi \psi J^{\alpha}(x) J^{\beta}(y) \rangle \nn \\
&=& - m \, e^2 \,  \int \frac{d^4 l}{(2 \pi)^4} \, tr \left\{\frac{i}{\lsl+\psl - m} \g^{\alpha} \frac{i}{\lsl-m} \g^{\beta} \frac{i}{\lsl-\qsl-m} \right\} + \textrm{exch.}
\eea
 A direct computation gives
\bea
\Lambda^{\alpha \beta}(p,q) = 
G_1(s;s_1,s_2,m^2) \, u^{\alpha \beta}(p,q) 
+ G_2(s;s_1,s_2,m^2) \, w^{\alpha \beta}(p,q),
\eea
where
\bea
3 \, s \, F_1(s;s_1,s_2,m^2) &=& G_1(s;s_1,s_2,m^2) - \frac{e^2}{6 \pi^2}  \\
3 \, s \, F_2(s;s_1,s_2,m^2) &=& G_2(s;s_1,s_2,m^2)
\eea
and 
\bea
G_1(s;s_1,s_2,m^2) &=& 
Ê\frac{ e ^2 \gamma Ê m^2}{ \pi^2 \sigma } +\frac{e^2\, \mathcal D_2(s,s_2,m^2)\, Ês_2 m^2}{ \pi^2 \sigma ^2} 
\left[s^2+4 s_1 s-2 s_2 s-5 s_1^2+s_2^2+4 s_1 Ê s_2\right]  Ê \nn \\
 Ê %
&& Ê \hspace{-3cm} 
- Ê \frac{e^2 \, \mathcal D_1 (s,s_1,m^2)\, s_1 m^2}{ \pi^2 \sigma ^2}
 Ê \left[-\left(s-s_1\right){}^2+5 s_2^2-4 \left(s+s_1\right)
 Ê s_2\right]  Ê\nn \\
 Ê %
 Ê&& \hspace{-3cm} Ê
 - e^2 \, \mathcal C_0 (s,s_1,s_2,m^2)\, 
 \left[ 
\frac{ m^2 \gamma}{2 \pi^2  \sigma^2} Ê \left[ \left(s-s_1\right){}^3-s_2^3+\left(3 s+s_1\right)Ê s_2^2+\left(-3 s^2-10 s_1 s+s_1^2\right) s_2 \right] -\frac{2 m^4 \gamma }{ \pi^2 \sigma }\right], \nn \\ \\ 
G_2(s;s_1,s_2,m^2) &=&
 - \frac{2 e^2 m^2}{ \pi^2 \sigma } 
 - \frac{2 e^2 \, \mathcal D_2(s,s_2,m^2) m^2}{\pi^2 \sigma Ê ^2} \, Ê\left[\left(s-s_1\right){}^2-2
 Ê s_2^2+\left(s+s_1\right) s_2\right]  Ê Ê Ê\nn \\
 Ê %
 Ê && \hspace{-3cm}
 Ê -\, \frac{2 \, e^2 \mathcal D_1 (s,s_1,m^2) \, m^2}{\pi^2 \sigma ^2}
 Ê \left[s^2+\left(s_1-2 s_2\right) s-2 s_1^2+s_2^2+s_1 s_2\right] \nn \\
 Ê %
 Ê Ê&& \hspace{-3cm}
 - \, e^2 \mathcal C_0 (s,s_1,s_2,m^2) \,
 Ê Ê\biggl[\frac{4 m^4}{ \pi^2 \sigma}
 Ê+\frac{m^2}{\pi^2 \sigma ^2} \, Ê\left[ s^3-\left(s_1+s_2\right) s^2
 Ê- \left(s_1^2-6 Ês_2 s_1+s_2^2\right) \right. s 
 Ê%
 Ê +\left. \left(s_1-s_2\right){}^2
 Ê \left(s_1+s_2\right)\right] \biggr], \nn \\
\eea
where $\gamma \equiv s -s_1 - s_2$, $\si \equiv s^2 - 2 (s_1+s_2)\, s + (s_1-s_2)^2$ and the scalar integrals $ \mathcal D_1 (s,s_1,m^2)$, $ \mathcal D_2 (s,s_1,m^2)$, $ \mathcal C_0 (s,s_1,s_2,m^2)$ for generic virtualities and masses are defined in Appendix \ref{P2scalars}. \\
We have checked that the final expressions of the form factors in the most general case, obtained either by imposing this condition on the 
energy momentum tensor or the Ward identity in the form given by Eq. (\ref{PNew1WIeq}) exactly coincide.

\section{Conformal anomaly poles}
 
There are close similarities between the effective action in the cases of the chiral and conformal anomalies, due to the presence of massless poles. In the previous chapter \cite{Armillis:2009sm} we have analyzed the fact that in the chiral case the anomaly is entirely generated by the longitudinal component $w_L$, which is indeed isolated for {\em any} configuration of the photon momenta and even for massive fermions. We also have discussed the coupling/decoupling properties of this structure in different kinematical regions.

To illustrate the emergence of a similar behaviour in the case of the conformal anomaly, it is sufficient to notice in the expression of 
$F_1$, given in Eq.(\ref{P1Fone}) in the most general form, the presence of the isolated contribution $( F_{1\,pole} = - e^2/(18 \pi^2 s))$ which survives in the massless limit but can be  isolated also in the massive case and for off-shell photons. This component, indeed, is responsible for the trace anomaly even though there appear extra corrections with mass-dependent terms. 
It is quite straightforward to figure out that the pole term $( F_{1\,pole})$ corresponds to a contribution to the gravitational effective action of the form (\ref{P2SSimple}), with a linearized scalar curvature. In fact we obtain
\bea
\Gamma_{anom}^{\mu\nu\alpha\beta}(p,q) = F_{1\, pole} \, t_1^{\mu\nu\alpha\beta} = - \frac{e^2}{18\pi^2} \frac{1}{k^2} \left(g^{\mu\nu}k^2 - k^{\mu}k^{\nu}\right)u^{\alpha\beta}(p,q)
\eea
which shows that the perturbative analysis is indeed in agreement with the variational solution given in Eq.(\ref{P2Gamanom}), as we have already expected.
Therefore, similarly to the case of the chiral anomaly, also in this case the anomaly is entirely given by $F_{1 pole}$, even in a configuration which is not obtained from a dispersive approach. The presence of mass corrections in 
$F_1$ is not a source of confusion, since there is a clear separation between anomaly and explicit breaking of the conformal symmetry.
Therefore we are entitled to disentangle the pole contribution, which describes the nonlocal contribution to the trace anomaly, from the rest $\tilde{\mathcal S}$, and rewrite the effective action as
\bea
\mathcal S =\mathcal S_{pole} +\tilde{\mathcal S}
\eea
with the pole part given by 
\beq
\mathcal{S}_{pole}= - \frac{e^2}{ 36 \pi^2}\int d^4 x d^4 y \left(\square h(x) - \partial_\mu\partial_\nu h^{\mu\nu}(x)\right)  \square^{-1}_{x\, y} F_{\alpha\beta}(x)F^{\alpha\beta}(y).
\eeq

Obviously also in this case, which is generic from the kinematical point of view, one can clearly show that the pole does not couple in the infrared if we compute the residue of the entire amplitude. The anomaly pole, in fact, couples in the spectral function only in a special kinematic configuration when the fermion-antifermion pair of the anomaly diagram is collinear, which is achieved for massless fermions and on-shell photons. This behaviour, which is in perfect analogy with the chiral anomaly case, is illustrated in detail in the next sections where the form factors $F_i$ are presented in some relevant kinematic configurations.

There is something to learn from perturbation theory: anomaly poles are not just associated to the collinear fermion-antifermion limit of the amplitude, but are also present in other, completely different kinematical domains where the collinear kinematics is not allowed and are not detected using a dispersive approach. They are present in the off-shell effective action as they are in the on-shell ones.

\section{Massive and massless contributions to anomalous Ward identities and the trace anomaly} 

Anomalous effects are usually associated in the literature with massless states, and for this reason, when we analyze the contribution to the anomaly for a massive correlator, we need to justify the distinction between massless and massive contributions.  
At nonzero momentum transfer $(k\neq 0)$ the second term on the right-hand side of Eq. (\ref{PNew1betaf}) is interpreted as
an anomalous contribution, proportional to the asymptotic $\beta$ function $(\beta_{as})$ of the theory, coming from the coefficient of the anomaly pole which appears in the form factor $F_1$. \\
The trace anomaly is connected with the regularization procedure involved in the computation of the diagrams. In our analysis we have used dimensional regularization (DR) and we have imposed conservation of the vector currents, the symmetry requirements on the correlator and the conservation of the energy momentum tensor. As we move from 4 to $d$ spacetime dimensions (before the renormalization of the theory), the anomaly pole term appears quite naturally in the expression of the correlator. This is not surprising, since QED in $d\neq 4$  dimensions is not even classically conformally invariant and the trace of the energy momentum tensor in the classical theory 
involves both a $F^2$ term ($\sim (d-4) F^2$) beside, for a massive correlator, a $\bar{\psi}\psi$ contribution
\bea
\label{PNew1TraceAnom}
g_{\mu\nu} T^{\mu\nu} = \frac{\beta_{as}(e)}{2 e^3} F^2 + (1 - \gamma_m) m \bar \psi \psi \,,
\eea
where $\beta_{as}$ and $\gamma_m$ are respectively the $\beta$ function and the anomalous mass dimension computed in a mass-independent renormalization scheme.\\
Let's summarize the basic features concerning the renormalization property of the correlator as they emerge from our direct computation. 
1) The trace Ward identity obtained by contracting the correlator ($\Gamma^{\mu\nu\alpha\beta}$) with $g_{\mu\nu}$ involves only the $F_1$ and $F_2$ 
form factors in the massive case; in the massless case the scale breaking appears uniquely due to $F_1$ via the term $-e^2/(18\pi^2)u^{\alpha\beta}(p,q)$, as pointed out before. This is interpreted as the anomalous breaking of scale invariance. The finiteness of the two form factors involved in the trace of the correlator is indeed evident.
2) The residue of the pole term in $F_1$ is affected by the renormalization of the entire correlator only by the re-definition of the bare coupling $(e^2)$ in terms of the renormalized coupling $(e_R^2)$ through the renormalization factor $Z_3$ (the form factor $F_{13}$ is the only one requiring renormalization).
At this point, the interpretation of the residue at the pole as a contribution proportional to the $\beta$ function of the theory is, in a way, ambiguous \cite{Manohar:1996cq}, since the $\beta$ function is related to a given renormalization scheme. We stress once more that  Eq. (\ref{PNew1betaf}) does not involve a renormalization scheme - which at this point has not yet been defined - but just a regularization. 
We have regulated the infinities of the theory but we have not specified a subtraction of the infinities. For this reason, the $\beta$ function appearing in Eq.(\ref{PNew1TraceAnom}), which identifies the anomalous trace term alone, without the inclusion of mass corrections, is always the asymptotic one $\beta_{as}$.

To fully understand the implications of Eq.(\ref{PNew1TraceAnom}) and the features of a massive fermion together with its decoupling properties, it is convenient to go back to the Ward identity 
(\ref{PNew1TWard}) and differentiate it with respect to the momentum $q$ and then set $p=-q$ $(k=0)$ by going to zero momentum transfer. 
One obtains the derivative Ward identity 
\beq
g_{\mu\nu}\Gamma^{\mu\nu\alpha\beta}(p,-p)= 2 p^2 \frac{d\Pi}{d p^2}(p^2)(p^2 g^{\alpha\beta}- p^{\alpha}p^{\beta}).
\label{PNew1unren}
\eeq
Notice that this result does not depend on the renormalization scheme due to the presence of the derivative with respect to $p^2$. 
The appearance of the derivative of the scalar self-energy of the photon on the right-had side of the previous equation allows to relate this expression to a particular $\beta$ function of the theory, which is not the asymptotic 
$\beta_{as}$ considered in Eq.(\ref{PNew1TraceAnom}), and includes also the mass corrections. In other words it takes into account all the effects described by Eq.(\ref{PNew1TraceAnom}) and not only the anomalous one. 

To illustrate this point we start from the expression of the scalar amplitude appearing in the photon self-energy in DR
\beq
\Pi(p^2,m)=\frac{e^2}{2 \pi^2}\left( \frac{1}{6 \epsilon} - \frac{\gamma}{6}- 
\int_0^1 dx x (1-x) \log\frac{m^2 - p^2 x(1-x)}{4 \pi \mu^2}\right)
\eeq
whose renormalization at zero momentum gives 
\beq
\Pi_R(p^2,m)=\Pi(p^2,m) - \Pi(0,m)=-\frac{e^2}{2 \pi^2}\int_0^1 x(1-x) 
\log\frac{ m^2 - p^2 x (1-x)}{m^2}.
\eeq
Using this expression, we can easily compute 
\beq
2 p^2\frac{d\Pi}{d p^2}=2 p^2\frac{d\Pi_R}{d p^2}= - \frac{e^2}{6\pi^2} + \frac{e^2\ m^2}{\pi^2}\int_0^1\,dx\ \frac{x(1-x)}
{m^2 - p^2 x(1-x) }.
\label{PNew1separated}
\eeq
Notice also that the $\beta$ function of the theory evaluated in the zero momentum subtraction scheme is exactly given by the right-hand side of the previous expression 
\beq
 \frac{\beta (e^2,p^2)}{e^2} = - 2 p^2\frac{d\Pi_R}{d p^2},
\label{PNew1defin}
\eeq
where $\beta(e^2,p^2)= 2 e \beta(e, p^2)$. 
Clearly, in the case of Eq.(\ref{PNew1defin}) all the mass contributions to Eq.(\ref{PNew1TraceAnom}) have been absorbed into the definition of this $\beta$ function. \\
Notice that if $ p^2 \ll m^2$ this $\beta$ function, after a rearrangement gives
\beq
- \frac{\beta (e^2,p^2)}{e^2}=\frac{e^2}{\pi^2}\int_0^1 dx \frac{p^2 x^2 (1-x)^2}{m^2 - p^2 x (1-x)}
\eeq
and therefore it vanishes as $\beta\sim O(p^2/m^2)$ for $p^2\to 0$. Equivalently, by taking the $m\to \infty$ limit we recover the expected decoupling of the fermion (due to a vanishing $\beta$)
since we are probing the correlator at a scale ($p^2$) which is not sufficient to resolve the contribution of the fermion loop. On the contrary, as $p^2\to \infty$, with $m$ fixed, the running of the $\beta$ function is the usual asymptotic one 
$\beta_{as}(e^2)\sim e^4/(6 \pi^2)$ modified by corrections $O(m^2/p^2)$. The UV limit is characterized by the same running typical of the massless case, as expected. 

It is worth to stress again that the right-hand side of Eq.(\ref{PNew1unren}) does not depend on the renormalization scheme, while the $\beta$ function does and Eq.(\ref{PNew1defin}) should be understood as a definition.

\section{The off-shell massless $TJJ$ correlator}
 
In the massless fermion case the scalar functions $F_i$ depend only on the kinematic invariants $s,s_1,s_2$ but we still retain the last entry of these functions and set it equal to $0$  for clarity, using the notation $F_i \equiv F_i (s;s_1,s_2,0)$. These new functions are computed starting from the massive ones and letting $m\rightarrow 0$ and $\mathcal A_0 (m^2) \rightarrow 0$, i.e. eliminating all the massless tadpoles generated in the zero fermion mass limit.

 The off-shell massless invariant amplitudes $F_i (s; s_1, s_2,0)$ are given in terms of the master integrals listed in Appendix \ref{P2scalars} with internal masses set to zero. We give here only the simplest invariant amplitudes, leaving the remaining ones to \cite{Armillis:2009pq} . 
The anomaly pole is clearly present in $F_1$, which is given by
\bea
%
 %
 F_{1} (s;\,s_1,\,s_2,\,0) &=& - \frac{e^2 }{18 \pi^2 s}, 
 \eea
while
  %
  \bea
 F_{2} (s;\,s_1,\,s_2,\,0) &=& 0. 
  \eea
 The complete $TJJ$ correlator is very complicated in this case as the long expressions of the form factors show, but a deeper analysis of its poles by computing the residue in $s=0$ can be useful to draw some conclusions.  The single pieces of $\Gamma^{\mu\nu\a\b}(s;\,s_1,\,s_2,\,0)$ indeed contribute as
\bea
\lim_{s\rightarrow0} \, s F_1 (s;\,s_1,\,s_2,\,0) \, t^{\mu\nu\a\b}_1 &=&
 - \frac{e^2}{18 \, \pi^2 } \, t^{\mu\nu\a\b}_1\big\vert_{s=0},\\
%
%
\lim_{s\rightarrow0} \, s F_3 (s;\,s_1,\,s_2,\,0) \, t^{\mu\nu\a\b}_3 &=&
\,\frac{e^2}{72 \, \pi^2 } \, t^{\mu\nu\a\b}_3 \big\vert_{s=0},\\
%
%
\lim_{s\rightarrow0} \, s F_5 (s;\,s_1,\,s_2,\,0) \, t^{\mu\nu\a\b}_5 &=&
\,\frac{e^2 }{72\, \pi^2} \, t^{\mu\nu\a\b}_5 \big\vert_{s=0}, \\
%
%
\lim_{s\rightarrow0} \, s F_7 (s;\,s_1,\,s_2,\,0) \, t^{\mu\nu\a\b}_7 &=&
\frac{e^2}{36\, \pi^2} \, t^{\mu\nu\a\b}_7 \big\vert_{s=0}.
\eea
The residues of the $F_i(s;\,s_1,\,s_2,\,0) $ not included in the equations above are all vanishing. Combining the results given above one can easily check that the entire correlator is completely free from coupled anomaly poles as
\bea
\lim_{s\rightarrow0} \, s \, \Gamma^{\mu\nu\a\b}(s;\,s_1,\,s_2,\,0) =0
\eea
in this rather general configuration. A similar result holds for the correlator responsible for the chiral anomaly and shows the decoupling of polar contributions in the infrared in the off-shell ($s_1, s_2 \neq 0$) case.

\section{The on-shell massive $TJJ$ correlator}

A particular  case of the $TJJ$ correlator is represented by its on-shell version with a
massive fermion in the loop.  If we contract $u^{	\a \b} (p,q)$ and $w^{	\a \b} (p,q)$ with the polarization vectors $\eps_\a (p)$ and $\eps_\b (q)$ requiring $\eps_\a (p) \, p^\a=0$, \mbox{${\eps_\b (p) \, p^\b=0}$ }, the first  tensor remains unchanged while  $w^{\a \b} (p,q)$ becomes $\widetilde w^{	\a \b} (p,q)= s_1 \, s_2 \, g^{\a \b}$. This will be carefully taken into account when computing the $s_1 \rightarrow 0$, $s_2 \rightarrow 0$ limit of the product of the invariant amplitudes $F_i$ with their corresponding tensors $t_i^{\,\mu\nu\a\b}$ ($i=1,\dots,13$).\\
The invariant amplitudes reported below describe $F_i(s;0,0,m^2)$ whose tensors $t_i^{\,\mu\nu\a\b}$  are also finite and non-vanishing. They are
\bea
F_1 (s;\,0,\,0,\,m^2) &=&
- \frac{e^2 }{18 \pi^2  s} \,  + \, \frac{e^2  m^2}{3 \pi^2 s^2} \, - \, \frac{e^2\, m^2} {3 \pi^2 s}\mathcal C_0 (s, 0, 0, m^2) 
\bigg[\frac{1}{2 \,  }-\frac{2 m^2}{ s}\bigg],  \nn \\
F_3 (s;\,0,\,0,\,m^2)  &=&
- \frac{e ^2}{144 \pi^2 s} - \frac{ e ^2 m^2}{12 \pi^2  s^2}
- \, \frac{e^2 \, m^2}{4 \pi^2  s^2} \mathcal D (s, 0, 0, m^2) \, 
\nn \\
&&
- \, \frac{e^2 \, m^2}{6 \pi^2 s } \mathcal C_0(s, 0, 0, m^2 )\, \left[ \frac{1}{2} + \frac{m^2}{s}\right], \nn \\
F_5 (s;\,0,\,0,\,m^2)  &=& F_3 (s;\,0,\,0,\,m^2) , \nn \\
F_7 (s;\,0,\,0,\,m^2)  &=&  - 4 \, F_3 (s;\,0,\,0,\,m^2) \nn \\
F_{13\, R} (s;\,0,\,0,\,m^2)  &=&
 \frac{11  e ^2}{144  \pi^2 }  +   \frac{ e ^2 m^2}{4 \pi^2 s}
 + \, e^2\mathcal C_0 (s, 0,0,m^2) \,\left[ \frac{m^4}{2 \pi^2 s}+\frac{ m^2}{4 \pi^2 }\right] \nn \\
 && + \,e^2  \mathcal D (s, 0, 0, m^2) \, \left[ \frac{5 m^2}{12 \pi^2  s} + \frac{1}{12}\right],
\label{PNew1masslesslimit}
\eea
where the on-shell scalar integrals $\mathcal D (s, 0, 0, m^2)$ and $\mathcal C_0 (s, 0,0,m^2)$ are defined in Appendix \ref{P2scalars}; here
$F_{13\, R}$ denotes the renormalized amplitude (in the on-shell renormalization scheme), obtained by first removing the UV pole present in the photon self-energy by the usual renormalization of the photon wavefunction and then taking the on-shell limit. The remaining  invariant amplitudes $F_i (s, 0,0,m^2)$ are zero or multiply vanishing  tensors in this kinematical configuration, therefore they do not contribute to the correlator. \\
Using the results given above, the full massive on-shell amplitude is given by
\bea
\Gamma^{\mu \nu \a \b} (s;0,0,m^2)  &=&
 F_1 \,(s; 0,0,m^2) \, \widetilde t_1^{\, \mu \nu \a \b}  \, +\,
F_3 \, (s; 0,0,m^2) \, (\widetilde t_3^{\, \mu \nu \a \b} \, + \,  \widetilde t_5^{\, \mu \nu \a \b}
- \, 4 \, \widetilde t_7^{\, \mu \nu \a \b}) \nn \\
 &+&  \, F_{\, 13,\, R\,} (s; 0,0,m^2) \, \widetilde t_{13}^{\, \mu \nu \a \b}  ,
\eea
so that the invariant amplitudes reduce from $13$ to $3$ and the three linear combinations of the tensors can be taken as a new basis
\bea
 \widetilde t_1^{\, \mu \nu \a \b} &=& \lim_{s_1,s_2 \rightarrow 0} \,  t_1^{\, \mu \nu \a \b} =
 (s \, g^{\mu\nu} - k^{\mu}k^{\nu}) \, u^{\a \b} (p,q)  
 \label{PNew1widetilde1}\\
 \widetilde t_3^{\, \mu \nu \a \b} \, + \,  \widetilde t_5^{\, \mu \nu \a \b}
- \, 4 \, \widetilde t_7^{\, \mu \nu \a \b} &=&  \lim_{s_1,s_2 \rightarrow 0}  \, (t_3^{\, \mu \nu \a \b} \, + \,  t_5^{\, \mu \nu \a \b}
- \, 4 \,  t_7^{\, \mu \nu \a \b}) = \nn  \\
&& \hspace{-0.5cm} - 2 \, u^{\a \b} (p,q) \left( s \, g^{\mu \nu} + 2 (p^\mu \, p^\nu + q^\mu \, q^\nu )
- 4 \, (p^\mu \, q^\nu + q^\mu \, p^\nu) \right)  \\
\widetilde{t}^{\, \mu \nu \alpha \beta}_{13} &=&  \lim_{s_1,s_2 \rightarrow 0} \,  t_{13}^{\, \mu \nu \a \b} =
\big(p^{\mu} q^{\nu} + p^{\nu} q^{\mu}\big)g^{\alpha\beta}
+ \frac{s}{2} \big(g^{\alpha\nu} g^{\beta\mu} + g^{\alpha\mu} g^{\beta\nu}\big) \nn \\
&&  \hspace{-0.5cm} - g^{\mu\nu} (\frac{s}{2} g^{\alpha \beta}- q^{\alpha} p^{\beta})
-\big(g^{\beta\nu} p^{\mu}
+ g^{\beta\mu} p^{\nu}\big)q^{\alpha}
 - \big (g^{\alpha\nu} q^{\mu}
+ g^{\alpha\mu} q^{\nu }\big)p^{\beta}, 
\label{PNew1widetilde13}
\nn \\
\eea
 as  previously done in the literature \cite{Berends:1975ah}.
If we extract the residue of the full amplitude we realize that even though some functions $F_i (s, 0,0,m^2) $ have kinematical singularities in $1/s$ this polar structure is no longer present in the complete massive correlator
\bea
\lim_{s \rightarrow 0} s \, \Gamma^{\mu \nu \alpha \beta}(s;0,0,m^2) = 0
\eea
showing that the $TJJ$ correlator exhibits no infrared poles in the presence of explicit conformal symmetry breaking terms.

\section{The general effective action and its various limits }

It is possible to identify the structure of the effective action in its most general form using the explicit expressions of the invariant amplitudes 
If we denote by $\mathcal{S}_i$ the contribution to the effective action due to each form factor $F_i$, then we can write it in the form 
\beq
\mathcal{S}_i= \int d^4 x \,d^4 y \,d^4 z\, \hat{t}_i^{\mu\nu\alpha\beta}(z,x,y) 
h_{\mu\nu}(z)A_\alpha(x) A_\beta(y)\int \frac{d^4 p\, d^4 q }{(2 \pi)^8}e^{-i p\cdot(x-z) -i q\cdot(y-z)} 
F_i(k,p,q)
\eeq
where $k\equiv p +q$. We have introduced the operatorial version of the tensor structures $t_i^{\mu\nu\alpha\beta}$, denoted by $\hat{t}_i$ that will be characterized below. Defining 
\beq
\hat{p}_x^\alpha\equiv i \frac{\partial}{\partial x_\alpha}, \qquad \hat{q}_y^\alpha\equiv i \frac{\partial}{\partial y_\alpha}, \qquad \hat{k}_z^\alpha\equiv - i \frac{\partial}{\partial z_\alpha}
\eeq
and using the identity
\beq
\hat{F}_i ( \hat{k}_z, \hat{p}_x, \hat{q}_y) \delta^4(x-z)\delta^4(y-z)
=\int \frac{d^4 p\, d^4 q }{(2 \pi)^8}e^{-i p\cdot(x-z) -i q\cdot(y-z)} 
F_i(k,p,q)
\eeq
where formally $\hat{F}_i$ is the operatorial version of $F_i$, 
we can arrange the anomalous effective action also in the form 
\beq
\mathcal{S}_i=\int d^4 x d^4 y d^4 z  \hat{F}_i ( \hat{k}_z, \hat{p}_x, \hat{q}_y)\left[\delta^4(x-z)\delta^4(y-z)\right]\hat{t}_i^{\mu\nu\alpha\beta}(z,x,y) h_{\mu\nu}A_\alpha(x)A_\beta(y).
\eeq
For instance we get
\bea
\hat{t}_1^{\mu\nu\alpha\beta}(z,x,y) h_{\mu\nu}A_\alpha(x)A_\beta(y)&=&
\frac{1}{2}\left(\square_z h(z) - 
\partial^z_\mu \partial^z_\nu h^{\mu\nu}(z)\right) F_{\alpha\beta}(x)F^{\alpha\beta}(y),\\
\hat{t}_2^{\mu\nu\alpha\beta}(z,x,y) h_{\mu\nu}A_\alpha(x)A_\beta(y) &=&
\left(\square_z h(z) - 
\partial^z_\mu \partial^z_\nu h^{\mu\nu}(z)\right) \partial_\mu F^{\mu}_\lambda(x)\partial_\nu F^{\nu\lambda}(y), \\
\hat{t}_3^{\mu\nu\alpha\beta}(z,x,y) h_{\mu\nu}A_\alpha(x)A_\beta(y) &=&
\frac{1}{2} h^{\mu\nu}(z)\left(\square_x g_{\mu\nu} - 
4 \partial^x_\mu \partial^x_\nu \right) F_{\alpha\beta}(x)F^{\alpha\beta}(y), \\
\hat{t}_4^{\mu\nu\alpha\beta}(z,x,y) h_{\mu\nu}A_\alpha(x)A_\beta(y) &=&
h^{\mu\nu}(z)\left(\square_x g_{\mu\nu}- 
4 \partial^x_\mu \partial^x_\nu \right) \partial_\mu F^{\mu}_\lambda(x)\partial_\nu F^{\nu\lambda}(y), \\
\hat{t}_5^{\mu\nu\alpha\beta}(z,x,y) h_{\mu\nu}A_\alpha(x)A_\beta(y) &=&
\frac{1}{2} h^{\mu\nu}(z)\left(\square_y 
g_{\mu\nu} - 
4 \partial^y_\mu \partial^y_\nu \right) F_{\alpha\beta}(x)F^{\alpha\beta}(y), \\
\hat{t}_6^{\mu\nu\alpha\beta}(z,x,y) h_{\mu\nu}A_\alpha(x)A_\beta(y) &=& 
h^{\mu\nu}(z)\left(\square_y g_{\mu\nu}- 
4 \partial^y_\mu \partial^y_\nu \right) \partial_\mu F^{\mu}_\lambda(x)\partial_\nu F^{\nu\lambda}(y), \\
\hat{t}_7^{\mu\nu\alpha\beta}(z,x,y) h_{\mu\nu}A_\alpha(x)A_\beta(y) &=&
 \frac{1}{2} h^{\mu\nu}(z)\left(
\partial^{x\,\lambda} \partial^y_\lambda g_{\mu\nu} - 
2 (\partial^y_\mu \partial^x_\nu  +  \partial^y_\nu \partial^x_\mu)\right)  F_{\alpha\beta}(x)F^{\alpha\beta}(y), \\
\hat{t}_8^{\mu\nu\alpha\beta}(z,x,y) h_{\mu\nu}A_\alpha(x)A_\beta(y) &=&
h^{\mu\nu}(z)\left(  \partial^{x\,\lambda} \partial^y_\lambda g_{\mu\nu} - 
2 (\partial^y_\mu \partial^x_\nu  +  \partial^y_\nu \partial^x_\mu)\right) \partial_\mu F^{\mu}_\lambda(x)\partial_\nu F^{\nu\lambda}(y)
\eea
and similar expressions for the remaining tensor structures. \\
However, the most useful forms of the effective action involve an expansion in the fermions mass, as in the $1/m$ formulation (the Euler-Heisenberg form) or for small $m$. In this second case the nonlocal contributions obtained from the anomaly poles appear separated from the massive terms, 
showing the full-fledged implications of the anomaly. This second formulation allows a smooth massless limit, where the breaking of the conformal anomaly is entirely due to the massless fermion loops.

In the ${1}/{m}$ case, for on-shell gauge bosons, the result turns out to be particularly simple. We obtain  
\bea
F_{1} (s, 0, 0, m^2) &=& 
\frac{7 e ^2 }{2160 \pi^2 } \frac{1}{m^2} + \frac{e ^2 s}{3024 \, \pi^2 }  \frac{1}{m^4} 
+O\left(\frac{1}{m^6}\right), \label{P11om}  \\
F_{3} (s, 0, 0, m^2) &=& F_{5} (s, 0, 0, m^2) = 
\frac{e^2 }{4320 \, \pi^2 }  \frac{1}{m^2} 
+ \frac{e^2 s}{60480 \, \pi^2} \frac{1}{m^4} +O\left(\frac{1}{m^6}\right), \\
F_{7} (s, 0, 0, m^2) &=&  - 4 \, F_{3} (s, 0, 0, m^2) \\
F_{13, R} (s, 0, 0, m^2) &=& 
\frac{11 e^2 s}{1440 \, \pi^2 } \frac{1}{m^2} + \frac{11 e^2 s^2}{20160 \, \pi^2 } \frac{1}{m^4} 
+ O\left(\frac{1}{m^6}\right),
\eea
which can be rearranged in terms of three independent tensor structures. Going to configuration space, the linearized expression of the contribution to the gravitational effective action due to the $TJJ$ vertex, in this case, can be easily obtained in the form
\bea
S_{TJJ} &=& \int d^4 x d^4 y d^4 z \, \Gamma^{\mu \nu \alpha\beta}(x,y,z) A_{\alpha}(x) \, A_{\beta}(y) \, h_{\mu \nu}(z) \nn \\
&=& \frac{ 7 \, e^2}{4320 \, \pi^2 \, m^2} \int d^4 x \left( \Box h - \partial^{\mu} \partial^{\nu} h_{\mu \nu}\right) F^2 \nn \\
&-& \frac{e^2}{4320 \, \pi^2 \, m^2} \int d^4 x \left( \Box h F^2 - 8 \partial^{\mu} F^{\alpha \beta} \partial^{\nu} F_{\alpha \beta} h_{\mu \nu} + 4 (\partial^{\mu} \partial^{\nu} F_{\alpha \beta})F^{\alpha \beta} h_{\mu \nu} \right) \nn \\
&+& \frac{11 \, e^2}{1440 \, \pi^2 \, m^2} \int d^4 x T^{\mu \nu}_{ph} \Box h_{\mu \nu}\,,
\eea
which shows three independent contributions linear in the (weak) gravitational field.
In this case there is no signature of the presence of pole singularities which are scaleless and described entirely the anomaly contribution. Of course $1/m$ expansions are legitimate, but there is no apparent sign left in Eq.(\ref{P11om}) of the presence of a massless term. This result is analog to the chiral case and it is obviously related to the IR decoupling in the presence of massive fermions which we have already discussed in the previous sections.

 Another important observation is that the contributions to the trace of the energy-momentum tensor, which is relevant also in the cosmological context \cite{Starobinsky:1980te, Shapiro:2006sy},  are all dominated by the pole term at high energy, since mass corrections contained in $F_1$ are clearly suppressed as $m^2/s$. Obviously, Eq.~(\ref{P11om}) differs systematically from the result obtained from the small $m$ expansion, where the nonlocality of the effective action and the presence of a massless scalar exchange, as a result of the conformal anomaly, 
is instead quite evident. We obtain in this second case $(s < 0)$
\bea
F_1(s,0,0,m^2) = F_{1 \, pole}   
+ \frac{ e^2 \, m^2}{12 \, \pi^2 \, s^2} 
\left[ 4 -  \log^2 \frac{m^2}{s} 	\right] + O \left( \frac{1}{s^3} \right)
\eea
where the anomalous form factor shows a massless pole beside some additional mass corrections.
This is an expansion, as in the case of the chiral anomaly, which describes the UV limit.

\section{The massless on-shell $TJJ$ correlator}
The nonlocal structure of the effective action, as we have pointed out in the previous sections, is not apparent 
within an expansion in $1/m$, nor this expansion has a smooth match with the massless case. 

The computation of the correlator $\Gamma^{\mu\nu\alpha\beta}  (s; 0,0,0) $ hides some subtleties in the  massless fermion limit  (with on-shell external photons), as the form factors $F_i$ and the tensorial structures $t_i$ both contain the kinematical invariants $s_1, s_2$.  For this reason the limit of both factors 
(form factor and corresponding tensor structure)  $F_i \, t_i^{\mu\nu\a\b}$ has to be taken carefully, starting from the expression of the massless $F_i (s; s_1,s_2,0) $ listed in \cite{Armillis:2009pq} and from the tensors $t_i^{\mu\nu\a\b}$ contracted with the physical polarization tensors. In this case only few form factors survive and in particular  
\bea
F_{1} (s; 0, 0, 0) &=& - \frac{e^2}{18 \pi^2  s}, \\
F_{3} (s; 0, 0, 0) &=&  F_{5} (s, 0, 0, 0) = - \frac{e^2}{144 \pi^2 \, s}, \\
F_{7} (s; 0, 0, 0) &=& -4 \, F_{3} (s, 0, 0, 0), \\
F_{13, R} (s; 0, 0, 0) &=& - \frac{e^2}{144 \pi^2} \, \left[ 12 \log \left(-\frac{s}{\mu^2}\right) - 35\right],
\eea
and hence the whole correlator with two onshell photons on the external lines is 
\bea
\Gamma^{\mu\nu\alpha\beta}  (s; 0,0,0) &=& 
F_{1} (s; 0, 0, 0) \, \widetilde{t}^{\, \mu \nu \alpha \beta}_{1} 
+ F_{3} (s; 0, 0, 0) \, \left( \widetilde{t}^{\, \mu \nu \alpha \beta}_{3}  + \widetilde{t}^{\, \mu \nu \alpha \beta}_{5}
- 4 	\, \widetilde{t}^{\, \mu \nu \alpha \beta}_{7}  \right) \nn \\
&+& 
F_{13, R}(s; 0, 0, 0) \, \widetilde{t}^{\, \mu \nu \alpha \beta}_{13}  \nn \\
&=&
- \frac{e^2}{48 \pi^2 \, s} \left[ 
\left(2 \, p^\b \, q^\a- s \, g^{\a\b}\right) \left( 2 \, p^\mu \, p^\nu + 2 \, q^\mu \, q^\nu - s \, g^{\mu\nu} \right)\right]
+ \, F_{13, R} \, \widetilde{t}^{\, \mu \nu \alpha \beta}_{13},
\label{PNew1gamma00}
\eea
where $\widetilde{t}^{\, \mu \nu \alpha \beta}_{i}$ are the tensors defined in Eqs.~(\ref{PNew1widetilde1}-\ref{PNew1widetilde13}).

The study of the singularities in $1/s$ for this correlator requires a different analysis for $F_1$ and the remaining form factors, as explicitly shown in Eq.~\ref{PNew1gamma00}, where $F_1$ has been kept aside from the others, even if it is proportional to $F_3$. Indeed $F_1$ is the only form factor multiplying a non zero trace tensor, $\widetilde{t}^{\, \mu \nu \alpha \beta}_{1}$, and responsible for the trace anomaly. If we take the residue of the onshell correlator for physical polarizations of the photons in the final state we see how the $4$ form factors and their tensors combine in such a way that the result is different from zero as 
\bea
\lim_{s\rightarrow 0 } \, s \, \Gamma^{\mu\nu\alpha\beta}  (s; 0,0,0)  = 
- \frac{e^2}{12 \pi^2 }  \, p^\b \, q^\a (p^\mu \, p^\nu + q^\mu \, q^\nu),
\eea
 where clearly each singular part in $1/s$ present in $F_1,F_3, F_5,F_7$ added up and the logarithmic behaviour in $s$ of $F_{13}$ has been regulated by the factor $s$ in front when taking the  limit. The result shows that the pole, in this case, is indeed coupled, as shown also by the dispersive analysis.

 \section{Conclusions}
We have presented a computation of the $TJJ$ correlator, responsible for the appearance of gauge contributions to the conformal anomaly in the effective action of gravity. We have used our results to present the general form of the gauge contributions to this action, in the limit of a weak gravitational field.   One interesting feature of this correlator is the presence of an anomaly pole \cite{Giannotti:2008cv}.

 Usually anomaly poles are interpreted as affecting the infrared region of the correlator and appear only in one special kinematical configuration, which requires massless fermions in the loop and on-shell conditions for the external gauge lines. In general, however, the anomaly pole affects the UV region even if it is not coupled in the infrared.  This surprising feature of the anomaly is present both in the case of the chiral anomaly \cite{Armillis:2009sm} {\em and} in the conformal anomaly. Here we have extracted explicitly this behaviour by a general analysis of the correlator, extending our previous study of the chiral anomaly. 

Indeed anomaly poles are the most interesting feature, at perturbative level, of the anomaly, being it conformal or chiral, and are described by mixed diagrams involving either a scalar (gravitational case) \cite{Giannotti:2008cv} or a pseudoscalar (chiral case) \cite{Armillis:2009sm,Coriano:2008pg}. The connection between the infrared and the ultraviolet, signalled by the presence of these contributions, should not be too surprising in an anomalous context. The pole-like behaviour of an anomalous correlator is usually ``captured" by a variational solution of a given anomaly equation, which implicitly assumes the presence of a pole term in the integrated 
functional \cite{Armillis:2009im}. By rediscovering the pole in perturbation theory, obviously, one can clearly conclude that variational solutions of the anomaly equations are indeed correct, although they miss homogeneous solutions  to the Ward identity, that indeed must necessarily be identified by an off-shell perturbative analysis of the correlators. This is the approach followed here and in \cite{Armillis:2009sm}.

We have also seen that  the identification of the massless anomaly pole allows to provide a  ``mixed" formulation of the effective action in which the pole is isolated from the remaining mass terms, 
extracted  in the $\tilde{\mathcal S}_{pole}$ part of the anomalous action, which could be used for further studies. 
We have also emphasized that a typical $1/m$ expansion of the anomalous effective action fails to convey fully the presence of scaleless contributions. 

There are various applications of our analysis which can be of interest for further studies. The first concerns the possible implication of these types of effective actions in cosmology, especially in inflationary scenarios where the coupling of gravity to matter via gauge interactions and the conformal anomaly plays an interesting phenomenological role. As we have stressed in the introduction, 
the local description of an anomalous effective action involves additional degrees of freedom which can be identified in the case of the gauge anomaly \cite{Coriano:2008pg} as well as in the conformal case \cite{Giannotti:2008cv}. In \cite{Giannotti:2008cv} the authors describe the role of the corresponding scalar degrees of freedom 
in the effective action emphasizing their meaning as possible composite. 
Of particular interest are the extensions of these analysis to the case of supersymmetric theories, in particular to $N=1$ superconformal theories, where the R-current, the supersymmetry current and the energy momentum tensor belong to the same supermultiplet, as are their corresponding anomalies. Clearly our computation is the first step in this direction, and can be extended with the inclusion of other types of fields in the perturbative expansion, reaching, as a starting point, all the relevant fields of the Standard Model.  In general, one could also use our approach to come with a complete 
description of the interplay between supersymmetry and the conformal anomaly, acting as a mediator of the gravitational interaction, which is of phenomenological interest. 

\chapter{The conformal anomaly and the gravitational coupling of QCD}
\label{Chap.GravitonQCD}

\section{Introduction}
In this chapter we move to the analysis of the anomalous $TJJ$ correlator in a non-abelian gauge theory which is the first step towards a full understanding of the gravitational effective action in the Standard Model.
The study of the effective action describing the coupling of a gauge theory to gravity via the trace anomaly \cite{Duff:1977ay} is an important aspect of quantum field theory, which is not deprived also of direct phenomenological implications. This coupling is mediated by the correlator involving the energy momentum tensor together with two vector currents (or $TJJ$ vertex), which describes the interaction of a graviton with two gauge bosons. 
At the same time, the vertex has been at the center of an interesting case study
of the renormalization properties of  composite operators in Yang Mills theories \cite{Collins:1994ee}, in the context of an explicit check of the violation of the Joglekar-Lee theorem \cite{Joglekar:1975nu} on the vanishing of S-matrix elements of BRST exact  operators.  In this second case it was computed on-shell, but only at zero momentum transfer.
In this chapter we are going to extend this computation in QCD and investigate the presence of massless singularities in its expression. These contribute to the trace anomaly and play a leading role in fixing the structure of the effective action that couples QCD to gravity. The analysis of \cite{Collins:1994ee}, which predates our study, unfortunately does not resolve the issue about the presence or the absence of the anomaly pole in the anomalous effective action of QCD
because of the restricted kinematics involved in that analysis of the $TJJ$ vertex, and for this reason we have to proceed with a complete re-computation.

As we have pointed out in the previous chapters, anomaly poles characterize quite universally (gravitational and chiral) anomalous effective actions, in the sense that account for their anomalies. They have been identified and discussed in the abelian case both by a dispersive analysis \cite{Giannotti:2008cv} and by a direct explicit computation of the related anomalous Feynman amplitudes, see chapter \ref{Chap.TJJQED} or \cite{Armillis:2009pq} for more details. Extensive analyses in the case of chiral theory have shown the close parallel in the solutions of the Ward identities, the coupling of the poles in the ultraviolet and in the infrared region with the gravity case \cite{Armillis:2008bg, Armillis:2009sm}.

It is therefore important to check whether similar contributions appear also in non-abelian gauge theories coupled to gravity. We recall that the same pole structure is found in the variational solution of the expression of the trace anomaly, where one tries to identify an action whose energy momentum tensor reproduces it. This action, found by Riegert long ago \cite{Riegert:1984kt}, is nonlocal and involves the Green function of a quartic (conformally covariant) operator. The action describes the structure of the singularities of anomalous correlators with any number of insertions of the energy momentum tensor and two vector fields ($T^n JJ$), which are expected to correspond both to single
and to higher order poles, for a sufficiently high $n$. For obvious reasons, explicit checks of this effective action using perturbation theory - as the number of external graviton lines grows -  becomes increasingly difficult to handle. The $TJJ$ correlator is the first (leading) contribution  to this infinite  sum of correlators in which the anomalous gravitational effective action is expanded.

Given the presence of a quartic operator in Riegert's nonlocal action, the proof that this action contains a single pole to lowest order (in the $TJJ$ vertex), once expanded around flat space, has been given in \cite{Giannotti:2008cv} by Giannotti and Mottola in QED, and provides the basis for the discussion of the anomalous effective action in terms of massless auxiliary fields contained in their work. The auxiliary fields are introduced in order to rewrite the action in a local form. We show by an explicit computation of the lowest order vertex that Riegert's action is indeed consistent in the non-abelian case as well, since its pole structure is recovered in perturbation theory, similarly to the abelian case. Therefore, one can reasonably conjecture the presence of anomaly poles in each gauge invariant subsets of the diagrammatic expansion, as the computation for the non-abelian $TJJ$ shows (here for the case of the single pole). In particular, this is in agreement with the result of \cite{Giannotti:2008cv},
where it is shown that, after expanding around flat spacetime, the quartic operator in Riegert's action becomes a simple $1/\square$ nonlocal interaction (for the $TJJ$ contribution), i.e. a pole term. We remark that the identification of an infrared coupled pole term in this and in others similar correlators, as we are going to emphasize in the following sections 
(at least in the case of QED and for the sector of QCD mediated by quark loops), requires an extrapolation to the massless fermion limit, and for this reason its interpretation as a long-range dynamical effect in the gravitational effective action requires some caution. In QCD, however, there is an extra sector that contributes to the 
same correlator, entirely due to virtual loops of gluons in the anomaly graphs, which remains unaffected by the massless fermion limit. 
In the gauge sector, being mass independent, the condition of infrared coupling of the corresponding pole only requires that the two external gluons are on-shell. This lend a special role to the exact non-abelian gauge theory, because it seems to imply the existence of an anomalous contribution in the $TJJ$ vertex unambiguously identifiable at any momentum scale, in the UV as well as in the IR. 
We also remark that the requirement of the on-shellness in the gluon sector may be contradicted by the condition of gluon confinement, which would indeed preclude the possibility of having an infrared coupled pole. Confinement, indeed, does not allow to have on-shell gluons and hence an infrared coupling of the anomaly in this sector.

This chapter is devoted to a presentation of a detailed study of the one-loop perturbative expansion of the off-shell $TJJ$ vertex together with its Ward identities which unambiguously define it. They have been derived from first principles and arise from the conservation equation of the energy-momentum tensor and from the BRST invariance of the Yang-Mills action. \\
The structure of the effective action and the characterization of its fundamental form factors at nonzero momentum transfer and its complete analytical structure is a novel result. In this respect, the classification of all the
relevant tensor structures which appear in the computation of this correlator is rather involved and has been performed
in the completely off-shell case.  We remark that the complexity of the final expression, in the off-shell case, prevents us from presenting
its form. For this reason we will give only the on-shell version of the complete vertex, which is expressed only in terms of three fundamental form factors.

Concerning the phenomenological relevance of this vertex, we just mention that it plays an essential role in the study of NLO corrections to processes involving a graviton exchange. In fact, in theories with extra dimensions,
gravitational propagation in the bulk naturally induces a Kaluza-Klein tower of spin-2 and graviscalar (also named dilaton) excitations in the effective four dimensional theory, lowering the gravity scale and enhancing the associated cross sections (with virtual or real graviton exchanges). 
In these scenarios
the vertex appears in the hard scattering of the corresponding factorization formula \cite{Mathews:2004xp,Kumar:2009nn,Agarwal:2010sp,Agarwal:2009xr} and has been computed in dimensional regularization. However, to our knowledge,  in all cases, there has been no separate discussion of the general structure of the vertex (i.e. as an amplitude) nor of its Ward identities, which, in principle, would require a more careful investigation because of the trace anomaly.

\section{The energy momentum tensor and the Ward identities}
We introduce the definition of the QCD energy-momentum tensor, which is given by
\bea
T_{\mu \nu} &=& -g_{\mu \nu} {\mathcal L}_{QCD}
-F_{\mu \rho}^a F_\nu^{a \rho} -{\frac{1} {\xi}}g_{\mu \nu}
\partial^\rho (A_\rho^a \partial^\sigma A_\sigma^a) +{\frac{1}{\xi}}(A_\nu^a \partial_\mu(\partial^\sigma A^a_\sigma)
  +A_\mu^a \partial_\nu(\partial^\sigma A_\sigma^a))
\nonumber\\
&+& {\frac{i}{4}} \Big[
  \overline \psi \gamma_\mu (\overrightarrow \partial_\nu
-i g T^a A_\nu^a)\psi  -\overline \psi (\overleftarrow \partial_\nu
+i g T^a A_\nu^a)\gamma_\mu\psi
 +\overline \psi \gamma_\nu (\overrightarrow \partial_\mu
-i g T^a A_\mu^a)\psi
\nonumber\\
&-& \overline \psi (\overleftarrow \partial_\mu
+i g T^a A_\mu^a)\gamma_\nu\psi \Big] +\partial_\mu \overline
\omega^a (\partial_\nu \omega^a -g f^{abc} A_\nu^c \omega^b)
+\partial_\nu \overline \omega^a (\partial_\mu \omega^a -g f^{abc}
A_\mu^c \omega^b), 
\label{P2EMT}
\eea
where $F_{\mu\nu}^a$ is the non-abelian field strength of the gauge field $A$
\bea
F_{\mu\nu}^a = \partial_{\mu}A_{\nu}^a - \partial_{\nu}A_{\mu}^a + g f^{abc}A^b_{\mu}A^c_{\nu}
\eea
and we have denoted with $\omega^a$ the Faddeev-Popov ghosts and with $\overline{\omega}^a$ the antighosts, while $\xi$ is the gauge-fixing parameter. The $T^a$'s are the gauge group generators in the fermion representation and $f^{abc}$ are the antisymmetric structure constants. For later use,
it is convenient to isolate the gauge-fixing and ghost dependent contributions from the entire tensor
\bea
T^{g.f.}_{\mu\nu} &=& {1 \over \xi}\left[A_\nu^a \partial_\mu(\partial \cdot A^a) +A_\mu^a \partial_\nu(\partial \cdot A^a)\right] -{1 \over \xi}g_{\mu \nu}
\left[- \frac{1}{2} (\partial \cdot A)^2 + \partial^\rho (A_\rho^a \partial \cdot A^a)\right], \\
T^{gh}_{\mu\nu} &=& \partial_{\mu}\bar\omega^a D^{ab}_{\nu}\omega^b + \partial_{\nu}\bar\omega^a D^{ab}_{\mu}\omega^b - g_{\mu\nu} \partial^{\rho}\bar\omega^a D^{ab}_{\rho}\omega^b.
\label{P2gfghost}
\eea

The coupling of QCD to gravity in the weak gravitational field limit is given by the interaction Lagrangian
\bea
\mathcal L_{int} = -\frac{1}{2} \kappa\, h^{\mu\nu} T_{\mu \nu}.
\eea
Notice that $T_{\mu \nu}$ as defined in Eq.~(\ref{P2EMT}) is symmetric, while traceless for a massless theory. The symmetric expression can be easily found as suggested in \cite{Freedman:1974gs}, by coupling the theory to gravity and then defining it via a functional derivative with respect to the metric, recovering (\ref{P2EMT}) in the flat spacetime case.

The conservation equation of the energy-momentum tensor takes the following form off-shell \cite{Nielsen:1975ph,Caracciolo:1989pt}
\bea
\partial^{\mu}T_{\mu\nu} &=& -\frac{\delta S}{\delta \psi} \partial_{\nu}\psi - \partial_{\nu}\bar \psi \frac{\delta S}{\delta \bar \psi} + \frac{1}{2}\partial^{\mu}\left( \frac{\delta S}{\delta \psi}\sigma_{\mu\nu}\psi - \bar \psi \sigma_{\mu\nu}\frac{\delta S}{\delta \bar \psi} \right) - \partial_{\nu}A_{\mu}^a \frac{\delta S}{\delta A_{\mu}^a} \nn\\
&+& \partial_{\mu}\left( A_{\nu}^a \frac{\delta S}{\delta A_{\mu}^a} \right) - \frac{\delta S}{\delta \omega^a} \partial_{\nu}\omega^a - \partial_{\nu}\bar \omega^a\frac{\delta S}{\delta \bar \omega^a} \,  \label{P2EMTdivergence}
\eea
where $\sigma_{\mu\nu}= \frac{1}{4}[\gamma_{\mu},\gamma_{\nu}]$. It is indeed conserved by using the equations of motion of the ghost, antighost and fermion/antifermion fields. The off-shell relation is particularly useful, since it can be inserted into the functional integral in order to derive some of the Ward identities satisfied by the correlator. In fact, the implications of the conservation of the energy-momentum tensor on the Green's functions can be exploited through the generating
functional, obviously defined as
\bea
&& Z[J,\eta,\bar\eta,\chi,\bar\chi,h] =
\int \mathcal D A \, \mathcal D \psi \, \mathcal D \bar\psi \,
\mathcal D \omega \, \mathcal D \bar \omega \, \exp\bigg\{ i \int d^4
x \left( \mathcal{L} + J_{\mu}A^{\mu} \right. \nn \\
&& \hspace{6cm}  \left.  + \bar \eta \psi +
\bar \psi \eta + \bar \chi \omega + \bar \omega \chi +
h_{\mu\nu}T^{\mu\nu}\right) \bigg\},
\eea
where $\mathcal{L}$ is the standard QCD action and we have added the coupling of the
energy-momentum tensor of the theory to the background gravitational field $h_{\mu\nu}$, which is the typical expression needed in the study of QCD coupled to gravity with a linear deviation from the flat metric. We have denoted with $J,\eta,\bar\eta,\chi,\bar\chi$ the
sources of the gauge field $A$ ($J$), the source of the fermion and antifermion fields ($\bar{\eta}, \eta$) and of the ghost and antighost fields ($\bar{\chi}$, $\chi$) respectively. The generating functional $W$ of the connected Green's functions is, as usual, denoted by \bea \exp i \,
W[J,\eta,\bar\eta,\chi,\bar\chi,h] =
\frac{Z[J,\eta,\bar\eta,\chi,\bar\chi,h]}{Z[0,0,0,0,0,0]}
\eea
(normalized to the vacuum functional) and the effective action is defined as the generating functional $\Gamma$ of the 1-particle irreducible and truncated amplitudes. This is obviously obtained from $W$ by a Legendre transformation respect to all the sources, except, in our case, $h_{\mu\nu}$, which is taken as a background external field
\bea
\Gamma[A_c,\bar \psi_c, \psi_c,
\bar \omega_c, \omega_c, h] = W[J,\eta,\bar\eta,\chi,\bar\chi,h] -
\int d^4 x \left( J_{\mu}A^{\mu}_c + \bar \eta \psi_c + \bar \psi_c \eta
+ \bar \chi \omega_c + \bar \omega_c \chi \right).
\label{P21PIfunctional}
\eea
The source fields are eliminated from the right hand side of Eq.~(\ref{P21PIfunctional}) inverting the relations
\bea
A^{\mu}_c = \frac{\delta W }{\delta J_{\mu}}, \qquad
\psi_c = \frac{\delta W }{\delta \bar \eta}, \qquad \bar \psi_c =
\frac{\delta W }{\delta \eta}, \qquad \omega_c = \frac{\delta W
}{\delta \bar \chi}, \qquad \bar \omega_c = \frac{\delta W }{\delta
\chi} \label{P2Legendre1}
\eea
so that the functional derivatives of the effective action $\Gamma$ with respect to its independent variables are
\bea
\frac{\delta\Gamma}{\delta A^{\mu}_c} = - J_{\mu}, \qquad
\frac{\delta\Gamma}{\delta \psi_c} = - \bar \eta, \qquad
\frac{\delta\Gamma}{\delta \bar \psi_c} = - \eta, \qquad
\frac{\delta\Gamma}{\delta \omega_c} = - \bar \chi, \qquad
\frac{\delta\Gamma}{\delta \bar \omega_c} = - \chi,
 \label{P2Legendre2}
\eea
and for the source $h_{\mu\nu}$ we have instead
\bea
\frac{\delta\Gamma}{\delta h_{\mu\nu}} = \frac{\delta W}{\delta h_{\mu\nu}}. \label{P2Legendre3}
\eea
The conservation of the energy-momentum tensor summarized in Eq.~(\ref{P2EMTdivergence}) in terms of classical fields, can be re-expressed
in a functional form by a differentiation of $W$ with respect to $h_{\mu\nu}$ and the use of Eq.~(\ref{P2EMTdivergence})
under the functional integral. We obtain
\bea
\partial_{\mu} \frac{\delta W}{\delta h_{\mu\nu}} &=&  \bar \eta \, \partial_{\nu} \frac{\delta W}{\delta \bar \eta} +
\partial_{\nu} \frac{\delta W}{\delta \eta} \eta - \frac{1}{2}\partial^{\mu}\left(\bar \eta \sigma_{\mu\nu} \frac{\delta W}{\delta \bar \eta}-
 \frac{\delta W}{\delta \eta}\sigma_{\mu\nu}\eta  \right) + \partial_{\nu} \frac{\delta W}{\delta J_{\mu}} J_{\mu}
 - \partial_{\mu} \left( \frac{\delta W}{\delta J_{\mu} }J_{\nu} \right) \nn \\
 &+& \bar \chi \partial_{\nu}\frac{\delta W}{\delta \bar \chi} + \partial_{\nu}\frac{\delta W}{\delta \chi} \chi \,,
\label{P2divW}
\eea
and finally, for the one particle irreducible generating functional, this gives
\bea
\partial_{\mu} \frac{\delta \Gamma}{\delta h_{\mu\nu}} &=& - \frac{\delta \Gamma}{\delta \psi_c}\partial^{\nu}\psi_c
- \partial^{\nu}\bar\psi_c \frac{\delta \Gamma}{\delta \bar \psi_c} + \frac{1}{2}\partial_{\mu}\left( \frac{\delta \Gamma}{\delta \psi_c} \sigma^{\mu\nu} \psi_c
- \bar \psi_c \sigma^{\mu\nu}\frac{\delta \Gamma}{\delta \bar \psi_c}  \right) \nn \\
&-& \partial^{\nu}A^{\mu}_c\frac{\delta \Gamma}{\delta A^{\mu}_c}
+\partial^{\mu}\left(A^{\nu}_c\frac{\delta \Gamma}{\delta A^{\mu}_c} \right) 
- \frac{\delta \Gamma}{\delta \omega_c}\partial^{\nu}\omega_c - \partial^{\nu}\bar \omega_c \frac{\delta \Gamma}{\delta \bar \omega_c}
\label{P2FuncWI},
\eea
obtained from Eq.~(\ref{P2divW}) with the help of Eqs.~(\ref{P2Legendre1}, (\ref{P2Legendre2}), (\ref{P2Legendre3}). \\
We summarize below the relevant Ward identities that can be used in order to fix the expression of the correlator.

\begin{itemize}
\item{\bf Single derivative general Ward identity}\\
The Ward identities describing the conservation of the energy-momentum tensor for the one-particle irreducible Green's functions with
an insertion of $T_{\mu\nu}$ can be obtained from the functional equation (\ref{P2FuncWI}) by taking functional derivatives with respect
to the classical fields. For example, the Ward identity for the graviton - gluon gluon vertex is obtained by differentiating Eq.~(\ref{P2FuncWI}) with respect to
$A_{c \, \alpha}^a(x_1)$ and $A_{c \, \beta}^b(x_2)$ and then setting all the external fields to zero
\bea
&& \partial^{\mu}\langle T_{\mu\nu}(x) A_\alpha^a (x_1)A_{\beta}^b (x_2)\rangle_{trunc}  = - \partial_{\nu}\delta^4(x_1-x) D^{-1}_{\alpha\beta}(x_2,x)
- \partial_{\nu}\delta^4(x_2-x) D^{-1}_{\alpha\beta}(x_1,x) \nn \\
&& + \partial^{\mu}\left( g_{\alpha \nu} \delta^4(x_1-x) D^{-1}_{\beta\mu}(x_2,x) + g_{\beta \nu} \delta^4(x_2-x) D^{-1}_{\alpha\mu}(x_1,x)\right) \label{P2WIcoord} 
\eea
where $D^{-1}_{\alpha\beta}(x_1,x_2)$ is the inverse gluon propagator defined as
\bea
D^{-1}_{\alpha\beta}(x_1,x_2) = \langle A_{\alpha}(x_1) A_{\beta}(x_2) \rangle_{trunc}  =  \frac{\delta^2 \Gamma}{\delta A^{\alpha}_c(x_1) \delta A^{\beta}_c(x_2) }
\eea
and where we have omitted, for simplicity, both the colour indices and the symbol of the $T$-product.
The first Ward identity (\ref{P2WIcoord}) becomes
\bea
k^{\mu}\langle T_{\mu\nu}(k) A_{\alpha}(p) A_{\beta}(q) \rangle_{trunc} =
q_{\mu} D^{-1}_{\alpha\mu}(p) g_{\beta\nu} + p_{\mu} D^{-1}_{\beta\mu}(q) g_{\alpha\nu}  - q_{\nu} D^{-1}_{\alpha\beta}(p) - p_{\nu} D^{-1}_{\alpha\beta}(q) \, . 
\label{P2WImom}
\eea

\item{\bf Trace Ward identity at zero momentum transfer}

It is possible to extract a Ward identity for the trace of the energy-momentum tensor for the same correlation function using just Eq.~(\ref{P2WImom}).
In fact, differentiating it with respect to $p_{\mu}$ (or $q_{\mu}$) and then evaluating the resulting expression at zero momentum transfer ($p = - q$) we obtain the Ward identity in $d$ spacetime dimensions
\bea
\langle T^{\mu}_{\mu}(0) A_{\alpha}(p) A_{\beta}(-p) \rangle_{trunc} = \left(2 -d + p\cdot \frac{\partial}{\partial p} \right) D^{-1}_{\alpha\beta}(p)
\label{P2WItrace}
\eea
where the number $2$ counts the number of external gluon lines. For $d = 4$ and using the transversality of the one-particle irreducible self-energy, namely
\bea
D^{-1}_{\alpha\beta}(p) = (p^2 g_{\alpha \beta} - p_{\alpha} q_{\beta}) \Pi (p^2),
\eea
the Ward identity in Eq.~(\ref{P2WItrace}) simplifies to
\bea
\langle T^{\mu}_{\mu}(0) A_{\alpha}(p) A_{\beta}(-p) \rangle_{trunc} = 2 p^2(p^2 g_{\alpha \beta} - p_{\alpha} q_{\beta}) \frac{d \Pi}{d p^2}(p^2).
\eea
The trace Ward identity in Eq.~(\ref{P2WItrace}) at  zero momentum transfer  can also be explicitly related to the $\b$ function and the anomalous dimensions of the renormalized theory. These enter through the renormalization group equation for the two-point function of the gluon.
Defining the beta function and the anomalous dimensions  as
\bea
\beta(g) = \mu \frac{\partial g}{\partial \mu}, \qquad \gamma(g) = \mu \frac{\partial}{\partial \mu}\log\sqrt{Z_A} , \qquad m\, \gamma_m(g) = \mu \frac{\partial m}{\partial \mu}
\eea
and denoting with $Z_A$ the wave function renormalization constant of the gluon field, with $g$  the renormalized coupling, and with $m$ the renormalized mass, the trace Ward identity can be related to these functions by the relation
\bea
\langle T^{\mu}_{\mu}(0) A_{\alpha}(p) A_{\beta}(-p) \rangle_{trunc} = \left[ \beta(g)\frac{\partial}{\partial g} - 2 \gamma(g) + m (\gamma_m(g) -1) \frac{\partial}{\partial m}   \right] D^{-1}_{\alpha\beta}(p).
\label{P2traceJJ}
\eea

\item{\bf Two-derivatives Ward identity via BRST symmetry}
\end{itemize}
We can exploit the BRST symmetry of the gauge-fixed lagrangian in order to derive some generalized Ward (Slavnov-Taylor) identities. We start by computing the BRST variation of the energy-momentum tensor, which is given by
\bea
\delta A_{\mu}^a &=& \lambda D_{\mu}^{ab} \omega^b, \label{P2brsA} \\
\delta \omega^a &=& - \frac{1}{2} g \lambda f^{abc}\omega^b \omega^c, \label{P2brsO} \\
\delta \bar \omega^a &=& -\frac{1}{\xi}(\partial^{\mu} A_{\mu}^a)\lambda, \label{P2brsOb} \\
\delta \psi &=&  i g \lambda \omega^a t^a \psi, \\
\delta \bar \psi &=& - i g \bar \psi t^a \lambda \omega^a,
\eea
where $\lambda$ is an infinitesimal Grassmann parameter.\\
A careful analysis of the energy-momentum tensor presented in Eq.~(\ref{P2EMT}) shows that the fermionic and the gauge part are gauge invariant and therefore invariant also under BRST. Instead the gauge-fixing and the ghost contributions must be studied in more detail. Using the transformation equations (\ref{P2brsA}) and (\ref{P2brsOb}) in (\ref{P2gfghost}) one can prove the two identities
\bea
\lambda \, T^{g.f.}_{\mu\nu} &=& - A^{a}_{\nu} \partial_{\mu} \delta \bar\omega^a - A^{a}_{\mu} \partial_{\nu} \delta \bar\omega^a + g_{\mu\nu} \left[\frac{1}{2}\partial\cdot A^a \delta \bar\omega^a + A_{\rho}^a\partial^{\rho} \delta\bar\omega^a \right], \label{P2BRS_Tgf}\\
\lambda \, T^{gh}_{\mu\nu} &=& - \partial_{\mu}\bar \omega^a \delta A^a_{\nu} - \partial_{\nu}\bar \omega^a \delta A^a_{\mu} + g_{\mu\nu} \partial^{\rho}\bar\omega^a \delta A_{\rho}^a,
\label{P2BRS_Tgh}
\eea
which show that the ghost and the gauge-fixing parts of the energy-momentum tensor (times the anticommuting factor $\lambda$) can be written as an appropriate BRST variation of ghost/antighost and gauge contributions.
Their sum, instead, can be expressed as the BRST variation of a certain operator plus an extra term which vanishes when using the ghost equations of motion
\bea
&& \lambda \left( T^{g.f.}_{\mu\nu} + T^{gh}_{\mu\nu} \right) = \delta \left[ - \partial_{\mu}\bar \omega^a A_{\nu}^a - \partial_{\nu}\bar \omega^a A_{\mu}^a + g_{\mu\nu} \left( A_{\rho}^a \partial_{\rho} \bar \omega^a + \frac{1}{2} \partial\cdot A^a \omega^a \right) \right]  \nn \\
&& + g_{\mu\nu} \frac{1}{2} \lambda \bar\omega^a \partial^{\rho} D^{ab}_{\rho}\omega^b, 
\eea
which shows explicitly the structure of the gauge-variant terms in the energy-momentum tensor. Using the nilpotency of the
BRST operator ($\delta^2=0$), the BRST variation of $T_{\mu\nu}$ is given by
\bea
\delta T_{\mu\nu} = \delta (T_{\mu\nu}^{g.f.} + T_{\mu\nu}^{gh}) = \frac{\lambda}{\xi}\left[ A_{\mu}^a \partial_{\nu} \partial^{\rho} D^{ab}_{\rho}\omega^b + A_{\nu}^a \partial_{\mu} \partial^{\rho} D^{ab}_{\rho}\omega^b - g_{\mu\nu}\partial^{\sigma}(A_{\sigma}^a \partial^{\rho}D^{ab}_{\rho}\omega^b ) \right],
\eea
where it is straightforward to recognize the equation of motion of the ghost field on its right-hand side.
Using this last relation, we are able to derive some constraints on the Green's functions involving insertions of the energy-momentum tensor. In particular, we are interested in some identities satisfied by the correlator $\langle T_{\mu\nu} A_{\alpha}^a A_{\beta}^b \rangle$ in order to define it unambiguously. For this purpose, it is convenient to choose an appropriate  Green's function, in our case this is given by $\langle T_{\mu\nu} \partial^{\alpha} A_{\alpha}^a \bar \omega^b \rangle$, and then exploit its BRST invariance to obtain
\bea
\delta \langle T_{\mu\nu} \partial^{\alpha} A_{\alpha}^a \bar \omega^b \rangle = \langle \delta T_{\mu\nu} \partial^{\alpha} A_{\alpha}^a \bar \omega^b \rangle + \lambda \langle T_{\mu\nu} \partial^{\alpha} D^{ac}_{\alpha}\omega^c \bar \omega^b \rangle - \frac{\lambda}{\xi} \langle T_{\mu\nu} \partial^{\alpha} A_{\alpha}^a \partial^{\beta} A_{\beta}^b \rangle = 0,
\label{P2BRSTward}
\eea
where the first two correlators, built with operators proportional to the equations of motion, contribute only with disconnected amplitudes, that are not part of the one-particle irreducible vertex function.
 From Eq.~(\ref{P2BRSTward}) we obtain the identity
 \bea
\partial_{x_1}^{\alpha}\partial_{x_2}^{\beta}\langle T_{\mu\nu}(x) A_{\alpha}^a(x_1)A_{\beta}^b(x_2)\rangle_{trunc} = 0,
\eea
which in momentum space becomes
\bea
p^{\alpha} q^{\beta} \langle T_{\mu\nu}(k) A_{\alpha}^a(p)A_{\beta}^b(q)\rangle_{trunc} = 0. \label{P2brstwi}
\label{P2pract1}
\eea
A subtlety in these types of derivations concerns the role played by the commutators, which are generated because of the T-product and can be ignored only if they vanish. In general, in fact, the derivatives
are naively taken out of the correlator, in order to arrive at Eq.~ (\ref{P2pract1}) and this can generate an error. In this case, due to the presence of an energy momentum tensor, the evaluation of these terms is rather involved. For this reason one needs to perform an explicit check of (\ref{P2pract1}) to ensure the consistency of the formal result in a suitable regularization scheme.
As we are going to show in the next sections, these three Ward identities turn out to be satisfied in dimensional regularization.

\section{The perturbative expansion}
\begin{figure}[t]
\begin{center}
\includegraphics[scale=0.9]{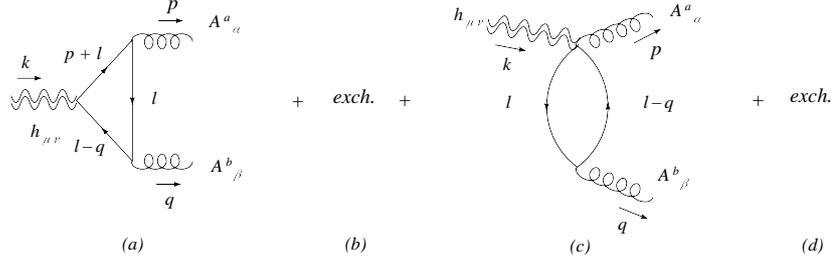}
\caption{\small The fermionic contributions  with a graviton $h_{\mu\nu}$ in the initial state and two gluons $A^a_\a, A^b_\b$ in the final state. }
\label{P2fermloop}
\end{center}
\end{figure}

\begin{figure}[t]
\begin{center}
\includegraphics[scale=0.9]{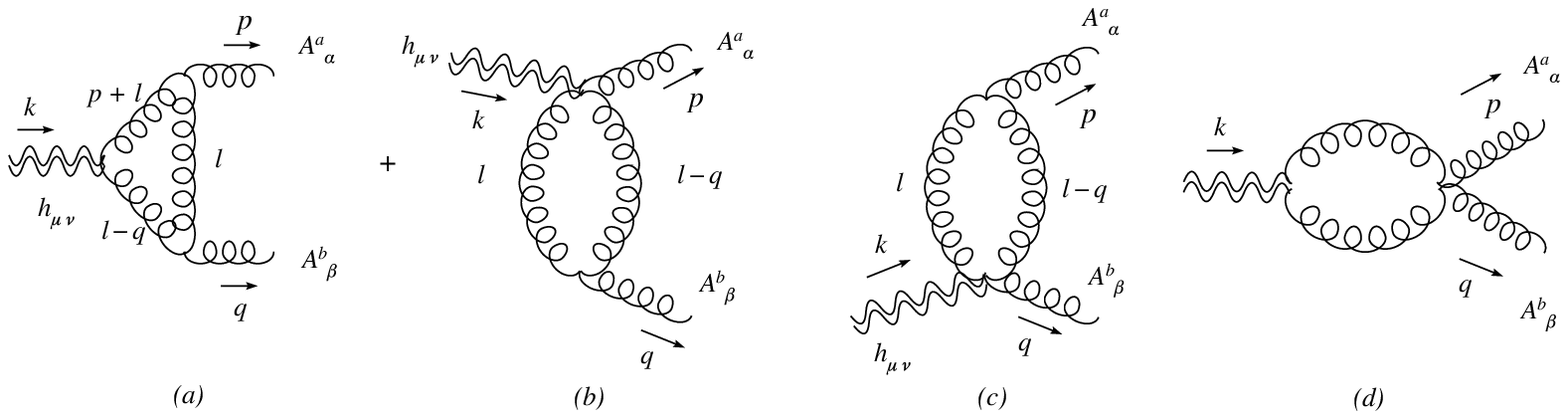}
\caption{\small The gauge contributions  with a graviton $h_{\mu\nu}$ in the initial state and two gluons $A^a_\a, A^b_\b$ in the final state. }
\label{P2gluonloop}
\end{center}
\end{figure}

\begin{figure}[t]
\begin{center}
\includegraphics[scale=0.9]{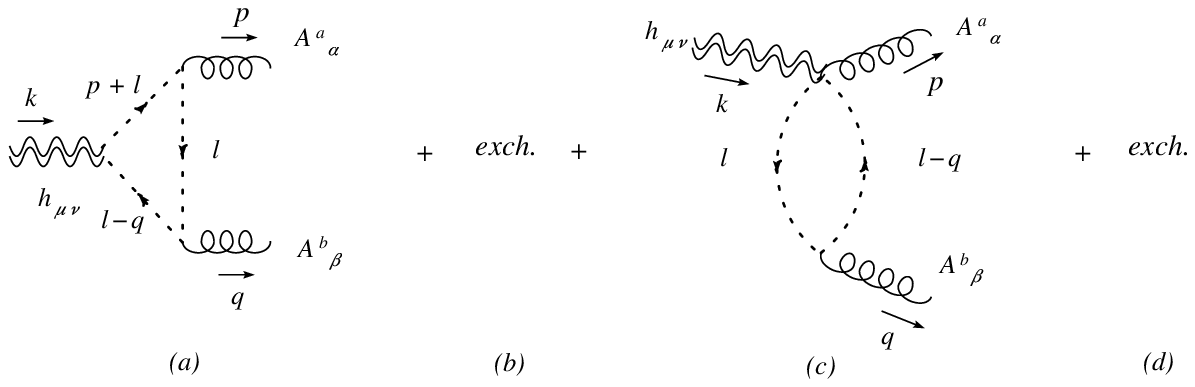}
\caption{\small The ghost contributions  with a graviton $h_{\mu\nu}$ in the initial state and two gluons $A^a_\a, A^b_\b$ in the final state. }
\label{P2ghostloop}
\end{center}
\end{figure}

The perturbative expansion is obtained by taking into account all the diagrams depicted in Figs.~\ref{P2fermloop}, \ref{P2gluonloop}, \ref{P2ghostloop}, where an incoming graviton appears in the initial state and two gluons with momenta $p$ and $q$ characterize the final state. The different contributions to the total amplitude are identified by the nature of the internal lines and are computed with the aid of the Feynman rules defined  in Appendix \ref{P2rules}. Each amplitude is denoted by $\Gamma$, with a superscript in square brackets indicating the figure of the corresponding diagram: the superscript $2$ refers to Fig.~\ref{P2fermloop}, $3$ refers to Fig.~\ref{P2gluonloop} and $4$ to Fig.~\ref{P2ghostloop}.

The contributions with  a massive fermion running in the loop are depicted in  Fig.~\ref{P2fermloop}; for the triangle in Fig.~\ref{P2fermloop}a we obtain
\bea
- i \frac{\kappa}{2}\, \Gamma^{{\bf[2a]} \,ab}_{\mu\nu\alpha\beta}(p,q) &=& \nn \\
&& \hspace{-3.cm} - \frac{\kappa}{2} g^2 \,{\rm Tr}(T^b T^a) \int\frac{d^4 l}{(2\pi)^4} {\rm tr}\left\{V^{\prime}_{\mu\nu}(l-q,l+p)\frac{1}{\lsl-\qsl-m}\gamma_{\beta}\frac{1}{\lsl-m}\gamma_{\alpha}\frac{1}{\lsl+\psl-m}  \right\} 
\eea
where the color factor is given by ${\rm Tr}(T^b T^a) = \frac{1}{2} \delta^{a b}$; the diagram in Fig.~\ref{P2fermloop}c contributes as
\bea
- i \frac{\kappa}{2} \, \Gamma^{{\bf[2c]} \, ab}_{\mu\nu\alpha\beta}(p,q) = - \frac{\kappa}{2} g^2 \,{\rm Tr}(T^a T^b) \int\frac{d^4 l}{(2\pi)^4} {\rm tr}\left\{W^{\prime}_{\mu\nu\alpha}\frac{1}{\lsl-\qsl-m}\gamma_{\beta}\frac{1}{\lsl-m} \right\},
\eea
with the vertices $V^{\prime}_{\mu\nu}(l-q,l+p)$ and $W^{\prime}_{\mu\nu\alpha}$ defined in Appendix \ref{P2rules}, Eqs.~(\ref{P2VGff}) and (\ref{P2WGffg}) respectively.
The remaining diagrams in Fig.~\ref{P2fermloop} are obtained by exchanging
$\alpha \leftrightarrow \beta$ and $p \leftrightarrow q$
\bea
- i \, \frac{\kappa}{2} \, \Gamma^{{\bf[2b]} \,ab}_{\mu\nu\alpha\beta}(p,q) &=&
- i \, \frac{\kappa}{2} \, \Gamma^{{\bf[2a]} \,ab}_{\mu\nu\alpha\beta}(p,q)\bigg|_{
\bmi[c]{.2\linewidth}
\footnotesize{$\alpha \leftrightarrow \beta$}
\vspace{-.2cm} \\
 \footnotesize {$p \leftrightarrow q$}
 \emi
}
\hspace{-2cm},
\\ \nn \\
 - i \, \frac{\kappa}{2} \, \Gamma^{{\bf[2d]} \,ab}_{\mu\nu\alpha\beta}(p,q) &=&
 - i \, \frac{\kappa}{2} \, \Gamma^{{\bf[2c]} \,ab}_{\mu\nu\alpha\beta}(p,q)\bigg|_{
 \bmi[c]{.2\linewidth}
 \footnotesize{$\alpha \leftrightarrow \beta$}
 \vspace{-.2cm} \\
  \footnotesize {$p \leftrightarrow q$}
  \emi
} \hspace{-2cm}.
\eea
Moving to the gauge sector we find the four contributions in Fig.~\ref{P2gluonloop}: the first one with a triangular topology is given by
\bea
- i \frac{\kappa}{2} \, \Gamma^{{\bf[3a]} \,ab}_{\mu\nu\alpha\beta}(p,q) = - \frac{\kappa}{2} g^2 f^{ade}f^{bde} \int\frac{d^4 l}{(2\pi)^4} \frac{1}{l^2\,(l+p)^2\,(l-q)^2} \bigg[V^{Ggg}_{\mu\nu\rho\sigma}(l-q,-l-p)\,\times  \nn \\
 V^3_{\tau\sigma\alpha}(-l,l+p,-p)\, \, V^3_{\rho\tau\beta}(-l+q,l,-q)\bigg],
\eea
where the color factor is $f^{ade}f^{bde} = C_A \, \delta^{a b}$. Those in Figs.~\ref{P2gluonloop}b and \ref{P2gluonloop}c, containing
gluon loops attached to the graviton vertex, are called  ``t-bubbles'' and can be obtained one from the other by the exchange of $\alpha \leftrightarrow \beta$ and $p \leftrightarrow q$. The first ``t-bubble'' is given by
\bea
- i \frac{\kappa}{2} \, \Gamma^{{\bf[3b]} \,ab}_{\mu\nu\alpha\beta}(p,q) = - \frac{1}{2}\frac{\kappa}{2} g^2 f^{ade}f^{bde} \int\frac{d^4 l}{(2\pi)^4} \frac{V^{Gggg}_{\mu\nu\rho\sigma\beta}(-l,l-p,-q)\,V^3_{\rho\alpha\sigma}(k,-p,-l+p)}{l^2\,(l-p)^2}
\eea
which is multiplied by an additional symmetry factor $\frac{1}{2}$. There is another similar contribution obtained from the previous one after exchanging $\alpha \leftrightarrow \beta$ and $p \leftrightarrow q$
\bea
- i \frac{\kappa}{2} \, \Gamma^{{\bf[3c]} \,ab}_{\mu\nu\alpha\beta}(p,q) =
- i \frac{\kappa}{2} \, \Gamma^{{\bf[3b]} \,ab}_{\mu\nu\alpha\beta}(p,q)\bigg|_{
 \bmi[c]{.25\linewidth}
 \footnotesize{$\alpha \leftrightarrow \beta$}
 \vspace{-.2cm} \\
  \footnotesize {$p \leftrightarrow q$}
  \emi
}\hspace{-3cm}.
\eea
The last diagram with gluons running in the loop is the one in Fig.~\ref{P2gluonloop}d which is given by
\bea
- i \frac{\kappa}{2} \, \Gamma^{{\bf[3d]} ab}_{\mu\nu\alpha\beta}(p,q) = \frac{1}{2}\frac{\kappa}{2} g^2 \int\frac{d^4 l}{(2\pi)^4} \frac{
V^{Ggg}_{\, \mu\nu\rho\sigma}(-l,l-p-q)\, \delta^{d f}\, \, V^{4\,abcd}_{\, \rho\alpha\sigma\beta} }{l^2\,(l-p-q)^2},
\label{P2gluond}
\eea
where $V^4$ is the four gluon vertex defined as
\bea
-i g^2 V^{4\,abcd}_{\mu\nu\rho\sigma} &=&
-i g^2\left[ f^{abe}f^{cde}(g_{\mu\rho}g_{\nu\sigma} - g_{\mu\sigma}g_{\nu\rho})
+ f^{ace}f^{bde}(g_{\mu\nu}g_{\rho\sigma} - g_{\mu\sigma}g_{\nu\rho})  \right. \nn  \\
&& \hspace{2cm} \left. + \, f^{ade}f^{bce}(g_{\mu\nu}g_{\rho\sigma} - g_{\mu\rho}g_{\nu\sigma}) \right]
\eea
and therefore
\bea
\delta^{d f}\,V^{4\,abcd}_{\rho\alpha\sigma\beta} &=& - C_A \delta^{a b} \tilde{V}^4_{\rho\alpha\sigma\beta} = - C_A \delta^{a b} \left( g_{\alpha\sigma}g_{\beta\rho} +  g_{\alpha\rho}g_{\beta\sigma} - 2 g_{\alpha\beta}g_{\sigma\rho} \right),
\eea
so that the amplitude in Eq.~(\ref{P2gluond}) becomes
\bea
- i \frac{\kappa}{2} \, \Gamma^{{\bf[3d]}\, ab}_{\mu\nu\alpha\beta}(p,q) = - \frac{1}{2}\frac{\kappa}{2} g^2 C_A \delta^{a b} \int\frac{d^4 l}{(2\pi)^4} \frac{V^{Ggg}_{\mu\nu\rho\sigma}(-l,l-p-q)\, \tilde{V}^4_{\rho\alpha\sigma\beta} }{l^2\,(l-p-q)^2}.
\eea
In the expression above we have explicitly isolated the color factor $C_A \delta^{a b}$ and the symmetry factor $\frac{1}{2}$.

Finally, the ghost contributions shown in Fig.~\ref{P2ghostloop} are  given by the sum of
\bea
- i \frac{\kappa}{2} \, \Gamma^{{\bf[4a]}\, ab}_{\mu\nu\alpha\beta}(p,q) = - \frac{\kappa}{2} g^2 f^{ade}f^{bde} \int\frac{d^4 l}{(2\pi)^4} \frac{C_{\mu\nu\rho\sigma}(l-q)^{\rho}(l+p)^{\sigma}l_{\alpha}(l-q)_{\beta} }{l^2\, (l+p)^2 \, (l-q)^2}
\label{P2ghosta}
\eea
for the triangle diagram in Fig.~\ref{P2ghostloop}a and
\bea
- i \frac{\kappa}{2} \, \Gamma^{{\bf[4b]} \, ab}_{\mu\nu\alpha\beta}(p,q) = \frac{\kappa}{2} g^2 f^{ade}f^{bde} \int\frac{d^4 l}{(2\pi)^4} \frac{C_{\mu\nu\alpha\sigma}l^{\sigma}(l-q)_{\beta} }{l^2 \, (l-q)^2}
\label{P2ghostb}
\eea
for the ``T-bubble" diagram shown in Fig.~\ref{P2ghostloop}c. The two exchanged diagrams are obtained from those in Eqs.~(\ref{P2ghosta}) and (\ref{P2ghostb}) with the usual replacement $\alpha \leftrightarrow \beta$ and $p \leftrightarrow q$.
\bea
- i \, \frac{\kappa}{2} \, \Gamma^{{\bf[4b]} \,ab}_{\mu\nu\alpha\beta}(p,q) &=&
- i \, \frac{\kappa}{2} \, \Gamma^{{\bf[4a]} \,ab}_{\mu\nu\alpha\beta}(p,q)\bigg|_{
\bmi[c]{.2\linewidth}
\footnotesize{$\alpha \leftrightarrow \beta$}
\vspace{-.2cm} \\
 \footnotesize {$p \leftrightarrow q$}
 \emi
}
\hspace{-2cm},
\\ \nn \\
 - i \, \frac{\kappa}{2} \, \Gamma^{{\bf[4d]} \,ab}_{\mu\nu\alpha\beta}(p,q) &=&
 - i \, \frac{\kappa}{2} \, \Gamma^{{\bf[4c]} \,ab}_{\mu\nu\alpha\beta}(p,q)\bigg|_{
 \bmi[c]{.2\linewidth}
 \footnotesize{$\alpha \leftrightarrow \beta$}
 \vspace{-.2cm} \\
  \footnotesize {$p \leftrightarrow q$}
  \emi
} \hspace{-2cm}.
\eea
Having identified the different sectors we obtain the total amplitude for quarks, denoted by a ``q'' subscript
\bea
\Gamma_{q, \, \mu\nu\a\b}^{ab}(p,q) =
\Gamma^{{\bf[2a]} \,ab}_{\mu\nu\alpha\beta}(p,q) +
\Gamma^{{\bf[2b]} \,ab}_{\mu\nu\alpha\beta}(p,q) +
\Gamma^{{\bf[2c]} \,ab}_{\mu\nu\alpha\beta}(p,q) +
\Gamma^{{\bf[2d]} \,ab}_{\mu\nu\alpha\beta}(p,q)
\label{P2Gammaq}
\eea
and the one for gluons and ghosts as
\bea
\Gamma_{g, \, \mu\nu\a\b}^{ab}(p,q) =
\sum_{j=3,4} \, \left[ \Gamma^{{\bf[ja]} \,ab}_{\mu\nu\alpha\beta}(p,q) +
\Gamma^{{\bf[jb]} \,ab}_{\mu\nu\alpha\beta}(p,q) +
\Gamma^{{\bf[jc]} \,ab}_{\mu\nu\alpha\beta}(p,q) +
\Gamma^{{\bf[jd]} \,ab}_{\mu\nu\alpha\beta}(p,q) \right].
\label{P2Gammag}
\eea
\section{The on-shell correlator, pole terms and form factors}
We proceed with a classification of all the diagrams contributing to the on-shell vertex, starting from the gauge invariant subset of diagrams that involve fermion loops and then moving to the second set, the one relative to gluons and ghosts. The analysis follows rather closely the method presented in the case of QED in previous works \cite{Giannotti:2008cv, Armillis:2009pq} and in the previous chapter, with a classification of all the relevant tensor structures which can be generated using the 43 monomials built out of the 2 of the 3 external momenta of the triangle diagram and the metric tensor $g_{\mu\nu}$. In general, one can proceed
with the identification of a subset of these tensor structure which allow to formulate the final expression in a manageable form. The fermionic triangle diagrams, which define one of the two gauge invariant subsets of the entire correlator, can be given in a simplified form also for off mass-shell external momenta, in terms of 13 form factors as in \cite{Giannotti:2008cv, Armillis:2009pq} while the structure of the gluon contributions are more involved. Some drastic simplifications take place in the on-shell case, where only 3 form factors - both in the quark and fermion sectors - are necessary to describe the final result.

We write the whole amplitude $\Gamma^{\mu\nu\a\b}(p,q)$ as
\bea
\Gamma^{\mu\nu\a\b}(p,q) =
\Gamma_q^{\mu\nu\a\b}(p,q) + \Gamma_g^{\mu\nu\a\b}(p,q),
\eea
 referring respectively to the contributions with quarks $(\Gamma_q)$ and with gluons/ghosts $(\Gamma_g)$ in Eqs.~(\ref{P2Gammaq}) and (\ref{P2Gammag}). We have omitted the color indices for simplicity. The amplitude $\Gamma$ is expressed in terms of 3 tensor structures and 3 form factors renormalized in the $\overline{MS}$ scheme
\beq
\Gamma^{\mu\nu\alpha\beta}_{q/g}(p,q) =  \, \sum_{i=1}^{3} \Phi_{i\,q/g} (s,0, 0,m^2)\, \delta^{ab}\, \phi_i^{\mu\nu\alpha\beta}(p,q)\,.
\label{P2Gamt}
\eeq
One comment concerning the choice of this basis is in order. The 3 form factors are more easily identified in the fermion sector after performing the on-shell limit of the off-shell amplitude, where the 13 form factors introduced in \cite{Giannotti:2008cv, Armillis:2009pq} for QED simplify into the 3 tensor structures that will be given below. It is then observed that the  tensor structure of the gluon sector, originally expressed in terms of the 43 monomials of  \cite{Giannotti:2008cv, Armillis:2009pq}, can be arranged consistently in terms of these 3 reduced structures.

The tensor basis on which we expand the on-shell vertex is given by
\bea
  \phi_1^{\, \mu \nu \a \b} (p,q) &=&
 (s \, g^{\mu\nu} - k^{\mu}k^{\nu}) \, u^{\a \b} (p,q),
 \label{P2widetilde1}\\
\phi_2^{\, \mu \nu \a \b} (p,q) &=& - 2 \, u^{\a \b} (p,q) \left[ s \, g^{\mu \nu} + 2 (p^\mu \, p^\nu + q^\mu \, q^\nu )
- 4 \, (p^\mu \, q^\nu + q^\mu \, p^\nu) \right],
\label{P2widetilde2} \\
\phi^{\, \mu \nu \alpha \beta}_{3} (p,q) &=&
\big(p^{\mu} q^{\nu} + p^{\nu} q^{\mu}\big)g^{\alpha\beta}
+ \frac{s}{2} \left(g^{\alpha\nu} g^{\beta\mu} + g^{\alpha\mu} g^{\beta\nu}\right) \nn \\
&&  \hspace{1cm} - g^{\mu\nu} \left(\frac{s}{2} g^{\alpha \beta}- q^{\alpha} p^{\beta}\right)
-\left(g^{\beta\nu} p^{\mu}
+ g^{\beta\mu} p^{\nu}\right)q^{\alpha}
 - \big (g^{\alpha\nu} q^{\mu}
+ g^{\alpha\mu} q^{\nu }\big)p^{\beta},
\label{P2widetilde3}
\eea
where $u^{\a \b} (p,q)$ has been defined in Eq.~(\ref{P2utensor}).
The form factors $\Phi_i(s,s_1,s_2,m^2)$ have as entry variables, beside $s=(p+q)^2$, the virtualities of the two gluons $s_1=p^2$ and $s_2=q^2$.\\
In the on-shell case only 3 invariant amplitudes contribute, which for the quark loop amplitude are given by
\bea
\Phi_{1\, q} (s,\,0,\,0,\,m^2) &=&
- \frac{g^2 }{36 \pi^2  s} \,  + \, \frac{g^2  m^2}{6 \pi^2 s^2} \, - \, \frac{g^2\, m^2} {6 \pi^2 s}\mathcal C_0 (s, 0, 0, m^2)
\bigg[\frac{1}{2 \,  }-\frac{2 m^2}{ s}\bigg],  \\
\Phi_{2\, q} (s,\,0,\,0,\,m^2)  &=&
- \frac{g ^2}{288 \pi^2 s} - \frac{ g ^2 m^2}{24 \pi^2  s^2}
- \, \frac{g^2 \, m^2}{8 \pi^2  s^2} \mathcal D (s, 0, 0, m^2) \,
\nn \\
&& \hspace{2cm}- \, \frac{g^2 \, m^2}{12 \pi^2 s } \mathcal C_0(s, 0, 0, m^2 )\, \left[ \frac{1}{2} + \frac{m^2}{s}\right],  \\
\Phi_{3\,q} (s,\,0,\,0,\,m^2) &=& \frac{11  g ^2}{288  \pi^2 }  +   \frac{ g ^2 m^2}{8 \pi^2 s}
 + \, g^2\mathcal C_0 (s, 0,0,m^2) \,\left[ \frac{m^4}{4 \pi^2 s}+\frac{ m^2}{8 \pi^2 }\right] \nn \\
 && \hspace{2cm} +  \frac{5 \,g^2 \, m^2}{24 \pi^2  s}  \mathcal D (s, 0, 0, m^2) + \frac{g^2}{24 \pi^2} \mathcal B_0^{\overline{MS}}(s, m^2),
\label{P2masslesslimit}
\eea
where the on-shell scalar integrals $\mathcal D (s, 0, 0, m^2)$, $\mathcal C_0 (s, 0,0,m^2)$ and $B_0^{\overline{MS}}(s, m^2)$ are computed in Appendix \ref{P2scalars}. In the massless limit the amplitude $\Gamma_q^{\mu\nu\a\b}(p,q)$ takes a simpler expression and the previous form factors become
\bea
\Phi_{1\,q} (s, 0, 0, 0) &=& - \frac{g^2}{36 \pi^2  s}, \\
\Phi_{2\,q} (s, 0, 0, 0) &=& - \frac{g^2}{288 \pi^2 \, s}, \\
\Phi_{3\,q} (s, 0, 0, 0) &=& - \frac{g^2}{288 \pi^2} \, \left[ 12 L_s - 35\right],
\eea
\begin{figure}[t]
\begin{center}
\includegraphics[scale=1.0]{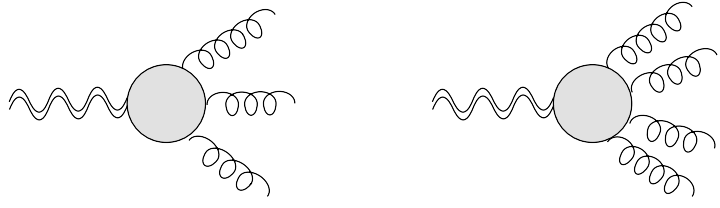}
\caption{\small Higher order contributions to the anomaly pole involved in the covariantization of the graviton/2-gluons amplitude. }
\label{P2additional}
\end{center}
\end{figure}

where
\bea
L_s \equiv \log \left( - \frac{s}{\mu^2} \right ) \qquad \qquad s<0.
\eea
In the gluon sector the computation of $\Gamma_g^{\mu\nu\a\b}(p,q)$ is performed analogously by using dimensional regularization with modified minimal subtraction ($\overline{MS}$) and we obtain for  on-shell gluons
\beq
\Gamma^{\mu\nu\alpha\beta}_{g}(p,q) =  \, \sum_{i=1}^{3} \Phi_{i\,g} (s,0,0)\,\delta^{ab}\, \phi_i^{\mu\nu\alpha\beta}(p,q)\, ,
\label{P2gl1}
\eeq
where the form factors obtained from the explicit computation are
\bea
\Phi_{1\,g}(s,0,0) &=& \frac{11 \, g^2}{72 \pi^2 \, s} \, C_A,\\
\Phi_{2\,g}(s,0,0) &=& \frac{g^2}{288 \pi^2 \, s} \, C_A, \\
\Phi_{3\,g}(s,0,0) &=& - \frac{g^2}{8 \pi^2} \, C_A \bigg[ \frac{65}{36} + \frac{11}{6} \mathcal B_0^{\overline{MS}}(s,0) - \mathcal B_0^{\overline{MS}}(0,0) +  s  \,\mathcal C_0(s,0,0,0) \bigg].
\label{P2gl2}
\eea
The renormalized scalar integrals can be found in Appendix \ref{P2scalars}.

The full on-shell vertex, which is the sum of the quark and pure gauge contributions, can be decomposed by using  the same three tensor structures $\phi_i^{\mu\nu\alpha\beta}$ appearing in the expansion of $\Gamma_q^{\mu\nu\a\b}(p,q)$ and $\Gamma_g^{\mu\nu\a\b}(p,q)$
\bea
\Gamma^{\mu\nu\alpha\beta}(p,q) =  \Gamma^{\mu\nu\alpha\beta}_g(p,q) + \Gamma^{\mu\nu\alpha\beta}_q(p,q) =  \sum_{i=1}^{3} \Phi_{i} (s,0, 0)\, \delta^{ab}\, \phi_i^{\mu\nu\alpha\beta}(p,q)\,,
\eea
with form factors defined as
\bea
\Phi_i(s,0,0) = \Phi_{i,\,g}(s,0,0) + \sum_{j=1}^{n_f}\Phi_{i, \,q}(s,0,0,m_j^2),
\eea
where the sum runs over the $n_f$ quark flavors. In particular we have
\bea
\Phi_{1}(s,0,0) &=& - \frac{g^2}{72 \pi^2 \,s}\left(2 n_f - 11 C_A\right) + \frac{g^2}{6 \pi^2}\sum_{i=1}^{n_f} m_i^2 \, \bigg\{ \frac{1}{s^2} \, - \, \frac{1} {2 s}\mathcal C_0 (s, 0, 0, m_i^2)
\bigg[1-\frac{4 m_i^2}{ s}\bigg] \bigg\}, \, \nn \\
\label{P2Phi1}\\
\Phi_{2}(s,0,0) &=& - \frac{g^2}{288 \pi^2 \,s}\left(n_f - C_A\right) \nn \\
&&- \frac{g^2}{24 \pi^2} \sum_{i=1}^{n_f} m_i^2 \, \bigg\{ \frac{1}{s^2}
+ \frac{ 3}{ s^2} \mathcal D (s, 0, 0, m_i^2)
+ \frac{ 1}{s } \mathcal C_0(s, 0, 0, m_i^2 )\, \left[ 1 + \frac{2 m_i^2}{s}\right]\bigg\},
\label{P2Phi2} \\
\Phi_{3}(s,0,0) &=& \frac{g^2}{288 \pi^2}\left(11 n_f - 65 C_A\right) - \frac{g^2 \, C_A}{8 \pi^2} \bigg[ \frac{11}{6} \mathcal B_0^{\overline{MS}}(s,0) - \mathcal B_0^{\overline{MS}}(0,0) +  s  \,\mathcal C_0(s,0,0,0) \bigg] \nn \\
&& + \, \frac{g^2}{8 \pi^2} \sum_{i=1}^{n_f}\bigg\{  \frac{1}{3}\mathcal B_0^{\overline{MS}}(s, m_i^2) + m_i^2 \, \bigg[
\frac{1}{s}   + \frac{5}{3 s}  \mathcal D (s, 0, 0, m_i^2)  \nn \\
&& + \mathcal C_0 (s, 0,0,m_i^2) \,\left[1 + \frac{2 m_i^2}{s}\right]
\bigg]\bigg\} ,
\label{P2Phi3}
\eea
with $C_A = N_C$ and the scalar integrals defined in Appendix \ref{P2scalars}.
Notice the appearance in the total amplitude of the $1/s$ pole in $\Phi_1$, which is present both in the quark and in the gluon sectors, and which saturates the contribution to the trace anomaly in the massless limit. In this case the entire trace anomaly is just proportional to this component, which becomes
\beq
\Phi_{1}(s,0,0) = - \frac{g^2}{72 \pi^2 \,s}\left(2 n_f - 11 C_A\right).
\label{P2polepole}
\eeq
The correlator $\Gamma^{\mu\nu\alpha\beta}(p,q)$, computed using dimensional regularization, satisfies all the Ward identities defined in the previous sections. Notice that the two-derivatives Ward identity introduced in Eq.~(\ref{P2brstwi})
\bea
p_{\a}  q_{\b} \, \Gamma^{\mu\nu\a\b}(p,q) = 0,
\eea
derived from the BRST symmetry of the QCD Lagrangian, is straightforwardly satisfied by the on-shell amplitude. This is easily seen from the tensor decomposition introduced in Eq.~(\ref{P2Gamt}) because all the tensors fulfill the condition
\bea
p_{\a}  q_{\b} \,  \phi_i^{\, \mu \nu \a \b} (p,q)  = 0.
\eea
Furthermore, we have checked at one-loop order the validity of the single derivative Ward identity given in Eq.~(\ref{P2WImom}) and describing the conservation of the energy-momentum tensor. Using the transversality of the two-point gluon function Eq.~(\ref{P2WImom}) this gives
\bea
k_{\mu} \, \Gamma^{\mu\nu\a\b}(p,q) =
\left( q^\nu \, p^\a \, p^\b - q^\nu \, g^{\a \b} \, p^2 + g^{\nu \b} \, q^\a \, p^2 - g^{\nu \b} \, p^\a \, p \cdot q \right) \, \Pi (p^2) \nn \\
+ \left( p^\nu \, q^\a \, q^\b - p^\nu \, g^{\a \b} \, q^2 + g^{\nu \a} \, p^\b \, q^2 - g^{\nu \a} \, q^\b \, p \cdot q \right) \, \Pi (q^2),
\eea
where the renormalized gluon self energies are defined as
\bea
\Pi (p^2) &=& \frac{g^2 \, C_A \, \d^{a b} \, }{144 \, \pi^2} \, \left( 15 \, \mathcal B_0^{\overline{MS}} (p^2, 0) -2 \right)  \nn \\
&+& \frac{g^2 \, \d^{ab}}{72 \, \pi^2 p^2}\sum_{i=1}^{n_f} \, \left[ 6 \, \mathcal A_0^{\overline{MS}} \,  (m_i^2) + p^2 - 6 \, m_i^2
- 3 \, \mathcal B_0^{\overline{MS}} (p^2, m_i^2) \left( 2 \, m_i^2 + p^2 \right)    \right]\, .
\label{P2Pigluon}
\eea
The QCD $\beta$ function can be related to the residue of the pole and can be easily computed starting from the amplitude  $\Gamma^{\mu\nu\alpha\beta}(p,q)$ for on-shell external lines and in the conformal limit
\bea
g_{\mu\nu}\,  \Gamma^{\mu\nu\alpha\beta}(p,q) =  3 \, s \, \Phi_1 (s; 0,0,0) \, u^{\a \b}(p,q) = - 2 \, \frac{\b (g)}{g}\, u^{\a \b}(p,q),
\eea
with the QCD $\b$ function given by
\bea
\b (g)= \frac{g^3}{16 \pi^2} \left (  - \frac{11}{3} \, C_A + \frac{2}{3} \, n_f \right) .
\eea
As we have already mentioned, after contracting the metric tensor $g_{\mu\nu}$ with the whole amplitude $\Gamma $, only the  tensor structure $\phi^{\mu\nu\a\b}_1(p,q)$ contributes to the anomaly,  being the remaining ones traceless, with a contribution entirely given by $\Phi_1|_{m=0}$  in Eq.~(\ref{P2Phi1}), i.e. Eq. (\ref{P2polepole}). In the massive fermion case, the anomalous contribution are corrected by terms proportional to the fermion mass $m$ and represent an explicit breaking of scale invariance. From a direct computation we can also extract quite straightforwardly the effective action, which is given by
\bea
S_{pole} &=&	- \frac{c}{6}\, \int d^4 x \, d^4 y \,R^{(1)}(x)\, \square^{-1}(x,y) \,  F^a_{\alpha \beta} \,  F^{a \, \alpha \beta} \nn\\
&=& \frac{1}{3} \, \frac{g^3}{16 \pi^2} \left (  - \frac{11}{3} \, C_A + \frac{2}{3} \, n_f \right)  \, \int d^4 x \, d^4 y \,R^{(1)}(x)\, \square^{-1}(x,y) \, F_{\alpha \beta}F^{\alpha \beta}
\eea
and is in agreement with Eq. (\ref{P2SSimple}), derived from the nonlocal gravitational action.
Here $R^{(1)}$ denotes the linearized expression of the Ricci scalar
\beq
 R^{(1)}_x\equiv \partial^x_\mu\, \partial^x_\nu \, h^{\mu\nu} - \square \,  h, \qquad h=\eta_{\mu\nu} \, h^{\mu\nu}
 \eeq
and the constant $c$ is related to the non-abelian $\beta$ function as
\beq
c= - 2 \, \, \frac{\beta (g)}{g}.
\eeq
Notice that the contribution coming from $TJJ$ generates the abelian part of the non-abelian field strength, while extra contributions (proportional to extra factors of $g$ and $g^2$) are expected from the $TJJJ$ and $TJJJJ$ diagrams (see Fig.~\ref{P2additional}). This situation is analogous to that of the gauge anomaly, where one needs to render gauge covariant the anomalous amplitude given by the triangle diagram. In that case the gauge covariant expression is obtained by adding to the $AVV$ vertex also the $AVVV$ and $AVVVV$ diagrams, with 3 and 4 external gauge lines, respectively. \\

The appearance of massless degrees of freedom in the effective action describing the coupling of gravity to the gauge fields
is rather intriguing, and is an aspect that will require further analysis. 

The nonlocal structure of the action that contributes to the trace anomaly, which is entirely reproduced, within the local description, by two auxiliary scalar fields, as pointed out in the previous chapter, seems to indicate that the effective dynamics of the coupling between gravity and matter might be controlled, at least in part, by these degrees of freedom. As we have just mentioned, however, this point requires a dedicated study and for this specific reason our conclusions remain open ended.

Our computation, however, being general, allows also the identification of other massless contributions to the effective action which are surely bound to play a role in the physical S-matrix. They appear in form factors such as $\Phi_2$ (Eq. \ref{P2Phi2}) 
and $\Phi_3$ (Eq. \ref{P2Phi3}) which do not contribute to the trace, but are nevertheless part of the 1-loop effective action mediated by the triangle graph.  

There are also some other comments, at this point, which are in order. Notice that while the isolation of the pole in the fermion sector indeed requires a massless fermion limit, as obvious from the structure of $\Gamma_q$, the other gauge invariant sector, described by $\Gamma_g$, is obviously not affected by this limit, being the corresponding form factors mass independent.

\section{Conclusions}
One of the standing issues of the anomalous effective action describing the interaction of a non-abelian theory to gravity is a test of its consistency with the standard perturbative approach. Thus, variational solutions of the effective action controlled by the trace anomaly should be reproduced by the perturbative expansion. Building on previous analysis in QED, reviewed in the previous chapter, here we have shown that also in the non-abelian case there is a perfect match between the two approaches. This implies that the interaction of gravity with a non-abelian gauge theory, mediated by the trace anomaly, indeed can be reformulated in terms of auxiliary scalar degrees of freedom, in analogy to the abelian case. We have proven this result by an explicit computation. Our findings indicate that
this feature is typical of each gauge invariant subsector of the non-abelian $TJJ$ amplitude, a result which is likely to hold also for singularities of higher order. These are expected to be present in correlators with a larger number of energy momentum insertions.  

\chapter{Gravity and the neutral currents in the electroweak theory}
\label{Chap.GravitonEW}

\section{Introduction}
In this chapter we present a complete study of the one graviton-two neutral gauge bosons vertex at one-loop level in the electroweak theory. This vertex provides the leading contribution to the interaction between the Standard Model and gravity, mediated by the trace anomaly, at first order in the inverse Planck mass and at second order in the electroweak expansion. 
We will show, in analogy with previous results in the QED and QCD cases, that the anomalous interaction between gravity and the gauge current of the Standard Model, due to the trace anomaly, is mediated, in each gauge invariant sector, by effective massless scalar degrees of freedom. 

%
The work presented in this chapter is the fourth in a sequence of investigations \cite{Armillis:2009pq,Armillis:2010qk,Armillis:2010pa}, motivated by 
the original analysis of \cite{Giannotti:2008cv}, aimed at studying the precise structure of the anomalous effective action which describes the anomalous breaking of scale invariance in the Standard Model (SM). Here we expand and fill in the details of a previous study \cite{Coriano:2011ti}. 
This breaking is induced by the trace anomaly \cite{Duff:1977ay,Duff:1993wm} and can be extracted from the exact computation of a set of diagrams involving
the graviton-gauge-gauge vertex. This work is a natural extension and an application of remarkable classical studies \cite{{Freedman:1974gs},Callan:1970ze,Adler:1976zt,Collins:1976yq} of the energy momentum tensor and of the corresponding trace anomaly in gauge theories. 

In the case of a gravitational background characterized by a small deviation with respect to the flat spacetime metric, this vertex is described by the correlation function containing one insertion of the energy momentum tensor (EMT) (denoted as $T$) on the correlation function of two gauge currents (denoted as $V, V'$). If we allow only conformally coupled scalars and
operators only up to dimension-4 in the Langrangian \cite{Freedman:1974gs} \cite{Callan:1970ze}, the EMT is uniquely defined by gravity and takes the form of a symmetric and (on-shell) conserved expression. In the massless limit, which in our case is equivalent to dealing with
an unbroken theory (i.e. before electroweak symmetry breaking) the EMT is classically (on-shell) traceless.

As remarked in \cite{Giannotti:2008cv} and in our previous studies in the context of QED \cite{Armillis:2009pq} and QCD \cite{Armillis:2010qk}, which have been extensively discussed in the previous chapters, the analysis of this correlator is interesting in several ways and allows to address some important issues concerning anomaly-mediated interactions between the SM and gravity. At the same time, this program is part of an attempt to characterize rigorously in quantum field theory the effective action which describes the interaction between matter and gravity beyond tree level, showing some interesting features, such as the appearance of effective massless scalar degrees of freedom as mediators of the breaking of scale invariance \cite{Giannotti:2008cv}, in close analogy with what found in the case of chiral gauge theories \cite{Armillis:2009sm,Armillis:2009im,Armillis:2008bg,Coriano:2008pg}. Beside these theoretical motivations, these corrections find also direct application in collider studies of low scale gravity.

In a theory such as the SM, the breaking of scale invariance is related both to the trace anomaly and to the spontaneous breaking of the gauge symmetry by the Higgs mechanism  \cite{Coriano:2011ti}, and both contributions may become significant in some specific scenarios. For example, the enduring discussion over the cosmological implications of the quantum breaking of scale invariance has spanned decades \cite{Dolgov:1993vg, Corradini:2007gd}, since the work of Starobinsky \cite{Starobinsky:1980te}, with his attempt to solve the problem of the cosmological "graceful exit" that predated inflationary studies. At the same time, the treatment of the trace anomaly using more refined
approaches such as the world-line formulation, has allowed for new ways to investigate the corresponding effective action
\cite{Bastianelli:2002qw}.

The computation of the effective action is, in principle,  rather challenging not only for the large number of diagrams involved, but also because of the need of a consistent way to define these interactions. The ambiguity
present in the definition of the fermion contributions, for instance, requires particular care, due to the presence
of axial-vector and vector currents in an external gravitational background. These have been analyzed building on the results of  \cite{Armillis:2010pa}, which provides the ground for the extensions contained in the present study. The current analysis is far more involved than the study discussed in the previous chapters, due to the appearance of a larger set of diagrams in the perturbative expansion. Their definition requires a suitable set of Ward and Slavnov-Taylor identities (STI's) which need to be identified from scratch and that we are going to discuss in fair detail. These are essential in order to establish the correctness of the computation and of the chosen regularization scheme, which is dimensional regularization with on-shell renormalization.

When we move from an exact gauge theory to a theory with spontaneous breaking of the gauge symmetry such as the SM, the contributions coming from the trace anomaly and from mass corrections are harder to disentangle, since the massless limit is not an option. However, even under these conditions, there are two possible ways of organizing the contributions to the one-loop effective action which may turn out handy. The first expansion, obviously, is the usual $1/m$ expansion, where $m$ is a large electroweak mass, valid below the electroweak scale. The second has been first discussed in a previous work \cite{Armillis:2009sm} and is characterized by the isolation of the anomalous massless pole contribution from the remaining subleading $O(m^2/s)$ corrections. These can be extracted from a complete computation.

The goal of this chapter is to discuss the role of the interactions mediated by the conformal anomaly  using as a realistic example the Lagrangian of the SM, by focusing our investigation on the neutral currents sector.  These contributions play a role, in general, also in scenarios of TeV scale gravity and as such are part of the radiative corrections to graviton-mediated processes at typical LHC energies. 

\subsection*{Description of the chapter contents}
The work presented in the following sections is organized as follows. In section \ref{P3EMT} we will provide the basic definition of the energy momentum tensor in a curved spacetime, followed by a direct computation of all of its components according to the Lagrangian of the SM
(section \ref{P3tmunusection}). 
In sections \ref{P3mastersection} and \ref{P3BRSTsection} we derive the fundamental Ward and Slavnov-Taylor identites which define the structure of the $TVV'$ vertex, expanded in terms of its $TAA$, $TAZ$ and $TZZ$ contributions, where $T$ couples to the graviton and $A$ and $Z$ are the photon and the neutral massive gauge boson, respectively. Complete results for all the amplitudes are given in section \ref{P3resultsection}, expressed in terms of a small set of form factors. As we are going to show, the contribution to the anomaly comes from a single form factor in each amplitude, multiplying a unique tensor structure. These form
factors are characterized by the appearance of a massless pole with a residue that can be related to the beta function of the theory and which is the signature of the anomaly \cite{Armillis:2009im}. We have extensively elaborated in previous works on the significance of such contributions in the ultraviolet region (UV) \cite{Coriano:2011ti}.

In the presence of spontaneous symmetry breaking the perturbative expansion of these form factors can be still arranged in the form of a $1/s$ contribution, with $s$ being the invariant mass of the graviton line, plus mass corrections of the form $v^2/s$, with $v$ being the electroweak vev.   The computation shows that the trace part of the amplitude is then clearly dominated at large energy (i.e for $s \gg v^2$) by the pole contribution, as we will discuss in section \ref{P3discussionsection}. Our conclusions and perspectives are given in section \ref{P3conclusions}. Several technical points omitted from the main sections have been included in the appendices to facilitate the reading of those more involved derivations.

\section{ The EMT of the Standard Model: definitions and conventions}
\label{P3EMT}
The expression of a symmetric and conserved EMT for the SM, as for any field theory
Lagrangian, may be obtained, more conveniently, by coupling the corresponding Lagrangian to the gravitational field, described
by the metric $g_{\mu\nu}$ of the curved background
\beq S = S_G + S_{SM} + S_{I}= -\frac{1}{\kappa^2}\int d^4 x \sqrt{-g}\, R + \int d^4 x
\sqrt{-g}\mathcal{L}_{SM} + \frac{1}{6} \int d^4 x \sqrt{-g}\, R \, \mathcal H^\dag \mathcal H      \, ,
\eeq
where $\kappa^2=16 \pi G_N$, with $G_N$ being the four dimensional Newton's constant and $\mathcal H$ is the Higgs doublet.
We recall that Einstein's equations take the form
\beq\label{P3Einsteinfunct.1}\,
\frac{\d}{\d g^{\mu\nu}(x)}S_G =- \frac{\d}{\d g^{\mu\nu}(x)}[ S_{SM} + S_I ]
\eeq
and the EMT in our conventions is defined as
\beq  T_{\mu\nu}(x)  = \frac{2}{\sqrt{-g(x)}}\frac{\d [S_{SM} + S_I ]}{\d g^{\mu\nu}(x)},
\eeq
or, in terms of the SM Lagrangian, as
\beq \label{P3TEI spaziocurvo}
\frac{1}{2} \sqrt{-g} T_{\mu\nu}{\equiv} \frac{\pd(\sqrt{-g}\mathcal{L})}
{\pd g^{\mu\nu}} - \frac{\pd}{\pd x^\s}\frac{\pd(\sqrt{-g}\mathcal{L})}{\pd(\pd_\s g^{\mu\nu})}\, ,
\eeq
which is classically covariantly conserved   ($g^{\mu\rho}T_{\mu\nu; \rho}=0$).  In flat spacetime, the covariant derivative is replaced by the ordinary derivative, giving the ordinary conservation equation ($ \pd_\mu T^{\mu\nu} = 0$).

We use the convention $\eta_{\mu\nu}=(1,-1,-1,-1)$ for the metric in flat spacetime, parameterizing its deviations from the flat case as
\beq\label{P3QMM} g_{\mu\nu}(x) = \h_{\mu\nu} + \kappa \, h_{\mu\nu}(x)\,,\eeq
with the symmetric rank-2 tensor $h_{\mu\nu}(x)$ accounting for its fluctuations.

 In this limit, the coupling of the Lagrangian to gravity is given by the term
\beq\label{P3Lgrav} \mathcal{L}_{grav}(x) = -\frac{\kappa}{2}T^{\mu\nu}(x)h_{\mu\nu}(x). \eeq
The corrections to the effective action describing the coupling of the SM to gravity that we will consider in our work
are those involving one external graviton and two gauge currents. These correspond to the leading contributions to the anomalous breaking of scale invariance of the effective action in a combined expansion in powers of $\kappa$ and of the electroweak coupling ($g_2$) (i.e. of $O(\kappa \, g_2^2)$).

Coming to the fermion contributions to the EMT, we recall that the fermions are coupled to gravity using the spin connection $\Omega$ induced by the
curved metric $g_{\mu\nu}$. This allows to define a spinor derivative $\mathcal{D}$ which transforms covariantly under local Lorentz transformations. If we denote with $\underline{a},\underline{b}$ the Lorentz indices of a local free-falling frame, and with
$\s^{\underline{a}\underline{b}}$ the generators of the Lorentz group in the spinorial representation, the spin connection takes the form
\beq
 \Omega_\mu(x) = \frac{1}{2}\s^{\underline{a}\underline{b}}V_{\underline{a}}^{\,\nu}(x)V_{\underline{b}\nu;\mu}(x)\, ,
\eeq
where we have introduced the vielbein $V_{\underline{a}}^\mu(x)$. The covariant derivative of a spinor in a given representation
$(R)$ of the gauge symmetry group, expressed in curved $(\mathcal{D}_{\mu})$ coordinates is then given by
\beq \mathcal{D}_{\mu} = \frac{\pd}{\pd x^\mu} + \Omega_\mu   + A_\mu,\eeq
where $A_\mu\equiv A_\mu^a\, T^{a  (R)}$ are the gauge fields and $T^{a (R)}$ the group generators,
giving a Lagrangian of the form
\beqa \mathcal{L}& = & \sqrt{-g} \bigg\{\frac{i}{2}\bigg[\bar\psi\g^\mu(\mathcal{D}_\mu\psi)
 - (\mathcal{D}_\mu\bar\psi)\g^\mu\psi \bigg] - m\bar\psi\psi\bigg\}.       \eeqa

\section{Contributions to $T_{\mu\nu}$}
\label{P3tmunusection}
In this section we proceed with a complete evaluation of the EMT for the SM Lagrangian coupled to gravity. We will do so for the entire quantum Lagrangian of the SM, which includes also the contributions from the ghosts and the gauge-fixing terms. Details on our conventions for this section have been collected in appendix (\ref{P3conventions}). \\
The full EMT is given by a minimal tensor $T^{Min}_{\mu\nu}$ (without improvement) and a term of improvement, $T^I_{\mu\nu}$, generated by the conformal coupling of the scalars
\bea
T_{\mu\nu} = T^{Min}_{\mu\nu} + T^I_{\mu\nu} \,,
\eea
where the minimal tensor is decomposed into
\bea
T^{Min}_{\mu\nu} = T^{f.s.}_{\mu\nu} + T^{ferm.}_{\mu\nu} + T^{Higgs}_{\mu\nu} + T^{Yukawa}_{\mu\nu} + T^{g.fix.}_{\mu\nu} + T^{ghost}_{\mu\nu}.
\eea
\subsection{The gauge and fermion contributions}
The contribution from the gauge kinetic terms derived from the field strengths of the SM is
\beqa
T^{f.s.}_{\mu\nu}
& = &
\eta_{\mu\nu}\frac{1}{4}\left[F^a_{\r\s}F^{a\,\r\s} + Z_{\r\s}Z^{\,\r\s}
       + F^A_{\r\s}F^{A\,\r\s} + 2 W^+_{\r\s}W^{-\,\r\s}\right] \nn \\
&-&   F^a_{\mu\r}F^{a\,\r}_\nu - F^A_{\mu\r}F^{A\,\r}_\nu
       - Z_{\mu\r}{Z_\nu}^{\rho} - W^+_{\mu\r}W^{-\,\r}_\nu - W^+_{\nu\r}W^{-\,\r}_\mu, \,
\eeqa
where $F^a_{\mu\nu}$, $F^A_{\mu\nu}$, $Z_{\mu\nu}$ and $W^{\pm}_{\mu\nu}$ are respectively the field strengths of the gluon, photon, $Z$ and $W^{\pm}$ fields defined in appendix (\ref{P3conventions}). The fermion contribution is rather lengthy and we give it here for a single fermion generation
\bea \label{P3TEIfermioni}
T^{ferm.}_{\mu\nu}
&=& - \h_{\mu\nu}\mathcal L_{ferm.} + \frac{i}{4}\bigg\{\bar\psi_{\nu_e}\g_\mu\stackrel{\rightarrow}{\pd}_\nu\psi_{\nu_e}
+ \bar \psi_e\g_\mu\stackrel{\rightarrow}{\pd}_\nu \psi_e  + \bar \psi_u\g_\mu\stackrel{\rightarrow}{\pd}_\nu \psi_u
+ \bar \psi_d\g_\mu\stackrel{\rightarrow}{\pd}_\nu \psi_d\nn\\
&-&  i\bigg[\frac{e}{\sqrt{2}\sin\th_W}\bigg(\bar\psi_{\nu_e} \g_\mu \frac{1-\g^5}{2}\psi_e\,W^+_\nu
+ \bar \psi_e\g_\mu\frac{1-\g^5}{2}\psi_{\nu_e}\,W^-_\nu\bigg)\nn\\
&+&  \frac{e}{\sin2\th_W}\bar\psi_{\nu_e} \g_\mu\frac{1-\g^5}{2}\psi_{\nu_e} Z_\nu
- \frac{e}{\sin2\th_W}\bar \psi_e\g_\mu\bigg(\frac{1-\g^5}{2} - 2\sin^2\th_W\bigg)\psi_e\,Z_\nu \nn\\
&+&  \frac{e}{\sqrt{2}\sin\th_W}\bigg(\bar \psi_u\g_\mu\frac{1-\g^5}{2}\psi_d\,W^+_\nu + \bar \psi_d\g_\mu
\frac{1-\g^5}{2}\psi_u\,W^-_\nu \bigg)\nn\\
&+&  \frac{e}{\sin2\th_W}\bar \psi_u\g_\mu\bigg(\frac{1-\g^5}{2} - 2\sin^2\th_W\frac{2}{3}\bigg)\psi_u\,Z_\nu \nn \\
&-& \frac{e}{\sin2\th_W}\bar \psi_d\g_\mu\bigg(\frac{1-\g^5}{2} - 2\sin^2\th_W\frac{1}{3}\bigg)\psi_d\,Z_\nu\nn\\
&+&  e A_\nu\bigg(- \bar \psi_e\g_\mu \psi_e + \frac{2}{3}\,\bar \psi_u\g_\mu \psi_u - \frac{1}{3}\,\bar \psi_d\g_\mu \psi_d
\bigg) + g_s G^a_\nu\bigg(\bar \psi_u\g_\mu t^a \psi_u + \bar \psi_d\g_\mu t^a \psi_d \bigg)\bigg]\nn\\
&-&  \bar\psi_{\nu_e}\g_\mu\stackrel{\leftarrow}{\pd}_\nu\psi_{\nu_e} -\bar \psi_e\g_\mu\stackrel{\leftarrow}{\pd}_\nu \psi_e
  - \bar \psi_u\g_\mu\stackrel{\leftarrow}{\pd}_\nu \psi_u - \bar \psi_d\g_\mu\stackrel{\leftarrow}{\pd}_\nu \psi_d\nn\\
&-&  i\bigg[\frac{e}{\sqrt{2}\sin\th_W}\bigg(\bar\psi_{\nu_e} \g_\mu \frac{1-\g^5}{2}\psi_e\,W^+_\nu
    + \bar \psi_e\g_\mu\frac{1-\g^5}{2}\psi_{\nu_e}\,W^-_\nu\bigg)\nn\\
&+&  \frac{e}{\sin2\th_W}\bar\psi_{\nu_e} \g_\mu\frac{1-\g^5}{2}\psi_{\nu_e} Z_\nu
- \frac{e}{\sin2\th_W}\bar \psi_e\g_\mu\bigg(\frac{1-\g^5}{2} - 2\sin^2\th_W\bigg)\psi_e\,Z_\nu \nn\\
&+&  \frac{e}{\sqrt{2}\sin\th_W}\bigg(\bar \psi_u\g_\mu\frac{1-\g^5}{2}\psi_d\,W^+_\nu + \bar \psi_d\g_\mu\frac{1-\g^5}{2}\psi_u\,W^-_\nu \bigg)\nn\\
&+& \frac{e}{\sin2\th_W}\bar \psi_u\g_\mu\bigg(\frac{1-\g^5}{2} - 2\sin^2\th_W\frac{2}{3}\bigg)\psi_u\,Z_\nu \nn \\
&-& \frac{e}{\sin2\th_W}\bar \psi_d\g_\mu\bigg(\frac{1-\g^5}{2} - 2\sin^2\th_W\frac{1}{3}\bigg)\psi_d\,Z_\nu\nn\\
&+&  e A_\nu\bigg(- \bar \psi_e\g_\mu \psi_e + \frac{2}{3}\,\bar \psi_u\g_\mu \psi_u - \frac{1}{3}\,\bar \psi_d\g_\mu \psi_d\bigg)
+ g_s G^a_\nu\bigg(\bar \psi_u\g_\mu t^a \psi_u + \bar \psi_d\g_\mu t^a \psi_d \bigg)\bigg] \nn \\
&+& (\mu \leftrightarrow \nu)\bigg\}   \,, 
\eea
where $\psi_{\nu_e}$, $\psi_e$, $\psi_u$ and $\psi_d$ are the Dirac spinors describing respectively the electron neutrino, the electron, the up and the down quarks while $\mathcal L_{ferm.}$ is given in appendix (\ref{P3conventions}).

\subsection{The Higgs contribution}
Coming to the contribution to the EMT from the Higgs sector, we recall that the scalar Lagrangian for the Higgs fields ($\mathcal H$) is given by
\beq \mathcal{L_{\mathcal H}} = (D^\mu \mathcal H)^\dagger(D_\mu \mathcal H)
+ \mu_{\mathcal H}^2 \mathcal H^\dagger \mathcal H - \lambda(\mathcal H^\dagger \mathcal H)^2\quad
\mu_{\mathcal H}^2,\l >0   \, ,\eeq
with the covariant derivative defined as
\beq D_\mu = \pd_\mu - i g W^a_\mu T^a - i g' B_\mu Y ,  \eeq
where, in this case, $T^a=\sigma^a/2$ are the generators of $SU(2)_L$, $Y$ is the hypercharge and the coupling constants $g$ and $g'$ are defined by
$e = g \, \sin \th_W = g' \cos \th_W$. As usual we parameterize
the vacuum $\mathcal H_0$ in the scalar sector in terms of the electroweak vev $v$ as
\beq \label{P3VEVHiggs}
\mathcal H_0 =
\left(\begin{array}{c} 0 \\ \frac{v}{\sqrt{2}} \end{array}\right)
\eeq
and we expand the Higgs doublet in terms of the physical Higgs boson $H$ and the two Goldstone bosons $\phi^{+}$, $\phi$ as
\bea
\label{P3Higgsparam}
\mathcal H = \left(\begin{array}{c} -i \phi^{+} \\ \frac{1}{\sqrt{2}}(v + H + i \phi) \end{array}\right),
\eea
then the masses of the Higgs ($m_H$) and of the $W$ and $Z$ gauge bosons are given by
\beq m_H =  \sqrt{2}\,\mu_{\mathcal H}\, ,\quad M_W  =  \frac{1}{2}g v\, , \quad M_Z = \frac{1}{2}\sqrt{g^2 +
g'^2}\,v.\,  \eeq
We obtain for the energy-momentum tensor of the Higgs contribution the following expression
\bea 
T^{Higgs}_{\mu\nu}
& = & -\h_{\mu\nu}\mathcal L_{Higgs} + \pd_\mu H\pd_\nu H + \pd_\mu \f\pd_\nu \f
      + \pd_\mu \f^+\pd_\nu \f^- + \pd_\nu \f^+\pd_\mu \f^- \nn\\
&+&  M_Z^2 Z_\mu Z_\nu + M_W^2(W^+_\mu W^-_\nu + W^+_\nu W^-_\mu)\nn\\
&+&  M_W\left(W^-_\mu\pd_\nu \f^+ + W^-_\nu\pd_\mu \f^+ + W^+_\mu\pd_\nu \f^- + W^+_\nu\pd_\mu \f^- \right)
+ M_Z(\pd_\mu \f Z_\nu + \pd_\nu \f Z_\mu)\nn\\
&+&  \frac{e M_W}{\sin\th_W}H\left( W^+_\mu W^-_\nu + W^+_\nu W^-_\mu \right)
    + \frac{e M_Z}{\sin 2\th_W}H \left(Z_\mu Z_\nu\right)\nn \\
&-&  \frac{e}{2\sin\th_W}\left[W^+_\mu\left(\f^-\stackrel{\leftrightarrow}{\pd}_\nu(H + i \f)\right)
     - W^+_\mu\left(\f^-\stackrel{\leftrightarrow}{\pd}_\nu(H + i \f)\right)\right]\nn \\
&-&  \frac{e}{2\sin\th_W}\left[W^-_\mu\left(\f^+\stackrel{\leftrightarrow}{\pd}_\nu(H - i \f)\right)
+ W^-_\nu\left(\f^+\stackrel{\leftrightarrow}{\pd}_\mu(H - i \f)\right)\right]\nn \\
&+&  i e \left( A_\mu + \cot 2\th_W Z_\mu \right)\left(\f^- \stackrel{\leftrightarrow}{\pd}_\nu \f^+\right)
+ i e \left( A_\nu + \cot 2\th_W Z_\nu \right)\left(\f^- \stackrel{\leftrightarrow}{\pd}_\mu \f^+\right)\nn \\
&-&  \frac{e}{\sin 2\th_W} \left[Z_\mu\left( \f\stackrel{\leftrightarrow}{\pd_\nu}H \right)
+ Z_\nu\left( \f\stackrel{\leftrightarrow}{\pd_\mu}H \right)\right]\nn 
\eea
\bea
&-&  i e M_Z\sin\th_W\left[ Z_\mu\left(W^+_\nu \f^- - W^-_\nu \f^+ \right)
+ Z_\nu\left(W^+_\mu \f^- - W^-_\mu \f^+ \right)\right]\nn \\
&-&  i e M_W\left[ A_\mu\left(W^-_\nu - W^+_\nu \f^-\right) + A_\nu\left(W^-_\mu - W^+_\mu \f^-\right)\right]\nn \\
&+&  \frac{e^2}{4\sin^2\th_W}H^2\left[\left(W^+_\mu W^-_\nu + W^+_\nu W^-_\mu + 2 Z_\mu Z_\nu\right)\right]\nn \\
&-&  \frac{i e^2}{2\cos\th_W} H \left[Z_\mu\left(W^+_\nu \f^- - W^-_\nu \f^+\right)
+ Z_\nu\left(W^+_\mu \f^- - W^-_\mu \f^+\right)\right]\nn \\
&+& \frac{e^2}{4\sin^2\th_W}\f^2\left(W^+_\mu W^-_\nu + W^+_\nu W^-_\mu + 2 Z_\mu Z_\nu\right)
+  \frac{e^2}{\sin\th^2_W}\f^+ \f^- \left(W^+_\mu W^-_\nu + W^+_\nu W^-_\mu\right)\nn \\
&-&  \frac{i e^2}{2\sin\th_W} H \left[A_\mu\left(W^-_\nu \f^+ - W^+_\nu \f^-\right)
+ A_\nu\left(W^-_\mu \f^+ - W^+_\mu \f^- \right)\right]\nn \\
&+&   \frac{e^2}{2\cos\th_W} \f \left[Z_\mu\left(W^+_\nu \f^- + W^-_\nu \f^+\right)
+ Z_\nu\left(W^+_\mu \f^- + W^-_\mu \f^+\right)\right] \nn \\
&-&   \frac{e^2}{2\sin\th_W} \f \left[ A_\mu\left(W^-_\nu \f^+ + W^+_\nu \f^-\right)
+ A_\nu\left(W^-_\mu \f^+ + W^+_\mu \f^- \right)\right]\nn \\
&+&  e^2\cot^2 2\th_W \f^+ \f^- Z_\mu Z_\nu + e^2 \f^+ \f^- A_\mu A_\nu
+   2 e^2\cot 2\th_W \f^+ \f^- \left(A_\mu Z_\nu + A_\nu Z_\mu \right)\, . 
\label{P3TEIHiggs}
\eea
In the Higgs Lagrangian $\mathcal L_{Higgs}$ and in the third line of the previous equation we have bilinear mixing terms involving the massive gauge bosons and their Goldstone. These terms will be canceled in the $R_{\xi}$ gauge by the EMT coming from the gauge-fixing contribution.

\subsection{Contributions from the Yukawa couplings}
The expression of the contributions coming from the Yukawa couplings are derived from the Lagrangian
\beq \mathcal{L}_{Yukawa} = \mathcal{L}^l_{Yukawa} + \mathcal{L}^q_{Yukawa}\, ,\eeq
where the lepton part is given by
\beq \label{P3termineYukawa}\mathcal{L}^l_{Yukawa} = -\lambda_e \bar L \, \mathcal H \, \psi_e^R - \lambda_e \, \bar\psi_e^R \, \mathcal H^\dag \, L,\,
\eeq
while the quarks give
\beq
\mathcal{L}^q_{Yukawa} =
- \lambda_d \, \bar Q \, \mathcal H \, \psi_d^R
- \lambda_d \, \bar \psi_d^R \, \mathcal H^\dag \, Q
- \lambda_u \, Q_i \, \e^{ij}\mathcal H^\ast_j \, \psi_u^R
- \lambda_u \, \bar \psi_u^R (\e^{ij}\mathcal H^\ast_j)^\dag \, Q_i \, .\eeq
In the previous expressions the coefficients $\lambda_e$, $\lambda_u$ and $\lambda_d$ are the Yukawa couplings, $L = (\psi_{\nu_e} \,\, \psi_e)_L$ and $Q = (\psi_u \, \, \psi_d)_L$ are the lepton and quark $SU(2)$ doublet while the suffix $R$ on the spinors identifies their right components.
The contribution from this sector to the total EMT is then given by
\bea
T_{\mu\nu}^{Yukawa}
& = & -\h_{\mu\nu}\mathcal{L}_{Yukawa} 
 =  \h_{\mu\nu}\bigg\{ m_e\bar \psi_e \psi_e + m_u \bar \psi_u \psi_u + m_d \bar \psi_d \psi_d \nn \\
      &+& i\,\frac{e}{\sqrt{2}\sin\th_W}\bigg[\frac{m_e}{M_W} \left(\f^-\bar \psi_e P_L \psi_{\nu_e} - \f^+\bar \psi_{\nu_e} P_R \psi_e\right)\nn\\
&+&     \frac{m_d}{M_W} \left(\f^-\bar \psi_d P_L \psi_u - \f^+\bar\psi_u P_R \psi_d \right)
      + \frac{m_u}{M_W} \left(\f^+\bar\psi_u P_L \psi_d - \f^-\bar\psi_d P_R \psi_u\right)\bigg]\nn
\eea
\bea
&+&     i\,\frac{e}{2\sin\th_W}\bigg[\frac{m_e}{M_W}\f\left(\bar \psi_e P_R \psi_e - \bar\psi_e P_L \psi_e \right)
      + \frac{m_d}{M_W}\f\left(\bar \psi_d P_R \psi_d - \bar \psi_d P_L \psi_d \right) \nn \\
&+& \frac{m_u}{M_W}\f\left(\bar \psi_u P_L \psi_u - \bar\psi_u P_R \psi_u \right)\bigg]
+     \frac{e \, H}{2\sin\th_W \, M_W} \bigg[m_e \bar \psi_e \psi_e + m_d \bar \psi_d \psi_d
      + m_u \bar \psi_u \psi_u\bigg] \bigg\}\, . 
       \label{P3TEIYukawa}
\eea
In the expression above we have used standard conventions for the chiral projectors $P_{R\,,L} = ({1 \pm \gamma^5})/{2}$. For simplicity we consider only one generation of fermions.
\subsection{Contributions from the gauge-fixing terms}
The contribution of the gauge-fixing Lagrangian can be computed is a similar way. We will work in the
$R_\xi$ gauge where we choose for simplicity the same gauge-fixing parameter $\xi$ for all the gauge sectors. In this case we obtain (see also appendix (\ref{P3conventions}))
\beqa\label{P3TEIgaugefixing}
T^{ g.fix. }_{\mu\nu}
& = & \frac{1}{\xi}\bigg\{G^a_\nu\pd_\mu(\pd^\si G^a_\si) + G^a_\mu\pd_\nu(\pd^\si G^a_\si)\nn\\
&+&   A_\nu\pd_\mu(\pd^\si A_\si) + A_\mu\pd_\nu(\pd^\si A_\si)
  +    Z_\nu\pd_\mu(\pd^\si Z_\si) + Z_\mu\pd_\nu(\pd^\si Z_\si)\nn \\
&+&  \frac{1}{2}\bigg[W^+_\mu\pd_\nu(\pd^\si W^-_\si) + W^+_\nu\pd_\mu(\pd^\si W^-_\si) +
       W^-_\mu\pd_\nu(\pd^\si W^+_\si) + W^-_\nu\pd_\mu(\pd^\si W^+_\si)\bigg] \bigg\}\nn \\
&-&  \h_{\mu\nu}\bigg\{-\frac{1}{2\xi}(\pd^\si A_\si)^2 - \frac{1}{2\xi}(\pd^\si Z_\si)^2
  -   \frac{1}{\xi}(\pd^\si W^+_\si)(\pd^\r W^-_\r) - \frac{1}{2\xi}(\pd^\s G^a_\s)^2\nn \\
&+&  \frac{1}{\xi}\pd^\r(A_\r\pd^\si A_\si) + \frac{1}{\xi}\pd^\r(Z_\r\pd^\si Z_\si)
  +   \frac{1}{\xi}\pd^\r\left[W^+_\r\pd^\si W^-_\si + W^-_\r\pd^\si W^+_\si\right]\nn\\
&+&  \frac{1}{\xi}\pd^\r(G^a_\r\pd^\s G^a_\s)\bigg\}
  +   \h_{\mu\nu}\frac{\xi}{2}M^2_Z \f\f + \h_{\mu\nu}\xi M^2_W \f^+ \f^- \nn \\
&-&   M_Z(Z_\mu \pd_\nu\phi + Z_\nu \pd_\mu\phi)
    - M_W(W^+_\mu\pd_\nu\phi^- + W^+_\nu\pd_\mu\phi^- + W^-_\mu\pd_\nu\phi^+ + W^-_\nu\pd_\mu\phi^+ )\, .
\eeqa
\subsection{The ghost contributions}
Finally, from the ghost Lagrangian one obtains the ghost contribution to the EMT, which is given by
\beqa\label{P3TEIGhost}
T^{ghost}_{\mu\nu}
& = &  -\h_{\mu\nu}\mathcal{L}_{ghost} +   \pd_\mu\bar{c}^a \left(\pd_\nu\d^{ac} + g_s f^{abc}G^b_\nu\right)c^c
  +    \pd_\nu\bar{c}^a \left(\pd_\mu\d^{ac} + \alpha_s f^{abc}G^b_\mu\right)c^c\nn \\
&+&   \pd_\mu\bar{\h}^Z\pd_\nu\h^Z + \pd_\nu\bar{\h}^Z\pd_\mu\h^Z
  +    \pd_\mu\bar{\h}^A\pd_\nu\h^A + + \pd_\nu\bar{\h}^A\pd_\mu\h^A\nn \\
&+&   \pd_\mu\bar{\h}^+\pd_\nu\h^- + \pd_\nu\bar{\h}^+\pd_\mu\h^-
  +    \pd_\mu\bar{\h}^-\pd_\nu\h^+ + \pd_\nu\bar{\h}^-\pd_\mu\h^+\nn \\
&+&   i g \bigg\{\pd_\mu\bar{\h}^+\left[W_\nu^+ (\cos\th_W \h^Z + \sin\th_W\h^A)
  -    (\cos\th_W Z_\nu + \sin\th_W A_\nu)\h^+ \right] \nn \\
&+&   \pd_\nu\bar{\h}^+\left[W_\mu^+ (\cos\th_W \h^Z + \sin\th_W\h^A)
  -    (\cos\th_W Z_\mu + \sin\th_W A_\mu)\h^+ \right]\nn \\
&+&   \pd_\mu\bar{\h}^-\bigg[\h^- (\cos\th_W Z_\nu + \sin\th_W A_\nu)
  -    (\cos\th_W\h_Z + \sin\th_W \h_A)W_\nu^-\bigg]\nn \\
&+&   \pd_\nu\bar{\h}^-\bigg[\h^- (\cos\th_W Z_\mu + \sin\th_W A_\mu)
  -    (\cos\th_W\h_Z + \sin\th_W \h_A)W_\mu^-\bigg]\nn \\
&+&   \pd_\mu(cos\th_W\bar{\h}^Z + \sin\th_W\bar{\h}^A )\left[W^+_\nu\h^-
  -    W^-_\nu\h^+\right]\nn \\
&+&   \pd_\nu(cos\th_W\bar{\h}^Z + \sin\th_W\bar{\h}^A )\left[W^+_\mu\h^-
  -    W^-_\mu\h^+\right]\bigg\}\, ,
\eeqa
where $c^a$, $\eta^A$, $\eta^Z$ and $\eta^{\pm}$ are respectively the ghost of the gluon, photon, $Z$ and $W^{\pm}$ bosons, while $\mathcal L_{ghost}$ is the SM Lagrangian for the ghost fields defined in appendix (\ref{P3conventions}).
\subsection{The EMT from the terms of improvement}
The terms of improvement contribute with an EMT of the form
\bea
T^I_{\mu\nu} = - \frac{1}{3} \bigg[ \partial_{\mu} \partial_{\nu} - \eta_{\mu\nu} \, \Box \bigg] \mathcal H^\dag \mathcal H = - \frac{1}{3} \bigg[ \partial_{\mu} \partial_{\nu} - \eta_{\mu\nu} \, \Box \bigg] \bigg( \frac{H^2}{2} + \frac{\phi^2}{2} + \phi^{+}\phi^{-} + v \, H \bigg).
\eea

\section{The master equation of the Ward identities}
\label{P3mastersection}
In this section we proceed with the derivation of the Ward identities describing the conservation of the EMT starting from the case of a simple model, containing a scalar, a gauge field and a single fermion in a curved spacetime and then moving to the case of the full SM Lagrangian. In both cases we start with the derivation of two master equations from which the Ward identities, satisfied by a specific correlator, can be extracted by functional differentiations.

We denote with $S[V_{\underline{a}}^\mu,\f,\psi,A_\mu]$ the action of the model. Its expression depends on the vielbein, the fermion field $\psi$, the complex scalar field $\phi$ and the abelian gauge field $A_\mu$. We can use this action and the vielbein to derive a useful form of the EMT
\beq \Theta^{\mu\nu} = - \frac{1}{V}\frac{\d S}{\d V^{{\underline{a}}}_\mu}V^{{\underline{a}}\nu},\eeq
in terms of the determinant of the vielbein $V(x){\equiv} \left|V^{\underline{a}}_\mu(x)\right|$.
Notice that this expression of the EMT is non-symmetric. The symmetric expression can be easily defined by the relation
 \beq T^{\mu\nu} = \frac{1}{2}(\Theta^{\mu\nu} + \Theta^{\nu\mu})\,
 \label{P3tmunu} \eeq
that will be used below.\\
We introduce the generating functional of the model, given by
\beqa\label{P3Z} Z[V,J,J^\dag,J^\mu,\c,\bar\c]
&=& \int \mD\f\mD\f^\dag\mD\psi\mD\bar{\psi}\mD A_\mu\,\mbox{exp}\bigg\{i S[V,\f,\psi,A_\mu]  + i\int d^4x\, \bigg[J^\dag(x)\f(x) \nn \\ 
&+& \f^\dag(x)J(x)
+ \bar{\c(x)}\psi(x) + \bar{\psi}(x)\c(x) + J^\mu(x)A_\mu(x)\bigg]\bigg\}\, ,
\eeqa
where we have denoted with $J(x)$, $J^\mu(x)$ and $\chi(x)$ the sources for the scalar, the gauge field and the spinor field respectively.
We will exploit the invariance of $Z$ under diffeomorphisms for the derivation of the corresponding Ward identities.
For this purpose we introduce a condensed notation to denote the functional integration measure of all the fields
\beqa
\label{P3SourcesToymodel}
\mD \Phi {\equiv} \mD\f\mD\f^\dag\mD\psi\mD\bar{\psi}\mD A_\mu\,
\eeqa
and redefine the action with the inclusion of external sources
\beqa
\label{P3ActionToyModel}
\tilde{S}&            =          & S + i\int d^4x\,\left( J^\mu A_\mu + J^\dag(x)\f(x) + \bar{\c}(x)\psi(x) + \textrm{h.c.}\right).\,
\eeqa
Notice that we have absorbed a factor $\sqrt{-g}$ in the definition of the sources, which clearly affects their transformation under changes of
coordinates (see also appendix (\ref{P3ward})).

The condition of diffeomorphism invariance of the generating functional $Z$ gives
\beq\label{P3covW}
Z[V,J,J^\dag,\c,\bar{\c},J^\mu] = Z[V',J',J^{'\,\dag},\c',\bar{\c}',J^{'\,\mu}]\, ,\eeq
where we have allowed an arbitrary change of coordinates $x^{'\,\mu} = F^\mu(x)$ on the spacetime manifold, which can be
parameterized locally as $x^{'\,\mu} = x^\mu + \e^\mu(x)$.
 The measure of integration is invariant under such changes $( \mD\Phi' = \mD\Phi)$ and we obtain to first order in $\e^\mu(x)$
\beqa\label{P3preWard}
\int \mD\Phi \, e^{i \tilde{S}}
& = & \int \mD\Phi \quad e^{i \tilde{S}}\,\bigg( 1  + i\int d^4xd^4y\, \bigg\{- V\Theta^\mu_{\,\,\,\underline{a}}\bigg[-\d^{(4)}(x-y)\pd_\nu V^{\underline{a}}_\mu(x) \nn\\
&& - [\pd_\mu\d^{(4)}(x-y)]V^{\underline{a}}_\nu\bigg]
- \pd_\nu[\d^{(4)}(x-y)J^\dag(x)]\f(x) - \f^\dag(x)\pd_\nu[\d^{(4)}(x-y)J(x)]\nn\\
&& - \pd_\nu[\d^{(4)}(x-y)\bar{\c}(x)]\psi(x) - \bar{\psi}(x)\pd_\nu[\d^{(4)}(x-y)\c(x)] \nn\\
&& - [\pd_\nu[J^\mu(x)\d^{(4)}(x-y)] + [\pd_\r\d^{(4)}(x-y)]\d^\mu_\nu J^\r(x)]A_\mu(x)\bigg\}\e^\nu(y)\bigg).
\eeqa
This expression needs some further manipulations in order to be brought into a convenient form for the perturbative test. Using some results of appendix \ref{P3ward} we rewrite it in an equivalent form and then perform the flat spacetime limit to obtain
\beqa\label{P3Ward}
 &&  \int\mD\Phi\, e^{i\tilde{S}}\, \bigg[\pd_\a T^{\a\b}(y) - J^\dag(y)\pd^\b\f(y)
            - \pd^\b\f^\dag(y)J(y)\nn  -  J^\a(y)\pd^\b A_\a(y) + \pd_\a[J^\a(y) A^\b(y)]\nn\\
  && -  \pd^\b\bar{\psi}(y)\c(y) - \bar{\c}(y)\pd^\b\psi(y)  -      \frac{1}{2}\pd_\a\bigg(\frac{\d S}{\d\psi(y)}\s^{\a\b}\psi(y)
            - \bar{\psi}(y)\s^{\a\b}\frac{\d S}{\d\bar{\psi}(y)}\bigg)\bigg] = 0\,.
\eeqa
A more general derivation is required in the case in which we have a theory which is SM-like, where we have more fields to consider. The master formula that one obtains is slightly more involved, but its structure is similar. Before specializing the derivation to the neutral sector of the SM we discuss the Ward identity for the amputated Green functions obtained from this functional integral.
\subsection{The master equation for connected and 1PI graphs}
We can extend the above analysis by deriving a different form of the master equation in terms of the generating functional of the connected graphs ($W$) or, equivalently, directly
in terms of the effective action ($\Gamma$), which collects all the 1-particle irreducible (1PI) graphs.  The Ward identities for the various correlators are then obtained starting from these master expressions via functional differentiation. For this purpose we extend the generating functional given in (\ref{P3Z}) by coupling the model to a weak external gravitational field $h_{\alpha\beta}$
\beqa
Z[J,J^\dag,J^{\mu},\c,\bar{\c},h_{\a\b}] =  \int \mD \Phi \, \exp\bigg\{i \tilde S - i \frac{\kappa}{2} \int d^4x \, h^{\alpha\beta}(x) \, T_{\alpha\beta}(x) \bigg\}.
\eeqa
The generating functional of connected graphs is then given by
\beq\label{P3W}\exp{\bigg\{i W[J,J^\dag,J^{\mu},\c,\bar{\c},h_{\a\b}]\bigg\}} =
\frac{Z[J,J^\dag,J^{\mu},\c,\bar{\c},h_{\a\b}]}{Z[0]}\,,
\eeq
normalized respect to the vacuum functional $Z[0]$. From this we obtain the relations
\beqa\label{P3Legendre1}
\phi_c(x) = \frac{\delta W }{\delta J^\dag(x)}\, , \quad \phi^\dag_c(x) = \frac{\delta W }{\delta J(x)}\, , \quad
\psi_c(x) = \frac{\delta W }{\delta \bar{\c}(x)}\, ,\quad
\bar{\psi}_c(x) = \frac{\delta W}{\delta\c(x)}\, ,\quad
A^{\mu}_c(x) = \frac{\delta W }{\delta J_{\mu}(x)} \nn \\
\eeqa
for the classical fields of the theory, identified by a subscript "c". The effective action is then defined via the usual Legendre transform of the fields except for the gravitational source $h_{\alpha\beta}$
\beqa\label{P3EffectiveAction}
\Gamma[\phi_c,\phi^\dag_c,A^{\mu}_c,\psi_c,\bar{\psi}_c,h_{\alpha\beta}]
&=& W[J,J^\dag,J^{\mu},\c,\bar{\c},h_{\a\b}] - \int d^4x\,\bigg[J^\dag(x)\phi_c(x) + \phi_c^\dag(x) J(x)\nn\\
&+& \bar{\psi}_c(x)\c(x) + \bar{\psi}_c(x)\c(x)  + J^{\mu}(x)A_{c\,\mu}(x) \bigg]
\eeqa
which satisfies the relations
\beqa\label{P3Legendre2}
&& \frac{\d\Gamma}{\d \phi_c(x)} =  - J^\dag(x) ,\quad
\frac{\d\Gamma}{\d \phi^\dag_c(x)} = - J(x) , \quad
\frac{\d\Gamma}{\d\psi_c(x)} = - \bar{\c}(x) , \nn \\
&& \frac{\d\Gamma}{\d \bar{\psi}_c(x)} = - \c(x) ,\quad
\frac{\d\Gamma}{\d A_{c\,\mu}(x)} = - J^{\mu}(x) .
\eeqa
Notice that the functional derivatives of both $W$ and $\Gamma$ respect to the classical background field $h_{\alpha\beta}$ coincide
\beq\label{P3deltah} \frac{\d W}{\d h_{\a\b}(x)} = \frac{\d\Gamma}{\d h_{\a\b}(x)}\, .\eeq
Therefore, the Ward identity (\ref{P3Ward}) can be rewritten in terms of the connected functional integral as
\beqa\label{P3WardW}
\pd_\a\frac{\d W}{\d h_{\a\b}}
&=& -\frac{\kappa}{2}\bigg\{J^\dag\pd^{\b}\frac{\d W}{\d J^\dag} + \pd^{\b}\frac{\d W}{\d J}J  + \pd^{\b}\frac{\d W}{\d J_{\mu}}J_{\mu}
    - \pd_{\a}\bigg( \frac{\d W}{\d J_{\b} }J^{\a}\bigg) \nn \\
&+& \bar{\c}\pd^{\b}\frac{\d W}{\d \bar{\c}} +
    \pd^{\b} \frac{\d W}{\d\c}\c - \frac{1}{2}\pd_{\a}\bigg(\bar \c \s^{\a\b} \frac{\d W}{\d
    \bar{\c}}- \frac{\d W}{\d\c}\s^{\a\b}\c\bigg)
 \bigg\}\, ,
\eeqa
or equivalently in terms of the 1PI generating functional
\beqa\label{P3WardGamma}
\pd_\a\frac{\d\Gamma}{\d h_{\a\b}}
&=& - \frac{\kappa}{2} \bigg\{- \frac{\d\Gamma}{\d \phi_c}\pd^{\b}\phi_c - \pd^{\b}\phi_c^\dag\frac{\d\Gamma}{\d \phi_c^\dag}
 - \frac{\d\Gamma}{\d A_{c\,\a}}\pd^{\b}A_{c\,\a}
+ \pd_{\a}\bigg(\frac{\d\Gamma}{\d A_{c\,\a}}
    A^{\b}_c\bigg)\nn\\
&-&  \pd^{\b}\bar{\psi}_c \frac{\delta\Gamma}{\d\bar{\psi}_c} -
        \frac{\d\Gamma}{\d\psi}_c\pd^{\b}\psi_c
        + \frac{1}{2}\pd_{\a}\bigg(\frac{\d\Gamma}{\d\psi}_c\s^{\a\b}\psi_c
    - \bar{\psi}_c\s^{\a\b}\frac{\d \Gamma}{\d\bar{\psi}_c}\bigg)
\bigg\}\, ,
\eeqa
having used  (\ref{P3Legendre1}), (\ref{P3Legendre2}), (\ref{P3deltah}),(\ref{P3WardW}).\\

\subsection{The Ward identity for $TVV'$}
In the case of the $TVV'$ correlator in the Standard Model the derivation of the Ward identity requires two functional differentiations of (\ref{P3WardGamma}) (extended to the entire spectrum of SM) respect to the classical fields $V^\a_c(x_1)$ and $V^{'\,\b}_c(x_2)$ where $V$ and $V'$ stand for the two neutral gauge bosons $A$ and $Z$, obtaining
\beqa
\label{P3WardTVV}
 -i\frac{\kappa}{2}\pd^{\mu}\langle T_{\mu\nu}(x) V_\a (x_1) V'_{\b} (x_2)\rangle_{amp}
&=& - \frac{\kappa}{2}\bigg\{- \pd_\nu\d^{(4)}(x_1-x) P^{-1\,VV'}_{\a\b}(x_2,x) \nn \\
&& \hspace{-6cm} - \pd_\nu\d^{(4)}(x_2-x) P^{-1\,V V'}_{\a\b}(x_1,x)
+ \pd^\mu[ \h_{\a\nu} \d^{(4)}(x_1-x) P^{-1\,V V'}_{\b\mu}(x_2,x) \nn \\ 
&& \hspace{-6cm} + \h_{\b\nu}\delta^{(4)}(x_2-x) P^{-1\,V V'}_{\a\mu}(x_1,x)]\bigg\}
\eeqa
where we have introduced the (amputated) mixed 2-point function
\beq P^{-1\,VV'}_{\a\b}(x_1,x_2) = \langle 0| T V_{\a}(x_1) V'_{\b}(x_2) | 0 \rangle_{amp}
= \frac{\delta^2 \Gamma}{\delta V^{\alpha}_c(x_1)\d V'^{\b}_c(x_2) }\, .\eeq
After a Fourier transform
\beqa
(2\pi)^4\d^{(4)}(k-p-q)\Gamma^{VV'}_{\mu\nu\a\b}(p,q)&=&
 -i\frac{\kappa}{2}\int d^4zd^4xd^4y\, \langle T_{\mu\nu}(z) V_{\a}(x) V'_{\b}(y)\rangle_{amp}\,
e^{-ikz + ipx + iqy}\, , \nn \\
\eeqa
Eq. (\ref{P3WardTVV}) becomes
\beqa\label{P3WardmomTVV}
k^{\mu}\Gamma^{VV'}_{\mu\nu\a\b}(p,q)
&=& - \frac{\kappa}{2}\bigg\{k^\mu P^{-1\,VV'}_{\a\mu}(p) \h_{\b\nu}
+ k^\mu P^{-1\,VV'}_{\b\mu}(q) \h_{\a\nu}  - q_\nu P^{-1\,VV"}_{\a\b}(p) - p_\nu P^{-1\,VV'}_{\a\b}(q)\bigg\} \, . \nn \\
\eeqa
The perturbative test of this relation, computationally very involved, as well as of all the other relations that we will derive in the next sections, is of paramount importance for determining the structure of the interaction vertex.

\section{BRST symmetry and Slavnov-Taylor identities}
\label{P3BRSTsection}
Before coming to the derivation of the STI's which will be crucial for a consistent definition of the $TVV$ correlator
for the Lagrangian of the SM, we give the BRST variation of the EMT in QCD and in the electroweak theory which will be used in the following. \\
The QCD sector gives
\beqa
\delta T_{\mu\nu}^{QCD} = \frac{1}{\xi}\left[ A_{\mu}^i \partial_{\nu} \partial^{\rho}
D^{ij}_{\rho}c^j + A_{\nu}^i \partial_{\mu} \partial^{\rho} D^{ij}_{\rho}c^j -
\eta_{\mu\nu}\partial^{\sigma}(A_{\sigma}^i \partial^{\rho}D^{ij}_{\rho}c^j ) \right] \,,
\eeqa
with $i,j$ being color indices in the adjoint representation of $SU(3)$, while
in the electroweak sector and in the interaction basis we have
\beqa\label{P3deltaTewbrstI}
\d T_{\mu\nu}^{e.w.}  =  \frac {1}{\xi}\bigg[W^r_\mu\pd_\nu\d \mathcal F^r
                           + W^r_\nu\pd_\mu\d \mathcal F^r
                           + B_\mu\pd_\nu\d \mathcal F^0
                           + B_\nu\pd_\mu \d \mathcal F^0\bigg]
                      - \eta_{\mu\nu}\frac{1}{\xi}\pd^\r\bigg[W^r_\r\d \mathcal F^r
                           + B_\r \d \mathcal F^0\bigg]. 
\eeqa
Here the indices $r$ and $0$ refer respectively to the $SU(2)$ and $U(1)$ gauge groups and can be expanded directly in the basis of the mass eigenstates (i.e. a=(+,-, A, Z)). We obtain
\beqa\label{P3deltaTewbrst}
\d T_{\mu\nu}^{e.w.}  &=&  \frac {1}{\xi}\bigg[W^+_\mu\pd_\nu\d \mathcal F^-
                           + W^-_\mu\pd_\nu \d \mathcal F^+
                           + A_\mu\pd_\nu\d \mathcal F^A + Z_\mu\pd_\nu \d \mathcal F^Z
                           + (\mu \leftrightarrow \nu) \bigg] \nn \\
                      &-& \frac {1}{\xi}\h_{\mu\nu}\pd^\r\bigg[W^+_\r\d \mathcal F^-
                           + W^-_\r \d \mathcal F^+
                           + A_\r\d \mathcal F^A + Z_\r\d \mathcal F^Z \bigg].
\eeqa

To proceed with the derivation of the STI's for the SM, we start introducing the generating functional of the theory in the presence of a background gravitational field $h_{\mu\nu}$ (also denoted as "$h$")
\beqa
Z(h,J) =  \int \mD \Phi \, \exp\bigg\{i \tilde S - i \frac{\kappa}{2} \int d^4x \, h^{\mu\nu}(x) \, T_{\mu\nu}(x) \bigg\}
\label{P3Zhj}
\eeqa
where $\tilde S$ denotes the action of the Standard Model $(S)$ with the inclusion of the external sources $(J, \omega, \xi)$ coupled to the SM fields
\beqa
\tilde S = S + \int d^4 x \, \left( J^{{a}}_\mu A^{\mu\,{a}} + \bar\w^{{a} }\eta^{{a}} + \bar\eta^{{a}} \w^{{a}} + \bar\xi^i \psi^i  + \bar \psi^i \xi^i \right),
\eeqa
with $a = A, Z, +, -$ and $i$ which run over the fermion fields.
We also define the functional describing the insertion of the EMT on the vacuum amplitude
\beqa
Z^T_{\mu\nu}(J; z) \equiv  \langle T_{\mu\nu}(z) \rangle_J =  \int \mD \Phi \, T_{\mu\nu}(z) \, \exp{i \tilde S}
\eeqa
where $Z^T_{\mu\nu}(J; z)$ is related to $Z(h,J)$ by
\beqa
-i \frac{\kappa}{2} Z^T_{\mu\nu}(J; z) = \frac{\delta}{\delta \, h^{\mu\nu}(z)} Z(h,J) \bigg |_{h = 0} \,.
\eeqa
 The STI's of the theory are obtained by using the invariance of the functional average under a change of integration variables
\beqa \label{P3ZT}
Z^T_{\mu\nu}(J; z) = \int \mD \Phi \, T_{\mu\nu}(z) \, \exp{i \tilde S} = Z^{T}_{\mu\nu}(J; z)^\prime = \int \mD \Phi' \, T'_{\mu\nu}(z) \, \exp{i \tilde S'}
\eeqa
which leaves invariant the quantum action $S$. These transformations, obviously, are the ordinary BRST variations of the fundamental fields of the theory. The integration measure is clearly invariant under these transformations and one obtains
\beqa \label{P3ST1}
\int \mD \Phi \, \exp{i \tilde S} \, \bigg\{ \d T_{\mu\nu}(z) + i \, T_{\mu\nu}(z) \int d^4 x \bigg[ J^{a}_\mu \d A^{\mu \, a} + \bar\w^{{a} } \d \eta^{{a}} + \d \bar\eta^{{a}} \w^{{a}} + \bar\xi^i \d \psi^i  + \d \bar \psi^i \xi^i \bigg] \bigg\} = 0 \,, 
\eeqa
where the operator $\d$ is the BRST variation of different fields, which is given in appendix (\ref{P3appendixBRST}).

The STI' s are then derived by a functional differentiation of the previous identity with respect to the sources. We just remark that since the BRST variations
increase the ghost number of the integrand by 1 unit, we are then forced to differentiate respect to the source of the antighost field in order go back to a zero ghost number in the integrand. This allows to extract correlation functions which are not trivially zero.  This procedure, although correct, may however generate STI' s among different correlators which are rather involved. For this reason we will modify the generating functional $Z^T_{\mu\nu}(J; z)$ by adding to the argument of the exponential extra contributions proportional to the product of the gauge fixing functions $\mathcal F^a(x)$ and of the
corresponding sources  $\chi^a(x)$. Therefore,  we redefine the action $\tilde{S}$ as $\tilde{S}_\chi$
\beqa
\tilde S_\chi \equiv \tilde S + \int d^4 x \,  \chi^{{a} }\mathcal F^{{a}} .
\eeqa
The condition of invariance of the generating functional that will be used below for the extraction of the STI's then becomes
\beq \label{P3ST2}
\int \mD \Phi \, \exp{i \tilde S} \, \bigg\{ \d T_{\mu\nu}(z) + i \, T_{\mu\nu}(z) \int d^4 x \bigg[ J^{a}_\mu \d A^{\mu \, a} + \bar\w^{{a} } \d \eta^{{a}} + \d \bar\eta^{{a}} \w^{{a}} + \bar\xi^i \d \psi^i  + \d \bar \psi^i \xi^i  + \chi^a \d \mathcal F^a \bigg] \bigg\} = 0 \,.
\eeq
The implications of BRST invariance on the correlator $TVV'$ are obtained
by functional differentiation of (\ref{P3ST2}) respect to the source $\chi^{a}(x)$ of the gauge-fixing function $\mathcal F^{a}$ and to the source  $\w^{a}(y)$ coupled to the antighost fields $\bar \eta^{a}$. For this reason in the following we set to zero the other external fields.
\subsection{STI for the $TAA$ correlator}
Eq. (\ref{P3ST2}) can be used in the derivation of the
STI's for the $TAA$ correlator by setting appropriately to zero all the components of the external sources except some of them. For instance, if only the sources in the photon sector $(\omega^A, \chi^A)$ are non-vanishing,
this equation becomes
\beq \label{P3STTA2}
 \int \mD \Phi \, \exp \left[i \, S + i \int d^4 x \left( \bar \eta^{A}\w^{A}  + \chi^{A} \mathcal F^{A} \right) \right]
 \bigg\{ \d T_{\mu\nu}(z) + i \, T_{\mu\nu}(z) \int d^4 x \, \bigg( \chi^{A} \mathcal E^{A} - \w^{A} \frac{1}{\xi}\mathcal F^{A}  \bigg)\bigg\}=0, \\
\eeq
where the function $\mathcal{E}^A$ denotes the finite part of the BRST variation (with the infinitesimal Grassmann parameter $\lambda$ removed) of the gauge-fixing function of the photon $\mathcal F^A$
\beqa
\label{P3photoneq}
\mathcal E^A(x) = \d \mathcal F^A(x) = \Box \, \eta^A + i\, e\, \partial^{\mu}\left( W^-_\mu \, \eta^+ - W^+_\mu \, \eta^- \right) \, .
\eeqa
Functional differentiating this relation with respect to $\chi^A(x)$ and $\w^A(y)$ and then setting to zero the external sources,
we obtain the STI for the  $\langle T A A \rangle$ correlator
\beqa \label{P3STTAA1}
\frac{1}{\xi}\langle T_{\mu\nu}(z)\pd^\a A_\a(x)\pd^\b A_\b(y) \rangle = \langle T_{\mu\nu}(z) \mathcal E^A(x) \bar{\h}^A(y) \rangle + \langle \d T_{\mu\nu}(z)\pd^\a A_\a(x)\bar{\h}^A(y) \rangle \,.
\eeqa
Its right-hand side can be simplified using the fields equation of motion. The BRST variation of $\mathcal F^A$, given by $\mathcal E^A$,  is indeed the equation of motion for the ghost of the photon.
This can be easily derived by computing the change of the action under a small variation of the antighost field of the photon
$\bar \eta^A$
\beqa \label{P3trasfbareta}
\bar \eta^A(x) \rightarrow \bar \eta^A(x) + \epsilon(x),
\eeqa
which gives, integrating by parts,
\beqa
\mathcal L \rightarrow  \mathcal L + ( \partial^{\mu} \epsilon) \left( \partial_{\mu} \eta^A + i \, e ( W^-_\mu \, \eta^+ - W^+_\mu \, \eta^- ) \right) = \mathcal L - \epsilon \, \d \mathcal F^A \,,
\eeqa
and the equation of motion $\d \mathcal F^A(x) = \mathcal E^A(x) = 0$.\\
The first correlator on the right hand side of Eq. (\ref{P3STTAA1}) can be expressed in terms of simpler correlation functions using the invariance of the generating functional $Z^T_{\mu\nu}(z)$ given in (\ref{P3ZT}) under the transformation (\ref{P3trasfbareta}). One obtains
\beqa \label{P3varbaretaZT1}
Z^T_{\mu\nu}(z)^\prime  =  \int\mD \Phi \, e^{i \, \tilde S} \exp \bigg\{i \int d^4x \, \epsilon(x) \bigg[ -\mathcal E^A(x) + \w^A(x) \bigg] \bigg\}\bigg(T_{\mu\nu}(z) + \d_{\bar \eta^A}T_{\mu\nu}(z)\bigg) =  Z^T_{\mu\nu}(z) 
\eeqa
where $\d_{\bar \eta^A}T_{\mu\nu}(z)$ denotes the variation of the EMT under the transformation (\ref{P3trasfbareta})
\beqa
\d_{\bar \eta^A} T_{\mu\nu}(z) &=& \pd_{\mu} \epsilon(z) [\pd_\nu\h^A + i e (W_\nu^-\h^+ - W_\nu^+\h^-)](z) +  (\mu \leftrightarrow \nu) \nn \\
&-& \eta_{\mu\nu}\pd^\r \epsilon(z) [\pd_\r\h^A + i e(W^-_\r\h^+ - W^+_\r\h^-)](z)\,.
\eeqa
This equation can be formally rewritten as an integral expression in the form
\beqa \label{P3varTbareta}
\d_{\bar \eta^A} T_{\mu\nu}(z) &=& \int d^4 x \,  \epsilon(x) \, \bar \delta_{\bar \eta^A} T_{\mu\nu}(z,x) \,,
\eeqa
where $\bar \delta T_{\mu\nu}(z,x)$ has been defined as
\beq
\bar{\d}T_{\mu\nu}(z,x) = \h_{\mu\nu}\pd_x^\r(\d^{(4)}(z - x)\,D_\r^{A}\h^A(x)) - \pd^x_\mu(\d^{(4)}(z - x)\, D^{A}_\nu \h^A(x))
- \pd^x_\nu(\d^{(4)}(z - x)\,D^{A}_\mu \h^A(x)) \,.
\label{P3deltaT}
\eeq
We have used the notation $D_\r^{A}\h^A$ to denote the covariant derivative of the ghost of the photon
\beqa
D_\r^{A}\h^A(x) = \pd_\r\h^A(x) + i e (W_\r^-\h^+ - W_\r^+\h^-)(x)
\eeqa
and its four-divergence equals the equation of  motion of the ghost $\eta^A$
\beqa
\partial^\r D_\r^{A}\h^A(x) = \mathcal E^A(x).
\eeqa
Using Eq. (\ref{P3varTbareta}) and expanding to first order in $\epsilon$, the identity in (\ref{P3varbaretaZT1}) takes the form \beqa \label{P3EM1}
\int\mD \Phi \, e^{i\, \tilde S}\bigg\{ T_{\mu\nu}(z) \bigg[-\mathcal E^A(x) + \w^A(x) \bigg] - i \, \bar{\d}T_{\mu\nu}(z,x)\bigg\}  =  0.
\eeqa
This relation represents the functional average of the equations of motion of the ghost $\eta^A$. As such, it can be used to derive the implications of the ghost equations on the correlation functions which are extracted from it.

For instance, to derive a relation for the first correlation function appearing on the rhs of Eq. (\ref{P3STTAA1}), it is sufficient to take a functional derivative of  (\ref{P3EM1}) respect to $\omega^A(y)$
\beqa
\langle T_{\mu\nu}(z) \mathcal E^A(x) \bar \eta^A(y) \rangle = - i \langle \bar \delta_{\bar \eta^A}
 T_{\mu\nu}(z,x) \bar \eta^A(y) \rangle - i \delta^{(4)}(x-y) \langle T_{\mu\nu}(z) \rangle.
\label{P3GreenF1}
\eeqa
Notice that the term proportional to $\delta^{(4)}(x-y)$ corresponds to a disconnected diagram and as such can be dropped
in the analysis of connected correlators. We can substitute in (\ref{P3GreenF1}) the explicit form of $\bar \delta T_{\mu\nu}(z,x)$, rewriting it in terms of the 2-point function of the covariant derivative of the ghost $\eta^A$ ($D^{A}_\r\eta^A$) and of the antighost $\bar\eta^A$
\beqa \label{P3Firstterm1}
\langle T_{\mu\nu}(z) \mathcal E^A(x) \bar \eta^A(y) \rangle & = &   -  i \bigg\{ 
\eta_{\mu\nu}\, \pd^\r_x \left[\d^{(4)}(z - x) \langle (D^{A}_\r\eta^A(x)\,\bar \eta^A(y) \rangle \right] \nn \\
&& - \left( \pd_\mu^x \left[ \d^{(4)}(z - x) \langle D^{A}_\nu\eta^A(x)\, \bar\eta^A(y)\rangle \right] + (\mu \leftrightarrow \nu) \right) \bigg\}.
\eeqa
The correlation functions involving the covariant derivative of the ghost and of the antighost, appearing on the right-hand side of
(\ref{P3Firstterm1}), are related - by some STI's - to derivatives of the photon 2-point function. We leave the proof of this point to appendix (\ref{P3appendixBRST}) and just quote the result. Then Eq. (\ref{P3Firstterm1}) becomes
\beqa \label{P3Firstterm2}
\langle T_{\mu\nu}(z) \mathcal E^A(x) \bar \eta^A(y) \rangle & = &   
- \frac{i}{\xi}\bigg\{ \eta_{\mu\nu} \, \pd^\r_x \left[\d^{(4)}(z - x) \partial^\alpha_y \langle A_\r(x) \, A_\alpha(y) \rangle \right] \nn \\
&& - \left( \pd_\mu^x \left[ \d^{(4)}(z - x) \partial^\alpha_y \langle A_\nu(x) \, A_\alpha(y)\rangle \right] + (\mu \leftrightarrow \nu) \right) \bigg\}.
\eeqa

\begin{figure}[t]
\centering
\includegraphics[scale=0.8]{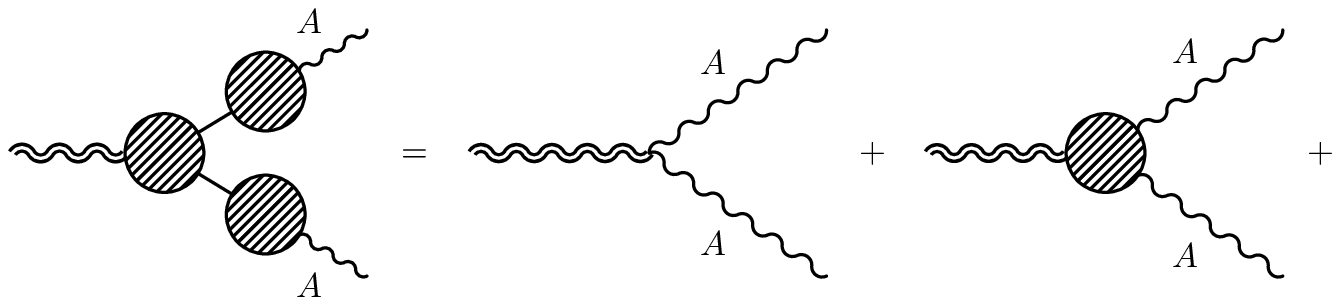}
\includegraphics[scale=0.8]{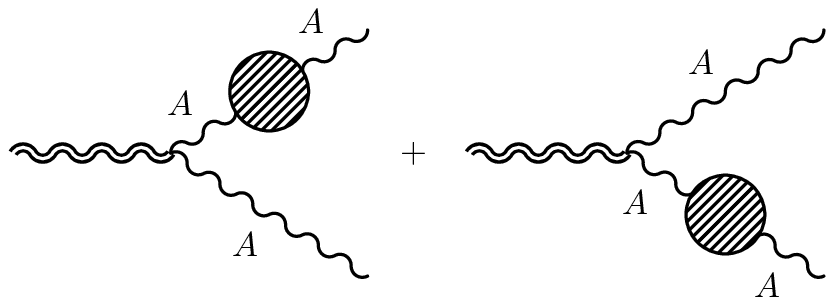}
\caption{One-loop loop decomposition of  $G_{\mu\nu\alpha\beta}^{AA}(p,q)$ in terms of the amputated function $\Gamma_{\mu\nu\alpha\beta}^{AA}(p,q)$ and of 2-point functions on the external legs. \label{P3Fig.greenAA}}
\end{figure}

Having simplified the first of the two functions on the right hand side of (\ref{P3STTAA1}), we proceed with the analysis of the second one, containing the BRST variation of the EMT, which can be expressed as a combination of BRST variations of the gauge-fixing functions $\mathcal F^{a}$
\beqa\label{P3deltaTewbrstM}
\d T_{\mu\nu} & = & \frac {1}{\xi}\bigg[W^+_\mu\pd_\nu\d F^- + W^-_\mu\pd_\nu \d F^+
                           + A_\mu\pd_\nu\d F^A + Z_\mu\pd_\nu \d F^Z + (\mu \leftrightarrow \nu) \bigg]\nn \\
                     && - \frac {1}{\xi}\h_{\mu\nu}\pd^\r\bigg[W^+_\r\d F^- + W^-_\r \d F^+ + A_\r\d F^A + Z_\r\d F^Z \bigg].
\eeqa
Similarly to the photon case, where $\d \mathcal F^A$ is proportional to the equation of motion of the corresponding ghost, also in this more general case we have
\beqa
\d \mathcal F^{{r} }= \mathcal E^{{r}}  \quad {r} = +,-,A,Z \,
\eeqa
and $\delta T_{\mu\nu}$ can be rewritten in the form
\beqa\label{P3deltaTewbrstM2}
\d T_{\mu\nu} & = & \frac {1}{\xi}\bigg[W^+_\mu\pd_\nu \mathcal E^- + W^-_\mu\pd_\nu \mathcal E^+
                           + A_\mu\pd_\nu \mathcal E^A + Z_\mu\pd_\nu \mathcal E^Z + (\mu \leftrightarrow \nu) \bigg]\nn \\
                     && - \frac {1}{\xi}\h_{\mu\nu}\pd^\r\bigg[W^+_\r \mathcal E^- + W^-_\r \mathcal E^+ + A_\r\mathcal E^A + Z_\r\mathcal E^Z \bigg].
\eeqa
The appearance of the operators $\mathcal E^{r}$ in the expression above suggests that Eq. (\ref{P3STTAA1}) can
be simplified if we derive STI's involving the equations of motion of the ghost fields. Therefore, we proceed with a functional average of the equation of motions of the ghosts
\beq\label{P3ghosteom}
\int \mD \Phi \, e^{i \, \tilde S_\chi}\bigg[- \mathcal E^{{r}}(z) + \w^{{r}}(z)\bigg] = 0 \quad r = +,-,A,Z \, .
\eeq
The terms appearing in Eq. (\ref{P3deltaTewbrstM2}) are obtained by acting on this generating functional with appropriate differentiations. For instance, to reproduce the term $\partial W^+ \mathcal E^-$ we take a functional derivative of (\ref{P3ghosteom}) with respect to the source $J^{a}_\r(z)$ followed by a differentiation respect to $z^\r$ obtaining
\beqa
&& \pd^\r_z\,\frac{\d}{\d J^{a}_\r(z)}  \int \mD \Phi \, e^{i \, \tilde S_\chi}\bigg[- \mathcal E^{{r}}(z) + \w^{{r}}(z)\bigg]  \nn \\
&& = i \int \mD \Phi \, e^{i\, \tilde S_\chi}\bigg[- \pd_z^\r \left(A^{a}_\r(z) \mathcal E^{{r}}(z) \right) +
\pd^\r_z \left( A^{a}_\r(z)\w^{{r}}(z) \right)\bigg] = 0\,. 
\label{P3eom1}
\eeqa
At this stage we need to take a derivative respect to the source $\chi^A(x)$ and to the source $\omega^A(y)$ of the antighost field $\bar\eta^A$
\beq
\label{P3eom2}
\int \mD \Phi \, e^{i\, \tilde S_\chi}\bigg[ \pd_z^\r \left( A^a_\r(z) \mathcal E^r(z) \right)  \partial^\alpha A_\alpha(x) \bar \eta^A(y) + i \, \delta^{{{r}}A}\pd^\r_z \left(A_\r(z) \d^{(4)}(z - y) \right)\pd^\a A_\a(x)\bigg] = 0\, .
\eeq
In the expression above the Kronecher $\delta^{{r}A}$  is 1 for ${r}=A$ and 0 for ${r}= +,-, Z$.
This shows that in $\delta T_{\mu\nu}$ in (\ref{P3deltaTewbrstM2}) only the photon contributes to the $\langle \d T_{\mu\nu}(z)\pd^\a A_\a(x)\bar{\h}^A(y) \rangle$ correlator and gives
\beqa \label{P3Secondterm}
\langle \d T_{\mu\nu}(z)\pd^\a A_\a(x)\bar{\h}^A(y) \rangle & = &  -\frac{i}{\xi}   \bigg\{\pd_\nu^z \d^{(4)}(z-y)  \pd_x^\a \langle A_\mu(z) A_\a(x) \rangle
 +  \pd_\mu^z \d^{(4)}(z-y)  \pd_x^\a \langle A_\nu(z) A_\a(x) \rangle  \nn\\
&& - \eta_{\mu\nu}\pd^\r_z  \left( \d^{(4)}(z-y) \pd_x^\a  \langle A_\r(z) A_\a(x) \rangle \right) \bigg\} \, .
\eeqa
Using the results of (\ref{P3Firstterm2}) and (\ref{P3Secondterm}) in (\ref{P3STTAA1}) we obtain a simple expression for the STI, just in terms of derivatives of the photon 2-point function
\beqa
 \frac{1}{\xi}\langle T_{\mu\nu}(z)\pd^\a A_\a(x)\pd^\b A_\b(y) \rangle &=& 
- \frac{i}{\xi}\bigg\{ \eta_{\mu\nu} \, \pd^\r_x \left[\d^{(4)}(z - x) \partial^\alpha_y \langle A_\r(x) \, A_\alpha(y) \rangle \right] \nn \\
&& \hspace{-4cm} - \eta_{\mu\nu}\pd^\r_z  \left[ \d^{(4)}(z-y) \pd_x^\a  \langle A_\r(z) A_\a(x) \rangle \right]
 - \bigg( \pd_\mu^x \left[ \d^{(4)}(z - x) \partial^\alpha_y \langle A_\nu(x) \, A_\alpha(y)\rangle \right]  \nn \\
&& \hspace{-4cm} -  \pd_\mu^z \d^{(4)}(z-y)  \pd_x^\a \langle A_\nu(z) A_\a(x) \rangle  + (\mu \leftrightarrow \nu)  \bigg)   \bigg\} \, ,
\eeqa
which in momentum space becomes
\beqa
p^\a \, q^\b \, G_{\mu\nu\alpha\beta}^{AA}(p,q) &=&  \frac{\kappa}{2} \, q^\a \, \bigg\{  p_\mu \, P^{AA}_{\nu\a}(q) + p_\nu \, P^{AA}_{\mu\a}(q)  - \eta_{\mu\nu} p^\r \, P^{AA}_{\r\a}(q) \bigg\} \nn \\
&+&  \frac{\kappa}{2} \, p^\a \, \bigg\{  q_\mu \, P^{AA}_{\nu\a}(p) + q_\nu \, P^{AA}_{\mu\a}(p)  - \eta_{\mu\nu} (p+q)^\r \, P^{AA}_{\r\a}(p) \bigg\}
\label{P3STAA0}
\eeqa
having defined
\beqa \label{P3STAA1}
(2 \pi)^4 \d^{(4)}(k-p-q) \, G_{\mu\nu\alpha\beta}^{AA}(p,q) &=& -i \frac{\kappa}{2} \int d^4 z \,d^4 x \,d^4 y \, \langle T_{\mu\nu}(z) A_{\alpha}(x) A_{\beta}(y) \rangle \, e^{-i k \cdot z + i p \cdot x + i q \cdot y} \,, \nn  \\
(2 \pi)^4 \d^{(4)}(p-q) \, P^{AA}_{\a\b}(p) & = &  \int d^4 x \,d^4 y \, \langle  A_{\alpha}(x) A_{\beta}(y) \rangle \, e^{i p \cdot x - i q \cdot y} \,.
\eeqa
The STI given in  (\ref{P3STAA0}) involves the Green function $G_{\mu\nu\alpha\beta}^{AA}(p,q)$ which differs from the vertex function $\Gamma_{\mu\nu\alpha\beta}^{AA}(p,q)$ for the presence of propagators on the external vector lines. In the one-loop approximation the decomposition of $G_{\mu\nu\alpha\beta}^{AA}(p,q)$ in terms of vertex and external lines corrections simplifies, as illustrated in Fig.(\ref{P3Fig.greenAA}). In momentum space this takes the form
\beqa \label{P3TAAGreen}
G_{\mu\nu\alpha\beta}^{AA}(p,q) &=&  V^{hAA}_{\mu\nu\sigma\rho}(p,q) \,
{P^{AA}_0}^{\sigma}_{\alpha}(p) \, {P^{AA}_0}^{\rho}_{\beta}(q)+ \Gamma_{\mu\nu\sigma\rho}^{AA, \,
1}(p,q) \, {P^{AA}_0}^{\sigma}_{\alpha}(p) \, {P^{AA}_0}^{\rho}_{\beta}(q) \nn\\
&+&  V^{hAA}_{\mu\nu\si\rho}(p,q) \, {P^{AA}_1}^{\sigma}_{\alpha}(p) \, {P^{AA}_0}^{\rho}_{\beta}(q)
+ V^{hAA}_{\mu\nu\si\rho}(p,q) \, {P^{AA}_0}^{\sigma}_{\alpha}(p) \, {P^{AA}_1}^{\rho}_{\beta}(q)
\, ,  \eeqa
where $V^{hAA}_{\mu\nu\sigma\rho}(p,q)$ is the tree level graviton-photon-photon interaction vertex defined in appendix \ref{P3FeynRules}.
The right-hand-side of Eq. (\ref{P3STAA0}) can be rewritten in the form
\beqa\label{P3STAA3}
p^\a \, q^\b \, G_{\mu\nu\alpha\beta}^{AA}(p,q)
&=& - i\frac{\kappa}{2}\frac{\xi}{q^2}\bigg\{ p_\mu q_\nu + p_\nu q_\mu - \eta_{\mu\nu} p \cdot q \bigg\}  - i \frac{\kappa}{2} \frac{\xi}{p^2}\bigg\{ q_\mu p_\nu + q_\nu p_\mu
      - \eta_{\mu\nu}( p \cdot q + p^2) \bigg\} \nn\\
&=& \frac{(-i \, \xi)^2}{p^2\, q^2} \, p^\a \, q^\b \, V^{hAA}_{\mu\nu\a\b}(p,q) \eeqa
which implies, together with (\ref{P3TAAGreen}), that
\beqa \label{P3STAA2}
p^\a \, q^\b \, \Gamma_{\mu\nu\a\b}^{AA, \, 1}(p,q)  = 0.
\eeqa
This is the Slavnov-Taylor identity satisfied by the one-loop vertex function.
\subsection{STI for the $T A Z $ correlator}\label{P3BRST TAZ}
The derivation of the STI for $TAZ $ follows a pattern similar to the $TAA$ case.
The starting point is the condition of BRST invariance of the generating functional given in Eq.(\ref{P3ST1}).
Also in this case we introduce some auxiliary sources $\chi^a(x)$ for the gauge-fixing terms, but we differentiate
(\ref{P3ST2}) with respect to
$\chi^A(x)$ and to the source $\w^Z(y)$ of the antighost $\bar\h^Z(y)$, and then set all the sources to zero.
We obtain a relation similar to Eq. (\ref{P3STTA2}), that is
\beqa \label{P3ST3AZ}
&& \int \mD \Phi \, \exp \left[i \, S + i \int d^4 x \left( \bar\eta^Z \w^Z
+ \chi^A \mathcal F^A \right) \right] \nn \\
&& \hspace{1.cm} \times \bigg\{ \d T_{\mu\nu}(z) + i \, T_{\mu\nu}(z) \int d^4 x \,
\bigg( - \w^Z \frac{1}{\xi}\mathcal F^Z + \chi^A \mathcal E^A \bigg)\bigg\} = 0 \, , 
\eeqa
where $\mathcal E^A(x)$, the operator describing the equation of motion of the photon, has been defined in
(\ref{P3photoneq}). Therefore, by taking a derivative with respect to $\chi^A(x)$ and to $\w^Z(y)$ we obtain

\beq\label{P3STAZ} \frac{1}{\xi}\langle T_{\mu\nu}(z\mathcal )F^A(x)\mathcal F^Z(y)\rangle =
\langle T_{\mu\nu}(z)\mathcal E^A(x) \bar\h^Z(y)\rangle
+ \langle \d T_{\mu\nu}(z)\mathcal F^A(x)\bar\h^Z(y)\rangle \, .\eeq
The right-hand-side of this equation can be simplified using the equation of motion for the ghost of the photon on
$Z^T_{\mu\nu}(J;z)$.

We start from the first of the two correlators $\langle T_{\mu\nu}(z)\mathcal E^A(x) \bar\h^Z(y)\rangle$.
Using the invariance of $Z^T_{\mu\nu}(J;z)$  respect to the variation (\ref{P3trasfbareta})
of the antighost of the photon $\bar\h^A$ and expressing $\d_{\bar\h^A}T_{\mu\nu}(z)$ as in Eqs. (\ref{P3varTbareta})
and (\ref{P3deltaT}), we obtain Eq. (\ref{P3EM1}). At this point we differentiate this relation respect to the source $\w^Z(y)$ obtaining
\beq\label{P3STAZfirstterm} \langle T_{\mu\nu}(z)\mathcal E^A(x) \bar\h^Z(y)\rangle
= -i \langle \bar\d_{\bar\h^A} T_{\mu\nu}(z,x)\bar\h^Z(y)\rangle
- i\d^{(4)}(x-y)\langle T_{\mu\nu}(z)\rangle\, .\eeq
As in the previous case, we omit the term which is proportional to the vev of the EMT, since this generates only disconnected diagrams. The explicit form of $\bar\d_{\bar\h^A} T_{\mu\nu}(z,x)$
allows to express Eq. (\ref{P3STAZfirstterm}) in the form
\beqa \label{P3STAZfirstterm1}
\langle T_{\mu\nu}(z) \mathcal E^A(x) \bar \eta^Z(y) \rangle
&=& - i \bigg\{\eta_{\mu\nu}\, \pd^\r_x \left[\d^{(4)}(z - x) \langle D^{A}_\r\eta^A(x)\,\bar
\eta^Z(y) \rangle \right] \nn \\
&& \hspace{-4cm} - \bigg(\pd_\mu^x \left[ \d^{(4)}(z - x) \langle D^{A}_\nu\eta^A(x)\,
\bar\eta^Z(y)\rangle \right]
 +  \pd_\nu^x \left[\d^{(4)}(z - x) \langle D^{A}_\mu\eta^A(x)\,
\bar\eta^Z(y)\rangle\right] \bigg) \bigg\}\, . \eeqa
To express
$\langle T_{\mu\nu}(z)\mathcal E^A(x) \bar\h^Z(y)\rangle$ in terms of 2-point functions and of their derivatives, we use the
identity
\beq \label{P3FZA}
\langle\bar\h^Z(y) D^A_\alpha\h^A(x)\rangle = \frac{1}{\xi}\langle \mathcal F^Z(y)A_\alpha(x)\rangle = 0\, ,
\eeq
which is proved in appendix (\ref{P3appendixBRST}).
This equation relates the correlators in Eq. (\ref{P3STAZfirstterm1}) to two-point functions involving the photon and the
gauge-fixing function of the $Z$ gauge boson $\mathcal F^Z$.
Using (\ref{P3FZA}), we then conclude that
\beq \langle T_{\mu\nu}(z) \mathcal E^A(x) \bar \eta^Z(y) \rangle = 0\, .\eeq
To complete the simplification of (\ref{P3STAZ}) we need to re-express
$\langle \d T_{\mu\nu}(z)\mathcal F^A(x)\bar\h^Z(y)\rangle$ in terms of 2-point functions. This correlation function involves
the BRST variation of the EMT, defined in (\ref{P3deltaTewbrstM2}), which contains a linear combination of operators proportional to the equations of motion of the ghosts. For this reason it is more convenient to start from the same equations
functionally averaged as in (\ref{P3ghosteom}), and then proceed with further differentiations, as shown in Eq. (\ref{P3eom1}). Finally, we perform a functional differentiation of (\ref{P3eom1}) respect to the sources
$\chi^A(x)$ and $\w^Z(y)$, analogously to Eq. (\ref{P3eom1}), thereby obtaining the relation
\beqa
\int \mD \Phi \, e^{i\, \tilde S_\chi}\,\bigg[ \pd_z^\r \left( A^a_\r(z) \mathcal E^r(z) \right)
\partial^\alpha A_\alpha(x) \bar \eta^Z(y) + i \, \delta^{rZ}\pd^\r_z \left(A_\r(z) \d^{(4)}(z - y)
\right)\pd^\a A_\a(x)\bigg] = 0 \, . 
\eeqa
Following this procedure for all the terms of $\d T_{\mu\nu}(z)$ we obtain
\beqa\label{P3SecondTermAZ} \langle\d T_{\mu\nu}(z)\pd^\a A_\a(x)\bh^Z(y)\rangle
&=& -\frac{i}{\xi} \bigg\{-\h_{\mu\nu}\pd^\s_z\bigg[\d^{(4)}(z-y)\langle
    Z_\s(z)\pd^\a A_\a(x)\rangle\bigg] \nn\\
&& \hspace{-3.cm}- \pd_\nu^z\d^{(4)}(z-y) \langle Z_\mu(z)\pd^\a A_\a(x)\rangle  - \pd_\mu^z\d^{(4)}(z-y) \langle Z_\nu(z)\pd^\a A_\a(x)\rangle \bigg\}\, . \eeqa
Given that this is the only non-vanishing correlator on the right-hand-side of Eq. (\ref{P3STAZ}),
we conclude that the BRST relation that we have been searching for can be expressed in the form
\beqa\label{P3BRSTfinalcoordAZ}
\frac{1}{\xi}\langle T_{\mu\nu}(z) \mathcal F^A(x)\mathcal F^Z(y)\rangle
&=& - \frac{i}{\xi}\bigg\{-\h_{\mu\nu}\pd^\s_z\bigg[\d^{(4)}(z-y)\langle
       Z_\s(z)\pd^\a A_\a(x)\rangle\bigg] \nn\\
&& \hspace{-3.cm} + \pd_\nu^z\d^{(4)}(z-y) \langle Z_\mu(z)\pd^\a A_\a(x)\rangle + \pd_\mu^z\d^{(4)}(z-y) \langle Z_\nu(z)\pd^\a A_\a(x)\rangle\bigg\}\, . \eeqa
Notice that on the left-hand-side of this identity, differently from the case of $T A A$, appear the gauge fixing functions of the photon and of the $Z$ gauge bosons
\beqa
\mathcal F^A  =  \pd^\s A_\s\, ,\qquad \qquad
\mathcal F^Z  =  \pd^\s Z_\s - \xi M_Z \f\, ,
\eeqa
which give
\beq
\langle T_{\mu\nu}(z) \mathcal F^A(x)\mathcal F^Z(y)\rangle =
\langle T_{\mu\nu}(z)\pd^\a A_\a(x)\pd^\b Z_\b(y)\rangle
-\xi M_Z\langle T_{\mu\nu}(z) \pd^\a A_\a(x) \phi(y)\rangle\, ,\eeq
where  $\phi$ is the Goldstone of the $Z$.
Going to momentum space, with the inclusion of an overall $-i{\kappa}/{2}$ factor we define
\beqa
\label{P3FourierAZ1}
(2 \pi)^4 \d^{(4)}(k-p-q) \, G^{AZ}_{\mu\nu\alpha\beta}(p,q) &=&
 -i \frac{\kappa}{2} \int d^4 z \,d^4 x \,d^4 y \,\langle T_{\mu\nu}(z)
A_{\alpha}(x) Z_{\beta}(y) \rangle \, e^{-i k \cdot z + i p \cdot x + i q \cdot y}\nn \\
(2 \pi)^4 \d^{(4)}(p-q) \, P^{AZ}_{\a\b}(p) &=& \int d^4 x \,d^4 y \,
\langle A_{\alpha}(x) Z_{\beta}(y)\rangle \, e^{- i p \cdot x + i q \cdot y}\label{P3FourierAZ2}\,, \nn\\
(2 \pi)^4 \d^{(4)}(k-p-q)\, G^{A\phi}_{\mu\nu\a}(p,q) &=& -i \frac{\kappa}{2}\int d^4 x \,d^4 y \,
\langle  T_{\mu\nu}(z)A_{\alpha}(x)\phi(y) \rangle \, e^{-i kz + i p \cdot x + i q \cdot y} \,,
\label{P3FourierAZ3}
\eeqa
\begin{figure}[t]
\centering
\includegraphics[scale=0.8]{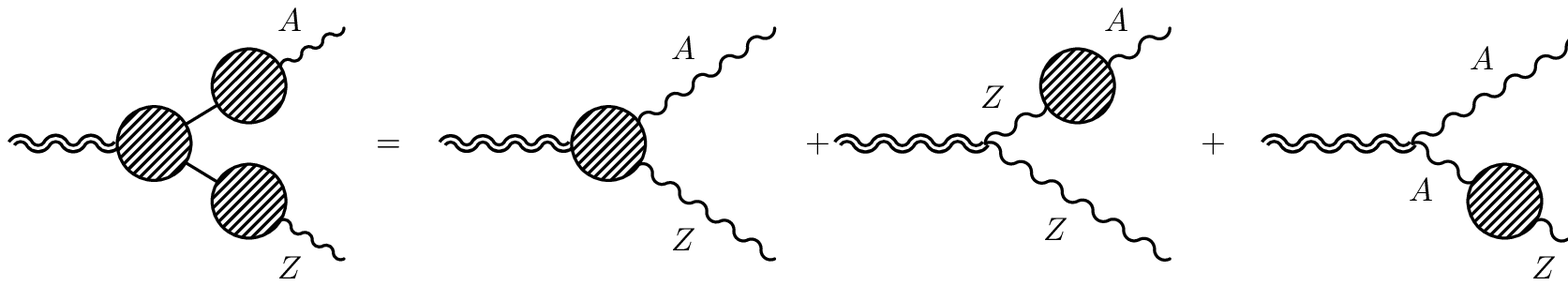}
\caption{One loop decomposition of  $G_{\mu\nu\alpha\beta}^{AZ}(p,q)$ in
terms of the amputated funtion $\Gamma_{\mu\nu\alpha\beta}^{AZ}(p,q)$ and of the corrections on the external lines. \label{P3Fig.greenAZ}}
\end{figure}
\begin{figure}[t]
\centering
\includegraphics[scale=0.8]{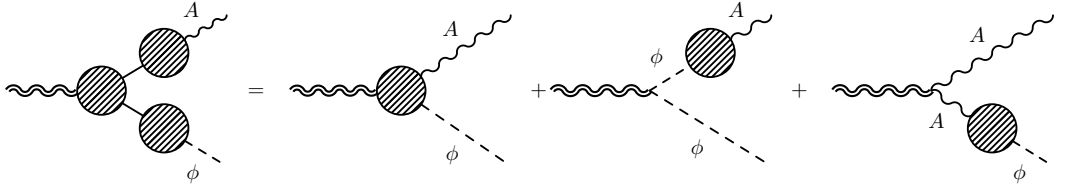}
\caption{Decomposition of $G_{\mu\nu\alpha}^{A\phi}(p,q)$ at 1-loop
in terms of the amputated correlator $\Gamma_{\mu\nu\alpha}^{A\phi,\,1}(p,q)$ and of the corrections on the external legs.\label{P3Fig.greenAphi}}
\end{figure}
and the final STI (\ref{P3BRSTfinalcoordAZ}) in momentum space reads as
\beqa\label{P3STAZ1}
p^\a q^\b G^{AZ}_{\mu\nu\a\b}(p,q) - i\xi M_Z p^\a G^{A\phi}_{\mu\nu\a}(p,q)
&=& \frac{\kappa}{2}\, p^\a\bigg\{q_\nu P^{ZA}_{\mu\a}(p) + q_\mu P^{ZA}_{\nu\a}(p) - \eta_{\mu\nu}(p+q)^\rho P^{ZA}_{\rho\a}(p)\bigg\}. \nn \\
\eeqa
At this point, we are interested in the identification of a STI for amputated Green functions. For this purpose we perform a decomposition on the left-hand-side of this equation similarly to Eq. (\ref{P3TAAGreen}) for $G^{AA}_{\mu\nu\a\b}(p,q)$, working in the
1-loop approximation.
In this case, the decomposition of the $G^{AZ}_{\mu\nu\a\b}(p,q)$ correlator, shown in Fig.
(\ref{P3Fig.greenAZ}), is given by
\beqa \label{P3TAZGreen}
G_{\mu\nu\alpha\beta}^{AZ}(p,q) &=&  \Gamma_{\mu\nu\sigma\rho}^{AZ, \, 1}(p,q) \,
{P^{AA}_0}^{\sigma}_{\alpha}(p) \, {P^{ZZ}_0}^{\rho}_{\beta}(q)  + V^{hZZ}_{\mu\nu\si\rho}(p,q) \, {P^{ZA}_1}^{\sigma}_{\alpha}(p) \, {P^{ZZ}_0}^{\rho}_{\beta}(q) \nn \\
&+& V^{hAA}_{\mu\nu\si\rho}(p,q) \, {P^{AA}_0}^{\sigma}_{\alpha}(p) \, {P^{AZ}_1}^{\rho}_{\beta}(q)
\, .
\eeqa
This decomposition, differently from the one in Eq. (\ref{P3TAAGreen}), does not contain a tree-level contribution $V^{hAZ}_{\mu\nu\a\b}(p,q)$ since this vertex is zero at the lowest order.\\
A similar procedure has to be followed for the correlator $G^{A\phi}_{\mu\nu\a}(p,q)$. Also in this case the vertices
 $V^{hA\phi}_{\mu\nu\a}(p,q)$,
$V^{hAZ}_{\mu\nu\a\b}(p,q)$ and $V^{h\phi Z}_{\mu\nu\b}(p,q)$ are zero at tree-level.
The 3-point function
$G^{A\phi}_{\mu\nu\a}(p,q)$, shown in Fig.(\ref{P3Fig.greenAphi}), is then decomposed into the form
\beqa
\label{P3TAGoldGreen}
G^{A\phi}_{\mu\nu\a}(p,q)
&=& \Gamma^{A\phi,\, 1}_{\mu\nu\rho}(p,q)\,{P^{AA}_0}^\rho_\a(p)\, P^{\phi\phi}_0(q)  + V^{hAA}_{\mu\nu\rho\si}\,{P^{AA}_0}^\rho_\a(p)\,{P^{A\phi}_1}^\si(q)
 +  V^{h\phi\phi}_{\mu\nu}\,{P^{\phi A}_1}_\a (p)\,P^{\phi\phi}_0(q) \, . \nn \\
 \eeqa
 The tree-level vertices used in Eq. (\ref{P3TAZGreen}) and (\ref{P3TAGoldGreen}) are defined in appendix \ref{P3FeynRules}.
The STI for this correlator is then obtained from (\ref{P3STAZ1}) using the decompositions in (\ref{P3TAZGreen}) and
(\ref{P3TAGoldGreen}). \\
One can show that the terms generated on the left-hand-side of (\ref{P3STAZ1}) by contracting tree-level vertices with the 1-loop insertions on the external legs, coincide with those generated from the right-hand side at the same order. For this one can use the expressions given in appendix (\ref{P3propagators}).
The result is summarized by the equation
\beq
p^\a q^\b \Gamma^{AZ,\,1}_{\mu\nu\a\b}(p,q)+ i\xi M_Zp^\a \Gamma^{A\phi,\, 1}_{\mu\nu\a}(p,q) = 0
\, , \eeq
which gives the STI at 1-loop for the amputated functions.

\subsection{STI for the $ T Z Z $ correlator}
The derivation of the STI for the $T Z Z$ follows a similar pattern.
We perform a functional derivative of (\ref{P3ST2}) respect to the source $\chi^Z(x)$ of
the gauge-fixing function $\mathcal F^Z(x)$ and to the source for the antighost $\bar\h^Z(y)$, which is $\w^Z(y)$.
We obtain a result quite similar to Eq.  (\ref{P3ST3AZ})
\beqa \label{P3ST3ZZ}
&& \int \mD \Phi \, \exp \left[i \, S + i \int d^4 x \left( \bar\eta^Z \w^Z
+ \chi^Z \mathcal F^Z \right) \right] \nn \\
&& \hspace{1.cm} \times \bigg\{ \d T_{\mu\nu}(z) + i \, T_{\mu\nu}(z) \int d^4 x \,
\bigg( - \w^Z \frac{1}{\xi}\mathcal F^Z + \chi^Z \mathcal E^Z \bigg)\bigg\} = 0\, . 
\eeqa
Here, clearly, $\mathcal E^Z(x)$ is the operator of the equations
of motion of the ghost
$\h^Z(x)$, derived from the BRST variation of the gauge-fixing function of the $Z$ gauge boson,
\bea \d\mathcal F^Z(x) &=& \mathcal E^Z(x)
= \Box\h^Z(x) + i e\frac{\cos\th_W}{\sin\th_W}\pd^\r(W^-_\r\h^+ - W^+_\r\h^-) \nn \\
&+& \frac{\xi e M_Z}{\sin2\th_W}[(v+H)\h^Z + i\cos\th_W(\f^+\h^- - \f^-\h^+)]  \nn \\
&=& \pd^\r D_\r^Z\h^Z(x)  + \frac{\xi e M_Z}{\sin2\th_W}[(v+H)\h^Z + i\cos\th_W(\f^+\h^-
- \f^-\h^+)],\,  \eea
where we have introduced, for convenience, the covariant derivative of the ghost $\h^Z(x)$, $D_\r^Z\h^Z(x)$,
which is given by
\beq
D^Z_\r\h^Z(x) = \pd_\r\h^Z(x) + ie\frac{\cos\th_W}{\sin\th_W}(W^-_\r\h^+ - W^+_\r\h^-)(x). \eeq
Performing a functional derivative of (\ref{P3ST3ZZ}) respect to $\chi^Z(x)$ and $\w^Z(y)$ we obtain the equivalent of Eq. (\ref{P3STAZ}), which is
\beq\label{P3STZZ}
\frac{1}{\xi}\langle T_{\mu\nu}(z\mathcal )F^Z(x)\mathcal F^Z(y)\rangle = \langle T_{\mu\nu}(z)\mathcal E^Z(x) \bar\h^Z(y)\rangle
+ \langle \d T_{\mu\nu}(z)\mathcal F^Z(x)\bar\h^Z(y)\rangle \, .
\eeq
At this point, the correlation functions on the right-hand-side of (\ref{P3STZZ}) must be re-expressed in terms of
2-point functions and of their derivatives. Also in this case we use a functional average of the equations of motion of the
ghost of the $Z$ gauge boson,  $\h^Z$, on the generating functional $Z^T_{\mu\nu}(J;z)$.
For this reason we start from the correlator $\langle T_{\mu\nu}(z)\mathcal E^Z(x) \bar\h^Z(y)\rangle$
and exploit the invariance of $Z^T_{\mu\nu}(J;z)$ under the BRST variation of the antighost field $\bar\h^Z(x)$,
\beq \label{P3trasfbaretaZ}\bar\h^Z(x) \rightarrow \bar\h^Z(x) + \e(x)\, ,\eeq
and express the variation of the EMT $\d_{\bar\h^Z}T_{\mu\nu}(z)$ as an integral,  having factorized the parameter $\e(x)$,
\beq \d_{\bar\h^Z} T_{\mu\nu}(z) = \int d^4 x\, \e(x)\bar\d_{\bar\h^Z} T_{\mu\nu}(z;x).\, \eeq
In this case
\beqa\label{P3bardeltaTEI ZZ}
\bar\d_{\bar\h^Z} T_{\mu\nu}(z,x)
&=& -\pd_\mu^x[\d^{(4)}(x-z)D^Z_\nu\h^Z(x)] - \pd_\nu^x[\d^{(4)}(x-z)D^Z_\mu\h^Z(x)]\nn\\
&& + \h_{\mu\nu}\mathcal \d^{(4)}(x-z)\mathcal E^Z(x) + \h_{\mu\nu}\pd^\r_x[\d^{(4)}(x-z)]D_\r^Z\h^Z(x)\,
.\eeqa
The equation obtained by the requirement of BRST invariance of  $Z^T_{\mu\nu}(J;z)$ is
\beq  \label{P3eq.motghZ}
\int\mD\Phi\,e^{i\tilde S}\bigg\{T_{\mu\nu}(z)\bigg[-\mathcal E^Z(x) + \w^Z(x)\bigg] - i\bar\d T_{\mu\nu}(z,x)\bigg\} = 0 \, .
\eeq
At this point we take a functional derivative of (\ref{P3eq.motghZ}) respect to $\w^Z(y)$ and then set all the sources to zero, obtaining
\beq
\label{P3STZZfirstterm}
\langle T_{\mu\nu}(z)\mathcal E^Z(x)\bar\h^Z(y)\rangle
= -i \langle \bar\d_{\bar\h^Z} T_{\mu\nu}(z,x)\bar\h^Z(y)\rangle
-i\d^{(4)}(x-y)\langle T_{\mu\nu}(z)\rangle  \, .
\eeq
Notice that if we are looking for a STI of connected graphs, then the term $-i\langle T_{\mu\nu}(z)\rangle$ does not contribute, being a disconnected part. Expressing $\bar\d_{\bar\h^Z}T_{\mu\nu}(z;x)$ according to (\ref{P3bardeltaTEI ZZ}), we conclude that Eq. (\ref{P3STZZfirstterm}) takes the form
\beqa
\langle T_{\mu\nu}(z)\mathcal E^Z(x)\bar\h^Z(y)\rangle
&=& - i\bigg\{\h_{\mu\nu}\langle\mathcal E^Z(x)\bar\h^Z(y)\rangle\d^{(4)}(x-z) + \h_{\mu\nu}\pd^\r_x[\d^{(4)}(x-z)]\langle D_\r^Z\h^Z(x)\bar\h^Z(y)\rangle\nn\\
&& \hspace{-3.cm} - \pd_\mu^x\bigg[\d^{(4)}(x-z)\langle D^Z_\nu\h^Z(x)\bar\h^Z(y)\rangle\bigg] - \pd_\nu^x\bigg[\d^{(4)}(x-z)\langle D^Z_\mu\h^Z(x)\bar\h^Z(y)\rangle\bigg]\bigg\} .\,\,
\eeqa
This equation can be simplified using the identities
\beqa
\langle\bar\h^Z(y)D^Z_\r\h^Z(x)\rangle = \frac{1}{\xi}\langle\mathcal F^Z(y)Z_\r(x)\rangle, \qquad \qquad
\langle\mathcal F^Z(x)\mathcal F^Z(y)\rangle = -i\xi\delta^{(4)}(x-y) \, ,\eeqa
which are proven in appendix (\ref{P3appendixBRST}) and we finally obtain the relation
\beqa\label{P3firsttermSTZZ}
\langle T_{\mu\nu}(z)\mathcal E^Z(x)\bar\h^Z(y)\rangle
&=& - \frac{i}{\xi}\bigg\{-i\xi^2\h_{\mu\nu}\d^{(4)}(x-y)\d^{(4)}(x-z)  + \h_{\mu\nu}\pd^\r_x[\d^{(4)}(x-z)]\langle Z_\r(x)\mathcal F^Z(y)\rangle\nn\\
&& \hspace{-1.cm} - \pd_\mu^x\bigg[\d^{(4)}(x-z)\langle Z_\nu(x)\mathcal F^Z(y)\rangle\bigg] - \pd_\nu^x\bigg[\d^{(4)}(x-z)\langle Z_\mu(x)\mathcal F^Z(y)\rangle\bigg]\bigg\}.\,
\eeqa
To complete the simplification of Eq. (\ref{P3STZZ}) an appropriate reduction of the correlator \\ $\langle \d T_{\mu\nu}(z)\mathcal F^Z(x)\bar\h^Z(y)\rangle$ is needed.
This can be achieved working as in the previous cases. We start from the equations of motion of the ghosts averaged with the functional integral $Z^T_{\mu\nu}$, and then take appropriate functional derivatives with respect to the sources in order to reproduce all the terms of Eq. (\ref{P3deltaTewbrstM2}) containing $\mathcal F^Z(x)$ ed $\bar\h^Z(y)$.  
We obtain the intermediate relation
\beq \int \mD \Phi \, e^{i\, \tilde S}\bigg[ \pd_z^\r \left( A^a_\r(z) \mathcal E^r(z) \right)
\mathcal F^Z(x) \bar \eta^Z(y)
+ i \, \delta^{rZ}\pd^\r_z \left(A_\r(z) \d^{(4)}(z - y)\right)\mathcal F^Z(x)\bigg] = 0\, ,\eeq
$ a = A, Z, +, -$,
while the final identity is given by
\beqa\label{P3secondtermSTZZ}
\langle\d T_{\mu\nu}(z)\mathcal F^Z(x)\bar\h^Z(y)\rangle
&=& -\frac{i}{\xi}\bigg\{\pd_\mu^z[\d^{(4)}(z-y)]\langle Z_\nu(z)\mathcal F^Z(x)\rangle + \pd_\nu^z[\d^{(4)}(z-y)]\langle Z_\nu(z)\mathcal F^Z(x)\rangle\nn\\
&-&  \h_{\mu\nu}\pd^\r_z\bigg[\d^{(4)}(z-y)\langle Z_\r(z)\mathcal F^Z(x)\rangle\bigg]\bigg\}\, .
\eeqa
Finally, inserting into (\ref{P3STZZ}) the results of (\ref{P3firsttermSTZZ}) and (\ref{P3secondtermSTZZ}), we obtain
\beqa\label{P3STZZfinalcoord}
\frac{1}{\xi}\langle T_{\mu\nu}(z)\mathcal F^Z(x)\mathcal F^Z(y)\rangle
&=& -\frac{i}{\xi}\bigg\{-i\xi^2\h_{\mu\nu}\d^{(4)}(x-y)\d^{(4)}(x-z) \nn \\ 
&& \hspace{-3.cm} + \h_{\mu\nu}\pd^\r_x\bigg[\d^{(4)}(x-z)\bigg]\langle Z_\r(x)\mathcal F^Z(y)\rangle
-  \pd_\mu^x\bigg[\d^{(4)}(x-z)\langle Z_\nu(x)\mathcal F^Z(y)\rangle\bigg] \nn \\
&& \hspace{-3.cm} - \pd_\nu^x\bigg[\d^{(4)}(x-z)\langle Z_\mu(x)\mathcal F^Z(y)\rangle\bigg]
 + \pd_\mu^z\bigg[\d^{(4)}(z-y)\bigg]\langle Z_\nu(z)\mathcal F^Z(x)\rangle \nn \\
&& \hspace{-3.cm} + \pd_\nu^z\bigg[\d^{(4)}(z-y)\bigg]\langle Z_\mu(z)\mathcal F^Z(x)\rangle
 - \h_{\mu\nu}\pd^\r_z\bigg(\d^{(4)}(z-y)\langle Z_\r(z)\mathcal F^Z(x)\rangle\bigg)\bigg\} \, .
\eeqa
We then move to momentum space introducing 2 and 3-point functions, generically defined as
\beqa
(2\pi)^4\d^{(4)}(p-q)P^{\phi_l \phi_m}(p) &=& \int d^4x d^4y\,\langle \phi_l(x)\phi_m(y)\rangle e^{-i px + i qy}\,, \\
(2\pi)^4\d^{(4)}(k-p-q)G^{\phi_l \phi_m}_{\mu\nu\a\b}(p,q) &=& -i\frac{\kappa}{2}\int d^4z d^4x d^4y \, \langle
T_{\mu\nu}(z)\phi_l(x)\phi_m(y)\rangle
e^{-i kz +i px +i qy}\, ,
\eeqa
for generic fields $\phi_l=(Z, \phi)$,  and rewrite (\ref{P3STZZfinalcoord}) in the form
\beqa
&& p^\a q^\b G^{ZZ}_{\mu\nu\a\b}(p,q) - i\xi M_Z p^\a G^{Z\f}_{\mu\nu\a}(p,q)
 - i\xi M_Z q^\b G^{\f Z}_{\mu\nu\b}(p,q) - \xi^2 M_Z^2 G^{\f\f}_{\mu\nu}(p,q) = \nn \\
&& \frac{\kappa}{2}\bigg\{
  i p_\mu[-i q^\b P^{ZZ}_{\nu\b}(q) - \xi M_Z P_\nu^{Z\f}(q)]
 +  i p_\nu[-i q^\b P^{ZZ}_{\mu\b}(q) - \xi M_Z P_\mu^{Z\f}(q)]\nn\\
&& + \, i q_\mu[- i p^\a P^{ZZ}_{\a\nu}(p) - \xi M_Z P^{Z\f}_\nu(p)]
 +  i q_\nu[- i p^\a P^{ZZ}_{\a\mu}(p) - \xi M_Z P^{Z\f}_\mu(p)]\nn\\
&& \, - i \h_{\mu\nu}k_\r[- iq_\b P^{\r\a}_{ZZ}(q) - \xi M_Z P^\r_{Z\f}(q)]
 -  i \h_{\mu\nu}k_\r[-ip_\a P^{\r\a}_{ZZ}(p) - \xi M_Z P^\r_{Z\f}(p)] - i\xi^2\h_{\mu\nu}\bigg\}. 
 \label{P3STZZmom}
\eeqa
\begin{figure}[t]
\centering
\includegraphics[scale=0.8]{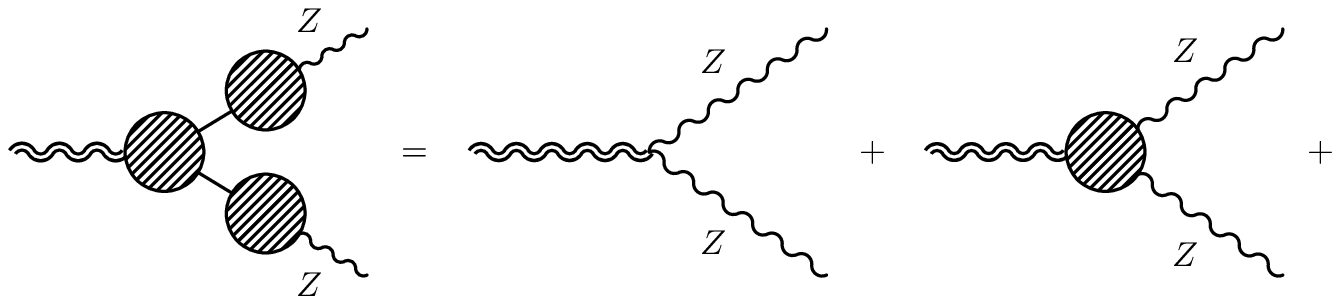}
\includegraphics[scale=0.8]{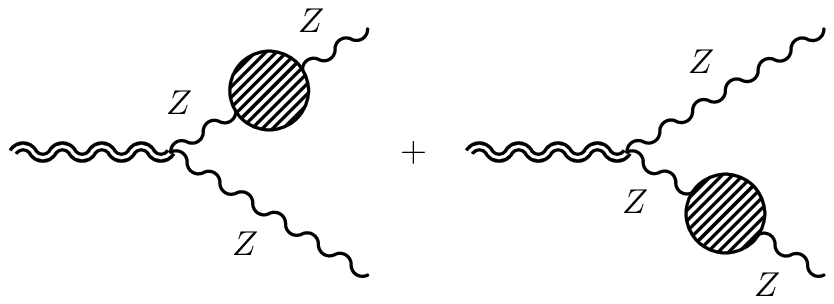}
\caption{Decomposition of $G_{\mu\nu\alpha\beta}^{ZZ}(p,q)$ in terms of the amputated $\Gamma_{\mu\nu\alpha\beta}^{ZZ}(p,q)$ and of the corrections on the external legs. \label{P3Fig.greenZZ}}
\end{figure}
\begin{figure}[t]
\centering
\includegraphics[scale=0.8]{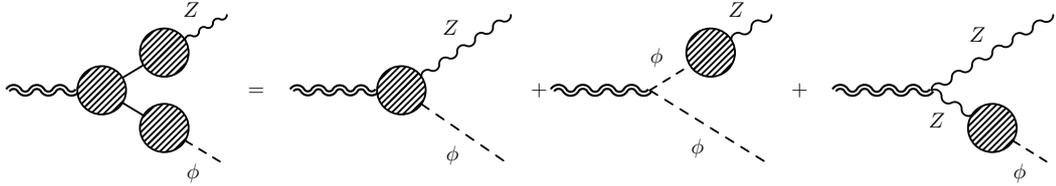}
\caption{Decomposizion of Green $G_{\mu\nu\alpha}^{Z\phi}(p,q)$ in
terms of the amputated function $\Gamma_{\mu\nu\alpha}^{Z\phi}(p,q)$ and of the corrections on the external lines \label{P3Fig.greenZphi}}
\end{figure}
\begin{figure}[t]
\centering
\includegraphics[scale=0.8]{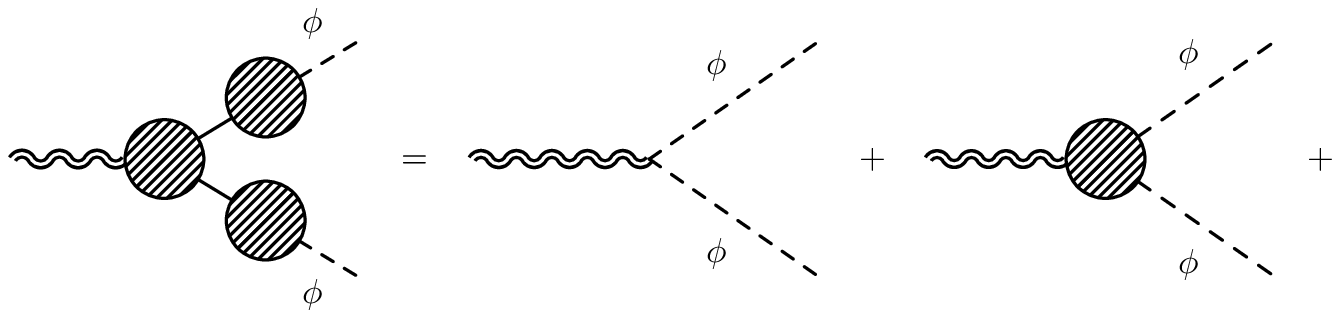}
\includegraphics[scale=0.8]{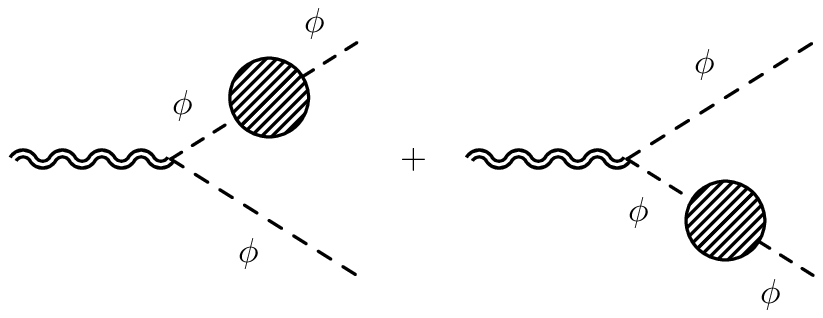}
\caption{One-loop decomposition of $G_{\mu\nu\alpha\beta}^{\phi\phi}(p,q)$
in terms of the amputated function $\Gamma_{\mu\nu}^{\phi\phi}(p,q)$ and of the corrections on the external lines.  \label{P3Fig.greenphiphi}}
\end{figure}
As in the cases of $ TAA$ and $ T A Z$, we are interested in deriving the form of the STI for amputated correlators.
From the left-hand-side of (\ref{P3STZZmom}) it is clear that there are 3 correlators which need to be decomposed, i.e. $G^{ZZ}_{\mu\nu\a\b}(p,q)$,
$G^{Z\phi}_{\mu\nu\a}(p,q)$ and $G^{\phi\phi}_{\mu\nu}(p,q)$. We have illustrated pictorially their decompositions at one loop order in
Figs. (\ref{P3Fig.greenZZ}), (\ref{P3Fig.greenZphi}) and (\ref{P3Fig.greenphiphi}),
while their explicit expressions are given by
\beqa
G_{\mu\nu\alpha\beta}^{ZZ}(p,q)
&=&  V^{hZZ}_{\mu\nu\sigma\rho}(p,q) \, {P^{ZZ}_0}^{\sigma}_{\alpha}(p) \,
     {P^{ZZ}_0}^{\rho}_{\beta}(q)+ \Gamma_{\mu\nu\sigma\rho}^{ZZ, \,1}(p,q) \,
     {P^{ZZ}_0}^{\sigma}_{\alpha}(p) \, {P^{ZZ}_0}^{\rho}_{\beta}(q) \nn\\
&+&  V^{hZZ}_{\mu\nu\si\rho}(p,q) \, {P^{ZZ}_1}^{\sigma}_{\alpha}(p) \, {P^{ZZ}_0}^{\rho}_{\beta}(q)
   + V^{hZZ}_{\mu\nu\si\rho}(p,q) \, {P^{ZZ}_0}^{\sigma}_{\alpha}(p) \, {P^{ZZ}_1}^{\rho}_{\beta}(q)  \, , \label{P3TZZGreen} \\
\nn \\
G_{\mu\nu\alpha}^{Z\phi}(p,q)
&=& \Gamma^{hZ\phi,\,1}_{\mu\nu\rho}(p,q)\,{P^{ZZ}_0}^\rho_\alpha(p)\,P^{\phi\phi}_0(q)
+  V^{hZZ}_{\mu\nu\rho}(p,q)\,{P^{ZZ}_0}^\rho_\alpha(p)\,P^{Z\phi}_1(q) \nn \\
&+&  V^{h\phi\phi}_{\mu\nu\rho}(p,q)\,{P^{\phi Z}_1}^\rho_\alpha(p)\,P^{\phi\phi}_0(q)\, , \label{P3TZphiGreen} \\
\nn \\
G_{\mu\nu}^{\phi\phi}(p,q)
&=& V^{h\phi\phi}_{\mu\nu}(p,q)\,P^{\phi\phi}_0(p)\,P^{\phi\phi}_0(q)
 +  \Gamma^{h\phi\phi,\,1}_{\mu\nu}(p,q)\,P^{\phi\phi}_0(p)\,P^{\phi\phi}_0(q)\nn\\
&& + V^{h\phi\phi}_{\mu\nu\rho}(p,q)\,P^{\phi\phi}_0(p)\,P^{\phi\phi}_1(q)
 +  V^{h\phi\phi}_{\mu\nu\rho}(p,q)\,P^{\phi\phi}_1(p)\,P^{\phi\phi}_0(q)\,. \label{P3TphiphiGreen}
 \eeqa
Eq. (\ref{P3STZZmom}), after the insertion of (\ref{P3TZZGreen}), (\ref{P3TZphiGreen}) and
(\ref{P3TphiphiGreen}), gives the STI for amputated functions that we have been looking for.
One can explicitly verify that the contributions on the left-hand-side of Eq. (\ref{P3STZZmom}) - generated both by the tree-level vertices and by the contraction of these with 1-loop 2-point functions on the external legs - are equal to the right-hand-side of the same equation. These checks are far from being obvious since they require a complete and explicit computation of all the correlators, as will be discussed next. Here we just conclude by quoting the STI for amputated functions, which takes the simpler form
\beq
p^\a q^\b \Gamma^{ZZ, \, 1}_{\mu\nu\alpha\b}(p,q) + i\xi M_Z p^\alpha \Gamma^{Z\phi, \, 1}_{\mu\nu\alpha}(p,q)
+ i\xi M_Z q^\beta \Gamma^{\phi Z, \, 1}_{\mu\nu\beta}(p,q) - \xi^2 M_Z^2 \Gamma^{\phi\phi, \, 1}_{\mu\nu}(p,q)
= 0\, .\eeq
This and the previous similar STI's  are fundamental relations which define consistently the coupling of one graviton to the neutral sector of the SM.

\section{ Perturbative results for all the correlators}
\label{P3resultsection}
\begin{figure}[t]
\centering
\subfigure[]{\includegraphics[scale=0.8]{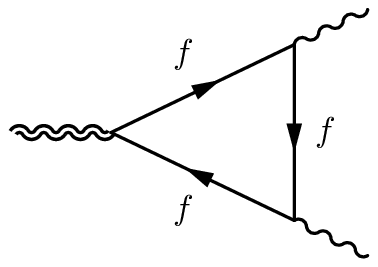}} \hspace{.5cm}
\subfigure[]{\includegraphics[scale=0.8]{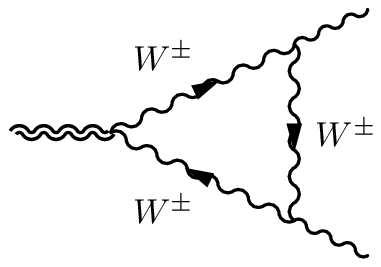}} \hspace{.5cm}
\subfigure[]{\includegraphics[scale=0.8]{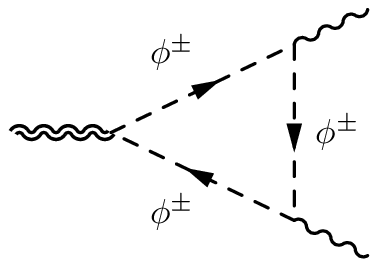}} \hspace{.5cm}
\subfigure[]{\includegraphics[scale=0.8]{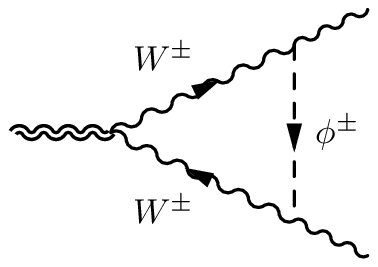}}
\\
\vspace{.5cm}
\subfigure[]{\includegraphics[scale=0.8]{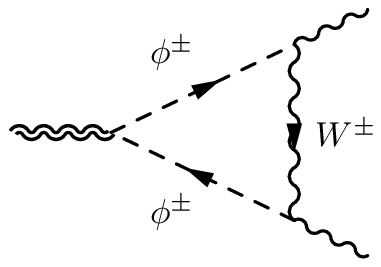}} \hspace{.5cm}
\subfigure[]{\includegraphics[scale=0.8]{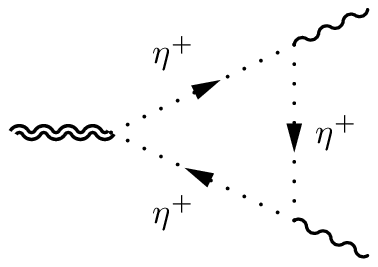}} \hspace{.5cm}
\subfigure[]{\includegraphics[scale=0.8]{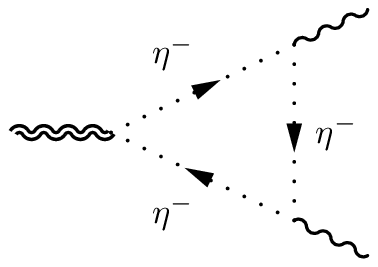}} \hspace{.5cm}
\caption{Amplitudes with the triangle topology for the three correlators $TAA$, $TAZ$ and $TZZ$. \label{P3triangles}}
\end{figure}

\begin{figure}[t]
\centering
\subfigure[]{\includegraphics[scale=0.75]{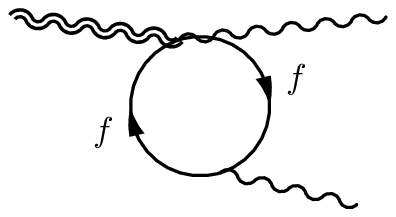}}\hspace{.5cm}
\subfigure[]{\includegraphics[scale=0.75]{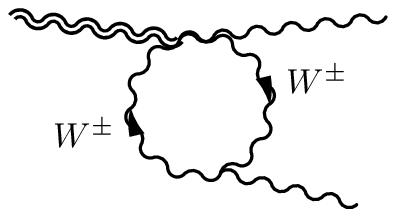}}\hspace{.5cm}
\subfigure[]{\includegraphics[scale=0.75]{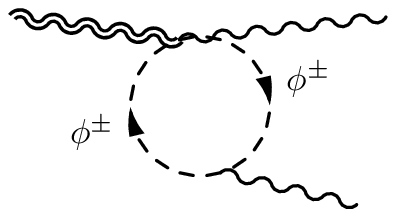}}\hspace{.5cm}
\subfigure[]{\includegraphics[scale=0.75]{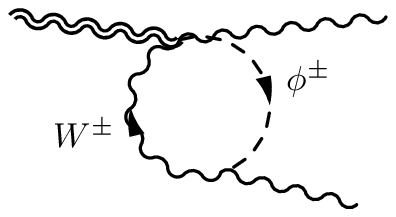}}
\\
\vspace{.5cm}
\subfigure[]{\includegraphics[scale=0.75]{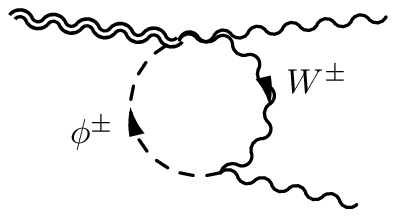}}\hspace{.5cm}
\subfigure[]{\includegraphics[scale=0.75]{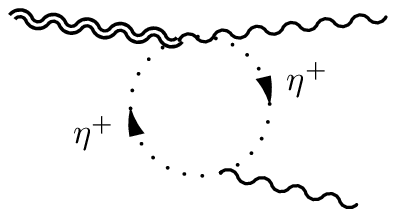}}\hspace{.5cm}
\subfigure[]{\includegraphics[scale=0.75]{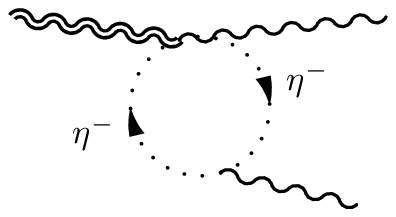}}\hspace{.5cm}
\caption{Amplitudes with t-bubble topology for the three correlators $TAA$, $TAZ$ and $TZZ$. \label{P3t-bubble}}
\end{figure}

\begin{figure}[t]
\centering
\subfigure[]{\includegraphics[scale=0.75]{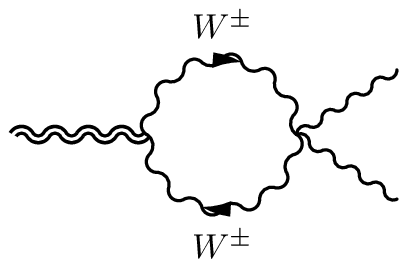}}\hspace{.5cm}
\subfigure[]{\includegraphics[scale=0.75]{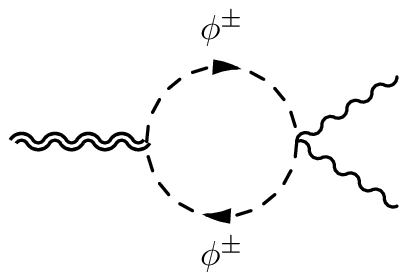}}
\caption{Amplitudes with s-bubble topology for the three correlators $TAA$, $TAZ$ and $TZZ$. \label{P3s-bubble}}
\end{figure}

\begin{figure}[t]
\centering
\subfigure[]{\includegraphics[scale=0.8]{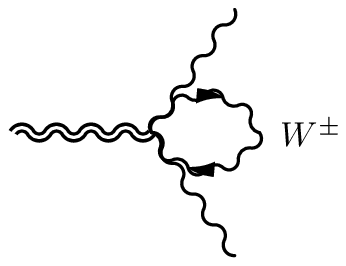}}\hspace{.5cm}
\subfigure[]{\includegraphics[scale=0.8]{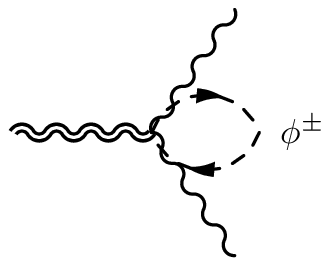}}
\caption{Amplitudes with the tadpole topology for the three correlators $TAA$, $TAZ$ and $TZZ$.\label{P3tadpoles}}
\end{figure}

\begin{figure}[t]
\centering
\subfigure[]{\includegraphics[scale=0.8]{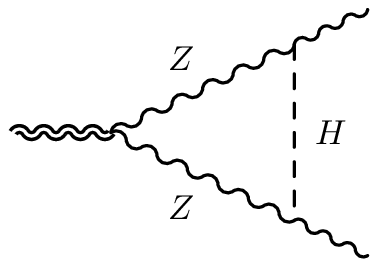}}\hspace{.5cm}
\subfigure[]{\includegraphics[scale=0.8]{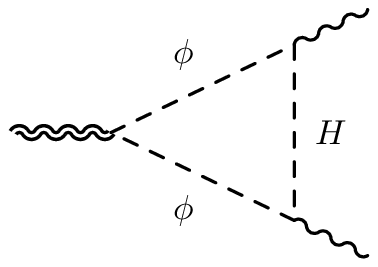}}\hspace{.5cm}
\subfigure[]{\includegraphics[scale=0.8]{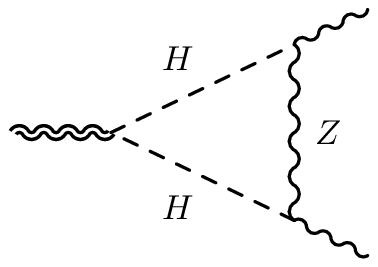}}\hspace{.5cm}
\subfigure[]{\includegraphics[scale=0.8]{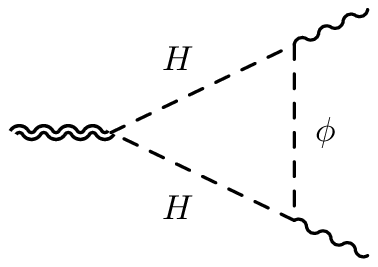}}
\caption{Amplitudes with the triangle topology for the correlator $TZZ$. \label{P3triangles1}}
\end{figure}

\begin{figure}[t]
\centering
\subfigure[]{\includegraphics[scale=0.75]{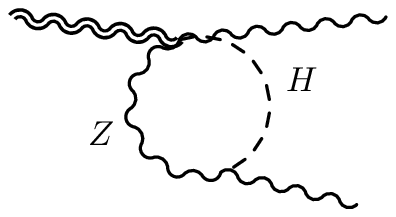}}\hspace{.5cm}
\subfigure[]{\includegraphics[scale=0.75]{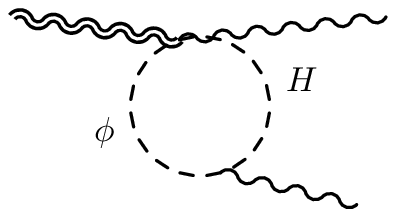}}
\caption{Amplitudes with the t-bubble topology for the correlator $TZZ$.\label{P3t-bubble1}}
\end{figure}

\begin{figure}[t]
\centering
\subfigure[]{\includegraphics[scale=0.75]{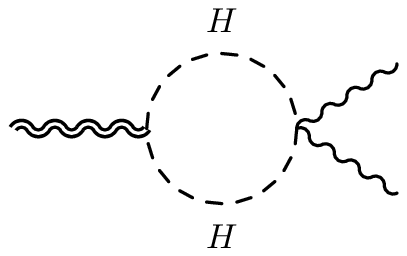}}\hspace{.5cm}
\subfigure[]{\includegraphics[scale=0.75]{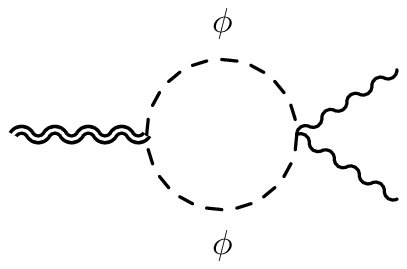}}
\caption{Amplitudes with the s-bubble topology for the correlator $TZZ$.\label{P3s-bubble1}}
\end{figure}

\begin{figure}[t]
\centering
\subfigure[]{\includegraphics[scale=0.8]{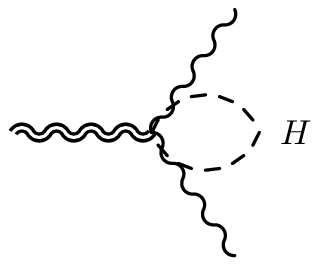}}\hspace{.5cm}
\subfigure[]{\includegraphics[scale=0.8]{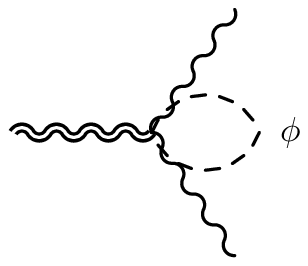}}
\caption{Amplitudes with the tadpole topology for the correlator $TZZ$.\label{P3tadpoles1}}
\end{figure}

\begin{figure}[t]
\centering
\subfigure[]{\includegraphics[scale=0.8]{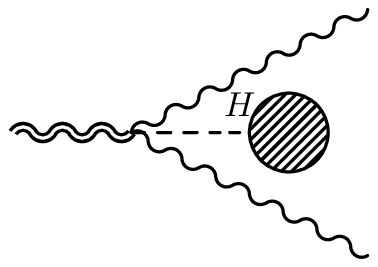}}\hspace{.5cm}
\subfigure[]{\includegraphics[scale=0.8]{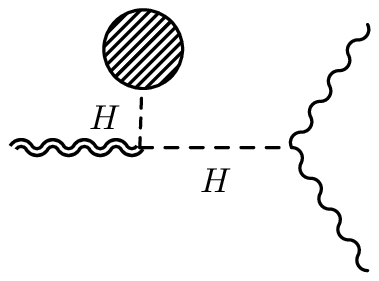}}
\caption{Amplitudes with the Higgs tadpole for the correlator $TZZ$ which vanish after renormalization.\label{P3tadpolesHiggs}}
\end{figure}

In this section we illustrate the various diagrammatic contributions appearing in the perturbative expansion of the $TV V'$ vertex.
We show in Figs. (\ref{P3triangles}-\ref{P3tadpolesHiggs}) all the basic diagrams involved, for which we are going to present explicit results. Figs. (\ref{P3triangles}) and (\ref{P3triangles1}) are characterized by a typical triangle topology, while (\ref{P3t-bubble}) and (\ref{P3t-bubble1}) denote typical terms where the point of insertion of the EMT coincides with that of a gauge current. We will refer to these last contributions with the term "t-bubbles", while those characterized by two gauge bosons emerging from a single vertex, such as in Figs. (\ref{P3s-bubble}) and (\ref{P3s-bubble1}), are called "s-bubble" diagrams. Other contributions are those with a topology of tadpoles, shown in Figs.
(\ref{P3tadpoles}), (\ref{P3tadpoles1}) and (\ref{P3tadpolesHiggs}).

The two sectors $TAA$ and $TAZ$ involve 32 diagrams each, while the $TZZ$ correlator includes 70 diagrams.
The computation of these diagrams is rather involved and has been performed in DR using the on-shell renormalization scheme  \cite{Ross:1973fp} and the t'Hooft-Veltman prescription for $\gamma_5$ matrix. We have used a
reduction of tensor integrals to the scalar form and checked explicitly all the Ward and STI's derived in the previous sections.  The reduction involves non-standard rank-4 integrals (due to the momenta coming from the insertion of the EMT on the triangle
topology) with 3 propagators.

One of the non trivial points of the computation concerns the treatment of diagrams containing fermion loops and insertions of the EMT on correlators with both vector ($J_V$) and axial-vector ($J_A$) currents. This problem has been analyzed and solved in
a related work \cite{Armillis:2010pa} to which we refer for more details. In particular, it has been shown that there are no mixed chiral and trace anomalies in diagrams of this type even in the presence of explicit mass corrections, due to the
vanishing of the $TJ_V J_A$ vertex mediated by fermion loops. This result has been obtained in a  simple $U(1)_V\times U(1)_A$ gauge model, with an explicit breaking of the gauge symmetry due to a fermion mass term. The result remains true both for global and local currents, being the gauge fields (vector and axial-vector) in the treatment of \cite{Armillis:2010pa} purely external fields. This preliminary analysis has been instrumental in all the generalizations discussed in this work.

At this point few more comments concerning the number of form factors introduced in our analysis are in order.
We recall, from a previous study \cite{Giannotti:2008cv}, that the number of original tensor structures which can be built out of the metric and of the two momenta $p$ and $q$ of the two gauge lines is 43  before imposing the Ward and the STI's of the theory. These have been classified in \cite{Giannotti:2008cv} and \cite{Armillis:2009pq}.
In particular, the form factors appearing in the fermion sector can be expressed (in the off-shell case) in terms of 13 tensor structures for the case of vector currents and of 22 structures for the axial-vector current, as shown in \cite{Armillis:2010pa}.

In the on-shell case, the fermion loops with external photons are parameterized just by 3 independent form factors. This analysis has been generalized more recently to QCD, with the computation of the graviton-gluon-gluon ($hgg$)
vertex in full generality \cite{Armillis:2009pq}. The entire vertex in the on-shell QCD case - which includes fermion and gluon loops -
is also parameterized just by 3 form factors. A similar result holds for the $TAA$ in the electroweak case. On the other hand the $TZZ$ and the $TAZ$ correlators have been expressed in terms of 9 form factors. A special comment deserves the handling of the symbolic computations. These have been performed using some software entirely written by us and implemented in the symbolic manipulation program {\em  MATHEMATICA}. This allows the reduction to scalar form of tensor integrals for correlators of rank-4 with the triangle topology. The software alllows to perform direct tests of all the Ward and Slavnov-Taylor identities on the correlator, 
which are crucial in order to secure the correctness of the result.

\subsection{ $\Gamma^{\mu\nu\alpha\beta}(p,q)$ and the terms of improvement}
Before giving the results for the anomalous correlators, we pause for some comments.

In our computations the gravitational field is non-dynamical and the analysis of the Ward and STI's shows that these can be
consistently solved only if we include the graviton-Higgs mixing on the graviton line. In other words, the graviton line is uncut. We will denote with $\Delta^{\mu\nu\alpha\beta}(p,q)$ these extra contributions and with $\Sigma^{\mu\nu\alpha\beta}(p,q)$ the completely cut vertex. These two contributions appear on the right-hand-side of the
expression of the correlation function $\Gamma^{\mu\nu\alpha\beta}(p,q)$
\bea
\Gamma^{\mu\nu\alpha\beta}(p,q) = \Sigma^{\mu\nu\alpha\beta}(p,q) + \Delta^{\mu\nu\alpha\beta}(p,q).
\eea
Finally, we just mention that we have excluded from the final expressions of the vertices all the contributions at tree-level. For this reason our results
are purely those responsible for the generation of the anomaly.

\begin{figure}[t]
\centering
\includegraphics[scale=0.8]{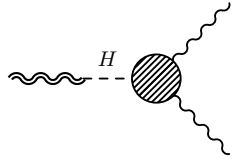}
\caption{Amplitude with the graviton - Higgs mixing vertex generated by the term of improvement. The blob represents the SM Higgs -VV' vertex at one-loop. \label{P3HVVImpr}}
\end{figure}
\begin{figure}[t]
\centering
\subfigure[]{\includegraphics[scale=0.8]{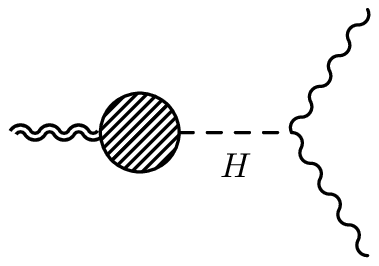}}\hspace{.5cm}
\subfigure[]{\includegraphics[scale=0.8]{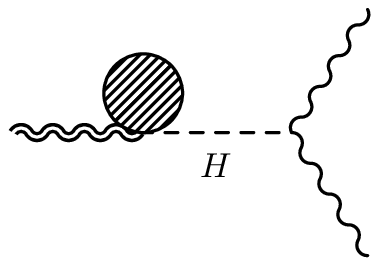}}\hspace{.5cm}
\subfigure[]{\includegraphics[scale=0.8]{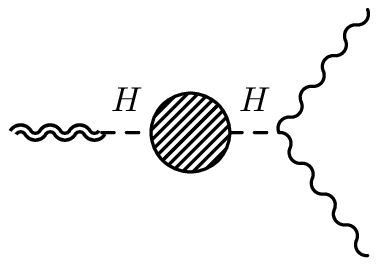}}\hspace{.5cm}
\caption{Leg corrections to the external graviton for the $TZZ$ correlator.\label{P3legcorr12}}
\end{figure}

\subsection{Results for the TAA correlator}
In this section we present the one-loop result of the computation of these correlators for on-shell vector bosons lines and discuss some of their interesting features, such as the appearance of massless anomaly poles in all the gauge invariants subsectors of the perturbative expansion.\\
We start from the case of the $TAA$ vertex and then move to the remaining ones.
In this case the full irreducible contribution $\Sigma^{\mu\nu\alpha\beta}(p,q)$ is written in the form
\bea
\Sigma ^{\mu\nu\a\b}(p,q) = \Sigma_{F}^{\mu\nu\a\b}(p,q) + \Sigma_{B}^{\mu\nu\a\b}(p,q) + \Sigma_{I}^{\mu\nu\a\b}(p,q),
\eea
where each term can be expanded in a tensor basis
\bea
\Sigma ^{\mu\nu\alpha\beta}_{F}(p,q) &=&  \, \sum_{i=1}^{3} \Phi_{i\,F} (s,0, 0,m_f^2) \, \phi_i^{\mu\nu\alpha\beta}(p,q)\,, \\
\Sigma ^{\mu\nu\alpha\beta}_{B}(p,q) &=&  \, \sum_{i=1}^{3} \Phi_{i\,B} (s,0, 0,M_W^2) \, \phi_i^{\mu\nu\alpha\beta}(p,q)\,, \\
\Sigma ^{\mu\nu\alpha\beta}_{I}(p,q) &=&  \Phi_{1\,I} (s,0, 0,M_W^2) \, \phi_1^{\mu\nu\alpha\beta}(p,q) + \Phi_{4\,I} (s,0, 0,M_W^2) \, \phi_4^{\mu\nu\alpha\beta}(p,q) \,.
\eea
The tensor basis on which we expand the on-shell vertex is given by
\bea
\label{P3phis}
  \phi_1^{\, \mu \nu \a \b} (p,q) &=&
 (s \, \eta^{\mu\nu} - k^{\mu}k^{\nu}) \, u^{\a \b} (p,q),
 \label{P3widetilde1} \nn \\
\phi_2^{\, \mu \nu \a \b} (p,q) &=& - 2 \, u^{\a \b} (p,q) \left[ s \, \eta^{\mu \nu} + 2 (p^\mu \, p^\nu + q^\mu \, q^\nu )
- 4 \, (p^\mu \, q^\nu + q^\mu \, p^\nu) \right],
\label{P3widetilde2} \nn \\
\phi^{\, \mu \nu \alpha \beta}_{3} (p,q) &=&
\big(p^{\mu} q^{\nu} + p^{\nu} q^{\mu}\big)\eta^{\alpha\beta}
+ \frac{s}{2} \left(\eta^{\alpha\nu} \eta^{\beta\mu} + \eta^{\alpha\mu} \eta^{\beta\nu}\right) - \eta^{\mu\nu}  \, u^{\a \b} (p,q) \nn \\
&&
-\left(\eta^{\beta\nu} p^{\mu}
+ \eta^{\beta\mu} p^{\nu}\right)q^{\alpha}
 - \big (\eta^{\alpha\nu} q^{\mu}
+ \eta^{\alpha\mu} q^{\nu }\big)p^{\beta}, \nonumber \\
\phi_4^{\mu\nu\alpha\beta}(p,q) &=& (s \, \eta^{\mu\nu} - k^{\mu}k^{\nu}) \, \eta^{\alpha\beta}\,,
\label{P3widetilde3}
\eea
where $u^{\a \b} (p,q)$ has been defined as
\beq
u^{\alpha\beta}(p,q) \equiv (p\cdot q) \,  \eta^{\alpha\beta} - q^{\alpha} \, p^{\beta},\,\\
\label{P3utensor}
\eeq
among which only  $\phi_1^{\, \mu \nu \a \b}$ shows manifestly a trace, the remaining ones being traceless.

The one loop vertex $\Sigma ^{\mu\nu\a\b}(p,q)$ with two on-shell photons is expressed as a sum of a fermion sector (F) (Fig. \ref{P3triangles}(a), Fig. \ref{P3t-bubble}(a)) ,
a gauge boson sector (B) (Fig. \ref{P3triangles}(b)-(g), Fig. \ref{P3t-bubble}(b)-(g), Fig. \ref{P3s-bubble}, Fig. \ref{P3tadpoles}) and a term of improvement denoted as $\Sigma^{\mu\nu\a\b}_{I}$. The contribution from the term of improvement is given by the diagrams depicted in Fig. \ref{P3triangles}(c), (d) and Fig. \ref{P3s-bubble}(b), with the graviton - scalar - scalar vertices determined by the $T_I^{\mu\nu}$. \\
The first three arguments of the form factors stand for the three independent kinematical invariants $k^2 = (p+q)^2 = s$, $p^2 = q^2 = 0$ while the remaining ones denote the particle masses circulating in the loop.

As already shown for QED and QCD, in the massless limit (i.e. before electroweak symmetry breaking), the entire contribution to the trace anomaly comes from the first tensor structure $\phi_1$ both for the fermion and for the gauge boson cases.\\
In the fermion sector the form factors are given by
\bea
\Phi_{1\, F} (s,\,0,\,0,\,m_f^2) &=& - i \frac{\kappa}{2}\, \frac{\alpha}{3 \pi \, s} \, Q_f^2 \bigg\{
- \frac{2}{3} + \frac{4\,m_f^2}{s} - 2\,m_f^2 \, \mathcal C_0 (s, 0, 0, m_f^2, m_f^2, m_f^2)
\bigg[1 - \frac{4 m_f^2}{s}\bigg] \bigg\}\, ,  \nn \\\\
\Phi_{2\, F} (s,\,0,\,0,\,m_f^2)  &=& - i \frac{\kappa}{2}\, \frac{\alpha}{3 \pi \, s} \, Q_f^2 \bigg\{
-\frac{1}{12} - \frac{m_f^2}{s} - \frac{3\,m_f^2}{s} \mathcal D_0 (s, 0, 0, m_f^2, m_f^2)  \nn \\
&-&  m_f^2 \mathcal C_0(s, 0, 0, m_f^2, m_f^2, m_f^2 )\, \left[ 1 + \frac{2\,m_f^2}{s}\right] \bigg\}\, , \\
\Phi_{3\,F} (s,\,0,\,0,\,m_f^2) &=&  - i \frac{\kappa}{2}\, \frac{\alpha}{3 \pi \, s} \, Q_f^2 \bigg\{
\frac{11\,s}{12}+ 3 m_f^2 +  \mathcal D_0 (s, 0, 0, m_f^2, m_f^2)\left[5 m_f^2 + s \right]\nn\\
&+&  s \, \mathcal B_0 (0, m_f^2, m_f^2) + 3 \, m_f^2 \, \mathcal C_0(s, 0, 0, m_f^2 , m_f^2, m_f^2) \left[s + 2m_f^2 \right] \bigg\}\, .
\eea
The form factor $\Phi_{1\, F}$ is characterized by the presence of an anomaly pole
\beq
\Phi^F_{1\, pole}\equiv i \kappa \frac{\alpha}{9 \pi \, s} \, Q_f^2
\eeq
which is responsible for the generation of the anomaly in the massless limit. This $1/s$ behaviour of the amplitude is also clearly identifiable in a $m_f^2/s$ (asymptotic) expansion ($s \gg m_f^2$), where $m_f$ denotes generically any fermion mass of the
SM.
In this second case, the scaleless contribution associated with the exchange of a massless state (i.e. the $1/s$ term) is corrected by other terms which are suppressed as powers of $m_f^2/s$. This pattern, as we are going to show, is general.

The other gauge-invariant sector of the $TAA$ vertex is the one mediated by the exchange of bosons and ghosts in the loop. In this
sector the form factors are given by
\bea
\Phi_{1\, B} (s,\,0,\,0,\,M_W^2) &=& - i \frac{\kappa}{2}\, \frac{\alpha}{\pi \, s} \bigg\{
\frac{5}{6} - \frac{2\,M_W^2}{s} + 2\,M_W^2 \, \mathcal C_0 (s, 0, 0, M_W^2, M_W^2, M_W^2)
\bigg[1 - \frac{2 M_W^2}{s}\bigg] \bigg\},  \nn \\\\
\Phi_{2\, B} (s,\,0,\,0,\,M_W^2)  &=& - i \frac{\kappa}{2}\, \frac{\alpha}{\pi \, s} \bigg\{
\frac{1}{24} + \frac{M_W^2}{2\,s} + \frac{3\,M_W^2}{2\,s} \mathcal D_0 (s, 0, 0, M_W^2, M_W^2)  \nn \\
&& \hspace{-1cm} + \frac{M_W^2}{2} \mathcal C_0(s, 0, 0, M_W^2, M_W^2, M_W^2 )\, \left[ 1 + \frac{2\,M_W^2}{s}\right] \bigg\}\, , \\
\Phi_{3\,B} (s,\,0,\,0,\,M_W^2) &=& - i \frac{\kappa}{2}\, \frac{\alpha}{\pi \, s} \bigg\{
-\frac{15\,s}{8}-\frac{3\,M_W^2}{2} - \frac{1}{2}\, \mathcal D_0 (s, 0, 0, M_W^2, M_W^2)\left[5 M_W^2+7\,s \right]\nn\\
&& \hspace{-1cm} - \frac{3}{4}s\, \mathcal B_0 (0, M_W^2, M_W^2) - \mathcal C_0(s, 0, 0, M_W^2 , M_W^2, M_W^2) \left[s^2 + 4 M_W^2\,s + 3\,M_W^4\right]
\bigg\}. \nn \\
\eea
As in the previous case, we focus our attention on $\Phi_{1\, B}$, which multiplies the tensor structure $\phi_1$, responsible for the generation of the trace of the EMT. The contribution of the anomaly pole is isolated in the form
\beq
\Phi_{1\, B, pole}\equiv - i \frac{\kappa}{2}\, \frac{\alpha}{\pi \, s} \frac{5}{6}\, .
\eeq
It is clear, also in this case, that in the massless limit ($M_W=0$), i.e. in the symmetric phase of the theory, this pole is completely responsible for the generation of the anomaly. At the same time, at high energy  (i.e. for $s\gg M_W^2$) the massless exchange can be easily exposed as a dominant contribution to the trace part of the correlator. Notice that, in general, the correlator has other $1/s$ singularities in the remaining form factors and even constant terms which are unsuppressed for a large $s$, but these are not part of the trace.

The contributions coming from the term of improvement are characterized just by two form factors
\bea
\Phi_{1\, I} (s,\,0,\,0,\,M_W^2) &=& - i \frac{\kappa}{2}\frac{\alpha}{3 \pi \, s} \bigg\{ 1 + 2 M_W^2 \,C_0 (s, 0, 0, M_W^2, M_W^2, M_W^2)\bigg\} \,,\\
\Phi_{4\, I} (s,\,0,\,0,\,M_W^2) &=&  i \frac{\kappa}{2}\frac{\alpha}{6 \pi }  M_W^2 \,C_0 (s, 0, 0, M_W^2, M_W^2, M_W^2) \,.
\eea
Now we consider the external graviton leg corrections $\Delta^{\mu\nu\alpha\beta}(p,q)$. In this case only the term of improvement contributes with the diagram depicted in Fig. \ref{P3HVVImpr}
\bea
\Delta^{\mu\nu\alpha\beta}(p,q) &=& \Delta^{\mu\nu\alpha\beta}_I (p,q) \nn \\
&&\hspace{-2.cm} = \Psi_{1\, I} (s,0, 0,m_f^2,M_W^2,M_H^2) \, \phi_1^{\mu\nu\alpha\beta}(p,q) + \Psi_{4 \, I} (s,0, 0,M_W^2)  \, \phi_4^{\mu\nu\alpha\beta}(p,q)\, .
\label{P3DAA}
\eea
This is built by combining the tree level vertex for graviton/Higgs mixing  - coming from the improved EMT -  and the Standard Model Higgs/photon/photon correlator at one-loop
\bea
\Psi_{1\, I} (s,\,0,\,0,\,m_f^2,M_W^2,M_H^2) &=& - i \frac{\kappa}{2} \frac{\alpha}{3 \pi \, s (s - M_H^2)} \nn \\
&& \hspace{-5cm} \times \bigg\{ 2 m_f^2 \, Q_f^2 \bigg[ 2 + (4 m_f^2 -s) C_0 (s, 0, 0, m_f^2, m_f^2, m_f^2) \bigg] \nn \\
&& \hspace{-5cm} + M_H^2 + 6 M_W^2 + 2 M_W^2 (M_H^2 + 6 M_W^2 - 4 s) C_0(s,0,0, M_W^2,M_W^2,M_W^2) \bigg\} \,, \\
\Psi_{4\, I} (s,\,0,\,0,M_W^2) &=& - \Phi_{4\, I} (s,\,0,\,0,\,M_W^2)  \, .
\eea
\subsection{Results for the TAZ correlator}
We proceed with the analysis of the $TAZ$ correlator, in particular we start with the irreducible vertex $\Sigma ^{\mu\nu\alpha\beta}(p,q)$ that can be defined, as in the previous case, as a sum of the three gauge invariant contributions: the fermion sector (F), (Fig. \ref{P3triangles}(a), Fig. \ref{P3t-bubble}(a)), the gauge boson sector (B), (Fig. \ref{P3triangles}(b)-(g), Fig. \ref{P3t-bubble}(b)-(g), Fig. \ref{P3s-bubble}, Fig. \ref{P3tadpoles}) and the improvement term (I) given by the diagrams depicted in Fig. \ref{P3triangles}(c), (d) and Fig. \ref{P3s-bubble}(b), with the graviton - scalar - scalar vertices determined by the $T_I^{\mu\nu}$
\bea
\Sigma ^{\mu\nu\a\b}(p,q) = \Sigma_{F}^{\mu\nu\a\b}(p,q) + \Sigma_{B}^{\mu\nu\a\b}(p,q) + \Sigma_{I}^{\mu\nu\a\b}(p,q).
\eea
Each of these terms can be expanded in the on-shell case ($p^2 = 0$, $q^2 = M_Z^2$) on a tensor basis $f_i^{\mu\nu\alpha\beta}(p,q)$
\bea
\Sigma ^{\mu\nu\alpha\beta}_{F}(p,q) &=&  \, \sum_{i=1}^{7} \Phi_{i\,F} (s,0, M_Z^2,m_f^2) \, f_i^{\mu\nu\alpha\beta}(p,q)\,, \\
\Sigma ^{\mu\nu\alpha\beta}_{B}(p,q) &=&  \, \sum_{i=1}^{9} \Phi_{i\,B} (s,0, M_Z^2,M_W^2) \, f_i^{\mu\nu\alpha\beta}(p,q)\,, \\
\Sigma ^{\mu\nu\alpha\beta}_{I}(p,q) &=&  \Phi_{1\,I} (s,0, M_Z^2,M_W^2) \, f_1^{\mu\nu\alpha\beta}(p,q) + \Phi_{8\,I} (s,0, M_Z^2,M_W^2) \, f_8^{\mu\nu\alpha\beta}(p,q) \,.
\eea
For the on-shell $TAZ$ correlator the tensor structures are explicitly defined as
\bea
f_1^{\mu\nu\alpha\beta}(p,q) &=&  (s \, \eta^{\mu\nu} - k^{\mu}k^{\nu}) [ \frac{1}{2}\left( s-M_Z^2 \right)\eta^{\alpha\beta} - q^\alpha p^\beta ]  \,,\nn \\
f_2^{\mu\nu\alpha\beta}(p,q) &=& p^{\mu} p^{\nu} [ \frac{1}{2}\left( s-M_Z^2 \right)\eta^{\alpha\beta} - q^\alpha p^\beta ]  \,,\nn \\
f_3^{\mu\nu\alpha\beta}(p,q) &=& (M_Z^2 \, \eta^{\mu\nu} - 4 q^{\mu} q^{\nu})[ \frac{1}{2}\left( s-M_Z^2 \right)\eta^{\alpha\beta} - q^\alpha p^\beta]  \,,\nn \\
f_4^{\mu\nu\alpha\beta}(p,q) &=& [ \frac{1}{2}\left( s-M_Z^2 \right) \eta^{\mu\nu} - 2 ( q^{\mu} p^{\nu} + p^{\mu} q^{\nu})][ \frac{1}{2}\left( s-M_Z^2 \right)\eta^{\alpha\beta} - q^\alpha p^\beta]  \,,\nn \\
f_5^{\mu\nu\alpha\beta}(p,q) &=& p^{\beta}[  \frac{1}{2}\left( s-M_Z^2 \right) (\eta^{\alpha\nu} q^{\mu} + \eta^{\alpha\mu} q^{\nu}) - q^{\alpha}(q^{\mu} p^{\nu} + p^{\mu} q^{\nu})] \,,\nn  \\
f_6^{\mu\nu\alpha\beta}(p,q) &=& p^{\beta}[ \frac{1}{2}\left( s-M_Z^2 \right) (\eta^{\alpha\nu} p^{\mu} + \eta^{\alpha\mu} p^{\nu}) -2 q^{\alpha} p^{\mu} p^{\nu}] \,,\nn \\
f_7^{\mu\nu\alpha\beta}(p,q) &=& (p^{\mu} q^{\nu} + p^{\nu} q^{\mu}) \eta^{\alpha \beta} +  \frac{1}{2}\left( s-M_Z^2 \right) (\eta^{\alpha\nu} \eta^{\beta\mu} + \eta^{\alpha\mu} \eta^{\beta\nu}) \nn \\ 
&-&  \eta^{\mu\nu}  [ \frac{1}{2}\left( s-M_Z^2 \right)\eta^{\alpha\beta} - q^\alpha p^\beta ] 
-  (\eta^{\beta\nu} p^{\mu} + \eta^{\beta\mu} p^{\nu}) q^{\alpha} - (\eta^{\alpha\nu} q^{\mu} + \eta^{\alpha\mu} q^{\nu}) p^{\beta} \,,\nn \\
f_8^{\mu\nu\alpha\beta}(p,q) &=&  (s \, \eta^{\mu\nu} - k^{\mu}k^{\nu}) \eta^{\alpha\beta}  \,,\nn \\
f_9^{\mu\nu\alpha\beta}(p,q) &=& q^{\alpha}[ 3 s (\eta^{\beta\mu}p^{\nu} + \eta^{\beta\nu}p^{\mu}) - p^{\beta}(s \, \eta^{\mu\nu} + 2 k^{\mu}k^{\nu} ) ].
\eea
We collect here just the form factors in the fermion and boson sectors which contribute to the trace anomaly, while the remaining ones are given in appendix \ref{P3formfactors}
\bea
\Phi_1^{(F)}(s,0,M_Z^2,m_f^2) &=& - i \frac{\kappa}{2} \frac{\alpha}{3 \pi s\, s_w \, c_w } C_v^f \, Q_f \bigg\{
-\frac{1}{3} +\frac{2 m_f^2}{s-M_Z^2} \nn \\
&& \hspace{-4cm} +\frac{2 m_f^2 \, M_Z^2 }{(s-M_Z^2)^2} \mathcal D_0\left(s,M_Z^2,m_f^2,m_f^2\right)
 - m_f^2  \bigg[1- \frac{4 m_f^2}{s-M_Z^2}\bigg] \mathcal C_0\left(s,0,M_Z^2,m_f^2,m_f^2,m_f^2\right)
  \bigg\}  \,, \\
\Phi_1^{(B)}(s,0,M_Z^2,M_W^2) &=& - i \frac{\kappa}{2} \frac{\alpha}{3 \pi s \, s_w \, c_w }  \bigg\{
\frac{1}{12}(37 - 30 s_w^2) - \frac{M_Z^2}{2(s-M_Z^2)}(12 s_w^4 - 24 s_w^2 + 11) \nn \\
&& \hspace{-4cm} -\frac{M_Z^2}{2 \left(s-M_Z^2\right)^2} \left(2 M_Z^2 \left(6 s_w^4-11 s_w^2+5\right)-2 s_w^2 s +s \right) \mathcal D_0\left(s,M_Z^2,M_W^2,M_W^2\right) \nn \\
&& \hspace{-4cm} -\frac{M_Z^2 c_w^2}{s-M_Z^2} \left(2 M_Z^2 \left(6 s_w^4-15 s_w^2+8\right)+s \left(6 s_w^2-5\right)\right) \mathcal C_0\left(s,0,M_Z^2,M_W^2,M_W^2,M_W^2\right) \bigg\} \, .
\eea
Moreover, the improvement term is defined by the following two form factors
\bea
\Phi_{1\,I} (s,0, M_Z^2,M_W^2) &=&- i \frac{\kappa}{2} \frac{\alpha \, (c_w^2 - s_w^2)}{6 \pi \, s_w \, c_w \, (s-M_Z^2)}\bigg\{ 1 + 2 M_W^2 \, \mathcal C_0\left(s,0,M_Z^2,M_W^2,M_W^2,M_W^2\right) \,, \nn \\
&+&  \frac{M_Z^2}{s-M_Z^2}  \mathcal D_0\left(s,M_Z^2,M_W^2,M_W^2\right)  \bigg\}\, , \\
\Phi_{2\,I} (s,0, M_Z^2,M_W^2) &=& - i \frac{\kappa}{2} \frac{\alpha}{6 \pi}  s_w^2 \, M_Z^2 \, \mathcal C_0\left(s,0,M_Z^2,M_W^2,M_W^2,M_W^2\right).
\eea
Now we consider the external graviton leg corrections $\Delta^{\mu\nu\alpha\beta}(p,q)$. In this case only the improvement term contributes with the diagram shown in Fig. \ref{P3HVVImpr}
\bea
\Delta^{\mu\nu\alpha\beta}(p,q) &=& \Delta^{\mu\nu\alpha\beta}_I (p,q) \nn \\
&& \hspace{-2.cm} = \Psi_{1\, I} (s,0, M_Z^2,m_f^2,M_W^2,M_H^2) \, \phi_1^{\mu\nu\alpha\beta}(p,q) + \Psi_{4 \, I} (s,0, M_Z^2,M_W^2)  \, \phi_4^{\mu\nu\alpha\beta}(p,q).
\label{P3DAZ}
\eea
This is built by joining the graviton/Higgs mixing tree level vertex - coming from the improved energy-momentum tensor - and the Standard Model Higgs/photon/Z boson one-loop correlator.
\bea
\Psi_{1\, I} (s,0, M_Z^2,m_f^2,M_W^2,M_H^2) &=& - i \frac{\kappa}{2} \frac{\alpha}{6 \pi \, s_w \, c_w \, (s-M_H^2)(s-M_Z^2)} \nn \\
&& \hspace{-4.5cm} \times \bigg\{ 2 m_f^2 \, C_v^f \, Q_f \bigg[ 2 + 2 \frac{M_Z^2}{s-M_Z^2} \, \mathcal D_0\left(s,M_Z^2,m_f^2,m_f^2\right)   \nn \\
&& \hspace{-4.5cm}  + (4 m_f^2 + M_Z^2 -s) \mathcal C_0\left(s,0,M_Z^2,m_f^2,m_f^2,m_f^2\right) \bigg] + M_H^2 (1 - 2 s_w^2) + 2 M_Z^2 (6 s_w^4 - 11 s_w^2 + 5) \nn \\
&&  \hspace{-4.5cm}  + \frac{M_Z^2}{s-M_Z^2} \left(M_H^2 (1 - 2 s_w^2) + 2 M_Z^2 (6 s_w^4 - 11 s_w^2 + 5)\right) \mathcal D_0\left(s,M_Z^2,M_W^2,M_W^2\right) \nn \\
&&  \hspace{-4.5cm}  + 2 M_W^2  \, \mathcal C_0\left(s,0,M_Z^2,M_W^2,M_W^2,M_W^2\right) \left( M_H^2 (1-2s_w^2) + 2 M_Z^2 (6 s_w^4 - 15 s_w^2 + 8) \right. \nn \\
&& \hspace{-4.5cm}  \left. + 2 s (4 s_w^2-3) \right) \bigg\},  \\
\Psi_{4\, I} (s,0, M_Z^2,M_W^2) &=& - i \frac{\kappa}{2} \frac{\alpha \, c_w}{6 \pi \, s_w} M_Z^2 \bigg\{ \frac{2}{s - M_H^2} \mathcal B_0\left( 0, M_W^2,M_W^2\right) \nn \\
&-&  s_w^2 \, \mathcal C_0\left(s,0,M_Z^2,M_W^2,M_W^2,M_W^2\right) \bigg\}. 
\eea
\subsection{Results for the TZZ correlator}

Our analysis starts with the irreducible amplitude and then we move to the insertions on the external graviton leg.\\
The irreducible vertex $\Sigma ^{\mu\nu\alpha\beta}(p,q)$ of the TZZ correlator for on-shell Z bosons can be separated into 
three contributions defined by the mass of the particles circulating in the loop, namely the fermion mass $m_f$ (fermion sector
(F)  with diagrams depicted in Fig. \ref{P3triangles}(a), Fig. \ref{P3t-bubble}(a)), the $W$ gauge boson mass $M_W$ (the $W$ gauge
boson sector (W) with diagrams Fig. \ref{P3triangles}(b)-(g), Fig. \ref{P3t-bubble}(b)-(g), Fig. \ref{P3s-bubble}, Fig.
\ref{P3tadpoles}),the $Z$ and the Higgs bosons masses, $M_Z$ and $M_H$ ($(Z,H)$ sector with the contributions represented in Figs.
\ref{P3triangles1} - \ref{P3tadpoles1}), which cannnot be separated because of scalar integrals with both masses in their internal
lines. There is also a diagram proportional to a Higgs tadpole (Fig. \ref{P3tadpolesHiggs}(a)) which vanishes after 
renormalization and so it is not included in the results given below. Finally there is the improvement term (I) given by 
the diagrams depicted in Fig. \ref{P3triangles}(c), (d), Fig. \ref{P3s-bubble}(b), Fig. \ref{P3triangles1}(b), (c), (d) and Fig.
\ref{P3s-bubble1} with the graviton - scalar - scalar vertices given by the $T_I^{\mu\nu}$. We obtain
\bea
\Sigma ^{\mu\nu\alpha\beta}(p,q) = \Sigma ^{\mu\nu\alpha\beta}_F (p,q)  + \Sigma ^{\mu\nu\alpha\beta}_W (p,q) + \Sigma
^{\mu\nu\alpha\beta}_{Z,H} (p,q) +  \Sigma ^{\mu\nu\alpha\beta}_I (p,q).
\eea
These four on-shell contributions can be expanded on a tensor basis given by 9 tensors
\bea
t_1^{\mu\nu\alpha\beta}(p,q) &=&  (s g^{\mu\nu} - k^{\mu}k^{\nu}) \left[ \left( \frac{s}{2}-M_Z^2 \right)g^{\alpha\beta} -
q^\alpha p^\beta \right] \,, \nn \\
t_2^{\mu\nu\alpha\beta}(p,q) &=&  (s g^{\mu\nu} - k^{\mu}k^{\nu}) g^{\alpha\beta} \,, \nn \\
t_3^{\mu\nu\alpha\beta}(p,q) &=&  g^{\mu\nu} g^{\alpha\beta} - 2 \left( g^{\mu\alpha} g^{\nu\beta} + g^{\mu\beta} g^{\nu\alpha}  \right) \,, \nn \\
t_4^{\mu\nu\alpha\beta}(p,q) &=& \left(p^{\mu}p^{\nu} + q^{\mu}q^{\nu} \right) g^{\alpha\beta} - M_Z^2 \left( g^{\mu\alpha} g^{\nu\beta} + g^{\mu\beta} g^{\nu\alpha}  \right) \,,\nn \\
t_5^{\mu\nu\alpha\beta}(p,q) &=&  \left(p^{\mu}q^{\nu} + q^{\mu}p^{\nu} \right) g^{\alpha\beta} - \left( \frac{s}{2} - M_Z^2 \right) \left( g^{\mu\alpha} g^{\nu\beta} + g^{\mu\beta} g^{\nu\alpha}  \right) \,, \nn \\
t_6^{\mu\nu\alpha\beta}(p,q) &=& \left(g^{\mu\alpha}q^{\nu} + g^{\nu\alpha}q^{\mu} \right) p^{\beta} +  \left(g^{\mu\beta}p^{\nu} + g^{\nu\beta}p^{\mu} \right) q^{\alpha} - g^{\mu\nu}p^{\beta}q^{\alpha} \,, \nn \\
t_7^{\mu\nu\alpha\beta}(p,q) &=& \left(g^{\mu\alpha}p^{\nu} + g^{\nu\alpha}p^{\mu} \right) p^{\beta} +  \left(g^{\mu\beta}q^{\nu} + g^{\nu\beta}q^{\mu} \right) q^{\alpha} \,, \nn \\
t_8^{\mu\nu\alpha\beta}(p,q) &=& \left[ 2 \left( p^{\mu} p^{\nu} + q^{\mu} q^{\nu} \right) - M_Z^2 g^{\mu\nu} \right] p^{\beta}q^{\alpha} \,, \nn \\
t_9^{\mu\nu\alpha\beta}(p,q) &=& \left[ 2 \left( p^{\mu} q^{\nu} + q^{\mu} p^{\nu} \right) - \left(\frac{s}{2}- M_Z^2 \right) g^{\mu\nu} \right] p^{\beta}q^{\alpha} \,,
\eea
and can be written in terms of form factors $\Phi_i$
\bea
\Sigma^{\mu\nu\alpha\beta}_F (p,q) &=& \sum_{i = 1}^9 {\Phi^{(F)}_i (s,M_Z^2,M_Z^2,m_f^2) \, t_i^{\mu\nu\alpha\beta}(p,q)} \,, \\
\Sigma^{\mu\nu\alpha\beta}_W (p,q) &=& \sum_{i = 1}^9 {\Phi^{(W)}_i (s,M_Z^2,M_Z^2,M_W^2) \, t_i^{\mu\nu\alpha\beta}(p,q)} \,, \\
\Sigma^{\mu\nu\alpha\beta}_{Z,H} (p,q) &=& \sum_{i = 1}^9 {\Phi^{(Z,H)}_i (s,M_Z^2,M_Z^2,M_Z^2,M_H^2) \, t_i^{\mu\nu\alpha\beta}(p,q)} \,, \\
\Sigma^{\mu\nu\alpha\beta}_{I} (p,q) &=& \Phi^{(I)}_1 (s,M_Z^2,M_Z^2,M_W^2,M_Z^2,M_H^2) \, t_1^{\mu\nu\alpha\beta}(p,q) \nn \\
&+&  \Phi^{(I)}_2 (s,M_Z^2,M_Z^2,M_W^2,M_Z^2,M_H^2) \, t_2^{\mu\nu\alpha\beta}(p,q) \,, 
\eea
where the first three arguments of the $\Phi_i$ represent the mass-shell and virtualities of the external lines $k^2 = s, \, p^2 = q^2 = M_Z^2$, while the remaining ones give the masses in the internal lines.\\ Moreover, we expand each form factor into a basis of independent scalar integrals.
\subsubsection{The fermion sector}
We start from the fermion contribution to $TZZ$ and then move to those coming from a $W$ running inside the loop (W loops) or a
$Z$ and a Higgs ($Z, H$ loops). We expand each form factor in terms of coefficients ${C_{(F)}}_j^i$
\bea
\Phi^{(F)}_i (s,M_Z^2,M_Z^2,m_f^2) = \sum_{j = 0}^4 {C_{(F)}}_j^i(s,M_Z^2,M_Z^2,m_f^2) \, \mathcal I^{(F)}_j
\label{P3fermionhZZ}
\eea
where $\mathcal I^{(F)}_j$ are a set of scalar integrals given by
\bea
\mathcal I^{(F)}_0 &=& 1 \, , \nn \\
\mathcal I^{(F)}_1 &=& \mathcal A_0(m_f^2) \, , \nn \\
\mathcal I^{(F)}_2 &=& \mathcal B_0(s, m_f^2,m_f^2) \, , \nn \\
\mathcal I^{(F)}_3 &=& \mathcal D_0(s, M_Z^2, m_f^2,m_f^2) \, , \nn \\
\mathcal I^{(F)}_4 &=& \mathcal C_0(s, M_Z^2,M_Z^2 ,m_f^2,m_f^2,m_f^2) \, . \label{P3IntF}
\eea
As in the previous case, only $\Phi^{(F)}_1$ contributes to the anomaly, and we will focus our attention only on this form factor. The expressions of all the coefficients ${C_{(F)}}_j^i$ for $(i\neq 1)$ can be found in appendix \ref{P3formfactors}. We obtain
\bea
{C_{(F)}}_0^1 &=&   -\frac{i  \kappa\, \alpha  \, m_f^2}{6 \pi  s^2 c_w^2 s_w^2 \left(s-4 M_Z^2\right)}
\left(s-2 M_Z^2\right) \left(C_a^{f \, 2}+C_v^{f\, 2}\right)
+ \frac{i \alpha \, \kappa}{36 \pi c_w^2 s_w^2 \, s}\left(C_a^{f \, 2}+C_v^{f \, 2}\right)\, ,  \nn \\
{C_{(F)}}_1^1 &=&   {C_{(F)}}_2^1 = 0 \, , \nn \\
{C_{(F)}}_3^1 &=&     -\frac{i \kappa\, \alpha  \, m_f^2}{3 \pi  s^2 c_w^2 \left(s-4 M_Z^2\right){}^2 s_w^2}
\left(\left(2 M_Z^4-3 s M_Z^2+s^2\right) C_a^{f \, 2}+C_v^{f \, 2} M_Z^2 \left(2 M_Z^2+s \right)\right) , , \nn \\
{C_{(F)}}_4^1 &=&   -\frac{i \kappa\, \alpha \, m_f^2}{12 \pi  s^2 c_w^2 \left(s-4 M_Z^2\right){}^2 s_w^2}
\left(s-2 M_Z^2\right) \bigg(\left(4 M_Z^4-2 \left(8 m_f^2+s \right) M_Z^2+s \left(4 m_f^2+s \right)\right) C_a^{f \, 2}\nn\\
&& +C_v^{f \, 2} \left(4 M_Z^4+2 \left(3 s-8 m_f^2\right) M_Z^2-s \left(s-4 m_f^2\right)\right)\bigg)\, .
\eea
The anomaly pole of $\Phi^{(F)}_1$ is entirely contained in ${C_{(F)}}_0^1$ and it is given by
\beq
\Phi^{(F)}_{1\, pole}\equiv\frac{i \alpha \, \kappa  \left(C_a^{f \, 2}+C_v^{f \, 2}\right)}{36 \pi c_w^2 s_w^2 \, s}\, .
\eeq
\subsubsection{The $W$ boson sector}
As we move to the contributions coming from loops of $W$'s, the 9 form factors are expanded as
\bea
\Phi^{(W)}_i (s,M_Z^2,M_Z^2,M_W^2) = \sum_{j = 0}^4 {C_{(W)}}_j^i(s,M_Z^2,M_Z^2,M_W^2) \, \mathcal I^{(W)}_j
\label{P3boson1hZZ}
\eea
where $\mathcal I^{(W)}_j$ are now given by
\bea
\mathcal I^{(W)}_0 &=& 1 \, , \nn \\
\mathcal I^{(W)}_1 &=& \mathcal A_0(M_W^2) \, , \nn \\
\mathcal I^{(W)}_2 &=& \mathcal B_0(s, M_W^2,M_W^2) \, , \nn \\
\mathcal I^{(W)}_3 &=& \mathcal D_0(s, M_Z^2, M_W^2,M_W^2) \, , \nn \\
\mathcal I^{(W)}_4 &=& \mathcal C_0(s, M_Z^2,M_Z^2 ,M_W^2,M_W^2,M_W^2) \, . \label{P3IntW}
\eea
The anomaly pole is extracted from the expansion of $\Phi^{(W)}_1 $, whose coefficients are
\bea
{C_{(W)}}_0^1 &=&  \frac{-i \kappa\, \alpha}{2 s_w^2 \, c_w^2 \, \pi \, s}
\bigg\{ \frac{M_Z^2}{6 s \left(s-4 M_Z^2\right)} \bigg[ 2 M_Z^2 \left(-12 s_w^6+32 s_w^4-29 s_w^2+9\right)\nn\\
&& +s \left(12 s_w^6-36 s_w^4+33 s_w^2-10\right)\bigg]+  \frac{(60s_w^4-148s_w^2+81)}{72}  \bigg\}\, , \nn \\
{C_{(W)}}_1^1 &=& {C_{(W)}}_2^1 = 0\, , \nn \\
{C_{(W)}}_3^1 &=&  \frac{- i \kappa \, \alpha \, M_Z^2}{ 12 s_w^2 \, c_w^2 \, \pi \, s^2 \, (s-4 M_Z)^2}   \bigg( 4 M_Z^4 (12 s_w^6 -32 s_w^4 +29 s_w^2 - 9) \nn \\
 && + 2 M_Z^2 s (s_w^2 - 2)(12 s_w^4 - 12 s_w^2 +1) + s^2 (-4 s_w^4+8s_w^2-5)   \bigg)\, , \nn \\
{C_{(W)}}_4^1 &=&   \frac{- i \kappa \, \alpha \, M_Z^2}{ 12 s_w^2 \, c_w^2 \, \pi \, s^2 \, (s-4 M_Z)^2}  \bigg(-4M_Z^6(s_w^2-1)(4 s_w^2-3)(12 s_w^4-20s_w^2+9) \nn \\
&& + 2 M_Z^4 s (18 s_w^4-34s_w^2 + 15)(4(s_w^2-3)s_w^2+7) - 2M_Z^2 s^2 (12 s_w^8-96s_w^6 \nn \\
&& +201s_w^4-157s_w^2+41)+ s^3(-12 s_w^6+32s_w^4-27s_w^2+7)  \bigg) \, .
\eea
As one can immediately see, the pole is entirely contained in ${C_{(W)}}_0^1$, and we obtain
\beq
\Phi^{(W)}_{1\, pole} \equiv  - i \frac{\kappa}{2} \frac{\alpha}{s_w^2 \, c_w^2 \, \pi \, s} \frac{(60s_w^4-148s_w^2+81)}{72}\, .
\eeq
\subsubsection{The $(Z,H)$ sector}
Finally, the last contribution to investigate in the $TZZ$ vertex is the one coming from a Higgs
($H$) or a Z boson ($Z$) running in the loops. Also in this case we obtain
\bea
\Phi^{(Z,H)}_i (s,M_Z^2,M_Z^2,M_Z^2,M_H^2) = \sum_{j = 0}^8 {C_{(Z,H)}}_j^i(s,M_Z^2,M_Z^2,M_Z^2,M_H^2) \, \mathcal I^{(Z,H)}_j
\label{P3boson2hZZ}
\eea
with the corresponding $\mathcal I^{(Z,H)}_j$ given by
\bea
\mathcal I^{(Z,H)}_0 &=& 1 \, , \nn \\
\mathcal I^{(Z,H)}_1 &=& \mathcal A_0(M_Z^2) \, , \nn \\
\mathcal I^{(Z,H)}_2 &=& \mathcal A_0(M_H^2) \, , \nn \\
\mathcal I^{(Z,H)}_3 &=& \mathcal B_0(s, M_Z^2,M_Z^2) \, , \nn \\
\mathcal I^{(Z,H)}_4 &=& \mathcal B_0(s, M_H^2,M_H^2) \, , \nn \\
\mathcal I^{(Z,H)}_5 &=& \mathcal B_0(M_Z^2, M_Z^2,M_H^2) \, , \nn \\
\mathcal I^{(Z,H)}_6 &=& \mathcal C_0(s, M_Z^2,M_Z^2 ,M_Z^2,M_H^2,M_H^2) \, , \nn \\
\mathcal I^{(Z,H)}_7 &=& \mathcal C_0(s, M_Z^2,M_Z^2 ,M_H^2,M_Z^2,M_Z^2) \, . \label{P3IntZH}
\eea
Again, as before, the contributions to $\Phi^{(Z,H)}_1$ are those responsible for a non vanishing trace in the massless limit. These are given by
\bea
{C_{(Z,H)}}_0^1 &=& \frac{i \kappa \, \alpha}{24 \pi  s^2 c_w^2 s_w^2 \left(s-4 M_Z^2\right)}
\left(M_H^2 \left(s-2 M_Z^2\right)+3 s M_Z^2- 2 M_Z^4\right) + \frac{7 i \alpha  \kappa }{144 \pi  s c_w^2 s_w^2}\, ,  \nn \\
{C_{(Z,H)}}_1^1 &=& \frac{i \kappa \, \alpha}{12 \pi  s^2 c_w^2 s_w^2 \left(s-4 M_Z^2\right)}
\left(M_Z^2-M_H^2\right) \, ,\nn \\
{C_{(Z,H)}}_2^1 &=& - {C_{(Z,H)}}_1^1  \, ,\nn \\
{C_{(Z,H)}}_3^1 &=&  \frac{i \kappa \, \alpha}{24 \pi  s^2 c_w^2 s_w^2 \left(s-4 M_Z^2\right){}^2}
   \left(2 M_H^2 \left(s M_Z^2-2 M_Z^4+s^2\right)+3 s^2 M_Z^2-14 s M_Z^4+8 M_Z^6\right)  \, ,\nn \\
{C_{(Z,H)}}_4^1 &=&  -\frac{i \kappa \, \alpha}{24 \pi  s^2 c_w^2 s_w^2 \left(s-4 M_Z^2\right){}^2}
   \left(2 M_H^2+s\right) \left(2 M_H^2 \left(s-M_Z^2\right)-3 s M_Z^2\right) \, ,  \nn \\
{C_{(Z,H)}}_5^1 &=&  \frac{i \kappa \, \alpha}{12 \pi  s^2 c_w^2 \left(s-4 M_Z^2\right){}^2 s_w^2}
\left(s M_H^4+6 \left(s-M_H^2\right) M_Z^4+\left(2 M_H^4-3 s M_H^2-3 s^2\right) M_Z^2\right)\, ,\nn \\
{C_{(Z,H)}}_6^1 &=&  \frac{i \kappa \, \alpha}{24 \pi  s^2 c_w^2 s_w^2 \left(s-4 M_Z^2\right){}^2}
\left(2 M_H^2+s\right) \bigg(M_Z^2 \left(-8 s M_H^2-2 M_H^4+s^2\right)\nn\\
&& +2 M_Z^4 \left(4 M_H^2+s\right)+2 s M_H^4\bigg) \, ,\nn  \\
{C_{(Z,H)}}_7^1 &=& \frac{i \kappa \, \alpha  M_H^2}{24 \pi  s^2 c_w^2 s_w^2
   \left(s-4 M_Z^2\right){}^2} \left(2 M_H^2 \left(s M_Z^2-2
   M_Z^4+s^2\right)-20 s M_Z^4+16 M_Z^6+s^3\right)\, , \nn \\
\eea
with the anomaly pole, extracted from ${C_{(Z,H)}}_0^1$, given by
\beq
\Phi^{(Z,H)}_{1\, pole} \equiv \frac{7 i \alpha  \kappa }{144 \pi  s c_w^2 s_w^2}\, .
\eeq
\subsubsection{ Terms of improvement and external leg corrections}
The expression of form factors $\Phi^{(I)}_1$ and $\Phi^{(I)}_2$ coming from the terms of improvement for the $\Sigma^{\mu\nu\alpha\beta}_I(p,q)$ vertex are given in appendix \ref{P3imprformfactors}.

The next task is to analyze the external leg corrections to the $TZZ$ correlator. This case is much more involved than the previous one because there are contributions coming from the minimal EMT (i.e. without the improvement terms) Fig. \ref{P3tadpolesHiggs}(b), Fig. \ref{P3legcorr12}(a)-(b) and from the improved $T^{\mu\nu}_I$. This last contribution can be organized into three sectors: the first is characterized by a contribution from the one-loop graviton/Higgs two-point function Fig. \ref{P3tadpolesHiggs}(b), Fig. \ref{P3legcorr12}(a). The second is constructed with the Higgs self-energy Fig. \ref{P3legcorr12}(c) and the last is built with the Standard Model Higgs/Z/Z one-loop vertex Fig. \ref{P3HVVImpr}. Furthermore, it is important to note that the diagram depicted in Fig. \ref{P3tadpolesHiggs}(b) is proportional to the Higgs tadpole and vanishes in our renormalization scheme. \\
The $\Delta^{\mu\nu\alpha\beta}(p,q)$ correlator is decomposed as
\bea
\Delta^{\mu\nu\alpha\beta}(p,q) &=& \bigg[ \Sigma^{\mu\nu}_{Min, \, hH}(k) + \Sigma^{\mu\nu}_{I, \, hH}(k) \bigg] \frac{i}{k^2 - M_H^2} V_{HZZ}^{\alpha\beta} \nn \\
&+& V_{I, \, hH}^{\mu\nu}(k)  \frac{i}{k^2 - M_H^2}  \Sigma_{HH}(k^2) \frac{i}{k^2 - M_H^2} V_{HZZ}^{\alpha\beta} 
+ \Delta^{\mu\nu\alpha\beta}_{I, \, HZZ}(p,q)
\eea
where $\Sigma_{HH}(k^2)$ is the Higgs self-energy given in appendix (\ref{P3propagators}) for completeness, $V_{HZZ}^{\alpha\beta}$ and $ V_{I, \, hH}^{\mu\nu}$ are tree level vertices defined in appendix (\ref{P3FeynRules}) and $\Delta^{\mu\nu\alpha\beta}_{I, \, HZZ}(p,q)$ is expanded into the two form factors of improvement as
\bea
\Delta^{\mu\nu\alpha\beta}_{I, \, HZZ}(p,q) &=& \Psi^{(I)}_1 (s,M_Z^2,M_Z^2,m_f^2,M_W^2,M_Z^2,M_H^2) \, t_1^{\mu\nu\alpha\beta}(p,q) \nn \\
&+&  \Psi^{(I)}_2 (s,M_Z^2,M_Z^2,m_f^2,M_W^2,M_Z^2,M_H^2) \, t_2^{\mu\nu\alpha\beta}(p,q) \, ,\nn
\eea
\bea
\Psi^{(I)}_i (s,M_Z^2,M_Z^2,m_f^2,M_W^2,M_Z^2,M_H^2)  &=& \sum^4_{j=0} \, {C^{(I)}_{(F)}}^i_j \left( s,M_Z^2,M_Z^2,m_f^2 \right) \, \mathcal I^{(F)}_j \nn \\
&& \hspace{-2.cm} + \sum^4_{j=0} \, {C^{(I)}_{(W)}}^i_j \left( s,M_Z^2,M_Z^2,M_W^2 \right) \, \mathcal I^{(W)}_j \nn \\
&& \hspace{-2.cm} + \sum^7_{j=0} \, {C^{(I)}_{(Z,H)}}^i_j \left( s,M_Z^2,M_Z^2,M_Z^2,M_H^2 \right) \, \mathcal J^{(Z,H)}_j
\eea
where the basis of scalar integrals $\mathcal I^{(F)}_j$ and $\mathcal I^{(W)}_j$ have been defined respectively in Eq. \ref{P3IntF} and \ref{P3IntW}. The $(Z,H)$ sector is expanded into a different set (instead of Eq. \ref{P3IntZH}) which is given by
\bea
\mathcal J^{(Z,H)}_0 &=& 1 \,, \nn \\
\mathcal J^{(Z,H)}_1 &=& \mathcal A_0 \left( M_Z^2 \right) \,, \nn \\
\mathcal J^{(Z,H)}_2 &=& \mathcal A_0 \left( M_H^2 \right) \,, \nn \\
\mathcal J^{(Z,H)}_3 &=& \mathcal B_0 \left( s, M_Z^2, M_Z^2 \right) \,, \nn \\
\mathcal J^{(Z,H)}_4 &=& \mathcal B_0 \left( s, M_H^2, M_H^2 \right) \,, \nn \\
\mathcal J^{(Z,H)}_5 &=& \mathcal B_0 \left( M_Z^2, M_Z^2, M_H^2 \right) \,, \nn \\
\mathcal J^{(Z,H)}_6 &=& \mathcal C_0 \left( s, M_Z^2, M_Z^2, M_Z^2,M_H^2,M_H^2 \right) \,, \nn \\
\mathcal J^{(Z,H)}_7 &=& \mathcal C_0 \left( s, M_Z^2, M_Z^2, M_H^2,M_Z^2,M_Z^2 \right) \,.
\eea
The expressions of these coefficients together with the graviton-Higgs mixing $\Sigma^{\mu\nu}_{Min, \, hH}(k)$,  $\Sigma^{\mu\nu}_{I, \, hH}(k)$ can be found in appendix \ref{P3externalleg}.

\section{Renormalization}
In this section we discuss the renormalization of the correlators. This is based on the identification of the
1-loop counterterms to the Standard Model Lagrangian which, in turn, allow to extract a counterterm vertex for the improved EMT.
We have checked that the renormalization of all the parameters of the Lagrangian is indeed sufficient to cancel all the singularities of all the vertices, as expected. We have used the on-shell scheme which is widely used in the electroweak theory. In this scheme the
renormalization conditions are fixed in terms of the physical parameters of the theory to all orders in perturbation theory. These are the masses of physical particles  $M_W, M_Z, M_H, m_f$, the electric charge $e$ and the quark mixing matrix $V_{ij}$. The renormalization conditions on the fields - which allow to extract the renormalization constants of the wave functions - are obtained by requiring a unit residue of the full 2-point functions on the physical particle poles.

We start by defining the relations
\bea
e_0 &=& (1+ \delta Z_e) e \,, \nn \\
M^2_{W,0} &=& M^2_W + \delta M^2_W \, , \nn \\
M^2_{Z,0} &=& M^2_Z + \delta M^2_Z \, , \nn \\
M^2_{H,0} &=& M^2_H + \delta M^2_H \, , \nn \\
\left( \begin{array}{c} Z_0 \\ A_0 \end{array} \right) &=& \left( \begin{array}{cc} 1+\frac{1}{2} \delta Z_{ZZ} & \frac{1}{2} \delta Z_{ZA} \\  \frac{1}{2} \delta Z_{AZ} & 1+ \frac{1}{2} \delta Z_{AA} \end{array}\right) \left( \begin{array}{c} Z \\ A \end{array} \right) \,, \nn \\
H_0 &=& \left( 1 + \frac{1}{2}\delta Z_H \right) H.
\eea
At the same time we need the counterterms for the sine of the Weinberg angle $s_w$ and of the vev of the Higgs field $v$
\bea
s_{w\,,0} = s_w + \delta s_w \,, \qquad v_0 = v + \delta v, \,
\eea
which are defined to all orders by the relations
\bea
s_w^2 = 1 - \frac{M_W^2}{M_Z^2} \,, \qquad v^2 = 4 \frac{M_W^2 \, s_w^2}{e^2} \,,
\eea
and are therefore linked to the renormalized masses and gauge couplings. Specifically, one obtains
\bea
\frac{\delta s_w}{s_w} = - \frac{c_w^2}{2 s_w^2} \left( \frac{\delta M_W^2}{M_W^2} - \frac{\delta M_Z^2}{M_Z^2} \right) \,, \qquad \frac{\delta v}{v} = \left( \frac{1}{2} \frac{\delta M_W^2}{M_W^2} + \frac{\delta s_w}{s_w} - \delta Z_e \right),
\eea
while electromagnetic gauge invariance gives
\bea
\delta Z_e = - \frac{1}{2}\delta Z_{AA} + \frac{s_w}{2 c_w}\delta Z_{ZA}.
\eea
We also recall that the wave function renormalization constants are defined in terms of the 2-point functions of the fundamental fields as
\bea
&& \delta Z_{AA} = - \frac{\partial \Sigma_T^{AA}(k^2)}{\partial k^2} \bigg |_{k^2=0} \,, \quad
\delta Z_{AZ} = - 2 Re \frac{\Sigma_T^{AZ}(M_Z^2)}{M_Z^2} \,, \quad
\delta Z_{ZA} = 2 \frac{\Sigma_T^{AZ}(0)}{M_Z^2} \,, \nn\\
&&\delta Z_{ZZ} = - Re \frac{\partial \Sigma_T^{ZZ}(k^2)}{\partial k^2} \bigg |_{k^2 = M_Z^2} \,, \quad
\delta Z_H = - Re \frac{\partial \Sigma_{HH}(k^2)}{\partial k^2} \bigg |_{k^2 = M_H^2} \,,\quad
\delta M_Z^2 = Re \, \Sigma_T^{ZZ}(M_Z^2) \,, \nn\\
&&\delta M_W^2 = \widetilde{Re} \, \Sigma_T^{WW}(M_W^2) \,, \quad \delta M_H^2 = Re \, \Sigma_{HH}(M_H^2). \,
\eea
From the counterterms Lagrangian defined in terms of the $Z_{V V'}$ factors given above, we compute the corresponding counterterm to the EMT $\delta T^{\mu\nu}$ and renormalized EMT
\bea
T^{\mu\nu}_0 = T^{\mu\nu} + \delta T^{\mu\nu}
\eea
which is sufficient to cancel all the divergences of the theory. One can also verify from the explicit computation that the terms of improvement, in the conformally coupled case, are necessary to renormalize the vertices containing an intermediate scalar with an external bilinear mixing (graviton/Higgs).
The vertices extracted from the counterterms are given by
\bea
\delta [TAA]^{\mu\nu\alpha\beta}(k_1,k_2) &=& - i \frac{\kappa}{2} \bigg\{ k_1 \cdot k_2 \, C^{\mu\nu\alpha\beta} + D^{\mu\nu\alpha\beta}(k_1,k_2)\bigg\} \, \delta Z_{AA} \,, \\
\delta [TAZ]^{\mu\nu\alpha\beta}(k_1,k_2) &=& - i \frac{\kappa}{2} \bigg\{ \left( \delta c_1^{AZ} \, k_1 \cdot k_2 + \delta c_2^{AZ} \, M_Z^2 \right) \, C^{\mu\nu\alpha\beta} + \delta c_1^{AZ} \, D^{\mu\nu\alpha\beta}(k_1,k_2) \bigg\} \,,  \\
\delta [TZZ]^{\mu\nu\alpha\beta}(k_1,k_2) &=& - i \frac{\kappa}{2} \bigg\{ \left( \delta c_1^{ZZ} \, k_1 \cdot k_2 + \delta c_2^{ZZ} \, M_Z^2 \right) \, C^{\mu\nu\alpha\beta} + \delta c_1^{ZZ} \, D^{\mu\nu\alpha\beta}(k_1,k_2) \bigg\} \,, 
\eea
where the coefficients $\delta c$ are defined as
\bea
\delta c_1^{AZ} = \frac{1}{2}\left( \delta Z_{AZ} + \delta Z_{ZA} \right) \,, \quad \delta c_2^{AZ} = \frac{1}{2} \delta Z_{ZA} \,, \quad
\delta c_1^{ZZ} = \delta Z_{ZZ} \,, \quad  \delta c_2^{ZZ} = M_Z^2 \, \delta Z_{ZZ} + \delta M_Z^2 \,. 
\eea
These counterterms are sufficient to remove the divergences of the completely cut graphs \\
 ($\Sigma^{\mu\nu\alpha\beta}(p,q)$) which do not contain a bilinear mixing, once we set
on-shell  the external gauge lines. This occurs both for those diagrams which do not involve the terms of improvement and for those involving $T_I$. Regarding those contributions which involve the bilinear mixing on the external graviton line, we encounter two situations. For instance, the insertion of the bilinear mixing on the $TAA$ vertex generates a reducible diagram of the form Higgs/photon/photon which does not require any renormalization, being finite. Its contribution has been denoted as $\Delta^{\mu\nu\alpha\beta}_I(p,q)$ in Eq. (\ref{P3DAA}). In the case of the $TAZ$ vertex the corresponding contribution is given in Eq. (\ref{P3DAZ}). In this second case the renormalization is guaranteed, within the Standard Model, by the use of the Higgs/photon/Z counterterm
\bea
\delta [HAZ]^{\alpha\beta} = i \frac{e \, M_Z}{2 s_w c_w} \delta Z_{ZA}\, \eta^{\alpha\beta} \,.
\eea
As a last case, we discuss the contribution to $TZZ$ coming from the bilinear mixing. The corrections on the graviton line involve the graviton/Higgs mixing $i \Sigma^{\mu\nu}_{hH}(k)$, the Higgs self-energy $i \Sigma_{HH}(k^2)$ and the term of improvement $\Delta^{\mu\nu\alpha\beta}_{I\,,HZZ}(p,q)$, which introduces the Higgs/Z/Z vertex (or $HZZ$) of the Standard Model. The Higgs self-energy and the $HZZ$ vertex, in the Standard Model, are renormalized with the counterterms
\bea
\delta [HH](k^2) &=& i (\delta Z_H \, k^2 - M_H^2 \delta Z_H - \delta M_H^2) \, ,\\
\delta [HZZ]^{\alpha\beta} &=& i \frac{e \, M_Z}{s_w \, c_w} \bigg[ 1 + \delta Z_e + \frac{2 s_w^2 - c_w^2}{c_w^2} \frac{\delta s_w}{s_w} + \frac{1}{2} \frac{\delta M_W^2}{M_W^2} + \frac{1}{2} \delta Z_H + \delta Z_{ZZ}  \bigg] \, \eta^{\alpha\beta} .
\eea
The self-energy $i \Sigma^{\mu\nu}_{hH}(k)$ is defined by the minimal contribution generated by $T_{\mu\nu}^{Min}$ and by a second term derived from $T_{\mu\nu}^I$. This second term is necessary in order to ensure the renormalizability of the graviton/Higgs mixing.
In fact, the use of the minimal EMT in the computation of this self-energy involves a divergence of the form
\bea
\delta [hH]^{\mu\nu}_{Min} = i \frac{\kappa}{2} \delta t \, \eta^{\mu\nu} \,, \label{P3CThH}
\eea
with $\delta t$ fixed by the condition of cancellation of the Higgs tadpole $T_{ad}$ ($\delta t + T_{ad} = 0$) and hence of any linear term in $H$ within the 1-loop effective Lagrangian of the Standard Model.
A simple analysis of the divergences in $i \Sigma^{\mu\nu}_{Min, \, hH}$ shows that the counterterm given in Eq. \ref{P3CThH} is not sufficient to remove all the singularities of this correlator unless we also include the renormalization of the term of improvement which is given by
\bea
\delta [hH]^{\mu\nu}_{I}(k) = - i \frac{\kappa}{2} \left( - \frac{1}{3} \right) i  \bigg[ \delta v + \frac{1}{2} \delta Z_H \bigg] v \, (k^2 \, \eta^{\mu\nu} - k^{\mu}k^{\nu}).
\eea
One can show explicitly that this counterterm indeed ensures the finiteness of $i \Sigma^{\mu\nu}_{hH}(k)$.
\section{Comments}
\label{P3discussionsection}
Before coming to our conclusions, we pause for some comments on the meaning and the implications of the current computation in a more general context. This concerns the superconformal anomaly and its coupling to supergravity. 

 The study of the mechanism of anomaly mediation between the Standard Model and gravity has several interesting features which for sure will require further analysis in order to be put on a more rigorous basis. However, here we have preliminarily shown that the perturbative structure of a correlator - obtained by the insertion of a gravitational field on 2-point  functions of gauge fields - can be organized in terms of a rather minimal set of fundamental form factors. Their expressions have been given in this work, generalizing previous results in the QED and QCD cases. The trace anomaly can be attributed, in all the cases, just to one specific tensor structure, as discussed in the previous analysis.

We have also seen that at high energy the breaking of conformal invariance, in a theory with a Higgs mechanism, has two sources, one of them being radiative. This can be attributed to the exchange of anomaly poles in each gauge invariant sector of the graviton/gauge/gauge vertex, while the second one is explicit. As discussed in \cite{Coriano:2011ti} this result has a simple physical interpretation,
since it is an obvious consequence of the fact that at an energy much larger than any scale of the theory, we should recover the role of the anomaly and its pole-like behaviour. 

In turn, this finding sheds some light on the significance of the anomaly cancellation mechanism in 4-dimensional field theory - discussed in the context of supersymmetric theories coupled to gravity - based on the subtraction of an anomaly pole in superspace  \cite{LopesCardoso:1991zt}. Let's briefly see why. 

The theory indeed becomes conformally invariant at high energy and, in presence of supersymmetric interactions, this invariance is promoted to a superconformal invariance.  In a superconformal theory, such as an ${\cal N}=1$ super Yang-Mills theory,  the superconformal anomaly multiplet, generated by the radiative corrections, puts on the same role the trace anomaly, the chiral anomaly of the corresponding $U(1)_R$ current and the gamma trace of the supersymmetric current. Notice that these three anomalies are "gauged" if they are coupled to a conformal gravity supermultiplet and all equally need to be cancelled. The role of the Green-Schwarz (GS) mechanism, in this framework, if realized as a pole subtraction, is, therefore, to perform a subtraction of these pole-like contributions which show up in the UV region, and has to be realized in superspace \cite{LopesCardoso:1991zt,Bagger:1999rd} for obvious reasons. Then, one can naturally ask what is the nature of the pole that is indeed cancelled by the mechanism, if this is acting in the UV. The answer, in a way, is obvious, since the mechanism works as an ultraviolet completion: the "poles" found in the perturbative analysis are a manifestation of the anomaly in the UV.

As we have explained at length in  \cite{Coriano:2011ti} these poles extracted in each gauge invariant sector do not couple in the infrared region, since the theory is massive and conformal invariance is lost in the broken electroweak phase. Looking for a residue of these poles in the IR, in the case of a massive theory, is simply meaningless. Indeed their role is recuperated in the UV, 
where they describe an effective  massless exchange present in the amplitude at high energy.

Therefore, the $1/s$ behaviour found in these correlators at high energy is the unique signature of the anomaly (they saturate the anomaly) in the same domain, and is captured within an asymptotic expansion in $v^2/s$  \cite{Coriano:2011ti}. Thus, the anomalous nature of the theory reappears as we approach a (classically) conformally invariant theory, with $s$ going to infinity.

 Obviously, this picture is only approximate, since the cancellation of the trace anomaly by the subtraction of a pole in superspace 
remains an open issue, given the fact that the trace anomaly takes contribution at all orders both in $G_N$ and in the gauge coupling.  
 The resolution of this point would require computations similar to the one that we have just performed for correlators of higher order. 
  Indeed, this is another aspect of the "anomaly puzzle" in supersymmetric theories when (chiral) gauge anomalies and trace anomalies appear on the same level, due to their coupling with gravity.

\section{Conclusions}
  \label{P3conclusions}
We have presented a complete study of the interactions between gravity and the fields of the Standard Model which are responsible for the generation of a trace anomaly in the corresponding effective action. The motivations in favour of these type of studies are several and cover both the cosmological domain and collider physics. In this second case these corrections are important especially in the phenomenological analysis of theories with a low gravity scale/large extra dimensions. We have defined rigorously the structure of these correlators, via an appropriate set of Ward and Slavnov-Taylor identities that we have derived from first principles. We have given the explicit expressions of these corrections, extending to the neutral current sector of the SM previous analysis performed in the QED and QCD cases.

\chapter{Standard Model corrections to flavor diagonal fermion-graviton vertices}
\label{Chap.GravitonFermions}

\section{Introduction}
The investigation of the perturbative couplings of ordinary field theories to gravitational backgrounds has received a certain attention 
in the past \cite{'tHooft:1974bx,Capper:1975ig,Berends:1974gk,Drummond:1979pp}.
While the smallness of the gravitational coupling may shed some doubts on the practical relevance of such corrections, with the advent of models on large (universal \cite{ArkaniHamed:1998rs, Antoniadis:1998ig, ArkaniHamed:1998nn} and warped \cite{Randall:1999ee,Randall:1999vf}) extra dimensions and, more in general, of models with a low gravity scale \cite{Dvali:2000hp} their case has found a new and widespread support. 
This renewed interest covers both theoretical and phenomenological aspects that
 have not been investigated in the past. They could play, for instance, a significant role in addressing issues such as the universality of the gravitational coupling to matter \cite{Degrassi:2008mw}, in connection with the Lagrangian of the Standard Model. On the more formal side, as pointed out in some studies \cite{Giannotti:2008cv,Mottola:2006ew,Armillis:2009im,Armillis:2009pq,Armillis:2010qk} and in the previous chapters, the structure of the effective action,
accounting for anomaly mediation between the Standard Model and gravity, 
shows, in its perturbative expansion, 
the appearance of new effective scalar degrees of freedom. These aspects went unnoticed before and, 
for instance, could be significant in a cosmological context. 
For this reason, we believe, they require further consideration.

All these perturbative analyses are usually performed at the leading order in the gravitational coupling. This is due to the rather involved 
expression of the operator responsible for such a coupling at classical and hence at quantum level: the energy-momentum tensor (EMT) of the Standard Model. Its expression in the electroweak theory is, indeed, very lengthy, and the classification of the several amplitudes (and form factors) in which it appears - at leading order in the electroweak expansion - requires a considerable effort.

Chapter \ref{Chap.GravitonFermions}, as the next chpater \cite{Coriano:2012cr, Coriano:2013msa}, are devoted to the analysis of the electroweak and strong one-loop corrections to the graviton-fermion vertex $Tf\bar f'$, which is another theoretically and phenomenologically relevant piece of the quantum effective action of the SM in a curved background. 
This work, based on the characterization of gravitational interaction within the flavor diagonal and off-diagonal sectors, extends a previous study \cite{Degrassi:2008mw} where these interactions have been analyzed for fermions of different flavors and at leading order in the external fermion masses. 
These off-diagonal (in flavor space) matrix elements are phenomenologically very intriguing, also because they provide the spin-2 counterpart of the standard electroweak spin-1 flavor-changing neutral currents (FCNC). As in the theory of weak interactions, the gravitational FCNC's are not present at the Born level but they emerge as a quantum effect. 

In \cite{Degrassi:2008mw} the authors present explicit analytic results at leading order in the external fermion masses. We have recomputed the off-diagonal $Tf\bar f'$ vertex integrating the previous studies, by including all the mass corrections which are necessary, for instance, for a phenomenological analysis of heavy quarks transitions (like $t \rightarrow b \, G$) in a gravitational background. As we already mentioned, these corrections could be accessible at collider energies in theories with a low-gravity scale. The complete analysis (including mass corrections) of the off-diagonal $Tf\bar f'$ vertex is provided in the following chapter, while here we discuss the flavor diagonal case, presenting the explicit results for the electroweak and strong corrections to the corresponding form factors.

A similar computation appeared in the past, limited to QED, in an old work of Berends and Gastmans \cite{Berends:1975ah}, where they evaluated the on-shell one-loop corrections to the graviton-photon and graviton-fermion vertices. Their studies showed that these corrections provide a modification of the Newton's potential for an electron in a gravitational field, with the emergence of a very tiny repulsive correction of the form $r_e/r^2$, where $r_e = \alpha/m_e$ is the electron radius, $\alpha$ the fine structure constant, and $m_e$ the electron mass. This term originates from the infrared behaviour of the $Tf\bar f$ matrix element in the limit of small electron mass, and emerges from a particular diagrammatic contribution characterized by the graviton-photon tree level coupling. Because there is no similar vertex in QED, the Coulomb potential is, instead, protected against $1/r^2$ corrections at one-loop order. We point out that the same situation is expected also in the QCD case, with a modification to the Newton's law enhanced by the appearance of the strong coupling constant and of the non-abelian color factors. On the other hand, the same analysis in the electroweak sector of the Standard Model leads to different conclusions, because the masses of the weak gauge bosons provide a natural cut-off in the low-energy limit, thus preventing any long-distance correction to the Newton's potential to play a role. 

We anticipate that these results are relevant for some phenomenological 
consequences that may affect studies on these interactions. One of the
 special contexts
 is represented by the neutrino sector \cite{Menon:2008wa}.
Here it is left open the possibility of extending our analysis to more general gravitational backgrounds, with the inclusion of a dilaton field.

\subsection*{Description of the chapter contents}
This chapter is organized as follows. 
In section \ref{P5Sec.PertExp} we illustrate the structure of the perturbative expansion, organized in terms of the various contributions to the EMT of the Standard Model. These are separated with respect to the particles running in the loops, which are the photon, the W and Z gauge bosons and the Higgs field.  We conclude this section with a classification of the relevant form factors. 

In section \ref{P5Sec.WardId} we briefly discuss the derivation of an important Ward identity involving the effective action which is crucial to test the correctness of 
our results and to secure their consistency. Section \ref{P5Sec.Ren} addresses the issue of the renormalization of the theory, 
which complements the analysis of \cite{Coriano:2012nm}. We recall that no new counterterms are needed - except for those of the Standard Model Lagrangian - to carry the perturbative expansion of EMT insertions on correlators of the Standard Model, under a certain condition. This condition requires that the non-minimal coupling ($\chi$) of the Higgs field be fixed at the conformal value $1/6$. We then proceed in section \ref{P5Sec.FormFac} with a description of the expression of the form 
factors for each separate gauge/Higgs contribution in the loop corrections. 

In section \ref{P5Sec.Infrared} we give a simple proof - at leading order in the gauge couplings - of the infrared finiteness of these loop corrections, once they are combined with the corresponding real emissions of massless gauge bosons, integrated over phase space. The proof is a generalization of the ordinary cancellation between real and virtual emissions, in inclusive cross sections, to the gravity case. Finally, in section \ref{P5Sec.Conclusions} we give our conclusions.

\section{The perturbative expansion}
\label{P5Sec.PertExp}
We will be dealing with the $Tf \bar{f}$ (diagonal) fermion case. The Standard Model Lagrangian and the associated EMT have been defined in the previous chapter. We refer to the latter for the various definitions and conventions. We introduce the following notation
\bea
\hat T^{\mu\nu} = i \langle p_2 | T^{\mu\nu}(0) | p_1 \rangle
\eea
to denote the general structure of the transition amplitude where the initial and final fermion states are defined with momenta $p_1$ and $p_2$ respectively. The external fermions are taken on mass shell and of equal mass $p_1^2 = p_2^2 = m^2$. We will be using the two linear combinations of momenta  $p = p_1 + p_2$ and $q = p_1 - p_2$  throughout this chapter in order to simplify the structure of the final result.

The tree-level Feynman rules needed for the computation of the $\hat T^{\mu\nu} $ vertex are listed in Appendix \ref{P5feynrules}, and its expression at Born level is given by
\bea
\label{P5treeT}
\hat T^{\mu\nu}_0 = \frac{i}{4} \bar u(p_2) \bigg\{ \gamma^{\mu} p^{\nu} + \gamma^{\nu} p^{\mu} \bigg\} u(p_1) \,.
\eea
Our analysis will be performed at leading order in the weak coupling expansion, and we will define a suitable set of independent tensor amplitudes 
(and corresponding form factors) to parameterize the result. \\
The external fermions can be leptons or quarks. In the latter case, since the EMT does not carry any non-abelian charge, the color matrix is diagonal and for notational simplicity, will not be included.

\begin{figure}[t]
\centering
\subfigure[]{\includegraphics[scale=0.8]{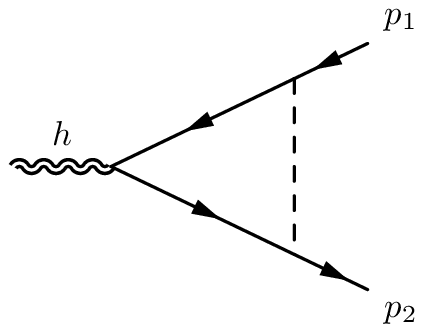}} \hspace{.5cm}
\subfigure[]{\includegraphics[scale=0.8]{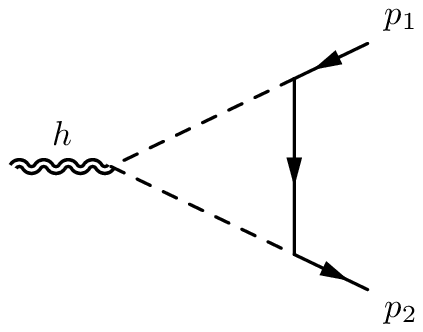}} \hspace{.5cm}
\subfigure[]{\includegraphics[scale=0.8]{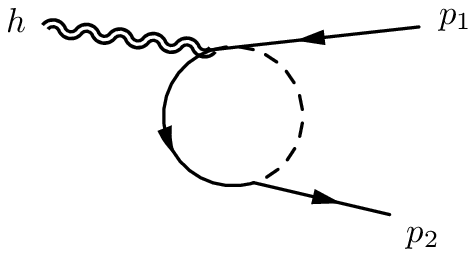}} \hspace{.5cm}
\subfigure[]{\includegraphics[scale=0.8]{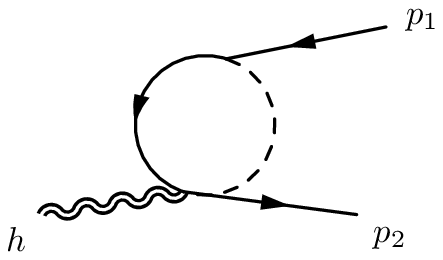}}
\caption{The one-loop Feynman diagrams of the graviton fermion vertex. The dashed lines can be gluons, photons, Higgs, $Z$ and $W$ bosons or their unphysical longitudinal parts. The internal fermion line can be of the same flavor of the external fermions if a neutral boson is exchanged in the loop, otherwise, for charged $W$ corrections, it can have different flavor because of the CKM matrix. \label{P5diagrams}}
\end{figure}

We decomposed the full matrix element into six different contribution characterized by the SM sectors running in the loop diagrams
\bea
\label{P5hatT}
\hat T^{\mu\nu} = \hat T^{\mu\nu}_{g} + \hat T^{\mu\nu}_{\gamma} + \hat T^{\mu\nu}_{h} + \hat T^{\mu\nu}_{Z} + \hat T^{\mu\nu}_{W} + \hat T^{\mu\nu}_{CT} 
\eea
where the subscripts stand respectively for the gluon, the photon, the Higgs, the $Z$ and the $W$ bosons and the counterterm contribution. Concerning the last term we postpone a complete discussion of the vertex renormalization to 
a follow-up section. \\
As we have already mentioned, we work in the $R_\xi$ gauge, where the sector of each massive gauge boson is always accompanied by the corresponding unphysical longitudinal part. This implies that the diagrammatic expansion of $\hat T^{\mu\nu}_{Z}$ and $\hat T^{\mu\nu}_{W}$ is characterized by a set of gauge boson running in the loops with duplicates obtained by replacing the massive gauge fields with their corresponding Goldstones. \\
The decomposition in Eq.(\ref{P5hatT}) fully accounts for the SM one-loop corrections to the flavor diagonal EMT matrix element with two external fermions. \\
The various diagrammatic contributions appearing in the perturbative expansion are shown in Fig.\ref{P5diagrams}. Two of them are characterized by a typical triangle topology, while the others denote terms where the insertion of the EMT and the fermion field occur on the same point. The computation of these diagrams is rather involved and has been performed in dimensional regularization using the on-shell renormalization scheme. We have used the standard reduction of tensor integrals to a basis of scalar integrals and we have checked explicitly the Ward identity coming from the conservation of the EMT, which are crucial to secure the correctness of the computation.  \\
\subsection{Tensor decompositions and form factors}
Now we illustrate in more detail the organization of our results. By using symmetry arguments and exploiting some consequences of the Ward identities, we have determined a suitable tensor basis on which our results are expanded. For massless vector bosons (gluons and photons), and for the Higgs field, because of the parity-conserving nature of their interactions, we have decomposed the matrix elements onto a basis of four tensor structures $O^{\mu\nu}_{V  k}$ with four form factors $f_k$ as

\bea
\hat T^{\mu\nu}_{g} &=& i \frac{\alpha_s }{4 \pi} C_2(N)  \sum_{k=1}^4 f_k(q^2) \, \bar u(p_2) \, O^{\mu\nu}_{V  k} \, u(p_1) \,, \\
\hat T^{\mu\nu}_{\gamma} &=& i \frac{\alpha }{4 \pi} Q^2  \sum_{k=1}^4 f_k(q^2) \, \bar u(p_2) \, O^{\mu\nu}_{V k} \, u(p_1) \,, \\
\hat T^{\mu\nu}_h &=& i  \frac{G_F}{16 \pi^2 \sqrt{2}}  \, m^2  \sum_{k=1}^4 f^h_k(q^2) \, \bar u(p_2) \, O^{\mu\nu}_{V  k} \, u(p_1) \,,
\eea
with the tensor basis defined as
\bea
\label{P5vectorbasis}
O^{\mu\nu}_{V  1} &=& \gamma^\mu \, p^\nu + \gamma^\nu \, p^\mu \,, \nn \\
O^{\mu\nu}_{V 2} &=& m \, \eta^{\mu\nu} \,, \nn \\
O^{\mu\nu}_{V  3} &=& m \, p^\mu \, p^\nu \,, \nn \\
O^{\mu\nu}_{V 4} &=& m \, q^\mu \, q^\nu \,.
\eea 
The form factors for the gluon and the photon contributions are identical, the only difference relies in the coupling constant and in the charge of the external fermions. The coefficient $C_2(N)$ is the quadratic Casimir in the $N$-dimensional fundamental representation, with $C_2 = 4/3$ for quarks and zero for leptons. $Q$ denotes the electromagnetic charge and $G_F$ the Fermi constant. Moreover, being the fermion-Higgs coupling proportional to the fermion mass, we have factorized $m^2$ in front of the Higgs form factors. Note that $O^{\mu\nu}_{V  2-4}$ are linearly mass suppressed, so that only $O^{\mu\nu}_{V  1}$ survives in the limit of massless external fermions. \\
Coming to the weak sector of our corrections, because of the chiral nature of the $Z$ and $W$ interactions, we have to decompose the matrix elements into a more complicated tensor basis of six elements as
\bea
\hat T^{\mu\nu}_Z &=& i \, \frac{G_F}{16 \pi^2 \sqrt{2}}   \sum_{k=1}^6  f^{Z}_k(q^2) \, \bar u(p_2) \, O^{\mu\nu}_{C  k} \, u(p_1) \,, \\ 
\hat T^{\mu\nu}_W &=& i \, \frac{G_F}{16 \pi^2 \sqrt{2}}  \sum_{k=1}^6  f^{W}_k(q^2) \, \bar u(p_2) \, O^{\mu\nu}_{C  k} \, u(p_1) \,, 
\eea
where we have defined
\bea
\label{P5chiralbasis}
O^{\mu\nu}_{C  1} &=& \left( \gamma^\mu \, p^\nu + \gamma^\nu \, p^\mu \right) P_L \,, \nn \\
O^{\mu\nu}_{C  2} &=& \left( \gamma^\mu \, p^\nu + \gamma^\nu \, p^\mu \right) P_R \,, \nn \\
O^{\mu\nu}_{C  3} &=& m \, \eta^{\mu\nu} \,, \nn \\
O^{\mu\nu}_{C  4} &=& m \, p^\mu \, p^\nu  \,, \nn \\
O^{\mu\nu}_{C  5} &=& m \, q^\mu \, q^\nu  \,, \nn \\
O^{\mu\nu}_{C  6} &=& m \, \left( p^\mu \, q^\nu + q^\mu \, p^\nu \right) \gamma^5 \,.
\eea
The most general rank-2 tensor basis that can be built with a metric tensor, two momenta ($p$ and $q$) and matrices $\gamma^\mu, \gamma^5$ has been given in \cite{Degrassi:2008mw}. The basis given in Eq. (\ref{P5chiralbasis}), compared to the flavor-changing case, is more compact.  We have imposed the  symmetry constraints on the external fermion states (of equal mass and flavor) and the conservation of the EMT, discussed in section \ref{P5Sec.WardId}.
For the form factors appearing in $\hat T^{\mu\nu}_W$ we introduce the notation  
\bea
\label{P5Ftof}
f^{W}_k(q^2) =  \sum_f V_{if}^* V_{fi} \, F^{W}_k(q^2, x_f) 
\eea
where $V_{i f}$ is the CKM mixing matrix and the index $i$ corresponds to the flavor of the external fermions. \\
We have extracted a single mass suppression factor coming from the contribution of the $O^{\mu\nu}_{C 3-6}$ operators. In the $\hat T^{\mu\nu}_Z$ matrix element, the leading terms in the small external fermion mass are given by the first two form factors. The situation is different for the $W$ case, in which only the first form factor is the leading term, being $f_2$ suppressed as $m^2$. We have decided not to factorize the $m^2$ term in order to make the notations uniform with the $Z$ case. \\
We remark that the expressions of the form factors is exact, having kept in the result the complete dependence from all the kinematic invariants and from the external and internal masses.

\section{The Ward identity from the conservation of the EMT}
\label{P5Sec.WardId}
In this section we simply quote the consequences of the conservation of the energy-momentum tensor that are contained in the Ward identities satisfied by the matrix elements defined above. As widely explained in the previous chapter, we can derive a master equation for the effective action $\Gamma$, the generating functional of all the 1-particle irreducible (1PI) graphs. 
The Ward identities for the various correlators are then obtained via functional differentiation. 
By a functional differentiation of Eq.(\ref{P3WardGamma}) with respect to the fermion fields and after a Fourier transform to momentum space we obtain
\bea
\label{P5WI}
q_{\mu} \, \hat T^{\mu\nu} = \bar u(p_2) \bigg\{ p_2^{\nu} \,  \Gamma_{\bar f f}(p_1) -  p_1^{\nu} \, \Gamma_{\bar f f}(p_2)
 + \frac{q_\mu}{2} \left( \Gamma_{\bar f f}(p_2) \, \sigma^{\mu\nu} - \sigma^{\mu\nu} \, \Gamma_{\bar f f}(p_1) \right) \bigg\} u(p_1) 
\eea
where $ \Gamma_{\bar f f}(p)$ is the fermion two-point function, diagonal in flavor space, given explicitly in Appendix \ref{P5selfenergies}. The perturbative test of this equation is of great importance for testing the correctness of our results, and in the following sections it will be used to relate different form factors, thus reducing the number of independent contributions to the $T f\bar{f}$ vertex.

\section{Renormalization}
\label{P5Sec.Ren}

The counterterms needed for the renormalization of the vertex can be obtained by promoting the counterterm Lagrangian to the curved background. The counterterm Feynman rules for the matrix element with the insertion of the EMT are easily extracted in the usual way and in our case, for a chiral fermion, we have
\bea
\label{P5TCT}
\hat T^{\mu \nu}_{CT} = i \langle p_2 | T^{\mu\nu}_{CT} (0) | p_1 \rangle =  \frac{i}{4} \bar u(p_2) \bigg\{ \delta Z_L \, O^{\mu\nu}_{C 1}   +  \delta Z_R \, O^{\mu\nu}_{C  2}  + 4 \frac{\delta m}{m} O^{\mu\nu}_{C  3}  \bigg\} u(p_1) \,,
\eea
with $\delta Z_L$, $\delta Z_R$ and $\delta m$ being the fermion wave function and the mass renormalization 
constants respectively, while $O^{\mu\nu}_{C 1-3}$ are defined in Eq.(\ref{P5chiralbasis}). \\
For vector-type interactions, in the gluon, photon and Higgs sector, $\delta Z_L = \delta Z_R$ and the expansion of the $\hat T^{\mu \nu}_{CT}$ matrix element naturally collapses to only two operators, $O^{\mu\nu}_{V  1} = O^{\mu\nu}_{C  1} + O^{\mu\nu}_{C  2}$ and $O^{\mu\nu}_{V  2} = O^{\mu\nu}_{C  3}$ of Eq.(\ref{P5vectorbasis}). \\
We have checked that the renormalization of the parameters of the SM Lagrangian is indeed sufficient to cancel all the singularities of the $\hat T^{\mu\nu}$ matrix element, as expected. As it can be easily seen from Eq.(\ref{P5TCT}), the form factors involved in the subtraction of infinities are just the first two for the gluon, the photon and the Higgs, and the first three for the massive gauge bosons. This is in agreement with simple power counting arguments. \\  
We have used the on-shell scheme, where the renormalization conditions are fixed in terms of the physical parameters of the theory to all orders in the perturbative expansion in the electroweak coupling constants. These are the masses of physical particles, the electric charge and the CKM mixing matrix. The renormalization conditions on the fields, which allow to extract the wave function renormalization constants, are satisfied by requiring a unitary residue of the full 2-point functions on the physical particle poles. For the fermion renormalization constants we obtain the following explicit expressions 
\bea
\delta Z_L &=&  - \tRE \, \Sigma^L (m^2) - m^2 \frac{\partial}{\partial p^2 } \tRE \bigg[ \Sigma^L(p^2) +  \Sigma^R(p^2) +  2 \Sigma^S(p^2) \bigg]_{p^2 = m^2}  \,, \\
\delta Z_R &=&  - \tRE \, \Sigma^R (m^2) - m^2 \frac{\partial}{\partial p^2 } \tRE \bigg[ \Sigma^L(p^2) +  \Sigma^R(p^2) +  2 \Sigma^S(p^2) \bigg]_{p^2 = m^2}  \,, \\
\delta m &=& \frac{m}{2} \tRE \bigg[ \Sigma^L(m^2) +  \Sigma^R(m^2) +  2 \Sigma^S(m^2)  \bigg] \,,
\eea
where the $\Sigma^{L, R, S}$ functions are the fermion self-energies defined in Appendix \ref{P5selfenergies}. The symbol $\tRE$ gives the real part of the scalar integrals appearing in the self-energies but it has no effect on the CKM matrix elements. If the mixing matrix is real $\tRE$ can be replaced with $\rm Re$.

\section{Form factors for the $\hat T^{\mu\nu}$ matrix element}
\label{P5Sec.FormFac}

\subsection{The massless gauge boson contribution}
We give the four form factors for the massless gauge boson cases, namely the gluon and the photon contributions. They depend on the kinematic invariant $q^2$, the square of the momenta of the graviton line, and from the dimensionless ratio $y = m^2 / q^2$. The form factors are expressed as a combination of one-, two- and three-point scalar integrals, which have been defined in Appendix. \ref{P5scalarint}, and are given by
\small
\bea
f_1(q^2) &=&
-\frac{4 y (2 y+1)}{3 (1- 4 y)^2}
+\frac{(8 y (7 y-4)+3) }{3 q^2 (1-4 y)^2 y} \mathcal A_0 \left(m^2\right)
-\frac{2 (y-1) }{3 (1-4 y)^2} \mathcal B_0 \left(q^2,0,0\right)      \nn \\
&&  +\frac{(17-44 y) }{48 y-12} \mathcal B_0 \left(q^2 ,m^2,m^2\right)
+\frac{1}{2} q^2 (2 y-1) \mathcal C_0 \left(0,m^2,m^2 \right) 
 +\frac{2 q^2 (1-2 y) y }{(1-4 y)^2} \mathcal C_0 \left(m^2,0,0\right) \,,
 \nn \\
f_2(q^2) &=&
\frac{4 (y (32 y-23)+3)}{3 (1-4 y)^2}
+\frac{2 \left(32 y^2-26 y+3\right) }{3 q^2 (1-4 y)^2 y} \mathcal A_0 \left( m^2 \right)
+ \frac{2 (10 y-1)}{3 (1-4 y)^2}  \mathcal B_0 \left(q^2,0,0 \right) \nn \\
&& +\frac{5}{3} \mathcal B_0 \left(q^2,m^2,m^2\right)
+\frac{8 q^2 y^2 }{(1-4 y)^2} \mathcal C_0 \left(m^2,0,0\right) \,,
\nn \\
f_3(q^2) &=&
\frac{4 y (8 y+19)-6}{3 q^2 (4 y-1)^3}
-\frac{4 (8 y-17)}{3 q^4 (4 y-1)^3}  \mathcal A_0 \left( m^2 \right)
-\frac{2 (26 y+1) }{3 q^2 (4 y-1)^3} \mathcal B_0 \left(q^2,0,0\right) \nn \\
&& +\frac{2 }{3 q^2 (4 y-1)} \mathcal B_0 \left(q^2,m^2,m^2 \right)
-\frac{8 y (y+1) }{(4 y-1)^3} \mathcal C_0 \left(m^2,0,0\right) \,,
 \nn \\
f_4(q^2) &=&
\frac{-80 y^2+68 y-9}{3 q^2 (1-4 y)^2}
+\frac{(20 y (4 y-1)+3) }{3 q^4 (1-4 y)^2 y} \mathcal A_0 \left(m^2\right)
+ \frac{(2-20 y)}{3 q^2 (1-4 y)^2}  \mathcal B_0 \left(q^2,0,0\right) \nn \\
&& -\frac{5 }{3 q^2} \mathcal B_0 \left(q^2,m^2,m^2\right)
-\frac{8 y^2 }{(1-4 y)^2} \mathcal C_0 \left(m^2,0,0\right) \,.
\eea
\normalsize
It is interesting to observe that not all the four form factors are independent because the Ward identity imposes relations among them. In fact, specializing Eq.(\ref{P5WI}) to the massless gauge bosons contributions, we obtain
\bea
f_2(q^2) + q^2 \, f_4(q^2) = - \frac{1}{2} \bigg[ \Sigma^L_{g / \gamma}(m^2) +  \Sigma^R_{g / \gamma}(m^2)+ 2  \Sigma^S_{g / \gamma}(m^2) \bigg],
\eea
as it can be checked from the explicit expressions given above. This relation can be used to test the correctness of our results and to reduce the number of independent form factors. We recall once more that the $\Sigma$'s denote the fermion self-energies which have been collected in Appendix \ref{P5selfenergies}. 

\subsection{The Higgs boson contribution}
In this section we present the results for the contribution of a virtual Higgs. As in the previous case, they are expanded in terms of scalar integrals and of the kinematic variables $q^2$, $y = m^2/q^2$ and $x_h = m^2/m_h^2$, where $m_h$ is Higgs mass. \\
As we have already mentioned, the conservation of the EMT induces a Ward identity on the correlation functions. 
This implies a relation between the form factors, which in the Higgs case becomes
\bea
f^h_2(q^2) + q^2 \, f^h_4(q^2) = - \frac{1}{2} \bigg[ \Sigma^L_{h}(m^2) + \Sigma^R_{h}(m^2) + 2 \Sigma^S_{h}(m^2) \bigg] \,.
\eea
Obviously, this equation has the same structure of the Ward identity found in the massless gauge bosons case, having expanded $\hat T^{\mu\nu}_h$ on the same tensor basis of Eq.(\ref{P5vectorbasis}). \\
Notice that the $f_2^h$ and $f_4^h$ form factors depend on the $\chi$ parameter of $\mathcal{S}_I$. This is expected, because the Higgs field can also couple to gravity with the EMT of improvement $T^{\mu\nu}_I$.  The Feynman rules for a graviton-two Higgs vertex are then modified with the inclusion of the $\chi$ dependence and affect the diagram represented in Fig.\ref{P5diagrams}(b), where this vertex appears. We obtain 
\small
\bea
f^h_1(q^2) &=& 
\frac{3 x_h-8 y -4 x_h y }{12 x_h(1- 4y)}
+\frac{2}{3 \, q^2 (1-4 y)} \bigg[  \mathcal A_0 \left(m_h^2 \right)   -  \mathcal A_0 \left(m^2 \right)    \bigg] \nn \\
&+&  \frac{1}{12 x_h^2 (1-4  y)^2}  \bigg[ x_h^2+8 (x_h (26 x_h+3)-3) y^2-2 (28 x_h+3) x_h y\bigg] \mathcal B_0 \left(q^2,m^2, m^2\right) \nn \\
&+& \frac{1}{6 x_h^2 (1-4  y)^2} \bigg[ x_h^2+4 (3-8 x_h) y^2+4 (2 x_h-1) x_h y\bigg]  \mathcal B_0 \left(q^2,m_h^2,m_h^2 \right) \nn \\
&+& \frac{y}{2 x_h (1-4 y)^2} \bigg[ x_h (4-32 y)+12 y+1\bigg]  \mathcal B_0 \left(m^2,m^2,m_h^2 \right)
- \frac{q^2 y}{2 x_h^3 (1-4 y)^2}  \bigg[ 4 x_h^3  \nn \\
&+& 4 (x_h (4 x_h-3) (4 x_h+1)+1) y^2 
+ (8 (1-4 x_h) x_h+3) x_h y\bigg]  \mathcal C_0 \left(m_h^2,m^2,m^2 \right) \nn \\
&+& \frac{q^2 y (x_h-2 y)}{x_h^3 (1-4 y)^2}   \left(x_h^2-4 x_h y+y\right)  \mathcal C_0 \left(m^2,m_h^2, m_h^2\right)
\,, \nn
\eea
\bea
f^h_2(q^2) &=&
\frac{40 x_h y-9 x_h-4 y}{3 x_h(1-4 y)}
+\frac{4}{3 q^2 (1-4 y)} \bigg[  \mathcal A_0 \left(m_h^2 \right) -  \mathcal A_0 \left(m^2 \right) \bigg] 
- \frac{2}{3 x_h^2 (1-4 y)^2} \bigg[ 2 x_h^2 (1-4 y)^2 \nn \\
&+& 9 x_h y (1-4 y)+6 y^2 \bigg]    \mathcal B_0 \left(q^2,m^2,m^2\right) 
+ \frac{1}{3 x_h^2 (1-4 y)^2} \bigg[ x_h^2+4 (3-20 x_h) y^2+8 (x_h+1) x_h y\bigg]  \nn \\
&\times& \mathcal B_0 \left(q^2,m_h^2, m_h^2\right) 
+ \frac{2}{x_h (1-4 y)^2} \bigg[ 4 (1-4 x_h) y^2+6 x_h y-x_h+y\bigg]   \mathcal B_0 \left(m^2, m^2, m_h^2 \right) \nn \\
&-& \frac{4 q^2 y (-4 x_h y+x_h+y)^2 }{x_h^3 (1-4 y)^2} \mathcal C_0 \left(m_h^2 ,m^2 ,m^2 \right) \nn \\
&-& \frac{q^2 (x_h (8 y-1)-2 y)}{x_h^3 (1-4 y)^2}  \bigg[ x_h^2 (2 y-1)+8 x_h y^2-2 y^2\bigg] \mathcal C_0 \left(m^2, m_h^2, m_h^2 \right) \nn \\
&+&  \chi \bigg\{
\frac{8}{1-4 y}  \bigg[ \mathcal B_0 (m^2 , m^2,m_h^2 ) - \mathcal B_0 (q^2,m_h^2,m_h^2 ) \bigg]
+ \frac{4 q^2 (x_h+2 y -8 x_h y ) }{x_h (1-4 y)} \mathcal C_0 (m^2 ,m_h^2, m_h^2 )
\bigg\} \,, \nn 
\eea
\bea
f^h_3(q^2) &=&
\frac{2 (x_h (22 y-3)-10 y)}{3 q^2 x_h (1-4 y)^2} 
+ \frac{2 (3-2 y) }{3 q^4 (1-4 y)^2 y} \bigg[  \mathcal A_0 \left(m_h^2 \right) - \mathcal A_0 \left(m^2 \right) \bigg] \nn \\
&+&  \frac{5 }{3 q^2 \, x_h^2 (4 y-1)^3} \bigg[ x_h^2+4 (4 (x_h-3) x_h+3) y^2+4 (3-2 x_h) x_h y\bigg] \mathcal B_0 \left(q^2 ,m^2, m^2 \right) \nn \\
&+& \frac{1}{3 q^2 \, x_h^2 (4 y-1)^3}  \bigg[ x_h^2 (7-88 y)+8 x_h y (26 y+1)-60 y^2\bigg] \mathcal B_0 \left(q^2,m_h^2,m_h^2 \right) \nn \\
&+& \frac{2}{q^2 x_h (4 y-1)^3} (x_h (2 (13-8 y) y-3)+8 (y-2) y+1)  \mathcal B_0 \left(m^2 ,m^2, m_h^2 \right) \nn \\
&+& \frac{10 y}{x_h^3 (4 y-1)^3}  (x_h (4 y-1)-2 y) (x_h (4 y-1)-y) \mathcal C_0 \left(m_h^2 ,m^2,m^2 \right) \nn \\
&+& \frac{1 }{x_h^3 (4 y-1)^3} \bigg[ x_h^3-2 \left(8 x_h^2-26 x_h+3\right) x_h y^2-2 (5 x_h+1) x_h^2 y \nn \\
&+& 4 (4 x_h-5) (4 x_h-1) y^3\bigg] \mathcal C_0 \left( m^2,m_h^2,m_h^2 \right) \,, \nn 
\eea
\bea
f^h_4(q^2) &=&
\frac{9 x_h+4 y -40 x_h y}{3 q^2 \, x_h (1-4  y)}
+\frac{(8 y- 3) }{3 q^4 y (1 -4 y)} \bigg[ \mathcal A_0 \left(m_h^2 \right) -  \mathcal A_0 \left(m^2 \right) \bigg] 
+  \frac{2}{3 q^2 x_h^2 (1-4  y)^2} \bigg[ 2 x_h^2 (1-4 y)^2 \nn \\
&+& 9 x_h y (1-4 y)+6 y^2\bigg]  \mathcal B_0 \left(q^2 ,m^2, m^2 \right) 
- \frac{1}{3 q^2 x_h^2 (1-4  y)^2}  \bigg[ x_h^2+4 (3-20 x_h) y^2 \nn \\
&+& 8 (x_h+1) x_h y\bigg]   \mathcal B_0 \left(q^2, m_h^2, m_h^2 \right) 
+ \frac{1}{q^2 x_h (1-4 y)^2}  \bigg[ x_h (4 (5-8 y) y-2)+2 y (4 y-5)+1\bigg] \nn \\ 
&\times& \mathcal B_0 \left(m^2 ,m^2, m_h^2 \right) 
+ \frac{4 y (x_h+y -4 x_h y )^2 }{x_h^3 (1-4 y)^2} \mathcal C_0 \left(m_h^2, m^2,m^2 \right) 
+ \frac{(x_h (8 y-1)-2 y)}{x_h^3 (1-4 y)^2}  \bigg[ x_h^2 (2 y-1) \nn \\
&+& 8 x_h y^2-2 y^2 \bigg]  \mathcal C_0 \left(m^2, m_h^2, m_h^2 \right) 
+  \chi \bigg\{
\frac{8 }{q^2( 1- 4  y)} \bigg[  \mathcal B_0 \left(q^2, m_h^2, m_h^2 \right) -  \mathcal B_0 \left(m^2, m^2, m_h^2 \right) \bigg]  \nn \\
&-& \frac{4 (x_h+2 y -8 x_h y) }{x_h (1 - 4 y)}  \mathcal C_0 \left( m^2, m_h^2, m_h^2\right)
\bigg\} \,.
\eea
\normalsize

\subsection{The $Z$ gauge boson contribution}
Coming to the form factors for the $Z$ boson contribution, which are part of $\hat T^{\mu\nu}_Z$, these are given in terms of the variables $q^2$, $y=m^2/q^2$ and $x_Z \equiv m^2/m_Z^2$, with the parameters $v$ and $a$ denoting the vector and axial-vector $Z$-fermion couplings. In particular we have
\bea
\label{P5Zfermconst}
v = I_3 - 2 s_W^2 Q\,, \qquad a = I_3 \,, \qquad c^2 = v^2 + a^2 \,,
\eea
where $I_3$ and $Q$ are, respectively, the third component of isospin and the electric charge of the external fermions, while $s_W$ is the sine of the weak angle. 

In this case, the structure of the Ward identity is more involved than the previous cases, being $\hat T^{\mu\nu}_Z$ expanded on a more complicated tensor basis, Eq.(\ref{P5chiralbasis}). We obtain two relations among the form factors that we have tested on our explicit computation, which are given by
\bea
\label{P5WIZ1}
f^Z_2 &=&  f^Z_1 + q^2 f^Z_6 + \frac{1}{4} \bigg[ \Sigma^R_Z(m^2) -  \Sigma^L_Z(m^2)  \bigg]  \,, \\
\label{P5WIZ2}
f^Z_3 &=& - q^2 f^Z_5  - \frac{1}{2} \bigg[ \Sigma^L_Z(m^2) + \Sigma^R_Z(m^2) + 2 \Sigma^S_Z(m^2) \bigg] \,.
\eea
Also in this case we have a dependence of the result on the parameter $\chi$, which appears in $f^Z_5$ and hence in $f^Z_3$. As for the Higgs field, also the gravitational coupling of the $Z$ Goldstone boson acquires a new contribution coming from the term of improvement $T_I$, shown by the Feynman diagram in Fig.\ref{P5diagrams}(b).

Here we present a list of the explicit expressions of $f^Z_1$, $f^Z_4$, $f^Z_5$ and $f^Z_6$ while $f^Z_2$ and $f^Z_3$ can be obtained using the Ward identity constraints of Eq.(\ref{P5WIZ1}) and Eq.(\ref{P5WIZ2}). We obtain 

\small
\bea
f^Z_1(q^2) &=&
\frac{q^2 y}{3 (4 y-1) x_Z^2}   \bigg[ x_Z \left(-4 y \left(5 a^2+5 a v+7 v^2\right)+a^2 (4 y-3) x_Z+6 (a+v)^2\right)+4 y (a+v)^2\bigg] \nn \\
&+&  \frac{4 y}{3 (1 - 4 y) x_Z} \left(2 a^2 x_Z+a^2-a v+v^2\right) \bigg[ \mathcal A_0 \left( m_Z^2 \right)  -   \mathcal A_0 \left( m^2 \right) \bigg] \nn \\
&+&  \frac{q^2 y}{6 (1-4 y)^2 x_Z^3}  \bigg[ x_Z \left((4 y-1) x_Z \left(-4 y \left(8 a^2+4 a v+11 v^2\right)+2 a^2 (4
   y-1) x_Z+17 (a+v)^2\right) \right. \nn \\
&+& \left.   6 y \left(4 y \left(5 a^2+8 a v+7 v^2\right)-7
   (a+v)^2\right)\right)-24 y^2 (a+v)^2\bigg]   \mathcal B_0 \left( q^2, m^2, m^2 \right)  \nn \\
&+&   \frac{2 q^2 y}{3 (1-4 y)^2 x_Z^3}   \bigg[ x_Z \left(x_Z \left(-2 y \left(15 a^2+20 a v+v^2\right)+a^2 (8 y+1)
   x_Z+64 a^2 y^2+2 (a+v)^2\right) \right. \nn \\
&+&  \left.   y \left(7 (a+v)^2-4 y \left(10 a^2+8 a v+13
   v^2\right)\right)\right)+6 y^2 (a+v)^2\bigg]   \mathcal B_0 \left( q^2, m_Z^2, m_Z^2 \right)  \nn \\
&+&  \frac{q^2 y}{(1-4 y)^2 x_Z^2}   \bigg[ 2 x_Z \left(-4 y^2 \left(a^2-4 a v-4 v^2\right)-y \left(a^2+4 a v+10
   v^2\right)-4 a^2 y x_Z+(a+v)^2\right) \nn \\
&+&y \left(4 y \left(3 a^2-4 a v+3 v^2\right)+a^2+6 a
   v+v^2\right)\bigg]    \mathcal B_0 \left( m^2, m^2, m_Z^2 \right)  \nn \\
&+&  \frac{4 q^4 y^2}{(1-4 y)^2 x_Z^4}   \bigg[ x_Z \left(x_Z \left(x_Z \left(a^2 x_Z-a^2-2 v y (4 a+v)+v^2\right)-y
   \left(4 a^2 (4 y-1)-6 a v  \right.\right.\right.   \nn \\ 
&+& \left.\left.\left.     v^2 (8 y+1)\right)\right)+y \left(2 y \left(4 a^2+4 a v+5
   v^2\right)-(a+v)^2\right)\right)-y^2 (a+v)^2\bigg]       \mathcal C_0 \left( m^2, m_Z^2, m_Z^2 \right)   \nn \\
&+&  \frac{q^4 y}{(1-4 y)^2 x_Z^4}   \bigg[ x_Z \left((4 y-1) x_Z \left((4 y-1) x_Z \left(2 y
   \left(a^2+v^2\right)-(a+v)^2\right)-2 y^2 \left(7 a^2+8 a v  10 v^2\right)  \right.\right.   \nn \\
&+&  \left.\left.   6 y
   (a+v)^2\right)+y^2 \left(4 y \left(7 a^2+12 a v+9 v^2\right)-9 (a+v)^2\right)\right)-4 y^3
   (a+v)^2\bigg]      \mathcal C_0 \left( m_Z^2, m^2, m^2 \right) \,, \nn
\eea

\bea
f^Z_4(q^2) &=&
-\frac{8 y}{3 x_Z^2 \left(1 -4 y \right)^2} \bigg[ a^2 x_Z \left((2 y-3) x_Z-14 y+6\right)-5 c^2 y \left(x_Z-1\right)\bigg] \nn \\
&+&  \frac{4 (2 y-3) \left(2 a^2 x_Z+c^2\right)}{3 q^2 (1-4 y)^2 x_Z} \bigg[ \mathcal A_0 \left( m^2 \right) - \mathcal A_0 \left( m_Z^2 \right)  \bigg] 
 + \frac{4 y}{3 (4  y-1)^3 x_Z^3}   \bigg[ x_Z \left((4 y-1) x_Z   \right.  \nn \\
&\times& \left.  \left(a^2 (1-4 y) x_Z+6 a^2 (8 y-3)+c^2 (4  y-1)\right)+3 y \left(4 a^2 (3-7 y)+7 c^2 (1-4 y)\right)\right)+30 c^2 y^2 \bigg]    \nn \\
&\times&    \mathcal B_0( q^2, m^2, m^2) 
+  \frac{4 y}{3 (4 y-1)^3 x_Z^3}   \bigg[ x_Z \left(y \left(12 a^2 (7 y-3)+c^2 (68 y+13)\right)-x_Z \left(a^2 (16 y+11) x_Z  \right. \right.  \nn \\
&+& \left.\left.  2 a^2 \left(64 y^2-34 y-3\right)+c^2 (26 y+1)\right)\right)-30 c^2 y^2 \bigg]   \mathcal B_0 \left( q^2, m_Z^2, m_Z^2 \right)   \nn \\
&+&  \frac{4 y}{(4 y-1)^3 x_Z^2}   \bigg[ 2 a^2 x_Z \left((6 y+1) x_Z-8 y^2+4 y-3\right)-c^2 (8 (y-2) y+1)  \left(x_Z-1\right)\bigg]  \nn \\
&\times& \mathcal B_0( m^2, m^2, m_Z^2) 
- \frac{4 q^2 y}{(4 y-1)^3 x_Z^4}   \bigg[ a^2 x_Z \left((6 y+1) x_Z^3+6 y (2 y-3) x_Z^2+2 y ((17-40 y) y+2) x_Z  \right.   \nn \\
&+& \left.  4  y^2 (7 y-3)\right)-c^2 y \left(x_Z \left(x_Z \left(-4 (y+1) x_Z+2 y (8 y+11)+1\right)-6 y  (6 y+1)\right)+10 y^2\right) \bigg]   \nn \\
&\times&   \mathcal C_0 \left( m^2, m_Z^2, m_Z^2 \right) 
+   \frac{4 q^2 y}{(4 y-1)^3 x_Z^4}   \bigg[ x_Z \left((4 y-1) x_Z \left(3 y \left(2 a^2 (5 y-2)+c^2 (4 y-1)\right)  \right.\right. \nn \\
&-& \left. \left.  2  a^2 (y (12 y-7)+1) x_Z\right)+4 y^2 \left(a^2 (3-7 y)+3 c^2 (1-4 y)\right)\right)+10 c^2  y^3\bigg]     \mathcal C_0 \left( m_Z^2, m^2, m^2 \right) \,, \nn
\eea

\bea
f^Z_5(q^2) &=&
\frac{2 y \left(2 a^2 x_Z \left((8 y-3) x_Z-44 y+12\right)+c^2 \left((32 y-9) x_Z+4  y\right)\right)}{3 (1-4 y) x_Z^2} \nn \\
&+&  \frac{2 (8 y-3) \left(2 a^2 x_Z+c^2\right)}{3 q^2 (1- 4 y) x_Z} \bigg[ \mathcal A_0 \left( m_Z^2 \right) - \mathcal A_0 \left( m^2 \right) \bigg] 
+  \frac{2 y}{3 (1-4 y)^2 x_Z^3}  \bigg[x_Z \left((4 y-1) x_Z  \right. \nn \\
&\times&  \left.  \left(4 a^2 (1-4 y) x_Z+12 a^2 (3 y-1)+5 c^2 (1-4 y)\right)+24 a^2 (1-3 y) y\right)+12 c^2 y^2\bigg]  \mathcal B_0( q^2, m^2, m^2) \nn \\
&+&  \frac{4 y}{3 (1-4 y)^2 x_Z^3}  \bigg[ a^2 x_Z \left(x_Z \left((16 y-7) x_Z+4 y (8 y-5)+6\right)+12 y (3 y-1)\right)+c^2 \left((4 y+5) y x_Z \right. \nn \\
&+&   \left. (1-10 y) x_Z^2-6 y^2\right)\bigg]   \mathcal B_0 \left( q^2, m_Z^2, m_Z^2 \right) 
+  \frac{2 y}{\left(x_Z-4 y x_Z\right)^2}   \bigg[ 2 a^2 x_Z \left((2-4 y) x_Z+2 (7-12 y) y  \right. \nn \\
&-& \left.  3\right)+c^2 \left(2 y \left(y \left(8 x_Z+4\right)-5\right)+1\right)\bigg]   \mathcal B_0 \left( m^2, m^2, m_Z^2 \right) 
- \frac{4 q^2 y}{(1-4 y)^2 x_Z^4}   \bigg[ a^2 x_Z \left(x_Z \left((2 y-1) x_Z  \right. \right.  \nn \\ 
&\times&  \left.\left.  \left(6 y-x_Z\right)+6 (3-8 y)  y^2\right)+4 (3 y-1) y^2\right)+c^2 y \left(x_Z \left(x_Z \left(4 y x_Z+2 y (8 y-7)+1\right) \right. \right. \nn \\
&+& \left. \left.  2 y (2 y+1)\right)-2 y^2\right)\bigg]    \mathcal C_0 \left( m^2, m_Z^2, m_Z^2 \right) \nn \\
&+& \frac{4 q^2 y^2}{(1-4 y)^2 x_Z^4}  \left((4 y-1) x_Z-y\right) \bigg[Êx_Z \left(4 a^2 (3 y-1)+c^2 (1-4 y)\right)-2 c^2 y \bigg]    \mathcal C_0 \left( m_Z^2, m^2, m^2 \right) \nn \\
&+&  \chi \, \frac{32 a^2 y}{1 - 4 y} \bigg\{  \mathcal B_0 \left( q^2, m_Z^2, m_Z^2 \right) -  \mathcal B_0 \left( m^2, m^2, m_Z^2 \right) - \frac{q^2 (2 y - x_Z)}{2x_Z} \mathcal C_0 \left( m^2, m_Z^2, m_Z^2 \right)
\bigg\} \,, \nn 
\eea

\bea
f^Z_6(q^2) &=&
\frac{2\, a \, v \, y}{3 (4 y-1) x_Z^2}  \left((8 y-9) x_Z-8 y\right)
+ \frac{2 \, a \, v (8 y- 3)}{3 q^2 (1-4 y) x_Z} \bigg[ \mathcal A_0 \left( m^2 \right) - \mathcal A_0 \left( m_Z^2 \right) \bigg] \nn \\
&+&  \frac{2 \, a \, v \, y}{3 (1-4 y)^2 x_Z^3}  \bigg[ 6 (7-16 y) y x_Z+(4 y-1) (8 y-17) x_Z^2+24 y^2\bigg]   \mathcal B_0 \left( q^2, m^2, m^2 \right) \nn \\
&+&  \frac{8 \, a \, v \, y}{3 (1-4 y)^2 x_Z^3}  \bigg[ (16 y-7) y x_Z+(20 y-2) x_Z^2-6 y^2\bigg]    \mathcal B_0 \left( q^2, m_Z^2, m_Z^2 \right) \nn \\
&-& \frac{2 \, a \, v \, y}{(1-4 y)^2 x_Z^2}  \bigg[ (8 y+2) x_Z-2 y+1\bigg]  \mathcal B_0 \left( m^2, m^2, m_Z^2 \right) \nn \\
&+&  \frac{16 \, a \, v \, q^2 \, y^3}{(1-4 y)^2 x_Z^4}  \bigg[ x_Z \left(x_Z \left(4 x_Z-3\right)-4 y+1\right)+y\bigg]    \mathcal C_0 \left( m^2, m_Z^2, m_Z^2 \right) \nn \\
&+&  \frac{4 \, a \, v \, q^2 \, y}{(1-4 y)^2 x_Z^4}  \left( (4 y-1) x_Z-y\right) \bigg[ (8 y-5) y x_Z+(4 y-1) x_Z^2-4 y^2\bigg]  \mathcal C_0( m_Z^2, m^2, m^2) \,.
\eea
\normalsize

\subsection{The $W$ gauge boson contribution}
Finally we collect here the results for the $\hat T^{\mu\nu}_W$ matrix element. They are expressed in terms of scalar integrals and of the kinematic invariants $q^2$, $y = m^2/q^2$, $x_W = m^2/m_W^2$ and $x_f = m^2/m_f^2$, where $m_f$ is the mass of the fermion of flavor $f$ running in the loop. \\
As in the $Z$ boson case the conservation equation for the EMT implies the following relations among the form factors
\bea
\label{P5WIW1}
f^W_2 &=&  f^W_1 + q^2 f^W_6 + \frac{1}{4} \bigg[ \Sigma^R_W(m^2) -  \Sigma^L_W(m^2)  \bigg]  \,, \\
\label{P5WIW2}
f^W_3 &=& - q^2 f^W_5  - \frac{1}{2} \bigg[ \Sigma^L_W(m^2) + \Sigma^R_W(m^2) + 2 \Sigma^S_W(m^2) \bigg] \,,
\eea
which we have tested explicitly. Also in this case, as for the form factors with the exchange of a $Z$ boson, $f_5^W$, and hence 
$f_3^W$, depends on the parameter $\chi$.

We recall that the $O^{\mu\nu}_{C  3-6}$ operators are characterized by a linear mass suppression in the limit of small external fermion masses, while $O^{\mu\nu}_{C  2}$, even if not explicitly shown, has a quadratic suppression, which is present only in the $W$ case. Therefore, the leading contribution, in the limit of external massless fermions, is then given by the first form factor $f^W_1$ alone. \\

We present the explicit results for the $F^W_1$ and $F^W_4$ to $F^W_6$ in Appendix \ref{P5FWappendix}, while $F^W_2$ and $F^W_3$ can be computed using the Ward identities of Eq.(\ref{P5WIW1}) and (\ref{P5WIW2}). The form factors $f^W_k$ are obtained from $F^W_k$ multiplying by the CKM matrix elements and then summing over the fermion flavors, as explained in Eq.(\ref{P5Ftof}).

\section{Infrared singularities and soft bremsstrahlung}
\label{P5Sec.Infrared}
Here we provide a simple proof of the infrared safety 
of the $Tf\bar{f}$ vertex against soft radiative corrections and emissions
of massless gauge bosons.

An infrared divergence comes from the topology diagram depicted in Fig.\ref{P5diagrams}(a) with a virtual massless gauge boson exchanged between the two fermion lines and it is contained in the three-point scalar integral $\mathcal C_0(0,m^2,m^2)$. If we regularize the infrared singularity with a small photon (or gluon) mass $\lambda$ the divergent part of the scalar integral becomes
\bea
\mathcal C_0 (0, m^2, m^2) = \frac{x_s}{m^2 (1 - x_s^2)} \bigg\{  - 2 \log \frac{\lambda}{m}  \log x_s + \ldots \bigg\}
\eea 
where the dots stand for the finite terms not 
proportional to $\log \frac{\lambda}{m}$, and $x_s = - \frac{1-\beta}{1+\beta}$ with $\beta = \sqrt{1 - 4 m^2 / q^2}.$ \\
In the photon case the infrared singular part of the matrix element is then given by
\bea
\hat T^{\mu\nu}_{\gamma} &=& i \frac{\alpha}{4 \pi} Q^2 \bigg[ - \frac{1}{y}(2 y -1) \frac{x_s}{1-x_s^2} \log \frac{\lambda}{m}  \log x_s \bigg] \, \bar u(p_2) \, O^{\mu\nu}_{V  1} u(p_1)  + \ldots  \nn \\
&=& \frac{\alpha}{4 \pi} Q^2 \bigg[ - \frac{4}{y}(2 y -1) \frac{x_s}{1-x_s^2} \log \frac{\lambda}{m}  \log x_s \bigg] \, \hat T^{\mu\nu}_0 + \ldots \,,
\eea
which is manifestly proportional to the tree level vertex. 
On the other hand, the gluon contribution is easily obtained from the previous equation by replacing $\alpha \, Q^2$ with $\alpha_s \, C_2(N)$.

For the massless gauge boson contributions there is another infrared divergence coming from the renormalization counterterm. Its origin is in the field renormalization constants of charged particles arising from photonic or gluonic corrections to the fermion self energies. For example, in the photon case we have
\bea
\delta Z_L \bigg|_{\gamma}^{IR} = \delta Z_R \bigg|_{\gamma}^{IR} = - \frac{\alpha}{4 \pi} Q^2 \bigg\{ 4 \log \frac{\lambda}{m}  +  \ldots \bigg\} \,.
\eea 

However the processes described by the $\hat T^{\mu\nu}$ matrix element alone are not of direct physical relevance, since they cannot be distinguished experimentally from those involving the emission of soft massless gauge bosons. Adding incoherently the cross sections of all the different processes with arbitrary numbers of emitted soft photons (or gluons) all the infrared divergences are expected to cancel, as in an ordinary gauge theory \cite{Bloch:1937pw,Kinoshita:1962ur,
Lee:1964is}. This cancellation takes place between the virtual and the real bremsstrahlung corrections, and is valid to all orders in perturbation theory. In our case one has to consider only radiation of a single massless gauge boson with energy $k_0 < \Delta E$, smaller than a given cutoff parameter.
\begin{figure}[t]
\centering
\subfigure[]{\includegraphics[scale=0.8]{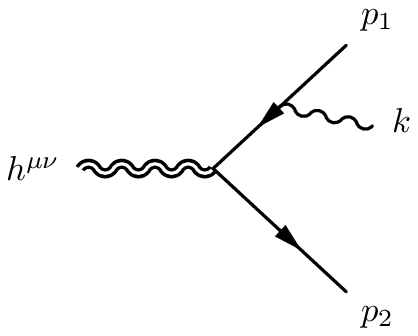}}  \hspace{2.5cm}
\subfigure[]{\includegraphics[scale=0.8]{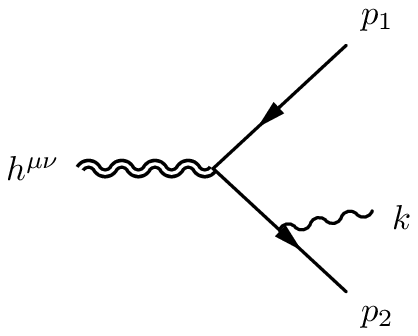}} 
\caption{Real emission diagrams of a massless gauge boson with momentum $k$. \label{P5realdiagrams}}
\end{figure}

For definiteness we consider the emission of a photon from the two external fermion legs. The gluon case, as already mentioned, is easily obtained from the final result with the replacement  $\alpha \, Q^2 \rightarrow \alpha_s \, C_2(N)$. In the soft photon approximation the real emission matrix element, corresponding to the sum of the two diagrams depicted in Fig. \ref{P5realdiagrams} is given by
\bea
\mathcal M_{soft} =   \mathcal M_0 \,  (e \, Q) \bigg[ \frac{\epsilon(k) \cdot p_1}{k \cdot p_1}  - \frac{\epsilon(k) \cdot p_2}{k \cdot p_2}  \bigg] \,,
\eea
where $k$ and $\epsilon(k)$ are the photon momentum and polarization vector respectively, while the sign difference between the two eikonal factors in the square brackets is due to the different fermion charge flow  of the diagrams. Here $\mathcal M_0$ is the Born amplitude which factorizes, in our case, as
\bea
\mathcal M_0 = A_{\mu\nu} \, \hat T_0^{\mu\nu} \,,
\eea
where $\hat T_0^{\mu\nu}$ is the tree level graviton vertex defined in Eq.(\ref{P5treeT}) and $A_{\mu\nu}$ is the remaining amplitude which does not participate to the soft photon emissions.

The cancellation of the infrared singularities occurs at the cross section level, therefore we have to square the soft photon matrix element, sum over the photon polarization and integrate over the soft photon phase space
\bea
d \sigma_{soft} = - d \sigma_0 \frac{\alpha}{2 \pi^2} Q^2 \int_{|\vec k | \le \Delta E} \frac{d^2 k}{2 k_0} \bigg[ \frac{p_1^2}{( k \cdot p_1)^2}  + \frac{p_1^2}{( k \cdot p_1)^2}  - 2 \frac{p_1 \cdot p_2}{k \cdot p_1 \, k \cdot p_2} \bigg]
\eea
where the infrared divergence is regularized by the photon mass $\lambda$ which appears through the photon energy $k_0 = \sqrt{|\vec k | + \lambda^2}$. \\
The generic soft integral
\bea
I_{i j} = \int_{|\vec k | \le \Delta E} \frac{d^2 k}{2 k_0}  \frac{2 p_i \cdot p_j}{k \cdot p_i \, k \cdot p_j} 
\eea
has been worked out explicitly in \cite{'tHooft:1978xw}, here we give only the infrared divergent parts needed in our case
\bea
I_{1 1} &=& I_{2 2} = 4 \pi \log \frac{\Delta E}{\lambda} + \ldots \,, \nn \\
I_{1 2} &=& - 8 \pi \left(1 - \frac{q^2}{2 m^2} \right) \frac{x_s}{1 - x_s^2}  \log \frac{\Delta E}{\lambda}  \log x_s + \ldots \,,
\eea
with $x_s = - \frac{1 - \beta}{1 + \beta}$ and $\beta = \sqrt{1 - 4m^2/q^2}$. \\
Using the previous results we obtain the infrared singular part of the soft cross section
\bea
d \sigma_{soft} = - d \sigma_0 \frac{\alpha}{4 \pi} Q^2 \bigg\{  8 \log \frac{\Delta E}{\lambda}  +  \frac{8}{y} (2 y -1)  \frac{x_s}{1-x_s^2}  \log \frac{\Delta E}{\lambda} \log x_s  + \ldots \bigg\} \,,
\eea
where dots stand for finite terms.

Exploiting the fact that the infrared divergences in the one-loop corrections and in the counterterm diagram multiply the tree level graviton vertex $\hat T^{\mu\nu}_0$, so that they contribute only with a term proportional to the Born cross section, we obtain
\bea
d \sigma_{virt} + d \sigma_{CT} + d \sigma_{soft} =   d \sigma_0  \, \frac{\alpha}{4 \pi} Q^2 \bigg[ 1  + \frac{1}{y} (2y-1) \frac{x_s}{1-x_s^2} \log x_s\bigg]  8 \log \frac{m}{\Delta E} + \ldots \,, 
\eea
for the photon case and an analog result for the gluon contribution. The sum of the renormalized virtual corrections with the real emission contributions is then finite in the limit $\lambda \rightarrow 0$.

\section{Conclusions} 
\label{P5Sec.Conclusions} 
We have computed  the one-loop electroweak and 
strong corrections to the 
flavor diagonal graviton-fermion vertices in the Standard Model.
The work presented in this chapter is an extension, to the flavor diagonal case, 
of previous related study in which 
only the flavor-changing fermion graviton interactions had been investigated.
The result of our computation has been expressed in terms of 
a certain numbers of on-shell form factors, 
which have been given at leading order in the electroweak expansion and 
by retaining the exact dependence on the fermion masses.
We have also  included in our analysis the contribution of a 
non-minimally coupled Higgs sector, 
with an arbitrary value of the coupling parameter.
All these results can be easily extended to theories with fermion 
couplings to massive graviton, graviscalar and dilaton fields.

Moreover, we proved the infrared safety of the fermion-graviton vertices against radiative corrections of soft photons and gluons, where 
the ordinary cancellation mechanism
between the virtual and real bremsstrahlung corrections have been generalized
to the fermion-graviton interactions.

There are several phenomenological implications of this study that one could consider. Beside the possible applications to models with a low gravity scale, which would make the corrections discussed here far more sizeable, 
one could consider, for instance,  the specialization of our results to the neutrino sector, a definitely appealing argument on the cosmological side. Another possible extension would be to include,
 as a gravitational background, also a dilaton field, generated, for instance, from metric compactifications.

\chapter{Standard Model corrections to flavor-changing fermion-graviton vertices}
\label{Chap.GravitonFermion2}

\section{Introduction}
In the previous chapter, following \cite{Degrassi:2008mw}, we have discussed the structure of the perturbative corrections to the graviton-fermion-antifermion 
($T f \bar{f}$) vertex in the Standard Model (SM), focusing our attention on the flavor diagonal sector. These studies address the structure of the interactions between the fermions of the Standard Model and gravity, beyond leading order in the weak coupling, 
which have never been presented before in their exact expressions. The choice of an external (classical) gravitational background allows to simplify 
the treatment of such interactions where the coupling is obtained by the insertion of the symmetric and improved energy-momentum tensor (EMT) into ordinary correlators of the Standard Model. 

We have addressed some of the main features of the perturbative structure of these corrections, presenting their explicit form, parameterized in terms of a certain set of form factors. We have also discussed some of their radiative properties in regard to their infrared finiteness and renormalizability, the latter being inherited directly from the Standard Model, when the coupling of the Higgs to the gravitational background is conformal. 

In general, one expects that such corrections are small,  although they could become more sizeable in theories with a low gravity scale 
 \cite{ArkaniHamed:1998rs, Antoniadis:1998ig, ArkaniHamed:1998nn, Randall:1999ee,Randall:1999vf, Dvali:2000hp}.  In particular, one can consider the possibility of including, in these constructions, backgrounds which are of dilaton type, with dilaton fields produced by metric compactifications. The same vertices characterize the interaction of a dilaton of a spontaneously broken dilatation symmetry with the ordinary fields of the Standard Model \cite{Goldberger:2007zk, Campbell:2011iw, Coriano:2012nm, Barger:2011hu}. This second possibility is particularly interesting, in view of the recent discovery of a Higgs-like scalar at the LHC and it will be addressed in the next chapter.
 
Perturbative studies of these vertices have their specific difficulties due to the proliferation of form factors, and the results have to be secured by consistency checks using some relevant Ward identities. These need to be derived from scratch using the full Lagrangian of the Standard Model, as discussed in chapter \ref{Chap.GravitonEW}. In this study we are going to reconsider the gravitational form factor of a Standard Model fermion in the presence of a background graviton in the off-diagonal flavor case, which had been discussed before \cite{Degrassi:2008mw}, extending that analysis. One of the goals of this re-analysis is to include all the mass corrections to the related form factors, which had not been given before. 
These corrections are important in order to proceed with a systematic phenomenological study. 
In this respect, mass corrections are important in order to extract the exact behaviour of these form factors in the infrared and ultraviolet limits, 
which may be of experimental interest. We have compared our new results against the previous ones given in \cite{Degrassi:2008mw} in the limit of massless external fermions and found complete agreement.

\section{The perturbative expansion}
\label{P6Sec.PertExp}
The interaction of one graviton with two fermions of different flavor is summarized by the vertex function
\bea
\hat T^{\mu\nu} \equiv  i \langle f_i, p_i| T^{\mu\nu}(0) | p_j, f_j\rangle
\eea
that we intend to study. Here $p_j$ ($f_j$) 
and $p_i$ ($f_i$) indicate the momenta (flavor) of initial and final 
fermions respectively. We will restrict to  the case of 
flavor-changing transitions, namely $f_i\neq f_j$.
In order to simplify the results we will also use the combinations of momenta $p = p_i + p_j$ and $q = p_j - p_i$. The external states are taken on their mass shell, $p_i^2 = m_i^2$ and $p_j^2=m_j^2$ 
and can be either leptons or quarks. From now on, we will assume that 
$m_i\neq m_j$. In the last case, since the EMT is diagonal in color space, the color structure is rather trivial and therefore we omit it.  

At tree level the flavor-changing gravitational interaction is absent so that the leading order contribution comes from the quantum corrections. 
At one loop level, instead, we decompose the $\hat T^{\mu\nu}$ matrix element as
\bea
\hat T^{\mu\nu} = \hat T^{\mu\nu}_W + \hat T^{\mu\nu}_{CT}
\eea
where the first term on the r.h.s represents the pure vertex corrections 
induced by the $W^\pm$ gauge boson and its Goldstone $\phi^\pm$ exchanges,
while the last term, $\hat T^{\mu\nu}_{CT}$, includes 
the usual counterterms (CT) coming from the wave-function renormalization 
insertions on the external legs.  
The inclusion of this last term  $\hat T^{\mu\nu}_{CT}$ is needed
in order to get finite results for the matrix element $\hat T^{\mu\nu}$, as it 
will be extensively  discussed in section \ref{P6Sec.Renorm}. 
The finiteness of the result is just a consequence of the
non-renormalization theorem of conserved currents, when applied to the case 
of a conserved EMT.

\begin{figure}[t]
\centering
\subfigure[]{\includegraphics[scale=0.7]{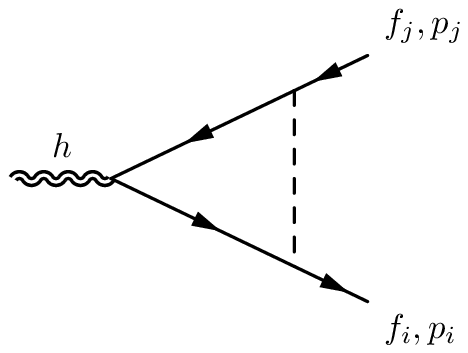}} \hspace{.5cm}
\subfigure[]{\includegraphics[scale=0.7]{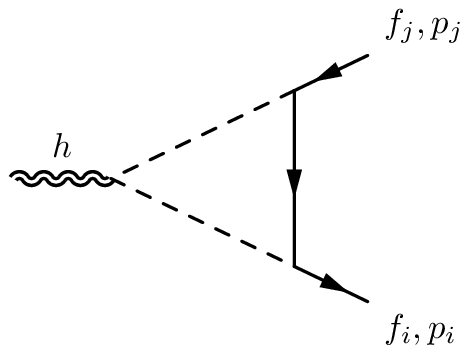}} \hspace{.5cm}
\subfigure[]{\includegraphics[scale=0.7]{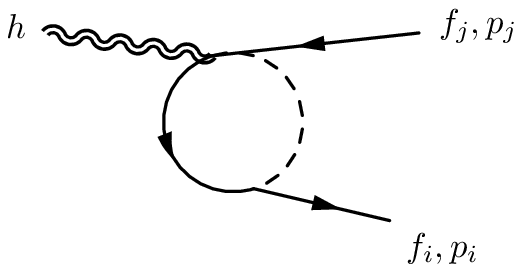}} \hspace{.5cm}
\subfigure[]{\includegraphics[scale=0.7]{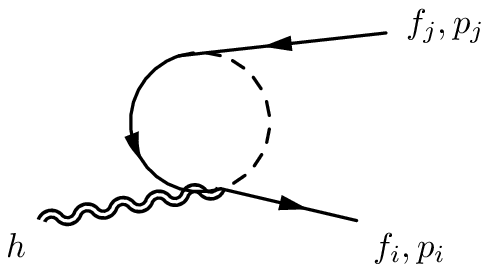}}
\caption{Diagrams of one-loop SM corrections to the flavor-changing graviton fermion vertex, where $f_{i,j}$ and $p_{i,j}$ specify the fermion flavors and corresponding momenta respectively, with $f_i \neq f_j$. \label{P6diagrams}}
\end{figure}

We choose to work in the $R_{\xi}$ gauge where every massive gauge field is always accompanied by its unphysical longitudinal part. The diagrammatic expansion of $\hat T^{\mu\nu}_W$ is depicted in Fig.\ref{P6diagrams} and is made of one contribution of triangle topology plus contact terms (see Fig. \ref{P6diagrams}  (c) and (d))  with a fermion and a graviton pinched on the same external point. The Feynman rules are listed in Appendix \ref{P6feynrules}. The computation of these diagrams has been performed in dimensional regularization using the on-shell renormalization scheme. To check the correctness of our results the Ward identity of the conservation of the EMT, which will presented in section \ref{P6Sec.WardId}, has been verified explicitly. 

Due to the chiral V-A nature of the $W$ interactions, we expand the flavor changing matrix element in terms of invariant amplitudes 
$f_i$ and tensor operators $O_i$ as 
\bea
\label{P6hatT}
\hat T^{\mu\nu} = i \frac{G_F}{16 \pi^2 \sqrt{2}} \sum_{k=1}^{12} f_k(p,q) \, \bar u_i(p_i) O^{\mu\nu}_k u_j(p_j)
\eea
with  the tensor operators given by
\bea
\begin{array}{ll}
O_1^{\mu\nu} = \left( \gamma^{\mu} p^{\nu}+\gamma^{\nu} p^{\mu}\right) P_L \qquad \qquad
& O_7^{\mu\nu}=  \eta^{\mu\nu} \, M_- \\
O_2^{\mu\nu}= \left( \gamma^{\mu} q^{\nu}+\gamma^{\nu} q^{\mu}\right) P_L 
& O_8^{\mu\nu}=  p^{\mu}p^{\nu} \, M_-  \\
 O_3^{\mu\nu}= \eta^{\mu\nu} \, M_+  
&O_9^{\mu\nu}=  q^{\mu}q^{\nu} \, M_-  \\
O_4^{\mu\nu}= p^{\mu}p^{\nu} \, M_+
&O_{10}^{\mu\nu}= \left(p^{\mu}q^{\nu}+q^{\mu}p^{\nu}\right) M_-\\
O_5^{\mu\nu}=  q^{\mu}q^{\nu} \, M_+
& O_{11}^{\mu\nu}= \frac{m_i m_j}{m_W^2}  \left( \gamma^{\mu} p^{\nu}+\gamma^{\nu} p^{\mu}\right) P_R \\
O_6^{\mu\nu}=  \left(p^{\mu}q^{\nu}+q^{\mu}p^{\nu}\right)\, M_+
&O_{12}^{\mu\nu}= \frac{m_i m_j}{m_W^2}  \left( \gamma^{\mu} q^{\nu}+\gamma^{\nu} q^{\mu}\right) P_R \end{array}
\label{P6basis}
\eea
where $P_{L,R}=(1\mp \gamma_5)/2$ and $M_{\pm}\equiv m_j P_R\pm m_i P_L$, and 
$u_{i,j}(p_{i,j})$ are the corresponding fermion
bi-spinor amplitudes in momentum space.\\ 
This is the most general rank-2 tensor basis that can be built out of two momenta, $p$ and $q$, a metric tensor and Dirac matrices $\gamma^\mu$ and $\gamma^5$. Its expression has been given in \cite{Degrassi:2008mw}. \\
For the form factors appearing in Eq.(\ref{P6hatT}) we use the following notation
\bea
f_k(p, q) = \sum_f  \lambda_f  \, F_{k}(p,q,m_f) \,,
\eea
where we have factorized the term $\lambda_f\equiv V_{f i} V^{*}_{f j}$
(the external fermions are assumed here to be quarks of down type), 
with $V_{ij}$ the corresponding CKM matrix element.

\section{Renormalization}
\label{P6Sec.Renorm}
The $\hat T^{\mu\nu}_W$ matrix element corresponding to the vertex corrections 
is ultraviolet divergent and, due to the non-renormalization
theorem of the conserved EMT, it is made finite by adding the contributions 
from the wave-function renormalization on the external legs, namely $\hat T^{\mu\nu}_{CT}$. 
In order to define the counterterm, as ilustrated in 
the previous chapter, we promote the counterterm SM Lagrangian to a curved background and then extract in the usual way the appropriate renormalized Feynman rules for single insertions of the EMT on the fields of the Standard Model. The metric is taken to be flat after all the functional differentiations. 
Then, for the off-diagonal flavor contributions $(i\neq j)$ to  
$\hat T^{\mu\nu}_{CT}$ we have
\bea
\label{P6ct}
\hat T^{\mu\nu}_{CT} &=& i \langle p_i, f_i | T^{\mu\nu}_{CT}(0) | p_j, f_j \rangle = 
\frac{i}{4} \bar u_i(p_i) \bigg\{ \left( \gamma^{\mu} p^{\nu}+\gamma^{\nu} p^{\mu}\right) \left( C^{L+}_{ij} P_L + C^{R+}_{ij} P_R \right) \nn \\
&& + \,2 \, \eta^{\mu\nu} \bigg[ C^{L-}_{ij}  \left( m_i P_L - m_j P_R \right) + C^{R-}_{ij}  \left( m_i P_R - m_j P_L \right) \bigg]
\bigg\} u_j(p_j) \,, 
\eea
where
\bea
C^{L \pm}_{ij} = \frac{1}{2} \left( \delta Z^L_{ij} \pm \delta Z^{L \dag}_{ij} \right) \,, \qquad   C^{R \pm}_{ij} = \frac{1}{2} \left( \delta Z^R_{ij} \pm \delta Z^{R \dag}_{ij} \right) \,,
\eea
with $\delta Z^{L,R}_{ij}$ being the fermion wave function renormalization constants. In the on-shell renormalization scheme, which we have chosen for our computation, the renormalization conditions are fixed in terms of the physical parameters of the Standard Model to all orders in the perturbative expansion. In particular for the fermion wave function renormalization constants with $i \ne j$ one obtains
\bea
\delta Z^L_{ij} &=& \frac{2}{m_i^2 - m_j^2} \tRE \bigg\{ m_j^2 \, \Sigma^L_{ij}(m_j^2) + m_i \, m_j \, \Sigma^R_{ij}(m_j^2) + \left( m_i^2 + m_j^2 \right) \Sigma^S_{ij}(m_j^2) \bigg\} \,, \nn \\
\delta Z^R_{ij} &=& \frac{2}{m_i^2 - m_j^2} \tRE \bigg\{ m_j^2 \, \Sigma^R_{ij}(m_j^2) + m_i \, m_j \, \Sigma^L_{ij}(m_j^2) + 2 \, m_i \, m_j \, \Sigma^S_{ij}(m_j^2) \bigg\} \,. 
\eea 
The symbol $\tRE$ gives the real part of the scalar integrals in the self-energies but it has no effect on the CKM matrix elements. Its presence yields $\delta Z^\dag_{ij} = \delta Z_{ij}\left( m_i^2 \leftrightarrow m_j^2 \right)$. Remember also that if the mixing matrix is real $\tRE$ can obviously be replaced with $\rm Re$. \\
For completeness we give the Standard Model flavor changing self-energies ($i \neq j $)
\bea
\label{P6selfenergies}
\Sigma^L_{ij}(p^2) &=& - \frac{G_F}{4 \pi^2 \sqrt{2}} \sum_f V_{if} V_{fj}^\dag \bigg[ \left( m_f^2 + 2 m_W^2 \right) \mathcal B_1(p^2, m_f^2, m_W^2) + m_W^2 \bigg] \,, \nn \\
\Sigma^R_{ij}(p^2) &=& - \frac{G_F}{4 \pi^2 \sqrt{2}} \, m_i \, m_j \sum_f V_{if} V_{fj}^\dag \, \mathcal B_1(p^2, m_f^2, m_W^2) \,, \nn \\
\Sigma^S_{ij}(p^2) &=& - \frac{G_F}{4 \pi^2 \sqrt{2}} \sum_f V_{if} V_{fj}^\dag \, m_f^2 \, \mathcal B_0(p^2, m_f^2, m_W^2) \,,
\eea
where
\bea
\mathcal B_1(p^2,m_0^2,m_1^2) = \frac{m_1^2 - m_0^2}{2 p^2} \bigg[ \mathcal B_0 (p^2,m_0^2,m_1^2) - \mathcal B_0(0,m_0^2,m_1^2) \bigg] - \frac{1}{2} \mathcal B_0(p^2,m_0^2,m_1^2) \,.
\eea
We have explicitly checked that the counterterm in Eq.(\ref{P6ct}) is indeed sufficient to remove all the ultraviolet divergences of the $\hat T^{\mu\nu}_W$ matrix element so that $\hat T^{\mu\nu}$ is finite, as expected.
\section{The Ward identity from the conservation of the EMT}
\label{P6Sec.WardId}
The conservation of the energy-momentum tensor constraints the $\hat T^{\mu\nu}$ matrix element reducing the 12 form factors defined above to a smaller subset of 6 independent contributions. We can derive the Ward identity by imposing the invariance of the 1-particle irreducible generating functional - which depends on the external gravitational metric - under a diffeomorphism transformation and then functional differentiating with respect to the fermion fields. We omit the details of this procedure, which has been discussed extensively in chapters \ref{Chap.GravitonEW} and \ref{Chap.GravitonFermions} for the $TVV'$ and the $Tf\bar{f}$ vertices respectively. The analysis, in this new case, follows similar steps.  
In momentum space, for the unrenormalized matrix element we obtain the Ward identity
\bea
\label{P6unrenormWI}
q_{\mu} \hat T^{\mu\nu}_W = \bar u_i(p_i) \bigg\{ p_i^\nu \Gamma_{ij}(p_i) - p_j^\nu \Gamma_{ij}(p_j) + \frac{q_\mu}{2} \left( \Gamma_{ij}(p_i) \sigma^{\mu\nu} - \sigma^{\mu\nu} \Gamma_{ij}(p_j) \right)\bigg\} u_j(p_j) \,,
\eea
where $\sigma^{\mu\nu}= [\gamma^{\mu},\gamma^{\nu}]/4$ and $\Gamma_{ij}(p)$ is the fermion two-point function which is given by
\bea
\Gamma_{ij}(p) = i \bigg[ \Sigma_{ij}^L(p^2) \, \psl \, P_L + \Sigma_{ij}^R(p^2) \, \psl \, P_R + \Sigma^S_{ij}(p^2) \left( m_i \, P_L + m_j \, P_R \right) \bigg] \,.
\eea
The off-diagonal (in flavor space) two-point form factors $\Sigma^{L,R,S}(p^2)$ are explicitly given in Eq.(\ref{P6selfenergies}). \\
The renormalized Ward identity is instead much simpler than Eq.(\ref{P6unrenormWI}) being just $q_{\mu} \hat T^{\mu\nu} = 0$. It implies a set of homogeneous equations \cite{Degrassi:2008mw} for the renormalized form factors $f_k(p,q)$
\bea
p \cdot q \, f_1(p,q) + q^2 f_2(p,q) &=& 0 \,, \nn \\
f_3(p,q) + q^2 f_5(p,q) + p \cdot q \, f_6(p,q) + \frac{p \cdot q}{2 m_W^2} f_{12}(p,q) &=& 0 \,, \nn \\
p \cdot q \, f_4(p,q) +q^2 f_6(p,q) + \frac{p \cdot q}{2 m_W^2} f_{11}(p,q) &=& 0 \,, \nn \\
f_2(p,q) + f_7(p,q) + q^2 f_9(p,q) + p \cdot q \, f_{10}(p,q) - \frac{p^2 + q^2}{4 m_W^2} f_{12}(p,q) &=& 0 \,, \nn \\
f_1(p,q) + p \cdot q \, f_8(p,q) + q^2 f_{10}(p,q) - \frac{p^2 + q^2}{4 m_W^2} f_{11}(p,q) &=& 0 \,, \nn \\
p \cdot q \, f_{11}(p,q) + q^2 f_{12}(p,q) &=& 0 \,,  
\eea
which provide a strong test on the correctness of our results and allow to reduce the number of independent contributions to the $\hat T^{\mu\nu}$ matrix element.

\section{Flavor-changing form factors}
In this section we present the explicit expressions of the renormalized form factors $F_k$ defined above. They have been computed in the on-shell case retaining the full dependence on the internal ($m_f, m_W$) and external masses ($m_i, m_j$) and on the virtuality, $q^2$, of the graviton line. They are expressed in terms of the dimensionless ratios $x_S = (m_i^2 + m_j^2)/q^2$, $x_D = (m_j^2 - m_i^2)/q^2$, $x_f = m_f^2/q^2$, $x_W = m_W^2/q^2$ and of the combination $\lambda = x_{D}^2-2 x_{S}+1$. We recall that $m_f$ is the mass of the fermion of flavor $f$ running in the loop. \\
Due to their complexity we expand our results onto a basis of massive one-, two- and three-point scalar integrals as
\bea
F_{k}(p,q,m_f) = \sum_{l=0}^{7} C_k^l \, I_l 
\label{P6Fi}
\eea
where
\bea
\begin{array}{ll}
I_0 = 1 \,, 	& 		I_4 = \mathcal B_0(q^2, m_f^2, m_f^2) \,, \\
I_1 =  \mathcal A_0(m_f^2) - \mathcal A_0(m_W^2) \,,   &   I_5 = B_0(q^2, m_W^2, m_W^2) \,,\\
I_2 =  \mathcal B_0(m_j^2, m_f^2, m_W^2) \,,        &   I_6 = \mathcal C_0(m_j^2, q^2, m_i^2, m_f^2, m_W^2, m_W^2) \,, \\
I_3 = \mathcal B_0(m_i^2, m_f^2, m_W^2) \,,      & I_7 = \mathcal C_0(m_j^2, q^2, m_i^2, m_W^2, m_f^2, m_f^2) \,.
\end{array}
\eea
We give the explicit results for the renormalized form factors $F_1$, $F_3$, $F_4$, $F_7$, $F_8$ and $F_{11}$ while the remaining six can be obtained exploiting the Ward identities derived in the previous section
\bea
\label{P6WIFF}
F_2 &=& - x_D \, F_1 \,, \nn \\
F_5 &=&  - \frac{1}{q^2} F_3 + x_D^2 \, F_4 + \frac{x_D^2}{m_W^2} F_{11} \,, \nn \\
F_6  &=& - x_D \,  F_4 - \frac{x_D}{2 m_W^2}  F_{11}  \,, \nn \\
F_9  &=& 2 \frac{x_D}{q^2} F_1 - \frac{1}{q^2} F_7 + x_D^2 \, F_8 - \frac{x_S \, x_D}{ m_W^2}  F_{11} \,, \nn \\
F_{10}  &=& - \frac{1}{q^2} F_1 - x_D \, F_8 + \frac{x_S}{2  m_W^2} F_{11}  \,, \nn \\
F_{12} &=& - x_D \, F_{11} \,. 
\eea
The coefficients $C_k^l$ defining $F_k$ in Eq.(\ref{P6Fi}) are given by
in the Appendix. \ref{P6formfactors}. \\
Finally we remark that the $F_3$, $F_5$, $F_7$ and $F_9$ form factors depend also from the parameter $\chi$ which appears in the gravitational coupling of the $\phi^\pm$ Goldstone bosons through the improved energy-momentum tensor.

\section{Conclusions}
We have presented the computation of the structure of the gravitational form factors of the Standard Model fermions in the  
off-diagonal flavor sector. The analysis has been developed according to the previous computation described in chapter \ref{Chap.GravitonFermions} where we have discussed the electroweak corrections in the flavor conserving case. The work presented in this chapter extends a previous investigation \cite{Degrassi:2008mw} of the same flavor-changing vertex in which the external mass dependence had not been included.

\chapter{An effective dilaton from a scale invariant extension of the Standard Model}
\label{Chap.Dilaton}

This chapter contains an extension of the analysis of the $TVV$ vertices of the Standard Model to the case of the $J_D VV$ correlator, where $J_D$ denotes the dilatation current. In particular we show that the anomaly poles extracted from the correlators involving the energy-momentum tensor insertions are also part of the $J_D VV$ vertices. It is natural to wonder what they mean. While in a gravitational context such poles describe effective degrees of freedom 
of gravity which are present in the anomaly action, a similar interpretation is also natural in the case of the dilatation current of the Standard Model.   
It does not require a large extrapolation to note that in this case these effective degrees of freedom can be identified with an effective dilaton. 
As we clarify, dilatons appear in certain field theories whenever we spontaneously break an underlying dilatation symmetry but they can also be part of gravity in the case of theories with extra dimensions.
It is, in general, expected that the breaking of a dilatation symmetry due to quantum effects should be associated with a Nambu-Goldstone mode that couples to the divergence of the current of the broken symmetry. Our perturbative analysis, which borrows from the previous chapters, shows that the signature of the anomalous breaking of the dilatation symmetry is in the emergence of an anomaly pole in the 
$J_D V V$ vertex, interpreted as the exchange of a composite state. We discuss the possible phenomenological implications of this result, which 
is the perturbative manifestation of an anomalous coupling. This causes a strong enhancement of certain decay channels of the associated resonance. We then proceed by showing how a mixing between this state and the Higgs of the Standard Model may take place. We also briefly investigate the difference between the Higgs and the dilaton couplings to the neutral currents in the various realization of the scale invariant extensions of the Standard Model Lagrangian, in particular if these are based on the notion either of a classical or of a quantum dilatation symmetry.

\section{Introduction} 

Dilatons are part of the low energy effective action of several different types of theories, from string theory to theories with compactified extra dimensions, but they may appear also in appropriate bottom-up constructions. For instance, in scale invariant extensions of the Standard Model, the introduction of a dilaton field allows to recover scale invariance, which is violated by the Higgs potential, by introducing a new, enlarged, Lagrangian. This is characterized both by a spontaneous breaking of the conformal and of the electroweak symmetries.  

In this case, one can formulate simple scale invariant extensions of the potential which can accomodate, via spontaneous breakings, two separate scales:
the electroweak scale ($v$), related to the vev of the Higgs field, and the conformal breaking scale ($\Lambda$), related to the vev of a new field $\Sigma=\Lambda + \rho$, with $\rho$ being the dilaton. The second scale can be fine-tuned in order to proceed with a direct phenomenological analysis and is, therefore, of outmost relevance in the search for new physics at the LHC.

In a bottom-up approach, and this will be one of the main points that we will address in our analysis, the dilaton of the effective scale invariant Lagrangian can also be interpreted as a composite scalar, with the dilatation current taking the role of an operator which interpolates between this state and the vacuum. We will relate this intepretation to the appearance of an anomaly pole in the correlation function involving the dilatation current ($J_D$)
and two neutral currents ($V, V'$) of the Standard Model, providing evidence, in the ordinary perturbative picture, in favour of such a statement.   

One of the main issues which sets a difference between the various types of dilatons is, indeed, the contribution coming from the anomaly, which is expected to be quite large. Dilatons obtained from compactifications with large extra dimensions and a low gravity scale, for instance, carry this coupling, which is phenomenologically relevant. 
The same coupling is present in the case of an effective dilaton, appearing as a Goldstone mode of the dilatation current, with some differences that we will specify in a second part of our work. The analysis will be carried out in analogy to the pion case, which in a perturbative picture is associated with the appearance of an anomaly pole in the $AVV$ diagram.

This chapter is organized as follows. In a first part we will characterize the leading one-loop interactions of a dilaton derived from a Kaluza-Klein compactification of the gravitational metric. The setup is analogous to that presented in 
\cite{Han:1998sg,Giudice:2000av} for a compactified theory with large extra dimensions and it involves all the neutral currents of the Standard Model. We present also 
a discussion of the same interaction in the QCD case for off-shell gluons.    

These interactions are obtained by tracing the $TVV'$ vertex, given in the previous chapters, with $T$ denoting the (symmetric and conserved) 
energy-momentum tensor (EMT) of the Standard Model. This study is accompanied by an explicit proof of the renormalizability of these interactions in the case of a conformally coupled Higgs scalar.

In a second part then we turn our discussion towards models in which dilatons are introduced from the ground up, starting with simple examples which should clarify - at least up to operators of dimension 4 - how one can proceed with the formulation of scale invariant extensions of the Standard Model. 
We illustrate the nature of the coupling of the dilaton to the mass dependent terms of the corresponding Lagrangian. The goal 
is to clarify that a fundamental (i.e. not a composite) dilaton, in a {\em classical} scale invariant extension of a given Lagrangian, does not necessarily couple to the anomaly, but only to massive states, exactly as in the Higgs case. For an effective dilaton, instead,  the Lagrangian, derived at tree level on the basis of classical scale invariance, as for a fundamental dilaton, which needs to be modified with the addition of an anomalous contribution, due to the composite nature of the scalar, in close analogy to the pion case.   

As we are going to show, if the dilaton is a composite state, identified with the anomaly pole of 
the $J_D VV$ correlator, an infrared coupling of this pole (i.e. a nonzero residue) is necessary in order to claim the presence of an anomaly enhancement in the $VV$ decay channel, with the $VV$ denoting on-shell physical asymptotic states, in a typical $S$-matrix approach. Here our reasoning follows quite closely the chiral anomaly case, where the anomaly pole of the $AVV$ diagram, which describes the pion exchange between the axial vector ($A$) and the vector currents, is infrared coupled only if $V$ denote physical asymptotic states. 
 
 Clearly, our argument relies on a perturbative picture and is, in this respect, admittedly limited, forcing this issue to be resolved at experimental level, as in the pion case. We recall that in the pion the enhancement is present in the di-photon channel and not in the 2-gluon decay channel. 
 
Perturbation theory, in any case, allows to link the enhancement of a certain dilaton production/decay channel, to the virtuality of 
the gauge currents in the initial or the final state.

\subsection{The energy momentum tensor}

We start with a brief summary of the structure of the Standard Model interactions with a $4D$ gravitational background, which is convenient in order to describe both the coupling of the dilaton, coming from the compactification of extra dimensions, and of a graviton at tree level and at higher orders.   
In the background metric $g_{\mu\nu}$ the action takes the form 
\beq S = S_G + S_{SM} + S_{I}= -\frac{1}{\kappa^2}\int d^4 x \sqrt{-g}\, R + \int d^4 x
\sqrt{-g}\mathcal{L}_{SM} + \chi \int d^4 x \sqrt{-g}\, R \, H^\dag H      \, ,
\eeq
where $\kappa^2=16 \pi G_N$, with $G_N$ being the four dimensional Newton's constant and $\mathcal H$ is the Higgs doublet.
We use the convention $\eta_{\mu\nu}=(1,-1,-1,-1)$ for the metric in flat spacetime, parameterizing its deviations from the flat case as
\beq\label{PNew2QMM} g_{\mu\nu}(x) \equiv \h_{\mu\nu} + \kappa \, h_{\mu\nu}(x)\,,\eeq
with the symmetric rank-2 tensor $h_{\mu\nu}(x)$ accounting for its fluctuations.
In this limit, the coupling of the Lagrangian to gravity is given by the term
\beq\label{PNew2Lgrav} \mathcal{L}_{grav}(x) = -\frac{\kappa}{2}T^{\mu\nu}(x)h_{\mu\nu}(x). \eeq
In the case of theories with extra spacetime dimensions the structure of the corresponding Lagrangian can be found in 
\cite{Han:1998sg,Giudice:2000av}. For instance, in the case of a compactification over a $S_1$ circle of a 5-dimensional theory to 4D, equation (\ref{PNew2Lgrav}) is modified in the form 
\beq
\label{PNew2Lgrav1} \mathcal{L}_{grav}(x) = -\frac{\kappa}{2} T^{\mu\nu}(x) \left(h_{\mu\nu}(x) + \rho(x) \,  \eta_{\mu\nu} \right) 
\eeq
which is sufficient in order to describe dilaton $(\rho)$ interactions with the fields of the Standard Model at leading order in $\kappa$, as in our case. In this case the graviscalar field $\rho$ is related to the $g_{55}$ component of the 5D metric and describes its massless Kaluza-Klein mode. The compactification generates an off-shell coupling of $\rho$ to the trace of the symmetric EMT. 

The derivation of the complete dilaton/gauge/gauge vertex in the Standard Model requires the computation of the trace of the EMT 
${T^\mu}_\mu$ (for the tree-level contributions), and of a large set of 1-loop 3-point functions. 
These are diagrams characterized by the insertion of the trace into 2-point functions of gauge currents. 
The full EMT is given by a minimal tensor $T_{Min}^{\mu\nu}$ (without improvement) and by a term of improvement, 
$T_I^{\mu\nu}$, originating from $S_I$. We refer to chapter \ref{Chap.GravitonEW} for the details.

\section{One loop electroweak corrections to dilaton-gauge-gauge vertices} 

In this section we will present results for the structure of the radiative corrections to the dilaton-gauge-gauge vertices in the 
case of two photons, photon/$Z$ and $Z Z$ gauge currents. The list of the relevant tree 
level interactions extracted from the SM Lagrangian,  
which have been used in the computation of these corrections, can be obtained from the vertices included in Appendix \ref{P3FeynRules} upon contraction with the metric tensor and with the replacement $\kappa/2 \to 1/\Lambda$.
We identify three classes of contributions, 
denoted as $\mathcal{A}$, $\Sigma$ and $\Delta$, with the $\mathcal A$-term coming from the conformal anomaly while the $\Sigma$ and 
$\Delta$ terms are related to the exchange of fermions, gauge bosons and scalars (Higgs/goldstones). The separation between the 
anomaly part and the remaining terms is typical of the $TVV'$ interaction. In particular one can check that in a 
mass-independent regularization scheme, such as Dimensional Regularization with minimal subtraction, this separation can be verified 
at least at one loop level and provides a realization of the (anomalous) conformal Ward identity 
\beq
\Gamma^{\alpha\beta}(z,x,y) 
\equiv \eta_{\mu\nu} \left\langle T^{\mu\nu}(z) V^{\alpha}(x) V^{\beta}(y) \right\rangle 
= \frac{\delta^2 \mathcal A(z)}{\delta A_{\alpha}(x) \delta A_{\beta}(y)} + \left\langle {T^\mu}_\mu(z) V^{\alpha}(x) V^{\beta}(y) \right\rangle,
\label{PNew2traceid1}
\eeq
where we have denoted by $\mathcal A(z)$ the anomaly and $A_{\alpha}$ the gauge sources coupled to the current $V^{\alpha}$. Notice that in the expression above $\Gamma^{\alpha\beta}$ denotes a generic 
dilaton/gauge/gauge vertex, which is obtained form the $TVV'$ vertex by tracing the spacetime indices $\mu\nu$. A simple way to test the validity 
of (\ref{PNew2traceid1}) is to compute the renormalized vertex $\langle T^{\mu\nu} V^\alpha V'^\beta\rangle$ (i.e. the graviton/gauge/gauge vertex) 
and perform afterwards its 4-dimensional trace. This allows to identify the left-hand-side of this equation. On the other hand, the insertion of 
the trace of $T^{\mu\nu}$ (i.e. $T^\mu_\mu$ )into a two point function $VV'$, allows to identify the second term on the 
right-hand-side of (\ref{PNew2traceid1}),  $\langle T^{\mu}_\mu(z) V^{\alpha}(x) V^{\beta}(y)\rangle$. 
The difference between the two terms so computed can be checked to correspond to the $\mathcal A$-term, obtained by two differentiations of the  anomaly functional $\mathcal A$. 
We recall that, in general, when scalars are conformally coupled, this takes the form
\beqa \label{PNew2TraceAnomaly}
\mathcal A(z)
&=& \sum_{i} \frac{\beta_i}{2 g_i} \, F^{\alpha\beta}_i(z) F^i_{\alpha\beta}(z) +... \,,
\eeqa
where $\beta_i$ are
clearly the mass-independent $\beta$ functions of the gauge fields and $g_i$ the corresponding coupling constants, while the ellipsis refer to 
curvature-dependent terms.%
We present explicit results starting for the $\rho VV'$ vertices ($V,V' = \gamma, Z$), denoted as $\Gamma_{VV'}^{\alpha \beta}$, which are decomposed in momentum space in the form 
\beq
\Gamma_{VV'}^{\alpha \beta}(k,p,q) = (2\, \pi)^4\, \delta(k-p-q) \frac{i}{\Lambda} 
\left( \mathcal A^{\alpha \beta}(p,q) + \Sigma^{\alpha \beta}(p,q) + \Delta^{\alpha \beta}(p,q)\right),
\eeq
where 
\beq
\mathcal A^{\alpha \beta}(p,q) = \int d^4 x\, d^4 y \, e^{i p \cdot x + i q\cdot y}\, 
\frac{\delta^2 \mathcal A(0)}{\delta A^\alpha(x)\delta A^\beta(y)}
\eeq
and 
\beq
 \Sigma^{\alpha \beta}(p,q) +  \Delta^{\alpha \beta}(p,q) = \int d^4 x\, d^4 y\, e^{ i p \cdot x + i q\cdot y}\, 
\left\langle {T^\mu}_\mu(0) V^\alpha(x) V^\beta(y) \right\rangle \,.
\eeq
We have denoted with $\Sigma^{\alpha \beta}$ the cut vertex contribution to $\Gamma^{\alpha\beta}_{\rho VV'}$, 
while $\Delta^{\alpha \beta}$ includes the dilaton-Higgs mixing on the dilaton line, as shown in Fig. \ref{PNew2figuremix}.
Notice that $\Sigma^{\alpha \beta}$ and $\Delta^{\alpha \beta}$ take contributions in two cases, specifically if the theory has an 
explicit (mass dependent) breaking and/or if the scalar - which in this case is the Higgs field - is not conformally coupled. 
The $\mathcal A^{\alpha\beta}(p,q)$ term represents the conformal anomaly while $\Lambda$ is dilaton interaction scale.

\subsection{The $\rho\gamma\gamma$ vertex}

The interaction between a dilaton and two photons is identified by the diagrams in Figs. \ref{PNew2figuretriangle},\ref{PNew2figuretadpole},\ref{PNew2figuremix} and is summarized by the expression 
\beqa
\Gamma_{\gamma \gamma}^{\alpha\beta}(p,q) = \frac{i}{\Lambda} \bigg[ \mathcal A^{\alpha\beta}(p,q) + 
\Sigma^{\alpha\beta}(p,q) + \Delta^{\alpha\beta}(p,q) \bigg] \,,\eeqa
with the anomaly contribution given by
\beqa
\label{PNew2Agammagamma}
\mathcal A^{\alpha\beta} =   \frac{\alpha}{\pi}\, \bigg[ -\frac{2}{3}\sum_{f} Q_{f}^2 + \frac{5}{2}  + 6\,\chi\bigg]\,
 u^{\alpha\beta}(p,q) \stackrel{\chi\rightarrow\frac{1}{6}}{=} - 2\, \frac{\beta_e}{e}\, u^{\alpha\beta}(p,q) \, ,
\eeqa
where
\bea
\label{PNew2utensor}
u^{\alpha\beta}(p,q) = (p\cdot q) \eta^{\alpha\beta} - q^{\alpha}p^{\beta} \,,
\eea
and the explicit scale-breaking term $\Sigma^{\alpha \beta}$ which splits into
\beqa
\label{PNew2Sigmagammagamma}
\Sigma^{\alpha\beta}(p,q) = \Sigma_F^{\alpha\beta}(p,q) +  \Sigma_B^{\alpha\beta}(p,q) +\Sigma_I^{\alpha\beta}(p,q) \,.
\eeqa
We obtain for the on-shell photon case ($p^2 = q^2 = 0$)
\beqa
\Sigma_F^{\alpha\beta}(p,q) 
&=& 
 \frac{\alpha}{\pi}\, \sum_f Q_f^2 m_f^2 \left[ \frac{4}{s} + 2 \left(\frac{4 
m_f^2}{s}-1\right) \mathcal C_0\left(s,0,0,m_f^2,m_f^2,m_f^2 \right)\right]\, u^{\alpha\beta}(p,q) \, , \nn \\
\Sigma_B^{\alpha\beta}(p,q) 
&=& 
 \frac{\alpha}{\pi}\, \left[ 6 M_W^2 \left(1-2\frac{M_W^2}{s}\right)  
\mathcal C_0(s,0,0,M_W^2,M_W^2,M_W^2) - 6 \frac{M_W^2}{s} - 1 \right]\, u^{\alpha\beta}(p,q) \, , \nn \\
\Sigma_I^{\alpha\beta}(p,q) 
&=& 
  \frac{\alpha}{\pi}\, 6 \chi \bigg[ 2 M_W^2 \mathcal C_0\left(s,0,0,M_W^2,M_W^2,M_W^2\right)\,
u^{\alpha\beta}(p,q) \nonumber \\
&-& 
M_W^2\, \frac{s}{2}\, \mathcal C_0(s,0,0,M_W^2,M_W^2,M_W^2) \, \eta^{\alpha \beta} \bigg]\, ,
\eeqa
while the mixing contributions are given by
\bea
\label{PNew2DeltaHgammagamma}
\Delta^{\alpha\beta}(p,q) 
&=&
\frac{\alpha}{\pi (s - M_H^2)} 6 \chi \bigg\{ 2 \sum_f Q_f^2 m_f^2  \bigg[ 2 + (4 m_f^2 -s ) \mathcal C_0(s,0,0,m_f^2,m_f^2,m_f^2) 
\bigg] \nn \\
&+& 
 M_H^2 + 6 M_W^2 + 2 M_W^2 (M_H^2 + 6 M_W^2 -4 s) \mathcal C_0(s,0,0,M_W^2,M_W^2,M_W^2)\bigg\} u^{\alpha \beta}(p,q) \nn \\
&+&  
\frac{\alpha}{\pi} 3 \chi s \, M_W^2 \, \mathcal C_0(s,0,0,M_W^2,M_W^2,M_W^2) \eta^{\alpha \beta} \, ,
\eea
with $\alpha$ the fine structure constant. The scalar integrals are defined in Appendix \ref{P2scalars}.
The $\Sigma$'s  and $\Delta$ terms are the contributions obtained from the insertion on the photon 2-point function of the trace of 
the EMT, ${T^\mu}_\mu$. Notice that $\Sigma_I$ includes all the trace insertions which originate from the terms of improvement $T_I$ 
except for those which are bilinear in the Higgs-dilaton fields and 
which have been collected in $\Delta$. The analysis of the Ward and Slavnov-Taylor identities for the graviton-vector-vector correlators shows that these can be consistently solved only if we include the graviton-Higgs mixing on the graviton line. 

We have included contributions proportional both to fermions ($F$) and boson ($B$) loops, beside the $\Sigma_I$.
A conformal limit on these contributions can be performed by sending to zero all the mass terms, which is equivalent to sending 
the vev $v$ to zero and requiring a conformal coupling of the Higgs $(\chi=1/6)$. 
In the $v\to 0$ limit, but for a generic parameter $\chi$, we obtain 
\beq
\lim_{v\to 0} \left( \Sigma^{\a\b} + \Delta^{\a\b} \right)
= \lim_{v\to 0} \left(\Sigma_{B}^{\a\b}  + \Sigma_{I}^{\a\b}\right) 
=  \frac{\alpha}{\pi} (6 \chi -1) u^{\alpha \beta}(p,q),
\eeq
which, in general, is non-vanishing.
Notice that, among the various contributions, only the exchange of a boson or the term of improvement contribute in this limit and 
their sum vanishes only if the Higgs is conformally coupled $(\chi = \frac{1}{6})$. \\

Finally, we give the decay rate of the dilaton into two on-shell photons in the simplified case in which we remove the term of improvement by sending $\chi \to 0$
\beqa
\Gamma(\rho \rightarrow \gamma\gamma) 
&=&
\frac{\alpha^2\,m_{\rho}^3}{256\,\Lambda^2\,\pi^3} \, \bigg| \beta_{2} + \beta_{Y} 
-\left[ 2 + 3\, x_W  +3\,x_W\,(2-x_W)\,f(x_W) \right]
+ \frac{8}{3} \, x_t\left[1 + (1-x_t)\,f(x_t) \right] \bigg|^2 \,, \nn \\
\label{PNew2PhiGammaGamma} 
\eeqa
where the contributions to the decay, beside the anomaly term, come from the $W$ and the fermion (top) loops
and $\beta_2 (= 19/6)$ and $\beta_Y (= -41/6)$ are the $SU(2)$ and $U(1)$ $\beta$ functions respectively.
Here, as well as in the other decay rates evaluated all through this chapter, the $x_i$ are defined as
\beq \label{PNew2x}
x_i = \frac{4\, m_i^2}{m^2_\rho} \, ,
\eeq
with the index "$i$" labelling the corresponding massive particle, and $x_t$ denoting the contribution from the top quark,
which is the only massive fermion running in the loop.
The function $f(x)$ is given by
\beqa
\label{PNew2fx}
f(x) = 
\begin{cases}
\arcsin^2(\frac{1}{\sqrt{x}})\, , \quad \mbox{if} \quad \,  x \geq 1 \\ 
-\frac{1}{4}\,\left[ \ln\frac{1+\sqrt{1-x}}{1-\sqrt{1-x}} - i\,\pi \right]^2\, , \quad \mbox{if} \quad \, x < 1.
\end{cases}
\eeqa
which originates from the scalar three-point master integral through the relation 
\beq \label{PNew2C03m}
C_0(s,0,0,m^2,m^2,m^2) = - \frac{2}{s} \, f(\frac{4\,m^2}{s}) \, .
\eeq

\subsection{The $\rho\gamma Z$ vertex} 
%
The interaction between a dilaton, a photon and a $Z$ boson is described by the $\Gamma^{\alpha \beta}_{\gamma Z}$ correlation 
function (Figs. \ref{PNew2figuretriangle},\ref{PNew2figuretadpole},\ref{PNew2figuremix}). In the on-shell case, with the kinematic defined by
\beq
p^2 = 0 \, \quad q^2 = M_Z^2 \, \quad k^2 = (p+q)^2 = s \,,
\eeq
the vertex $\Gamma^{\alpha \beta}_{\gamma Z}$ is expanded as 
\bea
\Gamma^{\alpha \beta}_{\gamma Z} &=& \frac{i}{\Lambda} \bigg[ \mathcal A^{\alpha\beta}(p,q)
+ \Sigma^{\alpha\beta}(p,q) + \Delta^{\alpha\beta}(p,q) \bigg] \nn \\
&=&
\frac{i}{\Lambda}\, \Bigg\{\,
\left[ \frac{1}{2}\,\left(s - M^2_Z\right)\,\eta^{\alpha\beta} - q^\alpha\,p^\beta\right] \, 
\left( \mathcal A_{\gamma Z} +  \Phi_{\gamma Z}(p,q)\right)
+\eta^{\alpha\beta}\, \Xi_{\gamma Z}(p,q) \Bigg\} \, .
\eea
The anomaly contribution is
\beq
\mathcal A_{\gamma Z} =  \frac{\alpha}{\pi\,s_w c_w}\,\left[-\frac{1}{3}\sum_{f}C^f_v\,Q_f + \frac{1}{12}\,(37-30s_w^2)
+ 3\, \chi\, (c_w^2 - s_w^2) \right]\, ,
\eeq
where $s_w$ and $c_w$ to denote the sine and cosine of the $\theta$-Weinberg angle.
Here $\Delta^{\alpha \beta}$ is the external leg correction on the dilaton line and the form factors $\Phi(p,q)$ and $\Xi(p,q)$ 
are introduced to simplify the computation of the decay rate and decomposed as
\bea
\Phi_{\gamma Z}(p,q)  &=& \Phi^{\Sigma}_{\gamma Z}(p,q) + \Phi^{\Delta}_{\gamma Z}(p,q) \,, \nn \\
\Xi_{\gamma Z}(p,q) &=& \Xi^{\Sigma}_{\gamma Z}(p,q) + \Xi^{\Delta}_{\gamma Z}(p,q) \,,
\eea
in order to distinguish the contributions to the external leg corrections ($\Delta$) from those to the cut vertex ($\Sigma$).
They  are given by
\beqa
\Phi^{\Sigma}_{\gamma Z}(p,q)
&=& 
\frac{\alpha}{\pi\, s_w\,c_w}\, \Bigg\{
\sum_f C_v^f \, Q_f \, \Bigg[\frac{2\, m_f^2}{s - M_Z^2} + \frac{2 m_f^2 \, M_Z^2}{(s - M_Z^2)^2}\, \mathcal D_0(s,M_Z^2,m_f^2,m_f^2)
\nonumber \\
&& \hspace{-20mm}
- m_f^2\, \left(1 - \frac{4\, m_f^2}{s - M_Z^2}\right)\, \mathcal C_0(s,0,M_Z^2,m_f^2,m_f^2,m_f^2) \Bigg]
-  \Bigg[ \frac{M_Z^2}{2\,\left(s - M_Z^2\right)}\, (12\, s_w^4 - 24\, s_w^2 + 11)
\nonumber\\
&& \hspace{-20mm}
\frac{M_Z^2}{2\,\left(s - M_Z^2 \right)^2} 
\left[2\, M_Z^2\,\Big(6\, s_w^4 - 11\, s_w^2 + 5 \Big)- 2\, s_w^2\, s + s \right]\, \mathcal D_0(s,M_Z^2,M_W^2,M_W^2)
\nonumber \\
&& \hspace{-20mm}
+ \frac{M_Z^2\, c_w^2}{s - M_Z^2}\,
\Big[2\, M_Z^2\, \left(6\, s_w^4 - 15\, s_w^2 + 8 \right) + s\, \left(6\, s_w^2 - 5\right)\Big]\,
\mathcal C_0(s,0,M_Z^2,M_W^2,M_W^2,M_W^2) \Bigg]
\nonumber \\
&& \hspace{-20mm}
+ \frac{3 \chi \,(c_w^2 - s_w^2)\,}{s - M_Z^2}\,
\bigg[M_Z^2 +  s\, \bigg( 2\, M_W^2 \, \mathcal C_0(s,0,M_Z^2,M_W^2,M_W^2,M_W^2) 
+ \frac{M_Z^2}{s-M_Z^2}\, \mathcal D_0(s,M_Z^2,M_W^2,M_W^2) \bigg) \bigg]
\Bigg\} \, ,
\nonumber
\eeqa
\beqa
\Xi^{\Sigma}_{\gamma Z}(p,q)
&=&
\frac{\alpha}{\pi}
\Bigg\{ - \frac{c_w\, M_Z^2}{ s_w} \mathcal B_0(0,M_W^2,M_W^2) + 3\, s\, \chi \, s_w^2\, M_Z^2 \, 
\mathcal C_0(s,0,M_Z^2,M_W^2,M_W^2,M_W^2) \Bigg\} \, ,  \nn \\
\Phi^{\Delta}_{\gamma Z}(p,q)
&=& 
\frac{3\,\alpha\,s\,\chi}{\pi s_w c_w (s-M_H^2)(s-M_Z^2)} 
\bigg\{ 2 \sum_f m_f^2 C_v^f Q_f \bigg[ 2 + 2 \frac{M_Z^2}{s-M_Z^2} \mathcal D_0(s,M_Z^2,m_f^2,m_f^2)  \nn \\
&& \hspace{-20mm}
+ (4 m_f^2 + M_Z^2 - s) \mathcal C_0(s,0,M_Z^2,m_f^2,m_f^2,m_f^2) \bigg]  + M_H^2(1-2 s_w^2) + 2 M_Z^2 (6 s_w^4 - 11 s_w^2 +5) \nn \\
&& \hspace{-20mm}
+ \frac{M_Z^2}{s-M_Z^2} ( M_H^2 (1-2 s_w^2) + 2 M_Z^2 (6 s_w^4 -11 s_w^2 + 5) ) \mathcal D_0(s,M_Z^2,M_W^2,M_W^2) \nn \\
&& \hspace{-20mm}
+ 2 M_W^2 \mathcal C_0(s,0,M_Z^2,M_W^2,M_W^2,M_W^2) ( M_H^2 (1-2 s_w^2) + 2 M_Z^2 (6 s_w^4 - 15 s_w^2 + 8) + 2 s (4 s_w^2-3)) \bigg\} 
\nn \\
\Xi^{\Delta}_{\gamma Z}(p,q)
&=&
\frac{3\,\alpha\,s\, \chi\, c_w}{\pi\, s_w}\, M_Z^2 \bigg\{ \frac{2}{s-M_H^2} \mathcal B_0(0, M_W^2,M_W^2) 
- s_w^2 \mathcal C_0(s,0,M_Z^2,M_W^2,M_W^2,M_W^2) \bigg\}\, .
\eeqa

As for the previous case, we give the decay rate in the simplified limit $\chi \to 0$ which is easily found to be
\beqa
\Gamma(\rho\rightarrow \gamma Z) 
&=&
\frac{9\,m_{\rho}^3}{1024\,\Lambda^2\,\pi} \, \sqrt{1-x_Z} \, \bigg( |\Phi^{\Sigma}_{\gamma Z}|^2(p,q)\,m_{\rho}^4\,(x_Z-4)^2  \nn \\
&+&  
48 \, Re\, \left\{\Phi^{\Sigma}_{\gamma Z}(p,q)\,\Xi^{\Sigma\,*}_{\gamma Z}(p,q) \, m_{\rho}^2\,(x_Z-4)\right\}
- 192 \, |\Xi^{\Sigma}_{\gamma Z}|^2(p,q)  \bigg) \, ,
\label{PNew2RateRhoGammaZ}
\eeqa
where $Re$ is the symbol for the real part.

\subsection{The $\rho Z Z$ vertex}

\begin{figure}[t]
\centering
\subfigure[]{\includegraphics[scale=.7]{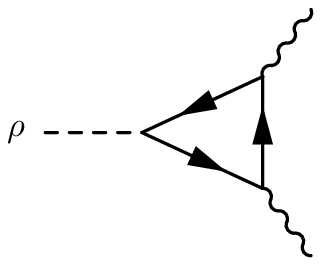}}
\hspace{.2cm}
\subfigure[]{\includegraphics[scale=.7]{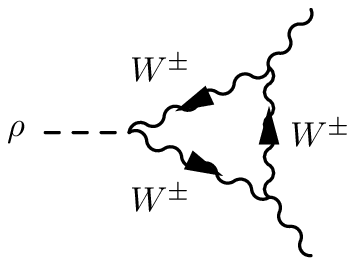}}
\hspace{.2cm}
\subfigure[]{\includegraphics[scale=.7]{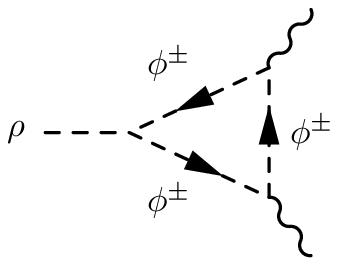}}
\hspace{.2cm}
\subfigure[]{\includegraphics[scale=.7]{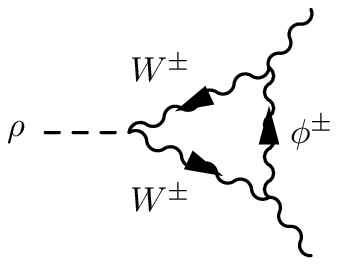}}
\hspace{.2cm}
\subfigure[]{\includegraphics[scale=.7]{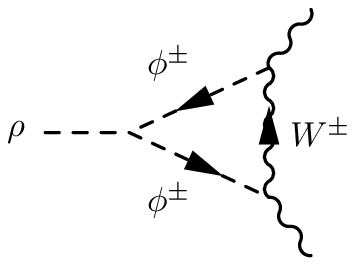}} \\
\hspace{.2cm}
\subfigure[]{\includegraphics[scale=.7]{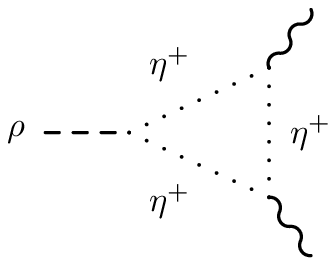}}
\hspace{.2cm}
\subfigure[]{\includegraphics[scale=.7]{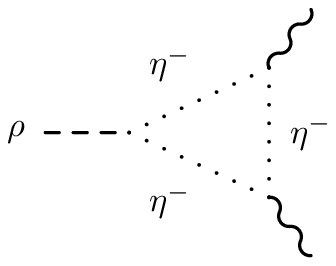}}
\hspace{.2cm}
\subfigure[]{\includegraphics[scale=.7]{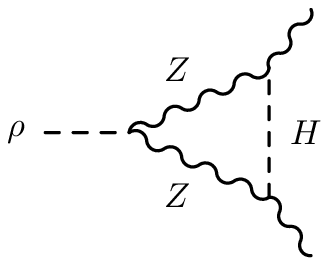}}
\hspace{.2cm}
\subfigure[]{\includegraphics[scale=.7]{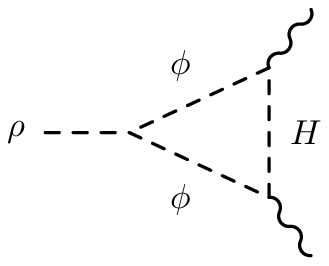}}
\hspace{.2cm}
\subfigure[]{\includegraphics[scale=.7]{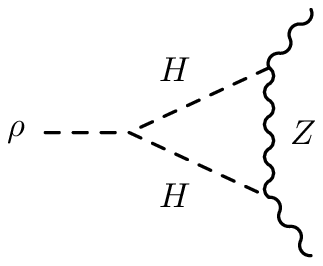}}
\hspace{.2cm}
\subfigure[]{\includegraphics[scale=.7]{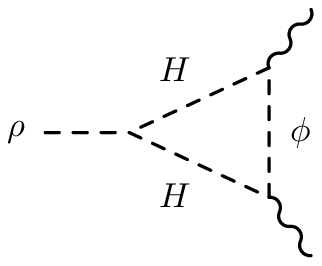}}  
\caption{Amplitudes of triangle topology contributing to the $\rho \gamma\gamma$, $\rho \gamma Z$ and $\rho ZZ$ interactions. They 
include fermion $(F)$, gauge bosons $(B)$ and contributions from the term of improvement (I). Diagrams (a)-(g) contribute to all the 
three channels while (h)-(k) only in the $\rho ZZ$ case.}
\label{PNew2figuretriangle}
\end{figure}

\begin{figure}[t]
\centering
\subfigure[]{\includegraphics[scale=.7]{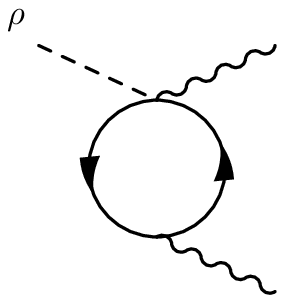}}
\hspace{.2cm}
\subfigure[]{\includegraphics[scale=.7]{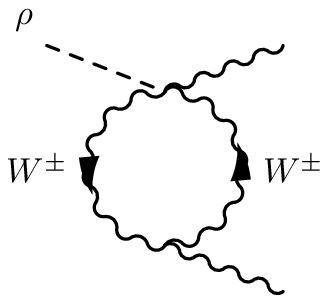}}
\hspace{.2cm}
\subfigure[]{\includegraphics[scale=.7]{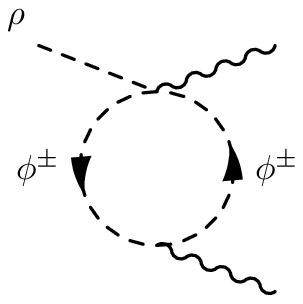}}
\hspace{.2cm}
\subfigure[]{\includegraphics[scale=.7]{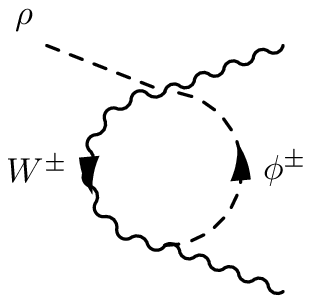}}
\hspace{.2cm}
\subfigure[]{\includegraphics[scale=.7]{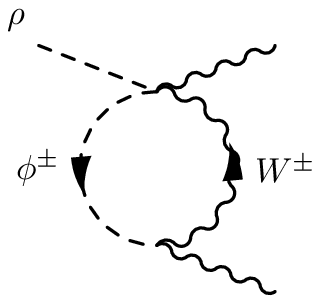}}
\hspace{.2cm}
\subfigure[]{\includegraphics[scale=.7]{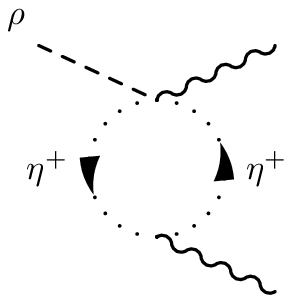}} \\
\subfigure[]{\includegraphics[scale=.7]{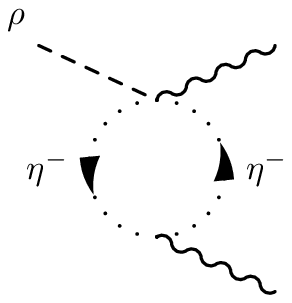}}
\hspace{.2cm}
\subfigure[]{\includegraphics[scale=.7]{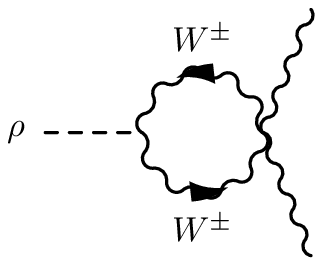}}
\hspace{.2cm}
\subfigure[]{\includegraphics[scale=.7]{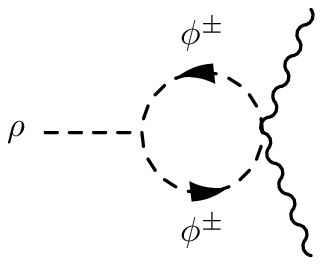}}
\hspace{.2cm}
\subfigure[]{\includegraphics[scale=.7]{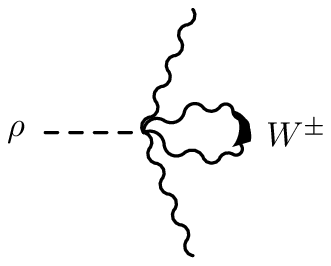}}
\hspace{.2cm}
\subfigure[]{\includegraphics[scale=.7]{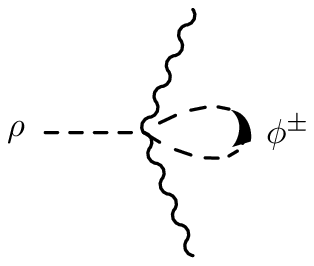}}
\hspace{.2cm}
\subfigure[]{\includegraphics[scale=.7]{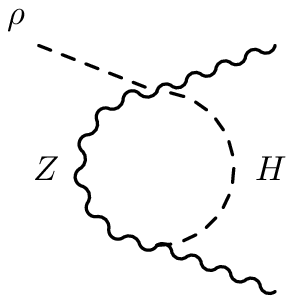}} \\
\hspace{.2cm}
\subfigure[]{\includegraphics[scale=.7]{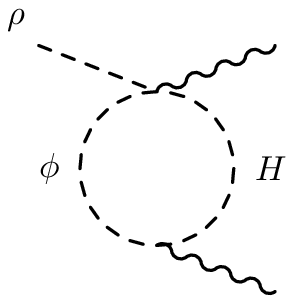}}  
\hspace{.2cm}
\subfigure[]{\includegraphics[scale=.7]{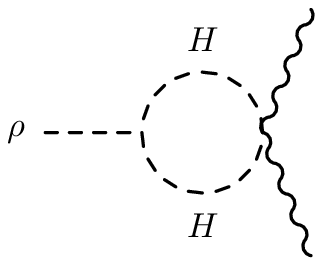}}
\hspace{.2cm}
\subfigure[]{\includegraphics[scale=.7]{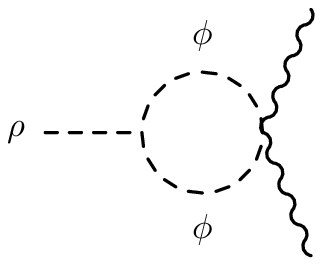}}  
\hspace{.2cm}
\subfigure[]{\includegraphics[scale=.7]{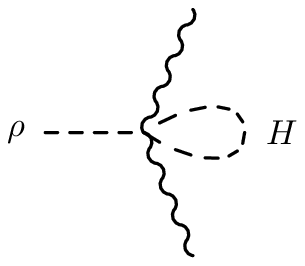}}
\hspace{.2cm}
\subfigure[]{\includegraphics[scale=.7]{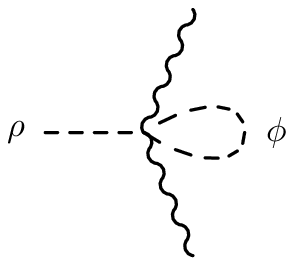}}  
\caption{Bubble and tadpole-like diagrams for $\rho \gamma\gamma$ $\rho \gamma Z $ and $\rho Z Z$. 
Amplitudes (l)-(q) contribute only in the $\rho ZZ$ channel.}
\label{PNew2figuretadpole}
\end{figure}

\begin{figure}[t]
\centering
\subfigure[]{\includegraphics[scale=.7]{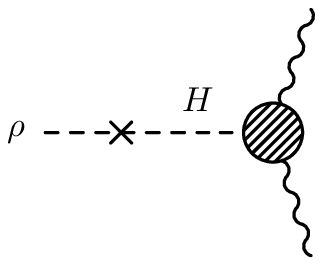}}
\hspace{.2cm}
\subfigure[]{\includegraphics[scale=.7]{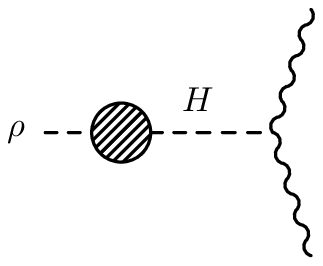}}
\hspace{.2cm}
\subfigure[]{\includegraphics[scale=.7]{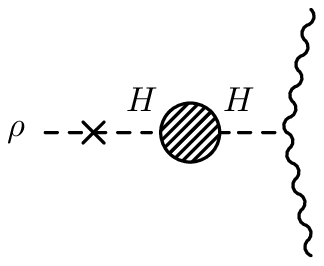}}
\caption{External leg corrections. Diagrams (b) and (c) appear only in the $\rho Z Z$ sector.}
\label{PNew2figuremix}
\end{figure}

The expression for the $\Gamma^{\alpha\beta}_{Z Z}$ vertex (Figs. \ref{PNew2figuretriangle},\ref{PNew2figuretadpole},\ref{PNew2figuremix}) defining 
the $\rho ZZ$ interaction is presented here in the kinematical limit given by $k^2 = (p+q)^2 = s$, $p^2 = q^2 = M_Z^2$ with two 
on-shell $Z$ bosons. The completely cut correlator takes contributions from a fermion sector, a $W$ gauge boson sector, a $Z-H$ 
sector together with a term of improvement.
There is also an external leg correction $\Delta^{\alpha \beta}$ on the dilaton line which is much more involved than in the previous 
cases because there are contributions coming from the minimal EMT and from the improved EMT . \\
At one loop order we have 
\bea \label{PNew21loopZZ}
&&
\Gamma_{Z Z}^{\alpha\beta}(p,q)
\equiv 
\frac{i}{\Lambda}\, \left[ \mathcal A^{\alpha\beta}(p,q) + \Sigma^{\alpha\beta}(p,q)+ \Delta^{\alpha\beta}(p,q) \right] \nn \\
&& = 
\frac{i}{\Lambda}\,
\Bigg\{
\left[\left(\frac{s}{2} - M^2_Z \right)\, \eta^{\alpha\beta} - q^\alpha\,p^\beta \right]\,
\left( \mathcal A_{ZZ} + \Phi^{\Sigma}_{ZZ}(p,q) + \Phi^{\Delta}_{ZZ}(p,q) \right)
+ \eta^{\alpha\beta}\, \left(\Xi^{\Sigma}_{ZZ}(p,q) + \Xi^{\Delta}_{ZZ}(p,q) \right) \Bigg\} ,\nn \\
\eea
where again $\Sigma$ stands for the completely cut vertex and $\Delta$ for the external leg corrections and
we have introduced for convenience the separation
\bea
\Phi^{\Sigma}_{ZZ}(p,q) &=& \Phi^{F}_{ZZ}(p,q) + \Phi^{W}_{ZZ}(p,q) + \Phi^{ZH}_{ZZ}(p,q) + \Phi^{I}_{ZZ}(p,q) \,, \nn \\
\Xi^{\Sigma}_{ZZ}(p,q) &=& \Xi^{F}_{ZZ}(p,q) + \Xi^{W}_{ZZ}(p,q) + \Xi^{ZH}_{ZZ}(p,q) + \Xi^{I}_{ZZ}(p,q)\, .
\eea
The form factors are given in \cite{Coriano:2012nm}, while here we report only the purely anomalous contribution
\beq
\mathcal A_{ZZ} = \frac{\alpha}{6 \pi c_w^2 s_w^2}\,
\left\{ -\sum_{f} \left({C_a^f}^2+{C_v^f}^2\right) + \frac{60\,s_w^6 - 148\,s_w^2 + 81}{4} - \frac{7}{4} 
+ 18\,\chi\,\left[ 1 - 2\,s_w^2\,c_w^2 \right]\right\}\, .
\eeq
Finally, we give the decay rate expression for the $\rho \to  Z Z$ process. 
At leading order it can be computed from the tree level amplitude
\beqa
\mathcal M^{\alpha\beta}(\rho \rightarrow ZZ) 
&=& 
\frac{2}{\Lambda}\,M_Z^2\, \eta^{\alpha\beta} \, , 
\eeqa
and it is given by  
\beqa
\Gamma(\rho \rightarrow ZZ) 
&=&
\frac{ m_{\rho}^3}{32\,\pi \Lambda^2} \, (1-x_Z)^{1/2}\, \left[ 1 - x_Z + \frac{3}{4}\,x_Z^2 \right].
\label{PNew2PhiZZ}
\eeqa
Including the one-loop corrections defined in Eq.(\ref{PNew21loopZZ}), one gets the decay rate at next-to-leading order
\beqa
 \Gamma(\rho\rightarrow ZZ) 
&=&
\frac{m_{\rho}^3}{32\,\pi\,\Lambda^2} \,\sqrt{1-x_Z}\, \bigg\{ 1 - x_Z + \frac{3}{4}\,x_Z^2 
+ \frac{3}{x_Z} \, \bigg[4\, Re\, \{\Phi^{\Sigma}_{ZZ}(p,q)\}(1-x_Z + \frac{3}{4}\,x_Z^2) 
\nn \\
&-& 
Re\, \{\Xi^{\Sigma}_{ZZ}(p,q)\}\,m_{\rho}^2\, \left( \frac{3}{4}\,x_Z^3 - \frac{3}{2}\,x_Z^2 \right) \bigg] \bigg\}\, .
\eeqa
%
%
%
%
%
%
%
%
\subsection{Renormalization of dilaton interactions in the broken electroweak phase}
\label{PNew2renorm}

In this section we address the renormalization properties of the correlation functions given above. Although the 
proof is quite cumbersome, one can check, from our previous results, that the 1-loop renormalization of the Standard Model 
Lagrangian is sufficient to cancel all the singularities in the cut 
vertices independently of whether the Higgs is conformally coupled or not. Concerning the uncut vertices, instead, the term of 
improvement plays a significant role in the determination of Green functions which are ultraviolet finite. In particular such a term 
has to appear with $\chi=1/6$ in order to guarantee the cancellation of a singularity present in the one-loop two-point function 
describing the Higgs dilaton mixing ($\Sigma_{\rho H}$).  
The problem arises only in the $\Gamma^{\alpha\beta}_{ZZ}$ correlator, where the $\Sigma_{\rho,H}$ two-point function is present as an 
external leg correction on the dilaton line. 

The finite parts of the counterterms are determined in the on-shell renormalization scheme which is widely used in the electroweak theory. 
From the counterterm Lagrangian we compute the corresponding counterterm to the trace of the EMT. As we have already mentioned, one can also verify from the 
explicit computation that the terms of improvement, in the conformally coupled case, are necessary to renormalize the vertices 
containing an intermediate scalar with an external bilinear mixing (dilaton/Higgs).
The counterterm vertices for the correlators with a dilaton insertion are
\beqa
\delta [\rho \gamma \gamma]^{\alpha\beta}  &=& 0
\\
\delta [\rho \gamma Z]^{\alpha\beta}  &=&  - \frac{i}{\Lambda}  \delta Z_{Z\gamma} \, M_Z^2  \, \eta^{\alpha\beta}  \, , 
\\
\delta [\rho Z Z]^{\alpha\beta}  &=&  - 2 \frac{i}{\Lambda}  (M_Z^2 \, \delta Z_{ZZ} + \delta M^2_Z)  \, \eta^{\alpha\beta} \, , 
\eeqa
where the counterterm coefficients are defined in terms of the 2-point functions of the fundamental fields as
\beqa
\delta Z_{Z \gamma} = 2 \frac{\Sigma_T^{\gamma Z}(0)}{M_Z^2} \,, \quad \delta Z_{ZZ} = - Re \frac{\partial 
\Sigma_T^{ZZ}(k^2)}{\partial k^2} \bigg |_{k^2 = M_Z^2} \,, \quad
\delta M_Z^2 = Re \, \Sigma_T^{ZZ}(M_Z^2) \,,
\eeqa
and are defined in Appendix \ref{P3propagators}.
It follows then that the $\rho \gamma \gamma$ interaction must be finite, as one can find by a direct inspection of the $\Gamma^{\alpha\beta}_{\gamma \gamma}$ vertex, while the others require the subtraction of their divergences.

These counterterms are sufficient to remove the divergences of the completely cut graphs which do 
not contain a bilinear mixing, once we set on-shell  the external gauge lines. This occurs both for those diagrams which do not 
involve the terms of improvement and for those involving $T_I$. 
Regarding those contributions which involve the bilinear mixing on the external dilaton line we encounter two different situations. \\
In the $\rho \gamma Z$ vertex the insertion of the bilinear mixing $\rho H$ generates a reducible diagram of the form Higgs/photon/Z 
whose renormalization is guaranteed, within the Standard Model, by the use of the Higgs/photon/Z counterterm
\beqa
\delta [H\gamma Z]^{\alpha\beta} = \frac{e \, M_Z}{2 s_w c_w} \delta Z_{Z\gamma}\, \eta^{\alpha\beta} \,.
\eeqa
As a last case, we discuss the contribution to $\rho ZZ$ coming from the bilinear mixing, already mentioned above. The corrections on 
the dilaton line involve the dilaton/Higgs mixing $\Sigma_{\rho H}$, the Higgs self-energy $\Sigma_{HH}$ and the term of improvement 
$\Delta^{\alpha\beta}_{I\,,HZZ}$, which introduces the Higgs/Z/Z vertex (or $HZZ$) of the Standard Model. 
The Higgs self-energy and the $HZZ$ vertex, in the Standard Model, are renormalized with the usual counterterms
\beqa
\delta [HH](k^2) 
&=& 
(\delta Z_H \, k^2 - M_H^2 \delta Z_H - \delta M_H^2) \, , 
\\
\delta [HZZ]^{\alpha\beta} 
&=& 
\frac{e \, M_Z}{s_w \, c_w} \bigg[ 1 + \delta Z_e + \frac{2 s_w^2 
- c_w^2}{c_w^2} \frac{\delta s_w}{s_w} + \frac{1}{2} \frac{\delta M_W^2}{M_W^2} + \frac{1}{2} \delta Z_H + \delta Z_{ZZ}  \bigg] 
\, \eta^{\alpha\beta} \,,
\eeqa
where
\bea
&& 
\delta Z_H = 
- Re \frac{\partial \Sigma_{HH}(k^2)}{\partial k^2} \bigg|_{k^2=M_H^2}\, ,\quad \delta M_H^2 = Re \Sigma_{HH}(M_H^2) \, , \quad
\delta Z_e = - \frac{1}{2} \delta Z_{\gamma \gamma} + \frac{s_w}{2 c_w} \delta Z_{Z \gamma} \,, \nn \\
&& \delta s_w = - \frac{c_w^2}{2 s_w} \left( \frac{\delta M_W^2}{M_W^2} - \frac{\delta M_Z^2}{M_Z^2} \right) \,, \quad \delta M_W^2 = 
Re \Sigma^{WW}_T(M_W^2) \,, \quad 
\delta Z_{\gamma \gamma} = - \frac{\partial \Sigma_T^{\gamma \gamma}(k^2)}{\partial k^2} \bigg|_{k^2 = 0} \,.
\eea
The self-energy $\Sigma_{\rho H}$ is defined by the minimal contribution generated by ${{T_{Min}}^\mu}_\mu$ 
and by a second term derived from ${{T_I}^\mu}_\mu$.
This second term, with the conformal coupling $\chi = \frac{1}{6}$, is necessary in order to ensure the renormalizability 
of the dilaton/Higgs mixing.
In fact, the use of the minimal EMT in the computation of this self-energy involves a divergence of the form
\beqa
\delta [\rho H]_{Min} = - 4 \frac{i}{\Lambda}\, \delta t \, , \label{PNew2CThH}
\eeqa
with $\delta t$ fixed by the condition of cancellation of the Higgs tadpole $T_{ad}$ ($\delta t + T_{ad} = 0$).
A simple analysis of the divergences in $\Sigma_{Min, \, \rho H}$ shows that the counterterm given in Eq. (\ref{PNew2CThH}) is not 
sufficient to remove all the singularities of this correlator unless we also include the renormalization originating from the term of improvement 
which is given by
\beqa
\delta [\rho H]_{I}(k) = - \frac{i}{\Lambda} 6 \, \chi \, v \bigg[ \delta v + \frac{1}{2} \delta Z_H \bigg]  k^2\,, \qquad 
\qquad \text{with} \quad \chi = \frac{1}{6} \,,
\eeqa
and
\bea
\delta v = v \bigg( \frac{1}{2} \frac{\delta M_W^2}{M_W^2} + \frac{\delta s_w}{s_w} - \delta Z_e \bigg) \,.
\eea
One can show explicitly that this counterterm indeed ensures the finiteness of $\Sigma_{\rho H}$.



\begin{figure}[t]
\centering
\subfigure[]{\includegraphics[scale=.7]{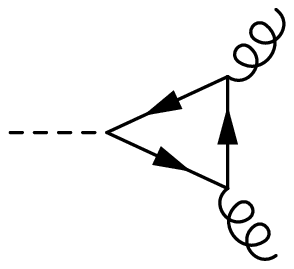}\label{PNew2QCD_NLOa}}
\hspace{.5cm}
\subfigure[]{\includegraphics[scale=.7]{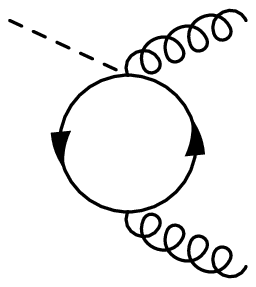}\label{PNew2QCD_NLOb}}
\hspace{.5cm}
\subfigure[]{\includegraphics[scale=.7]{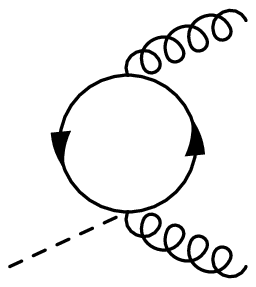}\label{PNew2QCD_NLOc}}
\hspace{.5cm}
\subfigure[]{\includegraphics[scale=.7]{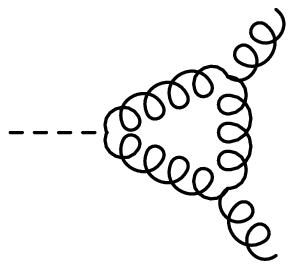}\label{PNew2QCD_NLOd}}
\hspace{.5cm}
\subfigure[]{\includegraphics[scale=.7]{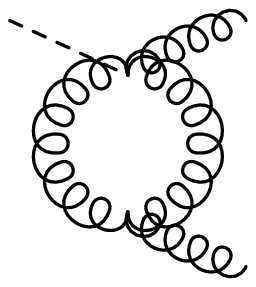}\label{PNew2QCD_NLOe}}
\hspace{.5cm}
\subfigure[]{\includegraphics[scale=.7]{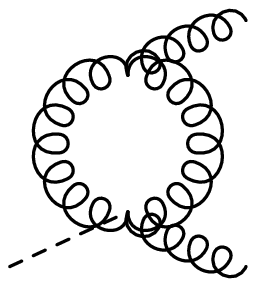}\label{PNew2QCD_NLOf}}\\
\hspace{.5cm}
\subfigure[]{\includegraphics[scale=.7]{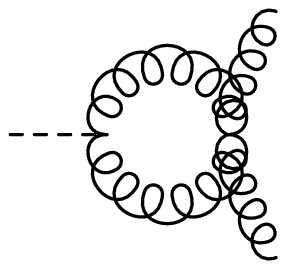}\label{PNew2QCD_NLOg}}
\hspace{.5cm}
\subfigure[]{\includegraphics[scale=.7]{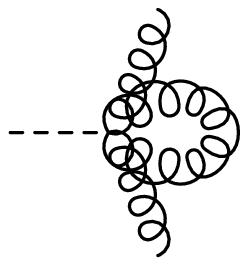}\label{PNew2QCD_NLOh}}
\hspace{.5cm}
\subfigure[]{\includegraphics[scale=.7]{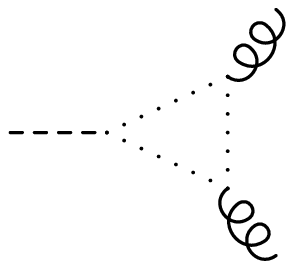}\label{PNew2QCD_NLOi}}
\hspace{.5cm}
\subfigure[]{\includegraphics[scale=.7]{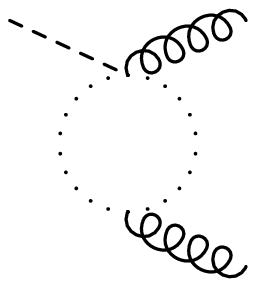}\label{PNew2QCD_NLOl}}
\hspace{.5cm}
\subfigure[]{\includegraphics[scale=.7]{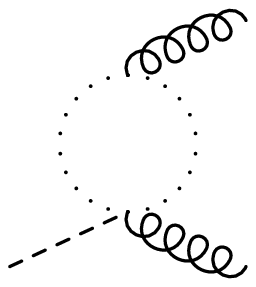}\label{PNew2QCD_NLOm}}
\caption{QCD vertices at next-to-leading order. In the on-shell gluon case only diagram (a)contributes.}
\label{PNew2QCD_NLO}
\end{figure}
%

\section{The off-shell dilaton-gluon-gluon vertex in QCD}

After a discussion of the leading corrections to the vertices involving one dilaton and two electroweak currents we investigate the 
interaction of a dilaton and two gluons beyond leading order, giving the expression of the full off-shell vertex. The corresponding 
interaction with two on-shell gluons has been computed in \cite{Giudice:2000av} and is simply given by the contribution of the 
anomaly and of the quark loop. 

We show in Fig. \ref{PNew2QCD_NLO} a list of the NLO QCD contributions to dilaton interactions. As we have just mentioned, in the 
two-gluon on-shell case one can show by an explicit computation that each of these contributions vanishes, except for diagram $(a)$, 
which is nonzero when a massive fermion runs in the loop. For this specific reason, in the parton model, the production of the 
dilaton in $pp$ collisions at the LHC is mediated by the diagram of gluon fusion, which involves a top quark in a loop.


We find convenient to express the result of the off-shell $\Gamma^{\alpha\beta}_{g g}$ vertex in the form 
\beqa \label{PNew2OffShellQCD}
\Gamma_{g g}^{\alpha \beta}(p,q) = 
\frac{i}{\Lambda}\, \bigg\{ A^{00}(p,q) \eta^{\alpha \beta} + A^{11}(p,q) p^{\alpha} p^{\beta} + A^{22}(p,q) q^{\alpha} q^{\beta} 
+ A^{12}(p,q) p^{\alpha}q^{\beta} + A^{21}(p,q) q^{\alpha}p^{\beta}\bigg\} \, ,
\eeqa
where $A^{ij}(p,q) = A^{ij}_g(p,q) + A^{ij}_q(p,q)$ which are diagonal ($\propto \delta^{a b}$) in color space.\\
After an explicit computation, we find
\beqa
A^{00}_g(p,q) 
&=& 
-\delta_{ab}\,\frac{g^2 \, N_C}{16 \pi^2} \bigg\{   
2 \left( p^2 +  q^2 + \frac{11}{3} p \cdot q \right) 
+ (p^2-q^2) \bigg[ \mathcal B_0(p^2,0,0) - \mathcal B_0(q^2,0,0)\bigg] 
\nonumber \\
&& 
+ \left(p^4 + q^4 - 2(p^2 + q^2) p \cdot q -6 p^2 q^2 \right) \mathcal C_0((p+q)^2, p^2, q^2,0,0,0)
\bigg\} \, , 
\nonumber 
\eeqa
\beqa
A^{11}_g(p,q) = A^{22}_g(q,p)
&=&   
\delta_{ab} \, \frac{g^2 \, N_C \,}{16\,\pi^2 }  \, 
\bigg\{ 2 + \frac{1}{p \cdot q^2 - p^2 \, q^2} \, 
\bigg[ (p+q)^2 \, p \cdot q \, \mathcal B_0((p+q)^2,0,0)
\nonumber \\
&& 
- p^2 \, (q^2 + p \cdot q) \, \mathcal B_0(p^2,0,0)  - (2 p \cdot q^2 - p^2 \, q^2 + p \cdot q \, q^2) \mathcal B_0(q^2,0,0) 
\nonumber \\
&& 
+ \left( p^2 \, q^2 ( 5 q^2 - p^2) + 2 p \cdot q^2 (p^2 + p \cdot q - 2 \, q^2 ) \right) \, 
\mathcal C_0((p+q)^2, p^2, q^2,0,0,0) \bigg] \bigg\}  \, , 
\nonumber 
\eeqa
\beqa
A^{12}_g(p,q) 
&=& 
\delta_{ab}\, \frac{g^2 \, N_C}{4 \pi^2} \, p \cdot q \, \mathcal C_0((p+q)^2, p^2, q^2,0,0,0)\, ,
\nonumber 
\eeqa
\beqa
A^{21}_g(p,q) 
&=& 
\delta_{ab}\, \frac{g^2 \, N_C}{24\,\pi^2 }  \, 
\bigg\{11 + \frac{3}{2}\,\frac{1}{p \cdot q^2 - p^2 \, q^2} (p^2 + q^2) \, \bigg[(p^2 + p \cdot q)\mathcal B_0(p^2,0,0) 
\nonumber \\
&& 
+ (q^2 + p \cdot q) \mathcal B_0(q^2,0,0) - (p+q)^2  \mathcal B_0((p+q)^2,0,0)  
\nonumber \\
&& 
- \left( p \cdot q (p^2 + 4 p \cdot q + q^2) - 2 \, p^2 \, q^2 \right) \mathcal C_0((p+q)^2, p^2, q^2,0,0,0) \bigg]\bigg\} \, ,
\nonumber 
\eeqa
\bea
A^{00}_q(p,q) &=&
\delta_{ab} \frac{g^2}{8 \pi^2} \sum_{i=1}^{n_f} \bigg\{ \frac{2}{3} p\cdot q - 2\, m_i^2 
+ \frac{m_i^2}{p \cdot q^2 - p^2 q^2} \bigg[
p^2 \left(p \cdot q +q^2\right)  \mathcal B_0(p^2,m_i^2,m_i^2 ) \nn \\
&+&
q^2 \left(p^2+p \cdot q \right) \mathcal B_0(q^2,m_i^2,m_i^2)  -
\left(p^2 \left(p \cdot q+2 q^2\right) + p \cdot q \, q^2\right) \mathcal B_0( (p+q)^2,m_i^2,m_i^2) \nn \\
&-&
\left(p^2 q^2 \left(p^2+q^2 -4 m_i^2 \right)+4 m_i^2 \, p \cdot q^2 +4 p^2 \, q^2 \, p \cdot q -2 \, p \cdot q^3\right)
  \mathcal C_0 ((p+q)^2,p^2,q^2,m_i^2,m_i^2,m_i^2) 
\bigg] \bigg\} , \nn
\eea
%
\bea
A^{11}_q(p,q) &=& A^{22}_q(q,p) =
\delta_{ab} \frac{g^2}{ 8\,\pi^2} \sum_{i=1}^{n_f} \frac{2 \, m_i^2 \, q^2}{ p \cdot q^2 - p^2 q^2} \bigg\{ -2 + \frac{1}{p \cdot q^2 - p^2 q^2} \bigg[
\left(q^2 \left(p^2+3 \, p \cdot q \right) \right. \nn \\
&+& \left. 2 \, p \cdot q^2\right) \mathcal B_0(q^2,m_i^2,m_i^2)
+ \left(p^2 \left(3 \, p \cdot q+q^2\right)+2 p \cdot q^2\right)  \mathcal B_0(p^2,m_i^2,m_i^2) \nn \\
&-& \left(p^2 \left(3 \, p \cdot q+2 q^2\right) + p \cdot q \left(4 \, p \cdot q+3 q^2\right)\right) \mathcal B_0((p+q)^2,m_i^2,m_i^2) \nn \\
&-& \left(2 \, p \cdot q^2 \left(2 m_i^2+p^2+q^2\right) + p^2 q^2 \left(p^2+q^2-4 m_i^2 \right)  \right.) \nn \\
&+& \left. 4 p^2 q^2 \, p \cdot q +2 \, p \cdot q ^3\right)
   \mathcal C_0((p+q)^2,p^2,q^2,m_i^2,m_i^2,m_i^2)
\bigg] \bigg\} , \nn
\eea
\bea
A^{12}_q(p,q) &=&
\delta_{ab} \frac{g^2}{ 8\,\pi^2} \sum_{i=1}^{n_f} \frac{2 \, m_i^2 \, p \cdot q^2}{ p \cdot q^2 - p^2 q^2} \bigg\{ 2 + \frac{1}{p \cdot q^2 - p^2 q^2} \bigg[
-\left(q^2 \left(p^2+3 \, p \cdot q \right)+ 2 \, p \cdot q^2\right) \mathcal B_0(q^2,m_i^2,m_i^2) \nn \\
&-&
\left(p^2 \left(3 \, p \cdot q +q^2\right)+ 2 \, p \cdot q^2\right) \mathcal B_0(p^2,m_i^2,m_i^2) +
\left(p^2 \left(3 \, p \cdot q +2 q^2\right)+p \cdot q \left(4 \, p \cdot q +3 q^2 \right)\right) \nn \\
&\times& \mathcal B_0( (p+q)^2,m_i^2,m_i^2)+
\left(2 \, p \cdot q^2 \left(2 m_i^2+p^2+q^2\right)+p^2 q^2 \left(p^2+q^2 -4 m_i^2 \right) \right. \nn \\
&+& \left.  4 \, p^2 \, q^2 \, p \cdot q +2\, p \cdot q^3\right)
 \mathcal C_0((p+q)^2,p^2,q^2,m_i^2,m_i^2,m_i^2)
\bigg] \bigg\} , \nn 
\eea
\bea
A^{21}_q(p,q)  &=& \delta_{ab} \frac{g^2}{ 8\,\pi^2} \sum_{i=1}^{n_f} \bigg\{ - \frac{2}{3} + \frac{2 \, m_i^2 \, p \cdot q}{p \cdot q - p^2 q^2} + \frac{m_i^2}{(p \cdot q -p^2 q^2)^2} \bigg[
- p^2 \left(q^2 \left(2 p^2+3 \, p \cdot q \right)+p \cdot q ^2\right)  \nn \\ 
&\times& \mathcal B_0(p^2, m_i^2,m_i^2) 
- q^2 \left(p^2 \left(3 \, p \cdot q+2 q^2\right)+p \cdot q^2\right) \mathcal B_0( q^2, m_i^2,m_i^2) 
+ \left(2 p^4 q^2+p^2 \left(6 \, p \cdot q \, q^2 \right. \right. \nn \\ 
&+& \left. \left. p \cdot q^2+2 q^4\right)+p \cdot q^2 q^2\right) \mathcal B_0( (p+q)^2,m_i^2,m_i^2)
+ p \cdot q \left(p^2 q^2 \left(3 p^2+3 q^2 - 4 m_i^2\right) \right. \nn \\
&+& \left. 4 \,   m_i^2 \, p \cdot q^2
 +8 p^2 \, q^2 \, p \cdot q -2 \, p \cdot q ^3\right)
   \mathcal C_0( (p+q)^2,p^2,q^2,m_i^2,m_i^2,m_i^2)
\bigg] \bigg\}, 
\eea
where $N_C$ is the number the color, $n_f$ is the number of flavor and $m_i$ the mass of the quark.
In the on-shell gluon case, Eq.(\ref{PNew2OffShellQCD}) reproduces the same interaction responsible for Higgs production at LHC augmented by an anomaly term.
This is given by
\beqa \label{PNew2OnShellQCD}
\Gamma^{\alpha\beta}_{gg}(p,q) =  
\frac{i}{\Lambda} \Phi(s) \, u^{\alpha\beta}(p,q) \, ,
\eeqa
with $u^{\alpha\beta}(p,q)$ defined in Eq.(\ref{PNew2utensor}), and with the gluon/quark contributions included in the $\Phi(s)$ form factor ($s = k^2 = (p+q)^2$) 
\beqa \label{PNew2OnShellPhi}
\Phi(s) 
= 
- \delta^{ab} \frac{g^2}{24\,\pi^2} \, \bigg\{
 \, \left(11\, N_C - 2\, n_f \right) + 12 \, \sum_{i=1}^{n_f} m_i^2 \, 
\bigg[ \frac{1}{s} \, - \, \frac{1} {2 }\mathcal C_0 (s, 0, 0, m_i^2, m_i^2, m_i^2) \bigg(1-\frac{4 m_i^2}{ s}\bigg) \bigg]
\bigg\} \, ,
\eeqa
where the first mass independent terms represent the contribution of the anomaly, while the others are the explicit mass corrections. \\ 
The decay rate of a dilaton in two gluons can be evaluated from the on-shell limit in Eq.(\ref{PNew2OnShellQCD}) and it is given by
\beqa
\Gamma(\rho \rightarrow gg) 
&=&
\frac{\alpha_s^2\,m_\rho^3}{32\,\pi^3 \Lambda^2} \, \bigg| \beta_{QCD} + x_t\left[1 + (1-x_t)\,f(x_t) \right] \bigg|^2 \,,
\label{PNew2Phigg}
\eeqa
where we have taken the top quark as the only massive fermion and $x_i$ and $f(x_i)$ are defined in Eq. (\ref{PNew2x}) and Eq. (\ref{PNew2fx}) respectively. 
Moreover we have set
$\beta_{QCD} = 11 N_C/3 - 2 \, n_f/3$ for the QCD $\beta$ function.

\section{Non-gravitational dilatons from scale invariant extensions of the Standard Model}
\label{PNew2NonGrav} 

As we have pointed out in the introduction, a dilaton may appear in the spectrum of different extensions of the Standard Model not 
only as a result of the compactification of extra spacetime dimensions, but also as an effective state, related to the breaking of a 
dilatation symmetry. In this respect, notice that in its actual formulation 
the Standard Model is not scale invariant, but can be such, at classical level, if we slightly modify the scalar potential with the 
introduction of a dynamical field $\Sigma$ that 
allows to restore this symmetry and acquires a vacuum expectation value. This task is accomplished by the replacement of every 
dimensionfull parameter $m$ according to $m \rightarrow m \frac{\Sigma}{\Lambda}$, where $\Lambda$ is the classical conformal 
breaking scale. 
In the case of the Standard Model, classical scale invariance can be easily accomodated with a simple change of the scalar potential. 

This is defined, obviously, modulo a constant, therefore we may consider, for instance, two equivalent choices 
\beqa
V_1(H, H^\dagger)&=& - \mu^2 H^\dagger H +\lambda(H^\dagger H)^2 =
\lambda \left( H^\dagger H - \frac{\mu^2}{2\lambda}\right)^2 - \frac{\mu^4}{4 \lambda}\nonumber \\
V_2(H,H^\dagger)&=&\lambda \left( H^\dagger H - \frac{\mu^2}{2\lambda}\right)^2
\eeqa
which gives two {\em different} scale invariant extensions 
\beqa
V_1(H,H^\dagger, \Sigma)&=&- \frac{\mu^2\Sigma^2}{\Lambda^2} H^\dagger H +\lambda(H^\dagger H)^2 \nonumber \\
V_2(H,H^\dagger, \Sigma)&=& \lambda \left( H^\dagger H - \frac{\mu^2\Sigma^2}{2\lambda \Lambda^2}\right)^2 \,,
\eeqa 
where $H$ is the Higgs doublet, $\lambda$ is its dimensionless coupling constant, while $\mu$ has the dimension of a mass and, 
therefore, is the only term involved in the scale invariant extension. More details of this analysis can be found in the following section.\\
The invariance of the potential under the addition of constant terms, typical of any Lagrangian, is lifted once we 
require the presence of a dilatation symmetry. Only the second choice $(V_2)$ guarantees the existence of a stable ground state 
characterized by a spontaneously 
broken phase. In $V_2$ we have parameterized the Higgs, as usual, around the electroweak vev $v$ as in Eq.(\ref{P3Higgsparam}), 
we have indicated with $\Lambda$ the vev of the dilaton field $\Sigma = \Lambda + \rho$, 
and we have set $\phi^+ = \phi = 0$ in the unitary gauge. \\
The potential $V_2$ has a massless mode due to the existence of a flat direction. 
Performing a diagonalization of the mass matrix we define the two mass eigenstates $\rho_0$ and $h_0$, which are given by 
\beq
 \left( \begin{array}{c}
 {\rho_0}\\
  h_0 \\
  \end{array} \right)
 =\left( \begin{array}{cc}
\cos\alpha & \sin\alpha \\
-\sin\alpha & \cos\alpha  \\
 \end{array} \right)
 \left( \begin{array}{c}
  \rho\\
 {h} \\
  \end{array} \right)
\eeq
with 
\beq
\cos\alpha=\frac{1}{\sqrt{1 + v^2/\Lambda^2}}\qquad \qquad  \sin\alpha=\frac{1}{\sqrt{1 + \Lambda^2/v^2}}.
\eeq
We denote with ${\rho_0}$ the massless dilaton generated by this potential, while 
$h_0$ will describe a massive scalar, interpreted as a new Higgs field, whose mass is given by  
\beq 
m_{h_0}^2= 2\lambda v^2 \left( 1 +\frac{v^2}{\Lambda^2}\right) \qquad \textrm{with} \qquad v^2=\frac{\mu^2}{\lambda},
\eeq
and with $m_h^2=2 \lambda v^2$ being the mass of the Standard Model Higgs.
The Higgs mass, in  this case, is corrected by the new scale of the spontaneous breaking of the dilatation symmetry ($\Lambda$), 
which remains a free parameter. 
 
The vacuum degeneracy of the scale invariant model  can be lifted by the introduction of 
extra (explicit breaking) terms which give a small mass to the dilaton field $\Sigma$.
To remove such degeneracy, one can introduce, for instance, the term
\beq
\mathcal{L}_{break} 
= \frac{1}{2} m_{\rho}^2 {\rho}^2 + \frac{1}{3!}\, {m_{\rho}^2} \frac{{\rho}^3}{\Lambda} + \dots \, ,
\eeq
where $m_{\rho}$ represents the dilaton mass.

It is clear that in this approach the coupling of the dilaton to the anomaly has to be added by hand.
The obvious question to address, at this point, is if one can identify in the effective action of the Standard Model 
an effective state which may interpolate between the dilatation current of the same model and the final state with two
neutral currents, for example with two photons. The role of the following sections will be to show 
rigorously that such a state can be identified in ordinary perturbation theory in the form of an anomaly pole.

We will interpret this scalar exchange as a composite state whose interactions with the rest of the Standard Model are 
defined by the conditions of scale and gauge invariance. In this respect, the Standard Model Lagrangian, enlarged by the introduction 
of a potential of the form $V_2(H,H^\dagger,\Sigma)$, which is expected to capture the dynamics of this pseudo-Goldstone mode, could
take the role of a workable model useful for a phenomenological analysis.  
We will show rigorously that this state couples to the conformal anomaly by a direct analysis of the $J_DVV$ correlator, 
in the form of an anomaly pole, with $J_D$ and $V$ being the dilatation and a vector current respectively.
Usual polology arguments support the fact that a pole  in a correlation function is there to indicate that a specific state can be created by 
a field operator in the Lagrangian of the theory, or, alternatively, as a composite particle of the same elementary fields.  

Obviously, a perturbative hint of the existence of such intermediate state does not correspond to a complete 
description of the state, in the same way as the discovery of an anomaly pole in the $AVV$ correlator of QCD (with $A$ being the 
axial current) is not equivalent to a proof of the existence of the pion. Nevertheless, massless poles extracted from the perturbative effective action do not appear for no reasons, 
and their infrared couplings should trigger further phenomenological interest.

\subsection{A classical scale invariant Lagrangian with a dilaton field}
\label{PNew2classical}

In this section we give an example in order to describe the construction of a scale invariant theory and to clarify some of the issues
concerning the coupling of a dilaton. In particular, the example has the goal to 
illustrate that in a classical scale invariant extension of a given theory, the dilaton couples only to operators which are mass dependent, 
and thus scale breaking, before the extension. We take the case of a fundamental dilaton field (not a composite) introduced 
in this type of extensions.

A scale invariant extension of a given Lagrangian can be obtained if we promote all the dimensionfull constants to dynamical 
fields. 
We illustrate this point in the case of a simple interacting scalar field theory incorporating the Higgs mechanism. 
At a second stage we will derive the structure of the dilaton interaction at order $1/\Lambda$, where 
$\Lambda$ is the scale characterizing the spontaneous breaking of the dilatation symmetry.

Our toy model consists in a real singlet scalar with a potential of the kind of $V_2(\phi)$ introduced in the previous section,
\beq
\label{PNew2original}
\mathcal L = \frac{1}{2} (\partial \phi)^2 - V_2(\phi) =
\frac{1}{2} (\partial \phi)^2 + \frac{\mu^2}{2}\, \phi^2 - \lambda\, \frac{\phi^4}{4} - \frac{\mu^4}{4\,\lambda}\, ,
\eeq
obeying the classical equation of motion
\beq \label{PNew2scalarEOM}
\square \phi = \mu^2\,\phi - \lambda\, \phi^3\, .
\eeq
Obviously this theory is not scale invariant due to the appearance of the mass term $\mu$. This feature is reflected in the trace of the EMT.
Indeed the canonical EMT of such a theory and its trace are
\bea
T^{\mu\nu}_{c}(\phi) 
&=& 
\partial^\mu \phi\, \partial^\nu \phi 
- \frac{1}{2}\,\eta^{\mu\nu} \bigg[ (\partial \phi)^2 + \mu^2 \,\phi^2 
-  \lambda\,\frac{\phi^4}{2} -  \frac{\mu^4}{2\,\lambda} \bigg] \, ,
\nn \\
{T_{c}^\mu}_\mu(\phi) &=& 
- (\partial\phi)^2 - 2\, \mu^2 \,\phi^2 +\lambda\, \phi^4 + \frac{\mu^4}{\lambda} \, .
\eea
It is well known that the EMT of a scalar field can be improved in such a way as to make its trace proportional only to the scale breaking parameter,
i.e. the mass $\mu$. This can be done by adding an extra contribution $T_I^{\mu\nu}(\phi, \chi)$ which is symmetric and conserved
\beq
T_I^{\mu\nu}(\phi,\chi)=\chi \left(\eta^{\mu\nu} \square \phi^2 - \partial^\mu \partial^\nu \phi^2 \right) \,,
\eeq
where the $\chi$ parameter is conveniently choosen.
The combination of the canonical plus the improvement EMT, 
$T^{\mu\nu} \equiv T_c^{\mu\nu} + T_I^{\mu\nu}$ has the off-shell trace
\beq
{T^\mu}_\mu(\phi,\chi)= 
(\partial\phi)^2\, \left( 6 \chi - 1 \right) - 2\, \mu^2\, \phi^2 
+ \lambda\, \phi^4 + \frac{\mu^4}{\lambda} + 6 \chi \phi\, \square \phi\, .
\eeq
Using the equation of motion (\ref{PNew2scalarEOM}) and chosing $\chi=1/6$ the trace relation given above 
becomes proportional uniquely to the scale breaking term $\mu$  
\beq \label{PNew2ImprovedTrace}
{T^\mu}_\mu(\phi,1/6) = - \mu^2 \phi^2 + \frac{\mu^4}{\lambda} \, .
\eeq
The scale invariant extension of the Lagrangian given in Eq.(\ref{PNew2original}) is achieved by promoting the mass terms to dynamical fields by the replacement 
\beq
\label{PNew2rep}
\mu \to \frac{\mu}{\Lambda} \, \Sigma,
\eeq
obtaining
\beq
\label{PNew2sigmaphi}
\mathcal L = 
\frac{1}{2}\, (\partial \phi)^2 +\frac{1}{2} (\partial \Sigma)^2 
+  \frac{ \mu^2}{2\,\Lambda^2}\, \Sigma^2\, \phi^2 - \lambda \frac{\phi^4}{4}
-  \frac{\mu^4}{4\,\lambda \, \Lambda^4}\, \Sigma^4
\eeq
where we have used Eq.(\ref{PNew2rep}) and introduced a kinetic term for the dilaton $\Sigma$. 
Obviously, the new Lagrangian is dilatation invariant, as one can see from the trace of the improved EMT 
\beq
{T^{\mu}}_{\mu}(\phi,\Sigma,\chi,\chi') = \left( 6\, \chi - 1 \right)\, (\partial\phi)^2
+ \left( 6 \chi^\prime -1\right)\, (\partial\Sigma)^2 
+ 6 \chi\, \phi\, \square \phi + 6 \chi^\prime\, \Sigma\, \square \Sigma 
- 2\, \frac{\mu^2}{\Lambda^2}\, \Sigma^2\, \phi^2 + \lambda\, \phi^4
+ \frac{1}{\lambda}\,\frac{\mu^4}{\Lambda^4}\,\Sigma^4 \,,
\eeq
which vanishes upon using the equations of motion for the $\Sigma$ and $\phi$ fields,
\bea
\square \phi &=& \frac{\mu^2}{\Lambda^2}\, \Sigma^2\, \phi -  \lambda\, \phi^3\, ,
\nn \\
\square \Sigma &=& \frac{\mu^2}{\Lambda^2}\, \Sigma\, \phi^2 - \frac{1}{\lambda}\, \frac{\mu^4}{\Lambda^4}\,\Sigma^3  \, ,
\eea
and setting the $\chi, \chi'$ parameters at the special value $\chi=\chi^\prime=1/6$. 

As we have already discussed in section \ref{PNew2NonGrav}, the scalar potential $V_2$ allows to perform 
the spontaneous breaking of the scale symmetry around a stable minimum point, 
giving the dilaton and the scalar field the vacuum expectation values $\Lambda$ and $v$ respectively
\beq
\Sigma  =  \Lambda + \rho \, , \quad \phi = v + h\, .
\eeq
For our present purposes, it is enough to expand the Lagrangian (\ref{PNew2sigmaphi}) around the vev for the dilaton field, 
as we are interested in the structure of the couplings of its fluctuation $\rho$
\beq \label{PNew2Manifest}
\mathcal L = \frac{1}{2}\, (\partial\phi)^2 + \frac{1}{2}\, (\partial\rho)^2 + \frac{\mu^2}{2}\, \phi^2 
- \lambda \, \frac{\phi^4}{4} - \frac{\mu^4}{4\,\lambda}
- \frac{\rho}{\Lambda}\,\left(- \mu^2\, \phi^2  + \frac{\mu^4}{\lambda}\right) + \dots\, ,
\eeq
where the ellipsis refer to terms that are higher order in $1/\Lambda$.
It is clear, from (\ref{PNew2ImprovedTrace}) and (\ref{PNew2Manifest}), that one can write an dilaton Lagrangian
at order $1/\Lambda$, as
\beq \label{PNew2RhoInteraction}
\mathcal L_{\rho} = (\partial\rho)^2 - \frac{\rho}{\Lambda}\, {T^{\mu}}_{\mu}(\phi,1/6) + \dots\, ,
\eeq
where the equations of motion have been used in the trace of the energy momentum tensor.
Expanding the scalar field around $v$ would render the previous equation more complicated and we omit it for definiteness.
We only have to mention that a mixing term $\sim \rho\, h$ shows up and it has to be removed diagonalizing 
the mass matrix, switching from interaction to mass eigenstates exactly in the way we discussed in the previous section. 

It is clear, from this simple analysis, that a dilaton, in general, does not couple to the anomaly,
but only to the sources of explicit breaking of scale invariance, i.e. to the mass terms.
The coupling of a dilaton to an anomaly is, on the other hand, necessary, 
if the state is interpreted as a composite pseudo Nambu-Goldstone mode of the dilatation symmetry.
Thus, this coupling has to be introduced by hand, in strict analogy with the chiral case.

\subsection{The $J_DVV$ and $TVV$ vertices}

This effective degree of freedom emerges both from the spectral analysis of the $TVV$ \cite{Giannotti:2008cv, Armillis:2009pq}, as we have illustrated in the previous chapters, and, 
as we are now going to show, of the  $J_D VV$ 
correlators, being the two vertices closely related. 
We recall that the dilatation current can be defined as 
\beq
J_D^\mu(z)= z_\nu T^{\mu \nu}(z) \qquad \textrm{with}  \qquad \partial\cdot J_D = {T^\mu}_\mu. 
\label{PNew2def}
\eeq
The $T^{\mu\nu}$ has to be symmetric and on-shell traceless for a classical scale invariant theory, while includes, at
quantum level, the contribution from the trace anomaly together with the additional terms describing the explicit breaking of the 
dilatation symmetry. 
The separation between the anomalous and the explicit contributions to the breaking of dilatation symmetry is present 
in all the analysis that we have performed on the $TVV$ vertex in dimensional regularization. 
In this respect, the analogy between these types of correlators and the $AVV$ diagram of the chiral anomaly goes 
quite far, since in the $AVV$ case such a separation has been shown to hold in the Longitudinal/Transverse 
(L/T) solution of the anomalous Ward identities \cite{Knecht:2003xy, Jegerlehner:2005fs, Armillis:2009sm}. 
This has been verified in perturbation theory in the same scheme.

We recall that the $U(1)_A$ current is characterized by an anomaly pole which describes the interaction between the 
Nambu-Goldstone mode, generated by the breaking of the chiral symmetry, and the gauge currents. 
In momentum space this corresponds to the nonlocal vertex 
\beq
\label{PNew2AVVpole}
V_{\textrm{anom}}^{\lambda \mu\nu}(k,p,q)=  \frac{k^\lambda}{k^2}\epsilon^{\mu \nu \alpha \beta}p_\alpha q_\beta +...
\eeq
with $k$ being the momentum of the axial-vector current and $p$ and $q$ the momenta of the two photons.
In the equation above, the ellipsis refer to terms which are suppressed at large energy. 
In this regime, this allows to distinguish the operator accounting for the chiral anomaly (i.e. $\square^{-1}$ in coordinate space)
from the contributions due to mass corrections. 
Polology arguments can be used to relate the appearance of such a pole to the pion state 
around the scale of chiral symmetry breaking. 

To identify the corresponding pole in the dilatation current of the $J_D VV$ correlator at zero momentum transfer, one can follow the 
analysis of \cite{Horejsi:1997yn}, where it is shown that the appearance of the trace anomaly is related to the presence of a
superconvergent sum rule in the spectral density of this correlator. At nonzero momentum transfer the derivation of a similar
behaviour can be obtained by an explicit computation of the spectral density of the 
$TVV$ vertex \cite{Giannotti:2008cv} or of the entire correlator, as illustrated in the previous chapters.

Using the relation between $J_D^\mu$ and the EMT $T^{\mu\nu}$ we introduce the $J_DVV$ correlator 
\beqa
\Gamma_D^{\mu\alpha\beta}(k,p)
&\equiv& 
\int d^4 z\, d^4 x\, e^{-i k \cdot z + i p \cdot x}\,
\left\langle J^\mu_D(z) V^\alpha(x)V^\beta(0)\right\rangle 
\label{PNew2gammagg}
\eeqa
which can be related to the $TVV$ correlator 
\beqa
\Gamma^{\mu\nu\alpha\beta}(k,p)&\equiv& \int d^4 z\, d^4 x\, e^{-i k \cdot z + i p \cdot x}\, 
\left\langle T^{\mu \nu}(z) V^\alpha(x) V^\beta(0)\right\rangle 
\eeqa
according to
\beqa
\Gamma_D^{\mu\alpha\beta}(k,p)&=& 
i \frac{\partial}{\partial k^\nu}\Gamma^{\mu\nu\alpha\beta}(k,p) \,.
\eeqa
As we have already mentioned, this equation allows us to identify a pole term in the $J_DVV$ diagram from the corresponding pole 
structure in the $TVV$ vertex. 
In the following we recall the emergence of the anomaly poles in the QED case. The same reasonings apply as well in all the gauge invariant sector of the Standard Model. 

\subsection{The dilaton anomaly pole in the QED case}
\label{PNew2SecPoleQED}

For definiteness, it is convenient to briefly review the characterization of the $TVV$ vertex in the QED case with a massive fermion 
which has been extensively discussed in chapter \ref{Chap.TJJQED}. \\
We recall that the first form factor, which introduces the anomalous trace contribution,
for two on-shell final state photons ($s_1=s_2=0$) and a massive fermion, is given by 
\beq
\label{PNew2oom}
{F_1 (s;\,0,\,0,\,m^2)} = 
F_{1\, pole} \,  + \, \frac{e^2 \,   m^2}{3 \, \pi ^2 \, s^2} \, - \frac{e^2 \, m^2}{3 \, \pi^2 \, s}  \, \mathcal C_0 (s, 0, 0, m^2,m^2,m^2) 
\bigg[\frac{1}{2}-\frac{2 \,m^2}{ s}\bigg],  \\
\eeq
where 
\beq F_{1\, pole}=- \frac{e^2 }{18 \, \pi^2 s} \,.
\eeq
In the massless fermion case two properties of this expansion are noteworthy: 1) the trace anomaly 
takes contribution only from a single tensor structure $(t_1)$ and invariant amplitude $(F_1)$ which coincides with the pole term; 
2) the residue of this pole as $s\to 0$ is nonzero, showing that the pole is coupled in the infrared. Notice that the form factor 
$F_2$, which in general gives a nonzero contribution to the trace in the presence of mass terms, is multiplied by a tensor 
structure ($t_2$) which {\em vanishes} when the two photons are on-shell.
Therefore, similarly to the case of the chiral anomaly, also in this case the anomaly is {\em entirely} given by the appearance 
of an anomaly pole. 
We stress that this result is found to be exact in dimensional regularization, which is a mass independent scheme:
at perturbative level, the anomalous breaking of the dilatation symmetry, related to an anomaly pole in the spectrum of all
the gauge-ivariant correlators studied in this work, is separated from the sources of {\em explicit} breaking. The latter are related to the mass parameters and/or to the gauge bosons virtualities $p^2$ and $q^2$. 

\begin{figure}[t]
\begin{center}
\includegraphics[scale=1.2]{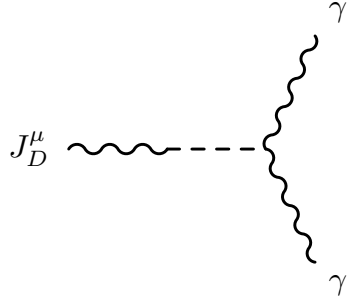}
\end{center}
\caption{Exchange of a dilaton pole mediated by the $J_D V V$ correlator. }
\label{PNew2dilatonpole}
\end{figure}

To analyze the implications of the pole behaviour discussed so far for the $TVV$ vertex and its connection with the $J_DVV$
correlator, we limit our attention on the anomalous contribution ($F_1\, t_1^{\mu\nu\alpha\beta}$), 
which we rewrite in the form
\beq
\Gamma_{pole}^{\mu\nu\alpha \beta}(k,p)\equiv
- \frac{e^2}{18\pi^2}\frac{1}{k^2}\left( \eta^{\mu \nu} k^2 - k^\mu k^\nu\right) u^{\alpha \beta}(p,q) \, , 
\qquad q = k - p \, .
\label{PNew2uref}
\eeq
This implies that the $J_D VV$ correlator acquires a pole as well
\beqa
\Gamma_{D\, pole}^{\mu\alpha\beta}
&=& 
- i \frac{e^2}{18 \pi^2}\frac{\partial}{\partial k^\nu}
\left[ \frac{1}{k^2}\, \left( \eta^{\mu\nu } k^2 - k^{\mu} k^{\nu}\right) u^{\alpha \beta}(p,k-p) \right]
\eeqa
and acting with the derivative on the right hand side we finally obtain 
\beq
\Gamma_{D \, pole}^{\mu\alpha\beta}(k,p)= 
i\, \frac{e^2}{6 \pi^2}\frac{k^\mu}{k^2}u^{\alpha \beta}(p,k-p) - i \frac{e^2}{18 \pi^2}\frac{1}{k^2}
\left( \eta^{\mu\nu } k^2 - k^{\mu} k^{\nu}\right)\frac{\partial}{\partial k_\nu}u^{\alpha \beta}(p,k-p).
\eeq
Notice that the first contribution on the right hand side of the previous equation corresponds to an anomaly pole, 
shown pictorially in Fig. \ref{PNew2dilatonpole}. In fact, by taking a derivative of 
the dilatation current only this term will contribute to the corresponding Ward identity
\beq
k_\mu \, \Gamma_D^{\mu\alpha\beta}(k,p) = i \frac{e^2}{6 \pi^2}u^{\alpha \beta}(p,k-p),
\label{PNew2res}
\eeq
which is the expression in momentum space of the usual relation $\partial J_D\sim FF$, while the second term trivially vanishes.  
Notice that the pole in (\ref{PNew2res}) has disappeared, and we are left just with its residue on the r.h.s., or, 
equivalently, the pole is removed in Eq. (\ref{PNew2uref}) if we trace the two indices $(\mu,\nu)$.

%

\subsection{Mass corrections to the dilaton pole}

The discussion of the mass corrections to the massless dilaton can follow quite closely the strategy adopted in 
the pion case using partially conserved axial currents (PCAC) techniques. Also in this case, as for PCAC in the past, 
one can assume a partially conserved dilaton current (PCDC) in order to relate the decay amplitude of the dilaton $f_\rho$ 
to its mass $m_\rho$ and to the vacuum energy. \\
For this goal we define the one-particle transition amplitudes for the dilatation current and the EMT between the vacuum and a dilaton 
state with momentum $p_\mu$
\beqa
\langle 0| J^\mu_D(x) |\rho, p \rangle 
&=& 
- i \, f_\rho \, p^\mu \, e^{-i p \cdot x} 
\nn \\
\langle 0| T^{\mu\nu}(x) |\rho, p \rangle 
&=&
\frac{f_\rho}{3} \, \left( p^\mu p^\nu - \eta^{\mu\nu} \, p^2\right)\, e^{-i p \cdot x},
\eeqa
both of them giving 
\beq
\label{PNew2rel1}
\partial_\mu \langle 0| J^\mu_D(x) |\rho, p\rangle = \eta_{\mu\nu} \langle 0| T^{\mu\nu}(x) |\rho, p \rangle = - f_\rho \, m_\rho^2 \, e^{-i p \cdot x}.
\eeq
We introduce the dilaton interpolating field $\rho(x)$ via a PCDC relation
\beq
\partial_\mu J^\mu_D(x) = - f_\rho \, m_\rho^2 \, \rho(x)
\label{PNew2pcdcrel}
\eeq
with 
\beq
\langle 0|\rho(x)|\rho, p\rangle = e^{-i p \cdot x}
\eeq
and the matrix element 
\beq
\mathcal A^\mu(q)= \int d^4 x \, e^{i q \cdot x} \,\langle 0| T \{ J^\mu_D(x) {T^\alpha}_\alpha(0) \} |0 \rangle,
\label{PNew2www}
\eeq
where $T\{\ldots\}$ denotes the time ordered product. \\
Using dilaton pole dominance we can rewrite the contraction of 	$q_\mu$ with this correlator as 
\bea
\label{PNew2inter}
\lim_{q_\mu \to 0} q_\mu \, \mathcal A^\mu(q)  = f_\rho \,\left\langle \rho, q=0| {T^\alpha}_\alpha(0) |0\right\rangle \,, 
\eea
where the soft limit $q_\mu \to 0$ with $q^2 \gg m_\rho^2 \sim 0$ has been taken. \\  
At the same time the dilatation Ward identity on the amplitude $\mathcal A^{\mu}(q)$ in Eq.(\ref{PNew2www}) gives 
\beqa
\label{PNew2WIJD}
q_\mu \mathcal A^\mu(q)
&=&
i \int d^4 x\,e^{i q \cdot x}\, \frac{\partial}{\partial x^\mu}\, 
\left\langle 0 \left| T\{ J^\mu_D(x) {T^\alpha}_\alpha(0) \} \right|0\right\rangle
\nn \\
&=&
i \int d^4 x \, e^{i q \cdot x} \,  
\left\langle 0 \left| T\{ \partial_\mu J^\mu_D(x) {T^\alpha}_\alpha(0)\} \right|0\right\rangle
+ i  \int d^4 x \, e^{i q \cdot x} \, \delta(x_0) \, 
\left\langle 0\left| \left[ J_D^0(x), {T^\alpha}_\alpha(0)\right] \right|0\right\rangle. \nn \\
\eeqa
The commutator of the time component of the dilatation charge density and the trace of the EMT can be rewritten as 
\beq
\left[ J_D^0(0, {\bf x}), {T^\alpha}_\alpha(0)\right] = -i \delta^3 ({\bf x})\left( 
d_T + x\cdot \partial\right){T^\alpha}_\alpha(0) 
\label{PNew2derr}
\eeq
where $d_T$ is the canonical dimension of the EMT $(d_T=4)$. 
Inserting Eq.(\ref{PNew2derr}) in the Ward identity (\ref{PNew2WIJD}) and neglecting the first term due to the nearly conserved dilatation current ($m_\rho \sim 0 $), 
we are left with 
\beq
q_\mu \mathcal A^\mu(q)= d_T \, \left\langle 0|{T^\alpha}_\alpha(0) |0\right\rangle.
\label{PNew2interdue}
\eeq
In the soft limit, with $q^2 \gg  m_\rho^2$, comparing Eq.(\ref{PNew2inter}) and Eq.(\ref{PNew2interdue}) we obtain 
\beqa
\lim_{q^\mu\to 0} q^\mu \mathcal{A}_\mu =
f_\rho \, \left\langle \rho, q=0| {T^\alpha}_\alpha(0) |0\right\rangle  = d_T \, \left\langle 0|{T^\alpha}_\alpha(0) |0\right\rangle.
\eeqa
Introducing the vacuum energy density $\epsilon_{vac}=\left\langle 0|T^0_0 |0\right\rangle = \frac{1}{4}  \left\langle 
0|T^\alpha_\alpha(0) |0\right\rangle$ and using the relation in Eq.(\ref{PNew2rel1}) we have 
\beq
\left\langle\rho, p=0| {T^\mu}_\mu |0\right\rangle =- f_\rho m_\rho^2 = \frac{d_T}{f_\rho} \epsilon_{vac}
\eeq
from which we finally obtain ($d_T = 4$)
\beq
f_\rho^2 m_\rho^2 = -16 \, \epsilon_{vac}. 
\eeq
This equation fixes the decay amplitude of the dilaton in terms of its mass and the vacuum energy. 
Notice that $\epsilon_{vac}$ can be related both to the anomaly and possibly to explicit contributions of the 
breaking of the dilatation symmetry since 
\beq
\label{PNew2beta}
\epsilon_{vac}= \frac{1}{4} \, \left\langle 0 \left| \frac{\beta(g)}{2 g} F_{\mu\nu}F^{\mu\nu} \right|0\right\rangle + ... 
\eeq
where the ellipsis saturate the anomaly equation with extra mass-dependent contributions.
In (\ref{PNew2beta}) we have assumed, for simplicity, the coupling of the pole to a single gauge field, with 
a beta function $\beta(g)$, but obviously, it can be generalized to several gauge fields.

\section{The infrared coupling of an anomaly pole and the anomaly enhancement} 

It is easy to figure out from the results of the previous sections that the coupling of a (graviscalar) dilaton to the anomaly 
causes a large enhancement of its 2-photons and 2-gluons decays. One of the features of the graviscalar interaction is that its coupling includes anomalous contributions which are part both of the two-photon and of the two-gluon cross sections. For this reason, 
if an enhancement respect to the Standard Model rates is found only in one of these two channels and it is associated to the exchange of a spin zero intermediate state, this result could be used to rule out the exchange of a graviscalar.  

On the other hand, for an effective dilaton, identified by an anomaly pole in the $J_D VV$ correlator of the Standard Model, the case is more subtle, since the coupling of this effective state to the anomaly has to be introduced by hand. This state should be identified, in the perturbative picture,  with the corresponding anomaly pole. The situation, here, is closely similar to the pion case, 
 where the anomalous contribution to the pion-photon-photon vertex is added - a posteriori - to a Lagrangian which is otherwise chirally symmetric. 
Also in the pion case the anomaly enhancement can be justified, in a  perturbative approach, by the infrared coupling of the anomaly pole of the 
$AVV$ diagram.

  In general, in the case of an effective dilaton, one is allowed to write down a Lagrangian which is assumed to be scale invariant and, at a second 
  stage, introduce the direct coupling of this state to the trace anomaly. The possibility of coupling such a state to the photon and to the gluons or just to the photons, for instance, is a delicate issue for which a simple perturbative approach is unable to offer a definitive answer. 
For instance, if we insist that confinement does not allow us to have, in any case, on-shell final state gluons, the two-gluon coupling of an effective dilaton, identified in the corresponding $J_D VV$ correlator, should not be anomaly enhanced. In fact, with one or two off-shell final state gluons, the residue of the anomaly pole in this correlator is zero. More comments on this issue can be found in chapter \ref{Chap.TJJQED} where the decoupling properties of the anomaly poles have been analyzed in detail.

We feel, however, that a simple perturbative analysis may not be completely sufficient to decide whether or not the coupling of such a state to the gluon anomaly takes place. On the other hand, there is no doubt, by the same reason, that such a coupling should occur in the 2-photon case, being the photons massless asymptotic states. In this case the corresponding anomaly pole of the $J_D \gamma\gamma$ vertex is infrared coupled.  

Similar enhancements are present in the case of quantum scale invariant extensions of the Standard Model \cite{Goldberger:2007zk},  where one assumes that the spectrum of the theory is extended with new massive states in order to set the $\beta$ functions of the gauge couplings to vanish. In a quantum scale invariant theory such as the one discussed in \cite{Goldberger:2007zk}, the dilaton couples only to massive states, but the heavy mass limit and the condition of the vanishing of the complete $\beta$ functions, leave at low energy a dilaton interaction proportional only to the $\beta$ functions of the massless low energy states. We have commented on this point in the following section. The "remnant" low energy interaction is mass-independent and coincides with that due to a typical anomalous coupling, although its origin is of different nature, since anomalous contributions are genuinely mass-independent. 

For this reason, the decays of a dilaton produced by such extensions carries "anomaly like" enhancements as in the graviscalar case. Obviously, such enhancements to the low energy states of the Standard Model would also be typical of the decay of a Higgs field, which couples proportionally to the mass of an intermediate state, if quantum scale invariance is combined with the decoupling of a heavy sector. This, in general, causes an enhancement of the Higgs decay rates into photons and gluons. A partial enhancement only of the di-photon channel could be accomplished, in this approach, by limiting the above quantum scale invariant arguments only to the electroweak sector.

\subsection{Quantum conformal invariance and dilaton couplings at low energy}
\label{PNew2quantum}

We consider the situation in which all the SM fields are embedded in a (quantum) Conformal Field Theory (CFT) extension \cite{Goldberger:2007zk}
and we discuss the (loop-induced) couplings of the dilaton to the massless gauge bosons. 
At tree level the dilaton of \cite{Goldberger:2007zk} couples to the SM fields only through their masses, as the fundamental dilaton which we have discussed previously, and, in this respect, it behaves like the SM Higgs, without scale anomaly contributions. 
For this reason the dilaton interaction with the massless gauge bosons is induced by quantum effects mediated by heavy particles
running in the loops (in this context heavier or lighter is referred to the dilaton mass), and not by anomalous terms.
When the mass $m_i$ of the particle running in the loop is much greater than the dilaton mass, the coupling to the massless 
gauge bosons becomes 
\bea
\mathcal L_{\rho} = \frac{\alpha_s}{8 \pi} \sum_i b_g^i \, \frac{\rho}{\Lambda} (F_{g \, \mu\nu}^a)^2 Ê 
+ Ê \frac{\alpha_{em}}{8 \pi} \sum_i b_{em}^i \, \frac{\rho}{\Lambda} (F_{\gamma \, \mu\nu})^2 Ê \,,
\eea
where $b_{em}^i$ and $b_{g}^i$ are the contributions of the heavy field $i$ to the one-loop $\beta$ function (computed in the 
$\overline{MS}$ scheme) for the electromagnetic and strong coupling constants respectively. The $\beta$ functions are normalized as
\bea
\beta_i = \frac{g^3}{16 \pi^2} b^i \,.
\eea
Note that this result is independent from the heavy mass $m_i$ as one can prove by analyzing the structure of the mass corrections 
of the dilaton coupling, which reads as
\beq
\Gamma_{\rho V V}\sim \frac{g^2}{\pi^2 \Lambda} \, m_i^2 \, 
\bigg[ \frac{1}{s}  -  \frac{1} {2 }\mathcal C_0 (s, 0, 0, m_i^2, m_i^2, m_i^2) \bigg(1-\frac{4 m_i^2}{ s}\bigg)\bigg] 
\sim 
\frac{g^2}{\pi^2 \Lambda} \, \frac{1}{6} + O\left( \frac{s}{m_i^2} \right)
\eeq
where $s=m_{\rho}^2$ is fixed at the dilaton mass and we have performed the large mass limit of the amplitude using 
\beqa
\mathcal C_0(s, 0, 0, m_i^2, m_i^2, m_i^2) \sim - \frac{1}{2 m_i^2} \left( 1 + \frac{1}{12} \frac{s}{m_i^2} + O(\frac{s^2}{m_i^4} ) \right)
\eeqa
valid for $m_i^2 \gg s = m_{\rho}^2$. This shows that in the case of heavy fermions, 
the dependence on the fermion mass cancels. 
Obviously, this limit generates an effective coupling which is proportional to the $\beta$ function related to the heavy flavours. 
The same reasonings can be employed to the Higgs case as well. It clear that this coupling to the massless gauge bosons is dependent 
from new heavy states and, therefore, from the UV completion of the SM. This is certainly the case for the Standard Model Higgs whose 
double photon decay is one of the most important decay channel for new physics discoveries. \\ 
For the dilaton case the situation is slightly different. Surely we do not understand the details of the CFT extension, nor its 
particle spectrum, but nevertheless we know that the conformal symmetry is realized at the quantum level. Therefore the complete 
$\beta$ functions, including the contribution from all states, must vanish
\bea
\beta = \frac{g^3}{16 \pi^2} Ê\bigg[ \sum_{i} b^i Ê+ \sum_{j} b^j \bigg] = 0 \,,
\eea
where $i$ and $j$ run over the heavy and light states respectively. Exploiting the consequence of the quantum conformal symmetry, the 
dilaton couplings to the massless gauge bosons become
\bea
\mathcal L_{\rho} = 
- \frac{\alpha_s}{8 \pi} \sum_j b_g^j \, \frac{\rho}{\Lambda} (F_{g \, \mu\nu}^a)^2 Ê - Ê \frac{\alpha_{em}}{8 \pi}
\sum_j b_{em}^j \, \frac{\rho}{\Lambda} (F_{\gamma \, \mu\nu})^2 Ê \,,
\eea
in which the dependence from the $\beta$ functions of the light states is now explicit. We emphasize that the appearance of the light 
states contributions to the $\beta$ functions is a consequence of the vanishing of the complete $\beta$, and, therefore, of the CFT 
extension and not the result of a direct coupling of the dilaton to the anomaly.

\section{Conclusions}
We have presented a general discussion of dilaton interactions with the neutral currents sector of the Standard Model. In the case of a 
fundamental graviscalar as a dilaton, we have presented the complete electroweak corrections to the corresponding interactions and we have discussed the renormalization properties of the same vertices. In particular, we have shown that the renormalizability of the dilaton vertices is inherited directly from that of the Standard Model only if the Higgs sector is characterized by a conformal coupling ($\chi$) fixed at the value 1/6.

Then we have moved to an analysis of the analytic structure of the $J_D VV$ correlator, showing that it supports an anomaly pole as an interpolating state, which indicates that such a state can be interpreted as the Nambu-Goldstone (effective dilaton) mode of the anomalous breaking of the dilatation symmetry. 

In fact, the trace anomaly seems to bring in some important information concerning the dynamics of the Standard Model, aspects that we have 
tried to elucidate. For this reason, we have extended a previous analysis of ours of the $TVV$ vertex, performed in the broken electroweak phase and in QCD, in order to 
characterize the dynamical behaviour of the analogous $J_D VV$ correlator. The latter carries relevant information on the 
anomalous breaking of the dilatation symmetry in the Standard Model. In fact, as we move to high energy, far above the electroweak scale, the Lagrangian of the Standard Model 
becomes approximately scale invariant. This approximate dilatation 
symmetry is broken by a quantum anomaly and its signature, as we have shown in our analysis, is in the appearance of 
an anomaly pole in the $J_DVV$ correlator. The same pole might appear in correlators with multiple insertions of $J_D$, but the proof 
of their existence is far more involved and requires further investigations. 
This pole is clearly massless in the perturbative picture, and accounts for the anomalous breaking 
of this approximate scale invariance.

\chapter{Conclusions and perspectives}
\label{Chap.Conclusions}

We have presented a detailed analysis of the quantum gravitational effective action in the Standard Model at one-loop order in perturbation theory, focusing on the contributions responsible for the appearance of the conformal anomaly. One of the main motivations of our work is to provide a more complete theoretical description of the interactions between gravity and the Standard Model spectrum. In this respect we have studied the correlation functions characterized by a single insertion of an energy momentum tensor (EMT) operator, describing the coupling of gravity with the QED, QCD and electroweak neutral gauge currents and with fermions in the diagonal and off-diagonal sectors of flavor space. \\
Much attention has been given to the derivation of a suitable set of Ward identities which exploit the invariance under diffeomorphism and BRST transformations. They have been an unavoidable tool to unambiguously define the correlators and to secure the correctness of our computation. Moreover, the renormalization properties of the same vertex functions have been discussed together with a complete determination of the one-loop renormalization counterterms.

One of the main features of the conformal anomalous correlators, which have been extensively studied in this thesis, is the emergence of anomaly poles in each gauge invariant sector of the Standard Model. In \cite{Giannotti:2008cv} this pole structure was identified for the first time in the $TJJ$ correlator in QED. As we have mentioned above, we have extended the analysis of \cite{Giannotti:2008cv} to the Standard Model and we have commented on the similarities with the chiral anomaly case. Indeed anomaly poles are the common signatures of chiral and conformal anomalies, sharing, for instance, the structure of the effective action and the decoupling property which we have discussed in this thesis. \\
The pole behaviour characterizing the anomalous correlators is clearly the signature of an effective degree of freedom interpolating between two external gauge fields and the anomalous current, the axial current or the energy momentum tensor for the chiral and the conformal anomalies, respectively. Concerning the conformal anomaly case, although the identification under general kinematical configurations of this pole structure is unambiguous, its interpretation is still open and may lead to an infrared modification of gravity and to macroscopic effects as discussed in \cite{Mottola:2006ew}. \\
We recall that the energy momentum tensor, besides its role in the definition of the interaction between the gravitational and the matter fields, has also an important meaning in flat theories with no gravity, being connected to the Noether's current of the dilatation symmetry. In fact, the pole term found in the $TJJ$ correlator is inherited by the dilatation current. Following the analogy with the chiral anomalous $AVV$ diagram of the strong interactions and the corresponding appearance of the pion particle as a Nambu Goldstone boson of the spontaneously broken chiral symmetry, we have speculated on the existence of a scalar mode, the dilaton, associated instead with the breaking of the scale symmetry. This interpretation is in agreement with a number of recently proposed low-energy effective models \cite{Goldberger:2007zk, Fan:2008jk} in which the breaking of the conformal and electroweak symmetries are simultaneously implemented.

Other possible applications of these studies concern the analysis of the anomaly supermultiplet in super Yang Mills theories. Our results clearly imply that two of three components of the Ferrara Zumino supermultiplet are saturated by anomaly poles. This suggests that an analogous pole should also appear in the gamma-trace of the third component, the supersymmetric current. These issues are relevant for a consistent definition of a mechanism of anomaly cancellation in supergravity. \\
Finally, we mention the important role played by correlators with the energy momentum tensor in CFT's, especially in the analyses of the irreversibility of the renormalization group flow  \cite{Komargodski:2011vj, Elvang:2012yc, Elvang:2012st, Luty:2012ww}, which has recently received a wide attention. We refer to \cite{Coriano:2012wp} for more details concerning the momentum space description of conformal correlators and their mapping from coordinate space, in the presence of anomalies.

\chapter*{Part two}
{\LARGE{ \textbf {Holographic non-gaussianities and applications \\ \\ of conformal field theory methods in momentum space}}}

\vspace{2cm}

In this second part of the thesis we discuss some applications of the methodology presented in the first part to two selected problems. 
The first concerns the use of conformal field theory techniques in the definition of scalar three-point functions, exploiting the constraints that come from this symmetry. The analysis is formulated in momentum space in which the constraints that characterize these correlators take the form of partial differential equations. We show the connection between their invariance under special conformal transformations and the appearance of a certain class of generalized hypergeometric functions. Moreover,
the method is applied to the calculation of a specific type of integrals which appear in higher-order perturbation theory, also called {\em master integrals}. They are the key elements in the computation of the radiative corrections to the perturbative expansion of massless theories. These integrals, whose calculation is usually based on Mellin Barnes techniques, is regained in an independent way using the corresponding conformal constraints. 

In the second contribution we discuss the computation of correlators involving multiple insertions of the energy momentum tensor, for some quantum field theories in $D=3$ spacetime dimensions. They play an important role in the characterization of the spectrum of the cosmological non-gaussianities in some specific cosmological models based on holographic dualities. This approach allows to relate the gravitational perturbations of the metric in a pre-inflationary phase of the early universe to the computation of a certain set of dual correlators in $D=3$. The model has been developed in a series of papers \cite{McFadden:2009fg, McFadden:2010vh, McFadden:2011kk, Bzowski:2011ab} to which we refer for additional information.

\chapter{Implications of conformal invariance in momentum space}

\section{Introduction} 

Conformal invariance plays an important role in constraining the structure of correlation functions of conformal field theories 
in any dimensions. It allows to fix the form of correlators - up to three-point functions - modulo a set of constants which are 
also given, once the field content of the underlying conformal field theory is selected \cite{Osborn:1993cr, Erdmenger:1996yc}.
The approach, which is largely followed in this case, is naturally formulated in position space, while the same conformal requirements, in 
momentum space, have been far less explored \cite{Coriano:2012wp}.\\
Conformal three-point functions have been intensively studied in the past, and a classification of their possible structures, in the presence of conformal anomalies, is 
available. Conformal anomalies emerge due to the inclusion of the energy momentum tensor in a certain correlator and, in some cases, find specific realizations in free field theories of 
scalars, vectors and fermions \cite{Osborn:1993cr, Erdmenger:1996yc}. Typical correlators which have been studied  are those 
involving the $TT, TOO, TVV$, and $TTT$, where $T$ denotes the energy momentum tensor, $V$ a vector and $O$ a generic 
scalar operator of arbitrary dimension.

The conformal constraints in position space, in this case, are combined with the Ward identities derived from the conservation of the energy momentum tensor and its 
tracelessness condition, valid at separate coordinate points, to fix the structure of each correlator. 
These solutions are obtained for generic conformal theories, with no reference to their Lagrangian realization which, in general, may not even exist. 
The solutions of the conformal constraints are then extended to include the contributions from the coincidence regions, where all the external points collapse to the same 
point. \\
Free field theory realizations of these correlators (for fermions, scalars and vectors) allow to perform a direct test of these results both in position 
and in momentum space, at least in some important cases, such as the $TVV$ or the $TTT$ 
(this latter only for $d=4$) \cite{Coriano:2012wp}, but obviously do not exhaust all possibilities.

Recently interest in the momentum space form of conformal correlators has arisen in the context of the study of anomalous conformal Ward identities, 
massless poles and scalar degrees of freedom associated with the trace anomaly \cite{Giannotti:2008cv, Mottola:2010, Armillis:2009pq, Armillis:2009im, Armillis:2010qk, Coriano:2011ti, Coriano:2011zk}, 
and because of their possible role in determining the form of conformal invariance in the non-Gaussian features of the Cosmic Microwave Background 
\cite{AntMazMott:1997, Antoniadis:2011ib}, or in inflation \cite{Maldacena:2011nz, Kehagias:2012pd}.
The possibility of retrieving information on conformal correlators in momentum space seems to be related, in one way or another, 
to the previous knowledge of the same correlators in configuration space, where the conformal constraints are easier to implement and solve. 
One question that can be naturally raised is if we are able to bypass the study of conformal correlators in position space, by fixing their structure directly in momentum space 
and with no further input. This approach defines an independent path which, as we are going to show, can be successful in some specific cases. 
We will illustrate the direct construction of the solution bringing the example of the scalar three-point correlator. 
The analysis in momentum space should not be viewed though as an unnecessary complication. 
In fact, the solution in the same space, if found, is explicit in the momentum variables and can be immediately compared with the integral representation of the position space solution, 
given by a generalized Feynman integral. As a corollary of this approach, in the case of three-point functions, we are able to determine the complete structure of such an integral, which 
is characterized by three free parameters related to the scaling dimensions of the original scalar operators, in an entirely new way. 
It is therefore obvious that this approach allows to determine the explicit form of an entire family of master integrals.   

In the scalar case we are able to show that the conformal conditions are equivalent to partial differential equations (PDE's) of 
generalized hypergeometric type, solved by functions of two variables, $x$ and $y$, which take the form of ratios of the external momenta.   
The general solution is expressed as a generic linear combination of four generalized hypergeometric functions of the same variables, or Appell's functions. 
Three out of the four constants of the linear combination can be fixed by the momentum symmetry. 
This allows to write down the general form of the scalar correlator in terms of a single multiplicative constant, 
which classifies all the possible conformal realizations of the scalar three-point function. 

In the final part of this chapter we go back to the analysis of the conformal master integrals, the Fourier transform of the scalar three-point correlators in position space.
We show that the usual rules of integration by parts satisfied by these integrals are nothing else but the requirement of scale invariance. 
Specifically, dilatation symmetry relates the master integral $J(\nu_1,\nu_2,\nu_3)$, labelled by the powers of the Feynman propagators $(\nu_1,\nu_2,\nu_3)$  - with $\nu_1 +\nu_2 +\nu_3=
\kappa$ - to those of the first neighboring plane $(\kappa\to \kappa +1)$.
On the other hand, special conformal constraints relate the integrals of second neighboring planes $(\kappa\to \kappa +2)$. 

\section{Conformal transformations}
%
In order to render our treatment self-contained, we present a brief review of the conformal transformations in $d > 2$ dimensions which identify, in Minkowski space, the conformal group $SO(2,d)$. \\
These may be defined as the transformations $x_\mu \rightarrow x'_\mu(x)$ that preserve the infinitesimal length up to a local factor 
\bea
d x_\mu d x^\mu \rightarrow d x'_\mu d x'^\mu = \Omega(x)^{-2} d x_\mu d x^\mu \,.
\eea
In the infinitesimal form, for $d > 2$, the conformal transformations are given by
\bea
\label{PNewCIxtransf}
x'_\mu(x) = x_\mu + a_\mu + {\omega_{\mu}}^{\nu} x_\nu + \lambda x_\mu + b_\mu x^2 - 2 x_\mu b \cdot x
\eea
with
\bea
\label{PNewCIOmega}
\Omega(x) = 1 -\sigma(x) \qquad \mbox{and} \qquad \sigma(x) = \lambda - 2 b \cdot x \,.
\eea
The transformation in Eq.(\ref{PNewCIxtransf}) is defined by translations ($a_\mu$), rotations ($\omega_{\mu\nu} = - \omega_{\nu\mu}$), dilatations ($\lambda$) and 
special conformal transformations ($b_\mu$). The first two define the Poincar\'{e} subgroup which leaves invariant the infinitesimal length ($\Omega(x) = 1$). \\
If we also consider the inversion
\bea
x_\mu \rightarrow x'_\mu = \frac{x_\mu}{x^2} \,, \qquad \qquad \Omega(x) = x^2 \,,
\eea 
we can enlarge the conformal group to $O(2,d)$. Special conformal transformations can be realized by a translation preceded and 
followed by an inversion.

Having specified the elements of the conformal group, we can define a quasi primary field $\mathcal O^i(x)$, where the index $i$ runs over a representation of the group $O
(1,d-1)$ to which the field belongs, through the transformation property under a conformal transformation $g$
\bea
\mathcal O^i(x) \stackrel{g}{\rightarrow} \mathcal O'^i(x') = \Omega(x)^\eta D^i_j(g) \mathcal O^j(x) \,,
\eea
where $\eta$ is the scaling dimension of the field and $D^i_j(g)$ denotes the representation of $O(1,d-1)$.
In the infinitesimal form we have
\bea
\delta_g \mathcal O^i(x) = - (L_g \mathcal O)^i(x) \,, \qquad \mbox{with} \qquad  L_g = v \cdot \partial + \eta \, \sigma + \frac{1}{2} \partial_{[ \mu} v_{\nu ]} \Sigma^{\mu\nu} \,,
\eea
where the vector $v_\mu$ is the infinitesimal coordinate variation $v_\mu = \delta_g x_\mu = x'_\mu(x) - x_\mu$ and 
$(\Sigma_{\mu\nu})^i_j$ are the generators of $O(1,d-1)$ in the representation of the field $\mathcal O^i$. The explicit form of the 
operator $L_g$ can be obtained from Eq.(\ref{PNewCIxtransf}) and Eq.(\ref{PNewCIOmega}) and is given by
\begin{align}
& \mbox{translations:} && L_g = a^\mu \partial_\mu \,, \nn \\
& \mbox{rotations:} && L_g = \frac{\omega^{\mu\nu}}{2} \left[ x_\nu \partial_\mu - x_\mu \partial_\nu - \Sigma_{\mu\nu} \right] \,, \nn \\
& \mbox{scale transformations :}    && L_g = \lambda \left[ x \cdot \partial +  \eta \right] \,, \nn  \\
& \mbox{special conformal transformations. :}    && L_g = b^\mu \left[ x^2 \partial_\mu - 2 x_\mu \, x \cdot \partial - 2 \eta \, x_\mu - 2 x_\nu {\Sigma_{\mu}}^{\nu} \right] \,.
\end{align}
Conformal invariant correlation functions of quasi primary fields can be defined by requiring that 
\bea
\sum_{r=1}^{n} \langle \mathcal O^{i_1}_1(x_1) \ldots \delta_g \mathcal O^{i_r}_r(x_r) \ldots \mathcal O^{i_n}_n(x_n) \rangle = 0 \,.
\eea
In particular, the invariance under scale and special conformal transformations, in which we are mainly interested, reads as
\bea
\label{PNewCIConformalEqCoord}
&& \sum_{r=1}^{n} \left( x_r \cdot \partial^{x_r} +  \eta_r  \right) \langle \mathcal O^{i_1}_1(x_1) \ldots   \mathcal O^{i_r}_r(x_r) 
\ldots  \mathcal O^{i_n}_n(x_n) \rangle = 0 \,, \nn \\
&& \sum_{r=1}^{n} \left( x_r^2 \partial^{x_r}_\mu - 2 x_{r \, \mu} \, x_r \cdot \partial^{x_r} - 2 \eta_r \, x_{r \, \mu} - 2 x_{r \, 
\nu} ( {\Sigma^{(r)}_{\mu}}^{\nu})^{i_r}_{j_r} \right) \langle \mathcal O^{i_1}_1(x_1) \ldots  \mathcal O^{j_r}_r(x_r) \ldots \mathcal 
O^{i_n}_n(x_n) \rangle = 0 \,. \nn \\
\eea
The constraints provided by conformal invariance have been solved in coordinate space and for arbitrary space-time dimension. One can show, 
for instance, that the two and three-point functions are completely determined by conformal symmetry up to a small number of independent
constants \cite{Polyakov:1970, Osborn:1993cr} . \\
Exploiting the same constraints in momentum space is somewhat more involved. 
In the following we assume invariance under the Poincar\'{e} group and we focus our attention on dilatations and special conformal transformations.\\
For this purpose we define the Fourier transform of a $n$ point correlation function as
\bea
&& (2 \pi)^d \, \delta^{(d)}(p_1 + \ldots + p_n) \, \langle \mathcal O^{i_1}_1(p_1) \ldots   \mathcal O^{i_n}_n(p_n) \rangle  \nn \\
&& \qquad \qquad \qquad = \int d^d x_1 \ldots d^d x_n \, \langle \mathcal O^{i_1}_1(x_1) \ldots    \mathcal O^{i_n}_n(x_n) \rangle e^{ 
i p_1 \cdot x_1 + \ldots + i p_n \cdot x_n},
\eea
where the correlation function in momentum space is understood to depend only on $n-1$ momenta, 
as the $n$-th one is removed using momentum conservation. \\
The momentum space differential equations describing the invariance under dilatations and special conformal 
transformations are obtained Fourier-transforming Eq.(\ref{PNewCIConformalEqCoord}). It is worth noting 
that some care must be taken, due to the appearance of derivatives on the delta function. As pointed 
out in \cite{Maldacena:2011nz}, these terms can be discarded and we are left with the two equations
\bea
\label{PNewCIConformalEqMom}
&& \bigg[ - \sum_{r=1}^{n-1} \left(  p_{r \, \mu} \, \frac{\partial}{\partial p_{r \, \mu}}  + d  \right) + \sum_{r=1}^{n} \eta_r 
\bigg] \langle \mathcal O^{i_1}_1(p_1) \ldots   \mathcal O^{i_r}_r(p_r) \ldots  \mathcal O^{i_n}_n(p_n) \rangle = 0 \,, \nn \\
&& \sum_{r=1}^{n-1} \left( p_{r \, \mu} \, \frac{\partial^2}{\partial p_{r}^{\nu} \partial p_{r \, \nu}}  - 2 \, p_{r \, \nu} \, 
\frac{\partial^2}{ \partial p_{r}^{\mu} \partial p_{r \, \nu} }    + 2 (\eta_r - d) \frac{\partial}{\partial p_{r}^{\mu}}  + 2 
(\Sigma_{\mu\nu}^{(r)})^{i_r}_{j_r} \frac{\partial}{\partial p_{r \, \nu}} \right) \nn \\
&& \hspace{7cm} \, \times \langle \mathcal O^{i_1}_1(p_1) \ldots  \mathcal O^{j_r}_r(p_r) \ldots \mathcal O^{i_n}_n(p_n) \rangle = 0 \, ,
\eea
which define an arbitrary conformal invariant correlation function in $d$ dimensions.
Note that we are dealing with a first and a second order partial differential equations in $n-1$ independent momenta. The choice of the momentum which is eliminated in the two equations given above is arbitrary. \\
Despite the apparent asymmetry in the definition of the special conformal constraint, due to the absence of the $n$-th scaling dimension $\eta_n$ and the $n$-th spin matrix $( \Sigma_{\mu\nu}^{(n)})^{i_n}_{j_n}$, the second of Eq.(\ref{PNewCIConformalEqMom}) does not depend on the specific momentum which is eliminated. We could have similarly chosen to express Eq.(\ref{PNewCIConformalEqMom}) in terms of the momenta $(p_1 \ldots p_{k-1}, p_{k+1}, \ldots p_n)$, with $p_k$ removed using the momentum conservation, and we would have obtained an equivalent relation. We have left to an appendix the formal proof of this point.
We have then explicitly verified the correctness of our assertion on the vector and tensor two-point correlators which have been discussed in the following section. In both cases, whatever momentum parameterization is chosen for Eq.(\ref{PNewCIConformalEqMom}), we have checked that special conformal constraints imply the equality of the scaling dimensions of the two (vector or tensor) operators, accordingly to the well-known result obtained in coordinate space. \\
We recall, anyway, that, 
apart from the following section, in which vector and tensor two-point functions are reviewed, the main results of this work, being focused on scalar operators, are free from all the complications arising from the presence of the spin matrices in the special conformal constraints.

\section{Two-point functions from momentum space and anomalies}
\label{PNewCITwoPointSection}

\subsection{General solutions of the scale and special conformal identities}

We start exploring the implications of these constraints on two-point functions. 
In particular, the quasi primary fields taken into account are 
scalar ($\mathcal O$), conserved vector ($V_\mu$) and conserved and traceless ($T_{\mu\nu}$) operators.

For the two-point functions the differential equations in Eq.(\ref{PNewCIConformalEqMom}) simplify considerably, being expressed in terms 
of just one independent momentum $p$, and take the form
\bea 
\label{PNewCIConformalEqMomTwoPoint}
&& \left( - p_{\mu} \, \frac{\partial}{\partial p_{\mu}}  + \eta_1 + \eta_2 - d \right) G^{ij}(p) = 0 \,, \nn \\
&& \left(  p_{\mu} \, \frac{\partial^2}{\partial p^{\nu} \partial p_{\nu}}  - 2 \, p_{\nu} \, \frac{\partial^2}{ \partial p^{\mu} 
\partial p_{\nu} }    + 2 (\eta_1 - d) \frac{\partial}{\partial p^{\mu}}  + 2 (\Sigma_{\mu\nu})^{i}_{k} \frac{\partial}{\partial 
p_{\nu}} \right)  G^{kj}(p)  = 0 \,,
\eea
where we have defined $G^{ij}(p) \equiv \langle \mathcal O_1^i(p) \mathcal O_2^j(-p) \rangle$. The 
first of Eq.(\ref{PNewCIConformalEqMomTwoPoint}) dictates the scaling behavior of the correlation function, while special conformal 
invariance allows a non zero result only for equal scale dimensions of the two operators $\eta_1 = \eta_2$, as we know from the 
corresponding analysis in coordinate space. We start by illustrating this point. \\
For the correlation function $G_S(p)$ of two scalar quasi primary fields the invariance under the Poincar\'{e} group obviously 
implies that $G_S(p) \equiv G_S(p^2)$, so that the derivatives with respect to the momentum $p_\mu$ can be easily recast in terms of 
the variable $p^2$. \\
The invariance under scale transformations implies that $G_S(p^2)$ is a homogeneous function of degree 
$\alpha = \frac{1}{2}(\eta_1 + \eta_2 - d)$. 
At the same time, it is easy to show that the second equation in (\ref{PNewCIConformalEqMomTwoPoint}) can be satisfied only if $\eta_1 = \eta_2$. 
Therefore conformal symmetry fixes the structure of the scalar two-point function up to an arbitrary overall constant $C$ as
\bea
\label{PNewCITwoPointScalar}
G_S(p^2) = \langle \mathcal O_1(p) \mathcal O_2(-p) \rangle = \delta_{\eta_1 \eta_2}  \, C\, (p^2)^{\eta_1 - d/2} \, .
\eea 
If we redefine
\bea
C=c_{S 12} \,  \frac{\pi^{d/2}}{4^{\eta_1 - d/2}} \frac{\Gamma(d/2 - \eta_1)}{\Gamma(\eta_1)} 
\eea
in terms of the new integration constant $c_{S 12}$, the two-point function reads as
\bea 
\label{PNewCITwoPointScalar2}
G_S(p^2) =  \delta_{\eta_1 \eta_2}  \, c_{S 12} \,  \frac{\pi^{d/2}}{4^{\eta_1 - d/2}} \frac{\Gamma(d/2 - \eta_1)}{\Gamma(\eta_1)} 
(p^2)^{\eta_1 - d/2} \,,
\eea
and after a Fourier transformation in coordinate space takes the familiar form
\bea
\langle \mathcal O_1(x_1) \mathcal O_2(x_2) \rangle \equiv \mathcal{F.T.}\left[ G_S(p^2) \right] =  \delta_{\eta_1 \eta_2} \,  c_{S 12} 
\frac{1}{(x_{12}^2)^{\eta_1}} \,,
\eea
where $x_{12} = x_1 - x_2$. 
The ratio of the two Gamma functions relating the two integration constants $C$ and $c_{S 12}$ correctly reproduces the ultraviolet singular behavior of the correlation function and plays a role in the discussion of the origin of the scale anomaly.

Now we turn to the vector case where we define $G_V^{\alpha \beta}(p) \equiv \langle V_1^\alpha(p) V_2^\beta(-p) \rangle$. If the 
vector current is conserved, then the tensor structure of the two-point correlation function is entirely fixed by the transversality 
condition, $\partial^\mu V_\mu = 0$, as 
\bea
\label{PNewCITwoPointVector0}
G_V^{\alpha \beta}(p) =  \pi^{\alpha\beta}(p) \, f_V(p^2)\,, \qquad \qquad \mbox{with} \qquad
\pi^{\alpha\beta}(p) = \eta^{\alpha \beta} -\frac{p^\alpha p^\beta}{p^2} 
\eea
where $f_V$ is a function of the invariant square $p^2$ whose form, as in the scalar case, is determined by the conformal constraints. 
Following the same reasonings discussed previously we find that
\bea 
\label{PNewCITwoPointVector}
G_V^{\alpha \beta}(p) = \delta_{\eta_1 \eta_2}  \, c_{V 12}\, 
\frac{\pi^{d/2}}{4^{\eta_1 - d/2}} \frac{\Gamma(d/2 - \eta_1)}{\Gamma(\eta_1)}\,
\left( \eta^{\alpha \beta} -\frac{p^\alpha p^\beta}{p^2} \right)\
(p^2)^{\eta_1-d/2} \,,
\eea
with $c_{V12}$ being an arbitrary constant. 
We recall that the second equation in (\ref{PNewCIConformalEqMomTwoPoint}) gives consistent results for the two-point function in Eq.(\ref{PNewCITwoPointVector}) only when the scale dimension $\eta_1 = d - 1$. We refer to appendix \ref{PNewCIAppTwoPoint} for more details. \\
To complete this short excursus, we present the solution of the conformal constraints for the two-point function built out of two energy momentum tensor operators which are symmetric, conserved and traceless
\bea
\label{PNewCIEMTconditions}
T_{\mu\nu} = T_{\nu\mu} \,, \qquad \qquad \partial^{\mu} T_{\mu\nu} = 0 \,, \qquad \qquad {T_{\mu}}^{\mu} = 0 \,.
\eea
Exploiting the conditions defined in Eq.(\ref{PNewCIEMTconditions}) we can unambiguously define the tensor structure of the correlation 
function $G^{\alpha\beta\mu\nu}_T(p) = \Pi_{d}^{\alpha\beta\mu\nu}(p) \, f_T(p^2)$ with
\bea 
\label{PNewCITT}
\Pi^{\alpha\beta\mu\nu}_{d}(p) = \frac{1}{2} \bigg[ \pi^{\alpha\mu}(p) \pi^{\beta\nu}(p) + \pi^{\alpha\nu}(p) \pi^{\beta\mu}(p) 
\bigg] 
- \frac{1}{d-1} \pi^{\alpha\beta}(p) \pi^{\mu\nu}(p) \,,
\eea
and the scalar function $f_T(p^2)$ determined as usual, up to a multiplicative constant, by requiring the invariance under 
dilatations and special conformal transformations. We obtain
\bea 
\label{PNewCITwoPointEmt}
G^{\alpha\beta\mu\nu}_T(p) = \delta_{\eta_1 \eta_2}  \, 
c_{T 12}\,\frac{\pi^{d/2}}{4^{\eta_1 - d/2}} \frac{\Gamma(d/2 - \eta_1)}{\Gamma(\eta_1)}\, 
\Pi^{\alpha\beta\mu\nu}_{d}(p) \, (p^2)^{\eta_1 - d/2} \,.
\eea
As for the conserved vector currents, also for the energy momentum tensor the scaling dimension is fixed by the second of Eq.(\ref{PNewCIConformalEqMomTwoPoint})
and it is given by $\eta_1 = d$. This particular value ensures that $\partial^\mu T_{\mu\nu}$ is also a quasi primary (vector) field. 
We have left to the appendix \ref{PNewCIAppTwoPoint} the details of the characterization of the vector and tensor two-point functions.\\
These formulae agree with those in the literature \cite{CapFriedLaT:1991}, and in particular those in Sec. 8 of Ref. \cite{Antoniadis:2011ib}
for the gravitational wave spectrum of the CMB.

\subsection{Divergences and anomalous breaking of scale identities}

The expressions obtained so far for the two-point functions in Eq.(\ref{PNewCITwoPointScalar2}),(\ref{PNewCITwoPointVector}) and (\ref{PNewCITwoPointEmt}),
allow to discuss very easily the question of the divergences and of the corresponding violations that these induce in the scale identities.
We can naturally see this noting that the Gamma function has simple poles for non positive integer arguments, which occur, in our case, when $\eta = d/2 + n$ with $n=0,1,2,\ldots$. \\
Working in dimensional regularization, we can parametrize the divergence through an analytic continuation of the space-time dimension, $d \to d - 2 \epsilon$, and, then, expand the product $\Gamma(d/2-\eta)\,(p^2)^{\eta - d/2}$, which appears in every two-point function, in a Laurent series around $d/2 - \eta = -n$. We obtain
\bea
\label{PNewCIexpansion}
\Gamma\left(d/2-\eta\right)\,(p^2)^{\eta-d/2} = \frac{(-1)^n}{n!} \left( - \frac{1}{\epsilon} + \psi(n+1)  + O(\epsilon) \right) (p^2)^{n + \epsilon} \,,
\eea
where $\psi(z)$ is the logarithmic derivative of the Gamma function, and $\epsilon$ takes into account the divergence of the two-point correlator for particular values of the scale dimension $\eta$ and of the space-time dimension $d$.

The singular behavior described in Eq.(\ref{PNewCIexpansion}) is responsible for the anomalous 
violation of scale invariance \cite{BrownCol:1980}, providing an extra contribution to the differential equation (\ref{PNewCIConformalEqMomTwoPoint}) 
obtained from the conformal symmetry constraints. Indeed, when $\eta = d/2 + n$, employing dimensional regularization, the first of 
Eq.(\ref{PNewCIConformalEqMomTwoPoint}) becomes
\bea
\label{PNewCIAnomScale1}
\left( p^2\, \frac{\partial}{\partial p^2} - n - \epsilon \right)\, G^{ij}(p^2) = 0\, , \qquad \mbox{with} \quad \eta_1 = \eta_2 
\equiv \eta
\eea
which is the Euler equation for a function $G^{ij}$ which behaves like $(p^2)^{n+ \epsilon}$. Due to the appearance of a divergence in $1/\epsilon$
in the correlation function, Eq.(\ref{PNewCIAnomScale1}) acquires an anomalous finite term 
in the limit $\epsilon \to 0$ and we obtain
\bea
\label{PNewCIAnomScale2}
\left( p^2\, \frac{\partial}{\partial p^2} - n  \right)\, G^{ij}(p^2) = G^{ij}_{sing}(p^2) \,, 
\eea
where $G^{ij}_{sing}(p^2)$ corresponds to the singular contribution in the correlation function, which we have decomposed according to
\bea
G^{ij}(p^2) = \frac{1}{\epsilon} G^{ij}_{sing}(p^2) + G^{ij}_{finite}(p^2)  \,.
\eea
As one can see from the r.h.s. of Eq.(\ref{PNewCIAnomScale2}), the coefficient of the divergence, $G^{ij}_{sing}(p^2)$, of the two-point function provides the source for its anomalous scaling. 

We illustrate the points discussed so far with some examples.
Consider, for instance, the scalar correlator in Eq.(\ref{PNewCITwoPointScalar2}) with scaling dimension $\eta_1 = \eta_2 \equiv \eta = d/2$. Due to the appearance of a pole in the Gamma function, the two-point correlator develops a divergence and becomes
\bea
\label{PNewCI2PFscalarDiv}
G_S(p^2) = - \, c_{S12} \, \frac{\pi^{d/2}}{\Gamma \left( d/2 \right)} \left[ \frac{1}{\bar \epsilon} + \log p^2 \right] \,,
\eea
where we have defined for convenience
\bea
\label{PNewCIepsbar}
\frac{1}{\bar \epsilon} = \frac{1}{\epsilon} + \gamma - \log(4 \pi)  \,,
\eea
with $\gamma$ being the Euler-Mascheroni constant. It is implicitly understood that the argument of the logarithm in Eq.(\ref{PNewCI2PFscalarDiv}) is made dimensionless, in dimensional regularization, by the insertion of a massive parameter. \\
As one can easily verify, the scalar two-point function given in Eq.(\ref{PNewCI2PFscalarDiv}) satisfies the anomalous scaling equation (\ref{PNewCIAnomScale2}) with a constant source term 
\bea
G_{S,sing}(p^2) = - \, c_{S12} \, \frac{\pi^{d/2}}{\Gamma \left( d/2 \right)} 
\eea
determined by the coefficient of the singularity. Note that the anomalous scaling behavior in Eq.(\ref{PNewCIAnomScale2}) is reproduced by the logarithmic contribution in Eq.(\ref{PNewCI2PFscalarDiv}). 

Now we turn to the discussion of a correlation function with two vector currents. As already mentioned, the scaling dimension of the conserved vector operator is fixed at the value $\eta = d-1$. In this case the divergences occur at $d = 2n + 2$ with $n= 0,1,\ldots \,$, so that, for $d>2$, the first singularity appears at $d = 4$.
Therefore the vector two-point function for $d=4$ is
\bea 
\label{PNewCI2PFvectorDiv}
G^{\alpha\beta}_V(p^2) = c_{V12}\,\frac{\pi^{2}}{8}\,
p^2\, \left[ \frac{1}{\bar\epsilon} - 1 + \log p^2 \right] \pi^{\alpha\beta}(p) \,,
\eea
with $\bar \epsilon$ defined in Eq.(\ref{PNewCIepsbar}).
As for the previous case, it is manifest that the two-point function in Eq.(\ref{PNewCI2PFvectorDiv}) satisfies the identity given in 
Eq.(\ref{PNewCIAnomScale2}), with the logarithm accounting for the source of the anomalous scaling behavior. 

Finally, we illustrate the case of the correlation function built with two (symmetric, conserved and traceless) energy momentum 
tensors with scale dimension $\eta=d$,
which is slightly more involved, as we have to pay attention to the fact that $\Pi_d^{\alpha\beta\mu\nu}(p)$ itself depends on the 
space-time dimension $d$. 
The singularities are generated when $d = 2 n$ with $n= 0,1,\ldots \,$, namely for even values of the space-time dimension.
For instance, the two-point function in $d=4$ is found to be given by
\bea
\label{PNewCI2PFtensorDiv}
G_T^{\alpha\beta\mu\nu}(p) = 
- c_{T12}\, \frac{\pi^2}{192}\, (p^2)^2\,  \bigg\{
\left[ \frac{1}{\bar\epsilon} - \frac{3}{2} + \log p^2  \right]\, \Pi^{\alpha\beta\mu\nu}_4(p)
- \frac{2}{9}\, \pi^{\alpha\beta}(p)\pi^{\mu\nu}(p)
\bigg\} \, .
\eea
As we have already discussed previously, the appearance of the singularity in the correlation function develops an anomalous term in 
the scale identity. Correspondingly, being the energy momentum tensor related to the dilatation current, $J_D^\mu = x_\nu 
T^{\mu\nu}$, it acquires an anomalous trace reflecting the violation of the scale symmetry. In this respect, the two-point function 
in Eq.(\ref{PNewCI2PFtensorDiv}) is characterized by a non vanishing trace
\bea \label{PNewCItrace}
\eta_{\mu\nu} \, G_T^{\alpha\beta\mu\nu}(p) =  c_{T12}\, \frac{\pi^2}{288}\, (p^2)^2\, \pi^{\alpha\beta}(p) \,,
\eea 
generated by the last term in Eq.(\ref{PNewCI2PFtensorDiv}) which, on the other hand, arises from the explicit dependence of the 
$\Pi_d^{\alpha\beta\mu\nu}(p)$ tensor on the space-time dimension. 
The non-zero trace of Eq.(\ref{PNewCItrace}) is the signature of a conformal or trace anomaly, whose coefficients are known for free fields 
\cite{Duff:1977ay,AndMolMott:2003}.

\section{Three-point functions for scalar operators}

In this section we turn to the momentum space analysis of conformal invariant three-point functions, by solving the constraints emerging 
from the invariance under the conformal group. %
We consider scalar quasi primary fields $\mathcal O_i$ with scale dimensions $\eta_i$ and define the three-point function
\bea
G_{123}(p_1,p_2) = \langle \mathcal O_1(p_1) \mathcal O_2(p_2) \mathcal O_3(-p_1 - p_2) \rangle \,.
\eea
The three-point correlator is a function of the two independent momenta $p_1$ and $p_2$, from which one can construct 
three independent scalar quantities, namely $p_1^2$, 
$p_2^2$ and $p_1 \cdot p_2$. We trade the last invariant for $p_3^2$  in order to manifest the symmetry properties of $G_{123}$ under
the exchange of any couple of operators. \\
We observe that scale invariance, the first equation in Eq.(\ref{PNewCIConformalEqMom}), implies that $G_{123}$ is a homogeneous 
function of degree $\alpha = -d + \frac{1}{2}(\eta_1 + \eta_2 + \eta_3)$. Therefore it can be written in the form
\bea
\label{PNewCIDilatationSol}
G_{123}(p_1^2, p_2^2, p_3^2) = (p_3^2)^{-d + \frac{1}{2}(\eta_1 + \eta_2 + \eta_3)} \, \Phi(x,y)  \qquad \mbox{with} \qquad x = 
\frac{p_1^2}{p_3^2} \,, \quad y = \frac{p_2^2}{p_3^2} \,,
\eea
where we have introduced the dimensionless ratios $x$ and $y$, which must not be confused with coordinate points. The dilatation 
equation only fixes the scaling behavior of the three-point correlator giving no further information on the dimensionless function
$\Phi(x,y)$. 

The last equation of (\ref{PNewCIConformalEqMom}), which describes the invariance under special conformal transformations, is the most
predictive one and, as we shall see, completely determines $\Phi(x,y)$ up to a multiplicative constant. \\
To show this, we start by rewriting Eq.(\ref{PNewCIConformalEqMom}) in a more useful form by introducing a change of variables from $(p_1^2,p_2^2,p_3^2)$ to 
$(x,y,p_3^2)$. The derivatives respect to the momentum components are re-expressed in terms of derivatives of the momentum invariants and their ratios as
\bea
\frac{\partial}{\partial p_{1}^{\mu}}  &=&   2 (p_{1\, \mu} + p_{2 \, \mu}) \frac{\partial}{\partial p_3^2} + \frac{2}{p_3^2}\left( 
(1- x) p_{1 \, \mu}  - x  \,  p_{2 \, \mu} \right) \frac{\partial}{\partial x} - 2  (p_{1\, \mu} + p_{2 \, \mu}) \frac{y}{p_3^2} 
\frac{\partial}{\partial y} \,, \nn \\
\frac{\partial}{\partial p_{2}^{\mu}}  &=& 2 (p_{1\, \mu} + p_{2 \, \mu}) \frac{\partial}{\partial p_3^2}   -   2  (p_{1\, \mu} + 
p_{2 \, \mu}) \frac{x}{p_3^2} \frac{\partial}{\partial x}   + \frac{2}{p_3^2}\left( (1- y) p_{2 \, \mu}  - y  \,  p_{1 \, \mu} 
\right) \frac{\partial}{\partial y}. \, 
\eea
Similar but lengthier formulas hold for second derivatives. Also notice that the derivatives with respect to $p_3^2$ can 
be removed using the solution of the dilatation constraint in Eq.(\ref{PNewCIDilatationSol}). Therefore we are left with a differential 
equation in the two dimensionless variables $x$ and $y$. \\
Due to the vector nature of the special conformal transformations, Eq.(\ref{PNewCIConformalEqMom}) can be projected out on the two 
independent momenta $p_1$ and $p_2$, obtaining a system of two coupled second order partial differential equations (PDE) for the 
function $\Phi(x,y)$. After several non trivial manipulations, these can be recast in the simple form
\bea
\label{PNewCIF4diff.eq}
\begin{cases}
 \bigg[ x(1-x) \frac{\partial^2}{\partial x^2} - y^2 \frac{\partial^2}{\partial y^2} - 2 \, x \, y \frac{\partial^2}{\partial x \partial y} +  \left[ \gamma - (\alpha + \beta + 1) x \right] \frac{\partial}{\partial x} \nn \\
\hspace{8cm} - (\alpha + \beta + 1) y \frac{\partial}{\partial y}  - \alpha \, \beta \bigg] \Phi(x,y) = 0 \,, \nn \\
 \bigg[ y(1-y) \frac{\partial^2}{\partial y^2} - x^2 \frac{\partial^2}{\partial x^2} - 2 \, x \, y \frac{\partial^2}{\partial x \partial y} +  \left[ \gamma' - (\alpha + \beta + 1) y \right] \frac{\partial}{\partial y} \nn \\
\hspace{8cm} - (\alpha + \beta + 1) x \frac{\partial}{\partial x}  - \alpha \, \beta \bigg] \Phi(x,y) = 0 \,, 
\end{cases} 
\\
\eea
with the parameters $\alpha, \beta, \gamma, \gamma'$ defined in terms of the scale dimensions of the three scalar operators as
\begin{align}
\label{PNewCIcoeffF4}
& \alpha = \frac{d}{2} - \frac{ \eta_1 + \eta_2 - \eta_ 3 }{2} \,,  && \gamma = \frac{d}{2} - \eta_1 + 1 \,, \nn \\
& \beta = d - \frac{\eta_1 + \eta_2 + \eta_3}{2}  \,, && \gamma' = \frac{d}{2} - \eta_2 + 1 \,.
\end{align}
It is interesting to observe that the system of equations in (\ref{PNewCIF4diff.eq}), coming from the invariance under special 
conformal transformations, is exactly the system of partial differential equations defining the hypergeometric Appell's function of 
two variables, $F_4(\alpha, \beta; \gamma, \gamma'; x, y)$, with coefficients given in Eq.(\ref{PNewCIcoeffF4}). 
The Appell's function $F_4$ is defined as the double series (see, e.g., \cite{AppellBook,BaileyBook,SlaterBook} for thorough 
discussions of the hypergeometric functions and their properties)
\bea
\label{PNewCIF4def}
F_4(\alpha, \beta; \gamma, \gamma'; x, y) = \sum_{i = 0}^{\infty}\sum_{j = 0}^{\infty} \frac{(\alpha)_{i+j} \, 
(\beta)_{i+j}}{(\gamma)_i \, (\gamma')_j} \frac{x^i}{i!} \frac{y^j}{j!} 
\eea
where $(\alpha)_i = \Gamma(\alpha + i)/ \Gamma(\alpha)$ is the Pochhammer symbol. \\ 
It is known that the system of partial differential equations (\ref{PNewCIF4diff.eq}), besides the hypergeometric function introduced in 
Eq.(\ref{PNewCIF4def}), has three other independent solutions given by
\bea
\label{PNewCIsolutions}
S_2(\alpha, \beta; \gamma, \gamma'; x, y) &=& x^{1-\gamma} \, F_4(\alpha-\gamma+1, \beta-\gamma+1; 2-\gamma, \gamma'; x,y) \,, \nn \\
S_3(\alpha, \beta; \gamma, \gamma'; x, y) &=& y^{1-\gamma'} \, F_4(\alpha-\gamma'+1,\beta-\gamma'+1;\gamma,2-\gamma' ; x,y) \,, \nn \\
S_4(\alpha, \beta; \gamma, \gamma'; x, y) &=& x^{1-\gamma} \, y^{1-\gamma'} \, 
F_4(\alpha-\gamma-\gamma'+2,\beta-\gamma-\gamma'+2;2-\gamma,2-\gamma' ; x,y) \, . \nn \\
\eea
Therefore the function $\Phi(x,y)$, solution of (\ref{PNewCIF4diff.eq}), is a linear combination of the four independent hypergeometric 
functions, i.e.
\bea
\label{PNewCISCTSol}
G_{123}(p_1^2, p_2^2, p_3^2) &=& (p_3^2)^{-d + \frac{1}{2}(\eta_1 + \eta_2 + \eta_3)} \, \Phi(x,y) \nn \\
&=& (p_3^2)^{-d + \frac{1}{2}(\eta_1 + \eta_2 + \eta_3)} \sum_{i=1}^{4} c_i(\eta_1,\eta_2,\eta_3) \, S_i(\alpha, \beta; \gamma, 
\gamma'; x, y) \,,
\eea
where we have denoted with $S_1$ the Appell's function $F_4$ given in Eq.(\ref{PNewCIF4def}), while the parameters 
$\alpha,\beta,\gamma,\gamma'$ are defined in Eq.(\ref{PNewCIcoeffF4}). The $c_i(\eta_1,\eta_2,\eta_3)$ appearing in the linear combination, 
are the arbitrary coefficients which may depend on the scale dimensions $\eta_i$ of the quasi primary fields and on the space-time 
dimension $d$. \\
The coefficients $c_i(\eta_1,\eta_2,\eta_3)$ can be determined, up to an overall multiplicative constant, by exploiting the symmetry 
of the correlation function under the interchange of two of the three scalar operators present in the correlator, which consists of the simultaneous exchange of momenta and 
scale dimensions $(p_i^2, \eta_i) \leftrightarrow (p_j^2,\eta_j)$. \\
Consider, for instance, the invariance of the three-point function under the exchange $\mathcal O_2(p_2) \leftrightarrow \mathcal 
O_3(-p_1-p_2)$, which is achieved by $(p_2^2, \eta_2) \leftrightarrow (p_3^2,\eta_3)$. Then Eq.(\ref{PNewCISCTSol}) becomes
\bea
G_{132}(p_1^2, p_3^2, p_2^2) = (p_2^2)^{-d + \frac{1}{2}(\eta_1 + \eta_2 + \eta_3)} \sum_{i=1}^{4} c_i(\eta_1,\eta_3,\eta_2) \, S_i(\tilde \alpha, \tilde \beta; \tilde \gamma, \tilde \gamma'; \frac{x}{y}, \frac{1}{y}) \,,
\eea
where
\begin{align}
& \tilde \alpha = \alpha(\eta_2 \leftrightarrow \eta_3) = \frac{d}{2} - \frac{ \eta_1 + \eta_3 - \eta_ 2 }{2} \,,  && \tilde \gamma = \gamma(\eta_2 \leftrightarrow \eta_3) = \frac{d}{2} - \eta_1 + 1 = \gamma \,, \nn \\
& \tilde \beta = \beta(\eta_2 \leftrightarrow \eta_3) = d - \frac{\eta_1 + \eta_2 + \eta_3}{2} = \beta \,, && \tilde \gamma' = \gamma'(\eta_2 \leftrightarrow \eta_3) = \frac{d}{2} - \eta_3 + 1 \,.
\end{align}
Note that the hypergeometric functions are now evaluated in $x/y$ and $1/y$. To reintroduce the dependence from $x$ and $y$, in order to exploit more easily the symmetry relation
\bea
G_{123}(p_1^2, p_2^2, p_3^2) = G_{132}(p_1^2, p_3^2, p_2^2) \,,
\eea 
we make use of the transformation property of $F_4$ \cite{AppellBook}
\bea
\label{PNewCItransfF4}
F_4(\alpha, \beta; \gamma, \gamma'; x, y) &=& \frac{\Gamma(\gamma') \Gamma(\beta - \alpha)}{ \Gamma(\gamma' - \alpha) \Gamma(\beta)} (- y)^{- \alpha} \, F_4(\alpha, \alpha -\gamma' +1; \gamma, \alpha-\beta +1; \frac{x}{y}, \frac{1}{y}) \nn \\
&+&  \frac{\Gamma(\gamma') \Gamma(\alpha - \beta)}{ \Gamma(\gamma' - \beta) \Gamma(\alpha)} (- y)^{- \beta} \, F_4(\beta -\gamma' +1, \beta ; \gamma, \beta-\alpha +1; \frac{x}{y}, \frac{1}{y}) \,.
\eea
After some algebraic manipulations, and repeating the procedure described so far for the other operator interchanges, the ratios between the coefficients $c_i$ take the simplified form
\bea
\frac{c_1(\eta_1, \eta_2, \eta_3)}{c_3(\eta_1, \eta_2, \eta_3)} &=& 
\frac{\Gamma \left(\eta _2-\frac{d}{2}\right) \Gamma
   \left(d-\frac{\eta _1}{2}-\frac{\eta _2}{2}-\frac{\eta
   _3}{2}\right) \Gamma \left(\frac{d}{2}-\frac{\eta
   _1}{2}-\frac{\eta _2}{2}+\frac{\eta _3}{2}\right)}{\Gamma \left(\frac{d}{2}-\eta _2\right) \Gamma
   \left(-\frac{\eta _1}{2}+\frac{\eta _2}{2}+\frac{\eta
   _3}{2}\right)  \Gamma
   \left(\frac{d}{2}-\frac{\eta _1}{2}+\frac{\eta
   _2}{2}-\frac{\eta _3}{2}\right)} \,, \nn \\
\frac{c_2(\eta_1, \eta_2, \eta_3)}{c_4(\eta_1, \eta_2, \eta_3)} &=& 
\frac{\Gamma \left(\eta
   _2-\frac{d}{2}\right)  \Gamma \left(\frac{\eta _1}{2}-\frac{\eta
   _2}{2}+\frac{\eta _3}{2}\right) \Gamma \left(\frac{d}{2}+\frac{\eta
   _1}{2}-\frac{\eta _2}{2}-\frac{\eta _3}{2}\right)}{ \Gamma \left(\frac{d}{2}-\eta _2\right) \Gamma
   \left(\frac{\eta _1}{2}+\frac{\eta _2}{2}-\frac{\eta
   _3}{2}\right) \Gamma
   \left(-\frac{d}{2}+\frac{\eta _1}{2}+\frac{\eta
   _2}{2}+\frac{\eta _3}{2}\right)} \,, \nn \\
\frac{c_1(\eta_1, \eta_2, \eta_3)}{c_4(\eta_1, \eta_2, \eta_3)} &=& 
\frac{\Gamma \left(\eta _1-\frac{d}{2}\right) \Gamma \left(\eta
   _2-\frac{d}{2}\right) \Gamma \left(d-\frac{\eta
   _1}{2}-\frac{\eta _2}{2}-\frac{\eta _3}{2}\right) \Gamma
   \left(\frac{d}{2}-\frac{\eta _1}{2}-\frac{\eta
   _2}{2}+\frac{\eta _3}{2}\right)}{\Gamma
   \left(\frac{d}{2}-\eta _1\right) \Gamma
   \left(\frac{d}{2}-\eta _2\right) \Gamma \left(\frac{\eta
   _1}{2}+\frac{\eta _2}{2}-\frac{\eta _3}{2}\right) \Gamma
   \left(-\frac{d}{2}+\frac{\eta _1}{2}+\frac{\eta
   _2}{2}+\frac{\eta _3}{2}\right)} \,,
\eea
and define $G_{123}(p_1^2, p_2^2, p_3^2)$ up to a multiplicative arbitrary constant $c_{123} \equiv c_{123}(\eta_1, \eta_2, \eta_3)$. This depends on the space-time dimension $d$, on the scale dimensions $\eta_i$ of the quasi primary fields and on their normalization. \\
The conformal invariant correlation function of three scalar quasi primary fields with arbitrary scale dimensions is then given by
\small
\bea
\label{PNewCIThreePointScalar}
G_{123}(p_1^2, p_2^2, p_3^2) &=&  \frac{c_{123} \,\, \pi^d \, 4^{d - \frac{1}{2}(\eta_1 + \eta_2 + \eta_3)} \, (p_3^2)^{-d + \frac{1}{2}(\eta_1 + \eta_2 + \eta_3)}}{\Gamma\left( \frac{\eta_1}{2} + \frac{\eta_2}{2} - \frac{\eta_3}{2}\right) \Gamma\left( \frac{\eta_1}{2} - \frac{\eta_2}{2} + \frac{\eta_3}{2}\right) \Gamma\left( - \frac{\eta_1}{2} + \frac{\eta_2}{2} + \frac{\eta_3}{2}\right) \Gamma\left( - \frac{d}{2} + \frac{\eta_1}{2} + \frac{\eta_2}{2} + \frac{\eta_3}{2}\right)} \bigg\{ \nn \\
&& \hspace{-2cm} \Gamma \left(\eta _1-\frac{d}{2}\right) \Gamma \left(\eta
   _2-\frac{d}{2}\right) \Gamma \left(d-\frac{\eta
   _1}{2}-\frac{\eta _2}{2}-\frac{\eta _3}{2}\right) \Gamma
   \left(\frac{d}{2}-\frac{\eta _1}{2}-\frac{\eta
   _2}{2}+\frac{\eta _3}{2}\right) \nn \\
&& \hspace{-2cm} \times \, F_4 \left( \frac{d}{2} - \frac{\eta_1 + \eta_2 - \eta_3}{2}, d - \frac{\eta_1 + \eta_2 + \eta_3}{2}; \frac{d}{2} - \eta_1 +1, \frac{d}{2} - \eta_2 +1; x, y \right) \nn \\
&& \hspace{-2cm} +\,   \Gamma \left(\frac{d}{2}-\eta _1\right) \Gamma
   \left(\eta _2-\frac{d}{2}\right) \Gamma \left(\frac{\eta _1}{2}-\frac{\eta _2}{2}+\frac{\eta
   _3}{2}\right) \Gamma
   \left(\frac{d}{2}+\frac{\eta _1}{2}-\frac{\eta
   _2}{2}-\frac{\eta _3}{2}\right) \nn \\
&& \hspace{-2cm} \times \, x^{\eta_1 - \frac{d}{2}} \, F_4\left( \frac{d}{2} - \frac{\eta_2 + \eta_3 - \eta_1}{2}, \frac{\eta_1 + \eta_3 - \eta_2}{2}; - \frac{d}{2} + \eta_1 +1, \frac{d}{2} - \eta_2 +1 ; x, y\right) \nn \\
&& \hspace{-2cm} + \, \Gamma \left(\eta _1-\frac{d}{2}\right) \Gamma
   \left(\frac{d}{2}-\eta _2\right) \Gamma \left(-\frac{\eta _1}{2}+\frac{\eta _2}{2}+\frac{\eta
   _3}{2}\right) \Gamma
   \left(\frac{d}{2}-\frac{\eta _1}{2}+\frac{\eta
   _2}{2}-\frac{\eta _3}{2}\right) \nn \\
&& \hspace{-2cm} \times \, y^{\eta_2 - \frac{d}{2}} \, F_4\left( \frac{d}{2} - \frac{\eta_1 + \eta_3 - \eta_2}{2} , \frac{\eta_2 + \eta_3 - \eta_1}{2}; 
\frac{d}{2} - \eta_1 +1, -\frac{d}{2} + \eta_2 +1 ; x, y\right) \nn \\
&& \hspace{-2cm} + \, \Gamma \left(\frac{d}{2}-\eta _1\right) \Gamma
   \left(\frac{d}{2}-\eta _2\right) \Gamma \left(\frac{\eta _1}{2}+\frac{\eta _2}{2}-\frac{\eta
   _3}{2}\right) \Gamma
   \left(-\frac{d}{2}+\frac{\eta _1}{2}+\frac{\eta
   _2}{2}+\frac{\eta _3}{2}\right) \nn \\
&& \hspace{-2cm} \times \,   x^{\eta_1 - \frac{d}{2}}  y^{\eta_2 - \frac{d}{2}} \, F_4 \left( -\frac{d}{2} + \frac{\eta_1+\eta_2+\eta_3}{2}, 
\frac{\eta_1+\eta_2-\eta_3}{2}; -\frac{d}{2} +\eta_1 +1, -\frac{d}{2} + \eta_2 +1; x, y \right) \bigg\} \,. \nn \\
\eea 
\normalsize
The convenient normalization employed in Eq.(\ref{PNewCIThreePointScalar}) for the three-point function reproduces, through the operator 
product expansion, as we are going to show next, the normalization of the two-point functions which we have chosen in 
Eq.(\ref{PNewCITwoPointScalar2}). 

As we shall identify the three-point correlator discussed in this section with specific Feynman amplitudes, this will fix the arbitrary constant $c_{123}$ using some information coming from the same operator product expansion analysis. This topic will be presented in section \ref{PNewCISection.Davy}. Indeed, the solution of the momentum space version of the conformal constraints provides an alternative computational tool for correlation functions with conformal symmetry.

It is worth to emphasize the connection between the invariance under special conformal transformations and appearance of the Appell's functions. Indeed we have shown how the constraints provided by the conformal group translate, in momentum space, in the well-known system of partial differential equations defining the hypergeometric series $F_4$. We have analyzed this connection in the case of a conformally invariant three-point function built with scalar operators in some detail. A similar correspondence should also hold for more complicated vector and tensor correlators.

\subsection{The Operator Product Expansion analysis}

In this section we show the consistency of our result with the operator product expansion (OPE) in conformal field theories in which
the structure of the Wilson's coefficients is entirely fixed by the scaling dimensions of the two operators. \\
Considering, for instance, the coincidence limit in the scalar case, one has
\bea
\mathcal O_i(x_1) \mathcal O_j(x_2) \sim \sum_k \frac{c_{ijk}}{(x_{12}^2)^{\frac{1}{2}(\eta_i + \eta_j - \eta_k)}} \mathcal O_k(x_2)  
\qquad \mbox{for} \quad x_1 \rightarrow x_2 \,,
\eea
where $x_{12} = x_1 - x_2$. It is worth noting that the coefficients $c_{ijk}$ are the same structure constants 
appearing in the three-point functions. \\
For the correlation function of three scalar operators the OPE implies the singular behavior
\bea
\label{PNewCIOPEcoord}
\langle \mathcal O_1(x_1) \mathcal O_2(x_2) \mathcal O_3(x_3) \rangle \stackrel{x_3 \rightarrow x_2}{\sim} 
\frac{c_{123}}{(x_{23}^2)^{\frac{1}{2}(\eta_2 + \eta_3- \eta_1)}} \langle \mathcal O_1(x_1) \mathcal O_2(x_2) \rangle \,,
\eea
with analogous formulae for the other coincidence limits. For the sake of simplicity, we choose a diagonal basis of quasi primary operators normalized as
\bea
\langle \mathcal O_{i}(x_1) \mathcal O_{j}(x_2) \rangle = \frac{\delta_{i j}}{(x_{12}^2)^{\eta_i}} \,. 
\eea
The momentum space version of the OPE in Eq.(\ref{PNewCIOPEcoord}) reads
\bea
\label{PNewCIOPEmom}
\langle \mathcal O_1(p_1) \mathcal O_2(p_2) \mathcal O_3(-p_1-p_2) \rangle \nn \\
&& \hspace{-4cm } \sim 
\frac{\pi^{d/2}}{4^{\frac{1}{2}(\eta_2+\eta_3-\eta_1)-\frac{d}{2}}} 
\frac{\Gamma(\frac{d}{2}-\frac{\eta_2+\eta_3-\eta_1}{2})}{\Gamma(\frac{\eta_2+\eta_3-\eta_1}{2})} 
\frac{c_{123}}{(p_3^2)^{\frac{d}{2}-\frac{1}{2}(\eta_2+\eta_3-\eta_1)}} \langle \mathcal O_1(p_1) \mathcal O_2(-p_1)\rangle\, ,
\eea
where the scalar two-point function is normalized as in Eq.(\ref{PNewCITwoPointScalar2}) with $c_{S12} = 1$. In the previous equation the 
symbol $\sim$ stands for the momentum space counterpart of the short distance limit $x_3 \rightarrow x_2$ which is achieved by the 
$p_3^2, p_2^2 \rightarrow \infty$ limit with $p_2^2 / p_3^2 \rightarrow 1$.  \\
The result for the scalar three-point function given in Eq.(\ref{PNewCIThreePointScalar}) is indeed in agreement, as expected, with the OPE 
analysis. This can be shown from Eq.(\ref{PNewCIThreePointScalar}) by a suitable expansion of the corresponding Appell's functions. In 
particular, in order to reproduce the momentum space singular behavior of Eq.(\ref{PNewCIOPEmom}), we need the hypergeometric leading 
expansion in the limit $x = p_1^2/p_3^2 \rightarrow 0$ and $y = p_2^2/p_3^2 \rightarrow 1$, which reads as \cite{AppellBook}
\bea
F_4(\alpha, \beta; \gamma, \gamma'; x, y) \sim \frac{\Gamma(\gamma') \Gamma(\gamma' -\alpha -\beta)}{\Gamma(\gamma' - \alpha) \Gamma(\gamma' - \beta)} \qquad \mbox{for} \quad x \rightarrow 0 \,, y \rightarrow 1 	\,.
\eea
In the previous equation we have retained only the terms with the correct power-law scaling in the $p_3^2$ variable, as dictated by the OPE analysis. In this case these contributions come from the terms of Eq.(\ref{PNewCIThreePointScalar}) which are proportional to the $S_2$ and $S_4$ solutions defined in Eq.(\ref{PNewCIsolutions}).
Analogously, in the limit $p_3^2, p_1^2 \rightarrow \infty$, with $p_1^2 / p_3^2 \rightarrow 1$, which is described in coordinate space by $x_3 \rightarrow x_1$, the leading behavior is extracted from $S_3$ and $S_4$. \\
The remaining coincidence limit $x_1 \rightarrow x_2$, corresponding to $p_1^2, p_2^2 \rightarrow \infty$ with $p_1^2 / p_2^2 \rightarrow 1$, is more subtle due to the apparent asymmetry in the momentum invariants $p_1^2, p_2^2, p_3^2$ of the three-point scalar correlator, as given in Eq.(\ref{PNewCIThreePointScalar}). In this case both $x$ and $y$ grow to infinity while their ratio $x/y \rightarrow 1$. Therefore it is necessary to apply the transformation defined in Eq.(\ref{PNewCItransfF4}) to each hypergeometric function appearing in Eq.(\ref{PNewCIThreePointScalar}). This can be viewed as an analytic continuation outside the domain of convergence $|\sqrt{x}| + |\sqrt{y}| < 1$, where the Appell's function is strictly defined as a double series.
The hypergeometric functions are then expanded according to
\bea
F_4(\alpha, \beta; \gamma, \gamma'; x, y) & \sim& (- y)^{- \alpha} \frac{ \Gamma(\gamma) \Gamma(\gamma') \Gamma(\beta - \alpha) 
\Gamma(\gamma  + \gamma' - 2 \alpha - 1)}{\Gamma(\beta)  \Gamma(\gamma -\alpha)  
\Gamma(\gamma' - \alpha)\Gamma(\gamma + \gamma' - \alpha - 1)}   \nn \\
&+& 
(- y)^{- \beta}   \frac{\Gamma(\gamma) \Gamma(\gamma') \Gamma(\alpha - \beta)  \Gamma(\gamma + \gamma' - 2 \beta - 1)  }{ 
\Gamma(\alpha)  \Gamma(\gamma' - \beta) \Gamma(\gamma - \beta)  \Gamma(\gamma + \gamma' - \beta - 1) }\, ,  \nn \\
&\mbox{for}& 
 \quad x, y \rightarrow \infty \,, \frac{x}{y} \rightarrow 1 \,. 
\eea
This completes the analysis of the OPE on the three-point scalar function in the three different coincidence limits.

\section{Feynman integral representation of the momentum space solution}
\label{PNewCISection.Davy}

We have seen in the previous sections that we can fix the explicit structure of the generic three-point scalar correlator 
in momentum space by solving the conformal constraints, which are mapped to a system of two hypergeometric differential equations of two variables. These variables take the form of 
two ratios of the external momenta. 
In particular we find that in any $d$ dimensional conformal field theory the solution of this system of PDE's is characterized by a single integration constant which depends on the 
specific conformal realization, as expected. 

In this section we want to point out the relationship between the scalar three-point functions studied so far and a certain class of Feynman master integrals.
These can be obtained by a Fourier transformation of the corresponding solution of the conformal constraints in coordinate space, which is well known to be
\bea \label{PNewCIOOO}
\langle \mathcal O_1(x_1) \mathcal O_2(x_2) \mathcal O_3(x_3)\rangle = \frac{c_{123}}{ \left(x_{12}^2\right)^{\frac{1}{2}(\eta_1 +\eta_2-\eta_3)}
\left(x_{23}^2\right)^{\frac{1}{2}(\eta_2 +\eta_3-\eta_1)} \left(x_{31}^2\right)^{\frac{1}{2}(\eta_3 +\eta_1-\eta_2)}}\, .
\eea
Transforming to momentum space, we find an integral representation, which necessarily has to coincide, up to an unconstrained overall constant, 
with the explicit solution found in the previous section, and reads as
\bea
\label{PNewCIdavy}
J(\nu_1,\nu_2,\nu_3) = \int \frac{d^d l}{(2 \pi)^d} \frac{1}{(l^2)^{\nu_3} ((l+p_1)^2)^{\nu_2} ((l-p_2)^2)^{\nu_1}}\, ,
\eea
with external momenta $p_1$, $p_2$ and $p_3$ constrained by momentum conservation $p_1 + p_2 + p_3 = 0$ and
the scale dimensions $\eta_i$ related to the indices $\nu_i$ as
\bea
\label{PNewCIetafromnu}
\eta_1 = d - \nu_2 - \nu_3 \,, \qquad
\eta_2 = d - \nu_1 - \nu_3 \,, \qquad
\eta_3 = d - \nu_1 - \nu_2 \,. 
\eea
This expression describes a family of master integrals which has been studied in \cite{Boos:1987bg, Davydychev:1992xr},
whose explicit relation with Eq.(\ref{PNewCIOOO}) is given by
\bea
&& \int \frac{d^d p_1}{(2\pi)^d} \frac{d^d p_2}{(2\pi)^d} \frac{d^d p_3}{(2\pi)^d} \, (2\pi)^d \delta^{(d)}(p_1 + p_2 + p_3) \, 
J(\nu_1,\nu_2,\nu_3) e^{- i p_1 \cdot x_1 - i p_2 \cdot x_2 - i p_3 \cdot x_3} \nn \\
&& = \frac{1}{4^{\nu_1+\nu_2+\nu_3} \pi^{3 d/2}}  \frac{\Gamma(d/2 - \nu_1) \Gamma(d/2 - \nu_2) \Gamma(d/2 - \nu_3)}{\Gamma(\nu_1) 
\Gamma(\nu_2) \Gamma(\nu_3)}  \frac{1}{(x_{12}^2)^{d/2- \nu_3} (x_{23}^2)^{d/2- \nu_1} (x_{31}^2)^{d/2- \nu_2}}\,, \nn \\
\eea
The integral in Eq.(\ref{PNewCIdavy}) satisfies the system of PDE's (\ref{PNewCIF4diff.eq}). 
Therefore, it can be expressed in terms of the general solution given in Eq.(\ref{PNewCIThreePointScalar}) which involves a linear 
combination of four Appell's functions, with the relative coefficients fixed by the symmetry conditions on the dependence from the external momenta.
Then Eq.(\ref{PNewCIThreePointScalar}) identifies $J(\nu_1,\nu_2,\nu_3)$ except for an overall  constant $c_{123}$ which we are now going to determine. This task can be 
accomplished, for instance, by exploiting some boundary conditions. \\
As for the OPE analysis discussed in the previous section, we may consider the large momentum limit in which the three-point integral 
collapses into a two-point function topology. 
Taking, for instance, the $p_2^2, p_3^2 \rightarrow \infty$ limit with $p_2^2/p_3^2 \rightarrow 1$ we have
\bea
\label{PNewCIdavylimit}
J(\nu_1,\nu_2,\nu_3) \sim \frac{1}{(p_2^2)^{\nu_1}} \int \frac{d^d l}{(2 \pi)^d} \frac{1}{(l^2)^{\nu_3} ((l + p_1)^2)^{\nu_2}} =  \frac{1}{(p_2^2)^{\nu_1}}  \frac{i^{1-d}}{(4 \pi)^{d/2}} \, G(\nu_2,\nu_3) \, (p_1^2)^{d/2 - \nu_2 - \nu_3} \,,
\eea
where
\bea
G(\nu, \nu') = \frac{\Gamma(d/2 - \nu) \Gamma(d/2 - \nu') \Gamma(\nu+\nu'-d/2)}{\Gamma(\nu) \Gamma(\nu') \Gamma(d - \nu -\nu')} \,.
\eea
Eq.(\ref{PNewCIdavylimit}) must be compared with the same limit taken on the explicit solution in Eq.(\ref{PNewCIThreePointScalar}), where the scale dimensions $\eta_i$ are replaced by $\nu_i$ through Eq.(\ref{PNewCIetafromnu}). This completely determines the multiplicative constant $c_{123}$ and the correct normalization of the three-point master integral, which is obtained by choosing 
\bea
\label{PNewCIc123}
c_{123} = \frac{i^{1-d}}{4^{\nu_1+\nu_2+\nu_3} \pi^{3 d/2}}  \frac{\Gamma(d/2 - \nu_1) \Gamma(d/2 - \nu_2) \Gamma(d/2 - \nu_3)}{\Gamma(\nu_1) \Gamma(\nu_2) \Gamma(\nu_3)}  \,.
\eea
Therefore the scalar master integral is given by
\bea
J(\nu_1,\nu_2,\nu_3) = G_{123}(p_1^2, p_2^2, p_3^2)
\eea
with scaling dimensions defined in Eq.(\ref{PNewCIetafromnu}) and the coefficient $c_{123}$ in Eq.(\ref{PNewCIc123}). Notice that this method allows us to bypass completely the Mellin-Barnes techniques which has been used previously in the analysis of the same integral.

\subsection{Recurrence relations from conformal invariance}

Having established the conformal invariance of the generalized three-point master integral $J(\nu_1,\nu_2,\nu_3)$, we can study the 
implications of the conformal constraints on the integral representation of Eq.(\ref{PNewCIdavy}). These are automatically satisfied by the explicit solution given in Eq.(\ref{PNewCIThreePointScalar}), but once that they are applied on 
$J(\nu_1,\nu_2,\nu_3)$, generate recursion relations among the indices of this family of integrals. Specifically, they relate integrals with $\nu_1+\nu_2+\nu_3=\kappa$ to those with $\nu_1+\nu_2+\nu_3=\kappa +1$ and 
$\nu_1+\nu_2+\nu_3=\kappa +2$. For instance, differentiating Eq.(\ref{PNewCIdavy}) under the integration sign according to the first of Eq.(\ref{PNewCIConformalEqMom}), which is the condition of scale invariance, 
we easily obtain the recursion relation
\bea 
\label{PNewCIdavyscale}
\nu_2\,p_1^2\,J(\nu_1,\nu_2+1,\nu_3) + \nu_1\,p_2^2\,J(\nu_1+1,\nu_2,\nu_3) 
&=&
\left(\nu_1 + \nu_2 + 2\, \nu_3 - d \right) \, J(\nu_1,\nu_2,\nu_3) \nn \\
&& \hspace{-4cm} + \,
\nu_2\, J(\nu_1,\nu_2+1,\nu_3-1) + \nu_1\, J(\nu_1+1,\nu_2,\nu_3-1) \, , 
\eea
together with the corresponding symmetric relations obtained interchanging 
$(p_1^2, \nu_1) \leftrightarrow (p_3^2, \nu_3)$ or $(p_2^2, \nu_2) \leftrightarrow (p_3^2, \nu_3)$. 
These equations link scalar integrals on two contiguous planes, as mentioned above. 
The recurrence relations obtained from scale invariance exactly correspond to those presented in \cite{Davydychev:1992xr}
and following from the usual integration-by-parts technique, which in this case is derived from the divergence theorem in dimensional regularization
\bea
\int \frac{d^d l}{(2\pi)^d}\, \frac{\partial}{\partial l_{\mu}}\,
\left\{ \frac{l_{\mu}}{(l^2)^{\nu_3}((l+p_1)^2)^{\nu_2}((l-p_2)^2)^{\nu_1}} \right\}= 0.
\label{PNewCIparts}
\eea
We can easily show the equivalence between Eq.(\ref{PNewCIparts}) and the first of Eq.(\ref{PNewCIConformalEqMom}) which is the constraint of scale invariance.  
In fact, the scale transformation acts on $J(\nu_1,\nu_2,\nu_3)$ in the form 
\bea \label{PNewCIdavyder}
\left[d - 2\,\left(\nu_1 + \nu_2 + \nu_3 \right)
- p_1\cdot \frac{\partial}{\partial p_1} - p_2\cdot \frac{\partial}{\partial p_2} \right]
\int d^d l   \frac{1}{(l^2)^{\nu_3}\,((l+p_1)^{2})^{\nu_2}\,((l-p_2)^{2})^{\nu_1}} = 0 \, .
\eea
Now we just invoke Euler's theorem on homogeneous functions on the integrand, which is of degree 
$- 2\,(\nu_1+\nu_2+\nu_3)$ in the momenta $p_1$, $p_2$ and $l$  and obtain the relation
\bea
&&
\left[ p_1\cdot \frac{\partial}{\partial p_1} + p_2\cdot \frac{\partial}{\partial p_2}
+ l\cdot \frac{\partial}{\partial l} \right] \frac{1}{(l^2)^{\nu_3}\,((l+p_1)^{2})^{\nu_2}\,((l-p_2)^{2})^{\nu_1}} \nn \\\
&& \hspace{8cm}
= \frac{-2 (\nu_1+\nu_2+\nu_3) }{(l^2)^{\nu_3}\,((l+p_1)^{2})^{\nu_2}\,((l-p_2)^{2})^{\nu_1}} \, .
\label{PNewCIeuler}
\eea
At this point, if we combine Eqs.(\ref{PNewCIdavyder}) and (\ref{PNewCIeuler}) and rewrite $d$ as $\frac{\partial}{\partial l} \cdot l$, 
we easily obtain the equivalence with Eq.(\ref{PNewCIparts}).

Other recursive relations can be found requiring Eq.(\ref{PNewCIdavy}) to satisfy the constraint of special conformal invariance 
which, from the second equation in Eq.(\ref{PNewCIConformalEqMom}), takes the form 
\beq
\left\{ p_{1\,\mu} \frac{\partial^2}{\partial p_1 \cdot \partial p_1} - 2\, p_{1\,\nu} \frac{\partial^2}{\partial p_{1}^{\mu}\partial p_{1\,\nu}}
- 2\,(\nu_2+\nu_3)\frac{\partial}{\partial p_{1}^{\mu}}+ (1 \leftrightarrow 2) \right\}\, J(\nu_1,\nu_2,\nu_3) = 0\, .
\label{PNewCIspecial}
\eeq
This is a vector condition which involves some tensor integrals of the same $J(\nu_1,\nu_2,\nu_3)$ family.  
Differentiating the integral $J(\nu_1,\nu_2,\nu_3)$ as in Eq.(\ref{PNewCIspecial}) and performing some standard manipulations one arrives at the implicit formula
\small
\bea
\label{PNewCIvectorid}
&& \hspace{-0.6cm}
\nu_2 \, p_{1\, \mu} \bigg[
\left(1 +\nu_2 + \nu_3-d/2\right) J(\nu_1,\nu_2+1,\nu_3)
+ (\nu_2+1) \left( J(\nu_1,\nu_2+2,\nu_3-1) - p_1^2\, J(\nu_1,\nu_2+2,\nu_3)\right)
\bigg] \nn \\
&&
\hspace{-0.6cm} 
+ \, \nu_1 \, p_{2 \, \mu} \bigg[
\left(1 +\nu_1 + \nu_3-d/2\right) J(\nu_1+1,\nu_2,\nu_3)
+ (\nu_1+1)\left( J(\nu_1+2,\nu_2,\nu_3-1) - p_2^2\, J(\nu_1+2,\nu_2,\nu_3)\right)
\bigg] \nn \\
&&
\hspace{-0.6cm} 
+ \, \nu_2 \, \bigg[ (\nu_3-1)\,  J_{\mu}(\nu_1,\nu_2+1,\nu_3) + (\nu_2+1) \left( J_{\mu}(\nu_1,\nu_2+2,\nu_3-1)
-  p_1^2\, J_{\mu}(\nu_1,\nu_2 + 2,\nu_3) \right) \bigg] \nn \\
&&
\hspace{-0.6cm} 
- \, \nu_1 \, \bigg[ (\nu_3-1)\,  J_{\mu}(\nu_1+1,\nu_2,\nu_3) + (\nu_1+1) \left( J_{\mu}(\nu_1+2,\nu_2,\nu_3-1) -  p_2^2\, J_{\mu}(\nu_1+2,\nu_2,\nu_3) \right) \bigg] = 0 \, , 
\eea
\normalsize
where the rank-$1$ tensor integral is defined as
\beq
J_{\mu}(\nu_1,\nu_2,\nu_3) = \int \frac{d^d l}{(2\pi)^d}\, 
\frac{l_{\mu}}{(l^2)^{\nu_3}((l+p_1)^2)^{\nu_2}((l-p_2)^2)^{\nu_1}}  = C_1(\nu_1,\nu_2,\nu_3)\, p_{1\,\mu} - C_2(\nu_1,\nu_2,\nu_3)\, p_{2\,\mu}  \, ,
\eeq
with the coefficients given by
\bea
C_1(\nu_1,\nu_2,\nu_3) 
&=& 
\frac{1}{(p_3^2 - p_1^2 - p_2^2)^2 - 4\, p_1^2\, p_2^2 }\, 
\bigg\{ (p_1^2+p_2^2-p_3^2)\, J(\nu_1-1,\nu_2,\nu_3) \nn \\
&& \hspace{-3cm}
- 2\, p_2^2\, J(\nu_1,\nu_2-1,\nu_3)
+
\left( -p_1^2 + p_2^2 + p_3^2 \right)\, J(\nu_1,\nu_2,\nu_3-1)
+ p_2^2\, \left( p_1^2 - p_2^2 + p_3^2 \right)\, J(\nu_1,\nu_2,\nu_3)
\bigg\} \nn \\
C_2(\nu_1,\nu_2,\nu_3)
&=& 
\frac{1}{(p_3^2 - p_1^2 - p_2^2)^2 - 4\, p_1^2\, p_2^2 }\, 
\bigg\{ (p_1^2+p_2^2-p_3^2)\, J(\nu_1,\nu_2-1,\nu_3) \nn \\
&& \hspace{-3cm}
- 2\, p_1^2\, J(\nu_1-1,\nu_2,\nu_3)
+
\left( p_1^2 - p_2^2 + p_3^2 \right)\, J(\nu_1,\nu_2,\nu_3-1)
+ p_1^2\, \left( -p_1^2 + p_2^2 + p_3^2 \right)\, J(\nu_1,\nu_2,\nu_3)
\bigg\}\, . 
\nn \\
\eea
Using the momentum expansion of the tensor integral defined above, we extract from Eq.(\ref{PNewCIvectorid}) the relation
\small
\bea
&& 
\hspace{-0.7cm}
\nu_2\, (\nu_3-1)\, C_1(\nu_1,\nu_2+1,\nu_3) 
+ \nu_2\,(\nu_2+1)\, \left( C_1(\nu_1,\nu_2+2,\nu_3-1) - p_1^2\, C_1(\nu_1,\nu_2+2,\nu_3) \right) \nn \\
&&
\hspace{-0.7cm}
- \nu_1\, (\nu_3-1)\, C_1(\nu_1+1,\nu_2,\nu_3) 
- \nu_1\,(\nu_1+1)\, \left( C_1(\nu_1+2,\nu_2,\nu_3-1) - p_2^2\, C_1(\nu_1+2,\nu_2,\nu_3) \right) \nn \\
&&
\hspace{-0.7cm}
+\, \nu_2 \bigg[ (\nu_2+1) \left(J(\nu_1,\nu_2+2,\nu_3-1) - p_1^2\, J(\nu_1,\nu_2+2,\nu_3)  \right)
+  \left(1+\nu_2+\nu_3 - d/2 \right)\, J(\nu_1,\nu_2+1,\nu_3) \bigg]= 0 \,, \nn \\
\label{PNewCISpecConf}
\eea
\normalsize
together with the corresponding symmetric equation obtained interchanging $(p_1^2, \nu_1) \leftrightarrow (p_2^2, \nu_2)$. \\
This result allows to express integrals in the plane $\nu_1+\nu_2+\nu_3 = \kappa+2$
in terms of those in the two lower ones. In fact, introducing in Eq.(\ref{PNewCISpecConf}) and in its symmetric one the explicit expressions for
$C_1$ and $C_2$ we get
\beq \label{PNewCIExpandingJ}
J(\nu_1+2,\nu_2,\nu_3) =
\frac{1}{\nu_1\,(\nu_1+1)\,(p_1^2 + p_2^2 - p_3^2)\,p_2^2\,p_3^2}\, \sum_{(a,b,c)} \mathcal C_{(a,b,c)} J(\nu_1+a,\nu_2+b,\nu_3+c)\, ,
\eeq
where the coefficients $\mathcal C_{(a,b,c)}$ are given by 
\bea
&&
\mathcal C_{(0,0,0)} = (\nu_3-1)\, \bigg( (\nu_1+\nu_2)\, p_1^2 - \nu_2\, p_3^2 \bigg)\, , 
\nn \\
&&\mathcal C_{(1,-1,0)} = \nu_1\, (\nu_3-1)\, (p_3^2-p_1^2) \, ,
\nn \\
&&
\mathcal C_{(-1,1,0)} = - \nu_2\, (\nu_3-1)\, p_1^2\, ,\nn
\eea
\bea
&&
\mathcal C_{(0,1,-1)} = \nu_2\, \bigg[ (\nu_2+1)\,p_1^2 - (2+\nu_2-\nu_3)\, p_3^2  \bigg]\, , 
\nn \\
&&
\mathcal C_{(1,0,-1)} = \nu_1 \, (p_1^2 (\nu_1 +1 ) - p_3^2 (\nu_3 -1)) \,,
\nn 
\eea
\bea
&&
\mathcal C_{(2,-1,-1)} = - \nu_1\,(\nu_1+1)\, (p_1^2 - p_3^2)\, , 
\nn \\
&&
\mathcal C_{(-1,2,-1)} = - \nu_2\, (\nu_2+1)\, p_1^2\, , \nn
\eea
\bea%
&&
\mathcal C_{(2,0,-2)} = - \nu_1\, (\nu_1+1)\, p_3^2\, , 
\nn \\
&&
\mathcal C_{(0,2,-2)} = \nu_2\, (\nu_2+1)\, p_3^2\, , 
\nn 
\eea
\bea
&&
\mathcal C_{(1,0,0)} = \nu_1\, \bigg[(p_1^2)^2\, \bigg(\frac{d}{2}-\nu_1-2 \bigg) - p_3^2\, (p_2^2-p_3^2)
\bigg(\frac{d}{2}-\nu_1-\nu_3-1\bigg)
\nn \\
&& \hspace{20mm}
+\, p_1^2\, \bigg( \bigg(1-\frac{d}{2}\bigg)\, p_2^2 +  p_3^2\, (2\, \nu_1 + \nu_3 + 3 - d)\bigg)\bigg]
\nn \\
&&
\mathcal C_{(0,1,0)} = \nu_2\, p_1^2\, \bigg[ (1-\frac{d}{2})\, \left(p_1^2-p_3^2\right) + (\frac{d}{2}-2-\nu_2)\, p_2^2 \bigg]\, ,
\nn \\
&&
\mathcal C_{(0,2,-1)} = - \nu_2\,(\nu_2+1)\, p_1^2\, p_3^2\, , 
\nn \\
&&
\mathcal C_{(-1,2,0)} = \nu_2\,(\nu_2+1)\, (p_1^2)^2\, ,
\nn \\
&&
\mathcal C_{(2,0,-1)} = \nu_1\,(\nu_1+1)\, p_3^2\, (p_1^2 + 2\, p_2^2 - p_3^2)\, , 
\nn \\
&&
\mathcal C_{(2,-1,0)} = \nu_1\, (\nu_1+1)\, p_2^2\, (p_1^2-p_3^2)\, .
\eea
Analogous results hold for $J(\nu_1,\nu_2+2,\nu_3)$ and $J(\nu_1,\nu_2,\nu_3+2)$ if we just make the usual exchanges
$(p_1^2,\nu_1) \leftrightarrow (p_2^2,\nu_2)$ and $(p_1^2,\nu_1) \leftrightarrow (p_3^2,\nu_3)$
both in the integrals and in the coefficients $\mathcal C_{(a,b,c)}$.

\section{Conclusions}

We have shown that the solution of the conformal constraints for a scalar three-point function can be obtained directly in momentum space by solving the differential equations 
following from them. This has been possible having shown that these constraints take the form of a system of two PDE's of generalized hypergeometric type. 
The solution is expressed as a linear combination of four independent Appell's functions. 
The use of the momentum symmetries of the correlator allows to leave free a single multiplicative integration constant to 
parameterize the general solution for any conformal field theory. If this solution is compared with the position space counterpart and its Fourier representation, which is given by a family of 
Feynman master integrals, we obtain the explicit expression of the same integrals in terms of special functions. Our solution coincides with the one found 
by Boos and Davydychev using Mellin-Barnes techniques, which in our case are completely bypassed. 
Having established this link, we have shown that by applying special conformal constraints on the master integral 
representation one obtains new recursion relations. \\
The momentum space approach discussed in this chapter can be used to treat more complicated correlators. For instance, this method can be employed in the analysis
of three-point functions involving the vector and the energy momentum tensor operators, like $VVV$, $TOO$, $TVV$ and $TTT$, as well as higher order ones,
such as the scalar four-point function, whose general structure has been known for a long time \cite{Polyakov:1970}.
Nevertheless, such a treatment is much more complicated, in the former case due to the tensor nature of the correlators, 
which implies a much more involved set of constraints, in the latter because of the increasing number of independent variables in the partial differential equations.

\chapter{Three and four point functions of stress energy tensors in $D=3$}
\label{Chap.NonGaussianities}

\section{Introduction} 
Conformal field theories in $D>2$ are significantly less known compared to their $D=2$ counterparts, where exact results stemming 
from the presence of an underlying infinite dimensional symmetry have allowed to proceed with their classification. In fact, as one 
moves to higher spacetime dimensions, 
the finite dimensional character of the conformal symmetry allows to fix, modulo some overall constants, only the structure of 2- and 
3-point functions \cite{Osborn:1993cr, Erdmenger:1996yc}. 
In 4-D, for instance, free field theory realizations of these specific correlators allow the identification of their explicit 
expressions, performing a direct comparison with their general form, which is predicted by the symmetry \cite{Coriano:2012wp}. 
Among these correlators, a special role is taken by those involving insertions of the energy momentum tensor (EMT), which can be 
significant in the context of several phenomenological applications. For instance, in $D=4$,  correlators involving insertions of the 
EMT describe the interaction of a given theory with gravity around the flat spacetime limit. Their study is quite involved due to the 
appearance of a trace anomaly \cite{Giannotti:2008cv, Armillis:2009pq, Coriano:2011ti, Coriano:2011zk}. They are part of the 
anomalous effective action of gravitons at higher order, but they also find application in the description of dilaton interactions 
and of the Higgs-dilaton mixing at the LHC \cite{Coriano:2012nm, Coriano:2012dg}.

In $D=3$ dimensions their computation simplifies considerably, due to the absence of anomalies, 
but it remains quite significant, especially in the context of the AdS/CFT correspondence \cite{Raju:2012zs} and supersymmetry in general \cite{Bastianelli:1999ab}. 
In particular, using a holographic approach, these correlators allow to describe the curvature gravitational perturbations   
in a pre-inflationary phase of the early universe characterized by strong gravity \cite{McFadden:2009fg, McFadden:2010vh, Bzowski:2011ab, Maldacena:2011nz}. 
Their imprints manifest through the non-gaussian behaviour of the bisprectrum and the trispectrum \cite{Mukhanov:1990me, Bartolo:2010qu} of scalar and tensor gravitational perturbations. A qualitative and quantitative estimation of such effects in holographic models brings us to the analysis of the $TTT$ and $TTTT$ correlators  in $D=3$. In this chapter we present a detailed computation of these correlation functions.
The possibility of using the perturbative analysis of these types of correlators in cosmology is what motivates our digression to $D=3$.

In holographic cosmological models the formulation of the correspondence between cosmological observables, like the correlation functions of gravitational perturbations, is a two-step process. The first one allows to map, through an analytic continuation, a $D=4$ cosmological background to a $D=4$ gravitational domain theory in the bulk, while at a second stage, standard gauge/gravity duality defines the mapping between the $D=4$ bulk and the appropriate boundary field theory, which is a $D=3$ non-abelian $SU(N)$ gauge theory in the large $N$ limit. Therefore scalar and tensor cosmological perturbations are related to correlators involving multiple insertions of the EMT's  computed in the boundary QFT. 
The boundary correlation functions associated to the bispectrum of scalar perturbations, of scalar-tensor and of tensor perturbations are, respectively,
the fully-contacted, partially-contracted and fully-uncontracted $TTT$. The determination of the trispectrum is, instead, associated to the $TTTT$ correlation function. 

In all the cases the calculation these cosmological perturbations involves correlation functions of EMT's in field theories including scalars $\phi^J$ (with $J=1,2,...n_\phi$), vectors 
$A_\mu^I$ (with $I=1,2,...n_A$) and fermions $\psi^L$ ( with $L=1,2,...n_\psi$) in the virtual 
corrections.  All the fields in such theories are in the adjoint of the gauge group $SU(N)$. 

The mapping is described, on the dual field theory side, in terms of an effective (t'Hooft) coupling constant $g_{eff}=g_{YM}^2 N/M$, with $g_{eff} <<1$, and requires a large$-N$ limit. $M$ is a typical momentum scale, related to the typical momenta of the correlators, and necessary in order to make $g_{eff}$ dimensionless. This implies that the gauge coupling ($g_{YM}/M$) has to be very weak, allowing an expansion of the dual theory in such a variable, which can be arrested to zeroth order, i.e. with free fields.

It has been pointed out in \cite{McFadden:2009fg, McFadden:2011kk} that the small amplitude ($O(10^{-9}$)) and the nearly scale invariant characters of the measured power spectrum, wich shows deviations from scale invariance which are of $O(10^{-3})$, are indeed predicted by such models. In particular, the large-$N$ limit, which is necessary in order to predict the small amplitude of the power spectrum, requires $N\sim 10^4$, given that its scales as $1/N^2$ in holographic models. At the same time, the small violation of scaling invariance in the same power spectrum is controlled by the t'Hooft coupling $g_{eff}$, therefore requiring that this coupling has to be small as well.

These considerations allow to simplify drastically the computation of these correlators on the dual side, as we have mentioned. In particular, we are allowed to deal with simple dual theories in order to identify the leading behaviour of the perturbative correlators which appear in the holographic formula for the perturbations. 

Henceforth, the non-abelian character of the dual gauge theory becomes unessential if we work at leading order in $g_{YM}$, and the vector contributions are proportional to those of a free abelian theory. 
This implies that all our computations can be and are performed 
in simple free field theories of scalars, abelian vectors and fermions, with the scalars and the fermions taken as gauge singlets. This choice simplifies the notations and allows us to obtain the correct result, to be used in the holographic mapping, just by introducing a correction factor, which will be inserted at the end.

The explicit form of the mapping has been given in \cite{McFadden:2011kk, Bzowski:2011ab}, for the 
$\langle \zeta \zeta\zeta\rangle$ (bispectrum) correlator, with $\zeta$ describing the gauge invariant curvature perturbation of the 
gravitational metric, which is mapped to the $TTT$. The same (uncontracted) 3-T correlator determines the bispectrum of more complex 
3-point functions, $\langle\zeta \zeta \gamma\rangle$, $\langle\zeta \gamma\gamma\rangle$ and $\langle\gamma\gamma\gamma\rangle$, 
involving tensor perturbations ($\gamma$) \cite{Bzowski:2011ab}. 

A similar, though more involved, mapping between the trispectrum correlator  $\langle
\zeta \zeta\zeta\zeta\rangle$ and the $TTTT$ 4-point function is expected to hold. The explicit form of this mapping is not yet available, since it involves a direct extension of the holographic approach developed in 
\cite{McFadden:2011kk, Bzowski:2011ab}. 

Even in the absence, at least at the moment, of a suitable generalization of the holographic expressions given in \cite{McFadden:2011kk, Bzowski:2011ab}, it is clear that a complete determination of the trispectrum in holographic models, given its complexity, is a two-stage process. This
requires 1) the explicit derivation of the holographic relation 
which maps the $\langle\zeta\zeta\zeta\zeta\rangle$ to the 4-T correlator, followed 2) by an explicit computation of these higher point functions via the dual mapping. 

For instance, in the case of the bispectrum ($\langle\zeta\zeta\zeta\rangle$), the holographic analysis has been put forward in \cite{McFadden:2011kk}, followed later on by an 
explicit computation of the relevant 3-D correlators $TTT$ given in \cite{Bzowski:2011ab}.
 
Our goal in the work presented in this chapter is to make one step forward in this program and present the explicit form of the $TTTT$ (fully traced)  correlator. The explicit form of the complete - uncontracted - correlator (which is of rank-8) is computationally very involved due to the higher 
rank tensor reductions and it will not be discussed here. 
  
At the same time we will proceed with an independent recomputation of the $TTT$ correlator in $D=3$, which has been investigated in 
\cite{Bzowski:2011ab, Maldacena:2011nz}. We anticipate that our analysis is in complete agreement with the result given in 
\cite{Bzowski:2011ab} for this correlator, and we will discuss the mapping between our approach and the one of \cite{Bzowski:2011ab}. This 
agreement allows to test our methods before coming to their generalization in the 4-T case. 

As we have already mentioned, the study of the 4-T correlator in $D=3$ is free of the complications present in their $D=4$
counterparts, which are affected by the scale anomaly and require renormalization. Checks of our computations have been performed at 
various levels. We secure the consistency of the result in the 4-T case by verifying the Ward identities which are expected to hold.

The chapter is organized as follows. After a summary section in which we outline our definitions and conventions, we move to a computation of the general form 
of the $TTT$ in our approach, followed by a brief section in which we provide a mapping between our result and those of \cite{Bzowski:2011ab}.  The Ward identities 
for the 3- and 4-point case, which are essential in order to test the consistency of our computations, are discussed together in a single section. We then move to the perturbative determination of the 4-T. We have collected in several 
appendices some of the technical aspects related to the diagrammatic expansion, specific to the $D=3$ case.  

\section{The search for non-gaussian fluctuations} 
Clues on the physics of the very early universe come from the analysis of the primordial gravitational fluctuations, which leave an
imprint on the cosmic microwave background (CMB) and on the evolution of large scale structures (see \cite{Mukhanov:1990me, Bartolo:2010qu}). So far, 
the cosmological data have shown to be compatible with the gaussian character of such fluctuations, which implies that they can be expressed in terms of 2-point functions. In Fourier space they define the so called 
"power spectrum" ($\Delta(k)$). In this case 3-point correlators associated to such fluctuations vanish, together with all the 
correlators containing an odd number of these fields.

Measurements of the power spectra of perturbations are not able to answer questions concerning the evolution and the interactions of 
quantum fields which generate such fluctuations in the very early universe. In fact, inflation models with different fields 
and interactions can lead to power spectra which are quite similar. For this reason, there is a justified hope that it will be 
possible to unveil, through the identification of a non-gaussian behaviour of such fluctuations, aspects of the physics of inflation 
which otherwise would remain obscure \cite{Komatsu:2009kd}. Tests of a possible non-gaussian behaviour of such perturbations can be performed using several 
observational probes, including analysis of the CMB, large scale structures and weak lensing, just to mention 
a few. 

One important result \cite{Maldacena:2002vr} in the study of the non-gaussian behaviour of single field inflation was the proof that 
in these models such fluctuations are small and, for this reason, the possible experimental detection of significant 
non-gaussianities would allow to rule them out.

\subsection{Domain-Wall/Cosmology correspondence and gauge/gravity duality}

An interesting approach \cite{McFadden:2010vh, Bzowski:2011ab, McFadden:2009fg} which allows to merge the analysis of fluctuations 
and of their quantization with ideas stemming from gauge/gravity duality, has been developed in the last few years. 
These formulations allow to define a
correspondence between two bulk theories, describing cosmological and domain wall gravitational backgrounds, and hence between their boundary duals, which are described by appropriate 3-D field theories.  
The two bulk metrics are related by an  
analytic continuation. Once that a cosmological model is mapped into a 4-D domain wall model, 
gauge/gravity duality can be used to infer the structure of the correlators in the bulk using a corresponding field theory on the boundary. Such a theory can 
be described by a combination of scalar, fermion and spin-1 sectors, formulated as simple field theories in flat 3-D backgrounds. 

Scalar and tensor fluctuations in domain wall backgrounds can then be described in terms of correlators involving multiple insertions 
of EMT's, computed in ordinary perturbation theory.  These result can be mapped back to describe the 
correspondence between bulk and boundary in the usual cosmological context, by an analytic continuation of the boundary correlators.  

In this framework, 
one can derive holographic formulas which allow to describe a primordial phase of strong gravity just by weakly coupled perturbations in the dual theory. We will present below, to make our discussion self-contained, the explicit expressions of one of these relations, which are of direct relevance for our analysis. 

We also mention that, in the conformal case, the 3-T and 4-T correlators in scalar and fermion free field theories provide a realization 
of the bispectrum and of the trispectrum of gravitational waves in De Sitter space \cite{Maldacena:2011nz}.
Discussions of the conformal properties of 3- and 4-point functions of primordial fluctuations can be found in \cite{ Kehagias:2012pd, Antoniadis:2011ib}. 

\section{Field theory realizations}

The correlation functions that we intend to study will be computed in four free field theories, namely
a minimally coupled and a conformally coupled scalar, a fermion and a spin 1 abelian gauge field.

If the classical theory is described by the action $\mathcal{S}$, the energy-momentum tensor (EMT) of the system is obtained by coupling it to a curved 3-D background metric $g_{\mu\nu}$ (with $\mathcal{S}\to \mathcal{S}[g]$) and functionally differentiating the action with respect to it. The formalism is similar to the ordinary one in the case of a 4-D gravitational spacetime  
\beq \label{P4EMT}
T^{\mu\nu}(z) = -\frac{2}{\sqrt{g_z}}\frac{\delta\,\mathcal{S}}{\delta g_{\mu\nu}(z)}\, ,
\eeq
and for this reason we will be using greek indices, with the understanding that they will run from 1 to 3. We will also set $\textrm{det}\, g_{\mu\nu}(z)\equiv g_z$ for the determinant of the 3-D metric.  

In the quantum theory, let $\mathcal W[g]$ be the euclidean generating functional depending on the classical background,
\beq\label{P4Generating}
\mathcal W[g] = \frac{1}{\mathcal{N}} \, \int \, \mathcal D\Phi \, e^{-\, \mathcal{S}} \, ,
\eeq
where $\mathcal{N}$ is a normalization factor, and $\Phi$ denotes all the quantum fields of the theory except the metric.
$\mathcal W[g]$ generates both connected and disconnected correlators of EMT's, which are 1-particle reducible. For notational simplicity we prefer to use this generating functional of the Green's function of the theory, rather than $\log \mathcal W$ and its Legendre transform. It is implicitly understood that, in the perturbative expansion of the corresponding correlators, we will consider only the connected components. In a 1-loop analysis the issue of 1-particle reducibility does not play any relevant role, and hence the use of $\mathcal W$ will make the manipulations more transparent.

Then it follows from (\ref{P4EMT}) that the quantum average of the EMT in the presence of the background source is given by
\beq 
< T^{\mu\nu}(z) >_g = 
\frac{2}{\sqrt{g_z}}\frac{\delta\, \mathcal W}{\delta\, g_{\mu\nu}(z)} \, .
\eeq
where the subscript $g$ indicates the presence of a generic metric background. 
Otherwise, all the correlators which do not carry a subscript $g$, are intended to be written in the flat limit. 
It is understood that the metric is generic while performing all the functional derivatives and ordinary differentiations of the correlators, and that the limit of flat space is taken only at the end. 

As we have mentioned above, we focus our analysis on the determination of the complete 3-T correlator of free (euclidean) field theories of scalar, vector and fermions in 3 space dimensions and on the 4-T fully traced correlator, which we are now going to introduce.


The actions for the scalar ($S$) and the chiral fermion field ($CF$), are respectively given by 
\beqa
\mathcal{S}_{S}
&=&
\frac{1}{2} \, \int d^3 x \, \sqrt{g}\,
\bigg[g^{\mu\nu}\,\nabla_\mu\phi\,\nabla_\nu\phi - \chi \, R \,\phi^2 \bigg]\, ,\label{P4scalarAction}\\
\mathcal{S}_{CF}
&=&
\frac{1}{2} \, \int d^3 x \, V \, {V_a}^\rho\,
\bigg[\bar{\psi}\,\gamma^a\,(\mathcal{D}_\rho\,\psi)
- (\mathcal{D}_\rho\,\bar{\psi})\,\gamma^a\,\psi \bigg] \, . \label{P4FermionAction}
\eeqa
Here $\chi$ is the parameter corresponding to the term of improvement obtained by coupling $\phi^2$ to the 3-D scalar curvature $R$.
We will be concerned with two cases for the scalar Lagrangian, the minimally and the conformally coupled ones.  $\chi = 0$ describes a minimally coupled scalar ($MS$).
In three dimensions, for $\chi = 1/8$ one has a classically conformal invariant theory (i.e. one with an EMT whose trace
vanishes upon use of the equations of motion), which is the second case that we will consider (denoted with the "conformal scalar" subscript, or $CS$). As we have mentioned, the absence of conformal anomalies guarantees that for any of these theories, those operators which are classically traceless, remain such also at quantum level.

The other conformal field theory which we will be concerned with, is the one describing the Lagrangian of a free chiral fermion $(CF)$  on a curved metric background \footnote{ Notice that even if our analysis is euclidean, the 3-D case that we discuss can be mapped straightforwardly to the analogous Minkowski one in $D=2+1$ by a simple analytic continuation.}. 
Here ${V_a}^\rho$ is the vielbein and $V (= \sqrt{g})$ its determinant, needed to embed
the fermion in such background, with its covariant derivative $\mathcal{D}_\mu$ defined as
\beq
\mathcal{D}_\mu = \pd_\mu + \Gamma_\mu =
\pd_\mu + \frac{1}{2} \, \Sigma^{bc} \, {V_b}^\sigma \, \nabla_\mu\,V_{c\sigma} \, .
\eeq
with $\Sigma^{ab} = \frac{1}{4}\, [\gamma^a,\gamma^b]$ in the fermion case.
Using the integrability condition of the vielbein
\beq
\nabla_{\mu}\, V_{b\sigma} = \pd_{\mu}\, V_{b\sigma} - \Gamma^\lambda_{\mu\sigma}\, V_{b\lambda} 
= \Omega_{ab,\mu}\, {V^a}_\sigma \, ,
\eeq
where $\Omega_{ab,\mu}$ is the spin connection, this can be expressed as
\beq \label{P4SpinCon}
\Omega_{a b ,\mu} = 
\frac{1}{2}\, V^{\lambda}_{a}\left( \pd_{\mu} V_{b \lambda} - \pd_{\lambda}V_{b \mu} \right)
- \frac{1}{2}\, V^{\lambda}_{b} \left( \pd_\mu V_{a \lambda} - \pd_\lambda V_{a \mu} \right) 
+ \frac{1}{2} \, V^{\rho}_{a}\, V^{\lambda}_{b}\, V^{h}_{\mu}\, \left(\pd_{\lambda}V_{h \rho}- \pd_{\rho}V_{h \lambda} \right). \, 
\eeq
After an expansion we can rewrite (\ref{P4FermionAction}) as
\beq
\mathcal S_{CF} 
= \frac{1}{2}\, \int d^3 x\, V\, {V_a}^\rho\, \bigg[\bar\psi\, \gamma^a\, (\pd_\rho\, \psi) - (\pd_\rho\,\bar{\psi})\,\gamma^a\,\psi
\bigg]
+\frac{1}{16}\, \int d^3 x\, V\, {V_a}^\rho\, \bar\psi\,\{\gamma^a,[\gamma^b,\gamma^c]\}\, \psi\, \Omega_{bc,\rho}\, .
\eeq
The functional derivative with respect to the metric appearing in (\ref{P4EMT}) is expressed in terms of the vielbein as
\beq\label{P4SymmVielbein}
\frac{\delta}{\delta g_{\mu\nu}(z)} = 
\frac{1}{4}\, \left( {V_a}^\nu(z)\, \frac{\delta}{\delta V_{a\mu}(z)} + {V_a}^\mu(z)\, \frac{\delta}{\delta V_{a\nu}(z)}\right)\, ,
\eeq
so that the EMT for a fermion field is defined by
\beq\label{P4SymmFerTEI}
T^{\mu\nu}_{CF}(z)
\stackrel{def}{\equiv} 
- \frac{1}{2\,V(z)}\bigg(V^{a\mu}(z) \, \frac{\delta}{\delta {V^a}_\nu(z)} + V^{a\nu}(z)\, \frac{\delta}{\delta {V^a}_\mu(z)}\bigg)\,
\mathcal S_{CF} \, .
\eeq
Finally, the action for the gauge field ($GF$) is given by
\beq
\mathcal{S}_{GF} = \mathcal{S}_M + \mathcal{S}_{gf} + \mathcal{S}_{gh}\, ,
\eeq
where the three contributions are the Maxwell (M), the gauge-fixing and the ghost actions
\beqa
\mathcal{S}_M    &=&   \frac{1}{4} \, \int d^3 x \, \sqrt{g} \, F^{\a\b} F_{\a\b}\, ,\\
\mathcal{S}_{gf} &=&   \frac{1}{2 \xi} \, \int d^3 x \, \sqrt{g} \, \left( \nabla_{\alpha}A^\alpha \right)^2\, ,\\
\mathcal{S}_{gh} &=& - \int d^3 x \, \sqrt{g}\, \partial^\a \bar{c} \, \partial_\a c\, .
\eeqa
The EMT's for the scalar and the fermion are
\beqa
T^{\mu\nu}_{S}
&=&
\nabla^\mu \phi \, \nabla^\nu\phi - \frac{1}{2} \, g^{\mu\nu}\,g^{\alpha\beta}\,\nabla_\alpha \phi \, \nabla_\beta \phi
+ \chi \bigg[g^{\mu\nu} \Box - \nabla^\mu\,\nabla^\nu + \frac{1}{2}\,g^{\mu\nu}\,R - R^{\mu\nu} \bigg]\, \phi^2 \,,
\label{P4ScalarEMT}\\
T^{\mu\nu}_{CF}
&=&
\frac{1}{4} \,
\bigg[ g^{\mu\rho}\,{V_a}^\nu + g^{\nu\rho}\,{V_a}^\mu - 2\,g^{\mu\nu}\,{V_a}^\rho \bigg]
\bigg[\bar{\psi} \, \gamma^{a} \, \left(\mathcal{D}_\rho \,\psi\right) -
\left(\mathcal{D}_\rho \, \bar{\psi}\right) \, \gamma^{a} \, \psi \bigg] \, ,
\label{P4FermionEMT}
\eeqa
while the energy-momentum tensor for the abelian gauge field is given by the sum
\beq
T^{\mu\nu}_{GF} = T^{\mu\nu}_M + T^{\mu\nu}_{gf} + T^{\mu\nu}_{gh}\, ,
\eeq
with
\beqa
T^{\mu\nu}_M
&=&
F^{\mu\a}{F^\nu}_{\a}  - \frac{1}{4}g^{\mu\nu}F^{\a\b}F_{\a\b} \, ,
\label{P4MaxwellEMT}
\\
T^{\mu\nu}_{gf}
&=&
\frac{1}{\xi}\{ A^\mu\nabla^\nu(\nabla_\r A^\r) + A^\nu\nabla^\mu(\nabla_\r A^\r ) -g^{\mu\nu}[ A^\r
\nabla_\r(\nabla_\s A^\s) + \frac{1}{2}(\nabla_\r A^\r)^2 ]\}\, ,
\label{P4GaugeFixEMT}
\\
T^{\mu\nu}_{gh}
&=&
g^{\mu\nu}\pd^{\r}\bar{c}\,\pd_{\r}c - \pd^\mu\bar{c}\, \pd^\nu c - \pd^\nu\bar{c}\, \pd^\mu c
\label{P4GhostEMT}\, .
\eeqa
The explicit expressions for the vertices involving one or more EMT's, which can be computed by functional differentiating the actions, 
have been collected in Appendix \ref{P4Vertices}. \\
We point out that in our computation of the contributions related to the gauge fields (GF), only the Maxwell
action $\mathcal{S}_M$ and the corresponding EMT, $T^{\mu\nu}_M$, are needed. One can check that there is a cancellation between the gauge-fixing and ghost contributions from $T_{gf}$ and $T_{gh}$. For those interested in a 
direct check of these results, we remark that the contributions generated from (\ref{P4ScalarEMT}) and those generated from (\ref{P4GhostEMT}) differ only by an overall sign factor, while the trilinear and quadrilinear vertices for the gauge-fixing part can be found in \cite{Coriano:2012wp}.

\section{Conventions and the structure of the correlators}
\begin{figure}[t]
\centering
\includegraphics[scale=0.8]{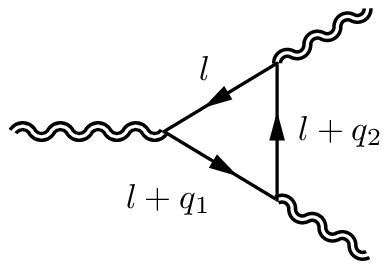}
\hspace{1cm}
\includegraphics[scale=0.8]{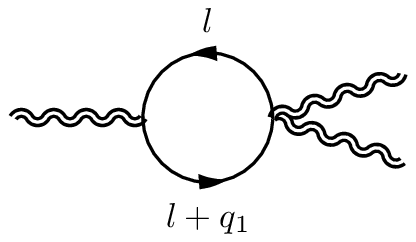}
\hspace{1cm}
\includegraphics[scale=0.8]{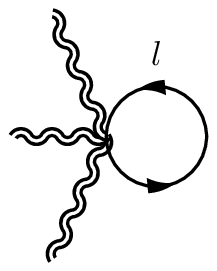}
\caption{Topologies appearing in the expansion of the $TTT$ correlator. Contributions involving coincident gravitons correspond to 
contact terms. \label{P4TTTtop} }
\end{figure}
We will be introducing two different notations for correlators involving the EMT's. The first is defined in terms of the symmetric 
$n-th$ order functional derivative of $\mathcal{W}$
\beqa \label{P4NPF}
< T^{\mu_1\nu_1}(x_1)...T^{\mu_n\nu_n}(x_n)> 
&=&
\bigg[\frac{2}{\sqrt{g_{x_1}}}...\frac{2}{\sqrt{g_{x_n}}} \,
\frac{\delta^n \mathcal{W}}{\delta g_{\mu_n\nu_n}(x_n) \ldots \delta g_{\mu_1\nu_1}(x_1)}\bigg]
\bigg|_{g_{\mu\nu} = \delta_{\mu\nu}} \nn \\
&=&
2^n\, \frac{\delta^n \mathcal{W}}{\delta g_{\mu_n\nu_n}(x_n) \ldots \delta g_{\mu_1\nu_1}(x_1)}\bigg|_{g_{\mu\nu} = 
\delta_{\mu\nu}} \, .
\eeqa
We will refer to this correlator as to a "symmetric" one. Notice that given the existence of an analytic continuation between the 3-D euclidean theory and the one in 2+1 dimensional spacetime, we will be referring to this vertex, for simplicity, as to the "n-graviton " vertex. 
 
This definition allows to leave the factor ${2}/{\sqrt{g}}$ outside of the actual differentiation in order to obtain symmetric expressions. We 
have denoted this correlator with a small angular brackets ($< \,\, >$) since these correlators include also 
contact terms.  Contact 
terms are easily identified in perturbation theory for bringing together at least two gravitons on the same spacetime point. Such terms are instead absent in the expression of correlation functions given by the (ordinary) expectation value of the product of $n$ EMT's, and which are denoted, in our case, with large angular brackets ($\langle\,\, \rangle$)  as in 
\beq
\langle T^{\mu_1\nu_1}(x_1) \, \ldots \, T^{\mu_n\nu_n}(x_n) \rangle =  \frac{1}{\mathcal N} \int \mathcal D \Phi \, 
T^{\mu_1\nu_1}(x_1) \, \ldots \, T^{\mu_n\nu_n}(x_n)  \, e^{-S} \bigg|_{g_{\mu\nu}=\delta_{\mu\nu}} \,.
\eeq
This second form of the correlator of EMT's will be referred to as "ordinary" or "genuine" $n$-point functions. 
It will also be useful to introduce the following notation to represent the functional derivative with respect to the background metric
\beqa
\label{P4funcder}
\left[f(x)\right]^{\muu\nuu\mud\nud\dots\mu_{n}\nu_{n}}(\xu,\xd,\dots,x_n) 
\equiv
\frac{\delta^n\, f(x)}{\delta g_{\mu_n\nu_n}(x_{n}) \, \ldots \, \delta g_{\mud\nud}(\xd) \, \delta g_{\muu\nuu}(\xu)}
\bigg|_{g_{\mu\nu}=\delta_{\mu\nu}} 
\eeqa
and take the flat spacetime limit at the end.
For later use we also define the notation with lower indices as
\beqa
\left[f(x)\right]_{\muu\nuu\ldots\mu_{n}\nu_{n}}(\xu,\xd,\dots,x_n) 
&\equiv& \nn \\
&& \hspace{-3cm} \delta_{\mu_1 \alpha_1} \delta_{\nu_1 \beta_1}\, \ldots\, \delta_{\mu_n \alpha_n}\delta_{\nu_n \beta_n}\,
\left[f(x)\right]^{\alpha_1 \beta_1 \alpha_2 \beta_2 \dots \alpha_{n}\beta_{n}}(\xu,\xd,\dots,x_n).
\eeqa
With this definition a single functional derivative of the action in a correlation function is always equivalent, modulo a factor, to an insertion of 
a $T^{\mu\nu}$ in the flat limit, since 
\beq
 \left[\mathcal S \right]^{\mu_1 \nu_1}(x_1) \equiv \frac{\delta \mathcal S}{\delta g_{\mu_1 \nu_1}(x_1)} \bigg|_{g_{\mu\nu}=\delta_{\mu\nu}} =  -\frac{1}{2}T^{\mu_1 \nu_1}(x_1) \,.
 \eeq
We can convert the two graviton (greek) indices $(\mu_1,\nu_1)$ into a latin index $(s_1)$ by contracting with a polarization vector 
of generic polarization $s_1$ as in 
\beq
\left[\mathcal S\right]^{(s_1)} = -\frac{1}{2}T^{(s_1)}\equiv\left[\mathcal S \right]^{\mu_1\nu_1}\epsilon_{\mu_1\nu_1}^{(s_1) \, *},
\eeq
with $s_1\equiv \pm$, being the two helicities, as explained below. Moreover the symbol $T$ will be used, in a correlator, to denote 
the trace of the EMT,  $T \equiv {T^\mu}_\mu$.
We will also stack the two $(\mu_1\nu_1)$ indices, one on top of the other, for simplicity, after a contraction, as in 
\beq 
\left[\mathcal S \right]^{\mu_1}_{\mu_1}\equiv  \left[\mathcal S \right]^{\mu_1\nu_1}\, 
\delta_{\mu_1 \nu_1}\qquad \textrm{or}
\qquad  
\left[\mathcal S \right]^{\mu_1\mu_2}_{\mu_1\mu_2} \equiv  \left[\mathcal S \right]^{\mu_1\nu_1\mu_2\nu_2}
\delta_{\mu_1 \nu_1} \delta_{\mu_2 \nu_2} 
\eeq
in order to make the tensorial expressions more compact. 

With the definition of Eq. (\ref{P4NPF}) the expansion of the $TT$ correlator becomes
\bea
< T^{\muu\nuu}(x_1)T^{\mud\nud}(x_2) >
= 4\, \bigg[ \langle \left[\mathcal S \right]^{\muu\nuu}(\xu)\left[\mathcal S \right]^{\mud\nud}(\xd) \rangle - \langle 
[S]^{\muu\nuu\mud\nud}(\xu,\xd) \rangle \bigg] \,.
\eea
The last term on the right hand side of the equation above, which is a massless tadpole, is scheme dependent and can be easily 
removed by a local finite counterterm. For this reason it can be set to zero and then the $TT$ correlation function, obtained by 
differentiation of the generating functional, coincides with the quantum average of two energy momentum tensors
\bea
< T^{\muu\nuu}(x_1)T^{\mud\nud}(x_2) > &=& 4  \langle \left[\mathcal S \right]^{\muu\nuu}(\xu)\left[\mathcal S 
\right]^{\mud\nud}(\xd) \rangle  
= \langle T^{\muu\nuu}(x_1)T^{\mud\nud}(x_2) \rangle \,.
\eea
This is not true for higher order correlation functions of $n$-gravitons, as we are going to show in a while, where contact terms also appear. \\
For the $TTT$ correlator the functional expansion is given by
\bea
\label{P43Tall}
&& < T^{\muu\nuu}(x_1)T^{\mud\nud}(x_2)T^{\mut\nut}(x_3) >
=
\langle T^{\muu\nuu}(\xu) T^{\mud\nud}(\xd)T^{\mut\nut}(\xt)\rangle 
\nn \\
&& 
-\, 4\, \bigg(\langle \left[\mathcal S\right]^{\muu\nuu\mud\nud}(\xu,\xd)\,T^{\mut\nut}(\xt)\rangle + 2\, \text{perm.} \bigg)
-\, 8\, \langle  \left[\mathcal S\right]^{\muu\nuu\mud\nud\mut\nut}(\xu,\xd,\xt)\rangle \bigg]   
\eea
whose right hand side is expressed in terms of ordinary 3-point correlators plus extra contact terms. The additional 
terms obtained by permutation are such to render symmetric the right hand side of the previous equation. 

We will present the expression of these contributions in the helicity basis for each sector (scalar, fermion and gauge field) in the 
sections below. Notice that the first term on the right hand side of Eq. (\ref{P43Tall}) corresponds to an ordinary (genuine) 3-point function, whose connected component is given, at 1 loop, by the triangle diagram of Fig. \ref{P4TTTtop},
while the last term is a massless tadpole (see the third diagram in Fig. 
\ref{P4TTTtop}), which can be set to zero
\beq
\langle \left[\mathcal S\right]^{\muu\nuu\mud\nud\mut\nut}(\xu,\xd,\xt)\rangle=0. 
\eeq
In the 3-T case, contact terms have the topology of a bubble, and are generated by correlators containing insertions of the second functional derivatives of the action respect to the metric 
(such as in $\left[S\right]^{\mut\nut\muq\nuq}$). One of them is shown in the second figure of Fig. \ref{P4TTTtop}, the others can be 
obtained by a reparameterization of the external momenta. These bubble terms are characterized by the insertion of two graviton lines 
on the same vertex. \\
Moving to the 4-T case, a similar expansion holds and is given by
\beqa
< T^{\muu\nuu}(x_1)T^{\mud\nud}(x_2)T^{\mut\nut}(x_3)T^{\muq\nuq}(x_4)>
&=&
\langle T^{\muu\nuu}(x_1)T^{\mud\nud}(x_2)T^{\mut\nut}(x_3)T^{\muq\nuq}(x_4)\rangle
\nn \\
&& \hspace{-75 mm}
-\, 4\, \bigg[\langle \left[\mathcal S\right]^{\muu\nuu\mud\nud}(\xu,\xd)T^{\mut\nut}(\xt) T^{\muq\nuq}(\xq)\rangle 
+ 5 \, \text{perm.} \bigg] \nn \\
&& \hspace{-75mm} + 16\, \bigg[\langle \left[\mathcal S\right]^{\muu\nuu\mud\nud}(\xu,\xd)\left[\mathcal 
S\right]^{\mut\nut\muq\nuq}(\xt,\xq)\rangle + 2 \, \text{perm.} \bigg]
\nn \\
&& \hspace{-75mm}
- \, 8\, \bigg[\langle \left[\mathcal S\right]^{\muu\nuu\mud\nud\mut\nut}(\xu,\xd,\xt)T^{\muq\nuq}(\xq)\rangle + 3 \, \text{perm.} \bigg] \nn \\
&& \hspace{-75mm} - \, 16\, \langle \left[\mathcal S\right]^{\muu\nuu\mud\nud\mut\nut\muq\nuq}(\xu,\xd,\xt,\xq)\rangle 
\label{P44PF}
\eeqa
with 
\beq
\langle \left[\mathcal S\right]^{\muu\nuu\mud\nud\mut\nut\muq\nuq}(\xu,\xd,\xt,\xq)\rangle=0, 
\eeq
being a massless tadpole contribution. Notice that the left hand side and the right hand side are both symmetric amplitudes, as they 
should. In this case the perturbative expansion in the three fundamental sectors - scalars, vector and fermion - generates diagrams of box type 
for the first 4-T correlator on the right hand side of (\ref{P44PF}), plus triangle, bubble and tadpole diagrams generated by the 
contact terms. The analysis of these contributions is more involved compared to the $TTT$ case, and will be performed in detail in 
the following sections. 
%

\section{Ward identities for the Green functions}

We proceed with a derivation of the relevant Ward identities satisfied by the 3- and 4-point functions of EMT's.

The diffeomorphism Ward identities are defined from the condition of general covariance on the generating functional $\mathcal W[g]$ 
\beq\label{P4masterWI0}
\nabla_{\nuu} < T^{\muu\nuu}(x_1) >_g
= \nabla_{\nuu} \bigg(\frac{2}{\sqrt{g_{x_1}}}\frac{\delta\mathcal W[g]}{\delta g_{\muu\nuu}(x_1)}\bigg)
= 0 \, .
\eeq
The Ward identities we are interested in are obtained by functional differentiation of Eq. (\ref{P4masterWI0}) and are given by
\bea
\pd_{\nuu} < T^{\muu\nuu}(x_1) T^{\mud\nud}(x_2) > &=& 0 \, , \label{P4WI2PFCoordinateFlat}
\eea
which shows that the two-point function is transverse, and by
\bea
\pd_{\nuu} < T^{\muu\nuu}(x_1)T^{\mud\nud}(x_2)T^{\mut\nut}(x_3) > 
&=& 
- 2\, \left[\Gamma^{\muu}_{\kappa\nuu}(x_1)\right]^{\mud\nud}(x_2) \langle T^{\kappa\nuu}(x_1)T^{\mut\nut}(x_3) \rangle \nn \\
&&
\hspace{-1cm} - 2\, \left[\Gamma^{\muu}_{\kappa\nuu}(x_1)\right]^{\mut\nut}(x_3) \langle T^{\kappa\nuu}(x_1)T^{\mud\nud}(x_2) \rangle 
\, ,  \label{P4WI3PFCoordinateFlat} \\
\pd_{\nuu} < T^{\muu\nuu}(x_1)T^{\mud\nud}(x_2)T^{\mut\nut}(x_3)T^{\muq\nuq}(x_4) > 
&=&    \nn \\
&& \hspace{-80mm} - 2\, \bigg[\left[\Gamma^{\muu}_{\kappa\nuu}(x_1)\right]^{\mud\nud}(x_2)
 < T^{\kappa\nuu}(x_1) T^{\mut\nut}(x_3)T^{\muq\nuq}(x_4) >
\nn \\
&& \hspace{-80mm}
+ \big( 2 \leftrightarrow 3, 2 \leftrightarrow 4 \big) \bigg]
- 4\, \bigg[ \left[\Gamma^{\muu}_{\kappa\nuu}(x_1)\right]^{\mud\nud\mut\nut}(x_2,x_3)
 \langle T^{\kappa\nuu}(x_1)T^{\muq\nuq}(x_4) \rangle 
+ \big( 2 \leftrightarrow 4, 3 \leftrightarrow 4 \big) \bigg] \,   \label{P4WI4PFCoordinateFlat} 
\eea
for the 3- and 4-point functions.
The functional derivatives of the Christoffel symbol, obtained from the expansion of the covariant derivative which appear in previous equations, are explicitly given in Appendix \ref{P4Vertices}. 

Before moving to momentum space, we define the Fourier transform using the following conventions
\beqa
&&
\int \, d^3 x_1 \,d^3 x_2 \, \ldots \,d^3 x_n \, 
\langle T^{\mu_1\nu_1}(x_1) \, T^{\mu_2\nu_2}(x_2) \, \ldots \, T^{\mu_n\nu_n}(x_n)\rangle \,
e^{-i(k_1\cdot x_1 + k_2 \cdot x_2 + \ldots + k_n \cdot x_n)} = 
\nn \\
&& \hspace{1cm}
(2\pi)^3\,\delta^{(3)}(\vec{k_1}+ \vec{k_2} + \ldots + \vec{k_n})\,\langle T^{\mu_1\nu_1}(\vec{k_1}) \, T^{\mu_2\nu_2}(\vec{k_2}) \, 
\ldots \, T^{\mu_n\nu_n}(\vec{k_n})\rangle \, , 
\label{P44PFMom}
\eeqa
where all the momenta are incoming. Similar conventions are chosen for the 2-, 3- and 4-point functions in all the expressions that 
follow. We will consider Fourier-transformed correlation functions with an implicit momentum conservation, i.e. we will omit the 
overall delta function. Tridimensional momenta in the perturbative expansions will be denoted as $\vec{k}$ with components $k_\mu$. The modulus of $\vec{k}$ will be simply denoted as $k$.

Going to momentum space the $TT$ correlator satisfies the simple Ward identity
\bea
k_{1\,\nuu} \langle T^{\muu\nuu}(\ku)T^{\mud\nud}(-\ku) \rangle = 0 \,,
\eea
while for the $TTT$ three-point function we obtain
\beqa
&& k_{1\,\nuu} < T^{\muu\nuu}(\ku) T^{\mud\nud}(\kd) T^{\mut\nut}(\kt) >
=
- k_3^{\muu} \langle T^{\mut\nut}(\kd)  T^{\mud\nud}(- \kd)  \rangle \nn \\
&& \hspace{1cm} - k_2^{\muu} \langle T^{\mud\nud}(\kt) T^{\mut\nut}(- \kt) \rangle \nn \\
&& \hspace{1cm}+ k_{3\, \nuu} \bigg[\delta^{\muu\nut} \langle T^{\nuu\mut}(\kd) T^{\mud\nud}(-\kd) \rangle
+ \delta^{\muu\mut} \langle T^{\nuu\nut}(\kd) T^{\mud\nud}(-\kd)  \rangle \bigg] 
\nn \\
&& \hspace{1cm}
+ k_{2\, \nuu} \bigg[\delta^{\muu\nud} \langle T^{\nuu\mud}(\kt) T^{\mut\nut}(-\kt)  \rangle
+ \delta^{\muu\mud}  \langle T^{\nuu\nud}(\kt) T^{\mut\nut}(-\kt) \rangle\bigg],
\eeqa
and finally, for the $TTTT$
\beqa
&&
k_{1\,\nuu} \,< T^{\muu\nuu}(\ku)T^{\mud\nud}(\kd)T^{\mut\nut}(\kt)T^{\muq\nuq}(\kq) >  \nn \\
&=& \bigg[- k_2^{\muu} \, < T^{\mud\nud}(\ku+\kd)T^{\mut\nut}(\kt)T^{\muq\nuq}(\kq) >
\nn \\
&+&
 k_{2\,\nuu}\, \bigg( \delta^{\muu\nud}\, < T^{\nuu\mud}(\ku+\kd)T^{\mut\nut}(\kt)T^{\muq\nuq}(\kq) > \nn \\
&+& \delta^{\muu\mud}\, < T^{\nuu\nud}(\ku+\kd)T^{\mut\nut}(\kt)T^{\muq\nuq}(\kq) >  \bigg)
+  \big( 2 \leftrightarrow 3, 2 \leftrightarrow 4\big) \bigg]  \nn \\
&+& \bigg[ 2\, k_{2\,\nuu}\, \bigg( 
\left[g^{\muu\mud}\right]^{\mut\nut}\, \langle T^{\nuu\nud}(\kq) T^{\muq\nuq}(-\kq) \rangle + 
\left[g^{\muu\nud}\right]^{\mut\nut}\, \langle T^{\nuu\mud}(\kq)T^{\muq\nuq}(-\kq) \rangle \bigg)
\nn \\
&+& \hspace{2mm}
 2\, k_{3\,\nuu}\, \bigg(
\left[g^{\muu\mut}\right]^{\mud\nud}\, \langle T^{\nuu\nut}(\kq) T^{\muq\nuq}(-\kq)  \rangle + 
\left[g^{\muu\nut}\right]^{\mud\nud}\, \langle T^{\nuu\mut}(\kq) T^{\muq\nuq}(-\kq)  \rangle \bigg)
\nn \\
&+&
 \bigg( k_2^{\nut} \delta^{\muu\mut} + k_2^{\mut} \delta^{\muu\nut}\bigg)\, \langle T^{\mud\nud}(\kq) T^{\mut\nut}(-\kq) \rangle \nn \\
&+& \bigg( k_3^{\nud} \delta^{\muu\mud} + k_3^{\mud} \delta^{\muu\nud}\bigg)\, \langle T^{\mut\nut}(\kq) T^{\mud\nud}(-\kq)  \rangle 
+
 \big( 2 \leftrightarrow 4, 3 \leftrightarrow 4 \big) \bigg] \,. \label{P4WI4PFMomentumFlat}
\eeqa
The functional derivatives of the metric tensor are computed using Eq. (\ref{P4funcder}) and given explicitly in Appendix
\ref{P4Vertices}.

For any conformal field theory in an odd dimensional spacetime the relation
\beq \label{P4TRACE}
g_{\mu\nu}\, < T^{\mu\nu} >_g = < {T^\mu}_\mu >_g = 0 \, 
\eeq
describes the invariance under scale transformations. This allows us to derive additional constraints on the fermion and on the conformally coupled scalar
correlators. Differentiating (\ref{P4TRACE}) with respect to the metric up to three times
and then performing the flat limit we obtain the three Ward identities
\beqa
&& < T(\vec{k_1})T^{\mud\nud}(\vec{-k_1})> = 0\, , \label{P4AWard2PF} \nn \\
&& < T(\vec{k_1})T^{\mud\nud}(\vec{k_2})T^{\mut\nut}(\vec{k_3}) > 
=  - 2 < T^{\mud\nud}(\vec{k_2})T^{\mut\nut}(\vec{-k_2})> - 2 < T^{\mud\nud}(\vec{k_3})T^{\mut\nut}(\vec{-k_3})>  \nn\\
&& < T(\vec{k_1})T^{\mud\nud}(\vec{k_2})T^{\mut\nut}(\vec{k_3})T^{\muq\nuq}(\vec{k_4}) > 
= 
-2\, < T^{\mud\nud}(\vec{k_2})T^{\mut\nut}(\vec{k_3})T^{\muq\nuq}(\vec{k_2}+\vec{k_3}) > \label{P4AWard3PF}
\nn \\
&&
-\,  2 \, < T^{\mud\nud}(\vec{k_3}+\vec{k_4})T^{\mut\nut}(\vec{k_3})T^{\muq\nuq}(\vec{k_4}) >
-2\, < T^{\mud\nud}(\vec{k_2})T^{\mut\nut}(\vec{k_2}+\vec{k_4})T^{\muq\nuq}(\vec{k_4}) >
\, .  \label{P4AWard4PF}
\eeqa
Tracing the last two equations with respect to the remaining two and three couples of indices respectively we get the constraints
\beqa
< T(\vec{k_1})T(\vec{k_2})T(\vec{k_3})> &=& 0 \, ,
\nn \\
< T(\vec{k_1})T(\vec{k_2})T(\vec{k_3})T(\vec{k_4}) > &=& 0
\label{P4TraceConstraints}
\eeqa
valid in the conformal case.

\section{Computation of $TTT$}

We begin this section recalling the results for the two-point function $TT$ in $D=3$, which takes the form 
\bea
\langle T^{\mu \nu}(\vec{k}) T^{\alpha \beta}(-\vec{k}) \rangle = 
A(k) \Pi^{\mu\nu\alpha\beta}(\vec{k}) + B(k) \pi^{\mu\nu}(\vec{k}) \pi^{\alpha\beta}(\vec{k})\, ,
\eea
where $\pi^{\mu\nu}(\vec{k})$ is a transverse projection tensor
\bea
\pi^{\mu\nu}(\vec{k}) = \delta^{\mu\nu} - \frac{k^{\mu} k^{\nu}}{k^2} \,,
\label{P4proj}
\eea
while $ \Pi^{\mu\nu\alpha\beta}(\vec{k})$ is transverse and traceless
\bea
\Pi^{\mu\nu\alpha\beta}(\vec{k}) = \frac{1}{2} \bigg[ \pi^{\mu\alpha}(\vec{k}) \pi^{\nu\beta}(\vec{k}) + 
\pi^{\mu\beta}(\vec{k}) \pi^{\nu\alpha}(\vec{k}) - \pi^{\mu\nu}(\vec{k}) \pi^{\alpha\beta}(\vec{k}) \bigg] \,.
\eea
The coefficients $A(k)$ and $B(k)$ for the minimal scalar case (MS), the conformally coupled scalar (CS), 
the gauge field (GF) and for the chiral fermion case (CF) are given by
\bea
&& 
A^{MS}(k) =  A^{CS}(k) =  A^{GF}(k) =  \frac{k^3}{256}\, , 
\qquad A^{CF}(k) =  \frac{k^3}{128} \,, \\
&& 
B^{MS}(k) =  B^{GF}(k) =  \frac{k^3}{256}\,, 
\qquad B^{CS}(k) = B^{CF}(k) = 0 \, .
\eea

At this point we proceed with the computation of the $TTT$ correlator. The contributions introduced in 
Eq. (\ref{P43Tall}) are built out of the vertices listed in the Appendix \ref{P4Vertices} and computed in terms of the usual tensor-to-scalar reductions of the loop integrals in $D=3$. The computations are finite at any stage and require only a removal of the massless tadpoles. Given the complex structure of the tensor result for the 3-T case, here we prefer to give its helicity projections instead of 
presenting it in an expansion on a tensor basis. We follow the same approach presented in \cite{Bzowski:2011ab}. \\
We define the helicity tensors as usual as
\bea
\epsilon^{(s)}_{\mu \nu}(\vec{k}) = \epsilon^{(s)}_{\mu}(\vec{k}) \, \epsilon^{(s)}_{\nu}(\vec{k}) \qquad \mbox{with} \quad s = \pm 1
\eea
where $\epsilon^{(s)}_{\mu}$ is the polarization vector for a spin 1 in $D=3$. They satisfy the standard orthonormality, traceless and transverse conditions
\bea
\epsilon^{(s)}_{\mu \nu}(\vec{k}) \epsilon^{(s') \, \, \mu \nu \, \, *}(\vec{k}) = \delta^{s s'} \,, \qquad \delta^{\mu \nu} \epsilon^{(s)}_{\mu \nu}(\vec{k}) = 0 \,, \qquad k^{\mu}  \epsilon^{(s)}_{\mu \nu}(\vec{k}) = 0 \,.
\eea
We consider a particular realization of the helicity basis choosing, without loss of generality, the three incoming momenta $\ku, \kd$ and $\kt$ to lie in $(x-z)$ plane
\bea
k_i^{\mu} = k_i (\sin \theta_i , 0, \cos \theta_i)\, , 
\eea
with the angles completely determined from the kinematical invariants as
\bea
&& \cos \theta_1 = 1 \,, \qquad   \cos \theta_2 = \frac{1}{2 k_1 k_2} (k_3^2-k_1^2-k_2^2) \,, \qquad \cos \theta_3 = \frac{1}{2 k_1 k_3}(k_2^2 - k_1^2 - k_3^2) \,, \nn \\
&& \sin \theta_1 = 0  \,, \qquad   \sin \theta_2 = \frac{\lambda(k_1,k_2,k_3)}{2 k_1 k_2} \,,  \qquad \sin \theta_3 = - \frac{\lambda(k_1,k_2,k_3)}{2 k_1 k_3} \,
\eea
and where 
\bea
\lambda^2(q_1,q_2,q_3) = (- q_1 + q_2 + q_3)(q_1 - q_2 + q_3) ( q_1 + q_2 - q_3)(q_1 + q_2 + q_3) \,.
\eea
Then the helicity tensors are explicitly given by
\bea
\epsilon^{(s)}_{\mu \nu}(\vec{k}) = \frac{1}{2} \left(
\begin{array}{ccc}
\cos^2 \theta_i & i s \cos \theta_i & - \sin \theta_i \cos \theta_i \\
i s  \cos \theta_i & -1 & - i s \sin \theta_i \\
- \sin \theta_i \cos \theta_i & - i s \sin \theta_i & \sin^2 \theta_i
\end{array}
\right) \,.
\eea 
Notice that $\epsilon_{\mu\nu}^{(s) \, *}(\vec{k}) = \epsilon_{\mu\nu}^{(s)}(- \vec{k})$, which turns useful in comparing our results with those of \cite{Bzowski:2011ab}. Notice also the different normalization of the polarization tensor $\epsilon_{\mu\nu}^{(s)}(\vec{k})$ used by us respect to \cite{Bzowski:2011ab}, by a factor of $\frac{1}{\sqrt{2}}$, which should be kept into account when comparing the results of each $\pm$ helicity projection. 
We are now ready to present in the following sections the $TTT$ correlator for the minimal and the conformally coupled scalar, the 
gauge field and the chiral fermion. All the other missing helicity amplitudes can be obtained from those given here by parity 
transformations or momentum permutations.

\subsection{Minimally coupled scalar}

We list the results for the $TTT$ correlator with a minimal scalar running in the loop.
They correspond to two of the three fundamental topologies appearing in the Fig. \ref{P4TTTtop}.  The different contributions in Eq. (\ref{P43Tall}) can be contracted with polarization 
tensors in order to extract the $\pm$ helicity amplitudes and the trace parts ($T\equiv T^\mu_\mu$). \\
The ordinary 3-point amplitudes are expressed only in terms of the Euclidean length of the external vectors and no relative angles. They are given by
\small
\bea
\langle T(\ku) T(\kd)T(\kt) \rangle_{MS}
&=&  
-\frac{1}{128} \bigg\{ k_1^3+\left(k_2+k_3\right) k_1^2+\left(k_2-k_3\right)^2 k_1+\left(k_2+k_3\right) \left(k_2^2+k_3^2\right) 
\bigg\} \,, \nn\\
\langle T(\ku)T(\kd) T^{(+)}(\kt) \rangle_{MS} 
&=& 
 \frac{1}{1024 k_3^2 (k_1+k_2+k_3)} (k_1 - k_2 - k_3) (k_1 + k_2 - k_3) (k_1 - k_2 + k_3)
\bigg\{ 3 k_1^3 \nn \\
&+& 
\left(5 k_2+6 k_3\right) k_1^2+\left(5 k_2^2+4 k_3 k_2+3 k_3^2\right) k_1
+ 3 k_2 \left(k_2+k_3\right)^2 \bigg\} \,, \nn \\
\langle T(\ku) T^{(+)}(\kd) T^{(-)}(\kt) \rangle_{MS} 
&=& 
-\frac{\left(k_1+k_2-k_3\right)^2 \left(k_1-k_2+k_3\right)^2}{4096 k_2^2 k_3^2} 
\bigg\{ 5 k_1^3-\left(k_2^2+4 k_3 k_2+k_3^2\right) k_1 \nn \\
&+& k_2^3+k_3^3\bigg\} \nn
\eea
\bea
\langle T(\ku) T^{(+)}(\kd) T^{(+)}(\kt) \rangle_{MS} 
&=& 
-\frac{\left(-k_1+k_2+k_3\right)^2}{4096 k_2^2 k_3^2 \left(k_1+k_2+k_3\right)^2} 
\bigg\{ k_3^7+\left(3 k_1+4
   k_2\right) k_3^6+2 \left(k_1^2+6 k_2 k_1 \right. \nn \\ 
&+& \hspace{-3cm} \left. 3 k_2^2\right)
   k_3^5 
+ \left(3 k_1^3+16 k_2 k_1^2+21 k_2^2 k_1+5 k_2^3\right)
   k_3^4+\left(17 k_1^4+36 k_2 k_1^3+38 k_2^2 k_1^2-8 k_2^3 k_1+5
   k_2^4\right) k_3^3 \nn \\
&+& \hspace{-3cm} \left(k_1+k_2\right)^2 \left(29 k_1^3+14
   k_2 k_1^2+9 k_2^2 k_1+6 k_2^3\right) k_3^2+4
   \left(k_1+k_2\right)^3 \left(5 k_1^3+k_2 k_1^2+k_2^3\right)
   k_3 \nn \\
&+& \hspace{-3cm} \left(k_1+k_2\right)^4 \left(5 k_1^3-k_2^2
   k_1+k_2^3\right) \bigg\} \nn\,,
\eea
\bea
\langle T^{(+)}(\ku) T^{(+)}(\kd) T^{(+)}(\kt) \rangle_{MS} 
&=&
-\frac{\left(k_1+k_2-k_3\right) \left(k_1-k_2+k_3\right)
   \left(-k_1+k_2+k_3\right)} {16384 k_1^2 k_2^2 k_3^2 \left(k_1+k_2+k_3\right)^3} 
 \bigg\{ 3 \left(k_1+k_2+k_3\right)^9 \nn \\
&-& \hspace{-3cm} 7
   \left(k_2 k_3+k_1 \left(k_2+k_3\right)\right)
   \left(k_1+k_2+k_3\right)^7+5 k_1 k_2 k_3
   \left(k_1+k_2+k_3\right)^6-64 k_1^3 k_2^3 k_3^3 \bigg\} \,, \nn 
\eea
\bea
\langle T^{(+)}(\ku) T^{(+)}(\kd) T^{(-)}(\kt) \rangle_{MS} 
&=& 
\frac{\left(k_1-k_2-k_3\right) \left(k_1+k_2-k_3\right)^3
   \left(k_1-k_2+k_3\right) } {16384 k_1^2 k_2^2 k_3^2 \left(k_1+k_2+k_3\right)}
\bigg\{ 3 k_3^5 + 4 \left(k_1+k_2\right) k_3^4 \nn \\
&+& \left(k_1^2+k_2^2\right) k_3^3 + \left(k_1^3+k_2^3\right) k_3^2 
+ 4 \left(k_1+k_2\right){}^2 \left(k_1^2 - k_2 k_1 +k_2^2\right) k_3 \nn \\
&+& 
\left(k_1+k_2\right)^3 \left(3 k_1^2 - k_2 k_1 + 3 k_2^2\right)\bigg\} \,. 
\eea
\normalsize
The helicity projections of the contact terms are
\small
\bea
\langle T(\ku) [\mathcal S]^{\mu_2 \mu_3}_ {\mu_2 \mu_3}(\kd,\kt) \rangle_{MS} 
&=& 
\frac{k_1^3}{256} \, , \nn \\
\langle T^{(s_1)}(\ku) [\mathcal S]^{\mu_2 \mu_3}_ {\mu_2 \mu_3}(\kd,\kt) \rangle_{MS} &=& 0 \, , \nn \\
\langle T(\ku) [\mathcal S]^{\mu_2 \, (s_3)}_ {\mu_2}(\kd,\kt) \rangle_{MS} 
&=& 
-  \frac{k_1 \, \lambda^2(k_1,k_2,k_3)}{4096 \, k_3^2} \, ,\nn  \\
\langle T(\ku) [\mathcal S]^{(s_2) \, (s_3)}(\kd,\kt) \rangle_{MS}
&=& 
- \frac{k_1 \, \lambda^2(k_1,k_2,k_3)}{4096 \, k_2^2 \, k_3^2}\bigg\{ - k_1^2 + k_2^2 + k_3^2 + 2 s_2 s_3 \, k_2 k_3\bigg\}\, ,
\nn
\eeqa
\beqa
\langle T^{(s_1)}(\ku) [\mathcal S]^{\mu_2 \, (s_3)}_ {\mu_2}(\kd,\kt) \rangle_{MS}
&=& 
\frac{k_1}{16384 \, k_3^2} \bigg\{ 
k_1^4 + k_2^4 + k_3^4 - 2 k_1^2 k_2^2 - 2 k_2^2 k_3^2 + 6 k_1^2 k_3^2 \nn \\
&&+ 
4 s_1 s_3 \, k_1 k_3  (k_1^2 -k_2^2 + k_3^2) \bigg\} \nn
\eeqa
\beqa
\langle T^{(s_1)}(\ku) [\mathcal S]^{(s_2) \, (s_3)}(\kd,\kt) \rangle_{MS}
&=& 
-\frac{k_1 \, \lambda^2(k_1,k_2,k_3)}{16384 \, k_2^2 k_3^2} \bigg\{ k_1^2 + k_2^2 + k_3^2  
+ 2 s_1 s_2 \, k_1 k_2 + 2 s_1 s_3 \, k_1 k_3 \nn \\
&&+  
2 s_2 s_3 \, k_2 k_3 \bigg\}\, . 
\eea
\normalsize
\subsection{Conformally coupled scalar}
In the case of a conformally coupled scalar the helicity amplitudes of the 3-point correlators are 
\small
\bea
\langle  T(\ku) T(\kd) T(\kt) \rangle_{CS} &=& 0 \,, \nn \\
\langle T(\ku) T(\kd) T^{(+)}(\kt) \rangle_{CS} &=& 0 \,, \nn \\
\langle T(\ku) T^{(s_2)}(\kd) T^{(s_3)}(\kt) \rangle_{CS} 
&=& 
\frac{1}{4} \langle T(\ku) T^{(s_2)}(\kd) T^{(s_3)}(\kt) \rangle_{CF} \,, \nn \\
\langle T^{(s_1)}(\ku) T^{(s_2)}(\kd) T^{(s_3)}(\kt) \rangle_{CS}
&=&  
\langle T^{(s_1)}(\ku) T^{(s_2)}(\kd) T^{(s_3)}(\kt) \rangle_{MS} \,, 
\eea
\normalsize
while the expressions of the contact terms take the form 
\small
\bea
\langle T(\ku) [\mathcal S]^{\mu_2 \mu_3}_ {\mu_2 \mu_3}(\kd,\kt) \rangle_{CS} &=& 0 \,, \nn \\
\langle T(\ku) [\mathcal S]^{\mu_2 \mu_3}_ {\mu_2 \mu_3}(\kd,\kt) \rangle_{CS} &=& 0 \,, \nn \\
\langle T(\ku) [\mathcal S]^{\mu_2 \, (s_3)}_ {\mu_2}(\kd,\kt) \rangle_{CS} &=& 0 \,, \nn \\
\langle T(\ku) [\mathcal S]^{(s_2) \, (s_3)}(\kd,\kt) \rangle_{CS} &=& 0 \,, \nn \\
\langle T^{(s_1)}(\ku) [\mathcal S]^{\mu_2 \, (s_3)}_ {\mu_2}(\kd,\kt) \rangle_{CS} 
&=&  
 \langle T^{(s_1)}(\ku) [\mathcal S]^{\mu_2 \, (s_3)}_ {\mu_2}(\kd,\kt) \rangle_{MS} \,, \nn   \\
\langle T^{(s_1)}(\ku) [\mathcal S]^{(s_2) \, (s_3)}(\kd,\kt) \rangle_{CS}
&=& 
\langle T^{(s_1)}(\ku) [\mathcal S]^{(s_2) \, (s_3)}(\kd,\kt) \rangle_{MS} \,.
\eea
\normalsize
\subsection{Gauge field}

Moving to the gauge contributions, a computation of the ordinary 3-point functions gives 
\small
\bea
\langle T(\ku) T(\kd) T(\kt) \rangle_{GF}
&=& 
-\frac{1}{128} \bigg\{ -3 k_1^3+\left(k_2+k_3\right)
   k_1^2+\left(k_2-k_3\right)^2
   k_1 \nn \\
&& -\left(k_2+k_3\right) \left(3 k_2^2-4 k_3 k_2+3 k_3^2\right) \bigg\}
\, , \nn
\eea
\bea
\langle T(\ku) T(\kd)T^{(+)}(\kt) \rangle_{GF} 
&=&
\frac{1}{1024 \,
  k_3^2 \left(k_1+k_2+k_3\right)} \left(k_1-k_2-k_3\right)
   \left(k_1+k_2-k_3\right) \left(k_1-k_2+k_3\right) \nn \\
&& \times   \bigg\{5 k_1^3 
+ \left(11 k_2+10 k_3\right)
   k_1^2+\left(11 k_2^2+12 k_3 k_2+5 k_3^2\right)
   k_1+5 k_2 \left(k_2+k_3\right)^2\bigg\} 
 \,, \nn 
\eea
\bea
\langle T(\ku)T^{(+)}(\kd) T^{(-)}(\kt) \rangle_{GF} 
&=&
\frac{\left(k_1^2- (k_2-k_3 )^2\right)^2 }{4096 \, k_2^2
   k_3^2} 
   \bigg\{7 k_1^3-\left(3 k_2^2+4 k_3 k_2+3
   k_3^2\right) k_1+k_2^3+k_3^3\bigg\} \,,  \nn \\
\langle T(\ku) T^{(+)}(\kd) T^{(+)}(\kt) \rangle_{GF} 
&=& 
\frac{\left(-k_1+k_2+k_3\right)^2}{4096 \, k_2^2 k_3^2
   \left(k_1+k_2+k_3\right)^2} 
   \bigg\{7 k_1^7 + 28 \left(k_2+k_3\right) k_1^6  \nn \\
&& \hspace{-2cm} + 
\left(39 k_2^2+88 k_3 k_2 + 39 k_3^2\right) k_1^5 + \left(k_2+k_3\right)
   \left(17 k_2^2+71 k_3 k_2+17 k_3^2\right)
   k_1^4 \nn \\
&&\hspace{-2cm} -
\left(k_2+k_3\right)^2 \left(7 k_2^2-34 k_3 k_2 + 7 k_3^2\right) k_1^3 
- 2 \left(k_2+k_3\right)^3 \left(3 k_2^2-5 k_3 k_2+3 k_3^2\right) k_1^2  \nn \\
&& \hspace{-2cm} + 
\left(k_2^6+4 k_3 k_2^5 + 7 k_3^2 k_2^4 + 40 k_3^3 k_2^3 + 7 k_3^4 k_2^2+4 k_3^5 k_2+k_3^6\right) k_1  \nn \\
&&\hspace{-2cm} +
\left(k_2+k_3\right)^5 \left(k_2^2-k_3 k_2+k_3^2\right) \bigg\}\, , \nn 
\eea
\bea
\langle T^{(+)}(\ku) T^{(+)}(\kd) T^{(+)}(\kt) \rangle_{GF} 
&=&
-\frac{\left(k_1+k_2-k_3\right) \left(k_1-k_2+k_3\right)
   \left(-k_1+k_2+k_3\right)} {16384 \, k_1^2 k_2^2 k_3^2 \left(k_1+k_2+k_3\right)^3} 
 \bigg\{ 3 \left(k_1+k_2+k_3\right)^9 \nn \\
&&\hspace{-2cm} - 7
   \left(k_2 k_3+k_1 \left(k_2+k_3\right)\right)
   \left(k_1+k_2+k_3\right)^7+5 k_1 k_2 k_3
   \left(k_1+k_2+k_3\right)^6 \nn \\
&& \hspace{-2cm} - 64 k_1^3 k_2^3 k_3^3 
-  4  \left(k_1+k_2+k_3\right)^6 \left(k_1^3+k_2^3+k_3^3\right)
\bigg\}\, ,\nn  
\eea
\bea
\langle T^{(+)}(\ku) T^{(+)}(\kd) T^{(-)}(\kt) \rangle_{GF} 
&=&
\frac{\left(k_1-k_2-k_3\right) \left(k_1+k_2-k_3\right)^3
   \left(k_1-k_2+k_3\right) } {16384 \, k_1^2 k_2^2 k_3^2 \left(k_1 + k_2 + k_3\right)}
\bigg\{ 3 k_3^5+4 \left(k_1+k_2\right) k_3^4 \nn \\
&&\hspace{-2cm} +
\left(k_1^2+k_2^2\right) k_3^3 + \left( k_1^3 + k_2^3 \right) k_3^2 
+ 4 \left(k_1+k_2\right){}^2 \left(k_1^2-k_2 k_1+k_2^2\right) k_3 	\nn \\
&& \hspace{-2cm} +
\left(k_1+k_2\right)^3 \left(3 k_1^2-k_2k_1+3 k_2^2\right) - 4 \left(k_1+k_2+k_3\right)^2 \left(k_1^3+k_2^3+k_3^3\right)
\bigg\}
 \,,
\eea
\normalsize
while the helicity projections of the contact terms are
\small
\bea
\langle T(\ku) [\mathcal S]^{\mu_2 \mu_3}_ {\mu_2 \mu_3}(\kd,\kt) \rangle_{GF} &=& \frac{3 \, k_1^3}{256} \,, \nn \\
\langle T^{(s_1)}(\ku) [\mathcal S]^{\mu_2 \mu_3}_ {\mu_2 \mu_3}(\kd,\kt) \rangle_{GF} &=& 0 \,, \nn \\
\langle T(\ku) [\mathcal S]^{\mu_2 \, (s_3)}_ {\mu_2}(\kd,\kt)\rangle_{GF} 
&=& 
- \frac{3 \, k_1 \, \lambda^2(k_1,k_2,k_3)}{4096 \, k_3^2} \,, \nn \\
\langle T(\ku) [\mathcal S]^{(s_2) \, (s_3)}(\kd,\kt) \rangle_{GF} 
&=& 
 \frac{k_1^3}{2048 \, k_2^2 \,  k_3^2} \left\{ k_1^4+k_2^4+k_3^4  - 2 k_1^2 k_2^2  - 2 k_1^2 k_3^2  + 6 k_2^2 k_3^2 
\right. \nn \\
&&+
\left.
4\, k_2 k_3  s_2 s_3   \left( k_2^2 + k_3^2 -  k_1^2  \right) \right\}\, , \nn \\
\langle T^{(s_1)}(\ku) [\mathcal S]^{\mu_2 \, (s_3)}_ {\mu_2}(\kd,\kt) \rangle_{GF} 
&=&  
 \frac{3 \, k_1}{16384 \, k_3^2} \bigg\{ 
k_1^4 + k_2^4 + k_3^4 - 2 k_1^2 k_2^2 - 2 k_2^2 k_3^2 + 6 k_1^2 k_3^2 \nn \\
&&+ 
4 s_1 s_3 \, k_1 k_3  (k_1^2 -k_2^2 + k_3^2) \bigg\}\, , \nn \\
\langle T^{(s_1)}(\ku) [\mathcal S]^{(s_2) \, (s_3)}(\kd,\kt) \rangle_{GF} &=& 0 \,.
\eea
\normalsize
\subsection{Chiral fermion}
The analysis can be repeated in the fermion sector. In this case we obtain  
\small
\bea
\langle T(\ku) T(\kd) T(\kt) \rangle_{CF} 
&=& 0 \,, \nn \\ 
\langle T(\ku) T(\kd) T^{(+)}(\kt) \rangle_{CF}
&=& 0 \,, \nn \\
\langle T(\ku) T^{(+)}(\kd) T^{(-)}(\kt) \rangle_{CF} 
&=& 
- \frac{k_2^3 + k_3^3}{1024 \, k_2^2 k_3^2} \bigg\{ k_1^2 - \left( k_2 - k_3 \right)^2 \bigg\}^2 \,, \nn \\
\langle T(\ku) T^{(+)}(\kd) T^{(+)}(\kt) \rangle_{CF} 
&=& 
- \frac{k_2^3 + k_3^3}{1024 \, k_2^2 k_3^2} \bigg\{ k_1^2 - \left( k_2 + k_3 \right)^2 \bigg\}^2  \,, \nn \\
\langle T^{(+)}(\ku) T^{(+)}(\kd) T^{(+)}(\kt) \rangle_{CF}
&=& 
-  \frac{\left(k_1+k_2-k_3\right) \left(k_1-k_2+k_3\right)
\left(-k_1+k_2+k_3\right)}{4096 \, k_1^2 k_2^2 k_3^2 \left(k_1+k_2+k_3\right)^3}
\bigg\{ 32 k_1^3 k_2^3 k_3^3 \nn \\
&& + \left(k_1+k_2+k_3\right)^9  -2 \left(k_2 k_3+k_1 \left(k_2+k_3\right)\right) \left(k_1+k_2+k_3\right)^7 \nn \\
&& + k_1 k_2 k_3 \left(k_1+k_2+k_3\right)^6 \bigg\}\, , \nn \\
\langle T^{(+)}(\ku) T^{(+)}(\kd) T^{(-)}(\kt) \rangle_{CF} 
&=& 
 \frac{\left(k_1-k_2-k_3\right) \left(k_1+k_2-k_3\right)^3 \left(k_1-k_2+k_3\right)}{4096\, k_1^2 k_2^2 k_3^2
\left(k_1+k_2+k_3\right)} \bigg\{ k_3^5+\left(k_1+k_2\right) k_3^4 \nn \\
&& \hspace{-2cm} - k_1 k_2 k_3^3 +\left(k_1+k_2\right)^2 \left(k_1^2-k_2 k_1+k_2^2\right) k_3+\left(k_1+k_2\right)^3 \left(k_1^2+k_2^2\right)\bigg\}\, ,
\eea
\normalsize
for the 3-point functions, while the helicity projections of the contact terms are
\small
\bea
\langle T(\ku) [\mathcal S]^{\mu_2 \mu_3}_ {\mu_2 \mu_3}(\kd,\kt) \rangle_{CF} &=& 0 \,, \nn \\
\langle T^{(s_1)}(\ku) [\mathcal S]^{\mu_2 \mu_3}_ {\mu_2 \mu_3}(\kd,\kt) \rangle_{CF} &=& 0 \,, \nn \\
\langle T(\ku) [\mathcal S]^{\mu_2 \, (s_3)}_ {\mu_2}(\kd,\kt) \rangle_{CF} &=& 0 \,, \nn \\
\langle T(\ku) [\mathcal S]^{(s_2) \, (s_3)}(\kd,\kt) \rangle_{CF} &=&  0  \,, \nn \\
\langle T^{(s_1)}(\ku) [\mathcal S]^{\mu_2 \, (s_3)}_ {\mu_2}(\kd,\kt) \rangle_{CF} &=& 0 \,, \nn \\
\langle T^{(s_1)}(\ku) [\mathcal S]^{(s_2) \, (s_3)}(\kd,\kt) \rangle_{CF} 
&=& 
- \frac{3 \, k_1 \, \lambda^2(k_1,k_2,k_3)}{32768 \, k_2^2 k_3^2} \bigg\{ k_1^2 + k_2^2 + k_3^2
+ 2 s_1 s_2 \, k_1 k_2 \nn \\ + 2 s_1 s_3 \, k_1 k_3 +  
2 s_2 s_3 \, k_2 k_3
\bigg\} \,. 
\eea
\normalsize

\subsection{Multiplicites in the non-abelian case} 
As we have mentioned in the introduction, the expressions of the correlation functions in the small t'Hooft limit in the dual theory can be obtained from the results presented in the previous sections, which are computed for simple free field theories with gauge singlet fields, by introducting some appropriate overall factors. The prescription is to introduce a factor $(N^2-1)$ in front of each of our correlators (and contact terms), with the addition of multiplicity factors $n_\phi, n_\phi', n_A$ and $n_\psi$ in each sector. These corresponds to the multiplicites of the conformal scalars, minimally coupled scalars, gauge fields and fermions, respectively. For instance, the scalar $\langle TTT \rangle$ correlator is obtained with the replacements 
\beqa
\label{P4repper}
\langle TTT \rangle&\rightarrow& (N^2-1)\left( n_\psi \langle  TTT \rangle_{CF} +n_\phi \langle  TTT \rangle_{CS} +n_\phi' \langle TTT \rangle_{MS} + n_A \langle TTT \rangle_{GF}\right),
 \eeqa
 and similarly for all the others. 
 Notice that this results is an exact one. It reproduces, in leading order in the gauge coupling, what one expects for these correlators in a non-abelian gauge theory. The large-$N$ limit, in this case, is performed by replacing the color factor $N^2-1$ in front with $N^2$.

%
%
\section{Mapping of our result to the holographic one}
In holographic cosmology, as we have briefly pointed out in the previous sections, the gravitational cosmological perturbations are expressed in terms of correlators of field theories living on its 3-D boundary, which for a large-$N$ and a small gauge coupling are approximated by free field theories. These theories are dual to the 4-D domain wall gravitational background. The analogous mapping between the cosmological background and the boundary theory requires an analytic continuation. This takes the form 
\bea
k_i\rightarrow -i k_i \qquad \qquad N \rightarrow  -iN, 
\label{P4cont}
\eea
in all the momenta of the 3-T correlators defined above, after the redefinition illustrated in (\ref{P4repper}),
with $N$ denoting the rank of the gauge group in \cite{Bzowski:2011ab}. From now on, in this section, we assume that in all the correlators computed in the previous sections we have done the replacement
(\ref{P4repper}), followed by the analytic continuation described by (\ref{P4cont}).
 
 The final formula for the gravitational perturbations then relates the scalar and tensor fluctuations of the metric to the imaginary parts of the redefined correlators\footnote{In the notation of \cite{Bzowski:2011ab} the euclidean momenta are denoted as $\bar{q}_i$, and correspond to our $k_i$ before the analystic continuation. The authors set $\bar q_i= - i q_i$ to define the euclidean momenta of the cosmological 
correlators in 3-D space, with $q_i$ the final momenta in this space}. The derivations of the 
holographic expressions for each type of perturbations are quite involved, but in the case of scalar cosmological perturbations, parameterized by the field $\zeta$, they take a slightly simpler form 
\bea
\label{P4holo_1}
\langle \zeta(q_1)\z(q_2)\z(q_3)\rangle 
 &=& -\frac{1}{256}\Big(\prod_i \Im [B(\bq_i)]\Big)^{-1}\times
\Im \Big[\langle T(\bq_1)T(\bq_2)T(\bq_3)\>\!\> + 4\sum_i B(\bq_i) \nn\\ 
&-& 2\Big( \langle T(\bq_1)\Upsilon(\bq_2,\bq_3)\rangle +\mathrm{cyclic\,perms.}\Big)\Big].
\eea
Similar formulas are given for the tensor and mixed scalar/tensor perturbations, which can be found in \cite{Bzowski:2011ab}.
In this expression we have omitted an overall factor of $(2\pi^3)$ times a delta function, for momentum conservation.
In \cite{Bzowski:2011ab} the authors use latin indices for the 3-D correlators and introduce the function 
\beq
\Upsilon_{ijkl}(x_1,x_2)\equiv \langle \frac{\delta T^{ij}(x_1)}{\delta g^{k l}(x_2)}\rangle
\label{P4yuppi}
\eeq
which characterises the contact terms, being proportional to a delta function ($\sim \delta(x_1-x_2)$). 
For the rest their conventions are 
\begin{align}
&T(\bq) = \delta_{ij} T_{ij}(\bq), \qquad
 \qquad \Upsilon(\bq_1,\bq_2) =\delta_{ij}\delta_{kl} \Upsilon_{ijkl}(\bq_1,\bq_2) \,.
\end{align}
 The coefficients  $B(\bq_i)$
are related to 2-point functions of the stress tensor.   $\Upsilon$ stands for
the trace of the $\Upsilon_{ijkl}$ tensor. Eq. (\ref{P4holo_1}) allows to map the computation of the bispectrum of the scalar perturbations in ordinary cosmology to a computation of correlation functions in simple free field 
theories. In this case the correlators on the right hand side are obtained by adding the scalar, fermion and gauge contributions. They correspond to fully traced 3-T correlators and contact terms whose imaginary parts are generated by the application of the replacements (\ref{P4cont}) on the diagrammatic results found in the previous sections.

In order to compare our results, expressed in terms of functional derivatives of $\mathcal{S}$ with those of \cite{Bzowski:2011ab} we define
\bea
\label{P4Upsilon}
\Upsilon^{\mu\nu\alpha\beta}(z,x) = \frac{\delta T^{\mu\nu}(z)}{\delta g_{\alpha \beta}(x)} \bigg|_{g_{\mu\nu} = \delta_{\mu\nu}} = - \frac{1}{2} \delta^{\alpha \beta}
\delta(z-x) T^{\mu\nu}(z) - 2 \, [S]^{\mu\nu\alpha\beta}(z,x)
\eea
which is the analog of $\Upsilon_{ijkl}(z,x)$ defined in Eq. (\ref{P4yuppi}).
Because the operations of raising and lowering indices do not commute with the metric functional derivatives, 
$\Upsilon ^{\mu\nu\alpha\beta}(z,x)$ and $\Upsilon_{\mu\nu\alpha\beta}(z,x)$ (we use greek indices) are not simply related
by the contractions with metric tensors. 
A careful analysis shows that the relation between the two quantities in the flat spacetime limit is
\bea
\Upsilon_{\mu\nu\alpha\beta}(z,x) =  \frac{\delta T_{\mu\nu}(z)}{\delta g^{\alpha \beta}(x)} \bigg|_{g_{\mu\nu} = \delta_{\mu\nu}} &=&
-\frac{1}{2} \delta(z-x) \bigg[ \delta_{\alpha \mu} T_{\beta \nu} + \delta_{\beta \mu} T_{\alpha \nu} 
+ \delta_{\alpha \nu} T_{\beta \mu} + \delta_{\beta \nu} T_{\alpha \mu} \bigg](z) \nn \\
&-& \delta_{\mu \rho} \delta_{\nu \sigma} \delta_{\alpha \gamma} \delta_{\beta \delta}   \Upsilon^{\rho\sigma\gamma\delta}(z,x) \,,
\eea
and taking in account Eq. (\ref{P4Upsilon}) we can finally map our contact terms with the expressions of \cite{Bzowski:2011ab}
\bea
&& \langle \Upsilon_{\mu\nu\alpha\beta}(z,x) T_{\rho\sigma}(y) \rangle = 
- \frac{1}{2} \delta(z-x)  \bigg[ \delta_{\alpha \mu} \langle T_{\beta \nu}(z) T_{\rho\sigma}(y)\rangle + \delta_{\beta \mu} \langle T_{\alpha \nu}(z) T_{\rho\sigma}(y)\rangle \nn \\
&& +  \delta_{\alpha \nu} \langle T_{\beta \mu}(z) T_{\rho\sigma}(y)\rangle 
+ 
\delta_{\beta \nu} \langle T_{\alpha \mu}(z) T_{\rho\sigma}(y)\rangle 
 -    \delta_{\alpha \beta}  \langle T_{\mu \nu}(z) T_{\rho\sigma}(y) \rangle  \bigg]  + 2 \,  \langle [\mathcal S]_{\mu \nu \alpha 
 \beta }(z,x) T_{\rho \sigma}(y) \rangle. \nn \\
\eea
This equation is sufficient in order to provide a complete mapping between our results and those of \cite{Bzowski:2011ab} in the $TTT$ case.

\section{Computation of $TTTT$}
\begin{figure}[t]
\label{P4top3T}
\centering
\includegraphics[scale=0.8]{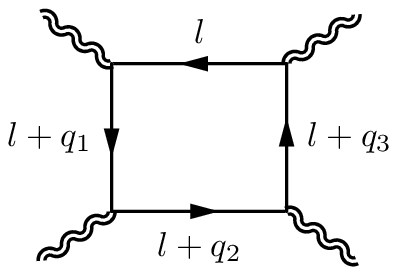}
\hspace{1cm}
\includegraphics[scale=0.8]{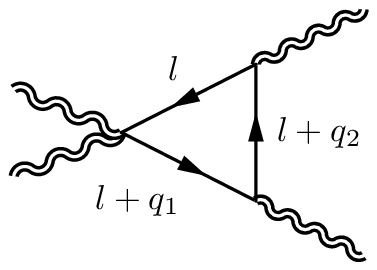}
\hspace{1cm}
\includegraphics[scale=0.8]{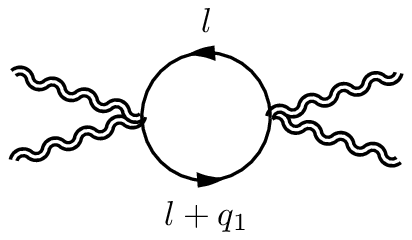}
\hspace{1cm}
\includegraphics[scale=0.8]{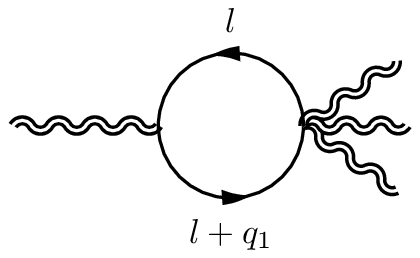}
\caption{Topologies appearing in the expansion of the $4-T$ correlator $\textrm{BoxTop}$, $\textrm{TriTop}$, $\textrm{BubTop}_{22}$ and $\textrm{BubTop}_{13}$. \label{P4TTTTtop}}
\end{figure}
\begin{figure}[t]
\centering
\includegraphics[scale=0.8]{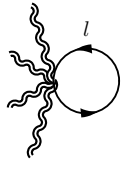}
\caption{Tadpole topology for the TTTT correlator. \label{P4TTTTtadtop}}
\end{figure}
The evaluation of the correlation function of 4 EMT' s is very involved, due to its tensor structure, which is of rank-8, but it becomes more manageable in the case of the scalar amplitude. This is obtained  by tracing all the indices pairwise, given by $\langle TTTT\rangle$ and all the corresponding contact terms, which are identified from the Feynman expansion of the 
related $<TTTT>$.  
The general structure of the 4-T correlator has been defined in full generality in Eq. (\ref{P44PF}) and the scalar component of this relation can be trivially extracted from the same equation. \\ 
We present the results for the minimally coupled scalar, the gauge field, the conformal scalar and the chiral fermion cases. The vertices with one, two and three EMT insertions used in the computation are given in the Appendix \ref{P4Vertices}. \\
We start analyzing in detail the minimally coupled scalar focusing on the classification of the contributions of different topologies. 
\subsection{$TTTT$ for the minimally coupled scalar case}
The first term on the right hand-side of Eq. (\ref{P44PF}), the "ordinary" $\langle TTTT \rangle$ correlator, 
is a four-point function with three box-like contributions which can be obtained, in momentum space, from each other with a suitable re-parameterization of the internal momenta circulating in the loop.  For this reason we compute just one box diagram $\textrm{BoxTop}^{MS}(\vec{q_1},\vec{q_2},\vec{q_3})$ with the internal momenta flowing in the loop as depicted in Fig. \ref{P4TTTTtop}, obtaining
\bea
\langle T(\ku)T(\kd)T(\kt)T(\kq)\rangle_{MS}  &=&16\left( \textrm{BoxTop}^{MS}(\ku,\ku+\kd,-\kq) \right. \nn \\
&& \hspace{-3cm} \left. + \textrm{BoxTop}^{MS}(\ku,\ku+\kd,-\kt) 
+ \textrm{BoxTop}^{MS}(\ku,\kd+\kt,-\kq)\right) \,.
\eea
The triangle terms in Eq. (\ref{P44PF}) are contact terms with the insertion of two external legs on the same vertex. There are six triangle diagrams of this type, characterized by the different couples of attached external momenta $\ku, \kd, \kt, \kq$. Each of these contributions is obtained from the triangle diagram $\textrm{TriTop}^{MS}(\vec{q_1},\vec{q_2})$, illustrated in Fig. \ref{P4TTTTtop}, with the following assignments of momenta
\bea
&& \textrm{TriTop}^{MS}(\ku , -\kq), \qquad \textrm{TriTop}^{MS}(\ku,-\kt), \qquad  \textrm{TriTop}^{MS}(\ku,\ku+\kd), \\ \nn
&& \textrm{TriTop}^{MS}(\ku + \kd, -\kq), \qquad \textrm{TriTop}^{MS}(\ku+\kt,-\kq) \qquad \textrm{TriTop}^{MS}(\ku+\kq,-\kt) \,.
\eea
Here we provide an example 
\bea
\langle \left[\mathcal S\right]^{\muu \, \mud}_{\muu \, \mud}(\ku,\kd)T(\kt) T(\kq)\rangle_{MS} =  4 \, \textrm{TriTop}^{MS}(\ku + \kd , -\kq) \,.
\eea
There are also two classes of contact terms with bubble topologies, namely $\textrm{BubTop}^{MS}_{22}(\vec{q_1})$ and $\textrm{BubTop}^{MS}_{13}(\vec{q_1})$ depicted in Fig. \ref{P4TTTTtop}. The former is defined by two vertices with double T insertions, while the latter is characterized by a three external leg insertion on the same point. In the first case there are three contributions coming from the distinct rearrangements of the external momenta into two pairs, given by
\bea
\textrm{BubTop}^{MS}_{22}(\ku + \kd) \qquad \textrm{BubTop}^{MS}_{22}(\ku + \kt) \qquad \textrm{BubTop}^{MS}_{22}(\ku + \kq) \,,
\eea 
while in the other topology class four terms appear which are given by
\bea
\textrm{BubTop}^{MS}_{13}(\ku) \qquad\textrm{BubTop}^{MS}_{13}(\kd) \qquad \textrm{BubTop}^{MS}_{13}(\kt) \qquad  \textrm{BubTop}^{MS}_{13}(\kq) \,.
\eea
From Eq. (\ref{P44PF}) we deduce that 
\bea
\langle \left[\mathcal S\right]^{\muu \, \mud}_{\muu \, \mud}(\ku,\kd) \left[\mathcal S\right]^{\mut \, \muq}_{\mut \, \muq}(\kt,\kq) \rangle_{MS} &=& \textrm{BubTop}^{MS}_{22}(\ku + \kd) \,,\\
\langle \left[\mathcal S\right]^{\muu \, \mud \, \mut}_{\muu \, \mud \, \mut}(\ku,\kd,\kt)  T(\kq)\rangle_{MS} &=& -2 \, \textrm{BubTop}^{MS}_{13}(\kq) \,,
\eea
and with similar expressions for the other contributions. Notice that there is also a massless tadpole in Fig. \ref{P4TTTTtadtop} which is scheme-dependent and can be set to zero.  \\
For bubble and triangular topologies, the results take a simple form 
\bea
\textrm{TriTop}^{MS}(\vec{q_1},\vec{q_2}) &=& \frac{1}{2048 \pi ^3}   \bigg\{ 
- \left( q_1^2 + q_{12}^2 + q_2^2 \right) \bigg[ 
q_1^2 \, \mathcal B_0 \left(q_1^2 \right)
+ q_{12}^2 \, \mathcal B_0 \left( q_{12}^2 \right)
+ q_2^2 \, \mathcal B_0 \left( q_2^2 \right) \bigg]  \nn \\
&+&
2 \, q_1^2 \, q_{12}^2  \, q_2^2 \, \mathcal C_0 \left( q_1^2, q_{12}^2, q_2^2 \right)
\bigg\} \,,  \\
\textrm{BubTop}^{MS}_{22}(\vec{q_1}) &=& \frac{q_1^4}{1024 \pi^3} \mathcal B_0(q_1^2) \,, \\
\textrm{BubTop}^{MS}_{13}(\vec{q_1}) &=& \frac{3 \, q_1^4}{1024 \pi^3} \mathcal B_0(q_1^2) \,,
\eea
where $q^2_{ij} = (q_i -q_j)^2$.\\
The diagram with box topology, $\textrm{BoxTop}^{MS}$, is more involved and it is expanded on a basis of scalar integrals $\mathcal I_i$ with coefficients $C_i^{MS}$. 
The basis is built from 2-, 3- and 4- point massless scalar integrals reported in Appendix \ref{P4scalarint}.
The elements of this basis are not all independents from each other because $\mathcal D_0$ can be removed using Eq. (\ref{P4D0toC0}). Nevertheless we show the results in this form in order to simplify their presentation. \\
The basis of scalar integrals is given by
\bea
\mathcal I_1(\vec{q_1},\vec{q_2},\vec{q_3}) &=& \mathcal D_0 \left(q_1^2,q_{12}^2,q_{23}^2,q_3^2,q_2^2,q_{13}^2\right) \,, \nn \\
\mathcal I_2(\vec{q_1},\vec{q_2},\vec{q_3}) &=& \mathcal C_0 \left(q_1^2,q_{12}^2, q_2^2 \right) \,, \nn \\
\mathcal I_3(\vec{q_1},\vec{q_2},\vec{q_3}) &=& \mathcal C_0 \left(q_1^2,q_{13}^2, q_3^2 \right) \,,  \nn \\
\mathcal I_4(\vec{q_1},\vec{q_2},\vec{q_3}) &=& \mathcal C_0 \left(q_2^2,q_{23}^2, q_3^2 \right) \,, \nn \\
\mathcal I_5(\vec{q_1},\vec{q_2},\vec{q_3}) &=& \mathcal C_0 \left(q_{12}^2,q_{23}^2, q_{13}^2 \right) \,, \nn \\
\mathcal I_6(\vec{q_1},\vec{q_2},\vec{q_3}) &=& \mathcal B_0 \left(q_1^2 \right) \,, \nn \\
\mathcal I_7(\vec{q_1},\vec{q_2},\vec{q_3}) &=& \mathcal B_0 \left(q_2^2 \right) \,, \nn \\
\mathcal I_8(\vec{q_1},\vec{q_2},\vec{q_3}) &=& \mathcal B_0 \left(q_3^2 \right) \,, \nn \\
\mathcal I_9(\vec{q_1},\vec{q_2},\vec{q_3}) &=& \mathcal B_0 \left(q_{12}^2 \right) \,, \nn \\
\mathcal I_{10}(\vec{q_1},\vec{q_2},\vec{q_3}) &=& \mathcal B_0 \left(q_{13}^2 \right) \,, \nn \\
\mathcal I_{11}(\vec{q_1},\vec{q_2},\vec{q_3}) &=& \mathcal B_0 \left(q_{23}^2 \right) \,, 
\eea
in terms of which the diagram with the box topology can be expressed as 
\bea
\textrm{BoxTop}^{MS}(\vec{q_1},\vec{q_2},\vec{q_3}) = \sum_{i=1}^{11} C^{MS}_i(\vec{q_1},\vec{q_2},\vec{q_3}) \, \mathcal I_i(\vec{q_1},\vec{q_2},\vec{q_3}).
\label{P4boxo}
\eea
The first few coefficients are given by
\small
\bea
C^{MS}_1(\vec{q_1},\vec{q_2},\vec{q_3})&=& \frac{q_1^2 \, q_{12}^2 \, q_{23}^2 \, q_3^2}{2048 \pi^3} \,, \nn 
\eea
\bea
C^{MS}_2(\vec{q_1},\vec{q_2},\vec{q_3}) &=& \frac{q_1^2 q_{12}^2}{2048 \pi^3 \, \lambda^2(q_1,q_{12},q_2)} \bigg[  q_3^2 q_1^4-\left(\left(2 q_3^2+2 q_{12}^2-q_{13}^2+q_{23}^2\right) q_2^2+q_{12}^2
   \left(q_3^2+q_{23}^2\right)\right) q_1^2 \nn \\
&+&  \left(q_2^2-q_{12}^2\right) \left(q_2^2 \left(q_3^2-q_{13}^2+q_{23}^2\right)-q_{12}^2 q_{23}^2\right) \bigg] \,, \nn
\eea
\bea
C^{MS}_3(\vec{q_1},\vec{q_2},\vec{q_3}) &=&  \frac{q_1^2 q_3^2}{2048 \pi^3 \, \lambda^2(q_1,q_{13},q_3)} \bigg[ q_{12}^2 q_1^4-\left(\left(q_{12}^2+2
   q_{13}^2+q_{23}^2\right) q_3^2+q_{13}^2 \left(-q_2^2+2 q_{12}^2+q_{23}^2\right)\right) q_1^2 \nn \\
&+& \left(q_3^2-q_{13}^2\right) \left(\left(q_2^2-q_{12}^2-q_{23}^2\right) q_{13}^2+q_3^2 q_{23}^2\right) \bigg] \,, \nn 
\eea
\bea
C^{MS}_4(\vec{q_1},\vec{q_2},\vec{q_3}) &=& \frac{q_3^2 q_{23}^2}{2048 \pi^3 \, \lambda^2(q_2,q_{23},q_3)} \bigg[ q_1^2 q_3^4-\left(\left(2
   q_1^2+q_{12}^2-q_{13}^2\right) q_2^2+\left(q_1^2+2 q_2^2+q_{12}^2\right) q_{23}^2\right) q_3^2 \nn \\
&+&  \left(q_2^2-q_{23}^2\right) \left(q_2^2 \left(q_1^2+q_{12}^2-q_{13}^2\right)-q_{12}^2 q_{23}^2\right) \bigg] \,, \nn \\
C^{MS}_5(\vec{q_1},\vec{q_2},\vec{q_3}) &=& \frac{q_{12}^2 q_{23}^2}{2048 \pi^3 \, \lambda^2(q_{12},q_{23},q_{13})} \bigg[ q_3^2
   q_{23}^4-\left(\left(q_1^2+q_3^2\right) q_{12}^2+\left(q_1^2-q_2^2+2 \left(q_3^2+q_{12}^2\right)\right) q_{13}^2\right) q_{23}^2 \nn \\
&+& \left(q_{12}^2-q_{13}^2\right) \left(q_1^2 q_{12}^2-\left(q_1^2-q_2^2+q_3^2\right) q_{13}^2\right) \bigg] \, ,
\eea
\normalsize
the remaining ones, having lengthier forms, have been collected in Appendix \ref{P4MSres}.

We remark, if not obvious, that being the $TTTT$ correlator in D=3 dimensions finite and hence (trace) anomaly free, the operation of tracing the indices of an energy-momentum tensor can be performed before or after the evaluation of the integrals appearing in the loops, with no distinction. In D=4 the two procedures are inequivalent, differing by the anomalous term. 

For this reason we have computed the $TTTT$ correlator in two distinct ways, obviously with the same result. In the first case we have traced all the four pairs of indices before computing the loop integrals. In this way we obtain directly the $TTTT$ correlator. In the second case, which is much more involved, we have calculated the $T^{\mu \nu}TTT$ correlation function with a pair of indices not contracted, we have evaluated the tensor integrals and then we have contracted the result with $\delta_{\mu\nu}$. This intermediate step is useful in order to test our computation with the diffeomorphism Ward identity given in Eq. (\ref{P4WI4PFMomentumFlat}). 

\subsection{$TTTT$ for the gauge field case}

As discussed in the previous section also in the case of an abelian gauge field the scalar component of the 4-T correlator can be decomposed in the sum of three box diagrams as
\bea
\langle T(\ku) T(\kd)T(\kt)T(\kq)\rangle_{GF}  &=& 16 \bigg( \textrm{BoxTop}^{GF}(\ku, \ku+\kd,-\kq) + \textrm{BoxTop}^{GF}(\ku,\ku+\kd,-\kt) \nn \\
 &+& \textrm{BoxTop}^{GF}(\ku,\ku+\kt,-\kq) \bigg) \,,
\eea
where the box diagram contributions can be written in terms of a minimal scalar box term (MS) and of extra terms as 
\bea
\textrm{BoxTop}^{GF}(\vec{q_1},\vec{q_2},\vec{q_3}) &=& \textrm{BoxTop}^{MS}(\vec{q_1},\vec{q_2},\vec{q_3})  \nn \\
&& \hspace{-2cm} + 
\frac{1}{2048 \pi^3}\bigg\{
\left(2 q_1^2-q_2^2-q_3^2-q_{12}^2-q_{13}^2\right) q_1^2
   \, \mathcal B_0\left(q_1^2\right) \nn \\
&& \hspace{-2cm} +  \left(2
   q_{23}^2 - q_2^2 - q_3^2 - q_{12}^2 - q_{13}^2\right) q_{23}^2
   \, \mathcal B_0\left(q_{23}^2\right)
+  \left(2 q_2^2 -q_1^2-q_3^2-q_{12}^2-q_{23}^2\right) q_2^2
 \,  \mathcal B_0\left(q_2^2 \right)  \nn \\
&& \hspace{-2cm}+ \left(2 q_3^2 -q_1^2-q_2^2-q_{13}^2-q_{23}^2\right) q_3^2
  \, \mathcal B_0 \left(q_3^2\right)
+  \left(2 q_{12}^2 -q_1^2-q_2^2-q_{13}^2-q_{23}^2\right) q_{12}^2
  \, \mathcal B_0 \left(q_{12}^2 \right) \nn \\
&& \hspace{-2cm} + \left(2 q_{13}^2 -q_1^2-q_3^2-q_{12}^2-q_{23}^2\right) q_{13}^2
 \, \mathcal B_0 \left(q_{13}^2\right) 
+ 2 \, q_2^2 q_{12}^2 q_1^2
  \, \mathcal C_0 \left(q_1^2,q_{12}^2,q_2^2 \right) \nn \\
&& \hspace{-2cm} + 2 q_3^2 q_{13}^2 q_1^2
  \,  \mathcal C_0 \left(q_1^2,q_{13}^2,q_3^2 \right)
+ 2 q_2^2 q_3^2 q_{23}^2
  \, \mathcal C_0 \left(q_2^2,q_{23}^2,q_3^2 \right)
+ 2 q_{12}^2 q_{13}^2 q_{23}^2
  \, \mathcal C_0 \left( q_{12}^2,q_{23}^2,q_{13}^2 \right)  \bigg\} \,. \nn \\
\eea
The reconstruction of the  
$< T(x_1)T(x_2)T(x_3)T(x_4)> $ amplitude can be obtained using the expression above plus the contributions of triangle and bubble topology, corresponding to the relative contact terms. In the case of the gauge field these are 
\bea
 \langle \left[\mathcal S\right]^{\muu \, \mud}_{\muu \, \mud}(\ku,\kd)T(\kt) T(\kq)\rangle_{GF} &=&  4 \, \textrm{TriTop}^{GF}(\ku + \kd , -\kq) \,, \\
 \langle \left[\mathcal S\right]^{\muu \, \mud}_{\muu \, \mud}(\ku,\kd) \left[\mathcal S\right]^{\mut \, \muq}_{\mut \, \muq}(\kt,\kq) \rangle_{GF} &=& \textrm{BubTop}^{GF}_{22}(\ku + \kd) \,,\\
 \langle \left[\mathcal S\right]^{\muu \, \mud \, \mut}_{\muu \, \mud \, \mut}(\ku,\kd,\kt)T(\kq)\rangle_{GF} &=& -2 \, \textrm{BubTop}^{GF}_{13}(\kq) \,,
\eea
and with similar expressions for the other contributions, obtained by a suitable reparameterization of the internal momenta. In this case the discussion is identical as in the minimally coupled scalar. The explicit results for these three topologies are given by
\bea
\textrm{TriTop}^{GF}(\vec{q_1},\vec{q_2}) &=& \frac{1}{2048 \pi ^3}   \bigg\{ 6 \, q_1^2 \, q_{12}^2  \, q_2^2 \, \mathcal C_0 \left( q_1^2, q_{12}^2, q_2^2 \right)
- 3 \, q_1^2 (- 3 q_1^2 + q_{12}^2 + q_2^2 ) \, \mathcal B_0(q_1^2) \nn \\
&-& 3 \, q_{12}^2 (q_1^2 - 3 q_{12}^2 + q_2^2 ) \, \mathcal B_0(q_{12}^2)  
- 3 \, q_2^2 ( q_1^2 + q_{12}^2 - 3 q_2^2 ) \, \mathcal B_0(q_2^2)
\bigg\} \,,   \\
\textrm{BubTop}^{GF}_{22}(\vec{q_1}) &=& \frac{9 \, q_1^4}{1024 \pi^3} \mathcal B_0(q_1^2) \,, \\
\textrm{BubTop}^{GF}_{13}(\vec{q_1}) &=& \frac{15 \, q_1^4}{1024 \pi^3} \mathcal B_0(q_1^2) \,.
\eea

\subsection{$TTTT$ for the conformally coupled scalar case}

As for the gauge fields, also for the conformally coupled scalar the totally traced component of the 4-T correlator is given by
\bea
\langle T(\ku) T(\kd) T(\kt)T(\kq)\rangle_{CS}  
&=& 
16 \bigg( \textrm{BoxTop}^{CS}(\ku, \ku+\kd,-\kq) \nn \\
&& \hspace{-3cm}+ \textrm{BoxTop}^{CS}(\ku,\ku+\kd,-\kt) 
+ 
\textrm{BoxTop}^{CS}(\ku,\ku+\kt,-\kq) \bigg) \,,
\eea
where the box diagram contributions can be written as
\bea
\textrm{BoxTop}^{CS}(\vec{q_1},\vec{q_2},\vec{q_3})
&=& 
\frac{1}{4096 \,\pi^3}\, \bigg\{
\mathcal B_0(q_{13}^2)\, \left(3 (\vec{q_1} \cdot \vec{q_3} )^2 + \vec{q_1} \cdot \vec{q_3} (2 q_2^2 - 3 \vec{q_2} \cdot \vec{q_3}) \right. \nn \\
&& \hspace{-4.5cm} \left. + q_1^2\, ( \vec{q_2} \cdot \vec{q_3} - q_3^2) 
+
\vec{q_1} \cdot \vec{q_2} (-3 \vec{q_1} \cdot \vec{q_3}  + q_3^2) \right)
- \mathcal B_0(q_2^2)\, \left(\vec{q_1} \cdot \vec{q_3} \, q_2^2 + 2\, q_1^2 (\vec{q_2} \cdot \vec{q_3} - q_3^2) 
\right.
\nn \\
&& \hspace{-4.5cm} + 
\left. \vec{q_1} \cdot \vec{q_2} \, (-3 \, \vec{q_2} \cdot \vec{q_3} + 2 q_3^2)\right) \bigg\}\, . 
\eea
Concerning the contact terms, the triangle and bubble topology contributions are  given by
\bea
\langle \left[\mathcal S\right]^{\muu \, \mud}_{\muu \, \mud}(\ku,\kd)T(\kt) T(\kq)\rangle_{CS} &=& 
\frac{1}{256 \, \pi^3} \vec{k_1} \cdot \vec{k_2} \, \vec{k_3} \cdot \vec{k_4} \, \mathcal B_0\left( (\vec{k_1} + \vec{k_2} )^2 \right)
\,, \\
\langle \left[\mathcal S\right]^{\muu \, \mud}_{\muu \, \mud}(\ku,\kd) \left[\mathcal S\right]^{\mut \, \muq}_{\mut \, \muq}(\kt,\kq) 
\rangle_{CS} 
&=& \frac{1}{1024 \, \pi^3} \vec{k_1} \cdot \vec{k_2} \,  \vec{k_3} \cdot \vec{k_4} \, \mathcal B_0\left( (\vec{k_1} + \vec{k_2} )^2 \right)   \,,\\
\langle \left[\mathcal S\right]^{\muu \, \mud \, \mut}_{\muu \, \mud \, \mut}(\ku,\kd,\kt)T(\kq)\rangle_{CS} &=& 0 \,.
\eea

\subsection{$TTTT$ for the chiral fermion case}

In the case of the chiral fermion field the ordinary EMT's correlation function is zero
\bea
\langle T(\ku) T(\kd) T(\kt)T(\kq)\rangle_{CF}  
&=& 
16 \bigg( \textrm{BoxTop}^{CF}(\ku, \ku+\kd,-\kq) \nn \\
&& \hspace{-3cm} + \textrm{BoxTop}^{CF}(\ku,\ku+\kd,-\kt) 
+ 
\textrm{BoxTop}^{CF}(\ku,\ku+\kt,-\kq) \bigg) = 0 \,,
\eea
despite the fact that the single box contribution is given by
\beqa
\textrm{BoxTop}^{CF}(\vec{q_1},\vec{q_2},\vec{q_3}) 
&=& 
\frac{1}{128 \,\pi^3}\, \bigg\{ 
B_0(q_2^2)\, (- q_2^2 \, \vec{q_1} \cdot \vec{q_3}   + \vec{q_1} \cdot \vec{q_2} \, \vec{q_2} \cdot \vec{q_3} ) 
\nn \\ 
&& \hspace{-4cm} + 
B_0(q_{31}^2)\, \left(- \vec{q_1} \cdot \vec{q_3} ( \vec{q_1} \cdot \vec{q_2} - \vec{q_1}\cdot \vec{q_3} + \vec{q_2} \cdot \vec{q_3}) 
+ q_1^2  (\vec{q_2} \cdot \vec{q_3} - q_3^2) + q_3^2 \, \vec{q_1} \cdot \vec{q_2} \right)
\bigg\} \,.
\eeqa
All the other topologies are identically zero,
\bea
\langle \left[\mathcal S\right]^{\muu \, \mud}_{\muu \, \mud}(\ku,\kd)T(\kt) T(\kq)\rangle_{CF} &=& 0 \,, \\
\langle \left[\mathcal S\right]^{\muu \, \mud}_{\muu \, \mud}(\ku,\kd) \left[\mathcal S\right]^{\mut \, \muq}_{\mut \, \muq}(\kt,\kq)  \rangle_{CF}  &=& 0 \,,\\
\langle \left[\mathcal S\right]^{\muu \, \mud \, \mut}_{\muu \, \mud \, \mut}(\ku,\kd,\kt)T(\kq)\rangle_{CF} &=& 0  \,,
\eea
with all the similar contributions obtained by exchanging the respective momenta.

\subsection{Relations between contact terms in the 4-T case} 
As we have mentioned in the introduction, the extension of the holographic formula for scalar and tensor perturbations remains to be worked out. This is expected to require a lengthy but straightforward extension of the methods developed for the analysis of the bispectrum. For this reason here we reformulate the structure of the contact terms, which are expected to be part of this extension, in a form similar to those presented in the previous section.  We recall that in our notations the contact terms are given as in Eq. (\ref{P44PF}). 

For example, extending the previous notations, the contact term with the triangle topology in the two formulations are related as
\bea \label{P4U1TT}
\langle \Upsilon_{\mu\nu\alpha\beta}(z,x) T_{\rho\sigma}(y) T_{\lambda\tau}(t) \rangle 
&=& 
- \frac{1}{2} \delta(z-x)  \bigg[ \delta_{\alpha \mu} \langle T_{\beta \nu}(z) T_{\rho\sigma}(y)T_{\lambda\tau}(t)\rangle
 + 
\delta_{\beta \mu} \langle T_{\alpha \nu}(z) T_{\rho\sigma}(y)T_{\lambda\tau}(t)\rangle \nn \\
&&  \hspace{-3cm} + 
\delta_{\alpha \nu} \langle T_{\beta \mu}(z) T_{\rho\sigma}(y)T_{\lambda\tau}(t)\rangle
+ \delta_{\beta \nu} \langle T_{\alpha \mu}(z) T_{\rho\sigma}(y)T_{\lambda\tau}(t)\rangle 
- \delta_{\alpha \beta}  \langle T_{\mu \nu}(z) T_{\rho\sigma}(y)T_{\lambda\tau}(t) \rangle  \bigg]  \nn \\
&& \hspace{-3cm} +
2 \,  \langle [\mathcal S]_{ \mu \nu \alpha \beta}(z,x) T_{\rho \sigma}(y) T_{\lambda\tau}(t) \rangle.
\eea
Similar relations hold for those contact terms of bubble topology  
\small
\bea \label{P4U1U1}
\langle \Upsilon_{\mu\nu\alpha\beta}(z,x) \Upsilon_{\rho\sigma\lambda\tau}(y,t) \rangle
&=&
\frac{1}{4}\, \delta(z-x) \delta(y-t)\, 
\bigg[
\bigg(
  \delta_{\alpha\mu}\delta_{\lambda\rho} \langle T_{\beta\nu}(z) T_{\tau\sigma}(x) \rangle 
+ \delta_{\alpha\mu}\delta_{\lambda\sigma} \langle T_{\beta\nu}(z) T_{\tau\rho}(x) \rangle 
\nn \\
&& \hspace{-45mm} 
+\,  \delta_{\alpha\mu}\delta_{\tau\rho}    \langle T_{\beta\nu}(z) T_{\lambda\sigma}(x)  \rangle
+    \delta_{\alpha\mu} \delta_{\tau\sigma} \langle T_{\beta\nu}(z) T_{\lambda\rho}(x) \rangle
+ ( \mu \leftrightarrow \nu )
\bigg)
+ (\alpha \leftrightarrow \beta)
\bigg]
\nn \\
&& \hspace{-45mm}
-\bigg\{ 
\delta(z-x)\, 
\bigg[
\frac{1}{4}\, \delta_{\lambda\tau}\, \delta(y-t)\, 
\bigg( 
  \delta_{\alpha\mu}\langle T_{\beta\nu}(z) T_{\rho\sigma}(y) \rangle 
+ \delta_{\alpha\nu}\langle T_{\beta\mu}(z) T_{\rho\sigma}(y) \rangle
+ \left( \alpha \leftrightarrow \beta \right) \bigg)
\nn \\
&& \hspace{-45mm} 
+ \bigg(
  \delta_{\alpha\mu} \langle T_{\beta\nu}(z) \left[\mathcal S\right]_{\rho\sigma\lambda\tau}(y,t) \rangle
+ \delta_{\alpha\nu} \langle T_{\beta\mu}(z) \left[\mathcal S\right]_{\rho\sigma\lambda\tau}(y,t) \rangle
+ \left(\alpha \leftrightarrow \beta \right) \bigg) \bigg]  \nn \\
&& \hspace{-45mm}
+\, \left(\mu,\nu,z,\alpha,\beta,x\right) \leftrightarrow \left(\rho,\sigma,y,\lambda,\tau,t\right) 
\bigg\}
\nn \\
&& \hspace{-45mm} 
+ \bigg( 
\frac{1}{4}\, \delta_{\alpha\beta}\, \delta_{\lambda\tau}\, \delta(z-x)\, \delta(y-t)\, \langle T_{\mu\nu}(z)T_{\rho\sigma}(y)\rangle
+ \, \delta_{\alpha\beta}\, \delta(z-x)\, 
\langle T_{\mu\nu}(z)\left[\mathcal S\right]_{\rho\sigma\lambda\tau}(y,t)\rangle
\nn \\
&& \hspace{-45mm}
+\, \delta_{\lambda\tau}\, \delta(y-t)\, \langle T_{\rho\sigma}(y) \left[\mathcal S\right]_{\mu\nu\alpha\beta}(z,x)\rangle
+ 4\, \langle \left[\mathcal S\right]_{\mu\nu\alpha\beta}(z,x) \left[\mathcal S\right]_{\rho\sigma\lambda\tau}(y,t)\rangle 
\bigg) \, .
\eea
\normalsize
Finally, we expect, in a possible generalization of the holographic formula for scalar and tensor perturbations at the trispectrum level, double functional derivatives of the EMT with respect to the metric 
\beq
\Upsilon_{\mu\nu\alpha\beta\rho\sigma}(z,x,y) = 
\frac{\delta^2 T_{\mu\nu}(z)}{\delta g^{\rho\sigma}(y)\delta g^{\alpha\beta}(x)}\bigg|_{g_{\mu\nu} = \delta_{\mu\nu}}\, .
\eeq
After some work, the expression of this object in terms of multiple functional derivatives of the action and of EMT's is found to be
\bea\label{P4U2}
\Upsilon_{\mu\nu\alpha\beta\rho\sigma}(z,x,y)
&=&
\frac{1}{2}\,
\delta(z-x)\delta(z-y)\, 
\bigg[
 \left[ g_{\mu\beta}g_{\alpha\kappa}g_{\nu\lambda} + g_{\mu\alpha}g_{\beta\kappa}g_{\nu\lambda}
+ (\mu \leftrightarrow \nu) - g_{\alpha\beta}g_{\mu\lambda}g_{\nu\kappa} \right]_{\rho\sigma}
\nn \\
&-&
\frac{1}{2} \delta_{\rho\sigma} \left( \delta_{\mu\beta}\delta_{\alpha\kappa}\delta_{\nu\lambda} 
+ \delta_{\mu\alpha}\delta_{\beta\kappa}\delta_{\nu\lambda}+ (\mu \leftrightarrow \nu) 
- \delta_{\alpha\beta}\delta_{\mu\lambda}\delta_{\nu\kappa} \right) \bigg]\, T_{\lambda\kappa}(z)
\nn \\
&-&
\delta(z-x)\, \left[ \delta_{\mu\beta}\delta_{\alpha\kappa}\delta_{\nu\lambda} 
+ \delta_{\mu\alpha}\delta_{\beta\kappa}\delta_{\nu\lambda}+ (\mu \leftrightarrow \nu) 
- \delta_{\alpha\beta}\delta_{\mu\lambda}\delta_{\nu\kappa} \right]
\,\left[\mathcal S\right]_{\lambda\kappa\rho\sigma}(z,y)
\nn \\
&+&
\delta(z-y)\, \left[\delta_{\rho\sigma}\, \left[\mathcal S\right]_{\mu\nu\alpha\beta}(z,x)
- 2\, \left[g_{a\alpha}g_{b\beta}g_{e\mu}g_{f\nu}\right]_{\rho\sigma}\, \left[\mathcal S\right]_{efab}(z,x)\right]
\nn \\
&-& 
2\,\left[\mathcal S\right]_{\mu\nu\alpha\beta\rho\sigma}(z,x,y) \, ,
\eea
where we refer to appendix \ref{P4Vertices} for a list of the derivatives of metric tensors.\\
We conclude by presenting, along the same lines, the expression of the last contact term which will be present in the holographic extension. 
This is related to our contact terms by the formula
\small
\bea \label{P4U2T}
\langle \Upsilon_{\mu\nu\alpha\beta\rho\sigma}(z,x,y)T_{\tau\omega}(t) \rangle 
&=&
\frac{1}{2}\,
\delta(z-x)\delta(z-y)\, 
\bigg[
 \left[ g_{\mu\beta}g_{\alpha\kappa}g_{\nu\lambda} + g_{\mu\alpha}g_{\beta\kappa}g_{\nu\lambda}
+ (\mu \leftrightarrow \nu) - g_{\alpha\beta}g_{\mu\lambda}g_{\nu\kappa} \right]_{\rho\sigma}
\nn \\
&& \hspace{-2cm}  -
\frac{1}{2} \delta_{\rho\sigma} \left( \delta_{\mu\beta}\delta_{\alpha\kappa}\delta_{\nu\lambda} 
+ \delta_{\mu\alpha}\delta_{\beta\kappa}\delta_{\nu\lambda}+ (\mu \leftrightarrow \nu) 
- \delta_{\alpha\beta}\delta_{\mu\lambda}\delta_{\nu\kappa} \right) \bigg]\, \langle T_{\lambda\kappa}(z)T_{\tau\omega}(t)\rangle
\nn \\
&& \hspace{-2cm} -
\delta(z-x)\, \left[ \delta_{\mu\beta}\delta_{\alpha\kappa}\delta_{\nu\lambda} 
+ \delta_{\mu\alpha}\delta_{\beta\kappa}\delta_{\nu\lambda}+ (\mu \leftrightarrow \nu) 
- \delta_{\alpha\beta}\delta_{\mu\lambda}\delta_{\nu\kappa} \right]
\,\langle \left[\mathcal S\right]_{\lambda\kappa\rho\sigma}(z,y)T_{\tau\omega}(t)\rangle
\nn \\
&& \hspace{-2cm} +
\delta(z-y)\, \left[\delta_{\rho\sigma}\, \langle \left[\mathcal S\right]_{\mu\nu\alpha\beta}(z,x) T_{\tau\omega}(t) \rangle
- 2\, \left[g_{a\alpha}g_{b\beta}g_{e\mu}g_{f\nu}\right]_{\rho\sigma}\, 
\langle \left[\mathcal S\right]_{efab}(z,x) T_{\tau\omega}(t) \rangle \right]
\nn \\
&& \hspace{-2cm} - 
2\, \langle \left[\mathcal S\right]_{\mu\nu\alpha\beta\rho\sigma}(z,x,y) T_{\tau\omega}(t) \rangle.
\eea
\normalsize
%
%


\section{Conclusions}
The study of holographic cosmological models is probably at its beginning and there is little doubt that the interest in these models will be growing in the near future. In these formulations, the metric perturbations of a cosmological inflationary phase characterized by strong gravity can be expressed in terms of correlation functions involving stress energy tensors 
in simple 3-D field theories. We have presented an independent derivation of all the amplitudes which are part of the 3-T correlators and extended the analysis to the fully traced component of the 4-T one. The analysis is rather involved and is based on an extension of the approach developed in \cite{Coriano:2012wp}, which dealt with the 3-T case in D=4. In D=3, the absence of anomalies simplifies considerably the treatment, but the perturbative expression of the 4-T amplitude carries the same level of difficulty of the 4-D case.

\appendix

\chapter{Appendix}

\section{Appendix. Feynman rules for the QCD sector}
\label{P2rules}

The Feynman rules used in the computation of the $TJJ$ correlator in QCD are collected here. The direction of each momentum is shown on the corresponding line.
\begin{itemize}
\item {Graviton - fermion - fermion vertex}
\\
\\
\bmi{95pt}
\includegraphics[scale=1.0]{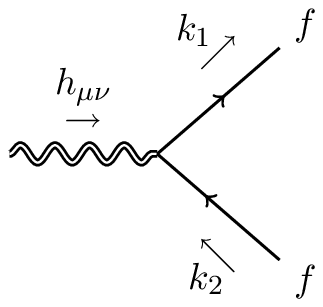}
\emi
\bmi{70pt}
\bann
&=& - i \, \frac{\kappa}{2} \, V^{\prime}_{\mu\nu}(k_1,k_2) \nn \\
&=& - i \, \frac{\kappa}{2} \, \bigg\{ \frac{1}{4} \left[\gamma_\mu (k_1 + k_2)_\nu
+\gamma_\nu (k_1 + k_2)_\mu \right]  \nn \\
&-& \frac{1}{2} g_{\mu \nu}
[\gamma^{\lambda}(k_1 + k_2)_{\lambda} - 2 m]  \bigg\} \nn \\
\eann
\emi
\bea
\label{P2VGff}
\eea
\item{Graviton - gluon - gluon vertex}
\\ \\
\bmi{110pt}
\includegraphics[scale=1.0]{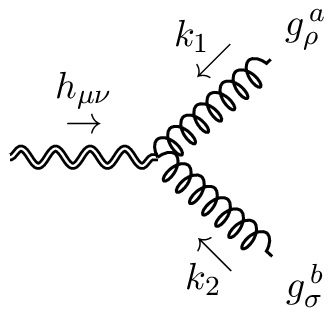}
\emi
\bmi{100pt}
\bann
&=& - i \, \frac{\kappa}{2} \, \delta_{a b} \, V^{Ggg}_{\mu\nu\rho\sigma}(k_1,k_2) \nn \\
 &= &- i \, \frac{\kappa}{2} \, \delta_{a b} \left\{ k_1\cdot k_2 \, C_{\mu\nu\rho\si} + D_{\mu\nu\rho\si}(k_1,k_2) + \frac{1}{\xi} \, E_{\mu\nu\rho\si}(k_1,k_2)  \right\}
\eann
\emi
\bea
\eea
\item{Graviton - ghost - ghost vertex}
\\ \\
\bmi{110pt}
\includegraphics[scale=1.0]{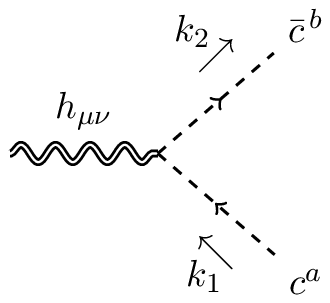}
\emi
\bmi{70pt}
\bann
& = & - i \,  \frac{\kappa}{2}\,  \delta^{a b} \, C_{\mu\nu\rho\sigma} \, k_{1\,\rho} \, k_{2\,\sigma}
\eann
\emi
\bea
\eea
%
%
\item{Graviton - fermion - fermion - gluon vertex}
\\ \\
\bmi{110pt}
\includegraphics[scale=1.0]{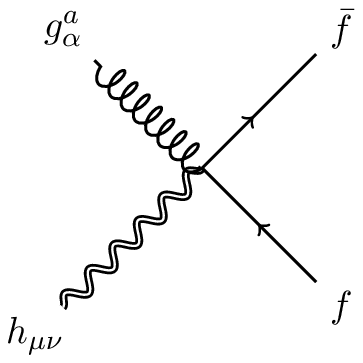}
\emi
\bmi{70pt}
\bann
&=& i g \, \frac{\kappa}{2} \,T^a\, W^{\prime}_{\mu\nu\alpha}
= i g \, \frac{\kappa}{2} \, T^a \left\{ -\frac{1}{2} (\gamma_\mu \, g_{\nu\alpha}
+\gamma_\nu \, g_{\mu\alpha}) +  g_{\mu \nu} \, \gamma_{\alpha} \right \}
\eann
\emi
\bea
\label{P2WGffg}
\eea
%
%
%
\item{Graviton - gluon - gluon - gluon  vertex}
\\ \\
\bmi{110pt}
\includegraphics[scale=1.0]{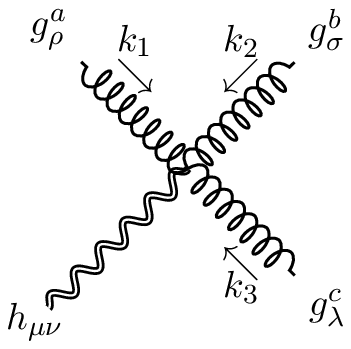}
\emi
\bmi{110pt}
\begin{eqnarray*}
&=&- g \frac{\kappa}{2} f^{a b c} V^{Gggg}_{\mu\nu\rho\sigma\lambda}(k_1,k_2,k_3) \nn \\
&=&  - g \frac{\kappa}{2} f^{a b c} \left\{ C_{\mu\nu\rho\sigma}(k_1-k_2)_{\lambda} + C_{\mu\nu\rho\lambda}(k_3-k_1)_{\sigma}   \right. \nn \\
&& \hspace{2.5cm}  + \left.  C_{\mu\nu\sigma\lambda}(k_2-k_3)_{\rho} + F_{\mu\nu\rho\sigma\lambda}(k_1,k_2,k_3)  \right\}
\hspace{1.7cm}
\end{eqnarray*}
\emi
\bea
\eea
%
%
\item{Graviton - ghost - ghost - gluon vertex}
\\ \\
\bmi{110pt}
\includegraphics[scale=1.0]{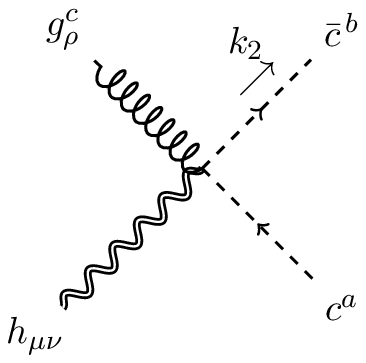}
\emi
\bmi{70pt}
\bann
&=& \frac{\kappa}{2} \, g \, f^{a b c} \, C_{\mu\nu\rho\sigma} \, k_{2}^{\sigma}
\eann
\emi
\bea
\eea
\end{itemize}
%
%
\bea
&& C_{\mu\nu\rho\sigma} = g_{\mu\rho}\, g_{\nu\sigma}
+g_{\mu\sigma} \, g_{\nu\rho}
-g_{\mu\nu} \, g_{\rho\sigma}\,
\nn \\
&& D_{\mu\nu\rho\sigma} (k_1, k_2) =
g_{\mu\nu} \, k_{1 \, \sigma}\, k_{2 \, \rho}
- \biggl[g^{\mu\sigma} k_1^{\nu} k_2^{\rho}
  + g_{\mu\rho} \, k_{1 \, \sigma} \, k_{2 \, \nu}
  - g_{\rho\sigma} \, k_{1 \, \mu} \, k_{2 \, \nu}
  + (\mu\leftrightarrow\nu)\biggr]\, \nn \\
&& E_{\mu\nu\rho\sigma} (k_1, k_2) = g_{\mu\nu} \, (k_{1 \, \rho} \, k_{1 \, \sigma}
+k_{2 \, \rho} \, k_{2 \, \sigma} +k_{1 \, \rho} \, k_{2 \, \sigma})
-\biggl[g_{\nu\sigma} \, k_{1 \, \mu} \, k_{1 \, \rho}
+g_{\nu\rho} \, k_{2 \, \mu} \, k_{2 \, \sigma}  +(\mu\leftrightarrow\nu)\biggr] \nn \\
&& F_{\mu\nu\rho\sigma\lambda} (k_1,k_2,k_3) =
g_{\mu\rho} \,  g_{\sigma\lambda} \, (k_2-k_3)_{\nu}
+g_{\mu\sigma} \, g_{\rho\lambda} \, (k_3-k_1)_{\nu}
+g_{\mu\lambda} \, g_{\rho\sigma}(k_1-k_2)_{\nu}
+ (\mu\leftrightarrow\nu) \nn \\
\label{CDEFtensors}
\eea
%
%
\section{Appendix. Feynman Rules for the EW sector}
\label{P3FeynRules}
\label{P5feynrules}
\label{P6feynrules}

We collect here all the Feynman rules used in the computation of the $TJJ$ in the electroweak theory. All the momenta are incoming

\begin{itemize}
\item{ graviton - gauge boson - gauge boson vertex}
\\ \\
\begin{minipage}{95pt}
\includegraphics[scale=1.0]{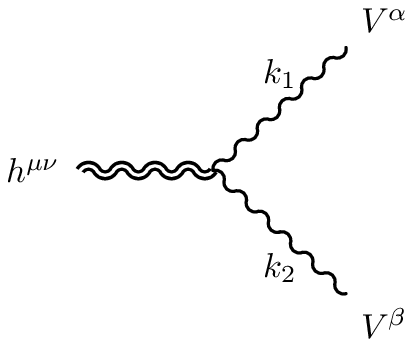}
\end{minipage}
\begin{minipage}{70pt}
\bea
= - i \frac{\kappa}{2} \bigg\{ \left( k_1 \cdot k_2  + M_V^2 \right) C^{\mu\nu\alpha\beta}
+ D^{\mu\nu\alpha\beta}(k_1,k_2) + \frac{1}{\xi}E^{\mu\nu\alpha\beta}(k_1,k_2) \bigg\}
\nn
\eea
\end{minipage}
\bea
\label{P3FRhVV}
\eea
where $V$ stands for the vector gauge bosons $A, Z$ and $W^{\pm}$.
\item{graviton - fermion - fermion vertex}
\\ \\
\begin{minipage}{95pt}
\includegraphics[scale=1.0]{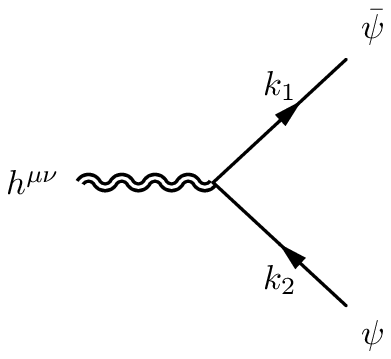}
\end{minipage}
\begin{minipage}{70pt}
\bea
=i \frac{\kappa}{8} \bigg\{ \gamma^\mu \, (k_1 - k_2)^\nu + \gamma^\nu \,(k_1 - k_2)^\mu - 2 \, \eta^{\mu\nu} \left( \ksl_1 - \ksl_2 + 2 m_f \right)\bigg\}
\nn
\eea
\end{minipage}
\bea
\label{P3FRhFF}
\eea
\item{graviton - ghost - ghost vertex }
\\ \\
\begin{minipage}{95pt}
\includegraphics[scale=1.0]{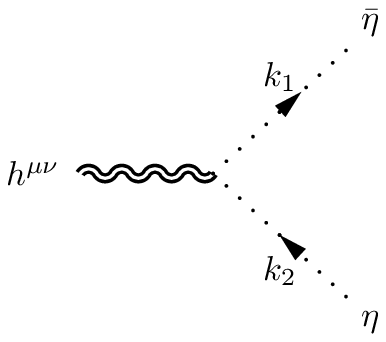}
\end{minipage}
\begin{minipage}{70pt}
\bea
=  i \frac{\kappa}{2} \bigg\{ k_{1\, \rho} \, k_{2 \, \sigma} \, C^{\mu\nu\rho\sigma}
- M_{\eta}^2 \, \eta^{\mu\nu} \bigg\}
\nn
\eea
\end{minipage}
\bea
\label{P3FRhUU}
\eea
where $\eta$ denotes the ghost fields $\eta^{+}$, $\eta^{-}$ ed $\eta^Z$.
\item{graviton - scalar - scalar vertex}
\\ \\
\begin{minipage}{95pt}
\includegraphics[scale=1.0]{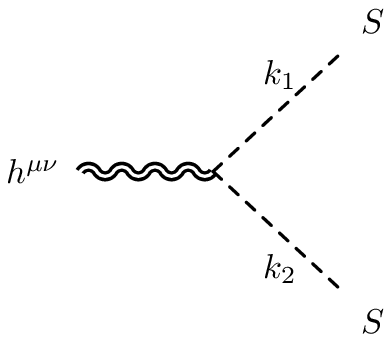}
\end{minipage}
\begin{minipage}{70pt}
\bea
&=&  i \frac{\kappa}{2} \bigg\{ k_{1\, \rho} \, k_{2 \, \sigma} \, C^{\mu\nu\rho\sigma}  - M_S^2 \, \eta^{\mu\nu} \bigg\} \nn \\
&=&  - i 2 \chi  \frac{\kappa}{2} \bigg\{ (k_1+k_2)^{\mu}(k_1+k_2)^{\nu} - \eta^{\mu\nu} (k_1+k_2)^2 \bigg\} \nn
\eea
\end{minipage}
\bea
\label{P3FRhSS}
\eea
where $S$ stands for the Higgs $H$ and the Goldstones $\phi$ and  $\phi^{\pm}$. The first expression is the contribution coming from the minimal energy-momentum tensor while the second is due to the term of improvement for a conformally coupled scalar.
\item{graviton - Higgs vertex}
\\ \\
\begin{minipage}{95pt}
\includegraphics[scale=1.0]{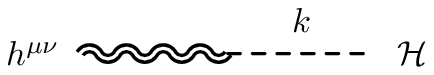}
\end{minipage}
\begin{minipage}{70pt}
\bea
\qquad &=&  i \frac{\kappa}{2} \frac{2 s_w M_W}{3 e} \bigg\{ k^\mu k^\nu - \eta^{\mu\nu}k^2 \bigg\} \nn
\eea
\end{minipage}
\bea
\label{P3FRhH}
\eea
This vertex is derived from the term of improvement of the energy-momentum tensor and it is a feature of the electroweak symmetry breaking because it is proportional to the Higgs vev.
\item{ graviton - three gauge boson vertex}
\\ \\
\begin{minipage}{95pt}
\includegraphics[scale=1.0]{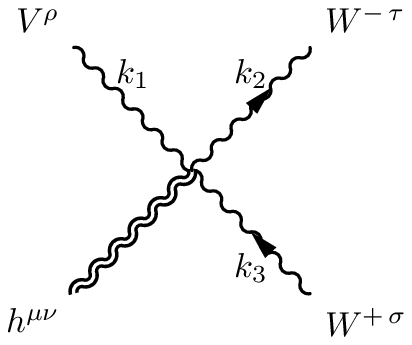}
\end{minipage}
\begin{minipage}{70pt}
\bea
= - i \, e \, \mathcal C_V \frac{\kappa}{2} &\bigg\{&  C^{\mu\nu\rho\sigma} \left(k_3^{\tau}
- k_1^{\tau} \right)  +   C^{\mu\nu\rho\tau} \left(k_1^{\sigma} - k_2^{\sigma} \right) \nn \\
&& + C^{\mu\nu\sigma\tau} \left(k_2^{\rho} - k_3^{\rho} \right)
+\,  F^{\mu\nu\rho\tau\sigma}(k_1,k_2,k_3) \bigg\}
\nn
\eea
\end{minipage}
\bea
\label{P3FRhVWW}
\eea
where $\mathcal C_A = 1$ and $\mathcal C_Z = \frac{c_w}{s_w}$.
\item{graviton - gauge boson - scalar - scalar vertex }
\\ \\
\begin{minipage}{95pt}
\includegraphics[scale=1.0]{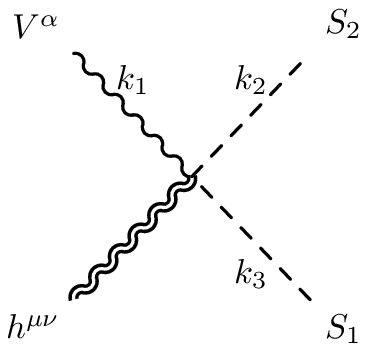}
\end{minipage}
\begin{minipage}{70pt}
\bea
= i \, e \, \mathcal C_{V S_1 S_2} \, \frac{\kappa}{2} \bigg\{ \left( k_{2\, \sigma} - k_{3 \,
\sigma} \right) C^{\mu\nu\alpha\si} \bigg\}
\nn
\eea
\end{minipage}
\bea
\label{P3FRhVSS}
\eea
with $\mathcal C_{V S_1 S_2}$ given by
\bea
\mathcal C_{A\phi^{+}\phi^{-}} = 1 \qquad
\mathcal C_{Z\phi^{+}\phi^{-}} = \frac{c_w^2 - s_w^2}{2 s_w \, c_w} \qquad
\mathcal C_{Z H \phi} =  \frac{i}{2 s_w \, c_w}. \nn
\eea
\item{graviton - gauge boson - ghost - ghost vertex}
\\ \\
\begin{minipage}{95pt}
\includegraphics[scale=1.0]{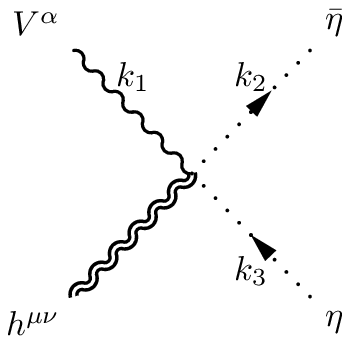}
\end{minipage}
\begin{minipage}{70pt}
\bea
= i \, e \, \mathcal C_{V \eta} \, \frac{\kappa}{2} \, \k_{2 \, \sigma} \,
C^{\mu\nu\alpha\sigma}
\nn
\eea
\end{minipage}
\bea
\label{P3FRhVUU}
\eea
where $V$ denotes the $A$, $Z$ gauge bosons and $\eta$ the two ghosts $\eta^{+}$
and $\eta^{-}$.
The coefficients $\mathcal C$ are defined as
\bea
\mathcal C_{A \eta^{+}} = 1 \qquad
\mathcal C_{A \eta^{-}} = -1 \qquad
\mathcal C_{Z \eta^{+}} =  \frac{c_w}{s_w} \qquad
\mathcal C_{Z \eta^{-}} =  -\frac{c_w}{s_w}. \nn
\eea
\item{graviton - gauge boson - gauge boson - scalar vertex}
\\ \\
\begin{minipage}{95pt}
\includegraphics[scale=1.0]{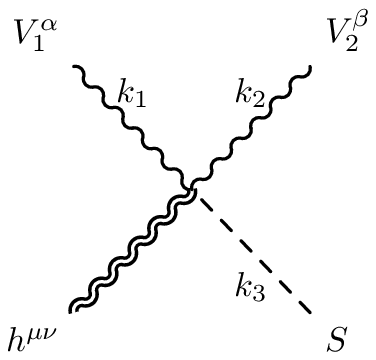}
\end{minipage}
\begin{minipage}{70pt}
\bea
= e \, \mathcal C_{V_1 V_2 S} \, \frac{\kappa}{2} \, M_W \, C^{\mu\nu\alpha\beta}
\nn
\eea
\end{minipage}
\bea
\label{P3FRhVVS}
\eea
where $V$ stands for $A$, $Z$ o $W^{\pm}$ and $S$ for $\phi^{\pm}$
and $H$. The coefficients are defined as

\bea
\mathcal C_{A W^{+} \phi^{-}} = 1 \qquad
\mathcal C_{A W^{-} \phi^{+}} = -1 \qquad
\mathcal C_{Z W^{+} \phi^{-}} = - \frac{s_w}{c_w} \qquad \nn \\
\mathcal C_{Z W^{-} \phi^{+}} = \frac{s_w}{c_w} \qquad
\mathcal C_{Z Z H} = - \frac{i}{s_w \, c_w^2} \qquad
\mathcal C_{W^{+} W^{-} H} = - \frac{i}{c_w}. \nn
\eea
\item{graviton - scalar - ghost - ghost vertex}
\\ \\
\begin{minipage}{95pt}
\includegraphics[scale=1.0]{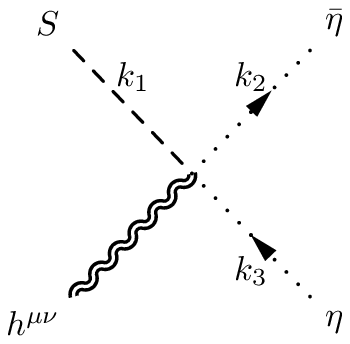}
\end{minipage}
\begin{minipage}{70pt}
\bea
= - i \, e \, \mathcal C_{S \eta} \, \frac{\kappa}{2} \, M_W \, \eta^{\mu\nu}
\nn
\eea
\end{minipage}
\bea
\label{P3FRhSUU}
\eea
where $S = H$ and $\eta$ denotes $\eta^{+}$, $\eta^{-}$ and
$\eta^{z}$. The vertex is defined with the coefficients
\bea
\mathcal C_{H \eta^{+}} = \mathcal C_{H \eta^{-}} = \frac{1}{2 s_w} \qquad \mathcal C_{H
\eta^{z}} = \frac{1}{2 s_w \, c_w}. \nn
\eea
\item{graviton - three scalar vertex}
\\ \\
\begin{minipage}{95pt}
\includegraphics[scale=1.0]{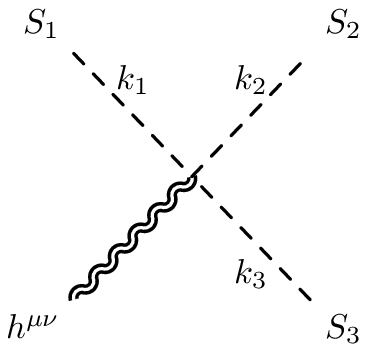}
\end{minipage}
\begin{minipage}{70pt}
\bea
= - i \, e \, \mathcal C_{S_1 S_2 S_3} \, \frac{\kappa}{2} \, \eta^{\mu\nu}
\nn
\eea
\end{minipage}
\bea
\label{P3FRhSSS}
\eea
with $S$ denoting $H$, $\phi$ and
$\phi^{\pm}$. We have defined the coefficients
\bea
\mathcal C_{H \phi \phi} = \mathcal C_{H \phi^{+} \phi^{-}} = \frac{1}{2 s_w \, c_w}
\frac{M_H^2}{M_Z} \qquad \mathcal C_{H H H} = \frac{3}{2 s_w \, c_w} \frac{M_H^2}{M_Z}. \nn
\eea
\item{graviton - scalar - fermion - fermion vertex}
\\ \\
\begin{minipage}{95pt}
\includegraphics[scale=1.0]{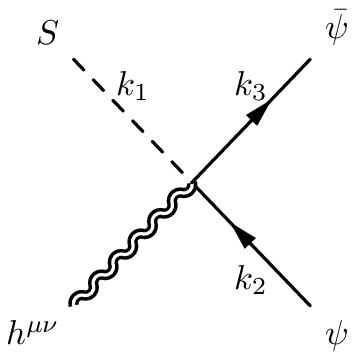}
\end{minipage}
\begin{minipage}{70pt}
\bea
=  \frac{\kappa}{2} \left( C^L_{S\bar \psi \psi} \, P_L + C^R_{S\bar \psi \psi} \, P_R \right) \, \eta^{\mu\nu}
\nn
\eea
\end{minipage}
\bea
\label{P5FRhSFF}
\eea
where  the coefficients are defined as
\bea
&& C^L_{h \bar \psi \psi}  = C^R_{h \bar \psi \psi} = - i \frac{e}{2 s_W} \frac{m}{m_W}  \,, \qquad
C^L_{\phi \bar \psi \psi}  = - C^R_{\phi \bar \psi \psi} =  i\frac{e}{2 s_W} \frac{m}{m_W} 2 I_3 \,, \nn \\
&& C^L_{\phi^+ \bar \psi \psi} = i \frac{e}{\sqrt{2} s_W} \frac{m_{\bar \psi}}{m_W} V_{\bar \psi \psi} \,, \qquad  C^R_{\phi^+ \bar \psi \psi} = - i \frac{e}{\sqrt{2} s_W} \frac{m_{\psi}}{m_W} V_{\bar \psi \psi} \,, \nn \\
&& C^L_{\phi^- \bar \psi \psi} = - i \frac{e}{\sqrt{2} s_W} \frac{m_{\bar \psi}}{m_W} V^*_{\bar \psi \psi} \,, \qquad  C^R_{\phi^- \bar \psi \psi} =  i \frac{e}{\sqrt{2} s_W} \frac{m_{\psi}}{m_W} V^*_{\bar \psi \psi} \,.
\eea
\item{graviton - gauge boson - fermion - fermion vertex}
\\ \\
\begin{minipage}{95pt}
\includegraphics[scale=1.0]{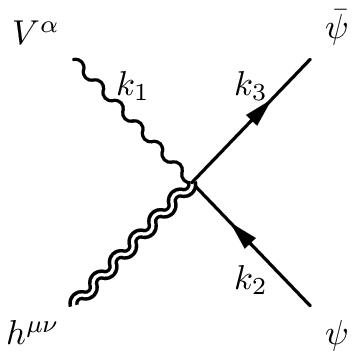}
\end{minipage}
\begin{minipage}{70pt}
\bea
= - \frac{\kappa}{2} \left( C^L_{V \bar\psi \psi} \, P_L + C^R_{V \bar\psi \psi} \, P_R\right) C^{\mu\nu\alpha\beta} \gamma_\beta
\nn
\eea
\end{minipage}
\bea
\label{P5FRhAFF}
\eea
with
\bea
&& C^L_{g \bar\psi \psi} = C^R_{g \bar\psi \psi} = i g_s T^a \,, \qquad  C^L_{\gamma \bar\psi \psi} = C^R_{\gamma \bar\psi \psi} = i e Q \,, \nn \\
&& C^L_{Z \bar\psi \psi} = i \frac{e}{2 s_W c_W} (v + a) \,, \quad C^R_{Z \bar\psi \psi} = i \frac{e}{2 s_W c_W} (v - a) \,, \nn \\
&& C^L_{W^+ \bar\psi \psi} = i \frac{e}{\sqrt{2} s_W} V_{\bar\psi \psi}  \,, \quad C^L_{W^- \bar\psi \psi} = i \frac{e}{\sqrt{2} s_W} V^*_{\bar\psi \psi}  \,, \quad C^R_{W^\pm \bar\psi \psi} = 0 \,,
\eea
and $v = I_3 - 2 s_W^2 Q$, $a = I_3$.
%
\\ \\
\begin{minipage}{95pt}
\includegraphics[scale=1.0]{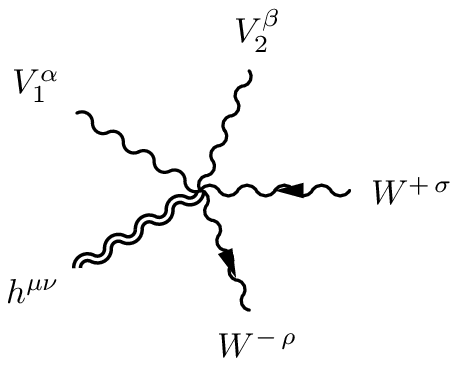}
\end{minipage}
\begin{minipage}{70pt}
\bea
\qquad \qquad = i \, e^2 \, \mathcal C_{V_1 V_2 }\frac{\kappa}{4}
\bigg\{G^{\mu\nu\alpha\beta\si\rho} + G^{\mu\nu\beta\alpha\si\rho} + G^{\mu\nu\alpha\beta\rho\si} + G^{\mu\nu\beta\alpha\rho\si}\bigg\}
\nn
\eea
\end{minipage}
\bea
\label{P3FRhVVWW}
\eea
where $V_1$ e $V_2$ denote $A$ or $Z$. The coefficients $\mathcal C$ are defined as
\bea
C_{A A} = 1 \qquad C_{A Z} = \frac{c_w}{s_w} \qquad C_{Z Z} = \frac{c_w^2}{s_w^2}. \nn
\eea
\item{graviton - gauge boson - gauge boson - scalar - scalar vertex}
\\ \\
\begin{minipage}{95pt}
\includegraphics[scale=1.0]{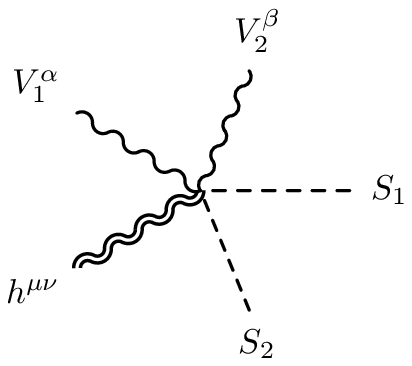}
\end{minipage}
\begin{minipage}{70pt}
\bea
 \qquad = - i \, e^2 \, \mathcal C_{V_1 V_2 S_1 S_2} \frac{\kappa}{2} \, C^{\mu\nu\alpha\beta}
\nn
\eea
\end{minipage}
\bea
\label{P3FRhVVSS}
\eea
where $V_1$ and $V_2$ denote the neutral gauge bosons $A$ and $Z$, while the possible scalars are
$\phi$, $\phi^{\pm}$ and $H$. The coefficients are
\bea
&& \mathcal C_{A A \phi^+ \phi^-} = 2
\qquad\mathcal C_{A Z \phi^+ \phi^-} = \frac{c_w^2 - s_w^2}{s_w \, c_w} \nn \\
&& \mathcal C_{Z Z \phi^+ \phi^-} = \frac{\left( c_w^2 - s_w^2 \right)^2}{2 s_w^2 \,c_w^2}
\qquad\mathcal C_{Z Z \phi \phi} =  \mathcal C_{Z Z H H} = \frac{1}{2 s_w^2 \, c_w^2} . \nn
\eea
\end{itemize}
The tensor structures $C$, $D$, $E$ and $F$ which appear in the Feynman rules defined above are given in Eq.(\ref{CDEFtensors}).

\section{Appendix. Scalar integrals for the QCD sector}
\label{P2scalars}
We collect in this appendix all the scalar integrals involved in the computation of the $TJJ$ vertex in QCD. To set all our conventions, we start with the definition of the one-point function, or massive tadpole $\mathcal  A_0 (m^2)$, the massive bubble $\mathcal B_0 (s, m^2) $  and  the massive three-point function $\mathcal C_0 (s, s_1, s_2, m^2)$
\bea
\mathcal A_0 (m^2) &=& \frac{1}{i \pi^2}\int d^n l \, \frac{1}{l^2 - m^2}
= m^2 \left [ \frac{1}{\bar \eps} + 1 - \log \left( \frac{m^2}{\mu^2} \right )\right],\\
 \mathcal B_0 (k^2, m^2) &=&  \frac{1}{i \pi^2} \int d^n l \, \frac{1}{(l^2 - m^2) \, ((l - k )^2 - m^2 )} \nn \\
 &=& \frac{1}{\bar \eps} + 2 - \log \left( \frac{m^2}{\mu^2} \right ) - a_3 \log \left( \frac{a_3+1}{a_3-1}\right), \\
\mathcal C_0 (s, s_1, s_2, m^2) &=&
 \frac{1}{i \pi^2} \int d^n l \, \frac{1}{(l^2 - m^2) \, ((l -q )^2 - m^2 ) \, ((l + p )^2 - m^2 )} \nn \\
&=&- \frac{1}{ \sqrt \sigma} \sum_{i=1}^3 \left[Li_2 \frac{b_i -1}{a_i + b_i}   - Li_2 \frac{- b_i -1}{a_i - b_i} + Li_2 \frac{-b_i +1}{a_i - b_i}  - Li_2 \frac{b_i +1}{a_i + b_i}
   \right], \nn \\
\label{P2C0polylog}
\eea
with
\bea
a_i = \sqrt {1- \frac{4 m^2}{s_i }} \qquad \qquad
b_i = \frac{- s_i + s_j + s_k }{\sqrt{ \sigma}},
\eea
where $s_3=s$ and in the last equation $i=1,2,3$ and $j, k\neq i$. \\
The one-point and two-point functions  written before in  $n=4 - 2 \, \eps$ are divergent in dimensional regularization with the singular parts given by
\bea
\mathcal A_0 (m^2) ^{sing.}  \rightarrow  \frac{1}{\bar \eps} \, m^2,  \qquad \qquad
\mathcal B_0 (s, m^2) ^{sing.}  \rightarrow  \frac{1}{\bar \eps} ,
\eea
with
\bea
\frac{1}{\bar \eps} = \frac{1}{\eps} - \g - \ln \pi
\label{P2bareps}
\eea
We use two finite combinations of scalar functions given by
\bea
&&  \mathcal B_0 (s, m^2) \, m^2 - \mathcal A_0 (m^2) =  m^2 \left[ 1 - a_3 \log \frac{a_3 +1}{a_3 - 1}  \right] , \\
&& \mathcal D_i \equiv \mathcal D_i (s, s_i,  m^2) =
\mathcal B_0 (s, m^2) - \mathcal B_0 (s_i, m^2) =  \left[ a_i \log\frac{a_i +1}{a_i - 1}
- a_3 \log \frac{a_3 +1}{a_3 - 1}  \right] \qquad i=1,2.
\label{P2D_i}
\nn \\
\eea
The scalar integrals $ \mathcal C_0 (s, 0,0,m^2) $ and $\mathcal D (s, 0, 0, m^2)$ are the $ \{ s_1 \rightarrow 0$, $s_2 \rightarrow 0 \} $ limits of the generic functions $\mathcal C_0(s,s_1,s_2,m^2)$ and $\mathcal D_1(s,s_1,m^2)$
\bea
\mathcal C_0 (s, 0,0,m^2) &=& \frac{1}{2 s} \log^2 \frac{a_3+1}{a_3-1}, \\
\mathcal D (s, 0, 0, m^2) &=& \mathcal D_1 (s,0,m^2)= \mathcal D_2(s,0,m^2) =
  \left[ 2 - a_3 \log \frac{a_3+1}{a_3-1}\right].
\eea
The singularities in $1/\bar \eps$ and the dependence on the renormalization scale $\mu$ cancel out when taking into account the difference of two functions $\mathcal B_0$, so that the $\mathcal D_i$'s  are well-defined; the three-point master integral is convergent.

The renormalized scalar integrals  in the modified minimal subtraction scheme named $\overline{MS}$ are defined as
\bea
\mathcal B_0^{\overline{MS}}(s,0) &=&  2 - L_s,
\label{P2B0masslessOS}\\
\mathcal B_0^{\overline{MS}}(0,0) &=&  \frac{1}{\omega},\\
\mathcal C_0(s,0,0,0) &=& \frac{1}{s}\left[ \frac{1}{\omega^2} + \frac{1}{\omega} L_s + \frac{1}{2} L_s^2 - \frac{\pi^2}{12}  \right],
\label{P2C0masslessOS}
\eea
where
\bea
L_s \equiv \log \left( - \frac{s}{\mu^2} \right ) \qquad \qquad s<0.
\eea
We have set  the spacetime dimensions to $n = 4 + 2 \omega$ with $\omega > 0$. The $1/\omega$ and $1/\omega^2$ singularities in Eqs.~(\ref{P2B0masslessOS}) and (\ref{P2C0masslessOS}) are infrared divergencies due to the zero mass of the gluons.
\section{Appendix. Scalar integrals for the EW sector}
\label{P5scalarint}
\label{P6scalarint}
We collect in this appendix the definition of the scalar integrals appearing in the computation of the correlators in the electroweak sector of the Standard Model.
One-, two- and three-point functions are denoted, respectively as $\mathcal A_0$, $\mathcal B_0$ and $\mathcal C_0$, with
\bea
\mathcal A_0 (m_0^2) &=& \frac{1}{i \pi^2}\int d^n l \, \frac{1}{l^2 - m_0^2} \,, \nn\\
\mathcal B_0 (k^2, m_0^2,m_1^2) &=&  \frac{1}{i \pi^2} \int d^n l \, \frac{1}{(l^2 - m_0^2) \, ((l + k )^2 - m_1^2 )} \,, \nn \\
\mathcal C_0 ((p_1+p_2)^2, p_1^2, p_2^2, m_0^2,m_1^2,m_2^2) &=& \frac{1}{i \pi^2} \int d^n l \, \frac{1}{(l^2 - m_0^2) \, ((l + p_1)^2 - m_1^2 ) \, ((l -p_2 )^2 - m_2^2 ) } \,. \nn \\
\eea
We have also used the finite combination of two-point scalar integrals
\bea
\mathcal D_0 (p^2, q^2, m_0^2,m_1^2) = \mathcal B_0 (p^2, m_0^2,m_1^2) - \mathcal B_0 (q^2, m_0^2,m_1^2) \,.
\eea
The explicit expressions of $A_0$, $B_0$ and $C_0$ in the most general case can be found in \cite{Denner:1991kt}. \\

In chapter \ref{Chap.GravitonFermions}, because the kinematic invariants on the external states of our computation are fixed, $q^2 = (p_1 - p_2)^2$, $p_1^2 = p_2^2 = m^2$, we have defined the shorter notation for the three-point scalar integrals
\bea
\mathcal C_0(m_0^2, m_1^2, m_2^2) = \mathcal C_0(q^2, m^2, m^2, m_0^2, m_1^2, m_2^2) \,,
\eea
with the first three variables omitted.

%
%


\section{Appendix. Conventions for the Standard Model Lagrangian}
\label{P3conventions}
We summarize here some of our conventions used in the computation of the various contributions to the total EMT of the SM.\\
The definitions of the field strengths are
\bea
F^a_{\mu\nu} & = & \pd_\mu G^a_\nu - \pd_\nu G^a_\mu + g_s f^{abc}G^b_\mu G^c_\nu \,, \\
F^A_{\mu\nu} & = & \pd_\mu A_\nu - \pd_\nu A_\mu + g\sin\th_W\c_{\mu\nu} \,, \\
Z_{\mu\nu} & = & \pd_\mu Z_\nu - \pd_\nu Z_\mu + g\cos\th_W\c_{\mu\nu} \,,\\
W^+_{\mu\nu} & = & \pd_\mu W^+_\nu - \pd_\nu W^+_\mu - i g\left[cos\th_W Z_\mu W^+_\nu + \sin\th_W A_\mu W^+_\nu - (\mu \leftrightarrow \nu)\right] \,, \\
W^-_{\mu\nu} & = & \pd_\mu W^-_\nu - \pd_\nu W^-_\mu + i g\left[cos\th_W Z_\mu W^-_\nu + \sin\th_W A_\mu W^-_\nu - (\mu \leftrightarrow \nu)\right] \,,
\eea
with $\c_{\mu\nu}$ given by  $\c_{\mu\nu} = i[W^-_\mu W^+_\nu - W^+_\mu W^-_\nu]$. As usual, we have denoted with $f^{abc}$ the structure constants of $SU(3)_C$, while $e = g \sin\th_W$.
The fermionic Lagrangian is
\small
\bea
\mathcal{L}_{ferm.}
&=& i\bar\psi_{\nu_e}\g^\mu\pd_\mu \psi_{\nu_e} + i\bar \psi_e\g^\mu\pd_\mu \psi_e
    + i\bar \psi_u\g^\mu\pd_\mu \psi_u + i\bar \psi_d\g^\mu\pd_\mu \psi_d  + \frac{e}{\sqrt{2}\sin\th_W}\bigg(\bar\psi_{\nu_e} \g^\mu \frac{1-\g^5}{2} \psi_e\, W^+_\mu \nn \\
&+& \bar \psi_e\g^\mu\frac{1-\g^5}{2}\psi_{\nu_e}\, W^-_\mu\bigg)
 + \frac{e}{\sin2\th_W}\bar\psi_{\nu_e} \g^\mu\frac{1-\g^5}{2}\psi_{\nu_e} Z_\mu  - \frac{e}{\sin2\th_W}\bar \psi_e\g^\mu\bigg(\frac{1-\g^5}{2} \nn \\
 &-& 2\sin^2\th_W\bigg)\psi_e\,Z_\mu
 + \frac{e}{\sqrt{2}\sin\th_W}\bigg(\bar \psi_u\g^\mu\frac{1-\g^5}{2}\psi_d\,W^+_\mu + \bar \psi_d\g^\mu
\frac{1-\g^5}{2}\psi_u\,W^-_\mu \bigg)\nn\\
&+&  \frac{e}{\sin2\th_W}\bar \psi_u\g^\mu\bigg(\frac{1-\g^5}{2} -
2\sin^2\th_W\frac{2}{3}\bigg)\psi_u\,Z_\mu
 - \frac{e}{\sin2\th_W}\bar \psi_d\g^\mu\bigg(\frac{1-\g^5}{2} -
2\sin^2\th_W\frac{1}{3}\bigg)\psi_d\,Z_\mu\bigg]\nn\\
&+& eA_\mu\bigg(-\bar \psi_e\g^\mu \psi_e + \frac{2}{3}\,\bar \psi_u\g^\mu \psi_u - \frac{1}{3}\,\bar \psi_d\g^\mu
\psi_d\bigg) + g_s G^a_\mu\bigg(\bar \psi_u\g^\mu t^a \psi_u + \bar \psi_d\g^\mu t^a \psi_d \bigg) \, .
\eea
\normalsize
The gauge-fixing Lagrangian is given by
\beq
\mathcal L_{g. fix.} = -\frac{1}{2\xi}(\mathcal F^A)^2 -\frac{1}{2\xi}(\mathcal F^Z)^2 -\frac{1}{\xi}(\mathcal F^+)(\mathcal F^-) - \frac{1}{2\xi}(\mathcal F^G)^2 \,,
\eeq
where the gauge-fixing functions in the $R_\xi$ gauge are defined by
\bea
\mathcal F^{G,i} & = & \pd^\s G^i_\s\, ,\nn\\
\mathcal F^A & = & \pd^\s A_\s\, ,\nn\\
\mathcal F^Z & = & \pd^\s Z_\s - \xi M_Z \f\, ,\nn\\
\mathcal F^+ & = & \pd^\s W^+_\s - \frac{1}{2}\xi g v \f^+\, ,\nn\\
\mathcal F^- & = & \pd^\s W^-_\s - \frac{1}{2}\xi g v \f^-\,,
\eea
and we have used for simplicity the same gauge-fixing parameter $\xi$ for all the gauge fields. Finally we give the ghost Lagrangian
\small
\bea
\mathcal{L}_{ghost}
& = & \pd^\mu\bar{c}^a \left(\pd_\mu\d^{ac} + g_s f^{abc}G^b_\mu\right)c^c
 + \pd^\mu\bar{\h}^Z\pd_\mu\h^Z + \pd^\mu\bar{\h}^A\pd_\mu\h^A + \pd^\mu\bar{\h}^+\pd_\mu\h^+ +   \pd^\mu\bar{\h}^-\pd_\mu\h^- \nn \\
&+&  i g \bigg\{\pd^\mu\bar{\h}^+\bigg[W_\mu^+ (\cos\th_W \h^Z + \sin\th_W\h^A) - (\cos\th_W Z_\mu + \sin\th_W A_\mu)\h^+ \bigg] \nn \\
&+& \pd^\mu\bar{\h}^-\bigg[\h^- (\cos\th_W Z_\mu + \sin\th_W A_\mu) - (\cos\th_W\h^Z + \sin\th_W \h^A)W_\mu^-\bigg]\nn \\
&+&  \pd^\mu(cos\th_W\bar{\h}^Z + \sin\th_W\bar{\h}^A )\bigg[W^+_\mu\h^- -   W^-_\mu\h^+\bigg]\bigg\}  - \frac{e \, \xi \, M_W}{\sin2\th_W} \bigg\{-i\f^+\bigg[\cos2\th_W\bar{\h}^+\h^Z \nn \\
&+&   \sin2\th_W\bar{\h}^+\h^A\bigg] + i \f^-\bigg[\cos2\th_W\bar{\h}^-\h^Z + \sin2\th_W\bar{\h}^-\h^A\bigg]\bigg\} - \frac{e\xi}{2\sin\th_W} M_W \bigg[(v + h + i \f)\bar{\h}^+\h^+ \nn \\
&+& (v + h - i \f) \bar{\h}^-\h^-\bigg] - i \frac{e\xi}{2\sin\th_W} M_Z ( - \f^-\bar{\h}^Z\h^+ + \f^+\bar{\h}^Z\h^-) - \frac{e\,\xi \, M_Z}{\sin2\th_W}(v + h)\bar{\h}^Z\h^Z\, .
\eea
\normalsize
\section{Appendix. Technical details of the derivation of the Ward identities}
\label{P3ward}
For the derivation of the Ward identities, the transformations of the fields are given by (we have absorbed a factor $\sqrt{-g}$ in their definitions)
\beqa
V^{'\,\underline{a}}_\mu(x)           &=& V^{\underline{a}}_\mu(x) -\int d^4y\,[\d^{(4)}(x-y)\pd_\nu V^{\underline{a}}_\mu(x)
                               + [\pd_\mu\d^{(4)}(x-y)]V^{\underline{a}}_\nu]\e^\nu(y)\, ,\nn\\
J'(x)                      &=& J(x) -\int d^4y\,\pd_\nu[\d^{(4)}(x-y)J(x)]\e^\nu(y)\, ,\nn\\
\c'(x)                     &=& \c(x) -\int d^4y\,\pd_\nu[\d^{(4)}(x-y)\c(x)]\e^\nu(y)\, .
\eeqa
The term which appears in the first line in the integrand of Eq. (\ref{P3preWard}) can be re-expressed in the following form
\bea \label{P3WIfirstterm}
&& - \int d^4 x \, V\Theta^\mu_{\,\,\,\underline{a}}\bigg[-\d^{(4)}(x-y)\pd_\nu V^{\underline{a}}_\mu(x) - [\pd_\mu\d^{(4)}(x-y)]V^{\underline{a}}_\nu\bigg] = - V\Theta^\mu_{\,\,\,\nu\,;\mu} + V\Theta^\mu_{\,\,\,\underline{a}}V^{\underline{a}}_{\mu\, ;\nu} \nn \\
&& =  - V\bigg[\Theta^\mu_{\,\,\,\nu\,;\mu} +  V_{\underline{a}\r} V^{\underline{a}}_{\mu\,;\nu}\frac{\Theta^{\mu\r} - \Theta^{\r\mu}}{2}\bigg] \,,
\eea
where in the last expression we used the covariant conservation of the metric tensor expressed in terms of the vierbein
\bea
g_{\mu\nu\,;\r} = 0 \Rightarrow V^{\underline{a}}_{\mu\,;\r}V_{\underline{a}\nu} = - V^{\underline{a}}_\mu V_{\underline{a}\nu\,;\r} = - V_{\underline{a}\mu} V^{\underline{a}}_{\nu\,;\r}.
\eea
Other simplifications are obtained using the invariance of the action under local Lorentz transformations \cite{Caracciolo:1989pt},
parameterized as
\beqa
\d V^{\underline{a}}_\mu = {\w^{\underline{a}}}_{\underline{b}} V^{\underline{b}}_\mu\, , \qquad
\d\psi     = \frac{1}{2}\s^{a b}\w_{{\underline{a}\underline{b}}}\psi\, , \qquad
\d\bar{\psi} = -\frac{1}{2}\bar\psi\s^{\underline{a} \underline{b}}\w_{\underline{a b}}\, ,
\eeqa
that gives, using the antisymmetry of $\w^{\underline{ a b}}$
\beq\label{P3vincolofermioni}
\frac{\d S}{\d\psi}\s^{\underline{a b}}\psi - \bar\psi\s^{\underline{a b}}\frac{\d S}{\d\bar{\psi}}
- \frac{\d S}{\d V^{\underline{b}}_\mu}V^{\underline{a}}_\mu + \frac{\d S}{\d V^{\underline{b}}_\mu} V^{\underline{a}}_\mu = 0\, .
\eeq
The previous equation can be reformulated in terms of the energy-momentum tensor $\Theta^{\mu\nu}$
\beq
V(\Theta^{\mu\r} - \Theta^{\r\mu}) = \bar{\psi}\s^{\mu\r}\frac{\d S}{\d\bar{\psi}}
- \frac{\d S}{\d\psi}\s^{\mu\r}\psi\,,
\eeq
which is useful to re-express Eq. (\ref{P3WIfirstterm}) in terms of the symmetric energy-momentum tensor $T^{\mu\nu}$ and to obtain finally, in the flat spacetime limit, Eq. (\ref{P3Ward}).
\section{Appendix. BRST transformations and Slavnov Taylor identities}
\label{P3appendixBRST}
Here we illustrate the derivation of some identities involving 2-point functions using the BRST invariance of
the generating functional
\beq Z[J, \mathcal{F}] = \int \mD\Phi\, e^{i\tilde S}\, ,\eeq
with
\beqa
\tilde S &=& S_{SM} \, + \, \int d^4x\bigg[ J_\mu(x)A^\mu(x) + \bar\eta^A(x)\w^A(x) + \ldots \nn\\
&+&  \chi^A(x)\mathcal F^A(x) + \chi^Z(x)\mathcal F^Z(x) + \chi^+(x)\mathcal F^+(x) + \chi^-(x)\mathcal F^-(x)\bigg]\, .
\eeqa
For convenience we have summarised the BRST transformation of the fundamental fields of the SM Lagrangian used in the derivations of the various STI's in section \ref{P3BRSTsection}
\beqa\label{P3brstQCD}
&& \d A^a_\mu  = \l D^{ab}_ \mu c^b, \qquad \d c^a   =  -\frac{1}{2}g \l f^{abc}c^b c^c, \qquad \d\bar{c}^a  =  -\frac{1}{\xi} \mathcal F^a \, \l = -\frac{1}{\xi}(\pd^\mu A^a_\mu)\l \,,\nn \\
&& \d\psi   =  i g \l c^a t^a\psi, \qquad \d\bar\psi   = - i g\bar\psi t^a\l c^a ,
\eeqa
for an unbroken non abelian gauge theory, and
\beqa
&& \d B_\mu        =  \l \, \pd_\mu\h_Y \qquad \d W^a_\mu    =  \l D^{ab}_{\mu}\eta^b_L = \l(\pd_\mu\h^a_L + \e^{abc}W^b_\mu\h^c_L), \nn \\
&& \d\bar{\eta}_Y   =  -\frac{\l}{\xi}\mathcal F^0, \qquad \d\bar{\eta^a_L}  =  -\frac{\l}{\xi}\mathcal F^a, \qquad \d\eta_Y  =  0, \qquad \d\eta^a_L  = \frac{\l}{2}g\e^{abc}\h^b_L\h^c_L,\nn \\
&& \d H   =   i g'Y H\l\h_Y + i g T^a H\l\h^a_L, \qquad \d H^\dag    =  -i g' H^\dag Y\l \h_Y - i g H^\dag T^a\l\h^a_L,
\eeqa
for the electroweak theory.\\
We require that $\delta_{BRST} Z[J,\mathcal{F}]=0$ under a variation of all the fields and gauge-fixing functions. We then differentiate the resulting equation with respect to the sources of the photon and of the antighost to obtain
\beq
 \label{P3dbrst}
 \frac{\d^2}{\d J^{A\,\mu}(x)\d\w^A(y)}\delta_{BRST} Z[J,\mathcal{F}] =\int\mD\Phi\, e^{i\tilde S}\bigg\{\bar\h^A(y)\d A_\mu(x)
+ \d\bar\eta^A(y)A_\mu(x)\bigg\} = 0\, .\eeq
Introducing the explicit BRST variation of the antighost field $\bar\eta^A(y)$ and of the gauge field $A_\mu(x)$ we obtain
\beq
\langle\bar\h^A(y) D^A_\mu\h^A(x)\rangle = \frac{1}{\xi}\langle\pd^\b A_\b(y) A_\mu(x)\rangle\, .
\eeq
Similarly, in the case of the $Z$ gauge boson, we take two functional derivatives of the condition of BRST invariance of $Z[J, \mathcal{F}]$, as in Eq. (\ref{P3dbrst}), but now respect to $J^{Z\,\mu}(x)$ and to $\w^Z(y)$, to obtain the relation
\beq
\langle\bar\h^Z(y)D_\r^Z\h^Z(x)\rangle = \frac{1}{\xi}\langle \mathcal F^Z(y)Z_\r(x)\rangle\,.
\eeq
On the other hand, two functional derivatives of the same invariance condition, now with respect to $J^{A\,\mu}(x)$ and to $\w^Z(y)$, give
\beq
\langle D^A_\r\h^A(x)\bar\h^Z(y)\rangle = \frac{1}{\xi}\langle\mathcal F^Z(y) A_\r(x)\rangle \, .
\eeq
\subsection{Identities from the ghost equations of motion}
A second class of identities is based on the equations of motion of the ghosts.
Differentiating $\delta_{BRST} Z[J,\mathcal F]$ respect to the source of the photon antighost
$\w^A(x)$ and to the source of the corresponding gauge-fixing function $\chi^A(y)$ gives
\beq
\frac{1}{\xi}\langle\pd^\a A_\a(x)\pd^\b A_\b(y)\rangle = \langle\bar\h^A(x)\mathcal E^A(y)\rangle\, .
\eeq
At this point we consider the functional average of the equation of motion of the ghost of the photon
\beq
\int\mD F\, e^{i\tilde S}\bigg\{-\mathcal E^A(y) + \w^A(y)\bigg\} = 0 \,
\eeq
and take a functional derivative of this expression respect to the source $\w^A(x)$ of the antighost  $\bar\h^A(x)$, obtaining the equation
\beq
\int \mD F\, e^{i\tilde S}\bigg\{-i \mathcal E^A(y)\bar\h^A(x) + \d^{(4)}(x-y)\bigg\} = 0\, ,
\eeq
or, in terms of Green's functions
\beq
\frac{1}{\xi}\langle\mathcal F^A(x)\mathcal F^A(y)\rangle
=\frac{1}{\xi}\langle\pd^\a A_\a(x)\pd^\b A_\b(y)\rangle
= \langle\bar\h^A(x)\mathcal E^A(y)\rangle = -i\d^{(4)}(x-y) \,,
\eeq
which involves the correlation function of the photon gauge-fixing function. \\
It is not hard to show, using the same method, the following identities
\bea
\langle \mathcal F^Z(x)\pd^\a A_\a(y)\rangle &=& 0  \Rightarrow \langle \mathcal F^Z(x) A_\a(y) \rangle = 0 \, , \nn \\
\langle \mathcal F^Z(x) \mathcal F^Z(y)\rangle  &=&  -i\xi\d^{(4)}(x-y)\,.
\eea
\section{Appendix. Standard Model propagators at one loop}
\subsection{Boson self-energies}
\label{P3propagators}
We report here the expressions of the self-energies appearing in Section \ref{P3BRSTsection}. They refer to the case of two vector
bosons ($V_1,V_2$), one vector boson and a scalar ($V S$) and two scalars ($S S$). The self-energies carrying Lorentz indices
are decomposed as
\bea
\Sigma^{V_1 V_2}_{\alpha\beta}(p) &=& -i \left( \eta_{\alpha\beta} - \frac{p_\alpha p_\beta}{p^2}
\right) \Sigma^{V_1 V_1}_T(p^2) - i \frac{p_\alpha p_\beta}{p^2} \Sigma^{V_1 V_1}_L(p^2) \, , \\
\Sigma^{V S}_{\alpha}(p) &=& p_\alpha \, \Sigma^{V S}_L(p^2) \,.
\eea
We denote with $\lambda$ the infrared regulator of the photon mass. We denote with $m_{l,i}$, $m_{u,i}$ and $m_{d,i}$ the masses of the lepton, u-type and d-type quarks of generation $i$ respectively.

The self-energies are then given by
\small
\bea
\Sigma^{AA}_T(p^2)
&=& - \frac{\alpha}{4 \pi} \bigg\{ \frac{2}{3} \sum_f \, N_C^f 2 Q_f^2 \bigg[
-(p^2 + 2 m_f^2)B_0(p^2, m_f^2, m_f^2)  +  2 m_f^2 B_0(0, m_f^2, m_f^2) + \frac{1}{3}p^2 \bigg] \nn \\
&+&  \bigg[ (3 p^2 + 4 M_W^2 ) B_0(p^2, M_W^2, M_W^2) - 4 M_W^2 B_0(0, M_W^2, M_W^2)\bigg]\bigg\}\, , \\
\Sigma^{AA}_L(p^2) &=& 0\, ,
\eea
\bea
\Sigma^{ZZ}_T(p^2)
&=& -\frac{\alpha}{4 \pi} \bigg\{ \frac{2}{3} \sum_f \,
N_C^f \bigg[ \frac{C_V^{f \, 2} + C_A^{f \, 2}}{2 s_w^2 c_w^2}
\bigg[ -(p^2 + 2m_f^2) B_0(p^2, m_f^2, m_f^2)  +  2 m_f^2 B_0(0, m_f^2, m_f^2) + \frac{1}{3}p^2 \bigg] \nn\\
&& \hspace{-2cm} + \frac{3}{4 s_w^2 c_w^2} m_f^2 B_0(p^2,m_f^2, m_f^2) \bigg]
+  \frac{1}{6 s_w^2 c_w^2}\bigg[ \bigg( (18 c_w^4 + 2 c_w^2 -\frac{1}{2})p^2
+ (24 c_w^4 + 16 c_w^2 -10)M_W^2 \bigg)\nn\\
&& \hspace{-2cm} \times B_0(p^2, M_W^2, M_W^2)
-  (24 c_w^4 - 8 c_w^2 + 2)M_W^2 B_0(0, M_W^2, M_W^2) + (4 c_w^2-1)\frac{p^2}{3}  \bigg] \nn \\
&& \hspace{-2cm} + \frac{1}{12 s_w^2 c_w^2} \bigg[ (2 M_H^2 -10 M_Z^2 - p^2) B_0(p^2, M_Z^2, M_H^2)
-  2 M_Z^2 B_0(0, M_Z^2, M_Z^2) - 2 M_H^2 B_0(0, M_H^2, M_H^2)  \nn \\
&& \hspace{-2cm} - \frac{(M_Z^2 - M_H^2)^2}{p^2}\left( B_0(p^2, M_Z^2, M_H^2)  - B_0(0, M_Z^2, M_H^2) \right) -
\frac{2}{3} p^2   \bigg] \bigg\}\, , \\
\Sigma^{ZZ}_L(p^2)
&=& - \frac{\alpha}{2 \pi s_w^2 c_w^2} \bigg\{
\sum_f \, N_C^2 \, C_A^{f \, 2} \, m_f^2 B_0(p^2, m_f^2, m_f^2)
+ M_W^2 (c_w^4 -s_w^4) B_0(p^2, M_W^2, M_W^2) \nn \\
&& \hspace{-2cm} - \frac{1}{4 p^2} \bigg[ \left( (M_Z^2-M_H^2)^2 - 4 M_Z^2 p^2\right) B_0(p^2, M_Z^2, M_H^2)
 + (M_Z^2 -M_H^2)\left( A_0(M_H^2 - A_0(M_Z^2))\right)\bigg] \bigg\}\, ,
\eea
\bea
\Sigma^{AZ}_T(p^2)
&=& \frac{\alpha}{4 \pi \, s_w \, c_w} \bigg\{ \frac{2}{3} \sum_f \, N_C^f \,
Q_f \, C_V^f \bigg[ (p^2 + 2m_f^2) B_0(p^2, m_f^2, m_f^2)
 - 2 m_f^2 B_0(0, m_f^2, m_f^2) -\frac{1}{3}p^2 \bigg] \nn \\
&& \hspace{-2cm} - \frac{1}{3} \bigg[ \left( (9 c_w^2 + \frac{1}{2})p^2
+ (12 c_w^2 + 4)M_W^2 \right) B_0(p^2, M_W^2, M_W^2)
-  (12 c_w^2 -2)M_W^2 B_0(0, M_W^2, M_W^2) + \frac{1}{3}p^2 \bigg]\bigg\}\, ,\nn\\ \\
\Sigma^{AZ}_L(p^2)
&=& -\frac{\alpha}{2 \pi \, s_w \, c_w} M_W^2 \, B_0(p^2, M_W^2, M_W^2)\, ,
\eea
\bea
\Sigma^{A\phi}_L(p^2)
&=& - \frac{\alpha}{2 \pi \, s_w} M_W^2 \, B_0(p^2, M_W^2, M_W^2)\, ,\\
\Sigma^{Z\phi}_L(p^2) &=& - \frac{\alpha}{2 \pi \, s_w^2 \, c_w^2} \bigg\{
C_A^{f \, 2} \frac{m_f^2}{M_Z} B_0(p^2, m_f^2, m_f^2)
+  \frac{M_W}{4} c_w (4 c_w^2 -3) B_0(p^2,M_W^2, M_W^2) \nn \\
&& \hspace{-2cm} + \frac{1}{8 M_Z \, p^2} \bigg[\left( (M_H^2 -M_Z^2)^2 - 3 M_Z^2 p^2 \right) B_0(p^2, M_Z^2, M_H^2)
+  (M_Z^2 -M_H^2)\left( A_0(M_H^2) \right)\bigg]\bigg\}\, ,
\eea
\bea
\Sigma^{\phi\phi}(p^2) &=&  i \frac{\alpha}{4 \pi \, s_w^2 \, c_w^2 \, M_Z^2} \bigg\{\sum_f \, N_C^f \,
C_A^f \, m_f^2 \bigg[  p^2 B_0(p^2, m_f^2,m_f^2) - 2 A_0(m_f^2)\bigg]  \nn \\
&+&  \frac{1}{8} \bigg[ \left(6 M_W^2 + M_H^2 \right) A0(M_W^2)
-  4 M_W^2 \left( p^2 B_0(p^2, M_W^2, M_W^2) + M_W^2 \right) \bigg] \nn \\
&+&     \frac{1}{16} \bigg[ 2 \left( (M_H^2 - M_Z^2)^2
- 2 M_Z^2 p^2 \right)  B_0(p^2, M_Z^2, M_H^2)\nn\\
&+&  (M_H^2 + 2M_Z^2)A_0(M_H^2) + (3 M_H^2 + 4 M_Z^2)A_0(M_Z^2)   - 4 M_Z^4 \bigg]\bigg\}\, ,
\eea
\bea
\Sigma_{HH}(p^2) &=& - \frac{\alpha}{4 \pi} \bigg\{ \sum_{f} N_C^f \frac{m_f^2}{2 s_w^2 M_W^2} \bigg[ 2 \mathcal A_0\left( m_f^2 \right) + (4 m_f^2 - p^2) \mathcal B_0 \left( p^2, m_f^2,m_f^2\right) \bigg] \nn \\
&-& \frac{1}{2 s_w^2} \bigg[ \left(6 M_W^2 - 2p^2 + \frac{M_H^4}{2 M_W^2} \right) \mathcal B_0 \left( p^2, M_W^2, M_W^2 \right) + \left( 3 + \frac{M_H^2}{2 M_W^2} \right) \mathcal A_0 \left( M_W^2 \right) - 6 M_W^2 \bigg] \nn \\
&-& \frac{1}{4 s_w^2 \, c_w^2} \bigg[ \left(6 M_Z^2 - 2p^2 + \frac{M_H^4}{2 M_Z^2} \right) \mathcal B_0 \left( p^2, M_Z^2, M_Z^2 \right) + \left( 3 + \frac{M_H^2}{2 M_Z^2} \right) \mathcal A_0 \left( M_Z^2 \right) - 6 M_Z^2 \bigg] \nn \\
&-& \frac{3}{8 s_w^2} \bigg[ 3 \frac{M_H^4}{M_W^2} \mathcal B_0 \left( p^2, M_H^2,M_H^2 \right) + \frac{M_H^2}{M_W^2} \mathcal A_0 \left( M_H^2 \right) \bigg] \bigg\}\, ,
\eea
\bea
\Sigma^{WW}_T(p^2) &=& -\frac{\alpha}{4 \pi} \bigg\{ \frac{1}{3 s_w^2} \sum_i \bigg[ \left( \frac{m_{l,i}^2}{2} - p^2 \right)
\mathcal B_0\left(p^2, 0, m_{l,i}^2 \right) + \frac{p^2}{3} + m_{l,i}^2 \mathcal B_0 \left( 0, m_{l,i}^2,m_{l,i}^2 \right)  \nn
\\
&& \hspace{-2cm} + \frac{m_{l,i}^4}{2 p^2} \left( \mathcal B_0\left(p^2, 0, m_{l,i}^2 \right)
- \mathcal B_0\left(0, 0, m_{l,i}^2 \right)\right) \bigg]
+ \frac{1}{s_w^2} \sum_{i,j}|V_{ij}|^2 \bigg[\left( \frac{m_{u,i}^2 + m_{d,j}^2}{2} - p^2\right)\nn\\
&& \hspace{-2cm} \times  \mathcal B_0 \left( p^2, m_{u,i}^2, m_{d,j}^2\right)
+ \frac{p^2}{3} + m_{u,i}^2 \mathcal B_0 \left( 0, m_{u,i}^2, m_{u,i}^2\right) + m_{d,j}^2 \mathcal B_0 \left( 0, m_{d,j}^2,
m_{d,j}^2\right)\nn\\
&&\hspace{-2cm} + \frac{(m_{u,i}^2 - m_{d,j}^2)^2}{2 p^2} \big( \mathcal B_0 \left( p^2, m_{u,i}^2, m_{d,j}^2\right)
- \mathcal B_0 \left( 0, m_{u,i}^2, m_{d,j}^2\right)\big) \bigg]\nn\\
&&\hspace{-2cm} + \frac{2}{3} \bigg[ (2 M_W^2 + 5 p^2) \mathcal B_0 \left( p^2, M_W^2, \lambda^2 \right)
- 2 M_W^2 \mathcal B_0 \left( 0, M_W^2, M_W^2\right) \nn \\
&&\hspace{-2cm} - \frac{M_W^4}{p^2} \big( \mathcal B_0\left( p^2, M_W^2, \lambda^2 \right) - \mathcal B_0 \left( 0,M_W^2, \lambda^2 \right)
\big) + \frac{p^2}{3}  \bigg] + \frac{1}{12 s_w^2} \bigg[
\big( (40 c_w^2 -1)p^2 \nn \\
&& \hspace{-2cm}+ (16 c_w^2 + 54 - 10 c_w^{-2}) M_W^2 \big) \mathcal B_0 \left(p^2, M_W^2, M_Z^2 \right) - (16 c_w^2 + 2) \big( M_W^2 \mathcal
B_0 \left( 0,M_W^2,M_W^2\right) \nn \\
&& \hspace{-2cm}+ M_Z^2 \mathcal B_0 \left( 0, M_Z^2, M_Z^2\right) \big) + (4 c_w^2 -1) \frac{2 p^2}{3} -  (8 c_w^2 +1) \frac{(M_W^2 -
M_Z^2)^2}{p^2} \big( \mathcal B_0 \left( p^2, M_W^2,M_Z^2\right) \nn \\
&&\hspace{-2cm}  -\mathcal B_0 \left(0, M_W^2,M_Z^2 \right) \big) \bigg] + \frac{1}{12 s_w^2} \bigg[ (2 M_H^2 - 10 M_W^2 - p^2) \mathcal B_0
\left(p^2, M_W^2,M_H^2 \right)- 2 M_W^2 \mathcal B_0 \left(0,M_W^2,M_W^2 \right) \nn\\
&&\hspace{-2cm} -2 M_H^2 \mathcal B_0 \left( 0, M_H^2,M_H^2\right) - \frac{(M_W^2
-M_H^2)^2}{p^2} \big( \mathcal B_0 \left( p^2, M_W^2, M_H^2\right)
- \mathcal B_0 \left( 0,M_W^2,M_H^2\right) \big) - \frac{2 p^2}{3}\bigg] \bigg\}\, . 
\eea
\normalsize
\subsection{Fermion self-energies}
\label{P5selfenergies}

The one-loop fermion two-point function, diagonal in the flavor space, is defined as
\bea
\Gamma_{\bar f f}(p) = i \bigg[ \psl P_L \, \Sigma^L(p^2) +  \psl P_R \, \Sigma^R(p^2)  + m \, \Sigma^S(p^2) \bigg]
\eea
where the three components $\Sigma^X(p^2)$, with $X = L,R,S$, are eventually given by the gluon, the photon, the Higgs, the Z and the W contributions
\bea
\Sigma^X(p^2) = \frac{\alpha_s}{4 \pi} C_2(N) \, \Sigma^X_g (p^2) + \frac{\alpha}{4 \pi} Q^2 \, \Sigma^X_{\gamma}(p^2) + \frac{G_F}{16 \pi^2 \sqrt{2}} \bigg[ m^2 \, \Sigma^X_h (p^2) +   \Sigma^X_Z (p^2) +  \Sigma^X_W(p^2) \bigg] \,.
\eea
The  $\Sigma^X(p^2)$ coefficients of the fermion self-energies are explicitly given by
\bea
&&  \Sigma^L_g (p^2)  = \Sigma^R_g (p^2) = \Sigma^L_\gamma (p^2)  = \Sigma^R_\gamma (p^2) =  - 2 \, \mathcal B_1 \left( p^2, m^2, 0 \right) - 1   \,, \nn   \\
&& \Sigma^S_g (p^2)  = \Sigma^S_\gamma (p^2) =  - 4 \, \mathcal B_0 \left( p^2, m^2, 0 \right) + 2     \,, \nn  \\
&& \Sigma^L_h (p^2) =  \Sigma^R_h (p^2) = - 2 \, \mathcal B_1 \left( p^2, m^2, m_h^2 \right)   \,, \nn \\
&& \Sigma^S_h (p^2) = 2 \, \mathcal B_0 \left( p^2, m^2, m_h^2 \right)   \,, \nn \\
&& \Sigma^L_W (p^2) = - 4 \sum_f V_{if}^* V_{f_i}  \bigg[ \left( m_f^2 + 2 m_W^2 \right) \mathcal B_1 \left( p^2, m_f^2, m_W^2 \right) + m_W^2 \bigg]   \,, \nn \\
&& \Sigma^R_W (p^2) = - 4 m^2 \sum_f V_{if}^* V_{f_i}  \,  \mathcal B_1 \left( p^2, m_f^2, m_W^2 \right)   \,, \nn  \\
&& \Sigma^S_W (p^2) = - 4  \sum_f V_{if}^* V_{f_i}  \, m_f^2 \, \mathcal B_0 \left( p^2, m_f^2, m_W^2 \right) \,, \nn \\
&& \Sigma^L_Z (p^2) = -2 m_Z^2 (v + a)^2 \bigg[ 2 \, \mathcal B_1 \left( p^2, m^2, m_Z^2 \right)  +1 \bigg]  - 2 m^2 \, \mathcal B_1 \left( p^2, m^2, m_Z^2 \right)  \,, \nn \\
&& \Sigma^R_Z (p^2) = -2 m_Z^2 (v - a)^2 \bigg[ 2 \, \mathcal B_1 \left( p^2, m^2, m_Z^2 \right)  +1 \bigg]  - 2 m^2 \, \mathcal B_1 \left( p^2, m^2, m_Z^2 \right)  \,, \nn \\
&& \Sigma^S_Z (p^2) = -2 m_Z^2 (v^2 -a^2) \bigg[ 4 \, \mathcal B_0 \left( p^2, m^2, m_Z^2 \right) -2 \bigg] - 2 m^2 \, \mathcal B_0 \left( p^2, m^2, m_Z^2 \right) \,,
\eea
where $v$ and $a$ are the vector and axial-vector $Z$-fermion couplings defined in Eq.(\ref{P5Zfermconst}) and
\bea
\mathcal B_1 \left( p^2, m_0^2, m_1^2 \right) = \frac{m_1^2 -m_0^2}{2 p^2} \bigg[ \mathcal B_0(p^2, m_0^2, m_1^2) -  \mathcal B_0(0, m_0^2, m_1^2) \bigg] -\frac{1}{2} \mathcal B_0(p^2, m_0^2, m_1^2) \,.
\eea


\section{Appendix. Explicit results for the $TJJ$ gravitational form factors in the SM}
\label{P3formfactors}
We give here the remaining coefficients appearing in the form factors of the $TAZ$ and $TZZ$ correlators.
\subsection{Form factors for the $TAZ$ vertex}
\small
\bea
%
%
\Phi_2^{(F)}(s,0,M_Z^2,m_f^2) &=& - i \frac{\kappa}{2} \frac{\alpha}{3 \pi s_w \, c_w } \frac{Q_f \, C_v^f }{s (s-M_Z^2)^3} \bigg\{\frac{1}{6}\big(12 m_f^2 \left(s M_Z^2+M_Z^4+s^2\right) \nn \\
&&\hspace{-3cm} -9 s^2 M_Z^2+12 s M_Z^4+2 M_Z^6+s^3\big)
+ 2 m_f^2 \left(s M_Z^2+M_Z^4+3 s^2\right) \mathcal D_0\left(s,0,m_f^2,m_f^2\right) \nn \\
&&\hspace{-3cm}   -\frac{M_Z^2[4 m_f^2(3 s M_Z^2+M_Z^4+9 s^2)+s(4 s M_Z^2+M_Z^4-3 s^2)]}{2(M_Z^2-s)} \mathcal D_0\left(s,M_Z^2,m_f^2,m_f^2\right) \nn \\
&&\hspace{-3cm}  +m_f^2 \left(4 m_f^2 \left(s M_Z^2+M_Z^4+s^2\right)+3 s^2 M_Z^2+6 s
   M_Z^4+M_Z^6+2 s^3\right) \mathcal C_0\left(s,0,M_Z^2,m_f^2,m_f^2,m_f^2\right)
\bigg\}\, , \nn \\
\Phi_3^{(F)}(s,0,M_Z^2,m_f^2) &=& - i \frac{\kappa}{2} \frac{\alpha}{12 \pi s_w \, c_w } \frac{Q_f \, C_v^f }{s (s-M_Z^2)} \bigg\{
- \frac{1}{6} \left(12 m_f^2+2 M_Z^2+s\right)
-2 m_f^2 \mathcal D_0\left(s,0,m_f^2,m_f^2\right) \nn \\
&& \hspace{-3cm} + \, \frac{[ 4 m_f^2(M_Z^2+2 s)+s M_Z^2 ]}{2 \left(M_Z^2-s\right)} \mathcal D_0\left(s,M_Z^2,m_f^2,m_f^2\right)
- m_f^2 \left(4 m_f^2+M_Z^2+2 s\right) \mathcal C_0\left(s,0,M_Z^2,m_f^2,m_f^2,m_f^2\right)
\bigg\}\, , \nn \\
\Phi_4^{(F)}(s,0,M_Z^2,m_f^2) &=& - \frac{2 (2s + M_Z^2)}{s-M_Z^2} \Phi_3^{(F)}(s,0,M_Z^2,m_f^2)\, , \nn \\
\Phi_5^{(F)}(s,0,M_Z^2,m_f^2) &=& - i \frac{\kappa}{2} \frac{\alpha}{6 \pi s_w \, c_w } \frac{Q_f \, C_v^f }{(s-M_Z^2)^2} \bigg\{
M_Z^2
-8 m_f^2 \mathcal D_0\left(s,0,m_f^2,m_f^2\right) \nn \\
&& \hspace{-3cm} + \, \frac{[s M_Z^2 - 8 m_f^2 (3 M_Z^2-s )]}{s-M_Z^2} \mathcal D_0\left(s,M_Z^2,m_f^2,m_f^2\right)
-6 m_f^2 \, M_Z^2 \, \mathcal C_0\left(s,0,M_Z^2,m_f^2,m_f^2,m_f^2\right)
 \bigg\}\, , \nn \\
\Phi_6^{(F)}(s,0,M_Z^2,m_f^2) &=& - i \frac{\kappa}{2} \frac{\alpha}{3 \pi s_w \, c_w } \frac{Q_f \, C_v^f }{(s-M_Z^2)^3} \bigg\{
M_Z^2 \left(6 m_f^2-M_Z^2+2 s\right)
+10 m_f^2 \, M_Z^2 \, \mathcal D_0\left(s,0,m_f^2,m_f^2\right) \nn \\
&&\hspace{-3cm} +  \, \frac{M_Z^2[ s(M_Z^2-3 s)-4 m_f^2(9 M_Z^2+4 s)]}{2 \left(M_Z^2-s\right)} \mathcal D_0\left(s,M_Z^2,m_f^2,m_f^2\right) \nn \\
&& \hspace{-3cm}+  \, 3 m_f^2 M_Z^2 \left(4 m_f^2+M_Z^2+3 s\right) \mathcal C_0\left(s,0,M_Z^2,m_f^2,m_f^2,m_f^2\right)
 \bigg\}\, , \nn \\
\Phi_7^{(F)}(s,0,M_Z^2,m_f^2) &=& - i \frac{\kappa}{2} \frac{\alpha}{12 \pi s_w \, c_w } \frac{Q_f \, C_v^f }{ (s-M_Z^2)} \bigg\{
\frac{1}{6} \left(36 m_f^2+M_Z^2+11 s\right)
+ \left(2 s-2 M_Z^2\right) \mathcal B_0\left(s,m_f^2,m_f^2\right) \nn \\
&& \hspace{-3cm} +2 m_f^2 \mathcal D_0\left(s,0,m_f^2,m_f^2\right)
+\left(4 M_Z^2 - \frac{2(m_f^2(M_Z^2+4 s)+M_Z^4)}{M_Z^2-s}\right) \mathcal D_0\left(s,M_Z^2,m_f^2,m_f^2\right) \nn \\
&& \hspace{-3cm} +6 m_f^2 \left(2 m_f^2+s\right) \mathcal C_0\left(s,0,M_Z^2,m_f^2,m_f^2,m_f^2\right)
\bigg\}\, ,
\eea
\bea
\Phi_2^{(B)}(s,0,M_Z^2,M_W^2) &=& - i \frac{\kappa}{2} \frac{\alpha}{3 \pi s_w \, c_w } \frac{1}{ s(s-M_Z^2)^3} \bigg\{
\frac{1}{12} \big(-6 s^2 M_Z^2 \left(12 \left(s_w^4+s_w^2\right)-25\right) \nn\\
&& \hspace{-3cm} + \, M_Z^6 \left(-72 s_w^4+174
   s_w^2-103\right)-36 s M_Z^4 \left(2 s_w^4-9 s_w^2+7\right)+s^3 \left(6
   s_w^2-5\right)\big) \nn \\
&& \hspace{-3cm} - \, 2 c_w^2 M_Z^2 \left(M_Z^2-s\right)^2 \mathcal B_0\left(s,M_W^2,M_W^2\right)
+ M_Z^2 c_w^2\left(6 s_w^2 \left(s M_Z^2+M_Z^4+3 s^2\right)-9 s M_Z^2-3 M_Z^4-13 s^2\right) \times \nn \\
&& \hspace{-3cm} \times \, \mathcal D_0\left(s,0,M_W^2,M_W^2\right)
+\frac{M_Z^2}{2
   \left(M_Z^2-s\right)} \big(2 s^2 M_Z^2 \left(54 s_w^4-115 s_w^2+61\right)+2 M_Z^6 \left(6 s_w^4-11
   s_w^2+5\right) \nn \\
&& \hspace{-3cm}   +2 s M_Z^4 \left(18 s_w^4-37 s_w^2+19\right)+s^3 \left(34
   s_w^2-35\right)\big) \mathcal D_0\left(s,M_Z^2,M_W^2,M_W^2\right) \nn \\
&& \hspace{-3cm}  - M_Z^2 c_w^2\big(2 s^2 M_Z^2 \left(6 s_w^4-3 s_w^2+4\right)+s M_Z^4 \left(12 \left(s_w^2-5\right) s_w^2+41\right)+2 M_Z^6 \left(6 s_w^4-15
   s_w^2+8\right) \nn \\
&& \hspace{-3cm}    -s^3 \left(6 s_w^2+5\right)\big) \mathcal C_0\left(s,0,M_Z^2,M_W^2,M_W^2,M_W^2\right)
\bigg\}\, , \nn
\eea
\bea
\Phi_3^{(B)}(s,0,M_Z^2,M_W^2) &=& - i \frac{\kappa}{2} \frac{\alpha}{12 \pi s_w \, c_w } \frac{1}{ s(s-M_Z^2)} \bigg\{
\frac{1}{12} \left(M_Z^2 \left(72 s_w^4-174 s_w^2+103\right)+s \left(5-6 s_w^2\right)\right) \nn \\
&& \hspace{-3cm} +2 M_Z^2 c_w^2 \mathcal B_0\left(s,M_W^2,M_W^2\right)
+3 M_Z^2 \left(2 s_w^4-3 s_w^2+1\right) \mathcal D_0\left(s,0,M_W^2,M_W^2\right) \nn \\
&& \hspace{-3cm} -\frac{M_Z^2 c_w^2}{s-M_Z^2} \left(M_Z^2 \left(6 s_w^2-5\right)+2 s \left(6 s_w^2-7\right)\right)\mathcal D_0\left(s,M_Z^2,M_W^2,M_W^2\right)\nn\\
&& \hspace{-3cm}+M_Z^2 c_w^2 \left(2 M_Z^2 \left(6 s_w^4-15 s_w^2+8\right)+s \left(7-6 s_w^2\right)\right) \mathcal C_0\left(s,0,M_Z^2,M_W^2,M_W^2,M_W^2\right)
\bigg\}\, , \nn \\
\Phi_4^{(B)}(s,0,M_Z^2,M_W^2) &=& - i \frac{\kappa}{2} \frac{\alpha}{6 \pi s_w \, c_w } \frac{1}{ s(s-M_Z^2)^2} \bigg\{
-\frac{1}{12} \left(M_Z^2+2 s\right) \big(M_Z^2 \left(72 s_w^4-174 s_w^2+103\right) \nn \\
&& \hspace{-3cm} +s \left(5-6 s_w^2\right)\big)
- 2 M_Z^2 c_w^2 \left(M_Z^2-4 s\right) \mathcal B_0\left(s,M_W^2,M_W^2\right)
+ 3 M_Z^2 c_w^2 \big(2 s_w^2 \left(M_Z^2+2 s\right) \nn \\
&& \hspace{-3cm}  -M_Z^2-6 s \big) \mathcal D_0\left(s,0,M_W^2,M_W^2\right)
+\frac{M_Z^2 \, c_w^2 }{s-M_Z^2} \left(M_Z^2+2 s\right) \left(M_Z^2 \left(6 s_w^2-5\right)+2 s \left(6 s_w^2-7\right)\right) \times \nn \\
&& \hspace{-3cm}  \times \, \mathcal D_0\left(s,M_Z^2,M_W^2,M_W^2\right)
- M_Z^2 c_w^2 \left(M_Z^2+2 s\right) \big(2 M_Z^2 \left(6 s_w^4-15 s_w^2+8\right) \nn \\
&& \hspace{-3cm} +s \left(7-6 s_w^2\right)\big) \mathcal C_0\left(s,0,M_Z^2,M_W^2,M_W^2,M_W^2\right)
\bigg\}\, , \nn \\
\Phi_5^{(B)}(s,0,M_Z^2,M_W^2) &=& - i \frac{\kappa}{2} \frac{\alpha}{6 \pi s_w \, c_w } \frac{1}{(s-M_Z^2)^2} \bigg\{
M_Z^2 \left(18 s_w^2-19\right)
+12 M_Z^2 c_w^2 \mathcal B_0\left(s,M_W^2,M_W^2\right) \nn \\
&& \hspace{-3cm} +8 M_Z^2 \left(3 s_w^4-4 s_w^2+1\right) \mathcal D_0\left(s,0,M_W^2,M_W^2\right)
- \frac{M_Z^2}{s-M_Z^2} \big(s \left(24 s_w^4-62 s_w^2+39\right) \nn \\
&& \hspace{-3cm}  -12 M_Z^2 \left(6 s_w^4-11 s_w^2+5\right)\big) \mathcal D_0\left(s,M_Z^2,M_W^2,M_W^2\right) \nn \\
&& \hspace{-3cm}  +6 M_Z^2 c_w^2 \left(M_Z^2 \left(2 s_w^2-1\right)-2 s\right) \mathcal C_0\left(s,0,M_Z^2,M_W^2,M_W^2,M_W^2\right)
\bigg\}\, , \nn \\
\Phi_6^{(B)}(s,0,M_Z^2,M_W^2) &=& - i \frac{\kappa}{2} \frac{\alpha}{3 \pi s_w \, c_w } \frac{1}{(s-M_Z^2)^3} \bigg\{
-\frac{1}{4} M_Z^2 \left(M_Z^2 \left(72 s_w^4-90 s_w^2+17\right)+s \left(53-54 s_w^2\right)\right) \nn \\
&& \hspace{-3cm} -5 M_Z^4 \left(6 s_w^4-11 s_w^2+5\right) \mathcal D_0\left(s,0,M_W^2,M_W^2\right)
-\frac{M_Z^2}{2 \left(s-M_Z^2\right)} \big(18 M_Z^4 \left(6 s_w^4-11 s_w^2+5\right) \nn \\
&& \hspace{-3cm} +s M_Z^2 \left(48 s_w^4-70 s_w^2+21\right)+24 s^2 c_w^2\big) \mathcal D_0\left(s,M_Z^2,M_W^2,M_W^2\right)
- 3 M_Z^2 c_w^2 \big(M_Z^4 \left(12 s_w^4-20 s_w^2+9\right) \nn \\
&& \hspace{-3cm} +s M_Z^2 \left(9-14 s_w^2\right)+2 s^2\big) \mathcal C_0\left(s,0,M_Z^2,M_W^2,M_W^2,M_W^2\right)
\bigg\}\, , \nn\eea
\bea
\Phi_7^{(B)}(s,0,M_Z^2,M_W^2) &=& - i \frac{\kappa}{2} \frac{\alpha}{6 \pi s_w \, c_w } \frac{1}{(s-M_Z^2)} \bigg\{
\frac{1}{24} \left(M_Z^2 \left(54 s_w^2 \left(7-4 s_w^2\right)-161\right)+s \left(270 s_w^2-277\right)\right) \nn \\
&& \hspace{-3cm} + \frac{1}{4} \left(M_Z^2 \left(43-42 s_w^2\right)+s \left(18 s_w^2-19\right)\right) \mathcal B_0\left(s,M_W^2,M_W^2\right)
+\frac{1}{2} c_w^2 \left(M_Z^2 \left(6 s_w^2-11\right)-6 s\right) \times \nn \\
&& \hspace{-3cm} \times \, \mathcal D_0\left(s,0,M_W^2,M_W^2\right)
- \frac{1}{4\left(s-M_Z^2\right)}\big(M_Z^4 \left(12 s_w^4+8 s_w^2-21\right)+2 s M_Z^2 \left(24 s_w^4-74 s_w^2+51\right) \nn \\
&& \hspace{-3cm}  +12 s^2c_w^2\big) \mathcal D_0\left(s,M_Z^2,M_W^2,M_W^2\right)
- 3 c_w^2  \big(M_Z^4 \left(6 s_w^4-11 s_w^2+5\right) \nn \\
&& \hspace{-3cm} +s M_Z^2 \left(6-8 s_w^2\right)+2 s^2\big) \mathcal C_0\left(s,0,M_Z^2,M_W^2,M_W^2,M_W^2\right)
\bigg\}\, , \nn \\
\Phi_8^{(B)}(s,0,M_Z^2,M_W^2) &=& \frac{i \alpha  \kappa  c_w M_Z^2 }{6 \pi  s s_w} \mathcal B_0(0,M_W^2,M_W^2)\, , \nn \\
\Phi_9^{(B)}(s,0,M_Z^2,M_W^2) &=& -\frac{i \alpha  \kappa  c_w M_Z^2 }{6 \pi  s s_w \left(s-M_Z^2\right)} \mathcal B_0(0,
M_W^2,M_W^2)\, .
\eea
\normalsize

\subsection{Form factors for the $TZZ$ vertex in the fermionic sector}
\label{P3fermionicFF}
The coefficients of Eq. (\ref{P3fermionhZZ}) are given by
\small
\bea
{C_{(F)}}_0^2 &=& \frac{i \kappa\, \alpha, m_f^2}{6 \pi  s^2 c_w^2 \left(s-4 M_Z^2\right) s_w^2}
\left(\left(2 M_Z^4-4 s M_Z^2+s^2\right) C_a^{f \, 2}+2 C_v^{f \, 2} M_Z^4\right)\, ,  \nn \\
{C_{(F)}}_1^2 &=&   0  \, ,\nn\\
{C_{(F)}}_2^2 &=&     \frac{i \kappa \, \alpha\,m_f^2}{6 \pi  s c_w^2 s_w^2}\,C_a^{f \, 2}  \, ,\nn\\
{C_{(F)}}_3^2 &=&   \frac{i \kappa \, \alpha\,  m_f^2 \,M_Z^2}{3 \pi  s^2 c_w^2 \left(s-4 M_Z^2\right){}^2 s_w^2}
\left(s^2 C_a^{f \, 2}-2 \left(C_a^{f \, 2}+C_v^{f \, 2}\right) M_Z^4
+2 s \left(C_v^{f \, 2}-C_a^{f \, 2}\right) M_Z^2\right) \, ,\nn\\
{C_{(F)}}_4^2 &=&   \frac{i \kappa \, \alpha \, m_f^2}{6 \pi  s^2 c_w^2 \left(s-4 M_Z^2\right){}^2 s_w^2}
\bigg(\bigg(4 M_Z^8-2 \left(8 m_f^2+5 s \right) M_Z^6+3 s \left(12 m_f^2+s \right) M_Z^4\nn\\
&-&16 s^2 m_f^2 M_Z^2+2 s^3 m_f^2\bigg) C_a^{f \,2}+C_v^{f \, 2} M_Z^4 \left(4 M_Z^4-2 \left(8 m_f^2+s \right) M_Z^2
+s \left(4 m_f^2+s \right)\right)\bigg)\, , \nn \\
\eea
\bea
{C_{(F)}}_0^3 &=& \frac{i \kappa \, \alpha}{192 \pi  c_w^2 \left(s-4 M_Z^2\right) s_w^2}
\bigg(4 \left(C_a^{f \, 2}+C_v^{f \, 2}\right) M_Z^4-2 \left(32 m_f^2 C_a^{f \, 2}+7 s \left(C_a^{f \, 2}+C_v^{f \,
2}\right)\right) M_Z^2\nn\\
&+& s\left(16 m_f^2 C_a^{f \, 2}+3 s \left(C_a^{f \, 2}+C_v^{f \, 2}\right)\right)\bigg)\, ,\nn\\
{C_{(F)}}_1^3 &=&     \frac{i \kappa \, \alpha}{48 \pi  c_w^2 s_w^2}  \left(C_a^{f \, 2}+C_v^{f \, 2}\right)\, , \nn\\
{C_{(F)}}_2^3 &=&     \frac{i \kappa \, \alpha}{48 \pi  c_w^2 s_w^2}\,\left(\left(3 m_f^2+s \right) C_a^{f \, 2}+C_v^{f \, 2} \left(s-m_f^2\right)\right)-\frac{i \alpha  \kappa}{24 \pi  c_w^2 s_w^2}\,\left(C_a^{f \, 2}+C_v^{f \, 2}\right) M_Z^2\, , \nn\\
{C_{(F)}}_3^3 &=&     \frac{i \kappa \, \alpha}{48 \pi  c_w^2 \left(s-4 M_Z^2\right){}^2 s_w^2}
\bigg(8 s^2 m_f^2 C_a^{f \, 2}+14 \left(C_a^{f \, 2}+C_v^{f \, 2}\right) M_Z^6
+\bigg(8 \left(5 C_a^{f \, 2}+C_v^{f \, 2}\right) m_f^2\nn\\
&-&17 s \left(C_a^{f \, 2}+C_v^{f \, 2}\right)\bigg) M_Z^4
+s \left(3 s \left(C_a^{f \, 2}+C_v^{f \, 2}\right)-2 \left(21 C_a^{f \, 2}+C_v^{f \, 2}\right) m_f^2\right) M_Z^2\bigg)\, ,
\nn\\
{C_{(F)}}_4^3 &=& \frac{i \kappa \, \alpha}{48 \pi  c_w^2 \left(s-4 M_Z^2\right){}^2 s_w^2}
\bigg(18 \left(C_a^{f \, 2}+C_v^{f \, 2}\right) M_Z^8-2 \left(8 \left(7 C_a^{f \, 2}+C_v^{f \, 2}\right) m_f^2+9 s \left(C_a^{f \, 2}+C_v^{f
\, 2}\right)\right) M_Z^6\nn\\
&+&\left(\left(160 m_f^4+116 s m_f^2+3 s^2\right) C_a^{f \, 2}+C_v^{f \, 2} \left(4 m_f^2+3 s \right) \left(8 m_f^2+s
\right)\right) M_Z^4\nn\\
&-&2 s m_f^2 \left(\left(40 m_f^2+21 s \right) C_a^{f \, 2} + C_v^{f \, 2} \left(8 m_f^2+5 s \right)\right)
M_Z^2+s^2 \left(5 C_a^{f \, 2}+C_v^{f \, 2}\right) m_f^2 \left(2 m_f^2+s \right)\bigg)\, , \nn \\
\eea
\bea
{C_{(F)}}_0^4 &=& -\frac{i \kappa \, \alpha}{144 \pi  s c_w^2 \left(s-4 M_Z^2\right){}^2 s_w^2}
\bigg(-44 \left(C_a^{f \, 2}+C_v^{f \, 2}\right) M_Z^6
+2 \bigg(96 \left(2 C_a^{f \, 2}+C_v^{f \, 2}\right) m_f^2 \nn \\
&+&  31 s \left(C_a^{f \, 2}+C_v^{f \, 2}\right)\bigg) M_Z^4
-s \left(48 \left(2 C_a^{f \, 2}+3 C_v^{f \, 2}\right) m_f^2+13 s
\left(C_a^{f \, 2}+C_v^{f \, 2}\right)\right) M_Z^2 \nn \\
&+&  s^2 \left(s C_a^{f \, 2}+C_v^{f \, 2} \left(24 m_f^2+s \right)\right)\bigg)
\, ,  \nn\\
{C_{(F)}}_1^4 &=&    \frac{i \kappa \, \alpha}{12 \pi  s c_w^2 \left(s-4
M_Z^2\right) s_w^2} \left(C_a^{f \, 2}+C_v^{f \, 2}\right) \left(s-3 M_Z^2\right) \, , \nn\\
{C_{(F)}}_2^4 &=& \frac{i \kappa \, \alpha \, m_f^2}{12 \pi  s c_w^2 \left(s-4 M_Z^2\right) s_w^2}\left(C_a^{f \, 2} \left(s-5 M_Z^2\right)-C_v^{f \, 2} \left(s-3 M_Z^2\right)\right)\, , \nn
\eea
\bea
{C_{(F)}}_3^4 &=&   \frac{i \kappa \, \alpha}{24 \pi  s c_w^2 \left(s-4 M_Z^2\right){}^3 s_w^2}
\bigg(-36 \left(C_a^{f \, 2}+C_v^{f \, 2}\right) M_Z^8+2 \bigg(56 \left(3 C_a^{f \, 2}-C_v^{f \, 2}\right) m_f^2\nn\\
&+&9 s \left(C_a^{f \, 2}+C_v^{f \, 2}\right)\bigg) M_Z^6
-2 s \left(\left(98 m_f^2+5 s \right) C_a^{f \, 2}+C_v^{f \, 2} \left(5 s-22 m_f^2\right)\right) M_Z^4\nn\\
&+&s^2 \left(C_a^{f \, 2}
+C_v^{f \, 2}\right) \left(12 m_f^2+s \right) M_Z^2+4 s^3 \left(C_a^f-C_v^f\right) \left(C_a^f+C_v^f\right) m_f^2\bigg)\, ,\nn\\
{C_{(F)}}_4^4 &=&    \frac{i \kappa \, \alpha}{12 \pi  s c_w^2 \left(s-4 M_Z^2\right){}^3 s_w^2}
\bigg(18 \left(C_a^{f \, 2}+C_v^{f \, 2}\right) M_Z^{10}-4 \bigg(4 \left(7 C_a^{f \, 2}+C_v^{f \, 2}\right) m_f^2\nn\\
&+&3 s \left(C_a^{f \, 2}+C_v^{f \, 2}\right)\bigg) M_Z^8 + 4 m_f^2 \left(\left(40 m_f^2+21 s \right) C_a^{f \, 2}+C_v^{f \, 2}
\left(8 m_f^2+11 s \right)\right) M_Z^6\nn\\
&-&2 s m_f^2 \left(24 \left(C_a^{f \,2}+C_v^{f \, 2}\right) m_f^2+s \left(7 C_a^{f \, 2}+13 C_v^{f \, 2}\right)\right) M_Z^4
+2 s^2 m_f^2 \bigg(C_v^{f \, 2} \left(9 m_f^2+4 s \right)\nn\\
&-&C_a^{f \, 2} \left(3 m_f^2+2 s \right)\bigg) M_Z^2+s^3 \left(C_a^f-C_v^f\right) \left(C_a^f+C_v^f\right) m_f^2 \left(2 m_f^2+s
\right)\bigg)\, ,
\eea
\bea
{C_{(F)}}_0^5 &=& -\frac{i \kappa \, \alpha}{288 \pi  s c_w^2 \left(s-4 M_Z^2\right){}^2 s_w^2}
\bigg(-88 \left(C_a^{f \, 2}+C_v^{f \, 2}\right) M_Z^6
+24 \bigg(16 \left(2 C_a^{f \, 2}+C_v^{f \, 2}\right) m_f^2 \nn \\
&+& 5 s \left(C_a^{f \,2}+C_v^{f \, 2}\right)\bigg) M_Z^4 
-12 s \left(8 \left(6 C_a^{f \, 2}+C_v^{f \, 2}\right) m_f^2+5 s \left(C_a^{f \,
2}+C_v^{f \,2}\right)\right) M_Z^2 \nn \\
&+& s^2 \left(96 m_f^2 C_a^{f \, 2}+7 s \left(C_a^{f \, 2}+C_v^{f \, 2}\right)\right)\bigg)\, ,
  \nn\\
{C_{(F)}}_1^5 &=&    \frac{i \kappa \, \alpha}{24 \pi  s c_w^2 \left(s-4
M_Z^2\right) s_w^2}\left(C_a^{f \, 2}+C_v^{f \, 2}\right) \left(s-6 M_Z^2\right) \, , \nn\\
{C_{(F)}}_2^5 &=&    -\frac{i \kappa \, \alpha}{24 \pi  s c_w^2 \left(s-4 M_Z^2\right) s_w^2}
\bigg(s \left(\left(s-3 m_f^2\right) C_a^{f \, 2}+C_v^{f \, 2} \left(m_f^2+s \right)\right)\nn\\
&-&2 \bigg(\left(3 C_v^{f \, 2}-5 C_a^{f \, 2}\right) m_f^2
+2 s \left(C_a^{f \, 2}+C_v^{f \, 2}\right)\bigg) M_Z^2\bigg) \, , \nn\\
{C_{(F)}}_3^5 &=&    -\frac{i \kappa \, \alpha}{24 \pi  s c_w^2 \left(s-4 M_Z^2\right){}^3 s_w^2}
\bigg(36 \left(C_a^{f \, 2}+C_v^{f \, 2}\right) M_Z^8+4 \bigg(\left(s-84 m_f^2\right) C_a^{f \, 2}\nn\\
&+&C_v^{f \, 2} \left(28 m_f^2+s\right)\bigg) M_Z^6
-3 s \left(44 \left(C_v^{f \, 2}-3 C_a^{f \, 2}\right) m_f^2+5 s \left(C_a^{f \, 2}+C_v^{f \, 2}\right)\right)
M_Z^4\nn\\
&+&2 s^2 \bigg(\left(s-79 m_f^2\right) C_a^{f \, 2}
+C_v^{f \, 2} \left(29 m_f^2+s \right)\bigg) M_Z^2+4 s^3 \left(5 C_a^{f \, 2}-2 C_v^{f \, 2}\right) m_f^2\bigg)\, , \nn\\
{C_{(F)}}_4^5 &=& -\frac{i \kappa \, \alpha}{24 \pi  s c_w^2 \left(s-4 M_Z^2\right){}^3 s_w^2}
\bigg(-36 \left(C_a^{f \, 2}+C_v^{f \, 2}\right) M_Z^{10}\nn\\
&+&2 \bigg(16 \left(7 C_a^{f \, 2}+C_v^{f \, 2}\right) m_f^2
+33 s \left(C_a^{f \,2}+C_v^{f \, 2}\right)\bigg) M_Z^8\nn\\
&-&2 \bigg(32 \left(5 C_a^{f \, 2}+C_v^{f \, 2}\right) m_f^4+4 s \left(51 C_a^{f \, 2}+5 C_v^{f \, 2}\right)m_f^2
+15 s^2 \left(C_a^{f \, 2}+C_v^{f \, 2}\right)\bigg) M_Z^6\nn\\
&+&s \bigg(384 m_f^4 C_a^{f \, 2}+16 s \left(16 C_a^{f \, 2}+C_v^{f \,2}\right)m_f^2
+3 s^2 \left(C_a^{f \, 2}+C_v^{f \, 2}\right)\bigg) M_Z^4\nn\\
&+&2 s^2 m_f^2 \bigg(C_v^{f \, 2} \left(6 m_f^2+s \right)
-C_a^{f \, 2} \left(66 m_f^2+35 s \right)\bigg) M_Z^2
+s^3 \left(7 C_a^{f \, 2}-C_v^{f \, 2}\right) m_f^2 \left(2 m_f^2+s \right)\bigg)\, , \nn \\
\eea
\bea
{C_{(F)}}_0^6 &=& \frac{i \kappa \, \alpha}{288 \pi  s c_w^2 \left(s-4 M_Z^2\right){}^2 s_w^2}
\left(C_a^{f \, 2}+C_v^{f \, 2}\right) \bigg(-72 M_Z^6+8 \left(48 m_f^2+19 s \right) M_Z^4\nn\\
&-&2 s \left(144 m_f^2+35 s \right) M_Z^2 +s^2 \left(48 m_f^2+11 s \right)\bigg)\, ,\nn\\
{C_{(F)}}_1^6 &=&    -\frac{i \kappa \, \alpha}{24 \pi  s c_w^2 \left(s-4 M_Z^2\right) s_w^2}
\left(C_a^{f \, 2}+C_v^{f \, 2}\right) \left(s-2 M_Z^2\right)\, , \nn\\
{C_{(F)}}_2^6 &=&   \frac{i \kappa\, \alpha}{24 \pi  s c_w^2 \left(s-4 M_Z^2\right) s_w^2}\,
\left(C_a^{f \, 2}+C_v^{f \, 2}\right) \left(s \left(m_f^2+s \right)-2 \left(m_f^2+2 s \right) M_Z^2\right) \, , \nn\\
{C_{(F)}}_3^6 &=&    \frac{i \kappa \, \alpha}{48 \pi  s c_w^2 \left(s-4 M_Z^2\right){}^3 s_w^2}
\bigg(24 \left(C_a^{f \, 2}+C_v^{f \, 2}\right) M_Z^8+24 \left(C_a^{f \, 2}+C_v^{f \, 2}\right) \left(s-4 m_f^2\right) M_Z^6\nn\\
&+&4 s \left(6\left(11 C_a^{f \, 2}+3 C_v^{f \, 2}\right) m_f^2-7 s \left(C_a^{f \, 2}+C_v^{f \, 2}\right)\right) M_Z^4\nn\\
&+&s^2\left(7 \left(s-20 m_f^2\right) C_a^{f \, 2}+C_v^{f \, 2} \left(7 s-44 m_f^2\right)\right) M_Z^2
+4 s^3 \left(5 C_a^{f \, 2}+2 C_v^{f \, 2}\right) m_f^2\bigg)\, ,\nn \\
{C_{(F)}}_4^6 &=&    \frac{i \kappa \, \alpha}{8 \pi  s c_w^2 \left(s-4 M_Z^2\right){}^3 s_w^2}
\bigg(\left(M_Z^4-4 m_f^2 M_Z^2+s m_f^2\right) \bigg(-4 M_Z^6+2 \left(8 m_f^2+9 s \right) M_Z^4\nn\\
&-&s \left(12 m_f^2+11 s \right) M_Z^2 +2 s^2\left(m_f^2+s \right)\bigg) C_a^{f \, 2}\nn\\
&+&C_v^{f \, 2} \bigg(-4 M_Z^{10}+2 \left(16 m_f^2+9 s \right) M_Z^8-\left(64 m_f^4+56 s m_f^2+11 s^2\right) M_Z^6\nn\\
&+& 2 s \left(32 m_f^4+16 s m_f^2+s^2\right) M_Z^4-s^2 m_f^2 \left(20 m_f^2+9 s \right) M_Z^2+s^3 m_f^2 \left(2 m_f^2+s
\right)\bigg)\bigg)\, ,
\eea
\bea
{C_{(F)}}_0^7 &=&      -\frac{i \kappa\, \alpha}{24 \pi  s c_w^2 \left(s-4 M_Z^2\right){}^2 s_w^2}
\left(C_a^{f \, 2}+C_v^{f \, 2}\right) M_Z^2 \left(6 M_Z^4-\left(32 m_f^2+7 s \right) M_Z^2+2 s \left(4 m_f^2+s \right)\right)
\, ,\nn\\
{C_{(F)}}_1^7 &=&    \frac{i \kappa \, \alpha}{12 \pi  s c_w^2 \left(s-4 M_Z^2\right) s_w^2}
\left(C_a^{f \, 2}+C_v^{f \, 2}\right) M_Z^2  \, ,\nn\\
{C_{(F)}}_2^7 &=&     -\frac{i \kappa \, \alpha}{12 \pi  s c_w^2 \left(s-4 M_Z^2\right) s_w^2}
\left(C_a^{f \, 2}+C_v^{f \, 2}\right) m_f^2 M_Z^2 \, ,\nn\\
{C_{(F)}}_3^7 &=&     -\frac{i \kappa \, \alpha}{48 \pi  s c_w^2 \left(s-4 M_Z^2\right){}^3 s_w^2}
\bigg(12 s^3 m_f^2 C_a^{f \, 2}-24 \left(C_a^{f \, 2}+C_v^{f \, 2}\right) M_Z^8\nn\\
&-& 4 \left(C_a^{f \, 2}+C_v^{f \, 2}\right) \left(s-24 m_f^2\right) M_Z^6
- 2 s \left(\left(s-20 m_f^2\right) C_a^{f \, 2}+C_v^{f \, 2} \left(76
m_f^2+s \right)\right) M_Z^4\nn\\
&+&s^2 \left(32 \left(C_v^{f \, 2}-2 C_a^{f \, 2}\right) m_f^2+3 s \left(C_a^{f \, 2}+C_v^{f \, 2}\right)\right)M_Z^2\bigg)
\, ,\nn\\
{C_{(F)}}_4^7 &=&    -\frac{i \kappa \, \alpha}{8 \pi  s c_w^2 \left(s-4 M_Z^2\right){}^3 s_w^2}
\bigg(\left(M_Z^4-4 m_f^2 M_Z^2+s m_f^2\right) \bigg(4 M_Z^6-16 m_f^2 M_Z^4\nn\\
&+&s \left(4 m_f^2-3s \right) M_Z^2+s^3\bigg) C_a^{f \, 2}
+C_v^{f \, 2} M_Z^2 \bigg(4 \left(M_Z^2-4 m_f^2\right){}^2 M_Z^4+8 s m_f^2
\left(5 M_Z^2-4 m_f^2\right) M_Z^2\nn\\
&+&s^3 \left(3 m_f^2+M_Z^2\right)+s^2 \left(4 m_f^4-20 M_Z^2 m_f^2-3 M_Z^4\right)\bigg)\bigg)\, ,
\eea
\bea
{C_{(F)}}_0^8 &=& \frac{i \kappa \, \alpha}{144 \pi  s^2 c_w^2 \left(s-4 M_Z^2\right){}^3 s_w^2}
\left(C_a^{f \, 2}+C_v^{f \, 2}\right) \bigg(-216 M_Z^8+8 \left(96 m_f^2+19 s \right) M_Z^6\nn\\
&-&24 s \left(16 m_f^2+s \right) M_Z^4+3 s^2 \left(3 s-16 m_f^2\right) M_Z^2+s^3 \left(24 m_f^2+s \right)\bigg) \, , \nn
\eea
\bea
{C_{(F)}}_1^8 &=&      -\frac{i \kappa \, \alpha}{12 \pi  s^2 c_w^2 \left(s-4 M_Z^2\right){}^2 s_w^2}
\left(C_a^{f \, 2}+C_v^{f \, 2}\right) \left(-6 M_Z^4+s M_Z^2+s^2\right)\, ,  \nn \\
{C_{(F)}}_2^8 &=&     \frac{i \kappa \, \alpha}{12 \pi  s^2 c_w^2 \left(s-4 M_Z^2\right){}^2 s_w^2}
\left(C_a^{f \, 2}+C_v^{f \, 2}\right) m_f^2 \left(-6 M_Z^4+s M_Z^2+s^2\right) \, , \nn \\
{C_{(F)}}_3^8 &=&     \frac{i \kappa \, \alpha}{12 \pi  s^2 c_w^2 \left(s-4 M_Z^2\right){}^4 s_w^2}
\bigg(36 \left(C_a^{f \, 2}+C_v^{f \, 2}\right) M_Z^{10}-4 \left(C_a^{f \, 2}+C_v^{f \, 2}\right) \left(20 m_f^2+3 s \right)
M_Z^8 \nn\\
&-&s \left(\left(20 m_f^2+s \right) C_a^{f \, 2}+C_v^{f \, 2} \left(s-108 m_f^2\right)\right) M_Z^6+s^2 \left(\left(34 m_f^2+3 s
\right)
C_a^{f \, 2}+C_v^{f \, 2} \left(3 s-62 m_f^2\right)\right) M_Z^4\nn\\
&+&s^3 \left(\left(s-22 m_f^2\right) C_a^{f \, 2}+C_v^{f \, 2} \left(2 m_f^2+s \right)\right) M_Z^2+2 s^4 \left(2 C_a^{f \, 2}+C_v^{f \, 2}\right) m_f^2\bigg) \, ,\nn \\
{C_{(F)}}_4^8 &=&      \frac{i \kappa \, \alpha}{12 \pi  s^2 c_w^2 \left(s-4 M_Z^2\right){}^4 s_w^2}
\bigg(-36 \left(C_a^{f \, 2}+C_v^{f \, 2}\right) M_Z^{12}+2 \left(C_a^{f \, 2}+C_v^{f \, 2}\right) \left(112 m_f^2+9 s \right) M_Z^{10}\nn\\
&+&4\left(-80 \left(C_a^{f \, 2}+C_v^{f \, 2}\right) m_f^4-2 s \left(15 C_a^{f \, 2}+31 C_v^{f \, 2}\right) m_f^2+3 s^2 \left(C_a^{f \, 2}
+C_v^{f \, 2}\right)\right) M_Z^8\nn\\
&+&s \left(256 \left(C_a^{f \, 2}+C_v^{f \, 2}\right) m_f^4+8 s \left(13 C_v^{f \, 2}-7 C_a^{f \, 2}\right)
m_f^2-9 s^2 \left(C_a^{f \, 2}+C_v^{f \, 2}\right)\right) M_Z^6\nn\\
&+&s^2 \bigg(-36 \left(C_a^{f \, 2}+C_v^{f \, 2}\right) m_f^4
+ 2s \left(23 C_a^{f \, 2}-13 C_v^{f \, 2}\right) m_f^2+3 s^2 \left(C_a^{f \, 2}+C_v^{f \, 2}\right)\bigg) M_Z^4\nn\\
&-&s^3 m_f^2 \left(10 \left(C_a^{f \, 2}+C_v^{f \, 2}\right) m_f^2+s \left(15 C_a^{f \, 2}+C_v^{f \, 2}\right)\right) M_Z^2\nn\\
&+&s^4 m_f^2 \left(2 \left(m_f^2+s \right) C_a^{f \, 2}+C_v^{f \, 2} \left(2 m_f^2+s \right)\right)\bigg)\, ,\eea
\bea
{C_{(F)}}_0^9 &=&    -\frac{i \kappa \, \alpha}{72 \pi  s^2 c_w^2 \left(s-4 M_Z^2\right){}^3 s_w^2}
\left(C_a^{f \, 2}+C_v^{f \, 2}\right) \bigg(108 M_Z^8-2 \left(192 m_f^2+83 s \right) M_Z^6\nn\\
&+&3 s \left(128 m_f^2+23 s \right) M_Z^4-6 s^2\left(28 m_f^2+s \right) M_Z^2+s^3 \left(24 m_f^2+s \right)\bigg)\, ,\nn\\
{C_{(F)}}_1^9 &=&    \frac{i \kappa \, \alpha}{6 \pi  s^2 c_w^2 \left(s-4 M_Z^2\right){}^2 s_w^2}
\left(C_a^{f \, 2}+C_v^{f \, 2}\right) \left(3 M_Z^4-3 s M_Z^2+s^2\right)\, ,\nn\\
{C_{(F)}}_2^9 &=&   -\frac{i \kappa \, \alpha}{6 \pi  s^2 c_w^2 \left(s-4 M_Z^2\right){}^2 s_w^2}
\left(C_a^{f \, 2}+C_v^{f \, 2}\right) m_f^2 \left(3 M_Z^4-3 s M_Z^2+s^2\right)  \, ,\nn\\
{C_{(F)}}_3^9 &=&     -\frac{i \kappa \, \alpha}{24 \pi  s^2 c_w^2 \left(s-4 M_Z^2\right){}^4 s_w^2}
\bigg(-72 \left(C_a^{f \, 2}+C_v^{f \, 2}\right) M_Z^{10}+4 \left(C_a^{f \, 2}+C_v^{f \, 2}\right)\left(40 m_f^2+21 s \right) M_Z^8\nn\\
&-& 4 s\left(2 \left(C_a^{f \, 2}+33 C_v^{f \, 2}\right) m_f^2+15 s \left(C_a^{f \, 2}+C_v^{f \, 2}\right)\right) M_Z^6 +4 s^2 \bigg(\left(5
s-18 m_f^2\right) C_a^{f \, 2}\nn\\
&+&5 C_v^{f \, 2} \left(6 m_f^2+s \right)\bigg) M_Z^4
+s^3 \left(s C_a^{f \, 2}+C_v^{f \, 2} \left(s-48 m_f^2\right)\right) M_Z^2+4 s^4 \left(C_a^{f \, 2}+2 C_v^{f \, 2}\right) m_f^2\bigg) \, ,\nn\\
{C_{(F)}}_4^9 &=&    -\frac{i \kappa\, \alpha}{12 \pi  s^2 c_w^2 \left(s-4 M_Z^2\right){}^4 s_w^2}
\bigg(36 \left(C_a^{f \, 2}+C_v^{f \, 2}\right) M_Z^{12}-16 \left(C_a^{f \, 2}+C_v^{f \,
2}\right) \left(14 m_f^2+3 s \right) M_Z^{10}\nn\\
&+&8 \left(40 \left(C_a^{f \, 2}+C_v^{f \, 2}\right) m_f^4+s \left(33 C_a^{f \, 2}+49 C_v^{f \,
2}\right) m_f^2+3 s^2 \left(C_a^{f \, 2}+C_v^{f \, 2}\right)\right) M_Z^8\nn\\
&-&s \bigg(352 \left(C_a^{f \, 2}+C_v^{f \, 2}\right) m_f^4
+20 s\left(5 C_a^{f \, 2}+13 C_v^{f \, 2}\right) m_f^2+9 s^2 \left(C_a^{f \, 2}+C_v^{f \, 2}\right)\bigg) M_Z^6\nn\\
&+&s^2 \left(180 \left(C_a^{f \,2}+C_v^{f \, 2}\right) m_f^4+8 s \left(C_a^{f \, 2}+10 C_v^{f \, 2}\right) m_f^2+3 s^2 \left(C_a^{f \,
2}+C_v^{f \, 2}\right)\right) M_Z^4 \nn\\
&-&s^3 m_f^2 \left(44 \left(C_a^{f \, 2}+C_v^{f \, 2}\right) m_f^2+s \left(3 C_a^{f \, 2}+17 C_v^{f \,2}\right)\right) M_Z^2\nn\\
&+&s^4 m_f^2 \left(\left(4 m_f^2+s \right) C_a^{f \, 2}+2 C_v^{f \, 2} \left(2 m_f^2+s \right)\right)\bigg)\, .
\eea
\normalsize

\subsection{Form factors for the $TZZ$ vertex in the $W$ sector}
The coefficients corresponding to Eq. (\ref{P3boson1hZZ}) are given by
\small
\bea
{C_{(W)}}_0^2 &=& \frac{-i \kappa \, \alpha \,  M_Z^2} {12 \, s_w^2 \, c_w^2 \, \pi \, s^2  (s-4M_Z^2)}   \bigg(  2 M_Z^4 \left(-12 s_w^6+32 s_w^4-29 s_w^2+9\right)\nn\\
&+&s M_Z^2 \left(4\left(s_w^4+s_w^2\right)-7\right)-2 s^2 \left(s_w^2-1\right) \bigg)\, , \nn \\
{C_{(W)}}_1^2 &=& 0 \, ,\nn\\
{C_{(W)}}_2^2 &=&  \frac{-i \kappa \, \alpha \, M_Z^2} {6 \, s_w^2 \, c_w^2 \, \pi \, s} (-2 s_w^4+3s_w^2-1)\, , \nn\\
{C_{(W)}}_3^2 &=& \frac{-i \kappa \, \alpha \, M_Z^2} {6\, s_w^2 \, c_w^2 \, \pi \, s^2 (s-M_Z^2)^2} \bigg( 2 M_Z^6 \left(12
s_w^6-32 s_w^4+29 s_w^2-9\right)  \nn \\ 
&+& s M_Z^4 \left(-24 s_w^6+92 s_w^4-110 s_w^2+41\right) 
+ s^2 M_Z^2 \left(-12 s_w^4+26 s_w^2-13\right)+ 2 s^3 \left(s_w^2-1\right)^2   \bigg)\, ,\nn\\
{C_{(W)}}_4^2 &=&  \frac{-i \kappa \, \alpha \, M_Z^2} {24 \, s_w^2 \, c_w^2 \, \pi \, s^2 (s-M_Z^2)^2} \bigg( -8 M_Z^8 \left(s_w^2 - 1
\right) \left(4 s_w^2-3\right) \left(12 s_w^4-20 s_w^2+9\right) \nn \\
&+&   4 s M_Z^6 \left(24 s_w^8-60 s_w^6+30 s_w^4+25 s_w^2-18\right)+ 2 s^2 M_Z^4 (-20 s_w^6+76 s_w^4 \nn \\
&-&   103 s_w^2+46) + s^3 M_Z^2 \left(-4 s_w^4+24 s_w^2-19\right) - 2 s^4 \left(s_w^2-1\right)   \bigg)\, ,\\
{C_{(W)}}_0^3 &=&  \frac{-i \kappa \, \alpha}{384 \, s_w^2 \, c_w^2 \, \pi (s-4 M_Z^2)}  \bigg(  4 M_Z^4 \left(124 s_w^4-228
s_w^2+101\right)\nn\\
&-&2 s M_Z^2 \left(412 s_w^4-836s_w^2+417\right)+ s^2 \left(172 s_w^4-356 s_w^2+181\right)\bigg) \, ,\nn \\
{C_{(W)}}_1^3 &=&  \frac{-i \kappa \, \alpha}{48\, s_w^2 \, c_w^2 \, \pi}   \bigg(6 s_w^4 - 10 s_w^2 + \frac{9}{2} \bigg)
\, ,\nn\\
{C_{(W)}}_2^3 &=&  \frac{-i \kappa \, \alpha}{48\, s_w^2 \, c_w^2 \, \pi}  \bigg(\frac{1}{2} M_Z^2 \left(3 s_w^2 \left(4
\left(s_w^2-7\right) s_w^2+43\right)-56\right)+s \left(9 s_w^4-19 s_w^2+\frac{39}{4}\right)  \bigg) \, ,\nn\\
{C_{(W)}}_3^3 &=&   \frac{-i \kappa \, \alpha}{96\, s_w^2 \, c_w^2 \, \pi (s-4M_Z^2)^2}  \bigg( 2 M_Z^6 \left(-48 s_w^6+196
s_w^4-304 s_w^2+151\right)+s M_Z^4 (24 s_w^6-148 s_w^4 \nn \\
&+&   342 s_w^2-209) -2 s^2 M_Z^2 \left(44 s_w^4-76 s_w^2+33\right) +24 s^3 \left(s_w^2-1\right)^2\bigg) \, ,\nn\\
{C_{(W)}}_4^3 &=&   \frac{-i \kappa \, \alpha}{48\, s_w^2 \, c_w^2 \, \pi (s-4 M_Z^2)^2}
\bigg(M_Z^8 \left(4 s_w^2-3\right) \left(48 s_w^6-36 s_w^4+16 s_w^2-31\right) \nn \\
&-& 2 s M_Z^6 \left(48 s_w^8+228 s_w^6-532 s_w^4+247 s_w^2+7\right)
+s^2 M_Z^4 (12 s_w^8+276 s_w^6  -407 s_w^4 \nn \\
&-& 10 s_w^2+128)-s^3 M_Z^2 \left(36 s_w^6+16 s_w^4-133 s_w^2+81\right)+12 s^4 \left(s_w^2-1\right)^2\bigg)\, ,
\eea
\bea
{C_{(W)}}_0^4 &=&  \frac{-i \kappa \, \alpha}{288\, s_w^2 \, c_w^2 \, \pi s(s-4M_Z^2)^2}  \bigg(   s^2 M_Z^2 \left(4 \left(72 s_w^4-93
s_w^2-35\right) s_w^2+225\right)
+4 M_Z^6 (576 s_w^6-1164 s_w^4\nn\\
&+&740 s_w^2-153)-2 s M_Z^4 \left(864 s_w^6-1332 s_w^4+100 s_w^2+369\right) +s^3 \left(-12 s_w^4+20 s_w^2-9\right)\bigg)
\, , \nn\\
{C_{(W)}}_1^4 &=&  \frac{-i \kappa \, \alpha \, (12 s_w^4-20s_w^2+9)(s-3 M_Z^2) }{24\, s_w^2 \, c_w^2 \, \pi s(s-4M_Z^2)}
\, ,\nn
\eea
\bea
{C_{(W)}}_2^4 &=&   \frac{-i \kappa \, \alpha \,   M_Z^2 \left(s_w^2-1\right) \left(M_Z^2 \left(-36 s_w^4+92 s_w^2-43\right)+s \left(12 s_w^4-28 s_w^2+13\right)\right)}{24\, s_w^2 \, c_w^2 \, \pi s(s-4M_Z^2)} \, , \nn\\
{C_{(W)}}_3^4 &=&    \frac{-i \kappa \, \alpha}{48 \, s_w^2 \, c_w^2 \, \pi s(s-4M_Z^2)^3 }  \bigg(
4 M_Z^8 \left(4 s_w^2 \left(84 s_w^4-375 s_w^2+452\right)-641\right)
+2 s M_Z^6 (-264 s_w^6+1396 s_w^4 \nn \\ 
&-&  1786 s_w^2+653)
-4 s^2 M_Z^4 \left(36 s_w^6+32 s_w^4-169 s_w^2+101\right)
+ s^3 M_Z^2 \left(48 s_w^6-60 s_w^4-40 s_w^2+51\right)\bigg) \, , \nn\\
{C_{(W)}}_4^4 &=&    \frac{-i \kappa \, \alpha}{24 \, s_w^2 \, c_w^2 \, \pi s(s-4M_Z^2)^3 }  \bigg(
2 M_Z^{10} \left(4 s_w^2-3\right) \left(48 s_w^6-36 s_w^4+16 s_w^2-31\right)
+ 2 s M_Z^8 \big(-288 s_w^8 \nn \\ 
&+& 312 s_w^6+52 s_w^4
+ 14 s_w^2-89\big) + 2 s^2 M_Z^6 (108 s_w^8 - 60 s_w^6-103 s_w^4-2 s_w^2+58)\nn\\
&+& s^3 M_Z^4 \left(-24 s_w^8-48 s_w^6+98 s_w^4+34 s_w^2-61\right)
+  s^4 M_Z^2 \left(12 s_w^6-16 s_w^4-5 s_w^2+9\right) \bigg)\, ,\\
{C_{(W)}}_0^5 &=&   \frac{-i \kappa \, \alpha}{576\, s_w^2 \, c_w^2 \, \pi s(s-4M_Z^2)^2}  \bigg(
8 M_Z^6 \left(576 s_w^6-1164 s_w^4+740 s_w^2-153\right)
-48 s M_Z^4 \big(24 s_w^6+92 s_w^4\nn\\
&-&246 s_w^2+127\big) + 24 s^2 M_Z^2 \left(156 s_w^4-316 s_w^2+157\right)
+s^3 \left(-492 s_w^4+1028 s_w^2-525\right) \bigg)\, ,\nn\\
{C_{(W)}}_1^5 &=&  \frac{- i \kappa \, \alpha}{24\, s_w^2 \, c_w^2 \, \pi s(s-4M_Z^2)}
\bigg( (6 s_w^4 -10 s_w^2 + \frac{9}{2})(s-6 M_Z^2)  \bigg) \, , \nn\\
{C_{(W)}}_2^5 &=& \frac{- i \kappa \, \alpha}{24\, s_w^2 \, c_w^2 \, \pi s(s-4M_Z^2)}   \bigg(   M_Z^4 \left(-36 s_w^6+128 s_w^4-135
s_w^2+43\right)\nn\\
&+&\frac{1}{2} s M_Z^2\left(12 s_w^6+24 s_w^4-99 s_w^2+61\right) +s^2 \left(-9 s_w^4+19 s_w^2-\frac{39}{4}\right)\bigg)\, , \nn\\
{C_{(W)}}_3^5 &=&   \frac{-i \kappa \, \alpha}{48\, s_w^2 \, c_w^2 \, \pi s(s-4M_Z^2)^3}   \bigg(
4 M_Z^8 \left(4 s_w^2 \left(84 s_w^4-375 s_w^2+452\right)-641\right)\nn\\
&+&4 s M_Z^6 (-396 s_w^6+1600 s_w^4- 1781 s_w^2+577)+s^2 M_Z^4 \left(696 s_w^6-2620 s_w^4+2758 s_w^2-839\right) \nn \\
&+&  2 s^3 M_Z^2 \left(-48 s_w^6+228 s_w^4-284 s_w^2+105\right)-24 s^4 \left(s_w^2-1\right)^2\bigg)\, ,\nn\\
{C_{(W)}}_4^5 &=&   \frac{-i \kappa \, \alpha}{24\, s_w^2 \, c_w^2 \, \pi s(s-4M_Z^2)^3} \bigg(
2 M_Z^{10} \left(4 s_w^2-3\right) \left(48 s_w^6-36 s_w^4+16 s_w^2-31\right)
+ s M_Z^8 (29  \nn \\ 
&-& 4 s_w^2 \left(540 s_w^4-995 s_w^2+458\right))
+ 2 s^2 M_Z^6 (-36 s_w^8+840 s_w^6-1243 s_w^4
+ 229 s_w^2+205) \nn \\ 
&+& s^3 M_Z^4 \left(12 s_w^8-372 s_w^6+245 s_w^4+508 s_w^2-391\right)
+ 8 s^4 M_Z^2 \left(3 s_w^6+10 s_w^4-28 s_w^2+15\right) \nn \\
&-& 12 s^5 \left(s_w^2-1\right)^2\bigg)\, ,
\eea
\bea
{C_{(W)}}_0^6 &=&  \frac{-i \kappa \, \alpha}{576 \, s_w^2 \, c_w^2 \, \pi s(s-4M_Z^2)^2} \bigg(
-24 M_Z^6 \left(16 s_w^2-13\right) \left(12 s_w^4-20 s_w^2+9\right)+32 s M_Z^4 (108 s_w^6-72 s_w^4\nn\\
&-&179 s_w^2+141) -8 s^2 M_Z^2 \left(72 s_w^6+276 s_w^4-790 s_w^2+435\right)+s^3 \left(540 s_w^4-1108 s_w^2+561\right)\bigg)
\, ,\nn\\
{C_{(W)}}_1^6 &=& \frac{ -i \kappa \, \alpha}{24\, s_w^2 \, c_w^2 \, \pi s(s-4M_Z^2)}  \bigg( (-6 s_w^4 +10 s_w^2 - \frac{9}{2})(s - 2 M_Z^2)\bigg)\, , \nn\\
{C_{(W)}}_2^6 &=& \frac{ -i \kappa \, \alpha}{24\, s_w^2 \, c_w^2 \, \pi s(s-4M_Z^2)}
\bigg( M_Z^4 \left(12 s_w^6-32 s_w^4+29 s_w^2-9\right)\nn\\
&-&\frac{1}{2} s M_Z^2 \left(12 s_w^6+40 s_w^4-123 s_w^2+69\right)
+s^2 \left(9 s_w^4-19 s_w^2+\frac{39}{4}\right) \bigg)\, , \nn 
\eea
\bea
{C_{(W)}}_3^6 &=& \frac{-i \kappa \, \alpha}{48\, s_w^2 \, c_w^2 \, \pi s(s-4M_Z^2)^3} \bigg(
12 M_Z^8 \left(4 s_w^2-3\right) \left(12 s_w^4-20 s_w^2+9\right)\nn\\
&+&12 s M_Z^6 \left(-36 s_w^6+40 s_w^4+s_w^2-7\right)
+s^2 M_Z^4 \left(264 s_w^6-260 s_w^4-102 s_w^2+111\right)\nn\\
&-&s^3 M_Z^2 (48 s_w^6+28 s_w^4 -180 s_w^2+105) +24 s^4 \left(s_w^2-1\right)^2 \bigg)\, , \nn \\
{C_{(W)}}_4^6 &=& \frac{-i \kappa \, \alpha}{8\, s_w^2 \, c_w^2 \, \pi s(s-4M_Z^2)^3} \bigg(
-2 M_Z^{10} \left(3-4 s_w^2\right){}^2 \left(12 s_w^4-20 s_w^2+9\right) \nn \\
&+& s M_Z^8 \left(4 s_w^2-3\right) \left(96 s_w^6-76 s_w^4-36 s_w^2+15\right)
-s^2 M_Z^6 (120 s_w^8+152 s_w^6 - 538 s_w^4+212 s_w^2+53)\nn\\
&+& s^3 M_Z^4 \left(12 s_w^8+120 s_w^6-163 s_w^4-61 s_w^2+92\right)
- 2 s^4 M_Z^2 \left(8 s_w^6+2 s_w^4-27 s_w^2+17\right) \nn \\
 &+& 4 s^5 \left(s_w^2-1\right)^2\bigg)\, , \\
{C_{(W)}}_0^7 &=&   \frac{-i \kappa \, \alpha \, M_Z^2}{384\, s_w^2 \, c_w^2 \, \pi s(s-4M_Z^2)} \bigg(
4 M_Z^2 \left(16 s_w^2-13\right) \left(12 s_w^4-20 s_w^2+9\right)\nn\\
&+&s \left(-372 s_w^4+748 s_w^2-375\right)
-\frac{5 s^2 \left(12 s_w^4-20 s_w^2+9\right)}{s-4M_Z^2} \bigg)\, ,\nn\\
{C_{(W)}}_1^7 &=&    \frac{-i \kappa \, \alpha  \, M_Z^2}{24\, s_w^2 \, c_w^2 \, \pi s(s-4M_Z^2)}( 12 s_w^4 -20 s_w^2+ 9 )
\, ,\nn\\
{C_{(W)}}_2^7 &=&     \frac{-i \kappa \, \alpha \,  M_Z^4 }{24\, s_w^2 \, c_w^2 \, \pi s(s-4M_Z^2)}   \bigg( 12 s_w^6 - 32 s_w^4 + 29 s_w^2 -9\bigg)\, ,\nn\\
{C_{(W)}}_3^7 &=&   \frac{-i \kappa \, \alpha \, M_Z^2}{24 \, s_w^2 \, c_w^2 \, \pi s(s-4M_Z^2)^3}  \bigg(
6 M_Z^6 \left(4 s_w^2-3\right) \left(12 s_w^4-20 s_w^2+9\right) +s M_Z^4 (-456 s_w^6+1348 s_w^4\nn\\
&-&1290 s_w^2+393 )+ s^2 M_Z^2 \left(96 s_w^6-220 s_w^4+108 s_w^2+15\right)
-6 s^3 \left(4 s_w^4-9 s_w^2+5\right) \bigg)\, ,\nn\\
{C_{(W)}}_4^7 &=&   \frac{-i \kappa \, \alpha \, M_Z^2}{8 \, s_w^2 \, c_w^2 \, \pi s(s-4M_Z^2)^3}  \bigg(
-2 M_Z^8 \left(3-4 s_w^2\right){}^2 \left(12 s_w^4-20 s_w^2+9\right) \nn \\
&+&   2 s M_Z^6 \left(96 s_w^8 -152 s_w^6+20 s_w^4+58 s_w^2-21\right)
-s^2 M_Z^4 (24s_w^8+112 s_w^6 \nn \\
&-& 330 s_w^4+254 s_w^2-59 )+ s^3 M_Z^2 \left(28 s_w^6-48 s_w^4+15 s_w^2+5\right)-4 s^4 \left(s_w^2-1\right)^2 \bigg)\, ,
\eea
\bea
{C_{(W)}}_0^8 &=&    \frac{-i \kappa \, \alpha}{288 \, s_w^2 \, c_w^2 \, \pi s^2(s-4M_Z^2)^3}  \bigg(
-24 M_Z^8 \left(32 s_w^2-23\right) \left(12 s_w^4-20 s_w^2+9\right)\nn\\
&+&8 s M_Z^6 (576 s_w^6-1068 s_w^4 + 484 s_w^2+15)
+24 s^2 M_Z^4 \left(24 s_w^6-116 s_w^4+166 s_w^2-73\right) \nn \\
&+&  3 s^3 M_Z^2 \left(-96 s_w^6+332 s_w^4-380 s_w^2+145\right)
+s^4 \left(12 s_w^4-20 s_w^2+9\right)\bigg)  \, , \nn\\
{C_{(W)}}_1^8 &=&   \frac{-i \kappa \, \alpha}{24\, s_w^2 \, c_w^2 \, \pi s^2(s-4M_Z^2)^2}
\bigg((-12 s_w^4 + 20 s_w^2 - 9)(s^2 + s M_Z^2 -6 M_Z^4) \bigg)\, ,\nn\\
{C_{(W)}}_2^8 &=&   \frac{-i \kappa \, \alpha \, M_Z^2}{24\, s_w^2 \, c_w^2 \, \pi s^2(s-4M_Z^2)^2}
\bigg( (-12 s_w^6 + 32 s_w^4 - 29 s_w^2 +9)(s^2 + s M_Z^2 - 6 M_Z^2) \bigg)  \, , \nn\\
{C_{(W)}}_3^8 &=&    \frac{-i \kappa \, \alpha}{48 \, s_w^2 \, c_w^2 \, \pi s^2(s-4M_Z^2)^4}  \bigg(
8 M_Z^{10} \left(20 s_w^2-11\right) \left(12 s_w^4-20 s_w^2+9\right)\nn\\
&-&8 s M_Z^8 (324 s_w^6-908 s_w^4 + 835 s_w^2-244)+2 s^2 M_Z^6 \left(744 s_w^6-1996 s_w^4+1626 s_w^2-375\right) \nn \\
&+& 4 s^3 M_Z^4 \left(-12 s_w^6+20 s_w^4+31 s_w^2-36\right)s^4 M_Z^2 \left(-48 s_w^6+172 s_w^4-208 s_w^2+85\right)\bigg)\, ,\nn
\eea
\bea
{C_{(W)}}_4^8 &=&   \frac{-i \kappa \, \alpha}{24 \, s_w^2 \, c_w^2 \, \pi s^2(s-4M_Z^2)^4}
\bigg(-4 M_Z^{12} \left(4 s_w^2-3\right) \left(20 s_w^2-11\right) \left(12 s_w^4-20 s_w^2+9\right) \nn \\
&+& 2 s M_Z^{10} \left(4 \left(384 s_w^6-876 s_w^4+691 s_w^2-218\right) s_w^2+93\right) 
- 4 s^2 M_Z^8 \left(108 s_w^8+156 s_w^6  \right. \nn \\
&-& \left. 695 s_w^4+564 s_w^2-130\right)-2 s^3 M_Z^6 (60 s_w^8
- 496 s_w^6+865 s_w^4-513 s_w^2+84)\nn\\
&+&2 s^4 M_Z^4 \left(12 s_w^8-48 s_w^6+41 s_w^4+17 s_w^2-21\right)
+  s^5 M_Z^2 \left(-12 s_w^6+40 s_w^4-45 s_w^2+17\right)\bigg)\, , \nn \\
{C_{(W)}}_0^9 &=& \frac{-i \kappa \, \alpha}{144 \, s_w^2 \, c_w^2 \, \pi s^2(s-4M_Z^2)^3}
\bigg(-12 M_Z^8 \left(32 s_w^2-23\right) \left(12 s_w^4-20 s_w^2+9\right)\nn\\
&+&2 s M_Z^6 (2304 s_w^6-4668 s_w^4
+ 2852 s_w^2-429)+3 s^2 M_Z^4 \left(-672 s_w^6+1356 s_w^4-812 s_w^2+113\right) \nn \\
&+& 3 s^3 M_Z^2 \left(96 s_w^6-212 s_w^4+148 s_w^2-31\right)+s^4 \left(-12 s_w^4+20 s_w^2-9\right) \bigg) \, ,\nn\\
{C_{(W)}}_1^9 &=&   \frac{-i \kappa \, \alpha}{12\, s_w^2 \, c_w^2 \, \pi s^2(s-4M_Z^2)^2}   \bigg( (12 s_w^4 - 20 s_w^2 + 9) (s^2 - 3 s M_Z^2 + 3 M_Z^4) \bigg) \, ,  \nn\\
{C_{(W)}}_2^9 &=&   \frac{-i \kappa \, \alpha \, M_Z^2}{12\, s_w^2 \, c_w^2 \, \pi s^2(s-4M_Z^2)^2}
\bigg( (12s_w^6 - 32 s_w^4 + 29 s_w^2 -9)(s^2 - 3 s M_Z^2 + 3 M_Z^4) \bigg)\, ,   \nn\\
{C_{(W)}}_3^9 &=&   \frac{-i \kappa \, \alpha}{24\, s_w^2 \, c_w^2 \, \pi s^2(s-4M_Z^2)^4}  \bigg(
4 M_Z^{10} \left(20 s_w^2-11\right) \left(12 s_w^4-20 s_w^2+9\right)\nn\\
&+&2 s M_Z^8 (-792 s_w^6+2020 s_w^4 - 1718 s_w^2+461)+6 s^2 M_Z^6 \left(120 s_w^6-260 s_w^4+158 s_w^2-13\right) \nn \\
&+& s^3 M_Z^4 \left(-288 s_w^6+588 s_w^4-316 s_w^2+9\right)+ s^4 M_Z^2 \left(2 s_w^2-1\right) \left(4 s_w^2-5\right)
\left(6 s_w^2-5\right) \bigg)  \, ,    \nn\\
{C_{(W)}}_4^9 &=&     \frac{-i \kappa \, \alpha}{12\, s_w^2 \, c_w^2 \, \pi s^2(s-4M_Z^2)^4}  \bigg(*
-2 M_Z^{12} \left(4 s_w^2-3\right) \left(20 s_w^2-11\right) \left(12 s_w^4-20 s_w^2+9\right) \nn \\
&+&   4 s M_Z^{10} \left(528 s_w^8-1188 s_w^6+892 s_w^4-227 s_w^2+3\right)-2 s^2 M_Z^8 (540 s_w^8-960 s_w^6 \nn \\
&+& 361 s_w^4+147 s_w^2-76)+s^3 M_Z^6 \left(264 s_w^8-368 s_w^6+26 s_w^4+90 s_w^2-3\right) \nn \\
&-&   2 s^4 M_Z^4 \left(12 s_w^8+12 s_w^6-55 s_w^4+35 s_w^2-3\right)s^5 M_Z^2 \left(12 s_w^6-28 s_w^4+21 s_w^2-5\right)\bigg)\, .
\eea
\normalsize
\subsection{Form factors for the $TZZ$ vertex in the $(Z,H)$ sector}
The coefficients corresponding to Eq. (\ref{P3boson2hZZ}) are given by
\small
\bea
{C_{(Z,H)}}_0^2  &=&   -\frac{i \kappa\, \alpha  M_Z^2}{12 \pi  s^2 c_w^2 \left(s-4 M_Z^2\right) s_w^2} \left(M_Z^4+M_H^2 M_Z^2-3 s M_Z^2+s^2\right)\, , \nn\\
{C_{(Z,H)}}_1^2  &=&    \frac{i \kappa\, \alpha}{24 \pi  s^2 c_w^2 \left(s-4 M_Z^2\right)
s_w^2}  \left(M_H^2-M_Z^2 \right)\left(s-2 M_Z^2\right)\, , \nn\\
{C_{(Z,H)}}_2^2  &=&  -{C_{(Z,H)}}_1^2 \, , \nn\\
{C_{(Z,H)}}_3^2  &=&  -\frac{i \kappa\, \alpha}{48 \pi  s^2 c_w^2 \left(s-4 M_Z^2\right){}^2 s_w^2}
\left(\left(8 M_Z^6+s^3\right) M_H^2+M_Z^2 \left(s-4 M_Z^2\right) \left(s-2 M_Z^2\right)\left(3 s-2 M_Z^2\right)\right)\, , \nn \\
{C_{(Z,H)}}_4^2  &=&    \frac{i \kappa\, \alpha}{48 \pi  s^2 c_w^2 \left(s-4 M_Z^2\right){}^2 s_w^2}
\left(2 \left(4 M_H^4-s^2\right) M_Z^4-s \left(2 M_H^2+s \right){}^2 M_Z^2+s^2 M_H^2\left(2 M_H^2+s \right)\right)\, ,\nn\\
{C_{(Z,H)}}_5^2  &=& \frac{i \kappa\, \alpha  M_Z^2}{6 \pi  s^2 c_w^2 \left(s-4 M_Z^2\right){}^2 s_w^2}
\left(-s M_H^4-\left(3 M_H^2+5 s \right) M_Z^4+\left(M_H^2+s \right)\left(M_H^2+2 s \right) M_Z^2\right) \, ,\nn
\eea
\bea
{C_{(Z,H)}}_6^2  &=&  -\frac{i \kappa\, \alpha  \left(2 M_H^2+s \right)}{48 \pi  s^2 c_w^2 \left(s-4 M_Z^2\right)^2 s_w^2}
\bigg(4 \left(7 s-4 M_H^2\right) M_Z^6+4 \left(M_H^2-s \right) \left(M_H^2+3 s \right) M_Z^4 \nn \\
&+& 2 s \left(-M_H^4-2 s M_H^2+s^2\right) M_Z^2+s^2 M_H^4\bigg)\, , \nn\\
{C_{(Z,H)}}_7^2  &=& -\frac{i \kappa\, \alpha}{48 \pi  s^2 c_w^2 \left(s-4 M_Z^2\right){}^2 s_w^2}
\bigg(\left(8 M_Z^6+s^3\right) M_H^4\nn\\
&+&4 M_Z^2 \left(s-4 M_Z^2\right) \left(2 M_Z^4-s M_Z^2+s^2\right) M_H^2+4 s M_Z^4 \left(s-4 M_Z^2\right){}^2\bigg)\, ,\\
{C_{(Z,H)}}_0^3  &=&  \frac{i \kappa\, \alpha}{384 \pi  c_w^2\left(s-4 M_Z^2\right) s_w^2}
\left(4 M_H^4+80 M_Z^4+3 s^2-2 \left(4 M_H^2+15 s \right) M_Z^2\right)\, ,\nn\\
{C_{(Z,H)}}_1^3  &=&  \frac{i \kappa\, \alpha}{192 \pi  c_w^2 \left(s-4 M_Z^2\right) s_w^2}  \left(4 M_H^2-s \right) \, ,\nn\\
{C_{(Z,H)}}_2^3  &=&   -\frac{i \kappa\, \alpha}{192 \pi  c_w^2 \left(s-4 M_Z^2\right) s_w^2}  \left(4 M_H^2-8 M_Z^2+s \right)
\, ,\nn\\
{C_{(Z,H)}}_3^3  &=&   \frac{i \kappa\, \alpha}{384 \pi  c_w^2 \left(s-4 M_Z^2\right){}^2 s_w^2}
\left(2 M_H^2-4 M_Z^2+s \right) \left(6 M_H^4+6 \left(s-4 M_Z^2\right) M_H^2  \right. \nn \\
&+& \left. \left(s-28 M_Z^2\right) \left(s-4
M_Z^2\right)\right)\, ,\nn\\
{C_{(Z,H)}}_4^3  &=&    -\frac{i \kappa\, \alpha}{384 \pi  c_w^2 \left(s-4 M_Z^2\right){}^2 s_w^2}\left(2 M_H^2-s \right)
\left(6 M_H^4-4 \left(2 M_Z^2+s \right) M_H^2 \right. \nn \\
&+& \left. \left(s-16 M_Z^2\right) \left(s-6 M_Z^2\right)\right)\, ,\nn\\
{C_{(Z,H)}}_5^3  &=&    \frac{i \kappa\, \alpha}{96 \pi  c_w^2 \left(s-4 M_Z^2\right){}^2 s_w^2}
\left(32 M_Z^6-2 \left(4 M_H^2+19 s \right) M_Z^4 \right. \nn \\
&+& \left. \left(6 M_H^4+11 s M_H^2+6 s^2\right) M_Z^2-6 s M_H^4\right)\, , \nn\\
{C_{(Z,H)}}_6^3  &=&   \frac{i \kappa\, \alpha} {96 \pi  c_w^2 \left(s-4 M_Z^2\right)^2 s_w^2} \bigg(3 M_H^8-4 \left(2 M_Z^2+s
\right) M_H^6+\left(32 M_Z^4+6 s M_Z^2+s^2\right) M_H^4 \nn \\
&-& 8 M_Z^4 \left(8 M_Z^2+s \right) M_H^2+s M_Z^2 \left(16 M_Z^4+3 s M_Z^2-s^2\right)\bigg) \, ,\nn\\
{C_{(Z,H)}}_7^3  &=&   \frac{i \kappa\, \alpha } {96 \pi  c_w^2 \left(s-4 M_Z^2\right){}^2 s_w^2} \bigg(3 M_H^8+6 \left(s-4
M_Z^2\right) M_H^6+4 \left(s-7 M_Z^2\right) \left(s-4 M_Z^2\right) M_H^4 \nn \\
&+& \left(s-16 M_Z^2\right) \left(s-4 M_Z^2\right)^2 M_H^2-2 M_Z^2 \left(s-4 M_Z^2\right){}^2 \left(3 s-4 M_Z^2\right)\bigg)\, ,
\eea
\bea
{C_{(Z,H)}}_0^4  &=&       \frac{i \kappa\, \alpha}{288 \pi  s c_w^2 \left(s-4 M_Z^2\right)^2 s_w^2}
\bigg(304 M_Z^6-10 s M_Z^4-13 s^2 M_Z^2+s^3\nn\\
&+&12 M_H^4 \left(M_Z^2+s \right)+12 M_H^2 \left(6 M_Z^4-8 s M_Z^2+s^2\right)\bigg)\, , \nn\\
{C_{(Z,H)}}_1^4  &=& \frac{i \kappa\, \alpha}{48\pi  s c_w^2 M_Z^2 \left(s-4 M_Z^2\right){}^2 s_w^2}
\left(-8 M_Z^6+5 s M_Z^4-2 s^2 M_Z^2+M_H^2 \left(-4 M_Z^4+2 s M_Z^2+s^2\right)\right)\, ,\nn\\
{C_{(Z,H)}}_2^4  &=&     -\frac{i \kappa\, \alpha}{48 \pi  s c_w^2 M_Z^2 \left(s-4 M_Z^2\right){}^2 s_w^2}
\left(16 M_Z^6-9 s M_Z^4+M_H^2 \left(-4 M_Z^4+2 s M_Z^2+s^2\right)\right)\, , \nn\\
{C_{(Z,H)}}_3^4  &=&     \frac{i \kappa\, \alpha  \left(2 M_H^2-4 M_Z^2+s \right)}{96 \pi  s c_w^2
\left(s-4 M_Z^2\right)^3 s_w^2} \bigg(6\left(M_Z^2+s \right) M_H^4+6 \left(s-4 M_Z^2\right) \left(M_Z^2+s \right) M_H^2 \nn \\
&+& M_Z^2 \left(4 M_Z^2-s \right) \left(28 M_Z^2+11 s \right)\bigg)\, ,\nn
\eea
\bea
{C_{(Z,H)}}_4^4  &=&    -\frac{i \kappa\, \alpha}{96 \pi  s c_w^2 \left(s-4 M_Z^2\right){}^3 s_w^2}
\left(2 M_H^2-s \right)\bigg(96 M_Z^6+26 s M_Z^4-5 s^2 M_Z^2\nn\\
&+&6 M_H^4 \left(M_Z^2+s \right)-4 M_H^2 \left(2 M_Z^4+7 s M_Z^2\right)\bigg)\, ,\nn\\
{C_{(Z,H)}}_5^4  &=&      \frac{i \kappa\, \alpha }{48 \pi  s c_w^2 M_Z^2 \left(s-4 M_Z^2\right)^3 s_w^2}
\bigg(448 M_Z^{10}-236 s M_Z^8-8 s^2 M_Z^6+6 s^3 M_Z^4 \nn \nn\\
&-& 2 M_H^2 \left(2 s-3 M_Z^2\right) \left(-24 M_Z^4-7 s M_Z^2+s^2\right) M_Z^2
+ M_H^4 \left(44 M_Z^6-16 s M_Z^4-14 s^2 M_Z^2+s^3\right)\bigg) \, ,\nn\\
{C_{(Z,H)}}_6^4  &=&     \frac{i \kappa\, \alpha}{48 \pi  s c_w^2 \left(s-4 M_Z^2\right)^3 s_w^2}  \bigg(6 \left(M_Z^2+s \right) M_H^8-4
\left(4 M_Z^4+10 s M_Z^2+s^2\right) M_H^6\nn\\
&+&(64 M_Z^6+92 s M_Z + 14 s^2 M_Z^2+s^3) M_H^4-16 \left(8 M_Z^8+5 s M_Z^6+2 s^2 M_Z^4\right) M_H^2\nn\\
&+& s M_Z^2 \left(32 M_Z^6+22 s M_Z^4+4 s^2 M_Z^2-s^3\right)\bigg)\, ,\nn \\
{C_{(Z,H)}}_7^4  &=&     \frac{i \kappa\, \alpha}{48 \pi  s c_w^2 \left(s-4 M_Z^2\right){}^3 s_w^2}
\bigg(6 \left(M_Z^2+s \right) M_H^8+12 \left(s-4 M_Z^2\right) \left(M_Z^2+s \right) M_H^6\nn\\
&+&\left(s-4 M_Z^2\right)\left(-56 M_Z^4 - 32 s M_Z^2+7 s^2\right) M_H^4\nn\\
&+&\left(s-16 M_Z^2\right)\left(s-4 M_Z^2\right){}^2 \left(2 M_Z^2+s\right) M_H^2 - 4 M_Z^4 \left(s-4M_Z^2\right){}^3\bigg)\, ,\\
{C_{(Z,H)}}_0^5  &=&    -\frac{i \kappa\, \alpha}{576 \pi  s c_w^2 \left(s-4 M_Z^2\right){}^2 s_w^2}
\bigg(-608 M_Z^6+12 \left(59 s-12 M_H^2\right) M_Z^4\nn\\
&-&24 \left(M_H^4+s M_H^2+7 s^2\right) M_Z^2+36 s M_H^4+11 s^3\bigg)\, ,\nn\\
{C_{(Z,H)}}_1^5  &=&   \frac{i \kappa\, \alpha}{96 \pi  s c_w^2 M_Z^2 \left(s-4 M_Z^2\right){}^2 s_w^2}
\left(M_Z^2 \left(s-2 M_Z^2\right) \left(8 M_Z^2+3 s \right)-4 M_H^2 \left(2 M_Z^4-2 s M_Z^2+s^2\right)\right)\, , \nn\\
{C_{(Z,H)}}_2^5  &=&   \frac{i \kappa\, \alpha}{96 \pi  s c_w^2 M_Z^2 \left(s-4 M_Z^2\right){}^2 s_w^2}
\left(-32 M_Z^6+18 s M_Z^4-5 s^2 M_Z^2+4 M_H^2 \left(2 M_Z^4-2 s M_Z^2+s^2\right)\right)\, , \nn\\
{C_{(Z,H)}}_3^5  &=&   -\frac{i \kappa\, \alpha  \left(2 M_H^2-4 M_Z^2+s \right)}
{192 \pi  s c_w^2 \left(s-4 M_Z^2\right){}^3 s_w^2} \bigg(6\left(3 s-2 M_Z^2\right) M_H^4 \nn \\
&+& 6 \left(s-4 M_Z^2\right) \left(3 s-2 M_Z^2\right) M_H^2+\left(s-4 M_Z^2\right) \left(56 M_Z^4-54 s M_Z^2+s^2\right)\bigg)
\, ,\nn \eea
\bea
{C_{(Z,H)}}_4^5  &=& \frac{i \kappa\, \alpha\left(2 M_H^2-s \right)} {192 \pi  s c_w^2 \left(s-4 M_Z^2\right)^3 s_w^2}
\bigg(-192 M_Z^6+4 \left(4 M_H^2+59 s \right) M_Z^4\nn\\
&-&12 \left(M_H^2+s \right) \left(M_H^2+3 s \right) M_Z^2 + s \left(18 M_H^4-4 s M_H^2+s^2\right)\bigg)\, ,\nn\\
{C_{(Z,H)}}_5^5  &=&   -\frac{i \kappa\, \alpha }{48 \pi  s c_w^2 M_Z^2 \left(s-4 M_Z^2\right){}^3 s_w^2} \bigg(-448 M_Z^{10}+4
\left(36 M_H^2+131 s \right) M_Z^8 -2(22 M_H^4 + 95 s M_H^2\nn\\
&+&107 s^2) M_Z^6 + s \left(62 M_H^4+93 s M_H^2+24 s^2\right) M_Z^4
-8 s^2 M_H^2 \left(4 M_H^2+s \right) M_Z^2+2 s^3 M_H^4\bigg)\, ,\nn\\
{C_{(Z,H)}}_6^5  &=&    \frac{i \kappa\, \alpha} {48 \pi  s c_w^2 \left(s-4 M_Z^2\right){}^3 s_w^2} \bigg(\left(6 M_Z^2-9 s
\right) M_H^8+8 \left(-2 M_Z^4+4 s M_Z^2+s^2\right) M_H^6\nn\\
&-&2(-32 M_Z^6+ 50 s M_Z^4+8 s^2 M_Z^2+s^3) M_H^4+8 \left(-16 M_Z^8+14 s M_Z^6+5 s^2 M_Z^4\right) M_H^2  \nn \\
&+& s M_Z^2 \left(32 M_Z^6-26 s M_Z^4-11 s^2 M_Z^2+2 s^3\right)\bigg)\, ,\nn
\eea
\bea
{C_{(Z,H)}}_7^5  &=&   -\frac{i \kappa\, \alpha}{48 \pi  s c_w^2 \left(s-4 M_Z^2\right){}^3 s_w^2}
\bigg(\left(9 s-6 M_Z^2\right) M_H^8+6 \left(s-4 M_Z^2\right) \left(3 s-2 M_Z^2\right) M_H^6\nn\\
&+&\left(s-4 M_Z^2\right)(56 M_Z^4 - 76 s M_Z^2+11 s^2) M_H^4+2 \left(s-16 M_Z^2\right)
\left(s-4 M_Z^2\right){}^2 \left(s-M_Z^2\right)M_H^2\nn\\
&-&2 M_Z^2 \left(s-4 M_Z^2\right){}^3 \left(3 s-2 M_Z^2\right)\bigg)\, ,\\
{C_{(Z,H)}}_0^6  &=&  -\frac{i \kappa\, \alpha}{576 \pi  s c_w^2 \left(s-4 M_Z^2\right){}^2 s_w^2}
\bigg(96 M_Z^6-172 s M_Z^4+80 s^2 M_Z^2-7 s^3 +12 M_H^4 \left(3 s-2 M_Z^2\right)\nn\\
&+&24 M_H^2\left(10 M_Z^4-9 s M_Z^2+s^2\right)\bigg)\, ,\nn\\
{C_{(Z,H)}}_1^6  &=&     -\frac{i \kappa\, \alpha}{96 \pi  s c_w^2 M_Z^2 \left(s-4 M_Z^2\right){}^2 s_w^2}
\left(-16 M_Z^6+6 s M_Z^4-3 s^2 M_Z^2+2 M_H^2 \left(4 M_Z^4+s^2\right)\right)\, , \nn\\
{C_{(Z,H)}}_2^6  &=&      \frac{i \kappa\, \alpha}{96 \pi  s c_w^2 M_Z^2 \left(s-4 M_Z^2\right){}^2 s_w^2}
\left(2 M_H^2 \left(4 M_Z^4+s^2\right)-s M_Z^2 \left(6 M_Z^2+s \right)\right) \, ,\nn\\
{C_{(Z,H)}}_3^6  &=&   \frac{i \kappa\, \alpha}{192 \pi  s c_w^2 \left(s-4 M_Z^2\right){}^3 s_w^2}
\bigg(\left(24 M_Z^2-36 s \right) M_H^6-6 \left(7 s-6 M_Z^2\right) \left(s-4 M_Z^2\right) M_H^4 \nn \\
&-& 2 \left(s-4 M_Z^2\right){}^2 \left(3 s-4 M_Z^2\right) M_H^2+\left(s-4 M_Z^2\right){}^2 \left(8 M_Z^4-30 s
M_Z^2+s^2\right)\bigg)\, ,\nn\\
{C_{(Z,H)}}_4^6  &=&    \frac{i \kappa\, \alpha}{192 \pi  s c_w^2 \left(s-4 M_Z^2\right){}^3 s_w^2}
\bigg(4 \left(40 M_H^4+2 s M_H^2-3 s^2\right) M_Z^4-4(6 M_H^6+49 s M_H^4 \nn \\
&-& 26 s^2 M_H^2+4 s^3) M_Z^2+s \left(36 M_H^6-6 s M_H^4-4 s^2 M_H^2+s^3\right)\bigg)\, ,\nn\\
{C_{(Z,H)}}_5^6  &=&   \frac{i \kappa\, \alpha}{48 \pi  s c_w^2 M_Z^2 \left(s-4 M_Z^2\right){}^3 s_w^2}
\bigg(12 \left(7 s-4 M_H^2\right)M_Z^8+6 \left(2 M_H^4+5 s M_H^2-5 s^2\right) M_Z^6 \nn \\
&+& s \left(-6 M_H^4-43 s M_H^2+6 s^2\right) M_Z^4+4 s^2 M_H^2 \left(4 M_H^2+s \right) M_Z^2-s^3 M_H^4\bigg)\, , \nn\\
{C_{(Z,H)}}_6^6  &=&  \frac{i \kappa\, \alpha}{16 \pi  s c_w^2 \left(s-4 M_Z^2\right){}^3 s_w^2}
\left(M_H^4-4 M_Z^2 M_H^2+s M_Z^2\right) \bigg(\left(2 M_Z^2-3 s \right) M_H^4\nn\\
&+&\left(-8 M_Z^4+8 s M_Z^2+s^2\right) M_H^2+s M_Z^2 \left(2 M_Z^2-3 s \right)\bigg)\, ,\nn \\
{C_{(Z,H)}}_7^6  &=&     -\frac{i \kappa\, \alpha}{16 \pi  s c_w^2 \left(s-4 M_Z^2\right){}^3 s_w^2}
\left(M_H^2-4 M_Z^2+s \right) \bigg(\left(3 s-2 M_Z^2\right) M_H^6\nn\\
&+&2 \left(4 M_Z^4-5 s M_Z^2+s^2\right) M_H^4+2 s M_Z^2 \left(s-4 M_Z^2\right){}^2\bigg)\, ,
\eea
\bea
{C_{(Z,H)}}_0^7  &=&    \frac{i \kappa\, \alpha}{96 \pi  s c_w^2 \left(s-4 M_Z^2\right){}^2 s_w^2}
\bigg(-16 M_Z^6+10 \left(s-4 M_H^2\right) M_Z^4 +\left(4 M_H^4+s^2\right) M_Z^2+4 s M_H^4\bigg)\, ,\nn\\
{C_{(Z,H)}}_1^7  &=&   \frac{i \kappa\, \alpha}{48 \pi  s c_w^2 \left(s-4 M_Z^2\right){}^2 s_w^2}
\left(8 M_Z^4-7 s M_Z^2+M_H^2 \left(6 s-4 M_Z^2\right)\right)\, ,\nn\\
{C_{(Z,H)}}_2^7  &=&     \frac{i \kappa\, \alpha}{48 \pi  s c_w^2 \left(s-4 M_Z^2\right){}^2 s_w^2}
\left(\left(4 M_Z^2-6 s \right) M_H^2+5 s M_Z^2\right)\, ,\nn\\
{C_{(Z,H)}}_3^7  &=&    \frac{i \kappa\, \alpha}{96 \pi  s c_w^2 \left(s-4 M_Z^2\right){}^3 s_w^2}
\bigg(12 \left(M_Z^2+s \right) M_H^6+6 \left(s-4 M_Z^2\right) \left(3 M_Z^2+2 s \right) M_H^4 \nn \\
&+&  \left(s-4 M_Z^2\right){}^2 \left(4 M_Z^2+s \right) M_H^2+M_Z^2 \left(s-4 M_Z^2\right){}^2 \left(4 M_Z^2+11 s \right)\bigg)
\, , \nn
\eea
\bea
{C_{(Z,H)}}_4^7  &=&   \frac{i \kappa\, \alpha}{96 \pi  s c_w^2 \left(s-4 M_Z^2\right){}^3 s_w^2}   \bigg(-12 \left(M_Z^2+s \right)
M_H^6+\left(80 M_Z^4+54 s M_Z^2+4 s^2\right) M_H^4 \nn \\
&-&  s \left(124 M_Z^4+10 s M_Z^2+s^2\right) M_H^2+s^2 M_Z^2 \left(26 M_Z^2+s \right)\bigg)\, ,\nn \\
{C_{(Z,H)}}_5^7  &=&  -\frac{i \kappa\, \alpha}{24 \pi  s c_w^2 \left(s-4 M_Z^2\right){}^3 s_w^2}
\bigg(6 \left(4 M_H^2+9 s \right) M_Z^6-\left(6 M_H^4+59 s M_H^2+18 s^2\right) M_Z^4\nn\\
&+&s \left(20 M_H^4+2 s M_H^2+3 s^2\right) M_Z^2+s^2 M_H^4\bigg)\, ,\nn\\
{C_{(Z,H)}}_6^7  &=&   \frac{i \kappa\, \alpha}{16 \pi  s c_w^2 \left(s-4 M_Z^2\right){}^3 s_w^2}
\left(M_H^4-4 M_Z^2 M_H^2+s M_Z^2\right) \bigg(2 \left(M_Z^2+s \right) M_H^4\nn\\
&-&\left(8 M_Z^4+4 s M_Z^2+s^2\right) M_H^2+2 s M_Z^2 \left(M_Z^2+s \right)\bigg) \, ,\nn \\
{C_{(Z,H)}}_7^7  &=&    \frac{i \kappa\, \alpha  M_H^2}{16 \pi  s c_w^2 \left(s-4 M_Z^2\right){}^3 s_w^2} \bigg(2 \left(M_Z^2+s
\right) M_H^6+\left(s-4 M_Z^2\right) \left(4 M_Z^2+3 s \right) M_H^4 \nn\\
&+& \left(s-4 M_Z^2\right){}^2 \left(2 M_Z^2+s \right) M_H^2+2 s M_Z^2 \left(s-4 M_Z^2\right){}^2\bigg)\, ,\\
{C_{(Z,H)}}_0^8  &=&   -\frac{i \kappa\, \alpha}{288 \pi  s^2 c_w^2 \left(s-4 M_Z^2\right){}^3 s_w^2}  \bigg(96 M_Z^8-100 s
M_Z^6+48 s^2 M_Z^4+15 s^3 M_Z^2+s^4 \nn \\
&+& 12 M_H^4 \left(-6 M_Z^4+3 s M_Z^2+4 s^2\right) +12 M_H^2 \left(44 M_Z^6-22 s M_Z^4-10 s^2 M_Z^2+s^3\right)\bigg)\, ,\nn\\
{C_{(Z,H)}}_1^8  &=&   \frac{i \kappa\, \alpha}{48 \pi  s^2 c_w^2 M_Z^2 \left(s-4 M_Z^2\right){}^3 s_w^2}
\bigg(16 M_Z^8+2 \left(4 M_H^2-9 s \right) M_Z^6\nn\\
&+& s \left(8 M_H^2+13 s \right) M_Z^4+2 s^2 \left(s-8 M_H^2\right) M_Z^2-s^3 M_H^2\bigg) \, ,\nn\\
{C_{(Z,H)}}_2^8  &=&   \frac{i \kappa\, \alpha}{48 \pi  s^2 c_w^2 M_Z^2 \left(s-4 M_Z^2\right){}^3 s_w^2}
\bigg(32 M_Z^8-2 s M_Z^6-19 s^2 M_Z^4\nn\\
&+&M_H^2 \left(-8 M_Z^6-8 s M_Z^4+16 s^2 M_Z^2+s^3\right)\bigg)\, , \nn\\
{C_{(Z,H)}}_3^8  &=&  -\frac{i \kappa\, \alpha} {96 \pi  s^2 c_w^2 \left(s-4 M_Z^2\right){}^4 s_w^2}
\bigg(12 \left(-6 M_Z^4+3 s M_Z^2+4 s^2\right) M_H^6+6 \left(s-4 M_Z^2\right)(-18 M_Z^4 \nn \\
&+&  5 s M_Z^2+8 s^2) M_H^4+4 \left(s-4 M_Z^2\right){}^2 \left(-8 M_Z^4+s M_Z^2+2 s^2\right) M_H^2 \nn \\
&+& M_Z^2 \left(s-4 M_Z^2\right)^3 \left(2 M_Z^2+7 s \right)\bigg) \, ,\nn \\
{C_{(Z,H)}}_4^8  &=&   -\frac{i \kappa\, \alpha}{96 \pi  s^2 c_w^2 \left(s-4 M_Z^2\right){}^4 s_w^2}
\bigg(4 \left(-88 M_H^4+42 s M_H^2+s^2\right) M_Z^6+4 \left(M_H^2-s \right)(18 M_H^4 +57 s M_H^2 \nn \\
&-& 17 s^2) M_Z^4+s \left(-36 M_H^6+218 s M_H^4-86 s^2 M_H^2+9 s^3\right) M_Z^2-2 s^2 M_H^2 \left(24 M_H^4-10 s
M_H^2+s^2\right)\bigg)\, ,\nn\\
{C_{(Z,H)}}_5^8  &=&    -\frac{i \kappa\, \alpha} {48 \pi  s^2 c_w^2 M_Z^2 \left(s-4 M_Z^2\right){}^4 s_w^2} \bigg(\left(-8 M_Z^8+60 s
M_Z^6-100 s^2 M_Z^4-22 s^3 M_Z^2+s^4\right) M_H^4 \nn \\
&+& \left(96 M_Z^{10}-228 s M_Z^8+262 s^2 M_Z^6+42 s^3 M_Z^4-4 s^4 M_Z^2\right) M_H^2
- 6 s M_Z^4 \left(-44 M_Z^6+38 s M_Z^4 \right. \nn \\
&-& \left. 2 s^2 M_Z^2+s^3\right)\bigg)\, , \nn \\
{C_{(Z,H)}}_6^8  &=&  -\frac{i \kappa\, \alpha} {48 \pi  s^2 c_w^2 \left(s-4 M_Z^2\right){}^4 s_w^2}
\bigg(-4 \left(80 M_H^4-24 s M_H^2+s^2\right) M_Z^8+2(112 M_H^6+44 s M_H^4 \nn \\
&-& 48 s^2 M_H^2+9 s^3) M_Z^6-6 \left(6 M_H^8+16 s M_H^6-38 s^2 M_H^4+20 s^3 M_H^2-3 s^4\right) M_Z^4 \nn \\
&+& s \left(18 M_H^8-144 s M_H^6+92 s^2 M_H^4-18 s^3 M_H^2+s^4\right) M_Z^2+2 s^2 M_H^4 \left(12 M_H^4-7 s
M_H^2+s^2\right)\bigg)\, ,\nn
\eea
\bea
{C_{(Z,H)}}_7^8  &=&  -\frac{i \kappa\, \alpha  M_H^2 }{48 \pi  s^2 c_w^2 \left(s-4 M_Z^2\right){}^4 s_w^2}
\bigg(6 \left(-6 M_Z^4+3 s M_Z^2+4 s^2\right) M_H^6+12 \left(s-4 M_Z^2\right)(-6 M_Z^4 \nn \\
&+& 2 s M_Z^2+3 s^2) M_H^4+2 \left(7 s-10 M_Z^2\right) \left(s-4 M_Z^2\right){}^2 \left(2 M_Z^2+s \right) M_H^2 \nn \\
&+& \left(s-4 M_Z^2\right){}^3 \left(-4 M_Z^4+4 s M_Z^2+s^2\right)\bigg)\, ,\\
{C_{(Z,H)}}_0^9  &=&   \frac{i \kappa\, \alpha}{144 \pi  s^2 c_w^2 \left(s-4 M_Z^2\right){}^3 s_w^2}
\bigg(-48 M_Z^8+26 s M_Z^6-3 s^2 M_Z^4+9 s^3 M_Z^2+s^4 \nn \\
&+& 12 M_H^2 \left(s-11 M_Z^2\right) \left(2 M_Z^4-2 s M_Z^2+s^2\right)+12 M_H^4 \left(3 M_Z^4-4 s M_Z^2+3 s^2\right)\bigg)
\, ,\nn\\
{C_{(Z,H)}}_1^9  &=&   \frac{i \kappa\, \alpha}{24 \pi  s^2 c_w^2 M_Z^2 \left(s-4 M_Z^2\right){}^3 s_w^2}
\left(8 M_Z^8-5 s M_Z^6-2 s^3 M_Z^2+M_H^2 \left(4 M_Z^6-10 s M_Z^4 \right. \right. \nn \\
&+& \left.\left. 7 s^2 M_Z^2+s^3\right)\right)\, , \nn\\
{C_{(Z,H)}}_2^9  &=&  -\frac{i \kappa\, \alpha}{24 \pi  s^2 c_w^2 M_Z^2 \left(s-4 M_Z^2\right){}^3 s_w^2}
\bigg(-16 M_Z^8+25 s M_Z^6-14 s^2 M_Z^4\nn\\
&+&M_H^2 \left(4 M_Z^6-10 s M_Z^4+7 s^2 M_Z^2+s^3\right)\bigg)\, ,\nn\\
{C_{(Z,H)}}_3^9  &=&   \frac{i \kappa\, \alpha}{48 \pi  s^2 c_w^2 \left(s-4 M_Z^2\right){}^4 s_w^2}  \bigg(12 \left(3 M_Z^4-4 s
M_Z^2+3 s^2\right) M_H^6+6 \left(s-4 M_Z^2\right)(9 M_Z^4 \nn \\
&-& 10 s M_Z^2+6 s^2) M_H^4+\left(s-4 M_Z^2\right)^2 \left(16 M_Z^4-18 s M_Z^2+7 s^2\right) M_H^2 \nn \\
&+&M_Z^2
\left(M_Z^2+4 s \right) \left(4 M_Z^2-s \right)^3\bigg) \, ,\nn \\
{C_{(Z,H)}}_4^9  &=&    -\frac{i \kappa\, \alpha}{48 \pi  s^2 c_w^2 \left(s-4 M_Z^2\right){}^4 s_w^2}
\bigg(12 \left(3 M_Z^4-4 s M_Z^2+3 s^2\right) M_H^6-2(88 M_Z^6-91 s M_Z^4 \nn \\
&+& 64 s^2 M_Z^2+8 s^3) M_H^4+s \left(84 M_Z^6-26 s M_Z^4+76 s^2 M_Z^2+s^3\right) M_H^2 \nn \\
&+& s^2 M_Z^2
\left(2 M_Z^4-21 s M_Z^2-8 s^2\right)\bigg)\, , \nn \\
{C_{(Z,H)}}_5^9  &=& \frac{i \kappa\, \alpha }{24 \pi  s^2 c_w^2 M_Z^2 \left(s-4 M_Z^2\right){}^4 s_w^2}  \bigg(-6 s \left(22
M_Z^4-16 s M_Z^2+7 s^2\right) M_Z^6+M_H^4(4 M_Z^8-12 s M_Z^6 \nn \\
&-& 23 s^3 M_Z^2+s^4)+M_H^2 \left(-48 M_Z^{10}+90 s M_Z^8-30 s^2 M_Z^6+58 s^3 M_Z^4-4 s^4 M_Z^2\right)\bigg)\, , \nn \\
{C_{(Z,H)}}_6^9  &=&    \frac{i \kappa\, \alpha}{24 \pi  s^2 c_w^2 \left(s-4 M_Z^2\right){}^4 s_w^2}  \bigg(6 \left(3 M_Z^4-4 s
M_Z^2+3 s^2\right) M_H^8-(112 M_Z^6-120 s M_Z^4 \nn \\
&+&  84 s^2 M_Z^2+11 s^3) M_H^6+2 \left(80 M_Z^8-46 s M_Z^6+24 s^2 M_Z^4+31 s^3 M_Z^2+s^4\right) M_H^4 \nn \\
&-&  3 s M_Z^2 \left(16 M_Z^6-8 s M_Z^4+16 s^2 M_Z^2+5 s^3\right) M_H^2+s^2 M_Z^2 \left(2 M_Z^6+9 s^2 M_Z^2+s^3\right)\bigg)
\, ,\nn\\
{C_{(Z,H)}}_7^9  &=&    \frac{i \kappa\, \alpha  M_H^2} {24 \pi  s^2 c_w^2 \left(s-4 M_Z^2\right){}^4 s_w^2}
\bigg(6 \left(3 M_Z^4-4 s M_Z^2+3 s^2\right) M_H^6+3 \left(s-4 M_Z^2\right)(12 M_Z^4 \nn \\
&-& 14 s M_Z^2+9 s^2) M_H^4+\left(s-4 M_Z^2\right){}^2 \left(20 M_Z^4-22 s M_Z^2+11 s^2\right) M_H^2 \nn \\
&+& \left(s-4 M_Z^2\right){}^3 \left(2 M_Z^4-5 s M_Z^2+s^2\right)\bigg)\, .
\eea
\normalsize

\subsection{The improvement contribution}
\label{P3imprformfactors}
The two form factors with the improvement contribution are given by
\small
\bea
 \Phi^{(I)}_1 (s,M_Z^2,M_Z^2,M_W^2,M_Z^2,M_H^2) &=& - i \frac{\kappa}{2} \frac{\alpha}{12 \pi \, s_w^2 \, c_w^2 \, s (s-4M_Z^2)^2} \bigg\{
 (c_w^2 - s_w^2)^2 \bigg[ s^2 - 6 M_Z^2 s + 8 M_Z^4 \nn \\
 && \hspace{-5cm} + 2 M_Z^2 (s + 2 M_Z^2) \, \mathcal D_0 \left( s, M_Z^2 , M_W^2, M_W^2 \right)   +  2 \left( c_w^2 M_Z^2 (8 M_Z^4 - 6 M_Z^2 s + s^2) - 2 M_Z^6 + 2 M_Z^4 s \right) \times \nn \\
 && \hspace{-5cm} \times \, \mathcal C_0 \left( s, M_Z^2, M_Z^2, M_W^2,M_W^2,M_W^2 \right) \bigg]  +  s^2 - 6 M_Z^2 s + 8 M_Z^4  + 2 M_Z^2 (s + 2 M_Z^2) \big[
 \mathcal B_0 \left(s, M_Z^2, M_Z^2 \right)  \nn \\
 && \hspace{-5cm} -  \mathcal B_0 \left(M_Z^2, M_Z^2, M_H^2 \right)  \big] + \left( 3 M_Z^2 s - 2 M_H^2 (s - M_Z^2) \right) \big[ \mathcal B_0 \left(s, M_H^2, M_H^2 \right) -  \mathcal B_0 \left(s, M_Z^2, M_Z^2 \right)\big] \nn \\
 && \hspace{-5cm} + M_H^2 \left( 2 M_H^2 (s-M_Z^2) + 8 M_Z^4 - 6 M_Z^2 s + s^2 \right) \mathcal C_0\left( s, M_Z^2,M_Z^2, M_H^2,M_Z^2,M_Z^2\right) \nn \\
 && \hspace{-5cm} + \left( 2 M_H^2 (M_H^2 - 4 M_Z^2)(s-M_Z^2) + s M_Z^2 (s+ 2 M_Z^2)\right) \mathcal C_0\left( s, M_Z^2,M_Z^2, M_Z^2,M_H^2,M_H^2\right)
 \bigg\} \, , \\
 \Phi^{(I)}_2 (s,M_Z^2,M_Z^2,M_W^2,M_Z^2,M_H^2) &=& i \frac{\kappa}{2} \frac{\alpha}{48 \pi \, s_w^2 \, c_w^2 (s-4 M_Z^2)^2} \bigg\{
 (c_w^2 -s_w^2)^2 \bigg[ 4 M_Z^4 (s - 4 M_Z^2) \nn \\
 && \hspace{-5cm} + 8 M_Z^4 (s-M_Z^2) \, \mathcal D_0 \left( s, M_Z^2, M_W^2, M_W^2\right) + 2 M_Z^4 \big( s^2 - 2 M_Z^2 s + 4 M_Z^4 + 4 c_w^2 M_Z^2 (s - 4 M_Z^2) \big) \times \nn \\
 && \hspace{-5cm}  \times \, \mathcal C_0 \left( s, M_Z^2, M_Z^2, M_W^2,M_W^2,M_W^2 \right) \bigg] + 4 M_Z^2 s_w^4 c_w^2 s (s - 4 M_Z^2)^2 \mathcal C_0 \left( s, M_Z^2, M_Z^2, M_W^2,M_W^2,M_W^2 \right)  \nn \\
 && \hspace{-5cm}  + 4 M_Z^2 (s-4 M_Z^2) + \big( M_Z^2 s (s + 2 M_Z^2) - M_H^2 (s^2 - 2 M_Z^2 s + 4 M_Z^4) \big) \big[ \mathcal B_0 \left(s, M_H^2, M_H^2 \right) -  \mathcal B_0 \left(s, M_Z^2, M_Z^2 \right) \big] \nn \\
 && \hspace{-5cm}  + 8 M_Z^4 (s - M_Z^2) \big[ \mathcal B_0 \left(s, M_Z^2, M_Z^2 \right) -  \mathcal B_0 \left(M_Z^2, M_Z^2, M_H^2 \right)  \big] +
 M_H^2 \big( 4 M_Z^4 (s-4 M_Z^2) \nn \\
 && \hspace{-5cm}  + M_H^2 (s^2 - 2 MZ^2 s + 4 M_Z^4)\big) \mathcal C_0\left( s, M_Z^2,M_Z^2, M_H^2,M_Z^2,M_Z^2\right) + \big( M_H^2(M_H^2 - 4 M_Z^2) (s^2 - 2 M_Z^2 s + 4 M_Z^4) \nn \\
 && \hspace{-5cm}  + 2 M_Z^2 s (s^2 - 6 M_Z^2 s + 14 M_Z^4)\big) \mathcal C_0\left( s, M_Z^2,M_Z^2, M_Z^2,M_H^2,M_H^2\right)
 \bigg\}\, .
\eea
\normalsize
\subsection{Coefficients of the external leg corrections}
\label{P3externalleg}
\small
\bea
{C^{(I)}_{(F)}}^1_0 &=&  \frac{i \, \kappa \, \alpha \, m_f^2 (C^{f \, 2}_v + C^{f \, 2}_a) (s- 2 M_Z^2)}{6 \pi s_w^2 c_w^2 s (s-M_H^2)(s-4M_Z^2)} \,, \nn\\
{C^{(I)}_{(F)}}^1_1 &=& {C^{(I)}_{(F)}}^1_2 = 0 \,, \nn\\
{C^{(I)}_{(F)}}^1_3 &=& \frac{i \, \kappa \, \alpha \, m_f^2}{3 \pi s_w^2 c_w^2 s (s-M_H^2)(s-4M_Z^2)^2 } \bigg[ M_Z^2 (C^{f \, 2}_v + C^{f \, 2}_a) (s+ 2 M_Z^2) + C^{f \, 2}_a (s-4M_Z^2) s \bigg] \,, \nn\\
{C^{(I)}_{(F)}}^1_4 &=& \frac{i \, \kappa \, \alpha \, m_f^2 (s-2 M_Z^2)}{12 \pi s_w^2 c_w^2 s (s-M_H^2)(s-4M_Z^2)^2} \bigg[ (C^{f \, 2}_v + C^{f \, 2}_a)
\left( 4 m_f^2 (s-4M_Z^2) + 4 M_Z^4 + 6 MZ^2 s - s^2 \right)\nn\\
&+& 2  C^{f \, 2}_a s (s-4 M_Z^2) \bigg]\, , \\
{C^{(I)}_{(F)}}^2_0 &=&  - \frac{i \, \kappa \, \alpha \, m_f^2 \, M_Z^2  (C^{f \, 2}_v + C^{f \, 2}_a) }{3 \pi s_w^2 c_w^2 s
(s-M_H^2)(s-4M_Z^2)} \,, \nn\\
{C^{(I)}_{(F)}}^2_1 &=& 0 \,, \nn\\
{C^{(I)}_{(F)}}^2_2 &=& - \frac{i \, \kappa \, \alpha \, m_f^2 \, C^{f \, 2}_a}{6 \pi s_w^2 c_w^2  (s-M_H^2)} \,,\nn \\
{C^{(I)}_{(F)}}^2_3 &=&  - \frac{i \, \kappa \, \alpha \, m_f^2 }{3 \pi s_w^2 c_w^2 s (s-M_H^2)(s-4M_Z^2)^2} \bigg[ 2 M_Z^4
(C^{f \, 2}_v + C^{f \, 2}_a) (s - M_Z^2) + C^{f \, 2}_a M_Z^2 s (s-4M_Z^2) \bigg]\,, \nn
\eea
\bea
{C^{(I)}_{(F)}}^2_4 &=& - \frac{i \, \kappa \, \alpha \, m_f^2}{6 \pi s_w^2 c_w^2 s (s-M_H^2)(s-4M_Z^2)^2} \bigg[
(C^{f \, 2}_v + C^{f \, 2}_a) M_Z^4 \left( 4 m_f^2 (s-4 M_Z^2) + 4 M_Z^4 - 2M_Z^2 s + s^2\right)\nn\\
&+& 2 C^{f \, 2}_a s (s-4M_Z^2)\left( m_f^2 (s-4M_Z^2)+M_Z^4\right)\bigg]\, , \\
{C^{(I)}_{(W)}}^1_0 &=& - \frac{i \, \kappa \, \alpha (s-2M_Z^2)}{24 \pi s_w^2 c_w^2 s (s-M_H^2)(s-4M_Z^2)}\bigg[ M_H^2 (1-2
s_w^2)^2 + 2 M_Z^2 (-12 s_w^6 + 32 s_w^4 - 29 s_w^2 +9)\bigg] \,, \nn\\
{C^{(I)}_{(W)}}^1_1 &=& {C^{(I)}_{(W)}}^1_2 = 0 \,, \nn\\
{C^{(I)}_{(W)}}^1_3 &=& - \frac{i \, \kappa \, \alpha \, M_Z^2}{12 \pi s_w^2 c_w^2 s (s-M_H^2)(s-4M_Z^2)^2} \bigg[
M_H^2 (1-2s_w^2)^2 (s+ 2 M_Z^2) \nn \\
&-&  2 (s_w^2-1)\bigg( 2 M_Z^4 (12 s_w^4 - 20 s_w^2 + 9) 
+ s M_Z^2 (12 s_w^4 - 20 s_w^2 + 1) + 2 s^2 \bigg) \bigg] \,, \nn\\
{C^{(I)}_{(W)}}^1_4 &=& - \frac{i \, \kappa \, \alpha \, M_Z^2}{12 \pi s_w^2 c_w^2 s (s-M_H^2)(s-4M_Z^2)^2} \bigg[
2 (s_w^2-1) \bigg( 2 M_Z^6 (4 s_w^2 -3) (12 s_w^4 - 20 s_w^2 + 9) \nn\\
&+& 2 M_Z^4 s (-36 s_w^6 + 148 s_w^4 - 163 s_w^2 + 54) + M_Z^2 s^2 (12 s_w^6 - 96 s_w^4 + 125 s_w^2 - 43) \nn \\
&+& 4 s^3 (2 s_w^4 - 3 s_w^2 + 1)\bigg)
- M_H^2 (1-2 s_w^2)^2 \left( M_Z^4 (8 s_w^2 - 6) + 2 M_Z^2 s (2-3 s_w^2) + s^2 (s_w^2 -1)\right)\bigg]\, , \nn \\ \\
{C^{(I)}_{(W)}}^2_0 &=& \frac{i \, \kappa \, \alpha \, M_Z^4}{ 12 \pi s_w^2 c_w^2 s (s-M_H^2)(s-4M_Z^2)}
\bigg[ M_H^2 (1-2 s_w^2)^2 + 2 M_Z^2 (-12 s_w^6 + 32 s_w^4 - 29 s_w^2 +9)\bigg] \,, \nn\\
{C^{(I)}_{(W)}}^2_1 &=& 0 \,, \nn\\
{C^{(I)}_{(W)}}^2_2 &=& - \frac{i \, \kappa \, \alpha \, M_Z^2}{24 \pi s_w^2 c_w^2 (s-M_H^2)} \bigg[ 8 s_w^4 - 13 s_w^2 + 5\bigg] \,, \nn\\
{C^{(I)}_{(W)}}^2_3 &=&  - \frac{i \, \kappa \, \alpha M_Z^2}{6 \pi s_w^2 c_w^2 s (s-M_H^2)(s-4M_Z^2)} \bigg[
M_H^2 M_Z^2 (1-2 s_w^2)^2 (M_Z^2 -s) -2 (s_w^2-1)\bigg( M_Z^6 (12 s_w^4 \nn \\
&-& 20 s_w^2 + 9)
- 3 M_Z^4 s \left( 4 (s_w^2 - 3)s_w^2 + 7 \right) + M_Z^2 s^2 (7 - 8 s_w^2) + s^3(s_w^2-1)\bigg)  \bigg]\,, \nn\\
{C^{(I)}_{(W)}}^2_4 &=& -\frac{i \, \kappa \, \alpha \, M_Z^4}{ 24 \pi s_w^2 c_w^2 s (s-M_H^2)(s-4M_Z^2)}\bigg[
M_H^2 \bigg( - 4 M_Z^6 (1-2 s_w^2)^2 (4 s_w^2 -3) \nn \\
&+&  2 M_Z^4 s (24 s_w^6 - 28 s_w^4 + 6 s_w^2 -1) 
+ M_Z^2 s^2 (-16 s^6 + 12 s_w^4 + 4 s_w^2 -1) + 2 s^3 s_w^4 (s_w^2-1)\bigg)\nn\\
&+& 2 (s_w^2 -1) \bigg( 4 M_Z^8 (4 s_w^2 - 3)(12
s_w^4 - 20 s_w^2 + 9) - 2 M_Z^6 s (24 s_w^6 - 52 s_w^4 + 6 s_w^2 + 15)\nn\\
&+& M_Z^4 s^2 (45 - 4 s_w^2 (s_w^2 +13)) + 2M_Z^2 s^3 (4 s_w^4 + 2 s_w^2 - 5) - s^4 (s_w^4 - 1) \bigg) \bigg]\, ,\\
{C^{(I)}_{(Z,H)}}^1_0 &=& - \frac{i \, \kappa \, \alpha (2 M_H^2 + M_Z^2) (s - 2 M_Z^2)}{24 \pi s_w^2 c_w^2 s (s-M_H^2)(s-4M_Z^2)} \,, \nn\\
{C^{(I)}_{(Z,H)}}^1_1 &=& - {C^{(I)}_{(Z,H)}}^1_2 = \frac{i \, \kappa \, \alpha (M_H^2 -M_Z^2)}{12 \pi s_w^2 c_w^2 s (s-M_H^2)(s-4 M_Z^2)} \,, \nn\\
{C^{(I)}_{(Z,H)}}^1_3 &=& - \frac{i \, \kappa \, \alpha}{24 \pi s_w^2 c_w^2 s (s-M_H^2)(s-4 M_Z^2)^2} \bigg[ 2 M_H^4 (s-M_Z^2) + 3 M_H^2 M_Z^2 s + 2 M_Z^2 (4 M_Z^4 - 9 M_Z^2 s + 2 s^2)\bigg] \,, \nn \\
{C^{(I)}_{(Z,H)}}^1_4 &=&  \frac{i \, \kappa \, \alpha}{8 \pi s_w^2 c_w^2 s (s-M_H^2)(s-4M_Z^2)^2} \bigg[ 2 M_H^4 (s-M_Z^2) - 3 M_H^2 M_Z^2 s \bigg] \,, \nn\\
{C^{(I)}_{(Z,H)}}^1_5 &=&  \frac{i \, \kappa \, \alpha}{12 \pi s_w^2 c_w^2 s (s-M_H^2)(s-4M_Z^2)^2} \bigg[ M_H^2 (s + 2 M_Z^2) (4 M_Z^2 - M_H^2 ) + 2 M_Z^2 s (s- 4 M_Z^2)\bigg] \,, \nn
\eea
\bea
{C^{(I)}_{(Z,H)}}^1_6 &=&  \frac{i \, \kappa \, \alpha \, M_H^2}{8 \pi s_w^2 c_w^2 s (s-M_H^2)(s-4M_Z^2)^2} \bigg[ 2 M_H^2 (s -M_Z^2) (4 M_Z^2 - M_H^2) - M_Z^2 s (s + 2 M_Z^2)\bigg] \,, \nn\\
{C^{(I)}_{(Z,H)}}^1_7 &=& - \frac{i \, \kappa \, \alpha \, M_H^2}{24 \pi s_w^2 c_w^2 s (s-M_H^2)(s-4M_Z^2)^2} \bigg[
2 M_H^4 (s - M_Z^2) + M_H^2 (4 M_Z^4 - 2 M_Z^2 s + s^2)\nn\\
&+& 2 M_Z^2 (8 M_Z^4 - 14 M_Z^2 s + 3 s^2 ) \bigg]\, , \\
{C^{(I)}_{(Z,H)}}^2_0 &=& \frac{i \, \kappa \, \alpha \, M_Z^4 (2 M_H^2 + M_Z^2)}{12 \pi s_w^2 c_w^2 s (s - M_H^2)(s-4 M_Z^2)}
\, ,\nn \\
{C^{(I)}_{(Z,H)}}^2_1 &=& - {C^{(I)}_{(Z,H)}}^2_2 = \frac{i \, \kappa \, \alpha (M_Z^2 - M_H^2)(s-2M_Z^2)}{24 \pi s_w^2 c_w^2 s
(s - M_H^2)(s -4 M_Z^2)} \, , \nn\\
{C^{(I)}_{(Z,H)}}^2_3 &=& \frac{i \, \kappa \, \alpha}{48 \pi s_w^2 c_w^2 s (s-M_H^2)(s-4M_Z^2)^2} \bigg[
M_H^4 (4 M_Z^4 - 2 M_Z^2 s + s^2) + M_H^2 M_Z^2 s (s+ 2 M_Z^2) \nn\\
&-& M_Z^2 (16 M_Z^6 - 28 M_Z^4 s + 18 M_Z^2 s^2 - 3 s^3) \bigg] \,,\nn\\
{C^{(I)}_{(Z,H)}}^2_4 &=& - \frac{i \, \kappa \, \alpha \, M_H^2}{16 \pi s_w^2 c_w^2 s (s- M_H^2)(s-4M_Z^2)^2} \bigg[
M_H^2 (4 M_Z^4 - 2 M_Z^2 s + s^2) - M_Z^2 s (s+ 2 M_Z^2) \bigg] \,, \nn\\
{C^{(I)}_{(Z,H)}}^2_5 &=&  \frac{i \, \kappa \, \alpha }{6 \pi s_w^2 c_w^2 s (s-M_H^2)(s-4M_Z^2)^2} \bigg[
M_Z^2 s (M_H^2 - 2 M_Z^2)^2 - M_H^2 M_Z^4 (M_H^2 - 4 M_Z^2) - M_Z^4 s^2 \bigg] \,, \nn\\
{C^{(I)}_{(Z,H)}}^2_6 &=& \frac{i \, \kappa \, \alpha \, M_H^2}{16 \pi s_w^2 c_w^2 s (s-M_H^2)(s-4M_Z^2)^2} \bigg[
M_H^2 (4 M_Z^4 - 2 M_Z^2 s + s^2) (M_H^2 - 4 M_Z^2) \nn \\
&+&  2 MZ^2 s (16 M_Z^4 - 6 M_Z^2 s + s^2)  \bigg] \,,\nn\\
{C^{(I)}_{(Z,H)}}^2_7 &=& \frac{i \, \kappa \, \alpha}{ 48 \pi s_w^2 c_w^2 s (s-M_H^2)(s-4M_Z^2)^2} \bigg[
M_H^6 (4 M_Z^4 - 2 M_Z^2 s + s^2) + 2 M_H^4 M_Z^2 (s^2 - 4 M_Z^4)\nn\\
&-& 4 M_H^2 M_Z^2 (8 M_Z^6 - 10 M_Z^4 s + 6 M_Z^2 s^2 -s^3) + 4 M_Z^4 s (s-4M_Z^2)^2 \bigg]\, .
\eea
\normalsize
The one - loop graviton - Higgs mixing amplitude is given by
\small
\bea
\Sigma^{\mu\nu}_{hH}(k) &=& \Sigma^{\mu\nu}_{Min, \, hH}(k) + \Sigma^{\mu\nu}_{I, \, hH}(k) =  i \frac{\kappa}{2} \frac{e}{288
\pi^2 \, s_w \, c_w \, M_Z \, s}
\bigg\{
2 m_f^2 \bigg[ 3 (s - 4 m_f^2) \mathcal B_0\left( s, m_f^2,m_f^2 \right) \nn \\
&+& 12 \mathcal A_0\left( m_f^2 \right)+ 2 s - 12 m_f^2 \bigg] +
6 ( 6 M_W^2 + M_H^2) \big( M_W^2 \mathcal \, B_0 \left( s, M_W^2 , M_W^2 \right) + M_W^2 -  \mathcal A_0\left( M_W^2 \right)\big)
\nn \\
&+& s\big( 18 M_W^2 \, \mathcal B_0 \left( s, M_W^2 , M_W^2 \right) + 18 M_W^2 + M_H^2 \big) + 3 M_Z^2 (M_H^2 + 6 M_Z^2 - 3 s)
\mathcal B_0 \left( s, M_Z^2, M_Z^2\right) \nn \\
&+& 9 M_H^4 \mathcal B_0 \left( s, M_H^2, M_H^2\right) - 3 (M_H^2 + 6 M_Z^2) \mathcal A_0\left( M_Z^2 \right) -9 M_H^2  \mathcal
A_0\left( M_H^2 \right) + 3 (3 M_H^4 + M_H^2 M_Z^2 \nn \\
&+& 6 M_Z^4) - s (2 M_H^2 + 9 M_Z^2)
\bigg\} \left( s \, \eta^{\mu\nu} - k^{\mu}k^{\nu} \right) \nn \\
&+&  i \frac{\kappa}{2} \frac{e}{16 \pi^2 \, s_w \, c_w } \bigg\{ 2 c_w
M_W \mathcal A_0\left( M_W^2 \right) + M_Z \mathcal A_0\left( M_Z^2 \right) \bigg\} \eta^{\mu\nu}\, .\nn\\
\eea
\normalsize

\section{Appendix. Explicit results for the flavor diagonal $T f \bar f$ form factors in the SM}
\label{P5FWappendix}

We collect in this appendix the expression of the form factors generated by the exchange of a W-boson in the loop in the flavor diagonal $Tf \bar f$ vertex. They are given by
\small
\bea
F^W_1(q^2) &=&
\frac{q^2 y \left(2 x_f^2 \left(3 x_W \left(y \left(x_W-10\right)+4\right)+8 y\right)+x_f x_W \left((6 y-3) x_W-8 y\right)-8 y \, x_W^2\right)}{6 (4 y-1) x_f^2 x_W^2}   \nn \\
&+&  \frac{y \left(x_f \left(3 x_W+2\right)+x_W\right)}{3 (1-4 y) x_f x_W} \bigg[  \mathcal A_0 \left(m_W^2 \right) -  \mathcal A_0 \left(m_f^2 \right) \bigg] 
+  \frac{q^2 y}{6 (1-4 y)^2 x_f^3 x_W^3} \bigg[ 2 x_f^3 \left(x_W \left(x_W \right. \right. \nn \\
&\times& \left.\left. \left(9 (2 y-1) y x_W-78 y^2+68 y-17\right)+42  y (2 y-1)\right)-24 y^2\right) 
+ x_f^2 x_W \left(2 (23  \right. \nn \\
&-& \left. 14 y) y x_W+(4 (5-12 y) y+1) x_W^2+72 y^2\right)+2 y (26 y-5) x_f x_W^3-24 y^2 x_W^3 \bigg]  \nn \\
&\times&   \mathcal B_0 \left( q^2, m_f^2, m_f^2 \right)
+  \frac{q^2 y}{3 (1-4 y)^2 x_f^3 x_W^3}  \bigg[ x_f^2 x_W \left(16 (y-1) y x_W+(4 y (3 y-1)+1) x_W^2 \right. \nn \\
&-& \left.  36 y^2\right) 
+ 4 x_f^3  \left(2 (y-1) x_W-3 y\right) \left((6 y-1) x_W-2 y\right)+12 (1-2 y) y x_f x_W^3+12 y^2  x_W^3 \bigg]   \nn \\ 
&\times& \mathcal B_0 \left( q^2, m_W^2, m_W^2 \right) 
+  \frac{q^2 y}{(1-4 y)^2 x_f^2  x_W^2}   \bigg[ x_f x_W \left((y (30 y-13)+4) x_f-2 y (y+1)\right) \nn \\
&-&  y x_W^2 \left(x_f \left((6 y-3) x_f-8 y+5\right) +2 (y+1)\right)+4 y (y+1) x_f^2 \bigg]     \mathcal B_0 \left( m^2, m_f^2, m_W^2 \right) \nn \\
&+&  \frac{2 q^4 y^2}{(1-4 y)^2 x_f^4 x_W^4} \bigg[  2 y x_f^3 x_W \left((8 y-2) x_f+5 y\right)+y x_f^2 x_W^2 \left(x_f \left(-4 (y-1) x_f-14 y+5\right) \right. \nn \\
&-&   \left. 6 y\right) 
 -2 y x_f x_W^3 \left(x_f \left(x_f \left(4 y x_f-3 y+3\right)-2 y+2\right)+y\right)-x_W^4 \left((2 y-1) x_f-2 y\right) \nn \\
&\times& \left(x_f \left(y x_f-2 y+1\right)+y\right)-4 y^2 x_f^4 \bigg]   \mathcal C_0 \left( m_f^2, m_W^2, m_W^2 \right) 
+ \frac{q^4 y}{(1-4 y)^2 x_f^4 x_W^4}   \bigg[ 2 y^2 x_f^3 x_W  \nn \\
&\times&   \left(9 (2 y-1) x_f+10 y\right)-y x_f^2 x_W^2 \left((34 y-25) y x_f+6 (y (9 y-8)+2) x_f^2+12 y^2\right) \nn \\
&-& y \left(x_f-1\right)^2 x_W^4  \left(\left(6 y^2-4 y+1\right) x_f^2+(2 y-1) y x_f-4 y^2\right)+2 x_f x_W^3 \left(4 (y-1) y^2 x_f  \right. \nn \\
&+& \left.  (2 y-1) \left(8 y^2-4 y+1\right) x_f^3+(3-2 y (y+2)) y x_f^2-2 y^3\right)-8 y^3 x_f^4 \bigg]   \mathcal C_0( m_W^2, m_f^2, m_f^2) \,, \nn
\eea

\bea
F^W_4(q^2) &=&
\frac{4 y}{3 (1-4 y)^2 x_f^2 x_W^2} \bigg[ x_f^2 \left((19 y-6) x_W-10 y\right)+x_f x_W \left((3-7 y) x_W+5 y\right)+5 y x_W^2\bigg] \nn \\
&+&  \frac{2 (2 y-3) \left(x_f \left(x_W+2\right)+x_W\right)}{3 q^2  (1-4  y)^2 x_f x_W} \bigg[ \mathcal A_0 \left( m_f^2 \right) - \mathcal A_0 \left( m_W^2 \right) \bigg] 
+  \frac{2 y}{3 (4 y-1)^3 x_f^3 x_W^3}  \bigg[ x_f^3 \left(x_W \right.  \nn \\
&\times& \left. \left(x_W \left(((5-6 y) y+1) x_W+y (108 y-77)+20\right)+6  (13-27 y) y\right)+60 y^2\right)+x_f^2 x_W \nn \\
&\times& \left((116 y-59) y x_W+2 (y (5 y-1)-1) x_W^2-90  y^2\right)+5 (1-10 y) y x_f x_W^3+30 y^2 x_W^3\bigg]    \nn \\
&\times&  \mathcal B_0 \left( q^2, m_f^2, m_f^2 \right)  
+  \frac{2 y}{3 (4 y-1)^3 x_f^3 x_W^3}  \bigg[ (x_W^3 \left(x_f \left(x_f \left((y-1) (6 y+1) x_f+14 (2   \right.\right.\right. \nn \\
&-& \left.\left.\left. 3 y) y  - 10\right)+3 y  (22 y-13)\right)-30 y^2\right)-(4 y-1) x_f^2 x_W^2 \left((7 y+4) x_f+25 y\right)   \nn
\eea
\bea
&+& 10 y x_f^2 x_W  \left((13 y-1) x_f+9 y\right)-60 y^2 x_f^3\bigg]   \mathcal B_0 \left( q^2, m_W^2, m_W^2 \right)  
+  \frac{2 y}{(4 y-1)^3 x_f^2 x_W^2}  \bigg[ (x_W^2  \nn \\
&\times& \left(\left(8 y^2-4 y  +  3\right) x_f-8 (y-2) y-1\right)-x_f x_W \left((24 (y-1) y+7) x_f+8 (y-2) y  \right.  \nn \\
&+&  \left. 1\right)+2 (8 (y-2) y   +  1) x_f^2\bigg]    \mathcal B_0 \left( m^2, m_f^2, m_W^2 \right)  
+   \frac{2 q^2 y}{(4 y-1)^3 x_f^4 x_W^4} \bigg[ -50 y^3 \left(x_f+1\right) \nn \\
&\times&   x_f^3 x_W+2 y x_f^2 x_W^2 \left(30 y^2 x_f   +  (y (11 y+5)-1) x_f^2+15 y^2\right)+y x_f x_W^3 \left(x_f \left(x_f \left((2 y (5 y \right.\right.\right.  \nn \\
&-&  \left.\left.\left.  6)+3) x_f+22 y^2-26 y+7\right)  +   6 (3-7 y) y\right)+10 y^2\right)-x_W^4 \left(4 (2-3 y) y x_f+(2 (y  \right. \nn \\ 
&-&  \left. 1) y+1) x_f^2+10 y^2\right) \left(x_f \left(y x_f-2 y   +  1\right)+y\right)+20 y^3  x_f^4\bigg]    \mathcal C_0 \left( m_f^2, m_W^2, m_W^2 \right)  \nn \\
&+&  \frac{2 q^2 y}{(4 y-1)^3 x_f^4 x_W^4}   \bigg[-2 y^2 x_f^3 x_W \left((37 y-18) x_f  
+  25 y\right)+6 y x_f^2 x_W^2  \left((16 y-9) y x_f   \right.  \nn \\
&+& \left.   3 (y (5 y-4)+1) x_f^2+5 y^2\right)+y \left(x_f-1\right)^2 x_W^4  \left((2 (y-1) y+1) x_f^2-10 y^2\right) \nn \\
&+&  x_f x_W^3 \left(6 (3-7 y) y^2 x_f+\left(-26 y^2+28  y-11\right) y x_f^2+\left(y \left(-38 y^2+34 y-11\right)+2\right) x_f^3 \right.  \nn \\
&+& \left.   10 y^3\right)+20  y^3 x_f^4\bigg]    \mathcal C_0 \left( m_W^2, m_f^2, m_f^2 \right)  \,, \nn \\
F^W_5(q^2) &=&
\frac{y}{3 (4 y-1) x_f^2 x_W^2}   \bigg[ x_f^2 \left(x_W+2\right) \left(3 (4 y-1) x_W-4 y\right)+x_f x_W \left((9-32 y) x_W+4 y\right) \nn \\
&+&   4 y x_W^2\bigg] 
+  \frac{(8 y-3) \left(x_f \left(x_W+2\right)+x_W\right)}{3 q^2 (1-4 y) x_f x_W} \bigg[  \mathcal A_0 \left( m_W^2 \right)  -  \mathcal A_0 \left( m_f^2 \right) \bigg] \nn \\
&+&  \frac{y}{3 (1-4 y)^2 x_f^3 x_W^3}    \bigg[x_f^3 \left(x_W+2\right) \left(12 y^2 \left(x_W-1\right)^2-4 y  \left(x_W-3\right) x_W+x_W^2\right)-x_f^2 x_W \nn \\
&\times&   \left(8 (2 y+1) y x_W+(4 y (11 y-7)+5)  x_W^2+36 y^2\right)+4 (2-11 y) y x_f x_W^3+12 y^2 x_W^3\bigg]   \nn \\
&\times&  \mathcal B_0( q^2, m_f^2, m_f^2) 
+  \frac{2 y}{3 (1-4 y)^2 x_f^3 x_W^3}   \bigg[  x_W^3 \left(x_f \left(x_f \left((y (6 y+5)-2) x_f+2 (10-9 y) y \right.\right.\right.  \nn \\
&-& \left.\left.\left.   5\right)+9 y  (2 y-1)\right)-6 y^2\right)+(4 y-1) x_f^2 x_W^2 \left((7 y-8) x_f+y\right)+2 y x_f^2 x_W  \left((13 y \right. \nn \\
&-& \left.  1) x_f+9 y\right)-12 y^2 x_f^3 \bigg]    \mathcal B_0 \left( q^2, m_W^2, m_W^2 \right)  
+  \frac{y}{(1-4 y)^2 x_f^2 x_W^2}    \bigg[ x_f^2 \left(x_W \left((1-2 y (4 y  \right.\right.  \nn \\
&+& \left.\left.   1)) x_W+2 (9-4 y) y-5\right)+4 y (4  y-5)+2\right)+x_f x_W \left(4 (1-2 y)^2 x_W+2 (5-4 y) y  \right.  \nn \\
&-&  \left. 1\right)+(2 (5-4 y) y-1) x_W^2\bigg]     \mathcal B_0( m^2, m_f^2, m_W^2)  
-  \frac{2 q^2 y}{(1-4 y)^2 x_f^4  x_W^4}   \bigg[ 10 y^3 \left(x_f+1\right) x_f^3 x_W  \nn \\
&-&    2 y x_f^2 x_W^2 \left((2 y+1) y   x_f+(y (3 y+4)-1) x_f^2+3 y^2\right)-y \left(x_f+1\right) x_f x_W^3 \left(2 (1 \right.    \nn \\
&-&  \left.     2 y) y  x_f+(2 (y-6) y+3) x_f^2+2 y^2\right)+x_W^4 \left(2 (1-2 y) y x_f+(2 (y-2) y+1) x_f^2   \right. \nn \\
&+&\left.    2  y^2\right) \left(x_f \left(y x_f-2 y+1\right)+y\right)-4 y^3 x_f^4\bigg]    \mathcal C_0( m_f^2, m_W^2, m_W^2) 
- \frac{2 q^2 y^2}{(1-4 y)^2  x_f^4 x_W^4}  \bigg[ 2 x_f x_W  \nn \\
&\times&   \left((y-1) x_f+y\right)+\left(x_f-1\right) x_W^2 \left((2  y-1) x_f-2 y\right)-4 y x_f^2\bigg] \bigg[ x_f^2 \left(y  \left(x_W-1\right)^2+x_W\right)   \nn \\
&-&   2 y x_f x_W \left(x_W+1\right)+y x_W^2\bigg]   \mathcal C_0 \left( m_W^2, m_f^2, m_f^2 \right) 
+  \chi \bigg\{   \frac{8 y \left(x_f+1\right)}{(1-4 y) x_f} \bigg[  \mathcal B_0 \left( q^2, m_W^2, m_W^2 \right)  \nn \\
&-&  \mathcal B_0 \left( m^2, m_f^2, m_W^2 \right)     \bigg] 
+  \frac{8 q^2 y}{(4 y-1) x_f^2 x_W}   \bigg[ y x_f \left(x_f+1\right)-x_W \left(x_f \left(y x_f-2   y+1\right)+y\right)\bigg]  \nn \\
&\times&  \mathcal C_0 \left( m_f^2, m_W^2, m_W^2 \right)  \bigg\} \,, \nn
\eea

\bea
F^W_6(q^2) &=&
\frac{y}{6 (1-4 y) x_f^2 x_W^2}  \bigg[-2 x_f x_W \left((4 y-9) x_f+4 y\right)+\left(x_f-1\right) x_W^2 \left(3 x_f+8
   y\right)+16 y x_f^2\bigg] \nn \\
&+&  \frac{(8 y-3) \left(x_f \left(x_W-2\right)-x_W\right)}{6 q^2 (1 - 4 y) x_f x_W} \bigg[ \mathcal A_0 \left( m_W^2 \right)  -  \mathcal A_0 \left( m_f^2 \right) \bigg] 
-   \frac{y}{6 (1-4 y)^2 x_f^3 x_W^3}    \bigg[\left(x_f-1\right) \nn \\
&\times&   x_W^3 \left(2 (5-8 y) y x_f+(2 y (12 y-7)-1) x_f^2+24  y^2\right)+2 x_f^2 x_W^2 \left(((53-48 y) y-17) x_f \right. \nn \\
&+& \left.  y (16 y+23)\right)+12 y x_f^2 x_W \left((10 y-7) x_f+6 y\right)-48 y^2 x_f^3\bigg]    \mathcal B_0 \left( q^2, m_f^2, m_f^2 \right)  \nn \\
&-&  \frac{y}{3 (1-4 y)^2 x_f^3 x_W^3}    \bigg[ -\left(x_f-1\right) x_W^3 \left(12 (1-2 y) y x_f+(4 y (3 y-1)+1) x_f^2+12 y^2\right) \nn \\
&+&   8 x_f^2 x_W^2 \left((4 (y-2) y+1) x_f-2 y (2 y+1)\right)-4 y x_f^2 x_W \left(7  (y-1) x_f+9 y\right)+24 y^2 x_f^3\bigg]    \nn \\
&\times&   \mathcal B_0 \left( q^2, m_W^2, m_W^2 \right) 
-  \frac{y}{2 (1-4 y)^2 x_f^2 x_W^2}   \bigg[ x_f^2 \left(x_W \left(2 y \left(x_W+7\right)+x_W+5\right)-4 y+2\right) \nn \\
&+&  x_f x_W  \left(-4 y x_W+2 y-1\right)+(2 y-1) x_W^2\bigg]     \mathcal B_0 \left( m^2, m_f^2, m_W^2 \right)  
-  \frac{2 q^2 y^2}{(1-4 y)^2 x_f^4 x_W^4}    \bigg[ 2 y x_f^3  \nn \\
&\times&    x_W \left((3 y-2) x_f+5 y\right)+y x_f^2 x_W^2 \left(x_f   \left((2 y+7) x_f+4 y+5\right)-6 y\right)-2 y x_f x_W^3 \left(x_f \left(x_f  \right.\right. \nn \\
&\times& \left.\left.   \left((3 y+1)  x_f-5 y+5\right)+y+2\right)+y\right)+\left(x_f-1\right) x_W^4 \left((2 y-1) x_f-2 y\right)  \left(x_f \left(y x_f   \right.\right.  \nn \\
&-& \left.\left.   2 y+1\right)+y\right)-4 y^2 x_f^4\bigg]   \mathcal C_0 \left( m_f^2, m_W^2, m_W^2 \right) 
-   \frac{q^2 y}{(1-4 y)^2 x_f^4 x_W^4}    \bigg[ 2 y^2 x_f^3 x_W \left((14 y-9) x_f \right. \nn \\
&+&  \left.  10 y\right)-y x_f^2 x_W^2 \left((16  y-25) y x_f+3 (y (12 y-13)+4) x_f^2+12 y^2\right)-y \left(x_f-1\right)^2 x_W^4 \nn \\
&\times&  \left(-y  x_f+(y (4 y-3)+1) x_f^2-4 y^2\right)+2 x_f x_W^3 \left(x_f \left(x_f \left((y (2 y (5  y-6)+5)-1) x_f \right.\right.\right. \nn \\
&-& \left.\left.\left.   6 y^3+3 y\right)-2 y^2 (y+2)\right)-2 y^3\right)-8 y^3 x_f^4\bigg]     \mathcal C_0 \left( m_W^2, m_f^2, m_f^2 \right) \,.
\eea
\normalsize

\section{Appendix. Explicit results for the flavor off-diagonal $T f \bar f$ form factors in the SM}
\label{P6formfactors}
We list the coefficients which appear in the expansion of the form factors $F_1$, $F_3$, $F_4$, $F_7$, $F_8$, $F_{11}$ for the off-diagonal, in flavor space, $T f \bar f$. The remaining ones, as already mentioned, can be computed using the Ward identities in Eq. (\ref{P6WIFF}). 
\small
\bea
C_1^0 &=& 
\frac{q^2}{12 \lambda} \bigg\{ 3 \left(x_D^2 - x_S^2 \right)+6(1-x_S) \left(x_f-6 x_W\right)+16
   \left(x_f-x_W\right) \left(x_f+2 x_W\right) \bigg\} \,, \nn \\
C_1^1 &=& -\frac{2 x_f-3 x_S+4 x_W+3}{6 \lambda} \,, \nn \\
C_1^2 &=& \frac{q^2}{8 \lambda ^2 \, x_D} 
\bigg\{x_D^4 \left(6 x_f+x_S-16 x_W-2\right)+x_D^3 \left(x_S \left(-10 x_f+18  x_W+3\right)+12 \left(x_f x_W+x_f^2  \right. \right. \nn \\
&+& \left. \left. x_f-2 x_W^2\right)-32 x_W-3\right)-x_D^2 \left(x_S^2 \left(20  x_f-46 x_W-6\right)+x_S \left(-28 x_f x_W-4 x_f \left(7 x_f+4\right) \right. \right. \nn \\
&+& \left. \left.  56 x_W^2+54 x_W+4\right)+24 \left(x_f-x_W\right) \left(x_f+2 x_W\right)+6 x_f+x_S^3-26 x_W-1\right) \nn \\
&+&x_D \left(x_S^2 \left(8 x_f+6 x_W+3\right)-2 x_S \left(2 x_f x_W+x_f \left(2 x_f+5\right)-4 x_W^2+x_W\right)-8 \left(x_f-x_W\right) \right. \nn \\
&\times& \left. \left(x_f+2 x_W\right)-3 x_S^3+10 x_W\right)-\left(1-2 x_S\right)^2  \left(2 x_W \left(2 x_f+x_S\right)+\left(x_S-2 x_f\right)^2-8 x_W^2\right)\bigg\} \,, \nn 
\eea
\bea
C_1^3 &=& \frac{q^2}{8 \lambda ^2 x_D} 
\bigg\{ x_D^4 \left(-\left(6 x_f+x_S-16 x_W-2\right)\right)+x_D^3 \left(x_S \left(-10 x_f+18 x_W+3\right)+12 \left(x_f x_W+x_f^2 \right.\right. \nn \\
&+& \left.\left. x_f-2 x_W^2\right)-32 x_W-3\right)+x_D^2 \left(x_S^2 \left(20 x_f-46 x_W-6\right)+x_S \left(-28 x_f x_W-4 x_f \left(7 x_f+4\right) \right.\right. \nn \\
&+& \left.\left. 56 x_W^2+54 x_W+4\right)+24  \left(x_f-x_W\right) \left(x_f+2 x_W\right)+6 x_f+x_S^3-26 x_W-1\right) \nn \\
&+& x_D \left(x_S^2 \left(8 x_f+6 x_W+3\right)-2 x_S \left(2 x_f x_W+x_f \left(2 x_f+5\right)-4 x_W^2+x_W\right)-8 \left(x_f-x_W\right) \right. \nn \\ 
&\times& \left. \left(x_f+2 x_W\right)-3 x_S^3+10 x_W\right)+\left(1-2 x_S\right)^2 \left(2 x_W \left(2 x_f+x_S\right)+\left(x_S-2 x_f\right)^2-8 x_W^2\right)\bigg\} \,, \nn \\
C_1^4 &=& \frac{q^2}{12 \lambda ^2} 
\bigg\{x_D^2 \left(3 \left(8 x_f-3\right) x_S-2 x_f \left(16 x_f+32 x_W+11\right)+10 x_W+9\right)-3 x_S^2 \left(8 x_f+26 x_W+3\right) \nn \\ 
&+& 4 x_S \left(-7 x_f x_W+x_f \left(13 x_f+5\right)+42 x_W^2+34 x_W\right)+24 \left(6 x_f-7\right) x_W^2+92 x_f x_W \nn \\
&-& 2 x_f \left(2 x_f+1\right) \left(12 x_f-1\right)+9 x_S^3-96 x_W^3-68 x_W\bigg\} \,, \nn \\
C_1^5 &=& \frac{q^2}{3 \lambda ^2}
\bigg\{ x_f \left(8 x_W \left(x_D^2+x_S-2\right)-2 x_D^2+3 x_S^2-2 x_S-36 x_W^2+1\right) \nn \\
&+& 4 x_W \left(x_D^2 \left(-3 x_S+4 x_W+5\right)+\left(x_S-3 x_W-2\right) \left(3 x_S-2
   x_W-1\right)\right)-12 x_f^2 \left(x_S-1\right)+12 x_f^3\bigg\} \,, \nn \\
C_1^6 &=& \frac{q^4}{2 \lambda ^2} 
\bigg\{ 8 x_W^3 \left(-2 x_D^2+5 x_f+4 x_S-2\right)-4 x_W^2 \left(x_D^2 \left(-2 x_f-2 x_S+3\right)+\left(x_f+x_S\right) \left(6 x_f+x_S\right) \right. \nn \\
&-& \left. 5 x_f-2 x_S\right)-2 x_W \left(-4 x_f^2 \left(x_D^2+x_S-2\right)+3 x_f \left(x_D^2 \left(2 x_S-3\right)-\left(x_S-2\right) x_S\right)+2 \left(x_D^2-x_S^2 \right) \right. \nn \\
&\times& \left. \left(x_D^2-x_S\right) +4 x_f^3\right)+x_f \left(2
   x_f-x_S+1\right) \left(-x_D^2+\left(x_S-2 x_f\right){}^2+4 x_f\right)-16 x_W^4 \bigg\} \,, \nn \\
C_1^7 &=& \frac{q^4}{8 \lambda ^2} 
\bigg\{ 2 x_D^2 \left(x_S \left(x_f \left(-4 x_f+32 x_W+7\right)-8 x_W-2\right)+\left(2-6 x_f\right) x_S^2+2 \left(3-16 x_f\right) x_W^2 \right. \nn \\
&+& \left. 4 x_f \left(4 x_f-9\right) x_W+x_f \left(2 x_f-1\right) \left(8 x_f+5\right)+8 x_W+1\right)+x_D^4 \left(4 x_f-1\right)+16 x_W^3 \left(10 x_f+9 x_S \right. \nn \\
&-& \left.  9\right)-4 x_W^2 \left(34 x_f x_S+24 x_f^2-50 x_f+3 x_S \left(9 x_S-16\right)+24\right)+8 x_W \left(-\left(x_f+8\right) x_S^2+\left(4 \left(x_f \right. \right. \right. \nn \\
&-& \left.\left.\left. 1\right) x_f+6\right) x_S+6 x_f-4 x_f^2 \left(x_f+2\right)+4 x_S^3-2\right)-\left(x_S-2 x_f\right)^2 \left(2 x_f \left(x_S-1\right)-8 x_f^2+3 x_S^2 \right. \nn \\
&-& \left. 4 x_S+2\right)-64 x_W^4\bigg\} \,, \\
C_{3}^0 &=& \frac{q^2}{6 \lambda } \bigg\{ -x_S \left(3 x_D^2+32 x_f-20 x_W+3\right)+2 x_f \left(7 x_D^2+4 x_W+9\right)-4 x_W \left(2 x_D^2+4 x_W+3\right) \nn \\
&+& 8 x_f^2+6 x_S^2\bigg\} \,, \nn \\
C_{3}^1 &=& \frac{1}{3 \lambda  \left(x_S^2-x_D^2\right)} \bigg\{ -x_D^2 \left(x_S \left(6 x_f+x_S+12 x_W+4\right)-4 x_f-8 x_W-3\right)+x_D^4 \nn \\
&+& x_S \left(4 x_S-3\right) \left(2 x_f+x_S+4 x_W\right) \bigg\} \,, \nn \\
C_{3}^2 &=& \frac{q^2}{4 \lambda ^2 \left(x_D+x_S\right)} 
\bigg\{ x_D^5 \left(-2 x_f+x_S+2 x_W+2\right)+x_D^4 \left(4 x_W \left(2 x_f+x_S+1\right)-2 x_f x_S+8 x_f^2 \right. \nn \\
&-& \left. 2 x_f-x_S^2+x_S-16 x_W^2\right)+x_D^3 \left(2 x_W \left(2 x_f \left(x_S+3\right)+x_S^2+x_S+2\right)+4 x_f^2 x_S+2 x_f x_S+12 x_f^2 \right. \nn \\
&-& \left. 2 x_f-8 \left(x_S+3\right) x_W^2-x_S^3-2 x_S^2-5 x_S+2\right)+x_D^2 \left(2 x_S^2 \left(-2 x_f+x_W-4\right)+x_S \left(2 x_f \left(2 x_W \right. \right. \right. \nn \\
&+& \left.\left. \left. 9\right)+4 x_f^2-2 x_W \left(4 x_W+9\right)+5\right)-4 x_f x_W-2 x_f \left(2 x_f+5\right)+x_S^3+8 x_W^2+12 x_W-1\right) \nn \\
&+& x_D \left(x_S^3 \left(-8 x_f+4 x_W-1\right)+2 x_S^2 \left(x_f \left(4 x_W+6\right)+4 x_f^2-x_W \left(8 x_W+9\right)+2\right) \right. \nn \\
&-& \left. 2 x_S \left(x_f \left(10 x_W-3\right)+10 x_f^2+\left(3-20 x_W\right) x_W+1\right)+2 \left(5-2 x_f\right) x_W-4 x_f \left(x_f+2\right)+2 x_S^4 \right. \nn \\
&+& \left. 8 x_W^2\right)+x_D^6+x_S^3 \left(-8 x_f+4 x_W+1\right)-2 x_S \left(2 x_f \left(5 x_W+2\right)+10 x_f^2-5 x_W \left(4 x_W+1\right)\right) \nn \\
&+& x_S^2 \left(8 x_f \left(x_W+2\right)+8 x_f^2-2 x_W \left(8 x_W+9\right)-1\right)+4 \left(x_f-x_W\right) \left(x_f+2 x_W\right)+2 x_S^4\bigg\} \nn \\
&+& \chi \frac{2 q^2}{\lambda}  \bigg\{ 2 \left(x_D+1\right) x_f-\left(x_D-1\right) \left(x_D+x_S\right)\bigg\} \,, \nn 
\eea
\bea
C_{3}^3 &=& -\frac{q^2}{4 \lambda ^2 \left(x_D-x_S\right)} 
\bigg\{ -x_D^5 \left(-2 x_f+x_S+2 x_W+2\right)+x_D^4 \left(4 x_W \left(2 x_f+x_S+1\right)-2 x_f x_S+8 x_f^2 \right. \nn \\ 
&-& \left. 2 x_f-x_S^2+x_S-16 x_W^2\right)+x_D^3 \left(-2 x_W \left(2 x_f \left(x_S+3\right)+x_S^2+x_S+2\right)-4 x_f^2 x_S-2 x_f x_S-12 x_f^2 \right. \nn \\
&+& \left. 2 x_f+8 \left(x_S+3\right) x_W^2+x_S^3+2 x_S^2+5 x_S-2\right)+x_D^2 \left(2 x_S^2 \left(-2 x_f+x_W-4\right)+x_S \left(2 x_f \left(2 x_W \right.\right.\right. \nn \\ 
&+& \left.\left.\left. 9\right)+4 x_f^2-2 x_W \left(4 x_W+9\right)+5\right)-2 x_f \left(2 x_W+5\right)-4 x_f^2+x_S^3+4 x_W \left(2 x_W+3\right)-1\right) \nn \\
&+& x_D \left(x_S^3 \left(8 x_f-4 x_W+1\right)-2 x_S^2 \left(x_f \left(4 x_W+6\right)+4 x_f^2-x_W \left(8 x_W+9\right)+2\right) \right. \nn \\
&+& \left. x_S \left(x_f \left(20 x_W-6\right)+20 x_f^2-40 x_W^2+6 x_W+2\right)+4 x_f \left(x_W+2\right)+4 x_f^2-2 x_S^4-2 x_W \left(4 x_W \right. \right. \nn \\
&+& \left.\left. 5\right)\right)+x_D^6+x_S^3 \left(-8 x_f+4 x_W+1\right)-2 x_S \left(2 x_f \left(5 x_W+2\right)+10 x_f^2-5 x_W \left(4 x_W+1\right)\right) \nn \\
&+& x_S^2 \left(8 x_f \left(x_W+2\right)+8 x_f^2-2 x_W \left(8 x_W+9\right)-1\right)+4 \left(x_f-x_W\right) \left(x_f+2 x_W\right)+2 x_S^4\bigg\} \nn \\
&-&\chi \frac{2 q^2}{\lambda} \bigg\{ 2 \left(x_D-1\right) x_f+\left(x_D+1\right) \left(x_D-x_S\right)\bigg\} \,, \nn \\
C_{3}^4 &=& \frac{q^2}{6 \lambda ^2} 
\bigg\{ x_D^2 \left(-12 x_S \left(2 x_f+x_W\right)+4 x_f \left(3-8 x_W\right)-28 x_f^2+3 x_S^2+4 x_W \left(3 x_W+5\right)+1\right) \nn \\
&+& x_D^4 \left(8 x_f-2\right)+x_S \left(4 x_f \left(4 x_W-7\right)+44 x_f^2+4 x_W \left(9 x_W-1\right)-1\right)+\left(22 x_f+2\right) x_S^2 \nn \\
&-& 8 \left(-x_f x_W \left(9 x_W+2\right)+3 x_f^3+2 x_f^2+6 x_W^2 \left(x_W+1\right)\right)+10 x_f-3 x_S^3-4 x_W\bigg\} \,, \nn \\
C_{3}^5 &=& -\frac{q^2}{6 \lambda ^2} 
\bigg\{ -x_D^2 \left(4 \left(x_f-10\right) x_W+2 x_f+15 x_S \left(2 x_W+1\right)+44 x_W^2-7\right)+x_D^4 \left(8 x_W+4\right) \nn \\
&+& x_S^2 \left(-18 x_f+28 x_W+5\right)+x_S \left(8 x_f \left(x_W+5\right)+36 x_f^2+26 x_W \left(2 x_W-3\right)-4\right) \nn \\
&-& 4 \left(x_f \left(-18 x_W^2+x_W+5\right)+6 x_f^3+9 x_f^2+2 x_W \left(6 x_W^2+x_W-4\right)\right)+3 x_S^3\bigg\} \nn \\
&+& \chi \frac{4 q^2}{\lambda } \left(x_D^2-2 x_f-x_S\right) \,, \nn \\
C_{3}^6 &=& \frac{q^4}{4 \lambda ^2} 
\bigg\{ 40 x_W^3 \left(-x_D^2+2 x_f+x_S\right)-4 x_W^2 \left(-x_D^2 \left(8 x_f+4 x_S+5\right)+2 x_D^4+4 \left(x_f+2\right) x_S \right. \nn \\
&+& \left. 4 \left(3 x_f^2+x_f-1\right)+3 x_S^2\right)+2 x_W \left(x_D^2-2 x_f-x_S\right) \left(5 x_D^2-4 \left(x_f+3\right) x_S+4 x_f \left(x_f+1\right)+x_S^2 \right. \nn \\
&+& \left. 6\right)+4 x_D^2 x_S-x_D^4-2 x_D^2-8 x_f x_S^3+24 x_f^2 x_S^2+24 x_f x_S^2-32 x_f^3 x_S-48 x_f^2 x_S-24 x_f x_S \nn \\ 
&+& 16 x_f^4+32 x_f^3+24 x_f^2+8 x_f+x_S^4-4 x_S^3+2 x_S^2-32 x_W^4\bigg\} \nn \\
&-& \chi \frac{2 q^4}{\lambda } \bigg\{ x_D^2 \left(2 x_W-1\right)-2 x_W \left(2 x_f+x_S\right)+\left(x_S-2 x_f\right)^2+4 x_f \bigg\} \,, \nn \\
C_{3}^7 &=& -\frac{q^4}{4 \lambda ^2} \bigg\{ -2 x_W \left(x_D^2+2 x_f+x_S-2\right)+\left(-2 x_f+x_S-1\right) \left(x_D^2+2 x_f-x_S\right)+8 x_W^2\bigg\} \nn \\
&\times& \bigg\{ x_D^2 \left(4 x_f-1\right)-4 x_S \left(x_f+x_W\right)+4 \left(\left(x_f-x_W\right)^2+x_W\right)+x_S^2\bigg\} \,, \\
C_{4}^0 &=& \frac{2}{3 \lambda ^2} \bigg\{ x_f \left(x_D^2-7 x_S+10 x_W+6\right)-x_W \left(7 x_D^2-19 x_S+20 x_W+12\right)+10 x_f^2\bigg\} \,, \nn \\
C_{4}^1 &=& \frac{2}{3 q^2 \lambda ^2 \left(x_D^2-x_S^2\right)}   
\bigg\{ x_D^2 \left(6 x_f x_S-10 x_f+12 x_S x_W-2 x_S^2+x_S-20 x_W-3\right)+2 x_D^4 \nn \\
&-& \left(x_S-3\right) x_S \left(2 x_f+x_S+4 x_W\right)\bigg\} \,, \nn
\eea
\bea
C_{4}^2 &=& \frac{1}{\lambda ^3 \left(x_D+x_S\right)}  
\bigg\{ x_D^4 \left(x_f \left(x_S+4 x_W+2\right)+4 x_f^2+x_W \left(-9 x_S-8 x_W+4\right)\right)+x_D^3 \left(x_f \left(2 \left(x_S \right. \right. \right. \nn \\
&+& \left. \left. \left. 6\right) x_W+x_S^2-2 x_S+7\right)+2 x_f^2 \left(x_S+6\right)-x_W \left(4 \left(x_S+6\right) x_W+7 x_S^2-10 x_S+5\right)\right) \nn \\
&+& 2 x_D^2 \left(x_f \left(2 x_S \left(-4 x_S+5 x_W+6\right)-7 x_W-4\right)+x_f^2 \left(10 x_S-7\right)+x_W \left(5 x_S \left(3 x_S-4 x_W-4\right) \right. \right. \nn \\
&+& \left.\left. 14 x_W+7\right)\right)+x_D \left(-x_f \left(x_S \left(x_S \left(12 x_S-16 x_W-15\right)+22 x_W+6\right)+8 x_W+3\right) \right. \nn \\
&+& \left. 2 x_f^2 \left(x_S \left(8 x_S-11\right)-4\right)+x_W \left(x_S \left(x_S \left(24 x_S-32 x_W-29\right)+44 x_W+2\right)+16 x_W+7\right)\right) \nn \\
&-& 2 x_D^5 x_W+x_W \left(2 x_f \left(2 \left(x_S-4\right) x_S+1\right)+x_S \left(6 \left(x_S-2\right) x_S+7\right)\right)+x_f \left(x_S \left(4 x_f \left(x_S-4\right) \right. \right. \nn \\
&-& \left.\left. 2 x_S^2+2 x_S-3\right)+2 x_f\right)-4 \left(2 \left(x_S-4\right) x_S+1\right) x_W^2 \bigg\} \,, \nn \\
C_{4}^3 &=& \frac{1}{\lambda ^3 \left(x_D-x_S\right)} 
\bigg\{-x_D^4 \left(x_f \left(x_S+4 x_W+2\right)+4 x_f^2+x_W \left(-9 x_S-8 x_W+4\right)\right)+x_D^3 \left(x_f \left(2 \left(x_S \right. \right. \right. \nn \\
&+&  \left. \left. \left.  6\right) x_W+x_S^2-2 x_S+7\right)+2 x_f^2 \left(x_S+6\right)-x_W \left(4 \left(x_S+6\right) x_W+7 x_S^2-10 x_S+5\right)\right) \nn \\
&+& 2 x_D^2 \left(x_f \left(2 x_S \left(4 x_S-5 x_W-6\right)+7 x_W+4\right)+x_f^2 \left(7-10 x_S\right)+x_W \left(5 x_S \left(-3 x_S+4 x_W+4\right) \right. \right. \nn \\
&-& \left.\left. 7 \left(2 x_W+1\right)\right)\right)+x_D \left(-x_f \left(x_S \left(x_S \left(12 x_S-16 x_W-15\right)+22 x_W+6\right)+8 x_W+3\right) \right. \nn \\
&+& \left. 2 x_f^2 \left(x_S \left(8 x_S-11\right)-4\right)+x_W \left(x_S \left(x_S \left(24 x_S-32 x_W-29\right)+44 x_W+2\right)+16 x_W+7\right)\right) \nn \\
&-& 2 x_D^5 x_W-x_W \left(2 x_f \left(2 \left(x_S-4\right) x_S+1\right)+x_S \left(6 \left(x_S-2\right) x_S+7\right)\right)+x_f \left(x_S \left(2 x_S \left(-2 x_f \right. \right. \right. \nn \\
&+& \left.\left.\left. x_S-1\right)+16 x_f+3\right)-2 x_f\right)+4 \left(2 \left(x_S-4\right) x_S+1\right) x_W^2 \bigg\} \,, \nn \\
C_{4}^4 &=& \frac{1}{6 \lambda ^3} \bigg\{ x_D^2 \left(x_S \left(36 x_f-6 x_W-9\right)-2 \left(2 x_f x_W+5 x_f \left(8 x_f+3\right)+6 x_W^2\right)+6 x_S^2+40 x_W+11\right) \nn \\
&-& 4 x_D^4 \left(2 x_f+1\right)-x_S^2 \left(10 x_f+108 x_W+5\right)+2 x_S \left(2 \left(-58 x_f x_W+25 x_f^2+x_f+81 x_W^2\right) + 77 x_W\right. \nn \\
&-& \left. 1\right)-4 \left(6 \left(13-15 x_f\right) x_W^2-59 x_f x_W+x_f \left(5 x_f \left(6 x_f+1\right)-2\right)+60 x_W^3+20 x_W\right)+3 x_S^3 \bigg\} \,, \nn \\
C_{4}^5 &=& \frac{1}{6 \lambda ^3} 
\bigg\{ -x_D^2 \left(-3 x_S \left(8 x_f-26 x_W+3\right)+50 x_f \left(2 x_W+1\right)+24 x_f^2+6 x_S^2-4 x_W \left(55 x_W+4\right) \right. \nn \\
&+& \left. 11\right)+4 x_D^4 \left(8 x_W+1\right)-20 x_W^2 \left(18 x_f+13 x_S-2\right)+2 \left(2 x_S-1\right) x_W \left(50 x_f+7 x_S+8\right) \nn \\
&+& 42 x_f x_S^2-132 x_f^2 x_S-56 x_f x_S+120 x_f^3+156 x_f^2+40 x_f-3 x_S^3+5 x_S^2+2 x_S+240 x_W^3 \bigg\} \,, \nn  \\
C_{4}^6 &=& \frac{q^2}{4 \lambda ^3} 
\bigg\{ x_D^4 \left(2 x_W \left(8 x_f-4 x_S+5\right)+4 x_f-2 x_S-56 x_W^2+3\right)+2 x_D^2 \left(x_S^2 \left(-6 x_f+13 x_W-2\right) \right. \nn \\
&+& \left. x_S \left(-5 \left(4 x_f+5\right) x_W+12 x_f^2+16 x_f+52 x_W^2+1\right)+2 \left(5 \left(12 x_f+1\right) x_W^2+\left(5-6 x_f\right) x_f x_W \right. \right. \nn \\
&-& \left.\left. x_f \left(4 x_f \left(x_f+3\right)+7\right)-50 x_W^3\right)+x_S^3+10 x_W-1\right)-2 x_S^3 \left(8 x_f+5 x_W+1\right)+x_S^2 \left(x_f \left(28 \right.\right. \nn \\
&-& \left.\left. 44 x_W\right)+72 x_f^2+4 \left(6-11 x_W\right) x_W+2\right)-4 x_S \left(6 x_f^2 \left(5-7 x_W\right)+x_f \left(60 x_W^2-26 x_W+4\right) \right. \nn \\
&+& \left. 32 x_f^3+x_W \left(10 \left(1-5 x_W\right) x_W+3\right)\right)+8 \left(50 x_f x_W^3+2 \left(1-15 x_f^2\right) x_W^2-x_f \left(2 x_f \left(5 x_f+9\right) \right. \right. \nn \\
&+& \left.\left. 7\right) x_W+x_f \left(x_f+1\right) \left(2 x_f \left(5 x_f+4\right)+1\right)-20 x_W^4\right)+x_S^4\bigg\} \,, \nn \\
C_{4}^7 &=& \frac{q^2}{4 \lambda ^3} 
\bigg\{x_D^4 \left(4 x_f \left(4 x_f-2 x_W+1\right)+2 x_S-2 x_W-3\right)-2 x_D^2 \left(x_S^2 \left(6 x_f-3 x_W-2\right) \right. \nn \\
&+& \left. x_S \left(10 x_f \left(2 x_f-4 x_W-1\right)+17 x_W+1\right)+6 \left(4 x_f-3\right) x_W^2+2 x_f \left(6 x_f+25\right) x_W-4 x_f \left(10 x_f^2 \right. \right. \nn \\
&+& \left.\left. x_f-2\right)+x_S^3+4 x_W^3-16 x_W-1\right)+8 x_W^3 \left(50 x_f+37 x_S-36\right)-12 x_W^2 \left(8 \left(4 x_f-3\right) x_S \right. \nn \\
&+& \left. 4 \left(x_f \left(5 x_f-9\right)+3\right)+15 x_S^2\right)+2 x_W \left(\left(26 x_f-34\right) x_S^2+\left(28 x_f \left(3 x_f-2\right)+22\right) x_S+44 x_f \right. \nn \\
&-& \left. 8 x_f^2 \left(5 x_f+9\right)+19 x_S^3-8\right)-\left(x_S-2 x_f\right)^2 \left(-20 x_f^2+\left(x_S-2\right) x_S+2\right)-160 x_W^4\bigg\} \,, \\
C_{7}^0 &=& \frac{q^2 x_D}{6 \lambda } \bigg\{ 3 x_D^2+x_S \left(4 x_f-4 x_W-6\right)-8 \left(x_f-x_W\right) \left(x_f+2 x_W\right)-4 x_f+4 x_W+3\bigg\} \,, \nn 
\eea
\bea
C_{7}^1 &=& \frac{2 x_D }{3 \lambda  \left(x_D^2-x_S^2\right)} 
\bigg\{ x_D^2 \left(-\left(x_f+x_S+2 x_W-1\right)\right)+x_S \left(x_S \left(-2 x_f+x_S-4 x_W-1\right) \right. \nn \\
&+& \left. 6 \left(x_f+2 x_W\right)\right)-3 \left(x_f+2 x_W\right)\bigg\} \,, \nn \\
C_{7}^2 &=& \frac{q^2}{4 \lambda ^2 \left(x_D+x_S\right)} 
\bigg\{ -x_D^4 \left(x_f \left(-2 x_S+4 x_W+2\right)+4 x_f^2+2 x_W \left(6 x_S-4 x_W+1\right)-1\right) \nn \\
&+& x_D^3 \left(3 x_S^2 \left(4 x_f-4 x_W-1\right)+x_S \left(34 x_W-6 \left(4 x_f x_W+4 x_f^2+x_f-8 x_W^2\right)\right)+16 \left(x_f-x_W\right) \right. \nn \\
&\times& \left. \left(x_f+2 x_W\right)+2 x_f-18 x_W+1\right)+x_D^2 \left(x_S^3 \left(8 x_f-8 x_W-2\right)+x_S^2 \left(-4 x_f \left(4 x_W+1\right)-16 x_f^2 \right. \right. \nn \\
&+& \left.\left. 4 x_W \left(8 x_W+9\right)+1\right)-2 x_S \left(3 x_f+x_W+1\right)+16 \left(x_f-x_W\right) \left(x_f+2 x_W\right)+6 x_f-10 x_W+1\right) \nn \\
&+& x_D \left(x_S \left(x_S \left(-24 x_f+8 x_W-7\right)+8 \left(x_f-x_W\right) \left(x_f+2 x_W\right)+22 x_f+6 x_S^2-2 x_W+2\right) \right. \nn \\
&-& \left. 2 \left(2 x_f+x_W\right)\right)+x_D^5 \left(-2 x_f-8 x_W+1\right)+\left(1-2 x_S\right){}^2 \left(-2 x_f+x_S+2 x_W\right) \left(x_S \right. \nn \\
&-& \left. 2 \left(x_f+2 x_W\right)\right)\bigg\} 
- \chi \frac{2 q^2}{\lambda} \bigg\{ 2 x_f \left(x_D+2 x_S-1\right)-x_D x_S+x_D^2+x_D-2 x_S^2+x_S\bigg\} \,, \nn \\
C_{7}^3 &=& \frac{q^2}{4 \lambda ^2 \left(x_D-x_S\right)} 
\bigg\{ -x_D^4 \left(x_f \left(-2 x_S+4 x_W+2\right)+4 x_f^2+2 x_W \left(6 x_S-4 x_W+1\right)-1\right) \nn \\
&+& x_D^3 \left(3 x_S^2 \left(-4 x_f+4 x_W+1\right)+x_S \left(6 \left(4 x_f x_W+4 x_f^2+x_f-8 x_W^2\right)-34 x_W\right)-16 \left(x_f-x_W\right) \right. \nn \\
&\times& \left. \left(x_f+2 x_W\right)-2 x_f+18 x_W-1\right)+x_D^2 \left(x_S^3 \left(8 x_f-8 x_W-2\right)+x_S^2 \left(-4 x_f \left(4 x_W+1\right)-16 x_f^2 \right. \right. \nn \\
&+& \left. \left. 4 x_W \left(8 x_W+9\right)+1\right)-2 x_S \left(3 x_f+x_W+1\right)+16 \left(x_f-x_W\right) \left(x_f+2 x_W\right)+6 x_f-10 x_W+1\right) \nn \\
&+& x_D \left(2 \left(2 x_f+x_W\right)-x_S \left(x_S \left(-24 x_f+8 x_W-7\right)+8 \left(x_f-x_W\right) \left(x_f+2 x_W\right)+22 x_f+6 x_S^2 \right. \right. \nn \\
&-& \left. \left. 2 x_W+2\right)\right)+x_D^5 \left(2 x_f+8 x_W-1\right)+\left(1-2 x_S\right)^2 \left(-2 x_f+x_S+2 x_W\right) \left(x_S-2 \left(x_f+2 x_W\right)\right)\bigg\} \nn \\
&+& \chi \frac{2 q^2}{\lambda }  \bigg\{ x_D \left(-2 x_f+x_S-1\right)+x_D^2-\left(2 x_S-1\right) \left(x_S-2 x_f\right)\bigg\} \,, \nn \\
C_{7}^4 &=& -\frac{q^2 x_D}{6 \lambda ^2} \left( -2 x_f+x_S-4 x_W-1\right) \bigg\{ x_D^2 \left(8 x_f-2\right)-2 x_S \left(2 x_f+6 x_W+1\right)+12 \left(x_f-x_W\right)^2 \nn \\
&-&4 x_f+3 x_S^2+12 x_W+1\bigg\} \,, \nn \\
C_{7}^5 &=& \frac{q^2 x_D}{6 \lambda ^2} \bigg\{ -48 x_W^3 + 4 x_W^2 \left(-8 x_D^2+18 x_f+7 x_S+1\right)+\left(-2 x_f+x_S-1\right) \left(x_D^2-4 \left(3 x_f+2\right) x_S \right. \nn \\
&+& \left. 4 \left(3 x_f \left(x_f+1\right)+1\right)+3 x_S^2\right)+2 \lambda \, x_W \left(-8 x_f+4 x_S+9\right)\bigg\}
+ \chi \frac{4 q^2 x_D}{\lambda } \left(2 x_f - x_S + 1\right) \,, \nn \\
C_{7}^6 &=& -\frac{q^4 x_D}{4 \lambda ^2} 
\bigg\{ 2 x_D^2 \left(-x_S \left(x_W \left(8 x_f+4 x_W+23\right)-2\right)+x_W \left(x_f \left(8 x_W-2\right)+8 x_f^2 \right. \right. \nn \\
&+& \left. \left. 2 \left(7-8 x_W\right) x_W+9\right)+2 x_S^2 x_W-1\right)+x_D^4 \left(12 x_W-1\right)-2 x_S^3 \left(4 x_f+3 x_W+2\right) \nn \\
&+& x_S^2 \left(4 \left(5 x_f x_W+6 x_f \left(x_f+1\right)+x_W^2\right)+42 x_W+2\right)-4 x_S \left(2 x_f^2 \left(x_W+6\right)+x_f \left(6-4 x_W \left(x_W \right. \right. \right. \nn \\
&+& \left.\left.\left. 1\right)\right)+8 x_f^3+x_W \left(2 \left(5-3 x_W\right) x_W+7\right)\right)+8 \left(\left(10 x_f+1\right) x_W^3-2 \left(x_f+1\right) \left(3 x_f-1\right) x_W^2 \right. \nn \\
&-& \left. x_f \left(2 x_f^2+x_f+2\right) x_W+x_f \left(x_f+1\right) \left(2 x_f \left(x_f+1\right)+1\right)-4 x_W^4\right)+x_S^4+4 x_W\bigg\} \nn \\
&+& \chi \frac{2 q^4 x_D}{\lambda }  \bigg\{ -x_D^2+\left(x_S-2 x_f\right) \left(-2 x_f+x_S+2 x_W\right)+4 x_f-2 x_W\bigg\} \,, \nn \\
C_{7}^7 &=& \frac{q^4 x_D}{4 \lambda ^2}  \left( -2 x_f+x_S-4 x_W-1\right) \left( 2 x_f+x_S-2 x_W-1\right) \bigg\{ x_D^2 \left(4 x_f-1\right)-4 x_S \left(x_f+x_W\right) \nn \\
&+& 4 \left(\left(x_f-x_W\right)^2+x_W\right)+x_S^2\bigg\} \,, \\
C_{8}^0 &=& \frac{10 x_D}{3 \lambda ^2} \left(x_f-x_W\right) \left(-2 x_f+x_S-4 x_W-1\right) \,, \nn 
\eea
\bea
C_{8}^1 &=& \frac{2 x_D}{3 q^2 \lambda ^2 \left(x_D^2-x_S^2\right)} \bigg\{ x_D^2 \left(4 x_f-5 x_S+8 x_W+5\right)+x_S \left(5 x_S \left(-2 x_f+x_S-4 x_W-1\right) \right. \nn \\
&+& \left.12 \left(x_f+2 x_W\right)\right)-6 \left(x_f+2 x_W\right)\bigg\} \,, \nn \\
C_{8}^2 &=& \frac{x_W-x_f}{\lambda ^3 \left(x_D+x_S\right)}  
\bigg\{ x_D^4 \left(4 x_f-x_S+8 x_W+6\right)+x_D^3 \left(9 x_S \left(2 x_f-x_S+4 x_W+2\right)-16 x_f-32 x_W \right. \nn \\
&-& \left. 7\right)-2 x_D^2 \left(-2 x_S^2 \left(3 x_f+6 x_W+2\right)+9 x_f+3 x_S^3+x_S+18 x_W+2\right)+x_D \left(x_S \left(x_S \left(24 x_f \right. \right. \right. \nn \\
&-&  \left.\left.\left. 12 x_S+48 x_W+17\right)-10 \left(3 x_f+6 x_W+1\right)\right)+4 x_f+8 x_W+1\right)+2 x_D^5 \nn \\
&-&\left(2 x_S-1\right)^3 \left(x_S-2 \left(x_f+2 x_W\right)\right)\bigg\} \,, \nn \\
C_{8}^3 &=& \frac{x_f-x_W }{\lambda ^3 \left(x_D-x_S\right)}  
\bigg\{ x_D^4 \left(-4 x_f+x_S-8 x_W-6\right)+x_D^3 \left(9 x_S \left(2 x_f-x_S+4 x_W+2\right)-16 x_f-32 x_W \right. \nn \\
&-& \left. 7\right)+2 x_D^2 \left(-2 x_S^2 \left(3 x_f+6 x_W+2\right)+9 x_f+3 x_S^3+x_S+18 x_W+2\right)+x_D \left(x_S \left(x_S \left(24 x_f \right. \right. \right. \nn \\
&-& \left.\left.\left. 12 x_S+48 x_W+17\right)-10 \left(3 x_f+6 x_W+1\right)\right)+4 x_f+8 x_W+1\right)+2 x_D^5 \nn \\
&+& \left(2 x_S-1\right)^3 \left(x_S-2 \left(x_f+2 x_W\right)\right)\bigg\} \,, \nn \\
C_{8}^4 &=& \frac{x_D}{6 \lambda ^3} 
\bigg\{x_D^2 \left(\left(13-16 x_f\right) x_S+32 x_f \left(x_f+2 x_W\right)+26 x_f-34 x_W-13\right)+x_S^2 \left(2 x_f+120 x_W+19\right) \nn \\
&-& 2 x_S \left(-56 x_f x_W+2 x_f \left(x_f+2\right)+150 x_W^2+86 x_W+3\right)+2 \left(-2 x_f \left(90 x_W^2+44 x_W+1\right) \right. \nn \\
&+& \left. 60 x_f^3-14 x_f^2+x_W \left(30 x_W \left(4 x_W+5\right)+43\right)+1\right)-15 x_S^3\bigg\} \,, \nn \\
C_{8}^5 &=& \frac{x_D}{6 \lambda ^3} 
\bigg\{ 4 x_W^2 \left(-16 x_D^2+90 x_f-13 x_S+29\right)+\left(-2 x_f+x_S-1\right) \left(-13 x_D^2-4 \left(15 x_f+1\right) x_S \right. \nn \\
&+& \left. 60 x_f \left(x_f+1\right)+15 x_S^2+2\right)+2 \lambda  x_W \left(-16 x_f+8 x_S-9\right)-240 x_W^3\bigg\} \,, \nn \\
C_{8}^6 &=& \frac{q^2 x_D}{4 \lambda ^3}  
\bigg\{ 2 x_D^2 \left(x_S \left(12 \left(2 x_f x_W-x_f+x_W^2\right)+7 x_W-4\right)-6 \left(4 x_f+3\right) x_W^2-6 x_f \left(4 x_f+5\right) x_W \right. \nn \\
&+& \left. 12 x_f \left(x_f+1\right)+x_S^2 \left(3-6 x_W\right)+48 x_W^3-5 x_W+2\right)+x_D^4 \left(4 x_W-1\right)+2 x_S^3 \left(20 x_f+7 x_W+4\right) \nn \\
&-& 2 x_S^2 \left(18 x_f \left(x_W+2\right)+60 x_f^2+\left(13-6 x_W\right) x_W+2\right)+4 x_S \left(-6 x_f^2 \left(x_W-8\right)+12 x_f \left(-3 x_W^2 \right. \right. \nn \\
&+& \left. \left. x_W+1\right)+40 x_f^3+x_W \left(2 \left(x_W-3\right) x_W+5\right)\right)+4 \left(-2 \left(50 x_f+13\right) x_W^3+6 \left(2 x_f \left(5 x_f+4\right)  \right. \right. \nn \\
&+& \left. \left.  1\right) x_W^2+\left(2 x_f^2 \left(10 x_f+9\right)-1\right) x_W-4 x_f \left(x_f+1\right) \left(5 x_f \left(x_f+1\right)+1\right)+40 x_W^4\right)-5 x_S^4\bigg\} \,, \nn \\
C_{8}^7 &=& \frac{q^2 x_D}{4 \lambda ^3} 
\bigg\{ 2 x_D^2 \left(x_S \left(-2 x_f \left(18 x_W+5\right)+15 x_W+4\right)+\left(6 x_f-3\right) x_S^2+6 \left(8 x_f-3\right) x_W^2 \right. \nn \\
&-&\left.  3 \left(2 x_f \left(4 x_f-9\right)+5\right) x_W-24 x_f^3+8 x_f-2\right)+x_D^4 \left(1-4 x_f\right)-2 x_S^3 \left(2 x_f+25 x_W+4\right) \nn \\
&+& 4 x_S \left(72 \left(x_f-1\right) x_W^2-3 \left(2 \left(x_f-3\right) x_f+5\right) x_W+4 x_f^3-2 x_f-70 x_W^3\right)+x_S^2 \left(x_f \left(8-36 x_W\right) \right. \nn \\
&+&\left. 90 x_W \left(2 x_W+1\right)+4\right)+8 \left(5 \left(7-10 x_f\right) x_W^3+6 \left(x_f-1\right) \left(5 x_f-3\right) x_W^2 \right. \nn \\
&+& \left. x_f \left(2 x_f+3\right) \left(5 x_f-3\right) x_W+2 \left(2-5 x_f\right) x_f^3+20 x_W^4\right)+5 x_S^4+20 x_W\bigg\} \,, \\
C_{11}^0 &=& \frac{q^2 x_W}{6 \lambda }  \bigg\{ 2 x_f-3 x_S+28 x_W+3 \bigg\} \,, \nn \\
C_{11}^1 &=& \frac{x_W}{3 \lambda  \left( x_D^2-x_S^2\right) } \bigg\{x_S^2 -x_D^2 + 6 \left(1- x_S \right) \left(x_f+2 x_W\right) \bigg\} \,, \nn
\eea
\bea
C_{11}^2 &=& \frac{q^2 x_W}{4 \lambda ^2 x_D \left(x_D+x_S\right)} 
\bigg\{ x_D^4 \left(-2 x_f+x_S+10 x_W+2\right)-x_D^3 \left(x_S \left(6 x_f-40 x_W-5\right)+8 \left(x_f-x_W\right) \right. \nn \\ 
&\times& \left.  \left(x_f+2 x_W\right)+x_S^2+28 x_W+2\right)-x_D^2 \left(2 x_W \left(2 x_f \left(x_S+4\right)-15 x_S^2+11 x_S+8\right)+4 x_f x_S^2 \right. \nn \\
&+& \left. 4 x_f^2 x_S-8 x_f x_S+16  x_f^2+6 x_f-8 \left(x_S+4\right) x_W^2+x_S^3+1\right)+x_D \left(x_S^2 \left(24 x_f-2 x_W+6\right) \right. \nn \\
&-& \left. 2 x_S \left(10 x_f x_W+x_f \left(10 x_f+11\right)-20 x_W^2+4 x_W+1\right)+4 \left(x_f x_W+x_f^2+x_f-2 x_W^2\right) \right. \nn \\
&-& \left. 7 x_S^3-2 x_W\right)+x_D^5-\left(1-2 x_S\right)^2 \left(2 x_W \left(2 x_f+x_S\right)+\left(x_S-2 x_f\right)^2-8 x_W^2\right)\bigg\} \,, \nn \\
C_{11}^3 &=& -\frac{q^2 x_W}{4 \lambda ^2 x_D \left(x_D-x_S\right)} 
\bigg\{ -x_D^4 \left(-2 x_f+x_S+10 x_W+2\right)-x_D^3 \left(x_S \left(6 x_f-40 x_W-5\right) \right. \nn \\
&+& \left. 8 \left(x_f-x_W\right) \left(x_f+2 x_W\right)+x_S^2+28 x_W+2\right)+x_D^2 \left(2 x_W \left(2 x_f \left(x_S+4\right)-15 x_S^2+11 x_S+8\right) \right. \nn \\
&+& \left. 4 x_f x_S^2+4 x_f^2 x_S-8 x_f x_S+16 x_f^2+6 x_f-8 \left(x_S+4\right) x_W^2+x_S^3+1\right)+x_D \left(x_S^2 \left(24 x_f-2 x_W \right. \right. \nn \\
&+& \left. \left. 6\right)-2 x_S \left(10 x_f x_W+x_f \left(10 x_f+11\right)-20 x_W^2+4 x_W+1\right)+4 \left(x_f x_W+x_f^2+x_f-2 x_W^2\right) \right. \nn \\
&-& \left. 7 x_S^3-2 x_W\right)+x_D^5+\left(1-2 x_S\right)^2 \left(2 x_W \left(2 x_f+x_S\right)+\left(x_S-2 x_f\right)^2-8 x_W^2\right)\bigg\} \,, \nn \\
C_{11}^4 &=& \frac{q^2 x_W}{6 \lambda ^2}  \bigg\{ x_D^2 \left(8 x_f-2\right)-2 x_S \left(2 x_f+15 x_W+1\right)+12 \left(x_f-4 x_W\right) \left(x_f-x_W\right) \nn \\
&-& 4 x_f+3 x_S^2+30 x_W+1\bigg\} \,, \nn \\
C_{11}^5 &=& \frac{q^2 x_W}{3 \lambda ^2} \bigg\{ -2 x_D^2 \left(8 x_W+1\right)-2 x_S \left(6 x_f-4 x_W+1\right)+12 \left(4 x_f x_W+x_f^2+x_f-5 x_W^2\right) \nn \\
&+& 3 x_S^2+8 x_W+1\bigg\} \,, \nn \\
C_{11}^6 &=& \frac{q^4 x_W}{2 \lambda ^2} 
\bigg\{ -12 x_W^2 \left(-2 x_D^2+6 x_f+x_S+1\right)-2 x_W \left(x_D^2 \left(4 x_f-2 x_S+3\right)+\left(4 x_f-2\right) x_S \right. \nn \\
&-& \left. 4 x_f \left(3 x_f+2\right)+x_S^2\right)+\left(-2 x_f+x_S-1\right) \left(x_D^2-\left(x_S-2 x_f\right)^2-4 x_f\right)+40 x_W^3\bigg\} \,, \nn \\
C_{11}^7 &=& \frac{q^4 x_W}{4 \lambda ^2} 
\bigg\{ 4 x_W \left(x_D^2 \left(4 x_f-1\right)+4 x_f \left(3 x_f+x_S\right)-8 x_f+x_S \left(3 x_S-4\right)+2\right)-\left(2 x_f+x_S-1\right) \nn \\
&\times& \left(x_D^2 \left(4 x_f-1\right)+\left(x_S-2 x_f\right)^2\right)-36 x_W^2 \left(2 x_f+x_S-1\right)+32 x_W^3\bigg\} \,.
\eea

\normalsize

\chapter{Appendix}

\section{Appendix. Invariance under the momenta parameterization}
\label{PNewCIappendix1}

In this appendix we show, exploiting the invariance under rotations and dilatations, that the second of Eq.(\ref{PNewCIConformalEqMom}) is independent of which momentum is eliminated. This implies that the two different parameterizations of the special conformal constraints
\bea
&& \sum_{r=1}^{n-1} \left( p_{r \, \mu} \, \frac{\partial^2}{\partial p_{r}^{\nu} \partial p_{r \, \nu}}  - 2 \, p_{r \, \nu} \, 
\frac{\partial^2}{ \partial p_{r}^{\mu} \partial p_{r \, \nu} }    + 2 (\eta_r - d) \frac{\partial}{\partial p_{r}^{\mu}}  + 2 
(\Sigma_{\mu\nu})^{i_r}_{j_r} \frac{\partial}{\partial p_{r \, \nu}} \right) \nn \\
&& \hspace{7cm} \, \times \langle \mathcal O^{i_1}_1(p_1) \ldots  \mathcal O^{j_r}_r(p_r) \ldots \mathcal O^{i_n}_n(p_n) \rangle = 0 \,, \label{PNewCIspecialconf1} \\
&& \sum_{\stackrel{r=1}{r \neq k}}^{n} \left( p_{r \, \mu} \, \frac{\partial^2}{\partial p_{r}^{\nu} \partial p_{r \, \nu}}  - 2 \, p_{r \, \nu} \, 
\frac{\partial^2}{ \partial p_{r}^{\mu} \partial p_{r \, \nu} }    + 2 (\eta_r - d) \frac{\partial}{\partial p_{r}^{\mu}}  + 2 
(\Sigma_{\mu\nu})^{i_r}_{j_r} \frac{\partial}{\partial p_{r \, \nu}} \right) \nn \\
&& \hspace{7cm} \, \times \langle \mathcal O^{i_1}_1(p_1) \ldots  \mathcal O^{j_r}_r(p_r) \ldots \mathcal O^{i_n}_n(p_n) \rangle = 0 \,, \label{PNewCIspecialconf2}
\eea
in which we have respectively removed the dependence on $p_n$ and $p_k$ in terms of the other momenta,
are indeed equivalent. \\
In order to simplify the presentation of the proof we introduce some convenient notations. We define
\bea
G^{i_1 \ldots i_n} &\equiv& \langle \mathcal O^{i_1}_1(p_1) \ldots  \mathcal O^{i_n}_n(p_n) \rangle \,, \nn \\
\mathcal R_{\mu\nu}(p_r) &\equiv& p_{r \, \nu} \frac{\partial}{\partial p_{r}^{\mu}} - p_{r \, \mu} \frac{\partial}{\partial p_{r}^{\nu}} \,, \nn \\
\mathcal D(p_{r}) &\equiv& - p_{r \, \nu} \frac{\partial}{\partial p_{r \, \nu}} - d \,, \nn \\
\mathcal K_{\mu}(p_{r}, \eta) &\equiv& p_{r \, \mu} \, \frac{\partial^2}{\partial p_{r}^{\nu} \partial p_{r \, \nu}}  - 2 \, p_{r \, \nu} \, 
\frac{\partial^2}{ \partial p_{r}^{\mu} \partial p_{r \, \nu} }    + 2 (\eta - d) \frac{\partial}{\partial p_{r}^{\mu}} \,
\eea
and preliminarily derive two constraints, which will be used in the following, emerging from the invariance under rotations and scale transformations respectively. \\
Using the same procedure described in section 2, we find the constraint coming from rotational invariance
\bea
\sum_{r = 1}^{n-1} \mathcal R_{\mu\nu}(p_r) G^{i_1 \ldots i_n} - \sum_{r = 1}^{n} \left( \Sigma_{\mu\nu}\right)^{i_r}_{j_r} G^{i_1 \ldots j_r \ldots i_n} = 0 \,,
\eea
from which, differentiating with respect to $p_{k \, \nu}$, we obtain
\bea
\label{PNewCIdiffRot}
\bigg[ \sum_{r = 1 \,, r \neq k}^{n-1} \mathcal R_{\mu\nu}(p_r) \frac{\partial}{\partial p_{k \, \nu}}  +   \mathcal F_{\mu}(p_k) \bigg] G^{i_1 \ldots i_n} 
- \sum_{r = 1}^{n} \left( \Sigma_{\mu\nu}\right)^{i_r}_{j_r} \frac{\partial}{\partial p_{k \, \nu}} G^{i_1 \ldots j_r \ldots i_n} = 0 \,, 
\eea
where $\mathcal F_{\mu}(p_k)$ is defined by
\bea
\label{PNewCItensorF}
\mathcal F_{\mu}(p_k) = (d-1) \frac{\partial}{\partial p_{k}^{\mu}} + p_{k}^\nu  \frac{\partial^2}{\partial p_{k}^{\nu} \, \partial p_{k}^{\mu}} - p_{k \, \mu}  \frac{\partial^2}{\partial p_{k}^{\nu} \, \partial p_{k \, \nu}} \,.
\eea
Another useful relation can be obtained differentiating the dilatation constraint in the first of Eq.(\ref{PNewCIConformalEqMom}) with respect to $p_{k\, \mu}$
\bea
\label{PNewCIdiffDil}
\bigg[ \sum_{r = 1\,, r \neq k}^{n-1} \left( \mathcal D(p_r) + \eta_r \right) \frac{\partial}{\partial p^\mu_k}  + (\eta_k + \eta_n - d- 1) \frac{\partial}{\partial p^\mu_k} - p_{k}^\nu \frac{\partial^2}{\partial p^\mu_k \partial p^\nu_k} \bigg] G^{i_1 \ldots  i_n} = 0 \,.
\eea
Having derived all the necessary relations, we can proceed with the proof of equivalence between Eqs.(\ref{PNewCIspecialconf1}) and (\ref{PNewCIspecialconf2}). We remove from Eq.(\ref{PNewCIspecialconf1}) the $k-th$ term containing the spin matrix $\left(\Sigma_{\mu\nu}\right)^{i_k}_{j_k}$ using Eq.(\ref{PNewCIdiffRot}) 
\bea
\label{PNewCIspecialconftemp1}
&& \bigg[ \sum_{r = 1}^{n-1} \mathcal K_\mu(p_r, \eta_r) + 2 \sum_{r=1 \,, r \neq k}^{n-1} \mathcal R_{\mu\nu}(p_r) \frac{\partial}{\partial p_{k \, \nu}} + 2 \mathcal F_\mu(p_k) \bigg] G^{i_1 \ldots  i_n} \nn \\
&& + \,  2 \sum_{r=1 \,, r \neq k}^{n-1} \left( \Sigma_{\mu\nu}\right)^{i_r}_{j_r} \left( \frac{\partial}{\partial p_{r \, \nu}} - \frac{\partial}{\partial p_{k \, \nu}}\right) G^{i_1 \ldots j_r \ldots i_n} 
- 2 \left( \Sigma_{\mu\nu} \right)^{i_n}_{j_n} \frac{\partial}{\partial p_{k \, \nu}} G^{i_1 \ldots  j_n}  = 0 \,,
\eea
and then we combine the $k-th$ operator $\mathcal K_\mu(p_k, \eta_k)$ with the $\mathcal F_\mu(p_k)$ contribution as
\bea
\mathcal K_{\mu}(p_k, \eta_k) + 2 \mathcal F_\mu(p_k) = - p_{k \, \mu} \frac{\partial^2}{\partial p_{k \, \nu} \partial p_k^\nu} + 2 (\eta_k - 1) \frac{\partial}{\partial p_k^\mu} \,.
\eea
Using Eq.(\ref{PNewCIdiffDil}) we rewrite the previous equation as
\bea
\mathcal K_{\mu}(p_k, \eta_k) + 2 \mathcal F_\mu(p_k) = - \mathcal K_\mu(p_k, \eta_n) - 2 \sum_{r = 1\,, r \neq k}^{n-1} \left( \mathcal D(p_r) + \eta_r \right) \frac{\partial}{\partial p_{k}^\mu}
\eea
so that Eq.(\ref{PNewCIspecialconftemp1}) can be recast in the following form
\bea
\label{PNewCIspecialconftemp2}
&& \bigg\{ \sum_{r=1 \,, r \neq k}^{n-1} \bigg[ \mathcal K_\mu(p_r , \eta_r) + 2 \, \mathcal R_{\mu\nu}(p_r) \frac{\partial}{\partial p_{k \, \nu}} - 2 \left( \mathcal D(p_r) + \eta_r \right) \frac{\partial}{\partial p_{k \, \mu}}  \bigg] - \mathcal K_\mu(p_k, \eta_n) \bigg\} G^{i_1 \ldots  i_n}   \nn \\
&&
+ \,  2 \sum_{r=1 \,, r \neq k}^{n-1} \left( \Sigma_{\mu\nu}\right)^{i_r}_{j_r} \left( \frac{\partial}{\partial p_{r \, \nu}} - \frac{\partial}{\partial p_{k \, \nu}}\right) G^{i_1 \ldots j_r \ldots i_n} 
- 2 \left( \Sigma_{\mu\nu} \right)^{i_n}_{j_n} \frac{\partial}{\partial p_{k \, \nu}} G^{i_1 \ldots  j_n}  = 0 \,.
\eea
In order to show the equivalence of Eq.(\ref{PNewCIspecialconftemp2}) with Eq.(\ref{PNewCIspecialconf2}) it is necessary to perform a change of variables from the independent set of momenta $(p_1, \ldots p_{n-1})$ to the new independent one $(p_1 \ldots p_{k-1}, p_{k+1}, \ldots p_n)$ from which $p_k$ has been removed using momentum conservation $p_k = - \sum_{r=1 \,, r \neq k}^ n p_r$. In this respect all the derivatives appearing in Eq.(\ref{PNewCIspecialconftemp2}) must be replaced according to
\bea
&& \frac{\partial}{\partial p_{r \, \mu}} \rightarrow \frac{\partial}{\partial p_{r \, \mu}} - \frac{\partial}{\partial p_{n \, \mu}}  \qquad \qquad \mbox{for} \quad r = 1, \ldots n-1 \quad \mbox{with} \quad r \neq k \,, \nn \\
&& \frac{\partial}{\partial p_{k \, \mu}} \rightarrow - \frac{\partial}{\partial p_{n \, \mu}} \qquad \qquad \mbox{for} \quad r = k \,.
\eea
It is just matter of tedious algebraic manipulations to show that the operators in curly brackets in Eq.(\ref{PNewCIspecialconftemp2}) simplify, after the change of variables, to give $\sum_{r = 1 \,, r \neq k}^n \mathcal K_\mu(p_r, \eta_r)$, while the two spin matrices sum up together and we are left with
\bea
\sum_{r = 1 \,, r \neq k}^n \mathcal K_\mu(p_r, \eta_r) G^{i_1, \ldots i_n} + 2 \sum_{r = 1 \,, r \neq k}^n \left( \Sigma_{\mu\nu} \right)^{i_r}_{j_r} \frac{\partial}{\partial p_{r \, \nu}} G^{i_1 \ldots j_r \ldots i_n} = 0
\eea
which is exactly Eq.(\ref{PNewCIspecialconf2}), where, now, $G^{i_1, \ldots i_n}$ is understood to be a function of the independent momenta $(p_1 \ldots p_{k-1}, p_{k+1}, \ldots p_n)$. This completes our derivation proving the independence of the special conformal constraints on which momentum is removed using the momentum conservation equation.

\section{Appendix. Conformal constraints on two-point functions}
\label{PNewCIAppTwoPoint}
In this appendix we provide some details on the solutions of the conformal constraints for the two-point functions with conserved vector and tensor operators. \\
In the first case the tensor structure of the two-point function is uniquely fixed by the transversality condition $\partial^\mu V_\mu$ as
\bea
G_V^{\alpha \beta}(p) = f(p^2) t^{\alpha \beta}(p)\,, \qquad \mbox{with} \quad t^{\alpha\beta}(p) = p^2 \eta^{\alpha\beta} - p^{\alpha} p^{\beta} \,.
\eea 
For the sake of simplicity, we have employed in the previous equation a slightly different notation with respect to Eq.(\ref{PNewCITwoPointVector0}), which, anyway, can be recovered with the identification $f(p^2) = f_V(p^2)/p^2$. \\
In order to exploit the invariance under scale and special conformal transformations it is useful to compute first and second order derivatives of the $t^{\alpha \beta}$ tensor structure. In particular we have
\bea
\label{PNewCITDerivatives}
t_1^{\alpha \beta, \mu}(p) &\equiv& \frac{\partial }{\partial p_\mu} t^{\alpha \beta}(p) = 2 \, p^{\mu} \eta^{\alpha \beta} - p^{\alpha} \eta^{\mu \beta} - p^{\beta} \eta^{\mu \alpha} \,, \nn \\
t_2^{\alpha \beta, \mu \nu}(p) &\equiv& \frac{\partial^2 }{\partial p_\mu \, \partial p_\nu} t^{\alpha \beta}(p) = 2 \, \eta^{\mu \nu} \eta^{\alpha \beta} - \eta^{\nu \alpha} \eta^{\mu \beta} - \eta^{\nu \beta} \eta^{\mu \alpha} \,, 
\eea
with the properties
\bea
&& p_\mu t_1^{\alpha \beta, \mu}(p) = 2 \, t^{\alpha \beta}(p) \,, \qquad     t_1^{\alpha \beta, \alpha}(p) = - (d - 1) p^\beta \,, \nn \\
&& p_\mu t_2^{\alpha \beta, \mu \nu}(p) = t_1^{\alpha \beta, \nu}(p) \,, \qquad    t_2^{\alpha \beta, \mu \mu}(p) = 2(d-1) \eta^{\alpha \beta} \,. 
\eea
As we have already mentioned, the invariance under scale transformations implies
\bea
\label{PNewCIFandLambda}
f(p^2) = (p^2)^\lambda \qquad \qquad \mbox{with} \quad \lambda = \frac{\eta_1 + \eta_2 - d}{2} -1 \,,
\eea
which can be easily derived from the first order differential equation in (\ref{PNewCIConformalEqMomTwoPoint}) using Eq.(\ref{PNewCITDerivatives}). Having determined the structure of the scalar function $f(p^2)$, one can compute the derivatives appearing in the second of Eq.(\ref{PNewCIConformalEqMomTwoPoint}), namely the constraint following from the invariance under the special conformal transformations
\bea
\label{PNewCIGDerivatives}
\frac{\partial}{\partial p_\mu} G_V^{\alpha \beta}(p) &=& (p^2)^{\lambda - 1} \bigg[ 2 \lambda \, p^\mu t^{\alpha \beta}(p) + p^2 \, t_1^{\alpha\beta, \mu}(p) \bigg] \,, \nn \\
\frac{\partial^2}{\partial p_\mu \, \partial p_\nu} G_V^{\alpha \beta}(p) &=&(p^2)^{\lambda - 2} \bigg[ 4 \lambda (\lambda -1) p^{\mu} p^{\nu} t^{\alpha \beta}(p)  + 2 \lambda p^2 \eta^{\mu\nu}   t^{\alpha \beta}(p) + 2 \lambda  p^2 p^{\mu} t_1^{\alpha\beta,\nu}(p)  \nn \\
&& \qquad +  \, 2 \lambda  p^2 p^{\nu} t_1^{\alpha\beta,\mu}(p) + (p^2)^2 t_2^{\alpha\beta, \mu\nu}(p) 
\bigg] \,,
\eea
where the definitions in Eq.(\ref{PNewCITDerivatives}) have been used.
Concerning the spin dependent part in Eq.(\ref{PNewCIConformalEqMomTwoPoint}), we use the spin matrix for the vector field, which, in our conventions, reads as
\bea
( \Sigma_{\mu\nu}^{(V)})^{\alpha}_{\beta} = \delta_{\mu}^{\alpha} \, \eta_{\nu \beta} - \delta_{\nu}^{\alpha} \, \eta_{\mu \beta} \,,
\eea
and obtain
\bea
\label{PNewCISigmaPart}
2( \Sigma_{\mu\nu}^{(V)})^{\alpha}_{\rho} \frac{\partial}{\partial p_\nu} G_V^{\rho\beta}(p) = - (p^2)^{\lambda- 1} \bigg[ 2 \lambda \, p^\alpha {t_{\mu}}^{\beta}(p) + (d-1)p^2 p^\beta \delta_\mu^\alpha  + p^2 {t_{1 \, \mu}}^{\beta, \alpha}(p) \bigg] \,.
\eea
Employing the results derived in Eq.(\ref{PNewCIGDerivatives}) and Eq.(\ref{PNewCISigmaPart}), we have fully determined the special conformal constraint on the two-point vector function.
Then we can project the second of Eq.(\ref{PNewCIConformalEqMomTwoPoint}) onto the three independent tensor structures $p_\mu \eta_{\alpha \beta}, p_\alpha \eta_{\mu \beta}, p_\beta \eta_{\alpha \mu}$, and setting $\lambda$ to the value given in Eq.(\ref{PNewCIFandLambda}), we finally obtain three equations for the scale dimensions $\eta_i$ of the vector operators
\bea
\begin{cases}
(\eta_1 - \eta_2) (\eta_1 + \eta_2 - d) = 0 \,, \nn \\
\eta_1 - d +1 = 0 \,, \nn \\
\eta_2 - d +1 = 0 \,. \nn
\end{cases}
\\
\eea
The previous system of equations can be consistently solved only for $\eta_1 = \eta_2 = d -1$, as expected. This completes our derivation of the vector two-point function which, up to an arbitrary multiplicative constant, can be written as in Eq.(\ref{PNewCITwoPointVector}).

The characterization of the two-point function with a symmetric, traceless and conserved rank-2 tensor follows the same lines of reasoning already explained in the vector case. These conditions (see Eq.(\ref{PNewCIEMTconditions})) fix completely the tensor structure of the two-point function as
\bea
G_T^{\alpha\beta\mu\nu}(p) = g(p^2) \, T^{\alpha\beta\mu\nu}(p) 
\eea
with
\bea
T^{\alpha\beta\mu\nu}(p) =  \frac{1}{2} \bigg[ t^{\alpha\mu}(p) t^{\beta\nu}(p) + t^{\alpha\nu}(p) t^{\beta\mu}(p) 
\bigg] 
- \frac{1}{d-1} t^{\alpha\beta}(p) t^{\mu\nu}(p) \,.
\eea
In order to recover the convention used in section \ref{PNewCITwoPointSection}, notice that $g(p^2) = f_T(p^2)/(p^2)^2$. \\
As in the previous case, we give the first and second order derivatives of the $T^{\alpha\beta\mu\nu}(p)$ tensor structure
\bea
\label{PNewCITTDerivatives}
T_1^{\alpha\beta\mu\nu, \rho}(p) &\equiv& \frac{\partial}{\partial p_\rho} T^{\alpha\beta\mu\nu}(p) = \frac{1}{2} \bigg[ t_1^{\alpha\mu, \rho}(p) t^{\beta\nu}(p) 
+ t^{\alpha\mu}(p) t_1^{\beta\nu, \rho}(p) + \left( \mu \leftrightarrow \nu \right) \bigg] \nn \\
&& - \frac{1}{d-1} \bigg[ t_1^{\alpha\beta, \rho}(p) t^{\mu\nu}(p) + t^{\alpha\beta}(p) t_1^{\mu\nu, \rho}(p) \bigg] \,, \nn \\
T_2^{\alpha\beta\mu\nu, \rho\sigma}(p) &\equiv& \frac{\partial}{\partial p_\rho \, \partial p_\sigma} T^{\alpha\beta\mu\nu}(p) = \frac{1}{2} \bigg[ 
t_2^{\alpha\mu, \rho \sigma}(p) t^{\beta\nu}(p) + t_1^{\alpha\mu, \rho}(p) t_1^{\beta\nu, \sigma}(p) + t_1^{\alpha\mu, \sigma}(p) t_1^{\beta\nu, \rho}(p) \nn \\
&& +  \, t^{\alpha\mu}(p) t_2^{\beta\nu, \rho \sigma}(p)+ \left( \mu \leftrightarrow \nu \right) \bigg]  \nn \\
&& - \frac{1}{d-1} \bigg[  t_2^{\alpha\beta, \rho\sigma}(p) t^{\mu\nu}(p) +   t_1^{\alpha\beta, \rho}(p) t_1^{\mu\nu, \sigma}(p) + (\mu\nu) \leftrightarrow (\alpha\beta)  \bigg] \,,
\eea
together with some of their properties
\bea
p_\rho T_1^{\alpha\beta\mu\nu, \rho}(p) = 4 \, T^{\alpha\beta\mu\nu}(p) \,, \qquad
p_\rho T_2^{\alpha\beta\mu\nu, \rho \sigma}(p) = 3 \, T_1^{\alpha\beta\mu\nu, \sigma}(p)  \,.
\eea
As we have already stressed, the first of Eq.(\ref{PNewCIConformalEqMomTwoPoint}) defines the scaling behavior of the two-point function, providing, therefore, the functional form of $g(p^2)$ which is given by
\bea
g(p^2) = (p^2)^\lambda \qquad \mbox{with} \quad \lambda = \frac{\eta_1 +\eta_2 -d}{2} -2 \,.
\eea
On the other hand, the second of Eq.(\ref{PNewCIConformalEqMomTwoPoint}), namely the constraint from the special conformal transformations, fixes the scaling dimensions of the tensor operators. In this case the spin connection is given by
\bea
(\Sigma_{\mu\nu}^{(T)})^{\alpha \beta}_{\rho \sigma } = \left( \delta_{\mu}^{\alpha} \, \eta_{\nu \rho} - \delta_{\nu}^{\alpha} \, \eta_{\mu \rho} \right) \delta_{\sigma}^{\beta}
+ \left( \delta_{\mu}^{\beta} \, \eta_{\nu \sigma} - \delta_{\nu}^{\beta} \, \eta_{\mu \sigma} \right) \delta_{\rho}^{\alpha} \,.
\eea
The algebra is straightforward but rather cumbersome due to the proliferation of indices. Here we give only the final result, which can be obtained projecting Eq.(\ref{PNewCIConformalEqMomTwoPoint}), making use of Eq.(\ref{PNewCITTDerivatives}), onto the different independent tensor structures
\bea
\begin{cases}
(\eta_1 - \eta_2) (\eta_1 + \eta_2 - d) = 0 \,, \nn \\
\eta_1 - d = 0 \,, \nn \\
\eta_2 - d = 0 \,, \nn
\end{cases}
\\
\eea
which implies the solution $\eta_1 = \eta_2 = d$, as described in Eq.(\ref{PNewCITwoPointEmt}).

\section{Appendix. Scalar integrals in $D=3$ dimensions}
\label{P4scalarint}
We give the expressions of the two, three and four point scalar integrals with internal masses set to zero in $D=3$. 
They are  defined as
\beqa
 \mathcal B_0(q_1^2) &=& \int d^3 l \frac{1}{l^2 \, (l+q_1)^2}  = \frac{\pi^3}{q_1}  \,, \nn \\
 \mathcal C_0(q_1^2, q_{12}^2, q_2^2) &=& \int d^3 l \frac{1}{l^2 \, (l+q_1)^2 \, (l+q_2)^2} = \frac{\pi^3}{q_1 \, q_{12} \, q_2}\,, \nn \\
 \mathcal D_0 \left(q_1^2,q_{12}^2,q_{23}^2,q_3^2,q_2^2,q_{13}^2\right) &=& \int d^3 l \frac{1}{l^2 \, (l+q_1)^2 \, (l+q_2)^2 \, (l+q_3)^2}
\eeqa
where $q_{ij}^2 = (q_i-q_j)^2$. \\
The box integral in $D=3$ in not independent from the 2- and 3- scalar point functions, indeed it is possible to show the following relation 
\small
\bea
&& \hspace{-0.5cm} \mathcal D_0 \left(q_1^2,q_{12}^2,q_{23}^2,q_3^2,q_2^2,q_{13}^2\right) =
\frac{1}{q_1^4 q_{23}^4-2 q_1^2 \left(q_3^2 q_{12}^2+q_2^2 q_{13}^2\right) q_{23}^2+\left(q_3^2
   q_{12}^2-q_2^2 q_{13}^2\right)^2} \times  \bigg\{  \nn \\
&& \hspace{-0.5cm}
 \bigg[q_2^2 q_{13}^4 - \left(\left(q_2^2+q_3^2-2 q_{23}^2\right) q_{12}^2+q_2^2 q_{23}^2\right) q_{13}^2 - q_1^2 q_{23}^2
   \left(q_{12}^2+q_{13}^2-q_{23}^2\right) - q_3^2 q_{12}^2 \left(q_{23}^2-q_{12}^2\right)\bigg]
   \mathcal C_0 \left(q_{12}^2,q_{23}^2,q_{13}^2 \right)       \nn \\
&& \hspace{-0.5cm}
+\bigg[q_2^2 q_{13}^4-\left(\left(q_2^2-2 q_3^2+q_{23}^2\right) q_1^2+q_3^2
   \left(q_2^2+q_{12}^2\right)\right) q_{13}^2+\left(q_1^2-q_3^2\right) \left(q_1^2 q_{23}^2-q_3^2 q_{12}^2\right)\bigg]
   \mathcal C_0 \left(q_1^2,q_3^2,q_{13}^2 \right) \nn \\
&& \hspace{-0.5cm}
+\bigg[ \left(2 q_2^2 q_{23}^2-q_{12}^2 \left(q_2^2-q_3^2+q_{23}^2\right)\right) q_3^2+q_1^2
   q_{23}^2 \left(-q_2^2-q_3^2+q_{23}^2\right)-q_2^2 q_{13}^2 \left(-q_2^2+q_3^2+q_{23}^2\right)\bigg]
   \mathcal C_0 \left(q_{23}^2,q_3^2,q_2^2\right) \nn \\
&& \hspace{-0.5cm}
+\bigg[q_{23}^2 q_1^4-\left(\left(q_{13}^2+q_{23}^2\right) q_2^2+q_{12}^2 \left(-2
   q_2^2+q_3^2+q_{23}^2\right)\right) q_1^2+\left(q_{12}^2-q_2^2\right) \left(q_3^2 q_{12}^2-q_2^2 q_{13}^2\right)\bigg]
   \mathcal C_0 \left(q_1^2,q_{12}^2,q_2^2 \right)
\bigg\}. \nn \\
\label{P4D0toC0}
\eea
\normalsize
\section{Appendix. The $TTTT$ for the minimal scalar case}
\label{P4MSres}
We give here the expressions of the remaining coefficients $C^{MS}_i$ of Eq. (\ref{P4boxo}). They are given by 
\small
\bea
&& C^{MS}_6(\vec{q_1},\vec{q_2},\vec{q_3}) = \frac{1}{2048\pi ^3} \bigg\{ \frac{4}{\lambda^2(q_1,q_{12},q_2)} \bigg[  -q_1^4 \left(\vec{q_1} \cdot \vec{q_2}
   \left(q_2^2+\vec{q_1} \cdot \vec{q_3}+2 \vec{q_2} \cdot \vec{q_3}\right)-q_2^2
   \left(q_2^2+\vec{q_1} \cdot \vec{q_3}\right) \right. \nn \\
 && \left. +\vec{q_1} \cdot \vec{q_2}^2\right) 
+ q_1^2
   \vec{q_1} \cdot \vec{q_2} \left(\vec{q_1} \cdot \vec{q_2}
   \left(\vec{q_1} \cdot \vec{q_3}+\vec{q_2} \cdot \vec{q_3} -q_2^2 \right)-2 q_2^2
   \vec{q_1} \cdot \vec{q_3}+\vec{q_1} \cdot \vec{q_2}^2\right)+q_1^6
   \left(q_2^2+\vec{q_2} \cdot \vec{q_3}\right) \nn \\
&& +\vec{q_1} \cdot \vec{q_2}^3 \vec{q_1} \cdot \vec{q_3} \bigg] 
 + \frac{4 \vec{q_1} \cdot \vec{q_3}^2}{\lambda^2(q_1,q_{13},q_3)} \left(q_1^2 \vec{q_2} \cdot \vec{q_3}-\vec{q_1} \cdot \vec{q_2}
   \vec{q_1} \cdot \vec{q_3}\right)+q_1^2 q_3^2 \bigg\} \,, \nn
\eea
\bea
&& C^{MS}_7(\vec{q_1},\vec{q_2},\vec{q_3}) =
\frac{1}{512 \pi ^3
   \lambda^2(q_1,q_{12},q_2)}
 \bigg\{ q_1^4 q_2^2 \vec{q_1} \cdot \vec{q_3}-\vec{q_2} \cdot \vec{q_3} \left(q_1^2
   \left(q_2^4-3 \vec{q_1} \cdot \vec{q_2}^2\right)-q_2^2
   \vec{q_1} \cdot \vec{q_2}^2 \right. \nn \\
&& \left. +q_1^4
   \left(q_2^2+\vec{q_1} \cdot \vec{q_2}\right)\right) \bigg\}
 - \frac{q_3^4}{512 \pi ^3 \lambda^2(q_2,q_{23},q_3)} 
   \left(\vec{q_1} \cdot \vec{q_2} \left(q_2^2+\vec{q_2} \cdot \vec{q_3}\right)-q_2^2
   \vec{q_1} \cdot \vec{q_3}\right)
+\frac{\left(q_1^2+q_2^2\right)
   \left(q_2^2+q_3^2\right)}{2048 \pi ^3}   \nn \\
&& + \frac{1}{256 \pi ^3 \lambda^2(q_1,q_{12},q_2) \lambda^2(q_2,q_{23},q_3)} 
\bigg\{3 \vec{q_1} \cdot \vec{q_2}^3 \vec{q_2} \cdot \vec{q_3}^2
   \left(\vec{q_2} \cdot \vec{q_3}-2 q_3^2\right) \nn \\
&& - q_2^2 \vec{q_1} \cdot \vec{q_2}
   \vec{q_2} \cdot \vec{q_3} \left(\vec{q_1} \cdot \vec{q_2}^2 \left(q_3^2+2
   \vec{q_2} \cdot \vec{q_3}\right)+\vec{q_1} \cdot \vec{q_2} \vec{q_1} \cdot \vec{q_3}
   \left(\vec{q_2} \cdot \vec{q_3}-4 q_3^2\right)+q_1^2 \vec{q_2} \cdot \vec{q_3}
   (-6 q_3^2   \right. \nn \\
&& - \left. 4 \vec{q_1} \cdot \vec{q_3}+\vec{q_2} \cdot \vec{q_3})\right)-q_2^4 \left(q_3^2
   \left(\vec{q_1} \cdot \vec{q_3} \left(4 q_1^2
   (\vec{q_1} \cdot \vec{q_2}+\vec{q_2} \cdot \vec{q_3})+\vec{q_1} \cdot \vec{q_2}^2\right)+q_1^2
   \vec{q_1} \cdot \vec{q_2} \vec{q_2} \cdot \vec{q_3}-2 \vec{q_1} \cdot \vec{q_2}^3\right)  \right. \nn \\
&& + \left. q_1^2
   \vec{q_2} \cdot \vec{q_3}^2 (\vec{q_1} \cdot \vec{q_3}-2 \vec{q_1} \cdot \vec{q_2})\right)+q_1^2
   q_3^2 q_2^6 (3 \vec{q_1} \cdot \vec{q_3}-2 \vec{q_1} \cdot \vec{q_2}) \bigg\} \,, \nn
\eea
\bea
&& C^{MS}_8(\vec{q_1},\vec{q_2},\vec{q_3}) = \frac{q_3^2}{128 \pi ^3
   \lambda^2(q_1,q_{13},q_3)\lambda^2(q_2,q_{23},q_3)} \bigg\{ \vec{q_1} \cdot \vec{q_2} \vec{q_1} \cdot \vec{q_3}^2 \left(q_3^2
   \left(q_2^2+2 \vec{q_2} \cdot \vec{q_3}\right)-q_3^4-2
   \vec{q_2} \cdot \vec{q_3}^2\right) \nn \\
&& + q_1^2 q_3^2 \vec{q_1} \cdot \vec{q_2}
   \left(\vec{q_2} \cdot \vec{q_3}-q_3^2\right)^2+\vec{q_1} \cdot \vec{q_3}^3
   \left(q_2^2-\vec{q_2} \cdot \vec{q_3}\right)
   \left(\vec{q_2} \cdot \vec{q_3}-q_3^2\right)+q_1^2 \vec{q_1} \cdot \vec{q_3}
   \left(q_3^4 \left(q_2^2-\vec{q_2} \cdot \vec{q_3}\right) \right. \nn \\
&& + \left. q_3^2
   \vec{q_2} \cdot \vec{q_3} \left(\vec{q_2} \cdot \vec{q_3}-2
   q_2^2\right)+\vec{q_2} \cdot \vec{q_3}^3\right)  \bigg\}
   +\frac{q_3^2}{2048 \pi ^3}
   \left(q_1^2+q_2^2+q_3^2-\vec{q_2} \cdot \vec{q_3}\right) \,, \nn
\eea
\bea
&& C^{MS}_9(\vec{q_1},\vec{q_2},\vec{q_3}) = \frac{q_{12}^2}{4096 \pi^3 \lambda^2(q_1,q_{12},q_2) \lambda^2(q_{12},q_{23},q_{13}) }
  \bigg\{ \left(\left(q_{12}^2-q_{13}^2\right)^2 -q_{23}^4-4 q_{12}^2
   q_{23}^2 \right)
   q_1^6+\left(-q_{23}^6 \right. \nn \\
&& +\left(3 q_2^2 +3 q_{13}^2  +  \left. 2
   \left(q_3^2+q_{12}^2\right)\right) q_{23}^4+\left(9
   q_{12}^4+8 q_2^2 q_{12}^2-3 q_{13}^4-2 \left(2
   q_3^2+q_{12}^2\right) q_{13}^2\right)
   q_{23}^2-\left(q_{12}^2-q_{13}^2\right){}^2 \left(3 q_2^2    \right. \right. \nn \\
&& \left. \left. - 2
   q_3^2 + 2 q_{12}^2-q_{13}^2\right)\right) q_1^4+\left(-4
   q_{12}^2 q_{23}^6+\left(-3 q_2^4-2 \left(2
   q_3^2+q_{12}^2\right) q_2^2+9 q_{12}^4+8 q_{12}^2
   q_{13}^2\right) q_{23}^4-4 \left(2 q_{12}^6  \right. \right. \nn \\
&& + \left. \left. 2 \left(q_3^2-2
   q_2^2\right) q_{12}^4+q_2^4 q_{12}^2+q_{13}^4 q_{12}^2-2
   \left(2 q_{12}^4+q_2^2 \left(q_3^2+2 q_{12}^2\right)\right)
   q_{13}^2\right) q_{23}^2-\left(-3 q_2^4+\left(4 q_3^2+6
   q_{12}^2\right) q_2^2 \right. \right. \nn \\
&& + \left. \left. q_{12}^4\right)
   \left(q_{12}^2-q_{13}^2\right){}^2\right)
   q_1^2+\left(q_2^2-q_{12}^2\right){}^2
   \left(q_{23}^6+\left(q_2^2+2 q_3^2-2 q_{12}^2-3
   q_{13}^2\right) q_{23}^4-\left(q_{12}^4-3 q_{13}^4 \right. \right.  \nn \\
&& + \left. \left. 2
   \left(2 q_3^2+3 q_{12}^2\right) q_{13}^2\right)
   q_{23}^2-\left(q_{12}^2-q_{13}^2\right){}^2
   \left(q_2^2+q_{13}^2-2
   \left(q_3^2+q_{12}^2\right)\right)\right) \bigg\} \,, \nn
\eea
\bea
&& C^{MS}_{10}(\vec{q_1},\vec{q_2},\vec{q_3}) = \frac{1}{16384 \pi^2  \lambda^2(q_1,q_{13},q_3) \lambda^2(q_{12},q_{23},q_{13}) }
\bigg\{ \left(3 q_{12}^6+\left(3 q_{23}^2-5
   q_{13}^2\right) q_{12}^4+\left(q_{13}^4-18 q_{23}^2 q_{13}^2\right. \right. \nn \\
&& \left. \left. -7 q_{23}^4\right) q_{12}^2  +  \left(q_{13}^2-q_{23}^2\right)^2
   \left(q_{13}^2+q_{23}^2\right)\right) q_1^6+\left(-\left(7
   q_3^2+q_{13}^2\right) q_{23}^6+\left(3 q_{13}^4+\left(2 q_2^2+5
   \left(q_3^2+q_{12}^2\right)\right) q_{13}^2  \right.  \right. \nn \\
&&  \left. \left. +41 q_3^2 q_{12}^2\right) q_{23}^4 + \left(-3 q_{13}^6+\left(-4 q_2^2+11
   q_3^2+34 q_{12}^2\right) q_{13}^4+q_{12}^2 \left(12 q_2^2+34
   q_3^2-15 q_{12}^2\right) q_{13}^2-37 q_3^2 q_{12}^4\right)
   q_{23}^2 \right. \nn \\
&& + \left. \left(q_{12}^2-q_{13}^2\right)^2
   \left(q_{13}^4+\left(2 q_2^2-9 q_3^2-5 q_{12}^2\right)
   q_{13}^2+3 q_3^2 q_{12}^2\right)\right) q_1^4+\left(\left(-7
   q_3^4-18 q_{13}^2 q_3^2+q_{13}^4\right) q_{12}^6+\left(-7
   q_{13}^6 \right. \right. \nn \\
&& + \left. \left. \left(-4 q_2^2+34 q_3^2+21 q_{23}^2\right)
   q_{13}^4+q_3^2 \left(12 q_2^2+5 q_3^2+34 q_{23}^2\right)
   q_{13}^2+41 q_3^4 q_{23}^2\right) q_{12}^4+\left(11 q_{13}^8+2
   \left(4 q_2^2-7 \left(q_3^2 \right. \right. \right. \right. \nn \\
&& + \left. \left. \left. \left. q_{23}^2\right)\right)
   q_{13}^6+\left(11 q_3^4+92 q_{23}^2 q_3^2+11 q_{23}^4-24 q_2^2
   \left(q_3^2+q_{23}^2\right)\right) q_{13}^4+2 q_3^2 q_{23}^2
   \left(17 \left(q_3^2 + q_{23}^2\right)-28 q_2^2\right) q_{13}^2 \right. \right. \nn \\
&& - \left. \left. 37
   q_3^4 q_{23}^4\right)
   q_{12}^2-\left(q_{13}^2-q_{23}^2\right)^2 \left(5
   q_{13}^6+\left(4 q_2^2+2 q_3^2+q_{23}^2\right) q_{13}^4+3 q_3^2
   \left(-4 q_2^2+3 q_3^2+6 q_{23}^2\right) q_{13}^2 \right. \right. \nn \\
&& - \left. \left. 3 q_3^4
   q_{23}^2\right)\right) q_1^2+\left(q_3^2-q_{13}^2\right){}^2
   \left(\left(q_3^2+q_{13}^2\right)
   q_{12}^6+\left(q_{13}^4+\left(2 q_2^2-q_3^2-9 q_{23}^2\right)
   q_{13}^2-7 q_3^2 q_{23}^2\right) q_{12}^4-\left(5
   q_{13}^6 \right. \right. \nn \\
&& + \left. \left. \left(4 q_2^2+q_3^2+2 q_{23}^2\right) q_{13}^4+3
   q_{23}^2 \left(-4 q_2^2+6 q_3^2+3 q_{23}^2\right) q_{13}^2-3
   q_3^2 q_{23}^4\right)
   q_{12}^2+\left(q_{13}^2-q_{23}^2\right){}^2 \left(3
   q_{13}^4 \right. \right. \nn \\
&& + \left. \left. \left(2 q_2^2+q_3^2+q_{23}^2\right) q_{13}^2+3 q_3^2
   q_{23}^2\right)\right) \bigg\} \,, \nn
\eea
\bea
&& C^{MS}_{11}(\vec{q_1},\vec{q_2},\vec{q_3}) = \frac{q_{23}^2}{4096 \pi^3 \lambda^2(q_2,q_{23},q_3) \lambda^2(q_{12},q_{23},q_{13}) }
\bigg\{ \left(-q_3^4-4 q_{23}^2
   q_3^2+\left(q_2^2-q_{23}^2\right)^2\right)
   q_{12}^6+\left(-q_3^6   \right. \nn \\
&&  \left.  + \left(2 q_{23}^2   +3
   \left(q_2^2+q_{13}^2\right)\right) q_3^4 - \left(3 q_2^4 - 9
   q_{23}^4 + 2 \left(q_2^2-4 q_{13}^2\right) q_{23}^2\right)
   q_3^2+\left(q_2^2-3 q_{13}^2-2 q_{23}^2\right)
   \left(q_2^2-q_{23}^2\right)^2\right) q_{12}^4  \nn \\
&& + \left(-4
   q_{23}^2 q_3^6+\left(-3 q_{13}^4+9 q_{23}^4-2 \left(q_{13}^2-4
   q_2^2\right) q_{23}^2\right) q_3^4-4 q_{23}^2 \left(q_2^4-4
   q_{23}^2 q_2^2+q_{13}^4+2 q_{23}^4 \right. \right. \nn \\
&& - \left. \left. 4 q_{13}^2
   \left(q_2^2+q_{23}^2\right)\right)
   q_3^2-\left(q_2^2-q_{23}^2\right){}^2 \left(-3 q_{13}^4+6
   q_{23}^2 q_{13}^2+q_{23}^4\right)\right)
   q_{12}^2-\left(q_{13}^2-q_{23}^2\right)^2 \left(-q_3^6+\left(3
   q_2^2 \right. \right. \nn \\
&& - \left. \left. q_{13}^2+2 q_{23}^2\right) q_3^4+\left(-3 q_2^4+6 q_{23}^2
   q_2^2+q_{23}^4\right) q_3^2+\left(q_2^2+q_{13}^2-2
   q_{23}^2\right) \left(q_2^2-q_{23}^2\right){}^2\right)+2 q_1^2
   \left(\left(q_2^4 \right. \right. \nn \\
&& - \left. \left. 2 \left(q_3^2+q_{23}^2\right)
   q_2^2+q_3^4+q_{23}^4\right) q_{12}^4-2 \left(2 q_3^2
   q_{23}^4+q_{13}^2 \left(q_2^4-2 \left(q_3^2+q_{23}^2\right)
   q_2^2+q_3^4+q_{23}^4\right)\right)
   q_{12}^2 \right. \nn \\
&& + \left. \left(q_{13}^2-q_{23}^2\right)^2 \left(q_2^4-2
   \left(q_3^2+q_{23}^2\right)
   q_2^2+q_3^4+q_{23}^4\right)\right) \bigg\} \,.
\eea
\normalsize
%
%

\section{Appendix. Multi-graviton interaction vertices}
\label{P4Vertices}

%
The computation of the vertices  can be done by taking at most three functional derivatives of the action
with respect to the metric, since the vacuum expectation values of the fourth order derivatives correspond to massless
tadpoles, which are set to zero, as explained in the previous sections. 
We keep the notation with square brackets to indicate the flat limit of the functional derivatives in momentum space,
showing explicitly the dependence on the momenta when this occurs (which is not always the case, as for the metric tensors). We have 
\beqa
\left[\sqrt{g}\right]^{\mud\nud}
&=& 
\left[V\right]^{\mud\nud}
=
\frac{1}{2}\, \delta^{\mud\nud} \, ,
\nn \\
\left[\sqrt{g}\right]^{\mud\nud\mut\nut}
&=&
\left[V\right]^{\mud\nud\mut\nut} 
=
\frac{1}{2}\, \left(\frac{1}{2}\, \delta^{\mud\nud}\, \delta^{\mut\nut} + \left[g^{\mud\nud}\right]^{\mut\nut}\right) \, ,
\nn \\
\left[\sqrt{g}\right]^{\mud\nud\mut\nut\muq\nuq}
&=& 
\left[V\right]^{\mud\nud\mut\nut\muq\nuq}
=
\left[\sqrt{g}\right]^{\muq\nuq}\, \left[\sqrt{g}\right]^{\mud\nud\mut\nut} \nn \\
&+& \frac{1}{2}\, \bigg( \frac{1}{2}\, \left[g^{\mud\nud}\, g^{\mut\nut}\right]^{\muq\nuq}
+ \left[g^{\mud\nud}\right]^{\mut\nut\muq\nuq} \bigg) \, ,
\nn \\
\left[ g_{\alpha\beta} \right]^{\mud\nud}
&=& 
\frac{1}{2}\left(\delta_\alpha^{\mud} \delta_\beta^{\nud} + \delta_\alpha^{\nud} \delta_\beta^{\mud} \right)
\label{P4Symm} \, , 
\nn \\
\left[ g^{\alpha\beta} \right]^{\mud\nud}
&=& 
- \frac{1}{2}\left(\delta^{\alpha\mud}\delta^{\beta\nud} + \delta^{\alpha\nud}\delta^{\beta\mud}\right)  \, ,
\nn \\
\left[ g^{\alpha\beta} \right]^{\mud\nud\mut\nut}
&=& 
- \frac{1}{2}\, \bigg[ \left[g^{\alpha\mud}\right]^{\mut\nut} \delta^{\beta\nud} 
                     + \delta^{\alpha\mud} \left[g^{\beta\nud}\right]^{\mut\nut}                     
                     + \left[g^{\alpha\nud}\right]^{\mut\nut} \delta^{\beta\mud} 
                     + \left[g^{\beta\mud}\right]^{\mut\nut} \delta^{\alpha\nud} \bigg],
\nn \\
\left[V_{a \lambda}\right]^{\mud\nud} 
&=& - \left[{V_a}^\lambda\right]^{\mud\nud} 
=
\frac{1}{4}\, \left({V_{a}}^{\mud}\,{\delta_\lambda}^{\nud} + {V_{a}}^{\nud}\,{\delta_\lambda}^{\mud} \right)\, , 
\nn \\
\left[V_{a \lambda}\right]^{\mud\nud\mut\nut}
&=& 
- \left[{V_a}^\lambda\right]^{\mud\nud\mut\nut} 
=
\frac{1}{4}\, \left(\left[{V_{a}}^{\mud}\right]^{\mut\nut}\,{\delta_\lambda}^{\nud} 
+ \left[{V_{a}}^{\nud}\right]^{\mut\nut}\,{\delta_\lambda}^{\mud} \right)\, , 
\nn \\
\left[V_{a \lambda}\right]^{\mud\nud\mut\nut\muq\nuq} 
&=&
-\left[{V_a}^\lambda\right]^{\mud\nud\mut\nut\mu\nuq} 
=
\frac{1}{4}\, \left(\left[{V_{a}}^{\mud}\right]^{\mut\nut\muq\nuq}\,{\delta_\lambda}^{\nud} 
+ \left[{V_{a}}^{\nud}\right]^{\mut\nut\muq\nuq}\,{\delta_\lambda}^{\mud} \right)\, .
\eeqa

For the Christoffel symbols, defined as
\beqa\label{P4Christoffel}
\Gamma^{\alpha}_{\beta\gamma}(z) 
&=& 
\frac{1}{2}\, g^{\alpha\kappa}(z)\, 
\left[-\pd_\kappa g_{\beta\gamma}(z) + \pd_\beta g_{\kappa\gamma}(z) + \pd_\gamma  g_{\kappa\beta}(z)\right]\, , 
\nn \\
\Gamma_{\alpha,\beta\gamma}(z) 
&=& 
\frac{1}{2}\, 
\left[-\pd_\alpha g_{\beta\gamma}(z) + \pd_\beta g_{\alpha\gamma}(z) + \pd_\gamma  g_{\alpha\beta}(z)\right]\, , 
\eeqa
we obtain
\beqa
\left[\Gamma^\alpha_{\beta\gamma}\right]^{\mud\nud}(\kd)
&=&
\frac{i}{2}\, \delta^{\alpha\lambda}\, \left(- \left[g_{\beta\gamma}\right]^{\mud\nud} k_{2\,\lambda} + 
\left[g_{\beta\lambda}\right]^{\mud\nud} k_{2\,\gamma} + \left[g_{\lambda\gamma}\right]^{\mud\nud} k_{2\,\beta} \right) \, ,
\nn \\
\left[\Gamma^\alpha_{\beta\gamma}\right]^{\mud\nud\mut\nut}(\kd,\kt)
&=&
\left[g^{\alpha\lambda}\right]^{\mud\nud}\, \left[\Gamma_{\lambda,\beta\gamma}\right]^{\mut\nut}(\kt)
\left[g^{\alpha\lambda}\right]^{\mut\nut}\, \left[\Gamma_{\lambda,\beta\gamma}\right]^{\mud\nud}(\kd) \, ,
\nn \\
\left[\Gamma^\alpha_{\beta\gamma}\right]^{\mud\nud\mut\nut\muq\nuq}(\kd,\kt,\kq)
&=&
\left[g^{\alpha\lambda}\right]^{\mud\nud}\, \left[\Gamma_{\lambda,\beta\gamma}\right]^{\mut\nut\muq\nuq}(\kt,\kq) \nn \\ 
&& \hspace{-3cm} +
\left[g^{\alpha\lambda}\right]^{\mut\nut}\, \left[\Gamma_{\lambda,\beta\gamma}\right]^{\mud\nud\muq\nuq}(\kd,\kq) 
+
\left[g^{\alpha\lambda}\right]^{\muq\nuq}\, \left[\Gamma_{\lambda,\beta\gamma}\right]^{\mud\nud\mut\nut}(\kd,\kt)\, .
\eeqa
It is straightforward, starting from these definitions, to derive the derivatives of the Ricci tensor
$\left[R_{\mu\nu}\right]^{\rho\sigma}(\vec{l})$ and $\left[R_{\mu\nu}\right]^{\rho\sigma\chi\omega}(\vec{l_1},\vec{l_2})$,
 defined as $R_{\mu\nu}(z) = {R^\lambda}_{\mu\lambda\nu}(z)$. We recall that in our conventions 
the Riemann tensor is defined as 
\beq
{R^\lambda}_{\mu\kappa\nu}(z) =
\pd_\nu \Gamma^\lambda_{\mu\kappa}(z) - \pd_\kappa \Gamma^\lambda_{\mu\nu}(z)
+ \Gamma^\lambda_{\nu\eta}(z)\Gamma^\eta_{\mu\kappa}(z) - \Gamma^\lambda_{\kappa\eta}(z)\Gamma^\eta_{\mu\nu}(z)\, .
\eeq

Next we list the interaction vertices for the dual theories. 
\begin{itemize}

\item \textbf{Scalar}
\small
\beqa
V_{S\phi\phi}^{\mu\nu}(\vec{p},\vec{q})
&=&
\frac{1}{2}\left(\delta^{\mu\alpha}\delta^{\nu\beta} - \frac{1}{2}\delta^{\mu\nu}\delta^{\alpha\beta}\right)\, 
\left(p_{\alpha} q_{\beta} + p_{\beta} q_{\alpha}\right)\, ,
\nn \\
&+& 
\chi\, \left(\delta^{\mu\nu}\delta^{\alpha\beta} - \delta^{\mu\alpha}\delta^{\nu\beta}\right)\, 
\left(p_{\alpha}p_{\beta} + p_{\alpha}q_{\beta} + q_{\alpha}p_{\beta}+ q_{\alpha}q_{\beta} \right) 
\nn \\
V_{SS\phi\phi}^{\mu\nu\rho\sigma}(\vec{p},\vec{q},\vec{l})
&=&
\frac{1}{2}\, \left( \left[\sqrt{g}\right]^{\rho\sigma}\,
\left(\delta^{\mu\alpha}\delta^{\nu\beta} - \frac{1}{2}\delta^{\mu\nu}\delta^{\alpha\beta}\right)
+ \left[g^{\mu\alpha}g^{\nu\beta} - \frac{1}{2}\, g^{\mu\nu}g^{\alpha\beta} \right]^{\rho\sigma}\,\right)
\left(p_{\alpha} q_{\beta} + p_{\beta} q_{\alpha}\right)\, ,
\nn \\
&& \hspace{-2cm}
+ \chi\,\left\{ \left( \left[\sqrt{g}\right]^{\rho\sigma}\,
\left(\delta^{\mu\nu}\delta^{\alpha\beta} - \delta^{\mu\alpha}\delta^{\nu\beta}\right)
+ \left[g^{\mu\nu}g^{\alpha\beta} - g^{\mu\alpha}g^{\nu\beta} \right]^{\rho\sigma}\,\right)
\left(p_{\alpha}p_{\beta} + p_{\alpha}q_{\beta} + p_{\beta}q_{\alpha} + q_{\alpha}q_{\beta}\right)
\right.
\nn \\
&& \hspace{-2cm}
+ \left.
\left(\delta^{\mu\nu}\delta^{\alpha\beta} - \delta^{\mu\alpha}\delta^{\nu\beta}\right)\,
\left[\Gamma^\lambda_{\alpha\beta}\right]^{\rho\sigma}(\vec{l})\, i\, (p_\lambda + q_\lambda)
-  \left(\frac{1}{2}\, \delta^{\mu\nu}\delta^{\alpha\beta} - \delta^{\mu\alpha}\delta^{\nu\beta}\right)\, 
\left[R_{\alpha\beta}\right]^{\rho\sigma}(\vec{l}) \right\} \, ,  
\nn
\eea
\bea
V_{SSS\phi\phi}^{\mu\nu\rho\sigma\chi\omega}(\vec{p},\vec{q},\vec{l_1},\vec{l_2})
&=&
\frac{1}{2}\bigg\{\left[\sqrt{g}\right]^{\rho\sigma\chi\omega}\, 
\left(\delta^{\mu\alpha}\delta^{\nu\beta} - \frac{1}{2}\, \delta^{\mu\nu}\delta^{\alpha\beta}\right)
\nn \\
&& \hspace{-2cm}
+ \left[\sqrt{g}\right]^{\rho\sigma}\, \left[g^{\mu\alpha}g^{\nu\beta} - \frac{1}{2}\, g^{\mu\nu} g^{\alpha\beta}\right]^{\chi\omega}	+
\left[\sqrt{g}\right]^{\chi\omega}\, \left[g^{\mu\alpha}g^{\nu\beta} - \frac{1}{2}\, g^{\mu\nu} g^{\alpha\beta}\right]^{\rho\sigma} 
\nn \\
&& \hspace{-2cm}
+ \left[g^{\mu\alpha}g^{\nu\beta} - \frac{1}{2}\, g^{\mu\nu} g^{\alpha\beta}\right]^{\rho\sigma\chi\omega} \bigg\}\, 
\left(p_{\alpha} q_{\beta} + p_{\beta} q_{\alpha}\right)\, .
\nn \\
&& \hspace{-2cm}
+ \chi\, \bigg\{\left[\sqrt{g}\right]^{\rho\sigma\chi\omega}\, 
\left(\delta^{\mu\nu}\delta^{\alpha\beta} - \delta^{\mu\alpha}\delta^{\nu\beta}\right)
\nn \\
&& \hspace{-2cm}
+ \left[\sqrt{g}\right]^{\rho\sigma}\, \left[g^{\mu\nu}g^{\alpha\beta} - g^{\mu\alpha}g^{\nu\beta}\right]^{\chi\omega}	+
\left[\sqrt{g}\right]^{\chi\omega}\, \left[g^{\mu\nu}g^{\alpha\beta} - g^{\mu\alpha}g^{\nu\beta}\right]^{\rho\sigma} 
\nn \\
&& \hspace{-2cm}
+ \left[g^{\mu\nu}g^{\alpha\beta} - g^{\mu\alpha}g^{\nu\beta}\right]^{\rho\sigma\chi\omega}
\bigg\}\, 
\left(p_\alpha p_\beta + p_\alpha q_\beta + q_\alpha p_\beta + q_\alpha q_\beta\right)
\nn \\
&& \hspace{-2cm}
+ \chi\, \bigg\{
\left(\left[\sqrt{g}\right]^{\chi\omega}\, 
\left[\delta^{\mu\nu}\delta^{\alpha\beta} - \delta^{\mu\alpha}\delta^{\nu\beta}\right] 
+ \left[g^{\mu\nu}g^{\alpha\beta} - g^{\mu\alpha}g^{\nu\beta}\right]^{\chi\omega} \right)\, 
\left[\Gamma^\lambda_{\alpha\beta}\right]^{\rho\sigma}(\vec{l_1}) 
\nn \\
&& \hspace{-2cm}
+ \left(\rho,\sigma,l_1\right) \leftrightarrow \left(\tau,\omega,l_2\right) 
+ \left(\delta^{\mu\nu}\delta^{\alpha\beta} - \delta^{\mu\alpha}\delta^{\nu\beta}\right)\, 
\left[\Gamma^\lambda_{\alpha\beta}\right]^{\rho\sigma\chi\omega}(\vec{l_1},\vec{l_2})
\bigg\}\, i\, \left(p_\lambda + q_\lambda\right)
\nn \\
&& \hspace{-2cm}
+ \chi \bigg\{ 
\left(\left[\sqrt{g}\right]^{\chi\omega}\, \left( \delta^{\mu\alpha}\delta^{\nu\beta} -
\frac{1}{2}\, \delta^{\mu\nu}\delta^{\alpha\beta}\right)+
\left[g^{\mu\alpha}g^{\nu\beta} - \frac{1}{2}\,g^{\mu\nu}g^{\alpha\beta}\right]^{\chi\omega} \right)
\, \left[R_{\alpha\beta}\right]^{\rho\sigma}(\vec{l_1})
\nn \\
&& \hspace{-2cm} 
+ \left(\rho,\sigma,l_1\right) \leftrightarrow \left(\tau,\omega,l_2\right) 
+\left(\delta^{\mu\alpha}\delta^{\nu\beta} - \frac{1}{2}\,\delta^{\mu\nu}\delta^{\alpha\beta}\right) 
\left[R_{\alpha\beta}\right]^{\rho\sigma\chi\omega}(\vec{l_1},\vec{l_2}) \bigg\}
\eeqa
\\ \\
\item \textbf{Fermion} 

\beqa
V^{\mu\nu}_{S\bar\psi\psi}(\vec{p},\vec{q}) 
&=&
\frac{1}{2}\, \left( \left[V\right]^{\mu\nu}\, {\delta_a}^{\lambda} + \left[{V_a}^\lambda\right]^{\mu\nu} \right)\, 
\gamma^a\, (p_\lambda - q_\lambda)\, ,
\nn \\
V^{\mu\nu\rho\sigma}_{SS\bar\psi\psi}(\vec{p},\vec{q},\vec{l_1})
&=&
\frac{1}{2}\, \bigg( \left[V\right]^{\mu\nu\rho\sigma}\, {\delta_a}^{\lambda}
+ \left[V\right]^{\mu\nu}\, \left[{V_a}^{\lambda}\right]^{\rho\sigma} 
+ \left[V\right]^{\rho\sigma}\, \left[{V_a}^{\lambda}\right]^{\mu\nu}
+ \left[{V_a}^{\lambda}\right]^{\mu\nu\rho\sigma}\bigg)\, \gamma^a\, (p_\lambda - q_\lambda)
\nn \\
&+&
\frac{1}{16}\, \bigg( \left[V\right]^{\mu\nu}\, {\delta_a}^\lambda + \left[{V_a}^\lambda\right]^{\mu\nu}\bigg)\,
\left\{\gamma^a,\left[\gamma^b,\gamma^c\right]\right\}\, \left[\Omega_{bc,\lambda}\right]^{\rho\sigma}(\vec{l_1})\, ,
\nn
\eeqa
\beqa
V^{\mu\nu\rho\sigma\chi\omega}_{SSS\bar\psi\psi}(\vec{p},\vec{q},\vec{l_1},\vec{l_2})
&=&
\frac{1}{2}\, \bigg( 
  \left[V\right]^{\mu\nu\rho\sigma\chi\omega}\, {\delta_a}^{\lambda}
+ \left[V\right]^{\mu\nu\rho\sigma}\, \left[{V_a}^{\lambda}\right]^{\chi\omega}
+ \left[V\right]^{\mu\nu\chi\omega}\, \left[{V_a}^{\lambda}\right]^{\rho\sigma} \nn \\
&+&
  \left[V\right]^{\rho\sigma\chi\omega}\, \left[{V_a}^{\lambda}\right]^{\mu\nu}
+ \left[V\right]^{\mu\nu}\, \left[{V_a}^{\lambda}\right]^{\rho\sigma\chi\omega}  
+ \left[V\right]^{\rho\sigma}\, \left[{V_a}^{\lambda}\right]^{\mu\nu\chi\omega} \nn \\
&+&
  \left[V\right]^{\chi\omega}\, \left[{V_a}^{\lambda}\right]^{\mu\nu\rho\sigma}
+ \left[{V_a}^{\lambda}\right]^{\mu\nu\rho\sigma\chi\omega}\bigg)\, \gamma^a\, (p_\lambda - q_\lambda) 
\nn \\
&+&
\frac{1}{16}\,\{\gamma^a,[\gamma^b,\gamma^c]\}\,
\bigg\{ \bigg(
\left[V\right]^{\mu\nu\rho\sigma}\, {\delta_a}^{\lambda}
+ \left[V\right]^{\mu\nu}\, \left[{V_a}^{\lambda}\right]^{\rho\sigma}
\nn \\
&+&
\left[V\right]^{\rho\sigma}\, \left[{V_a}^{\lambda}\right]^{\mu\nu}
+ \left[{V_a}^{\lambda}\right]^{\mu\nu\rho\sigma} \bigg)
\left[\Omega_{bc,\lambda}\right]^{\chi\omega}(\vec{l_2})
\nn \\
&+&
\bigg( \left[V\right]^{\mu\nu}\, {\delta_a}^\lambda + \left[{V_a}^\lambda\right]^{\mu\nu}\bigg)\,
\left[\Omega_{bc,\lambda}\right]^{\rho\sigma\chi\omega}(\vec{l_1},\vec{l_2})\bigg\}\, ,
\eeqa
where the spin connection was defined in (\ref{P4SpinCon}).
%
\item \textbf{Gauge field} 

%
\beqa
V^{\mu\nu\,\tau\vartheta}_{SAA}(\vec{p},\vec{q})
&=&
\frac{1}{2}\, \left(\delta^{\mu\lambda}\delta^{\alpha\kappa}\delta^{\nu\beta} 
+ \frac{1}{4}\delta^{\mu\nu}\delta^{\alpha\lambda}\delta^{\beta\kappa}\right)\,
\left[F_{\alpha\beta}F_{\lambda\kappa}\right]^{\tau\vartheta}(\vec{p},\vec{q})
\nn \\
V^{\mu\nu\rho\sigma\,\tau\vartheta}_{SSAA}(\vec{p},\vec{q})
&=&
\frac{1}{2}\, \left\{ \left[\sqrt{g}\right]^{\rho\sigma}\, 
\left(\delta^{\mu\lambda}\delta^{\alpha\kappa}\delta^{\nu\beta} 
+ \frac{1}{4}\delta^{\mu\nu}\delta^{\alpha\lambda}\delta^{\beta\kappa}\right)
\right.
\nn \\
&+&
\left.
\left[ g^{\mu\rho}g^{\alpha\sigma}g^{\nu\beta} + \frac{1}{4}\, g^{\mu\nu}g^{\alpha\beta}g^{\rho\sigma}\right]^{\rho\sigma}
\right\}\, \left[F_{\alpha\beta}F_{\lambda\kappa}\right]^{\tau\vartheta}(\vec{p},\vec{q})
\nn\\
V^{\mu\nu\rho\sigma\chi\omega\, \tau\vartheta}_{SSSAA}(\vec{p},\vec{q})
&=&
\frac{1}{2}\, \left\{
\left[\sqrt{g}\right]^{\rho\sigma\chi\omega}\,
\left(\delta^{\mu\lambda}\delta^{\alpha\kappa}\delta^{\nu\beta} 
+ \frac{1}{4}\delta^{\mu\nu}\delta^{\alpha\lambda}\delta^{\beta\kappa}\right)
\right.
\nn \\
&& \hspace{-2cm}
+ \left.
\left[\sqrt{g}\right]^{\rho\sigma}\, 
\left[g^{\mu\lambda}g^{\alpha\kappa}g^{\nu\beta} + \frac{1}{4}g^{\mu\nu}g^{\alpha\lambda}g^{\beta\kappa}\right]^{\chi\omega} +
\left[\sqrt{g}\right]^{\chi\omega}\, 
\left[g^{\mu\lambda}g^{\alpha\kappa}g^{\nu\beta} + \frac{1}{4}g^{\mu\nu}g^{\alpha\lambda}g^{\beta\kappa}\right]^{\rho\sigma}
\right.
\nn \\
&& \hspace{-2cm}
+ \left.
\left[ g^{\mu\lambda}g^{\alpha\kappa}g^{\nu\beta} 
+ \frac{1}{4}g^{\mu\nu}g^{\alpha\lambda}g^{\beta\kappa}\right]^{\rho\sigma\chi\omega}
\right\}\, \left[F_{\alpha\beta}F_{\lambda\kappa}\right]^{\tau\vartheta}(\vec{p},\vec{q})
\eeqa
where we have introduced
\beqa
\left[F_{\alpha\beta}F_{\lambda\kappa}\right]^{\tau\vartheta}(\vec{p},\vec{q})
&=& 
\int\, d^4x d^4y\, e^{- i\, p\cdot x -i\, q\cdot y}\, 
\frac{\delta^2 \left(F_{\alpha\beta}(0)F_{\lambda\kappa}(0)\right)}{\delta A_{\tau}(x)\delta A_{\vartheta}(y)}
\nn \\
&=&
-  \left(
\delta^{\tau}_{\lambda}\, \delta^{\vartheta}_{\alpha}\, p_{\kappa}\, q_{\beta}
- \delta^{\tau}_{\lambda}\, \delta^{\vartheta}_{\beta}\, p_{\kappa}\, q_{\alpha}
- \delta^{\tau}_{\kappa}\, \delta^{\vartheta}_{\alpha} p_{\lambda}\, q_{\beta}
+ \delta^{\tau}_{\kappa}\, \delta^{\vartheta}_{\beta}\, p_{\lambda}\, q_{\alpha}
+  \left(\tau,\vec{p}\right) \leftrightarrow  \left(\vartheta,\vec{q}\right)
\right) \,. \nn \\
\eeqa
\normalsize

\end{itemize}

\backmatter

\end{document}